\documentclass{dune} 
\pdfoutput=1            
\graphicspath{ {figures/}{executive-summary/figures/}{generated/} }

\def\expshort{DUNE\xspace}
\def\explong{The Deep Underground Neutrino Experiment\xspace}

\def\thedocsubtitle{Deep Underground Neutrino Experiment (DUNE)} 

\def\voltitlespfd{Volume 2: Single-Phase Module\xspace}

\def\voldpfd{\textbf{Dual-Phase Module\xspace} }

\newcommand{\nue}{$\nu_e$\xspace}

\newcommand{\ptoknubar}{$p \rightarrow K^+ \overline{\nu}$\xspace}

\newcommand{\kamland}{KamLAND\xspace}

\newcommand{\microboone}{MicroBooNE\xspace}
\newcommand{\minerva}{MINER$\nu$A\xspace}
\newcommand{\nova}{NO$\nu$A\xspace}

\newcommand{\lariat}{LArIAT\xspace}

\newcommand{\lartpc}{LArTPC\xspace}

\newcommand{\larsoft}{LArSoft\xspace}

\def\argon40{$^{40}$Ar}  
\def\Ar39{$^{39}$Ar}
\def\Cl40{$^{40}$Cl}
\def\K40{$^{40}$K}
\def\B8{$^{8}$B}

\def\fdfiducialmass{\SI{40}{\kt}\xspace}

\def\larmass{\SI{17.5}{\kt}\xspace} 
\def\tpcheight{\SI{12}{\meter}\xspace} 
\def\cryostatht{\SI{14.1}{\meter}\xspace} 
\def\cryostatlen{\SI{62.0}{\meter}\xspace} 
\def\cryostatwdth{\SI{14.0}{\meter}\xspace} 
\def\nominalmodsize{\SI{10}{kt}\xspace} 

\def\dunelifetime{\SI{20}{year}\xspace} 

\def\spmaxdrift{\SI{3.53}{\m}\xspace}
\def\spnumch{\num{384000}\xspace} 

\def\dpactivelarmass{\SI{12.096}{\kt}\xspace} 
\def\dpfidlarmass{\SI{10.643}{\kt}\xspace} 
\def\dpmaxdrift{\SI{12}{\m}\xspace} 
\def\dptpclen{\SI{60}{\meter}\xspace} 
\def\dptpcwdth{\SI{12}{\meter}\xspace} 
\def\dpswchpercrp{\num{36}\xspace} 
\def\dpnumswch{\num{2880}\xspace} 
\def\dptotcrp{\num{80}\xspace} 
\def\dpchpercrp{\num{1920}\xspace} 
\def\dpnumcrpch{\num{153600}\xspace} 
\def\dpchperchimney{\num{6400}\xspace} 
\def\dpnumpmtch{\num{720}\xspace} 
\def\dpstrippitch{\SI{3.125}{\milli\meter}\xspace} 
\def\dptargetdriftvoltpos{\SI{600}{kV}\xspace} 
\def\dptargetdriftvoltneg{\SI{-600}{kV}\xspace} 

\def\spreadout{\SI{5.4}{\ms}\xspace}
\def\dpreadout{\SI{16.4}{\ms}\xspace}
\def\snbtime{\SI{30}{\s}\xspace}
\def\snbpretime{\SI{10}{\s}\xspace}
\def\spsnbsize{\SI{45}{\PB}\xspace}

\def\offsitepbpy{\SI{30}{\PB/\year}\xspace}
\def\offsitegbyteps{\SI{1}{\GB/\s}\xspace}
\def\offsitegbps{\SI{8}{\Gbps}\xspace}

\def\surffnalbw{\SI{100}{\Gbps}\xspace}
\newcommand{\fnal}{Fermilab\xspace}
\newcommand{\surf}{SURF\xspace}

\newcommand{\efield}{E field\xspace}

\newcommand{\rms}{RMS\xspace} 
\newcommand{\threed}{3D\xspace}
\newcommand{\twod}{2D\xspace}

\newcommand{\detmodule}{detector module\xspace}
\newcommand{\dual}{DP\xspace}

\newcommand{\single}{SP\xspace}

\newcommand{\dpmod}{DP detector module\xspace}

\newcommand{\lar}{LAr\xspace}

\newcommand{\fdth}{feedthrough\xspace}

\newcommand{\pmt}{PMT\xspace}
\newcommand{\phel}{photoelectron\xspace}

\newcommand{\pwrsupps}{power supplies\xspace}

\DeclareSIUnit \s {\second}
\DeclareSIUnit \MB {\mega\byte}
\DeclareSIUnit \GB {\giga\byte}
\DeclareSIUnit \TB {\tera\byte}
\DeclareSIUnit \PB {\peta\byte}
\DeclareSIUnit \Mbps {\mega\bit/\s}
\DeclareSIUnit \Gbps {\giga\bit/\s}
\DeclareSIUnit \Tbps {\tera\bit/\s}
\DeclareSIUnit \Pbps {\peta\bit/\s}
\DeclareSIUnit \kton {\kilo\tonne} 
\DeclareSIUnit \kt {\kilo\tonne}
\DeclareSIUnit \Mt {\mega\tonne}
\DeclareSIUnit \eV {\electronvolt}
\DeclareSIUnit \keV {\kilo\electronvolt}
\DeclareSIUnit \MeV {\mega\electronvolt}
\DeclareSIUnit \GeV {\giga\electronvolt}
\DeclareSIUnit \m {\meter}
\DeclareSIUnit \cm {\centi\meter}
\DeclareSIUnit \in {\inchcommand}
\DeclareSIUnit \km {\kilo\meter}
\DeclareSIUnit \kV {\kilo\volt}
\DeclareSIUnit \kW {\kilo\watt}
\DeclareSIUnit \MW {\mega\watt}
\DeclareSIUnit \MHz {\mega\hertz}
\DeclareSIUnit \mrad {\milli\radian}
\DeclareSIUnit \year {year}
\DeclareSIUnit \POT {POT}
\DeclareSIUnit \sig {$\sigma$}
\DeclareSIUnit\parsec{pc}
\DeclareSIUnit\lightyear{ly}
\DeclareSIUnit\foot{ft}
\DeclareSIUnit\ft{ft}
\DeclareSIUnit \ppb{ppb}
\DeclareSIUnit \ppt{ppt}
\DeclareSIUnit \samples{S}

\usepackage[toc]{glossaries}
\makeglossaries

\newcommand{\dshort}[1]{\glsentrytext{#1}}

\newcommand{\dlong}[1]{\glsentrylong{#1}}

\newcommand{\dfirst}[1]{\glsfirst{#1}}
\newcommand{\dfirsts}[1]{\glsfirstplural{#1}}

\newcommand{\dword}[1]{\gls{#1}}
\newcommand{\dwords}[1]{\glspl{#1}}
\newcommand{\Dword}[1]{\Gls{#1}}
\newcommand{\Dwords}[1]{\Glspl{#1}}

\newcommand{\newduneword}[3]{
    \newglossaryentry{#1}{
        text={#2},
        long={#2},
        name={\glsentrylong{#1}},
        first={\glsentryname{#1}},
        firstplural={\glsentrylong{#1}\glspluralsuffix},
        description={#3}
    }
}

\newcommand{\newduneabbrev}[4]{
  \newglossaryentry{#1}{
    text={#2},
    long={#3},
    shortplural={{#2}\glspluralsuffix},
    longplural={{#3}\glspluralsuffix{}},
    name={\glsentrylong{#1}{} (\glsentrytext{#1}{})},
    first={\glsentryname{#1}},
    firstplural={\glsentrylong{#1}\glspluralsuffix{} (\glsentrytext{#1}\glspluralsuffix{})},
    description={#4}
  }
}

\newcommand{\newduneabbrevs}[5]{
  \newglossaryentry{#1}{
    text={#2},
    long={#3},
    plural={#4},
    shortplural={{#2}\glspluralsuffix},
    longplural={#4},
    name={\glsentrylong{#1}{} (\glsentrytext{#1}{})},
    first={#3 (#2)},
    firstplural={#4 (\glsentrytext{#1}\glspluralsuffix{})},
    description={#5}
  }
}

\newduneword{dword}{DUNE Word}{A term in the DUNE lexicon}

\newduneabbrev{nd}{ND}{near detector}{Refers to the detector(s) 
 installed close to the neutrino source at \fnal }

\newduneabbrev{fd}{FD}{far detector}{Refers to the \fdfiducialmass fiducial mass DUNE detector 
  to be installed at the far site at \surf in
  Lead, SD, to be composed of four \nominalmodsize modules}

\newduneabbrev{sp}{SP}{single-phase}{Distinguishes one of the DUNE far detector technologies by the fact that it operates using argon in its liquid phase only
  }

\newduneabbrev{dp}{DP}{dual-phase}{Distinguishes one of the DUNE far detector technologies by the fact that it operates using argon 
 in both gas and liquid phases}

\newduneabbrev{pds}{PDS}{photon detection system}{The detector 
  subsystem sensitive to light produced in the \lar }

\newduneabbrev{tpc}{TPC}{time projection chamber}{The portion of each
  DUNE \gls{detmodule} that records ionization electrons
  after they drift away from a cathode through the \lar, and also
  through gaseous argon in a \dual module. 
  The activity is recorded by digitizing the waveforms of current
  induced on the anode as the distribution of ionization charge passes by
  or is collected on the electrode}

\newduneabbrevs{apa}{APA}{anode plane assembly}{anode plane assemblies}{A unit of the \single
  detector module containing the elements sensitive to ionization in the \lar. 
  It contains two faces each of three planes of wires, and interfaces to the cold
  electronics and photon detection system} 

\newduneabbrev{cro}{CRO}{charge readout}{The system for detecting
  ionization charge distributions in a \dual detector module}

\newduneabbrev{lro}{LRO}{light readout}{The system for detecting
  scintillation photons in a \dual  detector module}

\newduneabbrev{fe}{FE}{front-end}{The front-end refers a point that is
  ``upstream'' of the data flow for a particular subsystem. 
  For example the front-end electronics is where the cold electronics
  meet the sense wires of the TPC and the front-end DAQ is where the
  DAQ meets the output of the electronics}

\newduneabbrev{adc}{ADC}{analog-to-digital converter}{A sampling of a voltage
  resulting in a discrete integer count corresponding in some way to
  the input}

\newduneabbrev{daq}{DAQ}{data acquisition}{The data acquisition system
  accepts data from the detector FE electronics, buffers
  the data, performs a \gls{trigdecision}, builds events from the selected
  data and delivers the result to the offline \gls{diskbuffer}}

\newduneword{detmodule}{detector module}{The entire DUNE far detector is
  segmented into four modules, each with a nominal \SI{10}{\kton}
  fiducial mass}

\newduneword{detunit}{detector unit}{A \gls{submodule} may be partitioned
  into a number of similar parts. 
  For example the single-phase TPC \gls{submodule} is made up of APA
  units}

\newduneword{diskbuffer}{secondary DAQ buffer}{A secondary
  \dshort{daq} buffer holds a small subset of the full rate as
  selected by a \gls{trigcommand}. 
  This buffer also marks the interface with the DUNE Offline}

\newduneabbrev{om}{OM}{online monitoring}{Processes that run inside
  the DAQ on data ``in flight,'' specifically before landing on the
  offline disk buffer, and that provide feedback on the operation of
  the DAQ itself and the general health of the data it is marshalling}

\newduneabbrev{dqm}{DQM}{data quality monitoring}{Analysis of the raw
  data to monitor the integrity of the data and the performance of the
  detectors and their electronics. This type of monitoring may be
  performed in real time, within the \gls{daq} system, or in later
  stages of processing, using disk files as input}

\newduneword{dumpbuffer}{DAQ dump buffer}{This \dshort{daq} buffer
  accepts a high-rate data stream, in aggregate, from an associated
  \gls{submodule} sufficient to collect all data likely relevant to
  a potential \gls{snb}}

\newduneabbrev{etl}{ETL}{external trigger logic}{Trigger processing
  that consumes \gls{detmodule} level \gls{trignote} information
  and other global sources of trigger input and emits
  \gls{trigcommand} information back to the \glspl{mtl}}

\newduneword{trignote}{trigger notification}{Information provided by
  \gls{mtl} to \gls{etl} about \gls{trigdecision} its processing}

\newduneword{trigprimitive}{trigger primitive}{Information derived by
  the DAQ \gls{fe} hardware that describes a region of space (e.g.,
  one or several neighboring channels) and time (e.g., a contiguous set
  of ADC sample ticks) associated with some activity}

\newduneword{externtrigger}{external trigger candidate}{Information
  provided to the \gls{mtl} about events external to a
  \gls{detmodule} so that it may be considered in forming
  \glspl{trigcommand}}

\newduneabbrev{daqoob}{OOB dispatcher}{out-of-band trigger command
  dispatcher}{This component is responsible for dispatching a \gls{snb} dump
  command to all \glspl{daqfer} in the \gls{detmodule}}

\newduneabbrev{mtl}{MTL}{module trigger logic}{Trigger processing
  that consumes \gls{detunit} level \gls{trigcommand} information
  and emits \glspl{trigcommand}. 
  It provides the \gls{etl} with \glspl{trignote} and receives back any
  \glspl{externtrigger}}

\newduneword{octant}{octant}{Any of the eight parts into which 4$\pi$
  is divided by three mutually perpendicular axes. 
  In particular in referencing the value for the mixing angle
  $\theta_{23}$}

\newduneword{submodule}{subdetector}{A detector unit of granularity less
  than one \gls{detmodule} such as the TPC of either a \single or \dual 
  module}

\newduneword{trigcandidate}{trigger candidate}{Summary information derived
  from the full data stream and representing a contribution toward
  forming a \gls{trigdecision}}

\newduneword{trigcommand}{trigger command}{Information derived from
  one or more \glspl{trigcandidate}  that directs elements of the
  \gls{detmodule} to read out a portion of the data stream}

\newduneabbrev{tcm}{TCM}{trigger command message}{A message flowing
  down the trigger hierarchy from global to local context.  Also see \gls{tpm}}

\newduneword{trigdecision}{trigger decision}{The process by which
  \glspl{trigcandidate} are converted into \glspl{trigcommand}}

\newduneabbrev{tpm}{TPM}{trigger primitive message}{A message flowing
  up the trigger hierarchy from local to global context.  Also see \gls{tcm}}

\newduneabbrev{eb}{EB}{event builder}{A software agent servicing one
  \gls{detmodule} by executing \glspl{trigcommand} by reading out
  the requested data}

\newduneabbrev{cob}{COB}{Cluster On Board}{An ATCA motherboard housing four RCEs}

\newduneabbrev{rce}{RCE}{reconfigurable computing element}{Data processor located outside of the cryostat on a \gls{cob} which contains \gls{fpga}, RAM and SSD resources, responsible for buffering data, producing trigger primitives, responding to triggered requests for data and sinking \gls{snb} dumps}

\newduneabbrev{bow}{BOW}{Bump On Wire}{A working name for the front-end readout computing elements used in the nominal DAQ design to interface the \dual  crates to the DAQ front-end computers}

\newduneabbrev{atca}{ATCA}{Advanced Telecommunications Computing
  Architecture}{An advanced computer architecture specification developed for the telecommunications, military, and aerospace industries that incorporates the latest trends high-speed interconnect technologies, next-generation processors, and improved reliability, availability and serviceability} 

\newduneabbrev{utca}{$\mu$TCA}{Micro Telecommunications Computing Architecture}{The computer architecture specification followed by the crates that house charge and light readout electronics in the dual-phase module} 

\newduneabbrev{udp}{UDP}{user datagram protocol}{A simple,
  connectionless Internet protocol that supports data integrity
  checksums, requires no handshaking, and does not guarantee packet delivery}

\newduneabbrev{amc}{AMC}{advanced mezzanine card}{Holds digitizing
  electronics and lives in \gls{utca} crates}

\newduneabbrev{rf}{RF}{radio frequency}{Electromagnetic emissions
  that are within the (radio) frequency band of sensitivity of the detector
  electronics}

\newduneabbrev{fpga}{FPGA}{field programmable gate array}{An
integrated circuit technology that allows the hardware to be reconfigured to
execute different algorithms after its manufacture and deployment}

\newduneabbrev{felix}{FELIX}{Front-End Link eXchange}{A
  high-throughput interface between front-end and trigger electronics
  and the standard PCIe computer bus}

\newduneword{daqpart}{DAQ partition}{A cohesive and
 coherent collection of DAQ hardware and software working together to trigger and read out some portion of one detector module; it consists of an integral number of \glspl{daqfrag}. 
 Multiple DAQ partitions may operate simultaneously, but each instance operates independently}

\newduneabbrev{fec}{FEC}{front-end computer}{The portion of one
  \gls{daqpart} that hosts the \gls{daqdr}, \gls{daqbuf} and
  \gls{daqds}.  It is connected to the \gls{daqfer} via fiber optic. 
Each \gls{detunit} of a  certain granularity, such as two \single \glspl{apa}, has one front-end computer
  that receives data from the readout hardware, hosts the primary DAQ
  memory buffer for that data, emits trigger candidates derived from
  that data, and satisfies requests for producing subsets of that data
  for egress}

\newduneword{daqfrag}{DAQ front-end fragment}{The portion of one
  \gls{daqpart} relating to a single \gls{fec} and corresponding to an
  integral number of \glspl{detunit}.  See also \gls{datafrag}}

\newduneword{datafrag}{data fragment}{A block of data read out from a single \gls{daqfrag} that
span a contiguous period of time as requested by a \gls{trigcommand}}

\newduneabbrev{daqfer}{FER}{DAQ front-end readout}{The portion of a
  \gls{daqfrag} that accepts data from the detector electronics and
  provides it to the \gls{fec}. 
  In the nominal design it is also responsible for generating channel
  level \glspl{trigprimitive}}

\newduneabbrev{daqdr}{DDR}{DAQ data receiver}{The portion of the
  \gls{daqfrag} that accepts data from the \gls{daqfer}, emits
  trigger candidates produced from the input trigger primitives, and
  forwards the full data stream to the \gls{daqbuf}}

\newduneabbrev{daqbuf}{primary buffer}{DAQ primary buffer}{The portion
  of the \gls{daqfrag} that accepts full data stream from the
  corresponding \gls{detunit} and retains it sufficiently long for it
  to be available to produce a \gls{datafrag}}

\newduneword{daqds}{data selector}{The portion of the \gls{daqfrag}
  that accepts \glspl{trigcommand} and returns the corresponding
  \gls{datafrag}}

\newduneabbrev{femb}{FEMB}{front-end mother board}{Refers a unit of
  the \gls{sp} cold electronics that contains the front-end amplifier
  and ADC ASICs covering 128 channels}

\newduneword{asic}{ASIC}{application-specific integrated circuit}

\newduneword{lv}{LV}{low voltage}

\newduneword{coldata}{COLDATA}{a 64-channel control and communications ASIC}

\newduneword{larasic}{LArASIC}{A 16-channel front-end ASIC that provides signal amplification and pulse shaping}

\newduneword{cmos}{CMOS}{Complementary metal-oxide-semiconductor}

\newduneword{enc}{ENC}{equivalent noise charge}

\newduneword{dre}{DRE}{dynamic range enhancement}

\newduneword{sar}{SAR}{successive approximation register}

\newduneword{protodune}{ProtoDUNE}{Either of the two DUNE prototype detectors constructed at CERN and operated in
  a CERN test beam (expected fall 2018). 
  One prototype implements \single and the other \dual technology}

\newduneword{pdsp}{ProtoDUNE-SP}{The \single ProtoDUNE detector}

\newduneword{pddp}{ProtoDUNE-DP}{The \dual ProtoDUNE detector}

\newduneword{wa105}{WA105 DP demonstrator}{The \SI[product-units=power]{3x1x1}{m} WA105 dual-phase prototype detector at CERN}

\newduneword{rawevent}{DAQ event block}{The unit of data output by the
  DAQ. 
  It contains trigger and detector data spanning a unique, contiguous
  time period and a subset of the detector channels.}

\newduneabbrev{ssd}{SSD}{solid-state disk}{Any storage device that
  may provide sufficient write throughput to receive, both collectively and
  distributed, the sustained full rate of data from a \gls{detmodule}
  for many seconds}

\newduneabbrev{hlt}{HLT}{high-level trigger}{A source of triggering at
  the module level}

\newduneabbrev{pid}{PID}{particle ID}{Particle identification}

\newduneword{readout window}{readout window}{A fixed, atomic and
  continuous period of time over which data from a \gls{detmodule}, in
  whole or in part, is recorded. 
  This period may differ based on the trigger that initiated the
  readout}

\newduneabbrev{zs}{ZS}{zero-suppression}{Used to delete some portion of a
  data stream that does not significantly deviate from zero or
  intrinsic noise levels. 
  It may be applied at different granularity from per-channel to per
  \dword{detunit}}

\newduneabbrev{rc}{RC}{run control}{The system for configuring,
  starting and terminating the DAQ}

\newduneabbrev{snb}{SNB}{supernova neutrino burst}{A prompt 
  increase in the flux of low-energy neutrinos emitted in the first few seconds of a core-collapse supernova.  It can also refer to a trigger command type that may be due to an SNB,
  or detector conditions that mimic its interaction signature}

\newduneabbrev{snble}{SNB/LE}{supernova neutrino burst and low
  energy}{Supernova neutrino burst and low-energy physics program}

\newduneabbrev{snews}{SNEWS}{SuperNova Early Warning System}{A global
  supernova neutrino burst trigger formed by a coincidence of SNB
  triggers collected from participating experiments}

\newduneabbrev{pps}{1PPS signal}{one-pulse-per-second signal}{An
  electrical signal with a fast rise time and that arrives in real
  time with a precise period of one second}

\newduneabbrev{sls}{SLS}{spill location system}{A system residing at
  the DUNE far detector site that provides information, possibly
  predictive, indicating periods of time when neutrinos are being
  produced by the \fnal Main Injector beam spills}

\newduneabbrev{wib}{WIB}{warm interface board}{Digital electronics
  situated just outside the \single cryostat that receives digital data
  from the FEMBs over cold copper connections and sends it to the RCE
  FE readout hardware}

\newduneabbrev{nic}{NIC}{network interface controller}{Hardware for controlling the interface to a communication network.  Typically, one that obeys the Ethernet protocol}

\newduneabbrev{wiec}{WIEC}{warm interface electronics crate}{Crates mounted on the signal flanges that contain the warm interface boards}

\newduneabbrev{ptc}{PTC}{power and timing cards}{Cards that provide further processing and distribution of the signals entering and exiting the \single cryostat}

\newduneabbrev{sipm}{SiPM}{silicon photomultiplier}{A solid-state
  avalanche photodiode sensitive to single \phel signals}

\newduneabbrev{cisc}{CISC}{cryogenic instrumentation and slow controls}{A DUNE
  consortium responsible for the cryogenic instrumentation and slow controls components}

\newduneword{consortium}{consortium}{A unit of organization in the
  DUNE project focused on one major component of the far detector}

\newduneword{art}{\textit{art}}{A software framework implementing an
  event-based execution paradigm} 

\newduneabbrev{sam}{SAM}{sequential
  access via metadata}{A data-handling system to store and retrieve
  files and associated metadata, including a complete record of the
  processing that has used the files}

\newduneword{artdaq}{\textit{art}DAQ}{A general-purpose \fnal data acquisition framework for distributed data combination and processing. 
It is designed to exploit the parallelism that is possible with
modern multi-core and networked computers}

\newduneword{order}{$\mathcal{O}(n)$}{of order $n$}

\newduneword{beamline}{beamline}{A sequence of control and monitoring devices used for the formation of a directed collection of particles}
\newduneabbrev{cdr}{CDR}{conceptual design report}{A formal project
  document 
   that describes the experiment
  at a conceptual level}

\newduneabbrev{cf}{CF}{conventional facilities}{Pertaining to
  construction and operation of buildings or caverns and conventional infrastructure}

\newduneabbrev{cp}{CP}{charge parity}{Product of charge and parity
  transformations}

\newduneword{cpt}{CPT}{product of charge, parity
  and time-reversal transformations}

\newduneabbrev{cpv}{CPV}{charge-parity symmetry violation}{Lack of
  symmetry in a system before and after charge and parity
  transformations are applied}

\newduneword{doe}{DOE}{U.S. Department of Energy}

\newduneword{dune}{DUNE}{Deep Underground Neutrino Experiment}

\newduneword{esh}{ES\&H}{Environment, Safety and Health}

\newduneabbrev{fgt}{FGT}{fine-grained tracker}{A near detector module ??}

\newduneabbrev{fscf}{FSCF}{far site conventional facilities}{The
  \gslpl{cf} at the DUNE far detector site, \surf}
  
\newduneabbrev{nscf}{NSCF}{near site conventional facilities}{The
  \gslpl{cf} at the DUNE near detector site, \fnal}

\newduneabbrev{gut}{GUT}{grand unified theory}{A class of theories that
  unifies the electro-weak and strong forces}

\newduneabbrev{lar}{LAr}{liquid argon}{The liquid phase of argon}

\newduneabbrev{lartpc}{LArTPC}{liquid argon time-projection chamber}{A
  class of detector technology that forms the basis for the DUNE far
  detector modules. 
  It typically entails observation of ionization activity by
  electrical signals and of scintillation by optical signals}

\newduneabbrev{lbl}{LBL}{long-baseline}{Refers to the distance between the 
  neutrino source  and the far detector.  It can also refer to the distance between the near and far detectors. 
  The ``long'' designation is an approximate and relative distinction. For DUNE, this distance  (between \fnal and \surf) is approximately \SI{1300}{km}}

\newduneabbrev{lbnf}{LBNF}{Long-Baseline Neutrino Facility}{The
  organizational entity responsible for developing the neutrino beam, the cryostats
  and cryogenics systems, and the conventional facilities for DUNE}

\newduneabbrev{mh}{MH}{mass hierarchy}{Describes the separation
  between the mass squared differences related to the solar and
  atmospheric neutrino problems}

\newduneabbrev{mi}{MI}{\fnal Main Injector}{An accelerator at
  \fnal that provides a beam of high-energy protons that upon
  striking a target produce secondaries that decay to provide the
  neutrinos directed toward the DUNE far detector}

\newduneabbrev{pot}{POT}{protons on target}{Typically used as a unit
  of normalization for the number of protons striking the neutrino
  production target}

\newduneabbrev{qa}{QA}{quality assurance}{The process by which quality is maintained so as to preserve
high availability and precise function}

\newduneabbrev{qc}{QC}{quality control}{A system of maintaining
  quality through testing products against a specification}

\newduneabbrev{sm}{SM}{Standard Model}{Refers to a theory describing
  the interaction of elementary particles}

\newduneabbrev{tdr}{TDR}{technical design report}{A formal project
  document 
  that describes the experiment at a technical level}

\newduneabbrev{tp}{IDR}{interim design report}{An intermediate
milestone on the path to a full \gls{tdr}} 

\newduneabbrev{ckm}{CKM matrix}{Cabibbo-Kobayashi-Maskawa
  matrix}{Refers to the matrix describing the mixing between mass and
  weak eigenstates of quarks}

\newduneabbrev{cl}{CL}{confidence level}{Refers to a probability
  used to determine the value of a random variable given its
  distribution}

\newduneabbrev{pmns}{PMNS matrix}{Pontecorvo-Maki-Nakagawa-Sakata
  matrix}{Describes the mixing between mass and weak eigenstates of
  the neutrino}

\newduneabbrevs{cpa}{CPA}{cathode plane assembly}{cathode plane assemblies}{The component of the \single detector module that provides the drift HV cathode}

\newduneabbrev{fc}{FC}{field cage}{The component of a LArTPC that contains and shapes the applied \efield}

\newduneabbrev{topfc}{top FC}{top field cage}{The horizontal portions of the \single FC on the top}

\newduneabbrev{botfc}{bottom FC}{bottom field cage}{The horizontal portions of the \single FC on the bottom}

\newduneabbrev{ewfc}{endwall FC}{endwall field cage}{The vertical portions of the \single FC near the wall}

\newduneabbrev{gp}{GP}{ground plane}{An electrode that is held to be
  electrically neutral relative to Earth ground voltage}

  \newduneword{gg}{ground grid}{An electrode that is held to be
  electrically neutral relative to Earth ground voltage  ??}

\newduneabbrev{alara}{ALARA}{as low as reasonably
  achievable}{Typically used with regard management of radiation
  exposure but may be used more generally. It means making every
  reasonable effort to maintain e.g., exposures, to as far below the
  limits as practical, consistent with the purpose for that the
  activity is undertaken}

\newduneabbrev{ecal}{ECAL}{electromagnetic calorimeter}{A detector
  component that measures energy deposition of traversing particles}

\newduneabbrev{hv}{HV}{high voltage}{Generally describes a voltage
  applied to drive the motion of free electrons through some media}

\newduneword{spmod}{SP module}{single-phase detector module}
\newduneword{dpmod}{DP module}{dual-phase detector module}

\newduneabbrev{tc}{TC}{technical coordination}{A dedicated DUNE
  project effort responsible for assuring proper interfaces between
  the collaboration consortia}

\newduneabbrev{cc}{CC}{charged current}{Refers to an interaction
  between elementary particles where a charged weak force carrier
  ($W^+$ or $W^-$) is exchanged}

\newduneabbrev{dis}{DIS}{deep inelastic scattering}{Refers to
  interaction of an elementary charged particle with a nucleus in an
  energy range where the interaction can be modeled as being with
  individual nucleons}

\newduneabbrev{fsi}{FSI}{final-state interactions}{Refers to
  interactions between elementary or composite particles subsequent to
  the initial, fundamental particle interaction, such as may occur as
  the products exit a nucleus}

\newduneabbrev{geant4}{Geant4}{GEometry ANd Tracking, version 4}{A
  software toolkit for the simulation of the passage of particles
  through matter using Monte Carlo methods}

\newduneabbrev{genie}{GENIE}{Generates Events for Neutrino Interaction
  Experiments}{Software providing an object-oriented neutrino
  interaction simulation resulting in kinematics of the products of
  the interaction}

\newduneabbrev{mc}{MC}{Monte Carlo}{Refers to a method of numerical
  integration that entails the statistical sampling of the integrand
  function. 
  Forms the basis for some types of detector and physics simulations}

\newduneabbrev{qe}{QE}{quasi-elastic}{Refers to interaction between
  elementary particles and a nucleus in an energy range where the
  interaction can be modeled as occurring between constituent quarks
  of one nucleon and resulting in no bulk recoil of the resulting
  nucleus}

\newduneabbrev{mou}{MOU}{memorandum of understanding}{A document
  summarizing an agreement between two or more parties}

\newduneabbrev{pip2}{PIP-II}{Proton Improvement Plan II}{A plan for
  improving the protons on target delivered for the DUNE neutrino
  production beam. 
  This is revision two of this plan and is planned to be followed by a PIP-III}

\newduneabbrev{sdsta}{SDSTA}{South Dakota Science and Technology
  Authority}{The legal entity that manages the Sanford Underground
  Research Facility, \surf, in Lead, S.D}

\newduneabbrev{wbs}{WBS}{work breakdown structure}{An organizational
  project management tool by which the tasks to be performed are
  partitioned in a hierarchical manner}

\newduneabbrev{br}{BR}{branching ratio}{A fractional probability for a
  decay of a composite particle to occur into some specified set or
  sets of products}

\newduneabbrev{dm}{DM}{dark matter}{The term given to the unknown
  matter or force that explains measurements of motion of galaxies
  that are otherwise inconsistent with the amount of mass associated
  with observed amount of photon production}

\newduneabbrev{dsnb}{DSNB}{Diffuse Supernova Neutrino Background}{The
  term describing the pervasive, constant flux of neutrinos due to all
  past supernova neutrino bursts}

\newduneabbrev{globes}{GLoBES}{General Long-Baseline Experiment
  Simulator}{A software package for simulating energy spectra of
  neutrino flux, interaction and measured (after application of some
  model of a detector response) energy spectra}

\newduneabbrev{nc}{NC}{neutral current}{Refers to an interaction
  between elementary particles where a neutrally charged weak force carrier
  ($Z^0$) is exchanged}

\newduneabbrev{nh}{NH}{normal hierarchy}{Refers to the neutrino mass
  eigenstate ordering whereby the sign of the mass squared difference
  associated with the atmospheric neutrino problem is positive}

\newduneabbrev{ih}{IH}{inverted hierarchy}{Refers to the neutrino mass
  eigenstate ordering whereby the sign of the mass squared difference
  associated with the atmospheric neutrino problem is negative}

\newduneabbrev{nsi}{NSI}{nonstandard interactions}{A general class of
  theory of elementary particles other than the Standard Model}

\newduneabbrev{msw}{MSW}{Mikheyev-Smirnov-Wolfenstein effect}{Explains
  the oscillatory behavior of neutrinos produced inside the sun as
  they traverse the solar matter}

\newduneabbrev{susy}{SUSY}{supersymmetry}{Theoretical symmetry between a fermion and a boson}

\newduneabbrev{wimp}{WIMP}{weakly-interacting massive particle}{A
  hypothesized particle that may be a component of dark matter}

\newduneabbrev{ce}{CE}{cold electronics}{Refers to readout electronics that operate at cryogenic temperatures}

\newduneabbrev{crp}{CRP}{charge-readout plane}{In the \dual technology, a  collection of
  electrodes in a planar arrangement placed at a particular voltage
  relative to some applied \efield such that drifting electrons
  may be collected and their number and time may be measured}

\newduneabbrev{dram}{DRAM}{dynamic random access memory}{A computer memory technology}

\newduneword{fermilab}{\fnal}{Fermi National Accelerator Laboratory (in Batavia, IL, the Near Site)}
\newduneword{fnal}{FNAL}{see \fnal}

\newduneabbrev{fs}{FS}{full stream}{Relates to a data stream that has not undergone selection, compression or other form of reduction}

\newduneabbrev{lem}{LEM}{large electron multiplier}{A micro-pattern detector suitable for use in ultra-pure argon vapor; LEMs consist of copper-clad PCB boards with sub-millimeter-size holes through which electrons undergo amplification}
\newduneabbrev{lng}{LNG}{liquefied natural gas}{Pertaining to natural gas in its liquid phase}

\newduneabbrev{mip}{MIP}{minimum ionizing particle}{Refers to a
  momentum traversing some medium such that the particle is losing
  near the minimum amount of energy per distance traversed} 

\newduneabbrev{pd}{PD}{photon detector}{Refers to the detector
  elements involved in measurement of number and arrival times of
  optical photons produced in a detector module} 

\newduneabbrev{pmt}{PMT}{photomultiplier tube}{A device that makes use
  of the photoelectric effect to produce an electrical signal from the
  arrival of optical photons}

\newduneabbrev{ppm}{ppm}{parts per million}{A number equal to $10^{-6}$}
\newduneabbrev{ppb}{ppb}{parts per billion}{A number equal to $10^{-9}$}
\newduneabbrev{ppt}{ppt}{parts per trillion}{A number equal to $10^{-12}$}

\newduneword{rio}{RIO}{reconfigurable input output}

\newduneword{rpc}{RPC}{resistive plate chamber}
\newduneword{s/n}{S/N}{signal-to-noise (ratio)}
\newduneword{ssp}{SSP}{SiPM signal processor}
\newduneword{sbn}{SBN}{Short-Baseline Neutrino program (at \fnal)}
\newduneword{stt}{STT}{straw tube tracker}
\newduneword{tr}{TR}{transition radiation}
\newduneword{wire board}{wire board}{At the head end of the APA in the \single TPC, stacks of electronics boards referred to as ``wire boards'' are arrayed to anchor the wires.  They also provide the connection between the wires and the cold electronics} 

\newduneabbrev{wls}{WLS}{wavelength shifting}{A material or process by
  which incident photons are absorbed by a material and photons are
  emitted at a different, typically longer, wavelength}

\newduneabbrev{tpb}{TPB}{tetra-phenyl butadiene}{A type of wavelength shifting material}

\newduneabbrev{sft}{SFT}{signal feedthrough}{A cryostat penetration allowing for the passage of cables or other extended parts}
\newduneabbrev{sftchimney}{SFT chimney}{signal feedthrough chimney}{In the \dual technology, a volume above the cryostat penetration used for a signal feedthrough}

\newduneword{catiroc}{CATIROC}{charge and time integrated readout chip}

\newduneabbrev{wr}{WR}{White Rabbit}{A component of the timing system that forwards clock signal and time-of-day reference data to the master timing unit}

\newduneabbrev{mch}{MCH}{MicroTCA Carrier Hub}{An network switching device}

\newduneabbrev{wrmch}{WR-MCH}{White Rabbit \gls{utca} Carrier Hub}{add def} 

\newduneword{cmp}{CMP}{configuration management plan}

\newduneword{qap}{QAP}{quality assurance plan} 

\newduneword{ieshp}{IESHP}{integrated environmental, safety and health plan}

\newduneword{dmp}{DMP}{data management plan} 

\newduneword{qam}{QAM}{quality assurance manager} 

\newduneabbrev{dss}{DSS}{detector support system}{The system used to support the \single detector within the cryostat}

\newduneabbrev{itf}{ITF}{integration and test facility}{A facility where various detector components will be tested prior to installation}

\newduneabbrev{tco}{TCO}{temporary construction opening}{An opening in the side of a cryostat through which detector elements are brought into the cryostat; utilized during construction and installation}

\newduneabbrev{uit}{UIT}{underground installation team}{An organizational unit responsible for installation in the underground area at the \surf site}

\newduneabbrev{pdr}{PDR}{preliminary design review}{A project management device by which an early design is reviewed}
\newduneabbrev{fdr}{FDR}{final design review}{A project management device by which a final design is reviewed}
\newduneabbrev{prr}{PRR}{production readiness review}{A project management device by which the production readiness is reviewed}
\newduneabbrev{orr}{ORR}{operational readiness review}{A project management device by which the operational readiness is reviewed}
\newduneabbrev{ppr}{PPR}{production progress review}{A project management device by which the progress of production is reviewed}
\newduneabbrev{edms}{EDMS}{electronic document management system}{A computerized system used at CERN by which documents are managed}

\newduneword{wrgm}{WR grandmaster}{White Rabbit grandmaster}

\newduneword{larsoft}{\larsoft}{Liquid Argon Software (\larsoft),  a shared base of physics software across \lartpc experiments}
\newduneword{nova}{\nova}{The \nova off-axis neutrino oscillation experiment at \fnal }
\newduneword{minerva}{\minerva}{The \minerva neutrino cross sections experiment at \fnal }
\newduneword{microboone}{\microboone}{The \lartpc-based \microboone neutrino oscillation experiment at \fnal }
\newduneword{wirecell}{wire-cell}{A tomographic automated \threed neutrino event reconstruction method for \lartpc{}s}
\newduneword{ftslite}{F-FTS-lite}{Light-weight version of the \fnal File Transfer system used for rapid data transfers out of the online systems}
\newduneword{fts}{FTS}{File Transfer System developed at \fnal to catalog and move data to permanent storage}

\newduneword{35t}{35 ton prototype}{The 35 ton prototype cryostat and \gls{sp} detector built at \fnal before the \gls{protodune} detectors}

\newduneabbrev{cuc}{CUC}{central utility cavern}{The central underground cavern  containing utilities such as central cryogenics and other systems, and the underground data center and control room}
\newduneabbrev{cfd}{CFD}{computational fluid dynamics}{High performance computer-assisted modeling of fluid dynamical systems}
\newduneword{vuv}{VUV}{vacuum ultra-violet}
\newduneword{tallbo}{TallBo}{A cylindrical cryostat at \fnal primarily used for developing scintillation light collection technologies for \lartpc detectors}

\newduneabbrev{eos}{EOS}{EOS}{The XRootD based distributed file system developed by CERN}
\newduneabbrev{ehn1}{EHN1}{Experiment Hall North One}{Location at CERN of the ProtoDUNE experiments}
\newduneword{led}{LED}{Light-emitting diode}
\newduneword{rtd}{RTD}{Cryostat level and temperature monitoring devices}
\newduneword{swc}{SWC}{Software \& Computing}
\newduneabbrev{las}{LAS}{LEM-anode Sandwich}{In the \dual technology, a \gls{lem} and its corresponding anode are mounted together in a module called a LEM-anode sandwich}

\newduneword{roi}{ROI}{region of interest}
\newduneword{hpc}{HPC}{high-performance computing facilities; generally computing facilities emphasizing parallel computing with aggregate power of more than a teraflop}

\newduneword{comfund}{common fund}{The shared resources of the collaboration}
\newduneabbrev{ims}{IMS}{integrated master schedule}{A project management device consisting of linked tasks and milestones}

\newduneword{hvdb}{HVDB}{dual-phase HV divider board}
\newduneword{sas}{SAS}{A pass-through chamber used to ensure safe transfer of materials, avoiding contamination in both directions}

\newduneword{fea}{FEA}{finite element analysis}

\newduneword{fss}{FSS}{field shaping strips}
\newduneword{lvds}{LVDS}{low-voltage differential signaling}

\hypersetup{
    pdftitle={\expshort TDR \thedocsubtitle},
    pdfauthor={\expshort Collaboration},
    final=true,
    colorlinks=false,
    linktocpage=true,
    linkbordercolor=blue,
    citebordercolor=green,
    urlbordercolor=magenta,
    filecolor=black,
    pdfpagemode=UseOutlines,
    pdfborderstyle={/S/U},  
}

\renewcommand\thedoctitle{\voldpfd} 
\begin{document}

\pagestyle{titlepage}
\includepdf[pages={-}]{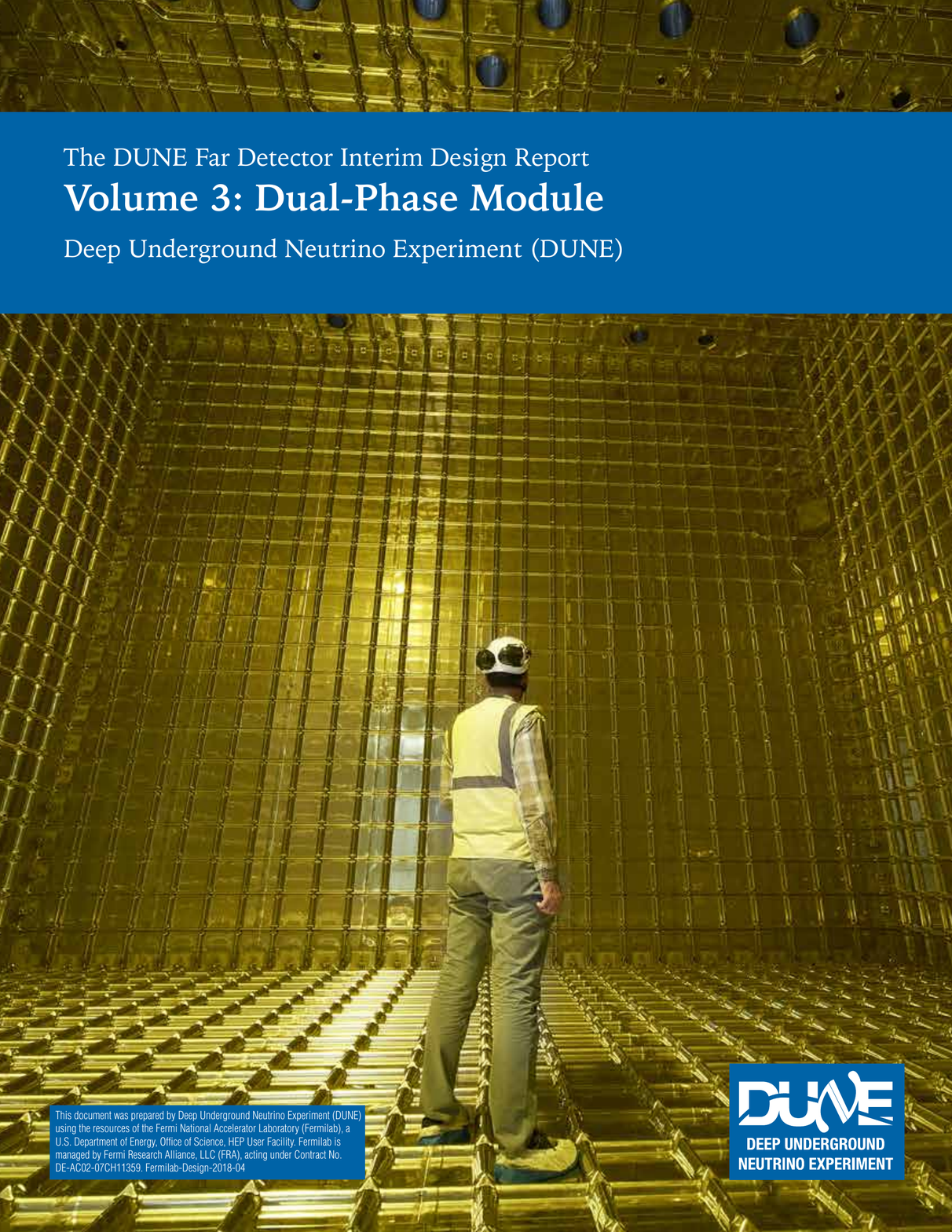}
\cleardoublepage



\cleardoublepage

\includepdf[pages={-}]{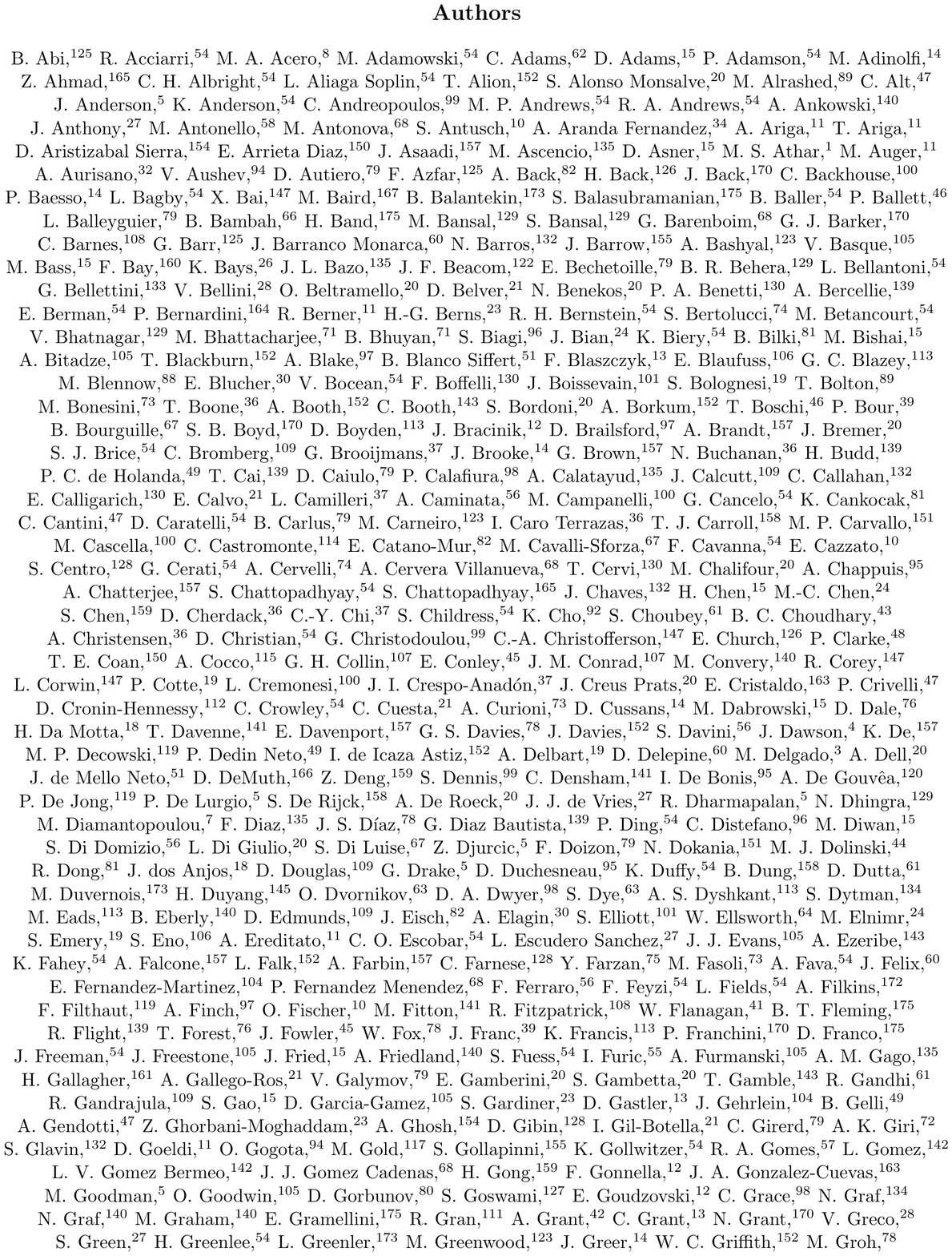}

\renewcommand{\familydefault}{\sfdefault}
\renewcommand{\thepage}{\roman{page}}
\setcounter{page}{0}

\pagestyle{plain} 


\textsf{\tableofcontents}

\textsf{\listoffigures}

\textsf{\listoftables}



\renewcommand{\thepage}{\arabic{page}}
\setcounter{page}{1}

\pagestyle{fancy}

\renewcommand{\chaptermark}[1]{%
\markboth{Chapter \thechapter:\ #1}{}}
\fancyhead{}
\fancyhead[RO,LE]{\textsf{\footnotesize \thechapter--\thepage}}
\fancyhead[LO,RE]{\textsf{\footnotesize \leftmark}}

\fancyfoot{}
\fancyfoot[RO]{\textsf{\footnotesize The DUNE Far Detector Interim Design Report}}
\fancyfoot[LO]{\textsf{\footnotesize \thedoctitle}}
\fancypagestyle{plain}{}

\renewcommand{\headrule}{\vspace{-4mm}\color[gray]{0.5}{\rule{\headwidth}{0.5pt}}}

\nocite{CD0}


\chapter{Design Motivation}
\label{ch:fddp-design}

\section{Introduction to the DUNE Dual-Phase Far Detector Module Design}
\label{sec:fddp-design-highlight}


The DUNE \dword{dp} \dword{detmodule} aims to open new windows of opportunity in the study of neutrinos, a goal it shares with the \dword{spmod}. 
 DUNE's rich physics program, with discovery potential for \dword{cp} in the neutrino sector, and its capability to make significant observations of nucleon decay and astrophysical events, is enabled by the exquisite resolution of the \lartpc detector technique, which the \dual design further augments. This design improves the \dword{s/n} ratio in the charge readout,  lowering the threshold for the smallest observable signals while also achieving a finer readout granularity.  The \dual technology enables the construction of larger drift volumes, thereby reducing 
the quantity of nonactive materials in the \lar. Although the physics requirements are identical for both the \single and \dual designs, 
 some aspects of the \dual design offer augmented performance. 


\section{Dual-Phase (DP) \lartpc Operational Principle}
\label{sec:fddp-operational-principle}


The basic operational principle is very similar to that of the \single design. 
Charged particles that traverse the active volume of the \lartpc ionize the medium, while also producing scintillation light.  The ionization electrons drift along an \efield towards a segmented anode where they deposit their charge, and \dwords{pd} pick up the scintillation light.
The precision tracking and calorimetry offered by the \dual
technology provides excellent capabilities for identifying interactions of interest while mitigating sources of background.  
Whereas the \single design has multiple drift volumes, the \dword{dpmod} design allows for a single, fully homogeneous \lar volume with a much longer drift length. This volume is surrounded by a \dword{fc} on the sides and a cathode at the bottom, which together define the drift field. 


The key differentiating concept of the \dual design is the amplification of the ionization signal in an avalanche process. In the \single design, charges drift horizontally to the anode, which consists of a set of induction and collection wire layers immersed in the \lar. In the \dual design, electrons drift vertically upward towards an extraction grid just below the liquid-vapor interface. After reaching the grid, an \efield stronger than the drift field extracts the electrons from the liquid up into the gas phase. Once in the gas, electrons encounter micro-pattern gas detectors with high-field regions, called \dwords{lem}. The \dwords{lem} amplify the electrons in avalanches that occur in these high-field regions. 
The amplified charge is then collected and recorded on a \twod anode
consisting of two sets of \SI{3.125}{mm}-pitch gold-plated copper strips that provide the $x$
and $y$ coordinates (and thus two views) of an event.



The extraction grid, \dword{lem} and anode are assembled into three-layered \textit{sandwiches} 
with precisely defined inter-stage distances and inter-alignment,  which are then connected together horizontally into
modular units of area \num{9}~m$^2$. These units are called \dwords{crp}.
The \dwords{crp} integrate the \dwords{lem} and anodes, and support the extraction grid. These units can be individually positioned in a horizontal plane a few millimeters beneath the liquid-gas interface, ensuring complete immersion of the extraction grid. 

The argon scintillation light, which at a wavelength of  \SI{127}{nm} is deep in the UV spectrum and it is recorded by an array of \dwords{pmt} located below the cathode.  
The \dwords{pmt}, coated with a wavelength-shifting material, shifts the light  closer to the visible spectrum and records the time and pulse characteristics of the incident light.


Two key factors affect the performance of the  \lartpc{}.  First, the \lar purity must be extremely high in order to achieve minimum charge attenuation over the longest drift lengths in the \lartpc{}.  This requires that the levels of electronegative contaminants (e.g., oxygen, water), be reduced and
maintained at $\sim$ppt levels.  The \dual and \single designs are subject to the same purity requirements. 
Second, the electronic readout
of the \lartpc{} requires very low noise levels in order that the signal from the drifting electrons
be clearly discernible over the baseline of the electronics.  This requires use of low-noise cryogenic electronics. 

The amplification of the electron signal in the gas phase mitigates the potential impact of these factors on the performance of the \dual design.  On the other hand this design requires  higher voltages on the cathode, relative to the \single, due to the longer drift field.

\begin{dunefigure}[Principle of the \dual readout]{fig:figure-label-DPprinciple}{Principle of the \dual readout}
\includegraphics[width=0.8\textwidth]{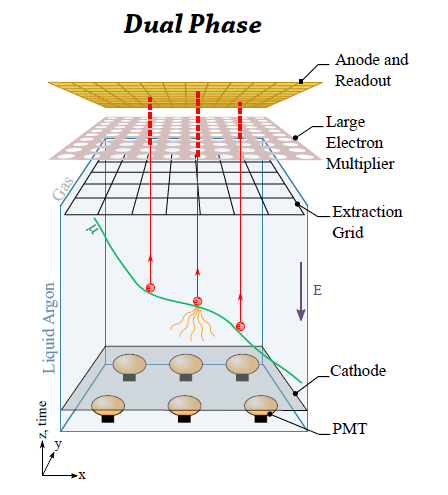}
\end{dunefigure}

\section{Motivation for the \dual Design} 
\label{sec:fddp-design-motivation}


The innovative \dual design is similar in many ways to the \single design, but implements some unique features and offers several advantages over it, providing an appealing and complementary approach, as summarized below:


\begin{itemize}
\item Gain on the ionization signal obtained in the gas phase:
\begin{itemize}
\item  leading to a robust and tunable \dword{s/n} and a lower detection threshold, and
\item  compensating for potential charge attenuation due to long drift paths; 
\end{itemize}
\item  Larger fiducial volume, enabling very long drift paths;
\item  Absence of dead material in the \lar drift volume;
\item  Finer readout pitch (\SI{3}{mm}), implemented in two identical collection views, $x$ and $y$;
\item  Fewer readout channels (\dpnumcrpch for \dual versus \spnumch for \single for a  \nominalmodsize module); 
\item  Fewer construction modules;
\item  Full accessibility and replaceability of the \dword{fe} electronics during the detector operation.

\end{itemize}

 The \dual design features maximize the capability of the experiment and are motivated to cope with the unprecedented scale of the \dwords{detmodule} and the deep underground location where construction will occur.

Among the features driven by the underground location of the experiment, all detector components are sized to fit within the constraints of the \surf shafts and access pathways. The \dword{crp} modules are essentially planar objects with a surface of \num{3}\,$\times$\,\SI{3}{m$^2$}. All the other detector 
components (\dword{fc} and cathode modules) are 
designed in order to stay within this envelope. The relatively small number of detector elements makes the underground installation easier.

A drift time of several milliseconds is typical for ionization charge to arrive at the anode wires after drifting several meters.  This lengthy duration, as well as aspects of the DUNE physics program looking for rare and low-energy processes, makes the deep underground location essential for the \dword{dpmod}.  The \SI{1.5}{km} overburden of earth will vastly reduce the rate of cosmic rays reaching the active volume of the \dword{dpmod}, greatly enhancing the ability to search for rare and low-energy signatures without the influence of cosmic-induced backgrounds.  


\section{Overview of Dual-Phase Design}
\label{sec:fddp-ov-description}

This \dual design implements a \dual liquid argon time projection chamber (\lartpc) augmented with a light-readout system.  \textit{\dual} refers to the extraction of ionization electrons at the interface between the liquid and gas argon and their amplification and collection in the gas phase.

The \dword{dpmod} features a  \dpactivelarmass active mass \lartpc, with all associated cryogenics, electronic readout, computing, and safety systems. The \dword{dpmod} is designed to maximize the active volume within the confines of the membrane cryostat while minimizing dead regions and the presence of dead materials in the drift region. The detector is built as a single active volume \dptpclen long, \dptpcwdth wide and \tpcheight high, with the anode at the top, the cathode near the bottom, and an array of \dwords{pmt} located  at the bottom 
underneath the cathode. The active volume (see Figure~\ref{fig:DPdet1}) is surrounded by the \dword{fc}. The ionization electrons in the liquid phase drift  in a uniform \efield towards the anode plane at the top of the active volume. This is made by an array of \num{80} independent \dword{crp} modules, \num{3}\,$\times$\,\SI{3}{m$^2$} each. The cryogenic \dword{fe} electronics is 
installed in the \dwords{sftchimney}
on the roof of the cryostat. There are no active electronics elements in the cryostat volume besides the \dword{pmt} bases.
The proposed design optimally exploits the cryostat volume of \cryostatwdth{}(w)\,$\times$\,\cryostatht{}(h)\,$\times$\cryostatlen{}(l) with an anode active area of \dptpcwdth{}\,$\times$\,\cryostatlen{} and a maximum drift length of \dpmaxdrift{}, corresponding to an active \lar mass of \dpactivelarmass  (\dpfidlarmass fiducial). 

The detector elements (\dwords{crp}, \dword{fc} and cathode) have been modularized such that their production can proceed in parallel with the construction of the DUNE caverns and cryostats, and sized so that they conform to the access restrictions for transport underground. Table~\ref{tab:dune-dp-parameters} summarizes 
the high-level parameters of the \dword{dpmod} while Figure~\ref{fig:DPdet1} shows 
the \dword{dpmod}'s main components.

\begin{dunetable}[Dual-phase module parameters]{lll}{tab:dune-dp-parameters}{\Dword{dpmod} parameters}
Parameter & Value & Note \\ \toprowrule
Cryostat \lar Mass & \larmass & \\ \colhline 
Active \lar Mass & \dpactivelarmass & \\  \colhline 
Active height & \tpcheight & \\  \colhline 
Active length & \dptpclen & \\  \colhline 
Maximum drift & \dpmaxdrift & \\ \colhline 
Number of \dwords{crp} &\dptotcrp & \\  \colhline 
Number of \dword{crp} channels & \dpnumcrpch & \\ \colhline 
Number of \dword{pmt} channels & \dpnumpmtch & \\ 
\end{dunetable}

\begin{dunefigure}[Diagram of the \dword{dpmod}]{fig:DPdet1}
  {The \dword{dpmod} with cathode, \dwords{pmt}, \dword{fc} and anode plane with \dwords{sftchimney}.}
  \includegraphics[width=0.9\textwidth]{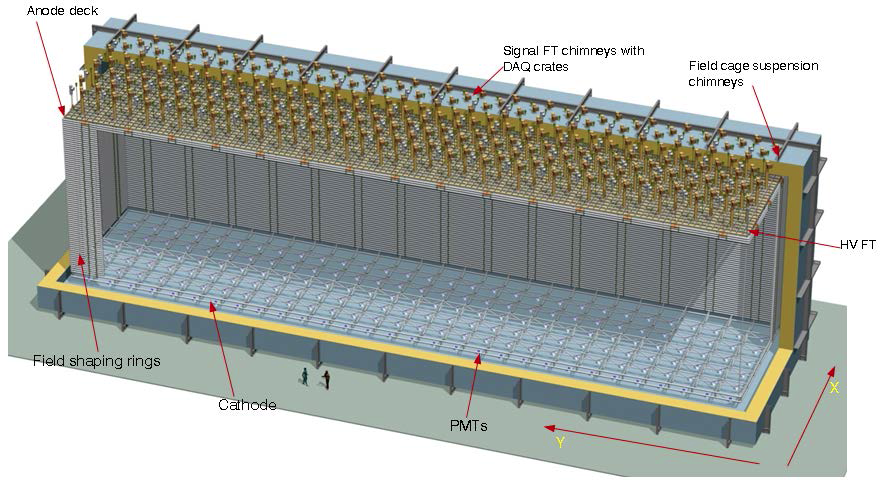}
\end{dunefigure}

The extraction of the electrons from the liquid to vapor phase is performed thanks to the submersed horizontal extraction grid, integrated in each \dword{crp} structure. A \dword{crp} unit includes \dpswchpercrp (0.5~m$\times$0.5~m) \dword{lem}/anode sandwiches, providing tunable amplification and charge collection on two independent views organized in strips of \SI{3}{m} length and \dpstrippitch pitch. There are \dpchpercrp readout channels for each \dword{crp}. Signals in each \dword{crp} unit are collected via three \dwords{sftchimney}
hosting the \dword{fe} cards with the cryogenic \dword{asic} amplifiers (\dpchperchimney channels/chimney) which are accessible and replaceable without contaminating the pure \lar volume. Each \dword{sftchimney} is coupled to a microTCA crate to provide for the signals' digitization and 
\dword{daq}. These crates are connected  via optical fiber links to the \dword{daq} back-end. The total number of readout channels  per \nominalmodsize module is \dpnumcrpch.

Each \dword{crp} unit is independently suspended by three stainless steel ropes. The vertical level of each \dword{crp} unit can be automatically adjusted with respect to the \lar level via three suspension \fdth{}s. The stainless steel ropes are operated by step motors located outside the suspension \fdth{}s. Slow-control \fdth{}s,  one per \dword{crp} unit, are used for level meters and temperature probes readout,
  for pulsing the calibration signals, and to apply the \dword{hv} bias on the two sides of the \dwords{lem} and on the extraction grid. The \dword{fc} and the anode are constructed of modules of similar dimensions as those in \dword{pddp}.

The number of components and corresponding parameters for the \dpactivelarmass \dword{dpmod} are summarized in Table~\ref{tab:DP_numbers}.

\begin{dunetable}[Quantities of items or parameters for the \dword{dpmod}]{ll}{tab:DP_numbers}{Quantities of items or parameters for the \dpactivelarmass  \dword{dpmod}}  Item & Number or Parameter    \\ \toprowrule
Anode plane size & W = \dptpcwdth, L = \dptpclen \\ \colhline
\dword{crp} unit size & W = \SI{3}{m}, L = \SI{3}{m}  \\ \colhline
\dword{crp} units & \num{4}\,$\times$\,\num{20} = \dptotcrp \\ \colhline
\dword{lem}-anode sandwiches per \dword{crp} unit & \dpswchpercrp \\ \colhline 
\dword{lem}-anode sandwiches (total) & \dpnumswch \\ \colhline
\dword{sftchimney} per \dword{crp} unit & \num{3} \\ \colhline
\dword{sftchimney} (total) & \num{240} \\ \colhline
Charge readout channels / \dword{sftchimney} & \num{640}  \\ \colhline
Charge readout channels (total) & \dpnumcrpch \\ \colhline
Suspension \fdth per \dword{crp} unit & \num{3}  \\ \colhline
Suspension \fdth{}s (total) & \num{240}  \\ \colhline
Slow Control \fdth per sub-anode & \num{1}  \\ \colhline
Slow Control \fdth{}s (total) & \num{80} \\ \colhline
\dword{hv} \fdth & \num{1}  \\ \colhline
\dword{hv} for vertical drift & \dptargetdriftvoltpos \\ \colhline
Voltage degrader resistive chains & \num{12} \\ \colhline
Cathode modules & \num{80}  \\ \colhline
Field cage rings & \num{197}     \\ \colhline
Field cage modules (\SI{3}{m}$\times$\SI{12}{m}) & \num{48}  \\ \colhline
\dwords{pmt} (total) & \dpnumpmtch (\num{1}/m$^2$) \\ 
\end{dunetable}

A number of factors make the \dual TPC concept, as described in this chapter, well suited to large detector sizes like the \dword{dpmod}.
In this design,the charge amplification in the \dwords{crp} compensates for the charge attenuation on the long drift paths.  This configuration also simplifies
construction by optimally exploiting the long vertical dimensions of the cryostat, providing a large homogeneous fiducial volume  free of embedded passive materials (effectively increasing the detector size), reducing the number of readout channels,  and ultimately lowering costs.

The \dwords{crp} collect the charge in a projective way,  with practically no dead region, and read the signals out  in two collection views, eliminating the need for  induction views, 
which  simplifies the reconstruction of complicated topologies. The tunable high \dword{s/n} provides operative margins with respect to the noise and electron lifetime conditions, and lowers the threshold on the minimal  detectable energy depositions .

The scope of a \dword{dpmod} includes the design, procurement, fabrication, testing, delivery, installation and
commissioning of the detector components, which is organized in detector consortia, specific to \dual or joint with \single. 

\begin{itemize}
\item \dword{crp}, including extraction grid, \dword{lem} and anode and readout planes (\dual consortium);
\item Analog and digital electronics and \dword{sftchimney} (\dual consortium); 
\item \dword{pds} (\dual Consortium);
\item Cathode, \dword{fc} and \dword{hv} system (joint \single{}-\dual consortium);  
\item Slow-control (joint \single{}-\dual consortium); 
\item Back-end \dword{daq} system (joint \single{}-\dual consortium).
\end{itemize}

\section{Detector systems}
\label{sec:fddp-ov-systems}
\subsection{Charge Readout Planes}
\label{v4:fddp-ov:crp}

An extraction efficiency of \num{100}\,\% of the electrons from the liquid to the gas phase is achieved with an \efield of the order of \SI{2}{kV/cm} across the liquid-gas interface, applied between an  extraction grid submersed in the liquid and charge amplification  devices situated in the ultra-pure argon gas. 

These amplification devices, called \dwords{lem}, are horizontally  oriented \SI{1}{mm}-thick printed  circuit boards with electrodes on the top and bottom surfaces. They are drilled through with many holes that collectively form a micro-pattern structure;  when a \SI{3}{kV} potential difference is applied across the electrodes the ionization electrons are amplified by avalanches (Townsend multiplication) occurring in the  pure argon gas in this micro-pattern structure due to the high \efield (\SI{30}{kV/cm}).

The use of avalanches to amplify the charges in the gas phase increases the \dword{s/n} ratio by at least one order of magnitude with a  gain between \numrange{20}{100}, significantly improving the event reconstruction quality. It also lowers the threshold for small energy depositions and provides a better resolution per volumetric pixel (voxel) compared to a \single \lartpc.  The charge is collected in a finely segmented \twod ($x$ and $y$) readout anode plane at the top of the gas volume and fed to the \dword{fe} electronics.   

The  collection, amplification and readout components are combined in an array of independent (layered) modules called \dwords{crp}. A \dword{crp} is  composed of several \num{0.5}\,$\times$\,\SI{0.5}{m$^2$} units, each of which is composed  of a \dword{lem}-anode \textit{sandwich}.  These units are embedded in a mechanically reinforced frame of FR-4 and iron-nickel invar alloy. This design guarantees the planarity requirements over the \dword{crp} span although the temperature gradient present in the gas phase and possible sagging effects with respect to the three suspension points. The \dword{crp} structure also integrates  the submersed extraction grid, which is an array of $x$ and $y$ oriented stainless steel wires, \SI{0.1}{mm} in diameter, with \dpstrippitch pitch. Thicknesses and possible biasing voltages for the different layers are indicated in Figure~\ref{fig:CRP_struct}.

\begin{dunefigure}[Thicknesses and HV values for electron extraction from liquid to gaseous Ar]{fig:CRP_struct}
{Thicknesses and \dword{hv} values for electron extraction from liquid to gaseous argon, their  multiplication by \dwords{lem} and their collection on the $x$ and $y$ readout anode plane. The \dword{hv} values are indicated for a drift field of \SI{0.5}{kV/cm} in \lar.}
\includegraphics[width=0.8\textwidth]{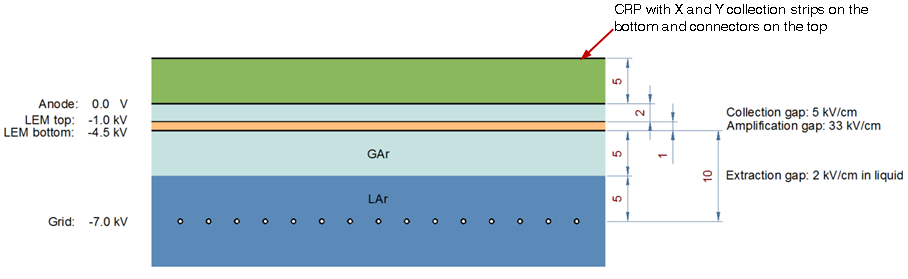}
\end{dunefigure}

Each \dword{crp} is independently suspended by three stainless-steel ropes linked to the top deck of the cryostat. This suspension system allows adjustment of the \dword{crp} distance and parallelism with respect to the \lar surface, and keeps the extraction grid immersed. A \dword{crp} provides an adjustable charge gain (with a minimal required gain of \num{20}) and two independent, orthogonal readout views, each with a pitch of \dpstrippitch.  The \dword{lem}/anode sandwiches  in the same \dword{crp} unit are interconnected with short flat cables so that each readout channel corresponds to a total strip length of \SI{3}{m}. Combined with the time information coming from the \lar scintillation readout by the \dword{pmt} arrays ($t_0$), a \dword{crp} provides \threed track imaging with $dE/dx$ information.  The \dwords{crp} and their components are described in Chapter~\ref{ch:fddp-CRP}.

The typical amplification achieved by this design, between \numrange{20}{100}, improves the \dword{s/n} ratio and thus  compensates for the charge losses that occur along the very long drift paths due to the presence of  electronegative impurities. Therefore, despite the longer drift length, this design requires no higher 
purity of the \lar than does the reference design, around \SI{0.1}{ppb} (or \SI{100}{ppt}) of oxygen equivalent, and yields a \SI{3}{ms} electron lifetime. The required level of purity can be reached by starting from  commercially available ppm-level bulk argon and filling a non-evacuated vessel~\cite{WA105-TDR}.


For instance, given an electron lifetime of \SI{3}{ms} corresponding to the minimal requirement,  a drift field of \SI{0.5}{kV/cm} and a \dword{lem} gain of \num{20}, \dword{s/n} ratio would be larger than  \num{40}:\num{1} for tracks up to \SI{6}{m} from the anode, while reaching  \num{11}:\num{1} for  \dword{mip} tracks that are \dpmaxdrift from the anode.

Other \fdth{}s than the signal chimney connected to the \dwords{crp} and the \dword{crp} slow control and \fdth{}s are planned for the cathode \dword{hv} connection, the \dwords{crp}' suspension and level adjustment, the \dword{hv} and signal readout of the \dwords{pmt}, and the monitoring instrumentation (level meters, temperature probes, strain gauges, etc.).

\subsection{Readout Electronics and Chimneys}
\label{v4:fddp-ov:electronics}

The electrical signals from the collected charges are passed to the outside of the tank via a set of dedicated \dword{sftchimney}. The chimney are pipes passing through the top layer of the cryostat insulation and closed at the top and at the bottom by ultra-high-vacuum flanges (warm and cold). The volume inside each \dword{sftchimney} is tight with respect to the cryostat inner volume and it is filled with nitrogen gas. The bottom (cold) flange of each \dword{sftchimney} is a short distance 
from the \dwords{crp} in the cryostat gas volume.

The cryogenic \dword{fe} electronics cards, housed at the bottom of the chimney are plugged to the top side of  the cold flange. The \dword{fe} cards are based on analog cryogenic preamplifiers implemented in \dword{cmos} \dword{asic} circuits for high integration and large-scale affordable production. 
The \dword{asic} circuits have been especially designed, following an R\&D process started in 2006, to match the signal dynamics of a \dword{dpmod}. Within the chimney, the cards are actively cooled to a temperature of about \SI{110}{K} and isolated with respect to the \lar vessel by the cold flange \fdth{}.  The bottom side of the cold flange is connected to the \dword{crp} via short flat cables (of length \SI{0.5}{m}) in order to minimize the input capacitance to the preamplifiers. Each \dword{sftchimney} collects \num{640} readout channels. 


The \dword{fe} cards are mounted on \SI{2}{m} long blades that slide on lateral guides that are integrated into the mechanical structure of the chimney, allowing access to and replacement of the analog \dword{fe} electronics from the outside without contaminating the \lar volume. 
Swapping the \dword{fe} electronics requires opening a flange at the top of the \dword{sftchimney} and feeding the \dword{sftchimney} with nitrogen in slight over-pressure with respect to the atmospheric pressure.  This operation quite rapid and straightforward, and can be performed while the \dword{detmodule} is taking data. The chimney act also as Faraday cages, thereby entirely decoupling the analog \dword{fe} electronics 
from possible noise pickup from the digital electronics.   


The digital electronics for the charge digitization system, also resulting from a long R\&D process started in 2006, is located 
on the roof the cryostat at room temperature. This makes it possible to use the low-cost, high-speed  networking technologies used in the telecommunication industries, such as \dword{utca}. 
Digitization cards in the \dword{amc} format read \num{64} channels per card. Each \dword{amc} card can digitize all  \num{64} channels at \SI{2.5}{MHz} and compress and transmit this continuous data stream, without zero skipping, over a network link operating at \SI{10}{Gbit/s}. Lossless data compression is particularly effective thanks to the high S/N ratio
 of \dual{}, which limits noise contributions at the level of one \dword{adc} count. Each \dword{sftchimney} is coupled to a \dword{utca} crate housing in \num{10} \dword{amc} digitization cards in order to read  \num{640} channels and transmit the data via the \dword{mch} switch via a \SI{10}{Gbit/s} optical link connected to the \dword{daq} back-end. It requires \num{240} \dword{utca} crates to read the entire \dword{detmodule}. The \dword{sftchimney} warm flange is used to connect the analog differential signals, via shielded VHDCI cables, to the \dword{amc} digitization cards and also to distribute the \dword{lv} and slow control signals to the analog \dword{fe} electronics.  

The light-readout digitization system is also based on \dword{utca} \dword{amc} cards derived from the design of the charge readout and hosting a specific circuitry, based on the CATIROC \dword{asic} for the light-readout triggering. By assuming a \dword{pmt} channel density similar to that in \dword{pddp}, five \dword{utca} crates are sufficient to read \dpnumpmtch \dwords{pmt}.

The timing synchronization is based on the \dword{wr} standard. Specifically developed timing \dword{mch} connected to a \dword{wr} network ensure the distribution of clock, absolute timing, and trigger information on the backplane of the \dword{utca} crates. The \dword{wrmch} are connected via \SI{1}{Gbit/s} optical fibers to a system of \dword{wr} switches which interconnect the \dword{wr} network. This ensures that the digitization performed by the various \dword{amc} cards is completely aligned and it also refers to the absolute UTC time. 
The \dword{wrgm} switch is connected to a GPS disciplined oscillator unit providing absolute time and the clock frequency reference to the system. The timing system includes \num{16} \dword{wr} switches and \num{240} (charge readout) + \num{5} (light readout) \dword{mch} units.    


The entire readout electronics system has been optimized to support the \dual features including: charge signal dynamics, readout organization via the chimney, the number of readout channels, the high \dword{s/n} ratio, and the possibility of performing continuous data streaming with zero losses to the \dword{daq} back-end. 


Employing cost-effective technologies and performing the corresponding R\&D needed to fully exploit these technologies are strategies that are in place since the beginning of the R\&D period in order reduce and optimize costs.   
This optimization adds to the fact that the number of readout channels is naturally lower for a \dword{dpmod} thanks to the long projective geometry: \dpnumcrpch channels for a DP module with 3 mm readout pitch to be compared to \spnumch channels for a \single module with 5 mm readout pitch. %


\subsection{Cathode, Field Cage and HV System}
\label{v4:fddp-ov:cathode}

The \dword{hv} system is designed by a common \single{}-\dual consortium.
The drift field (E ${\simeq}$ \SI{0.5}{kV/cm}) inside the fully active \lar volume is produced by applying \dword{hv} to the cathode plane at the bottom of the cryostat and is kept uniform by the \dword{fc}, a stack of \num{60} equally spaced field-shaping electrodes,  polarized at linearly decreasing voltages from the cathode  voltage to almost ground potential, reached at the level of the \dword{crp}. The electrodes are rectangles made of extruded aluminum profiles (vertical pitch \SI{60.6}{mm}) with rounded corners running horizontally (and stacked vertically) around the active volume. The aluminum profiles are supported and insulated by FRP supporting beams with a pattern of slots where the aluminum profiles can be mounted. Similarly as in \dword{pddp} the profiles are arranged in modules of \SI{3}{m} width including two FRP supporting columns. These modules are chained together and  are hanging from the cryostat roof. A chain of \num{6} modules covers the \dpmaxdrift drift. The aluminum profiles of the different modules are joined together with short clipping profiles in order to ensure the electrical continuity over the
\dptpclen long horizontal rings. 


The drift cage design shares common structure elements (aluminum profiles and FRP supporting beams) as the \single \dword{fc} design but with a different arrangement (vertically hung structure) in order to cope with the \dword{dpmod} geometry.

The cathode structure, constructed of a reinforced frame to guarantee its planarity, is suspended from the \dword{fc} and hangs near the 
bottom of the cryostat. It is a segmented structure of tubes of different sizes  arranged in a grid to minimize weight, limit sagging and avoid high \efield
regions in its proximity.  The segmented structure allows scintillation light to pass through and be detected by uniform arrays of \dwords{pmt} mounted \SI{1}{m} below it at the bottom of the tank. As in \dword{pddp}, the cathode is made out of modules \SI{3}{m} in size in order to allow for transportation and underground installation. 

\subsection{Photon Detection System}
\label{v4:fddp-ov:pd}

The \dword{pds} is based on an array of \dwords{pmt} uniformly distributed below the cathode. Assuming a similar channel density as in \dword{pddp}, this translates to \dpnumpmtch channels. The \dwords{pmt} have a \dword{tpb} coating on the photocathode's external glass surface that shifts the scintillation light from deep UV to visible light. The \dwords{pmt}  sit on the corrugated membrane cryostat floor, on 
mechanical supports that do not interfere with the membrane thermal contraction. 
A single cable provides both \dword{hv} and signal transmission by way of a positively biased base circuitry, thus reducing the required number of \fdth{} channels. Optical fibers provide the calibration system.

\subsection{Data Acquisition}
\label{v4:fddp-ov:daq}

The Ethernet-based \dword{daq} back-end system is designed by a joint \single{}-\dual consortium. It connects to 
\SI{10}{Gbits/s} optical links that provide continuous, lossless, compressed data streaming from the \dword{utca} crates, and it has the task of determining the trigger conditions for \textit{interesting} events (beam, cosmics, \dword{snb} neutrino interactions). This system also manages the recording of data to disk. 
The system can exploit the high \dword{s/n} ratio peculiar to the \dual design, the availability of the entire data stream without losses, and the possibility of going to lower detection thresholds for \dword{snb} events.

 It  is assumed that this \dword{daq} back-end system will be composed of a set of event-building and trigger machines, high-performance network elements, and a high-bandwidth distributed storage system based on an array of storage servers operating in parallel. In particular, the \dword{daq} system is expected to:

\begin{itemize}
\item Collect the high-bandwidth data volume coming from the data links of the \dword{fe} digitization crates; 
\item Put together the data streams from different crates in Regions Of Interest (ROI) or over the entire detector volume (A ROI is typically the size of a \dword{pddp} four-\dword{crp} surface, since events are contained in such a region.);
\item Process this data flow by an online trigger farm 
as a prelude to selecting relevant events to be recorded on disk (both neutrino beam and off-beam events);
\item Produce charge-readout triggers independently of the light-readout triggers and beam-spill information 
(In particular for \dword{snb} events, the trigger farm would issue triggers over a sliding timing window of about \SI{10}{s}  based on the presence of low-energy depositions; the entire content would be dumped to disk.).
\end{itemize}


The \dual readout architecture can be organized into \num{20} \dwords{roi}, each similar to the \dword{pddp} back-end architecture. Triggers are searched on the level-1 event builder machines, interconnecting multiple \dword{utca} crates, on a sliding windows of \SI{10}{s}  contained in the event builder RAM. 


The event builders combine the continuous lossless, compressed data (streaming from the charge readout) with beam data and light data in order to define the window $t_0$ and select disk streams from beam events, cosmics and \dword{snb} events. The data decompression is necessary on the event builders in order to perform the charge data analysis for the triggers definition. Compressed data are kept in memory, while the trigger definition analysis is performed, for further writing on disk from level-2 machines from the output streams: beam, cosmics and proton decay, and \dword{snb}. 
The level-1 event builders exchange trigger primitive data on the network with a global supervisor machine, which then decides what data to write onto disk.  The supervisor can order the dump onto disk of the event builder's memory  windows if a certain number of candidate energy depositions is found from the charge data. 
This scheme makes it possible to put selected portions of different \dwords{detmodule} in communication with one another.  
Typically for beam data and cosmic events, the amount of data written to disk can be limited to one or two \dwords{roi}; there are cases in which in case events occur at the border between two \dwords{roi} rather than in a single one.

\subsection{Cryogenics Instrumentation and Slow Controls}
\label{v4:fddp-ov:sc}
The \dword{cisc} system is designed by a joint \single{}-\dual consortium. 
This system controls the 
following items:

\begin{itemize}
\item Cryogenic instrumentation: measurements of temperatures (gas and liquid), pressure (gas), liquid level, purity monitors;
\item \dword{crp} instrumentation: temperatures, pulsing system, precision level meters, readout of \dword{crp} stepping motors;
\item Generation and control of \dword{hv} biasing of \dword{lem} and grid;
\item Generation and control of \dword{hv} biasing for the \dwords{pmt} and calibration (via optical pulsing);
\item Slow control of the \dword{utca} crates, analog \dword{fe}\dword{lv} control, charge injection control to preamplifiers;
\item Control of the cathode \dword{hv} biasing system;
\item Alignment survey of \dword{crp} position (via external reference points);
\item Control of the laser system;
\item Analysis of \lar purity and \dword{lem}  gain calibration.
\end{itemize}

\cleardoublepage

\chapter{Charge Readout Planes}
\label{ch:fddp-CRP}

\section{Charge Readout Planes (CRP) Overview}
\label{sec:fddp-crp-ov}

\subsection{Introduction}
\label{sec:fddp-crp-intro}

In the \dual \lartpc concept, the ionization electrons are multiplied in avalanches  occurring inside micro-pattern detectors, the \dfirsts{lem}, located in the argon gas phase above the \lar 
surface. The drift field of the TPC brings the electrons up to the \lar surface where they can  be    extracted into the gas using a 
\SI{2}{kV/cm} \efield defined across the liquid-gas interface.

This extraction field is defined by the potentials applied to submersed extraction grid (stainless steel wires tensioned in both $x$ and $y$ directions) and to the bottom side of the \dwords{lem}. The \dwords{lem} are printed circuit boards oriented horizontally, with conductive layers (electrodes) on the top and bottom surfaces, and many holes drilled through.  The holes form a micro-pattern structure within which the amplification occurs given the presence of a strong \efield.

By applying voltages across the two electrodes of the \dword{lem}, an \efield region (up to \SI{35}{kV/cm}) is defined in the holes, which produces an electronic signal gain exceeding \num{20} after the initial phase of charging-up the \dword{lem} dielectric material.  Electrons transiting these high \efield regions in the holes trigger Townsend multiplication in the pure argon gas.

The amplified charge is then collected and recorded on a \twod anode consisting of two sets of \dpstrippitch-pitch gold-plated copper strips that provide the $x$ and $y$ coordinates (and thus two orthogonal views) of the event. The strips are defined by a pattern of tracks on the bottom face of the anode printed circuit board. It is possible  to define two electrically insulated views of strips crossing each other orthogonally by ensuring continuity of the tracks with a set of vias and tracks extending to the top face of the anode printed circuit board.

Typical \efield{}s between each stage of the readout are
illustrated in Figure~\ref{fig:setup}. Table~\ref{tab:crp_dist} shows the inter-stage distance and the tolerances required to obtain uniformity of gain to within $\sim$5\%.

\begin{dunefigure}[Dual-phase readout]{fig:setup}
{Illustration of the \efield{}s in the amplification region of a \dual \lartpc. The simulated field lines in dark blue indicate the paths followed by the drifting charges (without diffusion).}
\includegraphics[width=.85\textwidth]{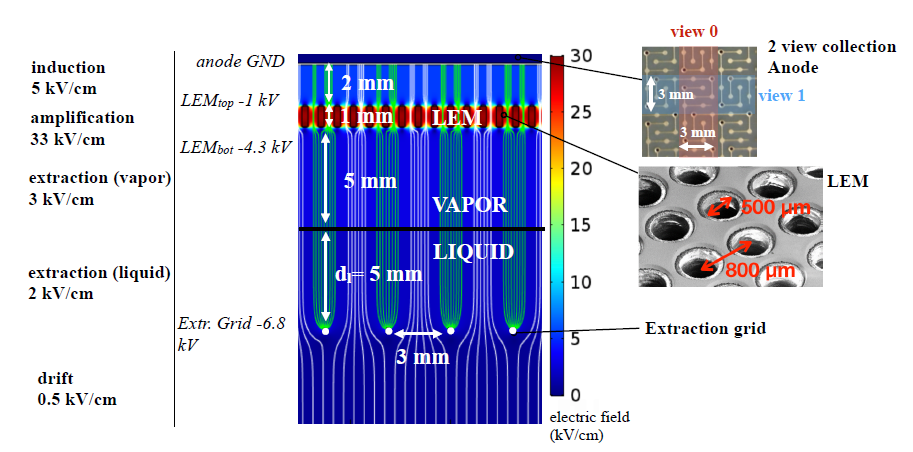}  
\end{dunefigure}
\begin{dunetable}[Interstage distances and \efield settings of the \dual readout components]{lp{2cm}p{2cm}l}{tab:crp_dist}{Interstage distances and \efield settings of the \dual readout components.} 
 Component & Distance [mm] & Tolerance [mm] & \efield [kV/cm]  \\ \toprowrule
 Anode-\dword{lem} top electrode  & \num{2} & \num{0.1} & \num{5}\\ \colhline
 \dword{lem} top-bottom electrode   & \num{1} & \num{0.01} & \numrange{30}{35}\\ \colhline
 \dword{lem} bottom electrode-grid        &\num{10} & \num{1} & \num{2} (in \lar) and \num{3} (in gaseous argon)\\
 \end{dunetable}

The extraction grid, \dword{lem} and anode are assembled into a three-layered unit, called a \dword{las}, with precisely defined inter-stage distances and inter-alignment. The \dwords{las}    are then connected together horizontally into modular units of area \SI{9}{m$^2$}. These units are called \dwords{crp}. Figure~\ref{fig:figure-label-crp} shows an 
engineering view of one \dword{crp} fully assembled.

\begin{dunefigure}
[View of a complete  \SI{9}{m$^{2}$} \dword{crp} module]
{fig:figure-label-crp}
{View of a complete  \SI{9}{m$^{2}$} \dword{crp} module.}
\includegraphics[width=0.75\textwidth]{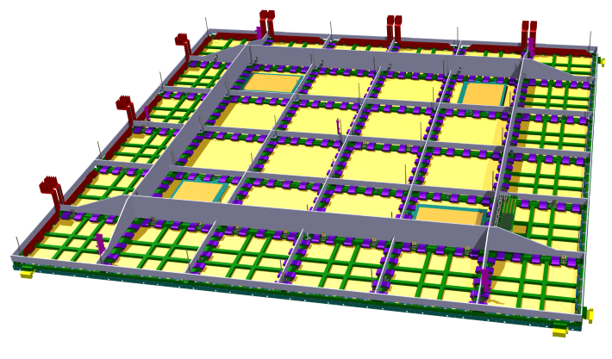}
\end{dunefigure}

The \dword{crp} mechanical structure provides the integration of the \dwords{las} over a large area by minimizing the dead spaces. The planarity of the active surface must be guaranteed despite possible sagging effects with respect to the three hanging points, differential thermal contraction effects on the various  components and the presence of a temperature gradient in the gas phase, which could induce different thermal contractions as a function of the distance from the liquid surface. In order to 
ensure the planarity, the design 
incorporates a supporting frame in Invar, which 
can provide a stiff supporting structure, extending vertically into the gas phase, with little sagging and 
very minor contraction effects. The \dword{lem} and anodes, which would be affected by a significant thermal contraction, are integrated into a G10 planar structure with similar contraction 
properties. The G10 structure is mechanically decoupled on the horizontal plane with respect to the Invar structure  and it is free to slide during thermal contraction. 

\subsection{Design Considerations}
\label{sec:fddp-crp-des-consid}

Each \dword{crp} is an independent detector element that performs charge
extraction, multiplication and collection, and has its own \dword{hv} system and independent signal \fdth{}s. The entire area of the \dword{lem} and anode in a \dword{crp} is active. The position and the parallelism with respect to the \lar surface can be individually adjusted for each \dword{crp}.

The \dword{lem} and corresponding anode are mounted into 
\num{50}$\times$\SI{50}{cm$^2$}  
\dwords{las} before being assembled with an extraction grid into a \dword{crp}. Each anode in a \dword{las} is segmented in \SI{50}{cm} long $x$ and $y$ strips . Adjacent \dword{las} anodes are bridged together to form readout strips of the required length by connecting short flat cables to KEL\footnote{KEL 8925E-068-1795 .27mm Pitch, 2 piece, IDC for 0.635mm FC Connector, Low profile type, KEL Corporation\texttrademark{}, \url{https://www.kel.jp/english/product/}.} connectors soldered onto the top sides of the anodes. The signals from the last anode in each  strip chain are brought to \fdth{}s mounted on the other side of the front-end electronics embedded inside dedicated signal-\fdth chimneys using \SI{50}{cm}-long flat cables.

Each \dword{crp} is independently hung from the vessel deck through its three
suspension \fdth{}s. It has its own \dword{hv} system and  independent signal and slow-control \fdth{}s.

Table~\ref{tab:crpphysicsparams} summarizes the set of requirements and parameters that drive the \dwords{crp} design. 

\begin{dunetable}
[Important parameters for the \dword{crp} system design]
{p{0.7\textwidth}r}
{tab:crpphysicsparams}
{Important parameters for the \dword{crp} system design}   
Parameter & Value \\ \toprowrule
 Planarity tolerance on the detection plane over \num{3}$\times$\SI{3}{m$^{2}$} & $\pm$\SI{0.5}{mm} \\ \colhline
 \dword{crp} vertical positioning precision & $\pm$\SI{0.5}{mm} \\ \colhline
 Range of vertical displacement& $\pm$\SI{20}{mm}\\ \colhline
 Lateral inter-\dword{crp} dead space & < \SI{10}{mm} \\\colhline
 Distance between the extraction grid wires and the \dword{lem} plane & 10mm\\ \colhline 
 \dword{hv} of the extraction grid wires in \lar &  \SI{6.8}{kV} \\ \colhline
 \dword{hv} of the \dword{lem} down surface in gas argon & \SI{4.5}{kV}\\ \colhline
 \dword{hv} of the \dword{lem} up surface in gas argon & \SI{1}{kV}\\ 
\end{dunetable}

The \dword{crp} design foreseen for the \dword{dpmod} is based on the one developed for \dword{pddp}.

\subsection{Scope}
\label{sec:fddp-crp-scope}

The scope of the \dwords{crp} includes the continued procurement of materials for, and the fabrication, testing, delivery and installation of the following systems: 

\begin{itemize}
\item  Production and \dword{qa} of the \dword{lem} and anodes;
\item  Production of the G10 and Invar frames;
\item Production of the suspension \fdth{}s and motorization;
\item Production of the extraction grid elements;
\item Production of the \dword{hv} distribution system associated with the \dword{crp} for applying voltages to the \dword{lem} and anodes;
\item Production of the temperature probe system associated with the \dword{crp};
\item Production of the level meter system associated with the \dword{crp};
\item Production of the strip-pulsing system  
associated with the \dwords{crp};
\item Production of the transportation and storage boxes for the \dwords{crp};
\item Assembly of the \dwords{las};
\item Assembly of the \dword{crp} structures, \dword{las}  grid elements, \dword{hv}, slow controls and cabling for each \dword{crp};
\item Testing of the assembled \dwords{crp};
\item Installation of the produced \dwords{crp} into the storage boxes;
\item Installation of the \dwords{crp} into the transportation boxes and delivery to \surf{}; 
\item Installation, cabling and test of the \dwords{crp} in the cryostat.
\end{itemize}

\section{CRP Design}
\label{sec:fddp-crp-design}

The complete \dword{crp} system includes, as illustrated in Figure~\ref{fig:figure-label-crp2}:
\begin{itemize}
\item mechanical frames,
\item the detection plane made of \dwords{lem} and anodes,
\item the extraction grid,
\item instrumentation devices: level meters, distance meters, anf temperature probes,
\item internal cabling: to patch panels (\dword{lem} \dword{hv}, slow control instruments),
\item suspension and control system.
\end{itemize}

\begin{dunefigure}[Main components of a \dword{crp} module of  \SI{9}{m$^{2}$}]{fig:figure-label-crp2}
{Main components of a \dword{crp} module of  \SI{9}{m$^{2}$}}
\includegraphics[width=0.8\textwidth]{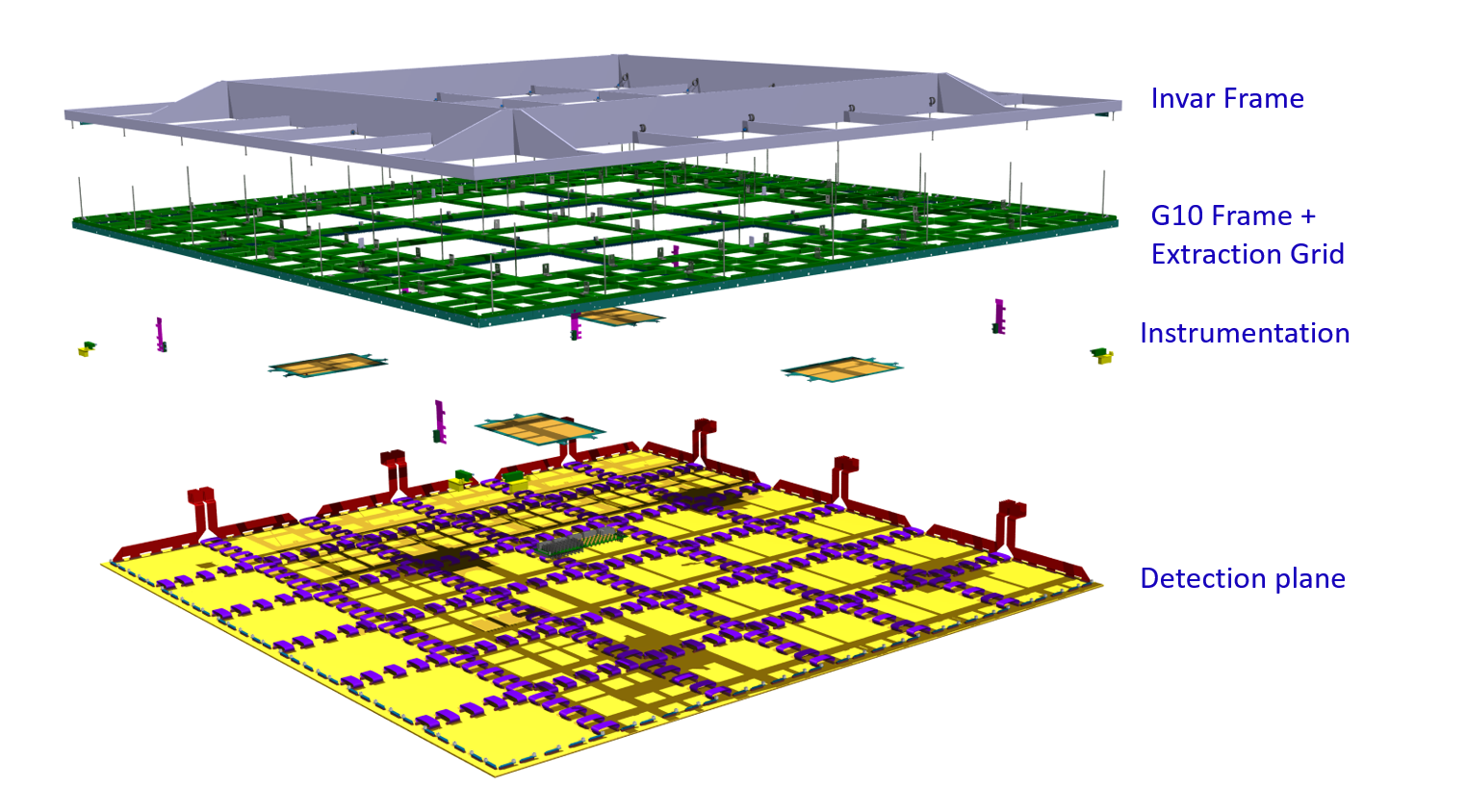}
\end{dunefigure}

\subsection{Mechanical structure}
\label{sec:fddp-crp-mechanics}
\subsubsection{Invar Frame}

The main mechanical supporting structure of the \dword{crp} is made of Invar nickel-iron alloy. This material is chosen for its low coefficient of thermal expansion leading to a very small deformation at cold temperatures, especially in the gas argon where the temperature gradient in the cryostat may be of the order of a few \si{K/cm}.
The structure is made of a grid of soldered Invar beams \SI{3000}{mm} long and \SI{6.5}{mm} wide. The four main ones have a height of \SI{150}{mm} while the internal ones are \SI{40}{mm} high, as are the four surrounding plates. The heights have been optimized to keep the stiffness as needed and to reduce the total frame weight to  
 about \SI{112}{kg}.
Figure~\ref{fig:invarframe} shows the soldering of one of the first \dword{crp} frame for \dword{pddp}.

\begin{dunefigure}[First \dword{crp} Invar frame under construction for \dword{pddp}]{fig:invarframe}
{First \dword{crp} Invar frame under construction for \dword{pddp}}
\includegraphics[width=0.95\textwidth]{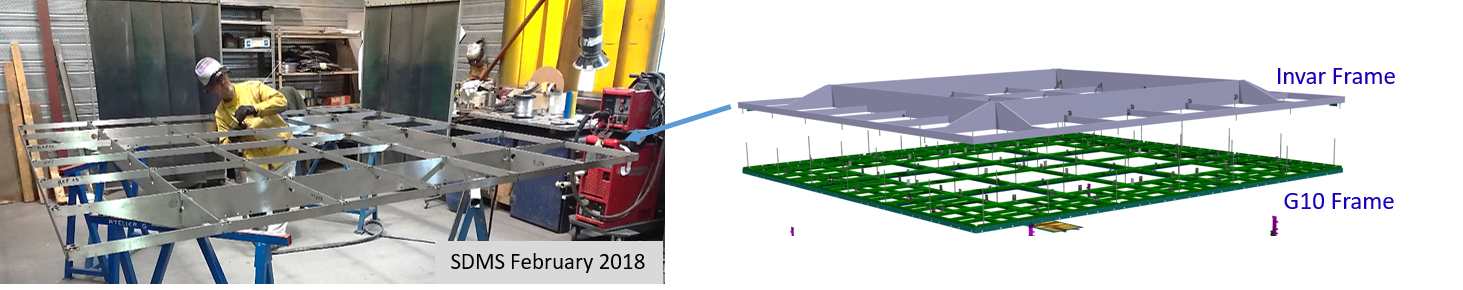}
\end{dunefigure}

\subsubsection{G10 Frames and Modules}
\label{sec:invar-frame}

The \num{3}$\times$\SI{3}{m$^{2}$}  G10 fiberglass structure used to attach the \num{36} \dwords{lem} and anodes, the extraction grid and the level meters is composed of an assembly of nine \num{1}$\times$\SI{1}{m$^{2}$} subframes. The choice of G10 is driven by the need to match the \dword{lem} and anode thermo-mechanical behavior and avoid over-stress due to differential thermal contraction. 
Figure~\ref{fig:crp-g10} shows the pattern of the nine G10 parts composing a full \dword{crp} frame as well as the supporting comb positioned at every meter, and the extraction grid support plates along the side of the \dword{crp}.

\begin{dunefigure}[G10 elements of a full \dword{crp} module]{fig:crp-g10}
{G10 elements of a full \dword{crp} module}
\includegraphics[width=0.8\textwidth]{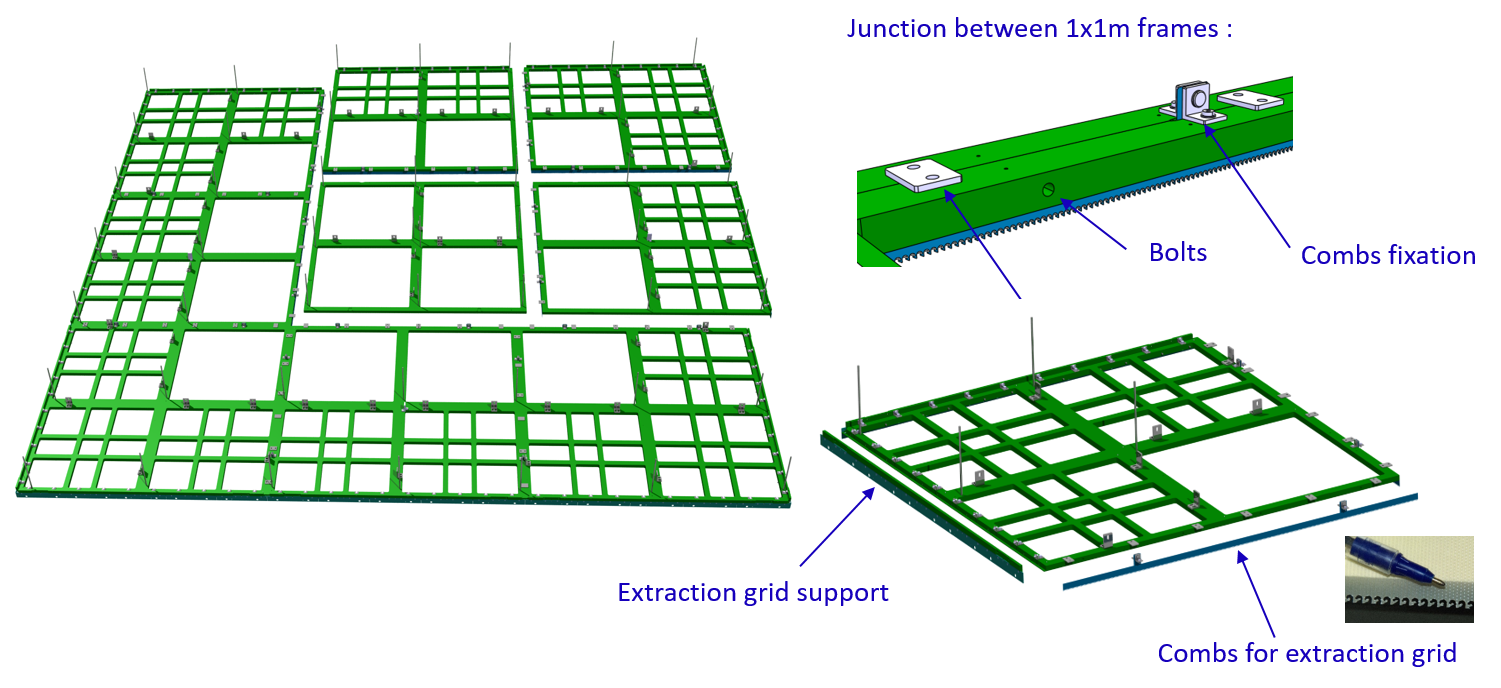}
\end{dunefigure}

Since G10 is a composite material created by stacking multiple layers of glass fibers,  the orientation of the stacking has to be taken into account prior to assembly.
At the construction level, the fiber directions are matched between the different subframes to ensure harmony in thermal shrinkage. Three different patterns have to be produced, one for the four angles, one for the four face centers and one for the central subframe.
Two versions of each pattern for the supporting extraction grid bars and the combs follow same rule.

The subframes have been designed to guarantee 
adequate stiffness in the  regions subject to larger tension while minimizing material in order to reduce the weight.
The G10 structure is \SI{15}{mm} thick and weights about \SI{68}{kg}. 

\subsubsection{Decoupling system}
During cooling, Invar's dimensions remain nearly unchanged whereas the G10 frame and \dwords{lem}-anodes contract in similar ways. Thermal decoupling allows a lateral sliding of the G10 frame, without changing the level. 
Dedicated decoupling systems are installed at each corner of the Invar frame (\num{50} systems by  \num{3}$\times$\SI{3}{m$^{2}$} module). One decoupling system that allows the G10 and \dword{lem}-anode elements to slide is shown in  Figure~\ref{fig:crp-decoupling}.

\begin{dunefigure}[Decoupling system attached to the Invar frame]{fig:crp-decoupling}
{Decoupling system attached to the Invar frame, with detailed view. Example of one of the systems built for \dword{pddp}.}
\includegraphics[width=0.8\textwidth]{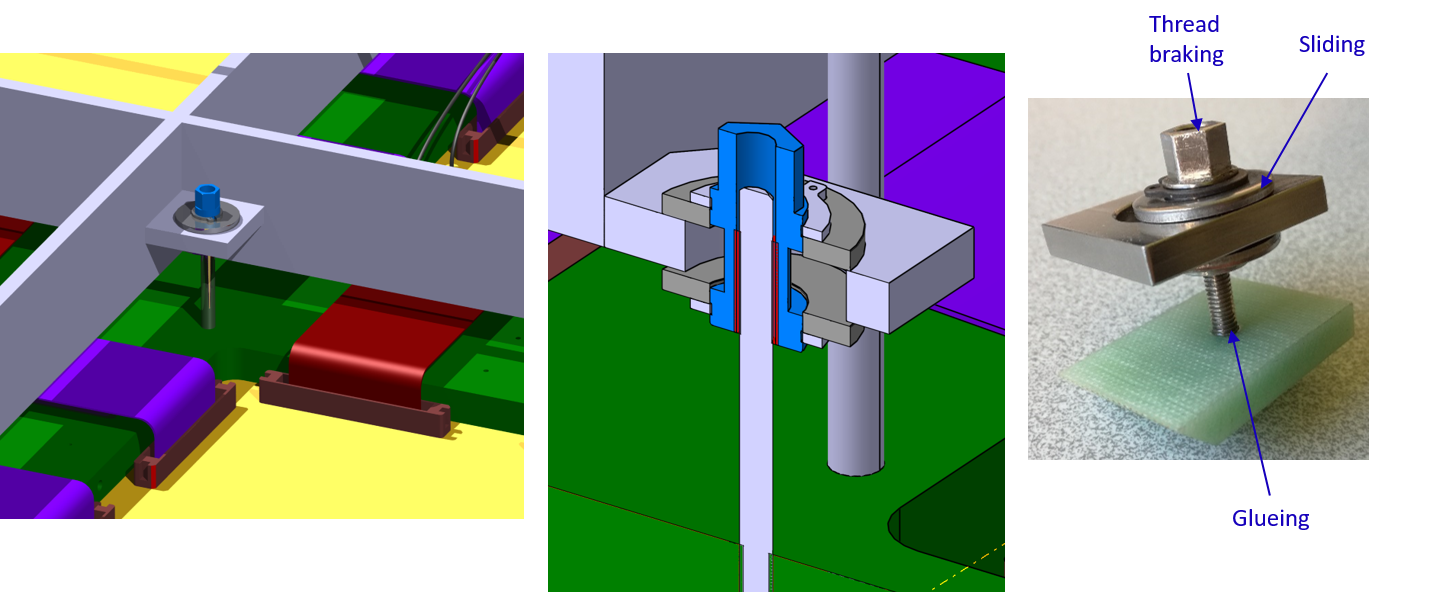}
\end{dunefigure}

The full weight of a \dword{crp} including the mass of \dwords{lem} and anodes is about \SI{330}{kg}.

\subsection{Extraction Grid}
\label{sec:fddp-crp-grid}
The extraction grid consists of \SI{100}{\micro\meter}-diameter stainless steel wires tensioned in both $x$ and $y$ directions over the entire \SI{3}{m} length and width of the \dword{crp} with \dpstrippitch pitch. They are soldered into groups of \num{64} on independent wire-tensioning pads oriented perpendicularly to the side of the \dword{crp} frame. Each wire-tensioning pad consists of a printed circuit board (PCB)  that is fixed very precisely to mechanical support beams screwed to the G10 frame of the \dword{crp}, as shown in Figure~\ref{fig:grid-parts}.
 
\begin{dunefigure}[Extraction grid components on the \dword{crp} structure]{fig:grid-parts}
{Extraction grid components on the \dword{crp} structure. On the left are the PCB plates and their supporting bars. On the right is the 
wire tensioning system with the pushing screws and calibrated wedges to keep the right distance.}
\includegraphics[width=0.8\textwidth]{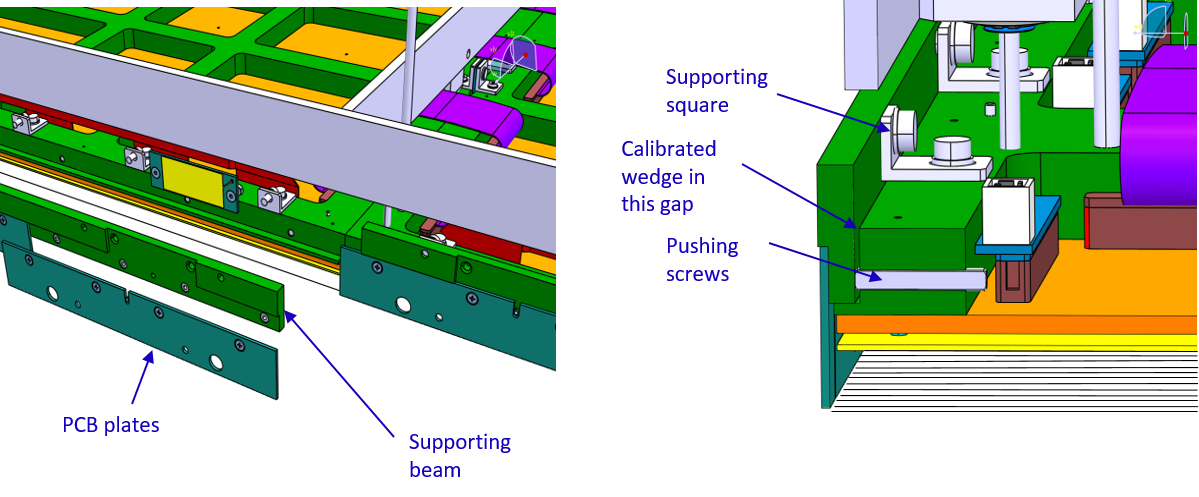}
\end{dunefigure}

The PCB has \num{64} soldering pads with \SI{200}{\micro\meter} grooves for precise positioning of the wires. During the 
soldering process each wire is tensioned and positioned in a groove. The PCB is then mounted on the G10 supporting bars and the tension of the group of \num{64} wires can be precisely adjusted by pushing the supporting bars against the \dword{crp}s G10 frame with screws. The tensioning is performed by tightening \textit{pushing screws}, adding a calibrated wedge, and locking the supporting square.
Supporting comb-teeth blades that are inserted between the \SI{1}{m$^2$} G10 subframes in both $x$ and $y$ restrict the sag to \SI{0.1}{mm} in the wires, which are \SI{3}{m} long in both directions. The array of blades penetrates the liquid surface and has the additional benefit of acting as a baffle against potential surface waves in the \lar.

The grid-\dword{hv} connection for a \dword{crp} is made through a varnished copper track into two of the \dword{crp}'s \num{60} PCB plates; a special isolated connection 
is made inside the \dword{crp} structure.

\subsection{Large Electron Multiplier (LEM)}
\label{sec:fddp-crp-lem}

Each \dword{lem} consists of a \SI{1}{mm}-thick,  \num{50}$\times$\SI{50}{cm$^{2}$} copper-clad standard PCB epoxy plate. Holes of \SI{500}{\micro\meter} diameter, through which electrons undergo amplification, are mechanically drilled in a hexagonal pattern with a pitch of \SI{800}{\micro\meter}, yielding about \num{180} holes per \si{cm$^2$}. In addition, each hole has a  \SI{40}{\micro\meter} dielectric rim, obtained by a chemical process, to prevent electrical discharges from occurring near the holes. The holes provide confinement for the UV photons produced during the avalanche process and thus act as a mechanical quencher to prevent photon feedback. This property makes the \dword{lem} suitable for operation in ultra-pure argon vapor without the addition of a quenching gas. The final gold-plated copper thickness of each \dword{lem} electrode is about  \SI{60}{\micro\meter}. Twenty peripheral and nine central \SI{2.2}{mm}-diameter holes are used for assembling the \dword{lem} module together with its anode on the G10 frame. Figure~\ref{fig:LEM_CFR-35} shows a picture of a \dword{lem} module used for  \dword{pddp}. To prevent \dword{hv} discharges near or across the edges, the \dword{lem} module has a  \SI{10}{mm} border free from metallization and another \SI{5}{mm} copper guard ring. Similar copper guard rings are located around the \SI{2.2}{mm} diameter holes and the \dword{hv} connectors. The latter are made with \SI{1.2}{mm} diameter male pins (Deutsch\footnote{Deutsch\texttrademark{} \url{http://www.deutsch.net/}.} 6860-201-22278.) that are soldered onto specifically designed pads imprinted on the \dword{lem} electrodes. The pins are insulated with circular tubes made in MACOR and a  \SI{10}{mm} circular clearance around each pin. 

The total active area of a \dword{lem} module used for  \dword{pddp} represents about \SI{86}{\%} of the \num{50}$\times$\SI{50}{cm$^{2}$} area. The choice of the \dword{lem} design for  \dword{pddp} was made in order to achieve stable operation conditions up to \SI{3.5}{kV} in \dual \lar (D\lar{}) mode, corresponding to amplification gains larger than \num{30} (\num{100}) after (before) charging-up of the \dword{lem} dielectric material. For the \dword{dpmod}, a further optimization of the current \dword{lem} design will  be performed in order  to find the best compromise between the detector active area and the operation stability.

\begin{dunefigure}
[Picture of a \dword{lem} module used for  \dword{pddp}]
{fig:LEM_CFR-35}
{Picture of a \dword{lem} module used for  \dword{pddp}}
\includegraphics[width=.8\textwidth]{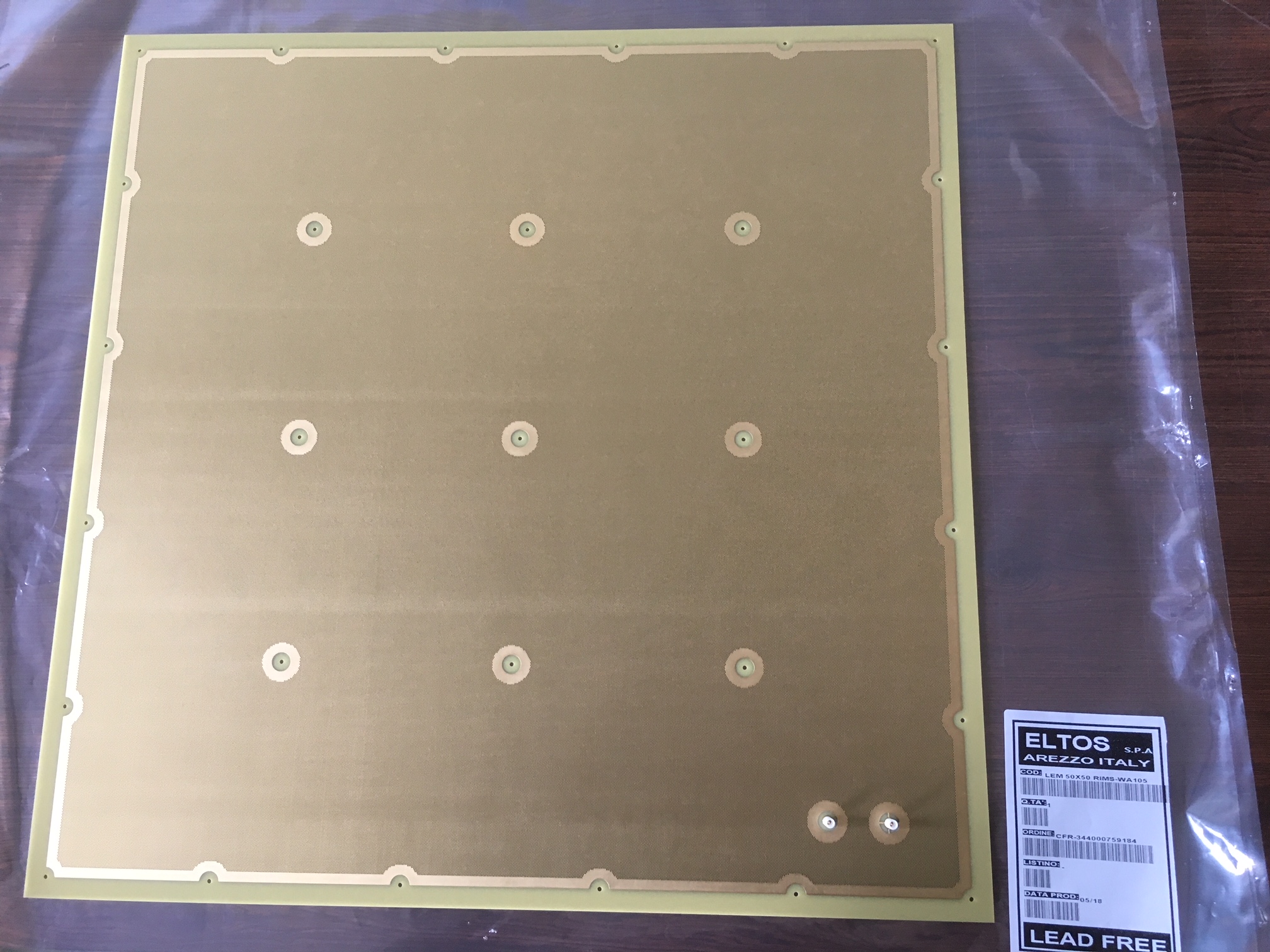}
\end{dunefigure}

\subsection{Anode}
\label{sec:fddp-crp-anode}
 
The anode is a four-layer PCB having a set of orthogonal strips with a \dpstrippitch pitch that provide the two views of the collected charge. The area of the anode is the same as that of the \dword{lem},  \num{50}$\times$\SI{50}{cm$^2$}. Twenty nine holes of \SI{2.2}{mm} diameter matching the \dword{lem} holes are used for the \dword{lem} and anode assembly on their G10 frame. The \SI{2}{mm} distance between the anode and the \dword{lem} is ensured by \num{29} precisely machined spacers made of polyetheretherketone (PEEK). 

The pattern of readout strips, printed on the bottom PCB layer and used for charge collection, is optimized such that the charge is evenly split between both views (Figure~\ref{fig:Anode}). Electrical insulation in the locations where orthogonal tracks would superimpose is achieved by 
using a system of vias between the top and bottom layers of the PCB to allow the tracks to cross over and under one another. 
Each strip, made of thin gold-plated copper tracks, has a capacitance per unit length to ground of about 
\SI{160}{pF/m}. The readout strips are routed to the top layer towards \num{68}-pin female connectors (KEL 8925E-068-179-F) soldered on the anode periphery. Each connector reads \num{32} strips; its \num{36} remaining pins are connected to the detector ground via a copper strip that runs around the periphery of the top layer of the anode (see Figure \ref{fig:Anode_CTest}). 

\begin{dunefigure}
[Picture of the anode symmetric \twod strip design.]
{fig:Anode}
{Picture of the anode symmetric \twod strip design.}
  \includegraphics[width=.8\textwidth]{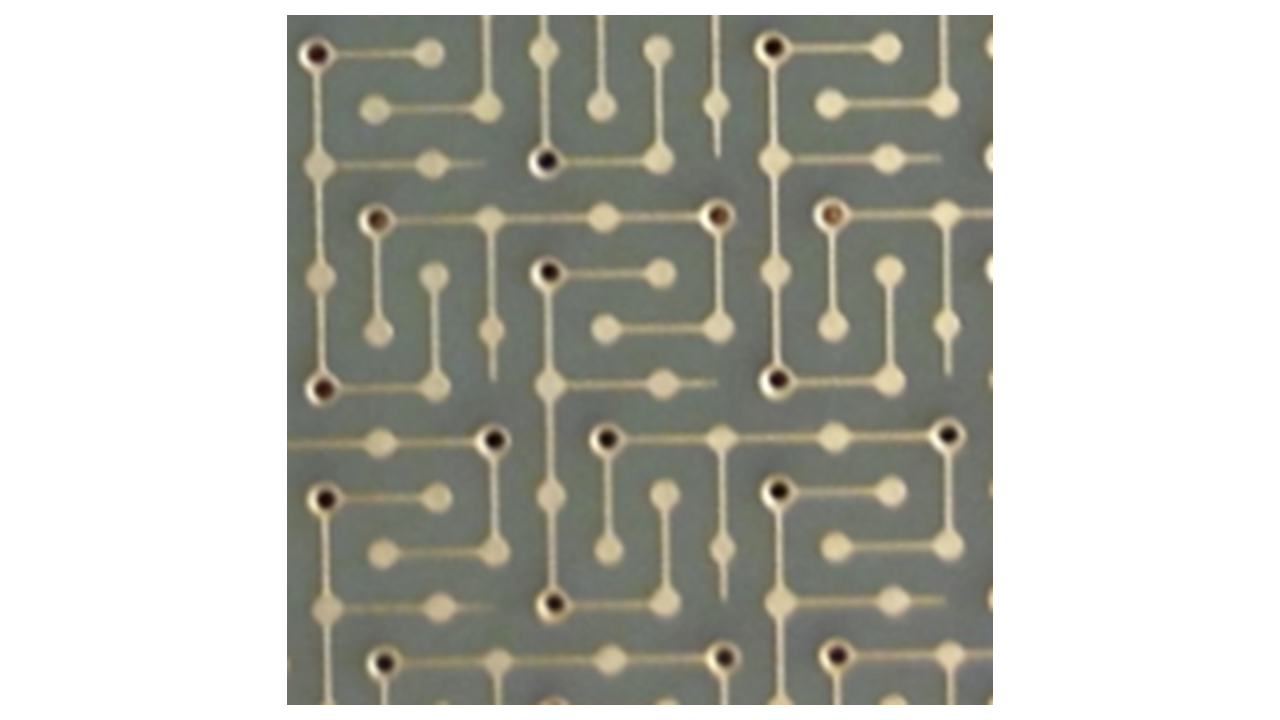}
\end{dunefigure}

\subsection{Instrumentation}
\label{sec:fddp-crp-instr}

\subsubsection{Distance meters and level meters}

The vertical and lateral positions of the \dwords{crp} are measured by different means. 
It is possible to know the vertical position of the \dword{crp} both from the suspension system and from dedicated capacitive measurements.
The  suspension system and its motorization on three points provides an accurate measurement for each \dword{crp} of order \SI{0.1}{mm} due to the accuracy of the steps motors and encoders. 
In order to exploit the information from the encoders,  the anchor points on the \dword{crp} must be surveyed during construction and the absolute positions of the suspensions must be surveyed from the roof of the cryostat during the installation of the \fdth.

There will be also the possibility to get relative measurements of the \dword{crp} position with respect to the \lar level using capacitive level meters put on the sides of the \dword{crp}s located at the periphery of the detection plane.  The  measurements of the capacitance between the \dword{lem} and the extraction grid, which depends on the height of the liquid above the grid, can also provide a relative measurement  of the \dword{crp}'s vertical position with respect to the liquid surface. This method is used for the \dword{crp}s which are not located at the periphery, while the ones at the periphery can exploit both the  \dword{lem}-grid capacitance measurements and the capacitive level meters.
 
 
The measurement of the distance between each \dword{crp} pair is performed by using capacitive devices called \textit{distance meters} made of two parallel plates. 
This measurement enables positioning of the \dwords{crp} with the correct  inter-\dword{crp} distance all along the detection plane. These devices do not require any contact among  \dword{crp}s and give an accuracy of the order of \SI{0.1}{mm} on the  inter-\dword{crp} distance.  Four devices per \dword{crp} side are embedded in the G10 frame side as can be seen on Figure~\ref{fig:distancemeter}.

\begin{dunefigure}[View of the distance meter plates on the sides of a \dword{crp}]{fig:distancemeter}
{CAD views (left, center) of the distance meter plates, represented by the yellow rectangles, on the sides of a \dword{crp}. Two different plate sizes are used in pairs, installed on facing sides of neighboring \dwords{crp}, 
to increase the measurement accuracy of the overlapping surface which translates in the capacitance to be measured. A set of five distance meter plates of the two sizes (right) received from the production factory}
\includegraphics[width=0.85\textwidth]{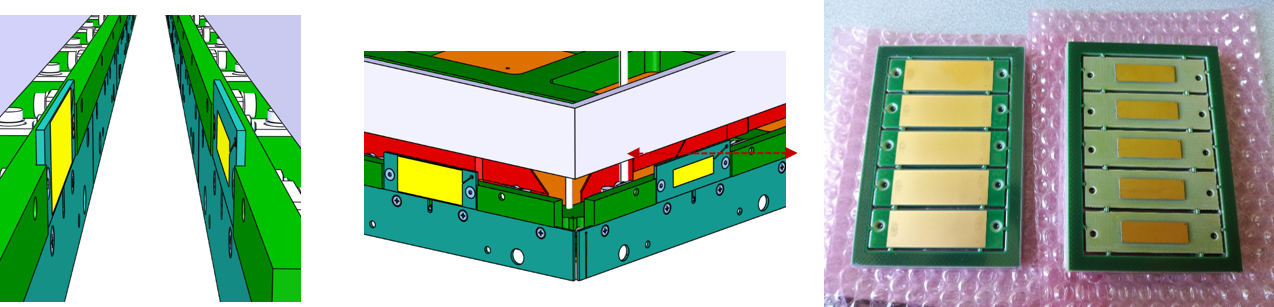}
\end{dunefigure}


\subsubsection{Thermometers}

The temperature in the gas above the anode plane is monitored at different heights with resistance thermometers (Pt sensors) soldered on  six PCB boards distributed over the full surface of the \dword{crp}, six 
sensors per PCB. The configuration and the Pt positions are shown in Figure~\ref{fig:ptsensor}.

\begin{dunefigure}[View of the thermometer board]{fig:ptsensor}
{View of the thermometer board and the positions of the Pt sensors along the PCB plate.}
\includegraphics[width=0.55\textwidth]{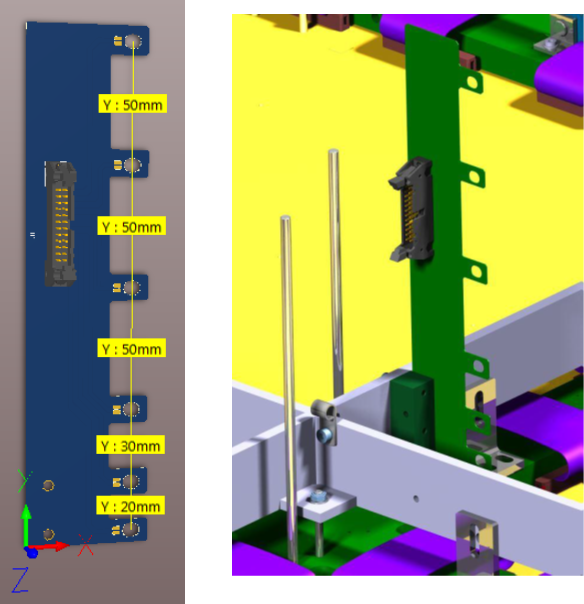}
\end{dunefigure}

\subsection{Suspension System and Drive}
\label{sec:fddp-crp-suspension}

Three suspension \fdth{}s are arranged in an equilateral triangle whose barycenter coincides with that of the \dword{crp}; an automated system is used to suspend the \dword{crp} at the required position and precisely adjust the \dword{crp} level with respect to the \lar surface.

Figure~\ref{fig:spft} shows the design of the suspension \fdth including the bellows and the motors. There are three \fdth{}s per \dword{crp}.

\begin{dunefigure}[Suspension \fdth and various assembly details]{fig:spft}
{Suspension \fdth and various assembly details.}
\includegraphics[width=0.7\textwidth]{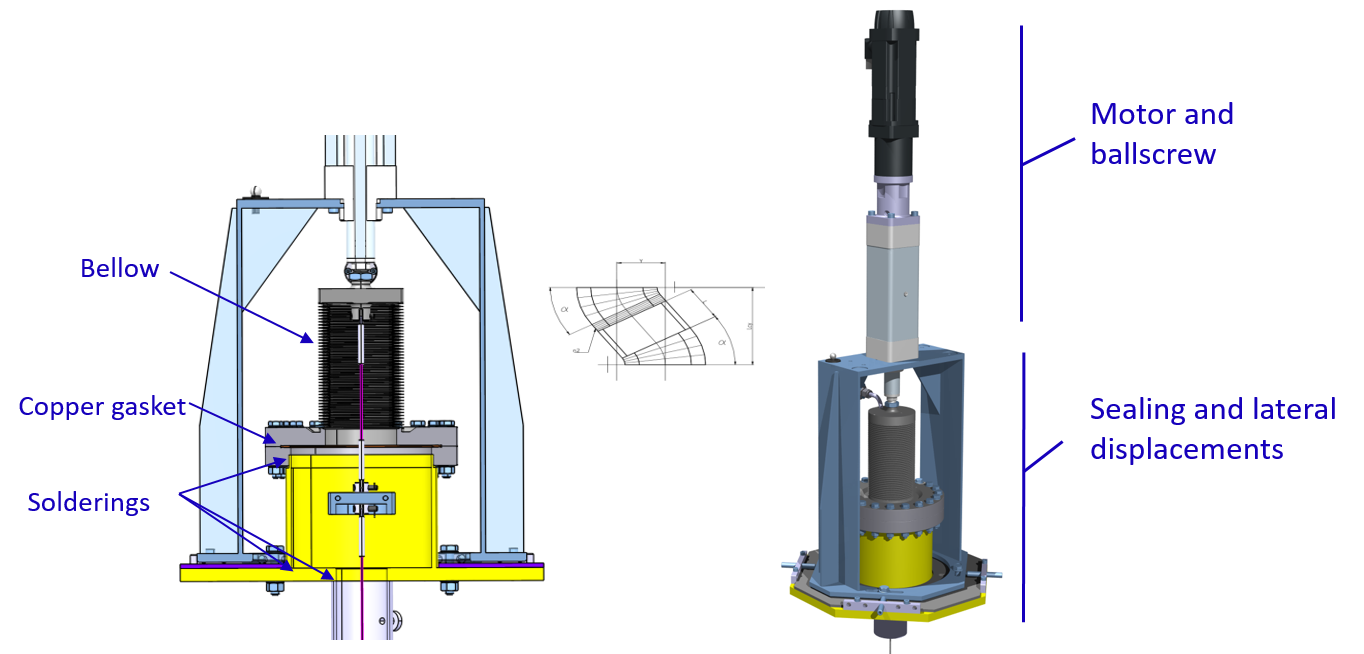}
\end{dunefigure}

At the level of the flanges on the crysotat roof the adjusting table on which the suspension \fdth is screwed allows a lateral stroke of \SI{26}{mm}. Any tranverse movement is absorbed by lateral deformation of the bellows.
The vertical stroke available with the bellows size is $\pm$\SI{20}{mm}.
The system incorporates a  mechanical stop and a simple obstruction of the chimney for maintenance purpose or bellows replacement.
At the top pf the suspension \fdth there is a special slot to position a laser tracker target
such that the \fdth position can be precisely surveyed during installation.

The suspension cables anchoring system on the \dword{crp} is shown in Figure~\ref{fig:anchor}. 
In case of variation of the cryostat pipes' verticality, this system allows to change an anchoring point on a module, in warm conditions. In cold conditions, the transverse movement is done with the suspension \fdth position adjustment.
\begin{dunefigure}[Anchoring system of the suspension cable on the \dword{crp} frame]{fig:anchor}
{Anchoring system of the suspension cable on the \dword{crp} frame.}
\includegraphics[width=0.6\textwidth]{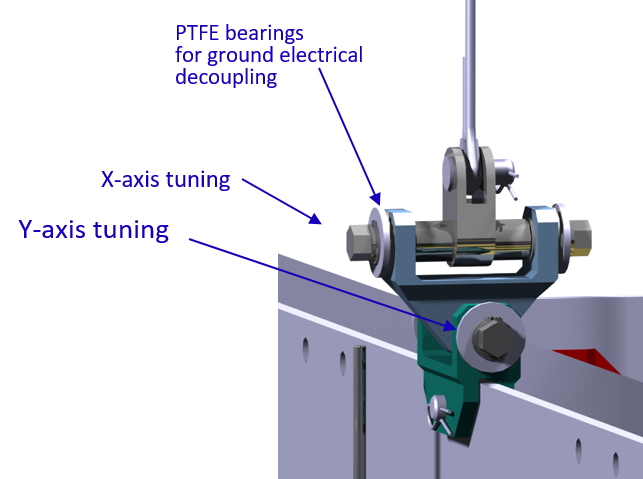}
\end{dunefigure}

Each motor has independent controls for tuning the horizontality of the plane using the  level information output from the 
sensors. In standard, stable running  conditions several \dword{crp} modules are associated together as a global system and controlled via an automatic process.



\section{Production and Assembly}
\label{sec:fddp-crp-prod-assy}

\subsection{G10 and Invar Frame Production}
\label{sec:fddp-crp-frame}

The G10 and Invar frames are both produced by industry. For the \dword{dpmod}, 
\dptotcrp Invar frames and \dpnumpmtch 
G10 subframes of \SI{1}{m$^2$} are required. 
For \dword{pddp}, the process and \dword{qc} procedures have been defined.
For the Invar structure, SDMS\footnote{SDMS\texttrademark{}, \url{www.sdms.fr/en}.}, the company chosen for \dword{pddp}'s four \dwords{crp},  
has produced them within the specifications without identified problems.
They had to organize the procurement of raw Invar plates, which depends on the available world stock. For the \dword{dpmod} it is important to plan ahead, 
 for this procurement. 
The manufacturing process followed by the manufacturer (SDMS) included eight steps:\\
\begin{itemize}
\item plate rectification,
\item  laser cutting,
\item  assembly,
\item  welding,
\item  geometrical controls,
\item  washing,
\item  packing, and
\item  shipping.
\end{itemize}

During the process a special validation step was included  for a full-size test frame to verify all the technical aspects and conformity to the requirements. 
The acceptance tests are performed on the geometry, the welding quality and the proper position of the different holes made in the Invar beams.

For \dword{pddp}, fabrication of the G10 frames for the four \dwords{crp} was done by a company with expertise in composite materials. The manufacturing process includes the 
production of frames with fiber layers in different orientations as explained in Section~\ref{sec:invar-frame}, the drilling of more than \num{100} screw threads per square meter of G10 frames, among them \num{80} with a \SI{2}{mm} diameter.
The acceptance tests are mostly based on geometrical and dimension criteria, and quality of the threads.  
The G10 production is longer and more difficult than that for the Invar structure. G10 production would require a minimum of two production sites, while one company would suffice for the Invar frame production.

\subsection{LEM and Anode Production}
\label{sec:fddp-crp-LASprod}

The construction of a  
\dword{dpmod} involves the production of \dpnumswch charge readout modules made of a \dword{lem} and its anode. For such a large production, it is desirable to use more than one manufacturer in order to mitigate risks and costs. The ELTOS\footnote{ ELTOS\texttrademark{} \url{www.eltos.com}.} company in Italy is making the \dword{lem} modules and the anodes for  \dword{pddp} and as such, has successfully worked out the \dword{qa} and \dword{qc} processes based on requirements and specifications agreed with the \dword{wa105} collaboration. So far, the fraction of rejected \dword{lem} modules produced by ELTOS, after the final phase of tests, has proven to be small, less than one \dword{lem} per \dword{crp}. The rejection factor for the anodes is, however, higher and corresponds to about \SI{10}{\%}, due to the very thin and fragile conducting strips forming the anode plane and also to the close proximity of the outer strips to the PCB edges. 


Small modifications to the anode design will be necessary for the \dword{dpmod} in order to increase the robustness of the manufacturing process. It is believed that such modifications are possible without  
significantly affecting the performance of the anode.   

A second manufacturer, ELVIA\footnote{ELVIA\texttrademark{} \url{https://www.pcb-elvia.com/}.} in France, is also known to have the capability for large-scale \dword{lem} and anode production with the requested specifications. Recently, several \dword{lem} prototypes have indeed been built for CEA/Irfu 
by ELVIA and extensive tests have 
demonstrated the same  \dword{lem} performance as the ELTOS products achieve.
 One key issue for a very large production, like the one necessary for DUNE, is to establish thorough \dword{qa} and \dword{qc} processes that can be applied throughout the whole \dword{lem} production. The definition of the \dword{lem} manufacturing requirements and specifications is naturally an important part of the tendering process. As far as the anodes are concerned, several prototypes have already been manufactured by ELVIA with satisfactory results. Despite the large size (\num{50}$\times$\SI{50}{cm$^2$}) of the anode, its fabrication follows a rather standard process mastered by several PCB manufacturers in Europe and around the world.     

\subsubsection{LEM production and QA and QC}
\label{sec:fddp-crp-LEMprod}

In the following scheme, an effective production time of \num{40} weeks per year and a total duration of two years is assumed. This corresponds to an average \dword{lem} production rate of one \dword{crp} or \num{36} \dword{lem} modules per effective week. The main limitation to the manufacturing speed comes from the \dword{lem} drilling process. While it is likely that ELTOS and ELVIA will have the capability each to produce \num{18}  \dword{lem} modules per week by assigning dedicated drilling machines, it would still be highly advisable to identify, well before the start of the \dword{lem} production phase, additional manufacturers capable of large-scale production. 

Based on the experience gained with  \dword{pddp}, the  \dword{qa} and \dword{qc} requirements during the \dword{lem} manufacturing process include the selection of the base material for the \dword{lem} PCB, thickness measurements of the dielectric material and copper layers before and after the process, a control of the size of the \dword{lem} holes, rims and outer dimensions, as 
well as a measurement of the electrical insulation across the two faces of the PCB.  
Table~\ref{tab:LEM_Tolerance} gives, as an example, the requested tolerance values on the various parameters of the \dword{lem} detectors for  \dword{pddp}. In addition, pre-series productions of several \dword{lem} modules from each manufacturing site will be necessary in order to validate the complete fabrication process prior to the start of the full production.

\begin{dunetable}[Tolerance values on various \dword{lem} parameters]
{p{.4\textwidth}p{.30\textwidth}}
{tab:LEM_Tolerance}
{Tolerance values on various \dword{lem} parameters} 
Parameter & Value and tolerance \\ \toprowrule
Dielectric thickness & \num{1.00}$^{+0.00}_{-0.05}$\,mm \\ \colhline
Average total thickness & \num{1.20}$^{+0.00}_{-0.06}$\,mm \\ \colhline
Dimensions & \num{499.5}$^{+0.00}_{-0.30}$\,mm$\times$499.5$^{+0.00}_{-0.30}$\,mm \\ \colhline
Final PCB thickness & \num{1.10}$^{+0.02}_{-0.05}$\,mm \\ \colhline
Active hole diameter & \num{0.50}$^{+0.00}_{-0.01}$\,mm \\ \colhline
Rim size & \num{40}$\pm$4\,$\mu$m \\ \colhline
Electrical insulation & >\SI{1}{\giga\ohm} \\
 \end{dunetable}

Once manufactured, the \dword{lem} modules are shipped to one or several collaboration sites for their final characterization and validation. Several infrastructures are necessary in order to perform these tasks, e.g., a survey bench, clean rooms and storage rooms, cleaning stations and \dword{hv} test setups. Here is a list of tasks to be performed, in sequential order:

\begin{itemize}
\item {\bf Visual inspection and survey.} Upon receipt of the \dword{lem} modules from the manufacturing sites, the first operations to be carried out are the visual inspection to examine the quality of the \dword{lem} surfaces, 
followed by the detector survey. The parameters that determine the \dword{lem} amplification gain are the thickness of the PCB and the geometry of the holes. It is therefore important to assess the uniformity of these parameters over the entire area of a \dword{lem} module. For  \dword{pddp}, this is performed on a sampling basis with the use of a confocal laser scanning microscope (CLSM). Several hundred measurements are done in each of \num{25} predefined locations, distributed uniformly over the \dword{lem} surface. With such an optical system, the total \dword{lem} and copper layer thicknesses as well as the rim size can be measured with a precision of a few microns.  
For the magnitude of production required for a \dword{dpmod}, the development of a fully automated survey system is mandatory. 

\item {\bf \dword{hv} connection.} The next step consists of soldering the \dword{hv} connection pins on the two \dword{lem} copper surfaces as well as gluing the MACOR insulators around the connectors.

\item {\bf Cleaning and polymerization.} The cleaning operation is an important phase of the \dword{lem} preparation. Following a procedure defined by CERN/EP-DT-EF-MP and CEA/Irfu,
 this is done using an ultrasonic bath at \SI{65}{C} with a micro-finishing solution (NGL 17.40 Sp ALU III) to clean the gold-plated copper surfaces of the \dword{lem}. It is followed by a rinsing phase with water and then with a spray of pressurized deionized water 
 ($<$\,30\,bar). The \dword{lem} is then dried in an oven 
at \SI{60}{C}  for several hours and then baked for three hours up to \SI{160}{C}, 
a temperature near the glass transition point of the dielectric material. From the cleaning operation on, each \dword{lem} is handled via an aluminum frame on which it is mounted in order to avoid any contact with the PCB surfaces. They are also to be 
handled and operated in a clean environment with the use, for example, of a laminar flux.  

\item {\bf \dword{hv} tests.} The final validation of a \dword{lem} is obtained after  successful and stable operation in pure argon at room temperature and an absolute pressure of about \numrange{3.2}{3.3}\,bar. A high-pressure vessel is used for this purpose.  The argon pressure value is precisely adjusted as a function of the argon temperature inside the vessel in order to reach the same gas density as the one existing in D\lar mode. In such a way, the \dword{lem} modules are tested at the same Townsend avalanche operation point as in cold. Figure~\ref{fig:HP} shows the \SI{360}{L} high-pressure vessel used at CEA/Irfu for the characterization of the  \dword{pddp} \dword{lem} modules. Up to nine \dword{lem} modules can be stacked inside this chamber for the \dword{hv} tests. The HP vessel is also instrumented with \dword{fe} electronics and a \dword{daq} system for gain measurements in pure argon. In this configuration, a single \dword{lem} is installed with its \twod charge-collecting anode inside 
a \num{50}$\times$\num{50}$\times$\SI{5}{cm$^3$} 
TPC together with a collimated $^{241}$Am open alpha source mounted on 
the cathode. Figure~\ref{fig:HP_ED} shows an event display of a \SI{5.5}{MeV} alpha track observed in pure argon gas at an 
absolute pressure of \num{1}\,bar.
During the \dword{lem} validation \dword{hv} tests, a two-step procedure is followed. After pumping for 
about \num{60} hours down to a residual pressure of \num{10}$^{-4}$\,mbar, the chamber is filled with dry synthetic air and 
\dword{hv}  up to \SI{4.5}{kV} is applied across the \dword{lem} in order to 
burn possible residual dust. Then, the vessel is pumped again and pure argon (graded \num{5.7})
is introduced at an absolute pressure of about \numrange{3.2}{3.3}\,bar. Each \dword{lem} module is tested 
and validated so see that it can reach the value of \SI{3.5}{kV} across the two faces, consistent with amplification gains higher than \num{100} before charging up, with no occurrence of discharges. The amount of time needed to perform the \dword{hv}  tests is 
typically one week for an entire batch of nine \dword{lem} modules. This last figure can probably be increased to \num{12} with a slightly larger volume vessel. 
\end{itemize}

While the tasks related to the \dword{lem} visual inspection, survey and the implementation of the \dword{hv}  connections can be done at 
a single participating institution 
site, the subsequent operations could be dispatched among several sites. In that case, it is  important that each 
site 
have the necessary infrastructure for 
both the cleaning phase and the \dword{hv} tests, as iterations are sometimes necessary in order to fully validate the 
\dword{lem} modules. With the assumption of \num{36} \dword{lem} modules produced per effective week, a reasonable number of setups needed for the \dword{hv}  tests, together with their associated instrumentation and infrastructure, could amount to three or four. Finally, based on the experience gained in  \dword{pddp}, the processing time needed for the preparation, characterization and tests of a batch of nine to twelve \dword{lem} modules is estimated to be about three weeks.  

\begin{dunefigure}
[High-pressure vessel for the characterization of the  \dword{pddp} \dword{lem} modules.]
{fig:HP} 
{High-pressure vessel for the characterization of the  \dword{pddp} \dword{lem} modules.}
  \includegraphics[width=.8\textwidth]{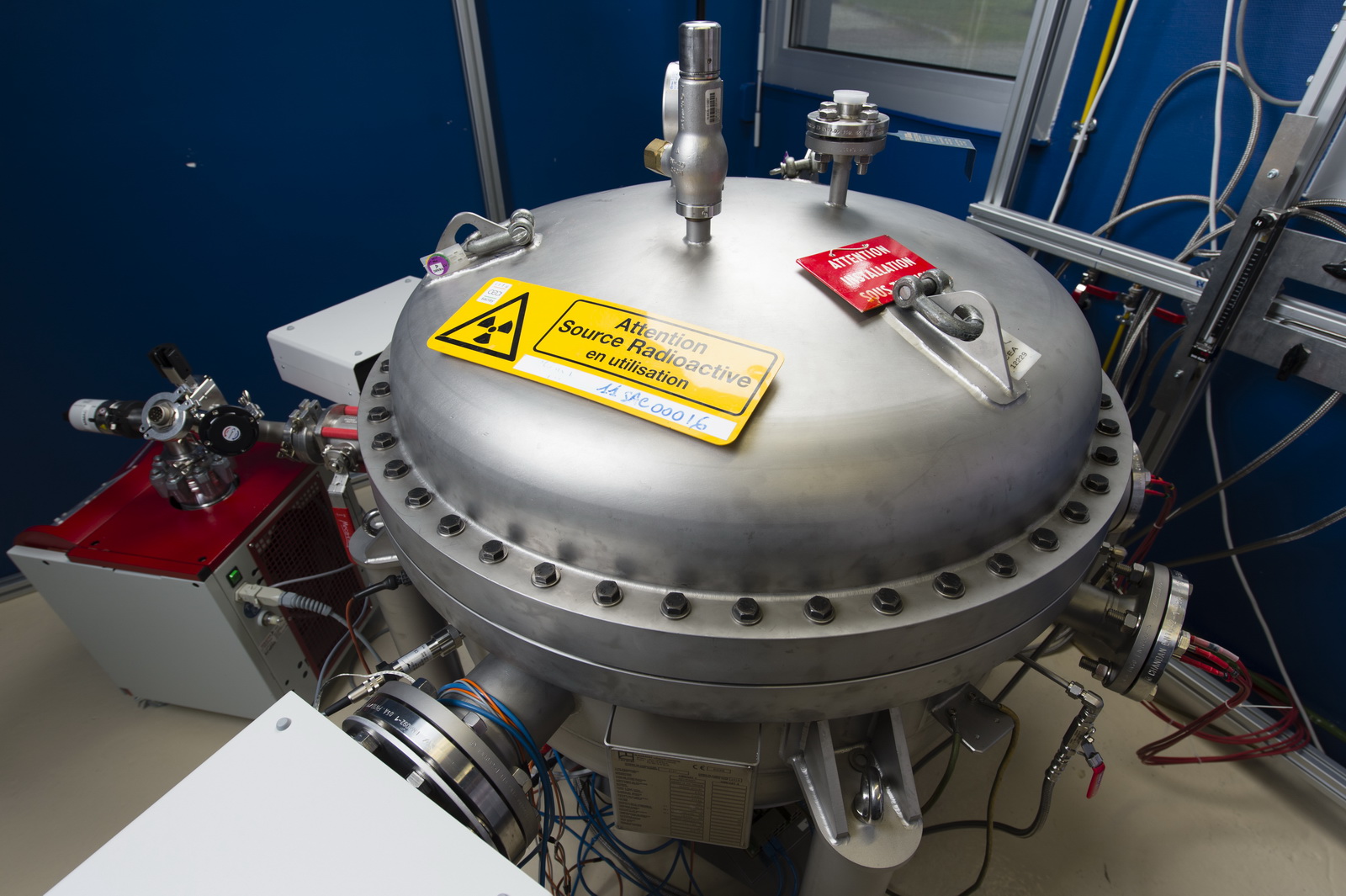}
\end{dunefigure}

\begin{dunefigure}
[Event display of a \SI{5.5}{MeV} alpha track in argon gas at \num{1}\,bar.]
{fig:HP_ED}
{Event display of a \SI{5.5}{MeV} alpha track in argon gas at \num{1}\,bar.  Left: $x$-View. Right: $y$-View. Top: Pulse-height distributions of hits in \dword{adc} counts. Bottom: Hit time [300\,ns bins] \textit{vs} strip number.}
\includegraphics[width=.8\textwidth]{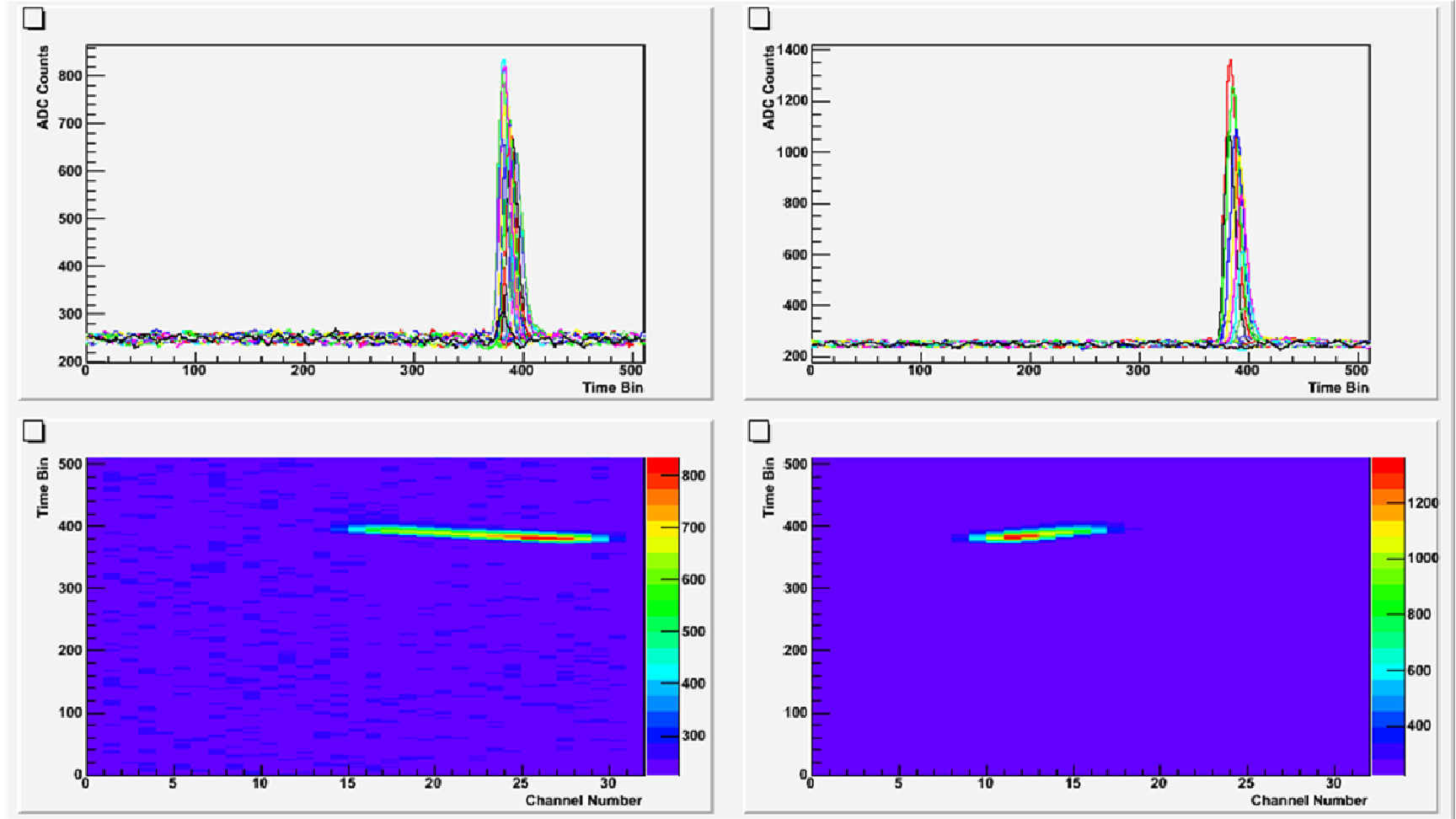}
\end{dunefigure}

\subsubsection{Anode production and QA/QC}
\label{sec:fddp-crp-ANODEprod}
It is realistic to assume that the lead time in the manufacturing of the anodes will be compatible with the production time
of the \dword{lem} modules. Table~\ref{tab:ANODE_Tolerance} gives the anode specifications for the dimensions of the PCB. In addition, visual inspection and continuity tests are requested at the manufacturing site after the soldering of the signal connectors. Upon receipt at the collaborating institution site(s) where the \dword{crp} assembly will take place, continuity and short circuit tests of the anode strips are performed (see Figure \ref{fig:Anode_CTest}). This task is rather quick and should not take more than one day per \dword{crp}.

\begin{dunetable}[Specifications for the anode dimensions]
{p{.4\textwidth}p{.30\textwidth}}
{tab:ANODE_Tolerance}
{Specifications for the anode dimensions} 
 Parameter & Value and tolerance\\ \toprowrule
Dimensions & 499.5$^{+0.2}_{-0.0}$\,mm$\times$499.5$^{+0.2}_{-0.0}$\,mm \\ \colhline
PCB thickness & 3.5$\pm$0.05\,mm \\ \colhline
PCB sagitta & < 1\,mm \\
 \end{dunetable}

\begin{dunefigure}
[Test of anode strip continuity]
{fig:Anode_CTest} 
{Test of anode strip continuity.}
 \includegraphics[width=.8\textwidth]{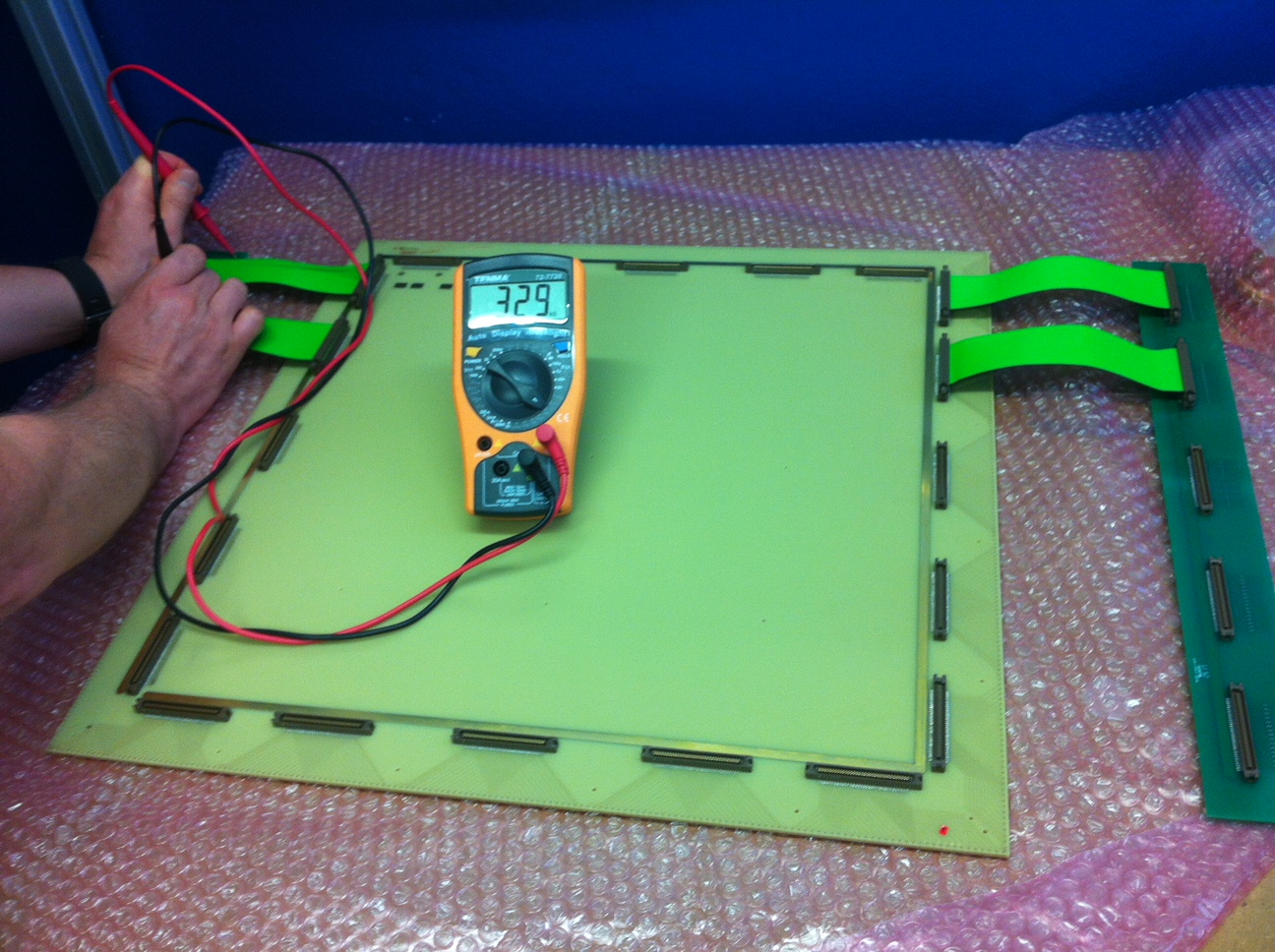}
\end{dunefigure}

\subsection{Tooling}
\label{sec:fddp-crp-tooling}

Once the anodes, the \dwords{lem}, the mechanical Invar and G10 frames are produced and delivered, the \dwords{crp} production site should be equipped with several toolings.
The \dwords{crp} construction activity requires a clean room large enough to accommodate the different work spaces. Separate work areas are required in this clean room for: 
\begin{itemize}
\item{a \SI{10}{m$^2$} table for assembling the G10 subframes and for surveying the whole structure;}
\item{a supporting structure to hang the Invar frame and to allow coupling of the G10 frame and the installation of the anodes, \dword{lem} and extraction grid modules;}
\item{storage of the anodes and \dwords{lem} on special shelves;}
\item{construction of the extraction grid modules and storage of the modules before mounting on the \dword{crp} structure.}
\end{itemize}
The required tooling includes: two mobile cranes (\SI{4}{m} span) to manipulate the Invar frame, an aluminum support frame and parts of the transport boxes, a large assembly table, storage shelves for the \dword{lem} and anodes and for cables, and a fabrication bench for the extraction grid modules.
Figure~\ref{fig:gridtooling} shows the \SI{3}{m} long wiring bench with details and the real tooling at CERN used for the \dword{crp} in \dword{pddp}.

\begin{dunefigure}[Extraction grid construction bench details]{fig:gridtooling}
{Extraction grid construction bench details and picture in the clean room 185 at CERN.}
\includegraphics[width=0.65\textwidth]{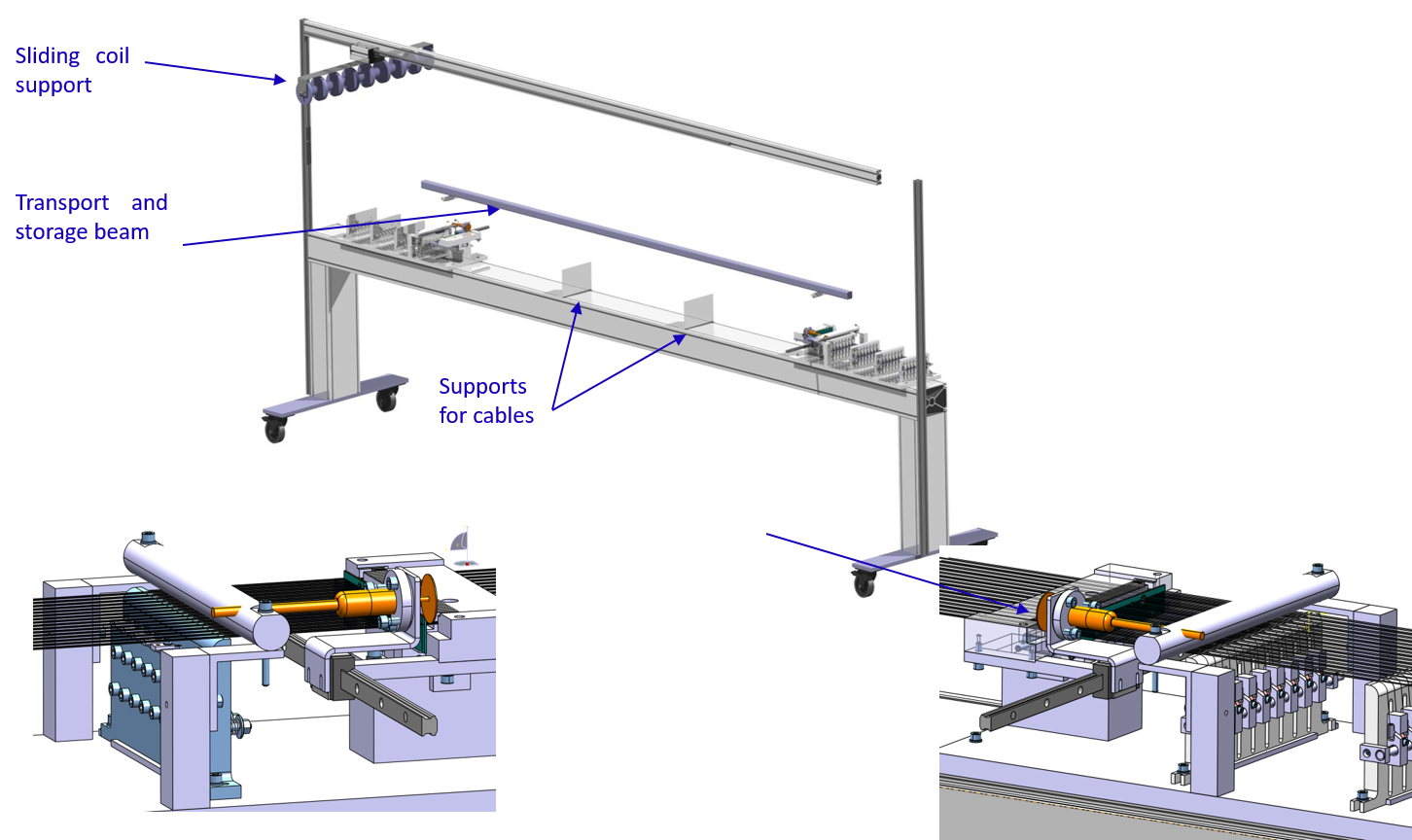}
\includegraphics[width=0.3\textwidth]{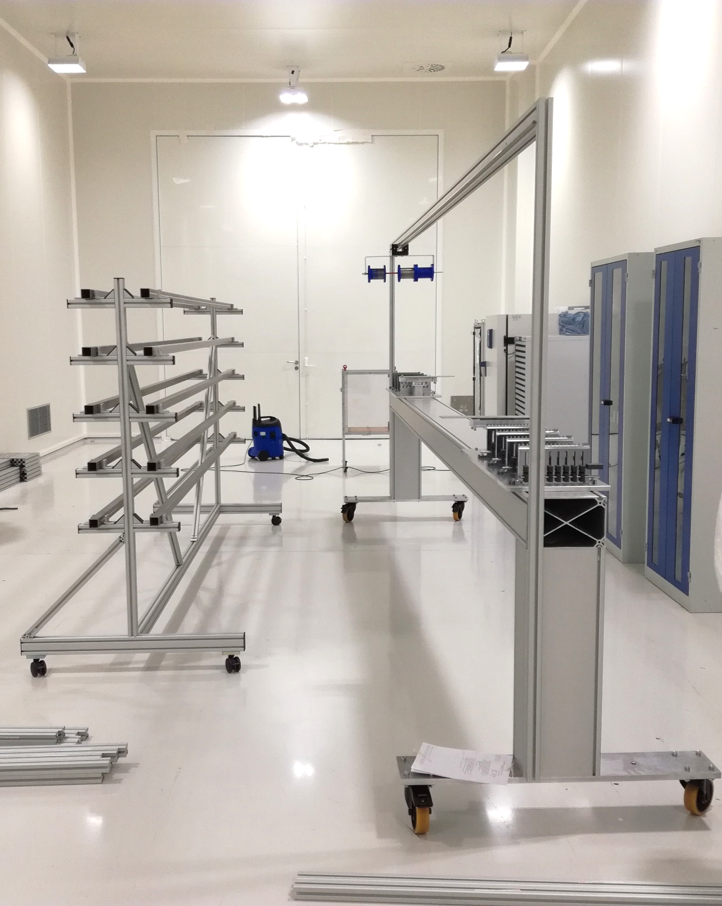}
\end{dunefigure}

\subsection{Assembly Procedures}
\label{sec:fddp-crp-assy}

The \dword{crp} assembly production activity will span two years 
prior to \dword{crp} installation underground at \surf. The \dwords{crp} will be integrated in at least two different production sites and stored locally in temporary storage boxes before being shipped to the \dword{itf}. 
The \dword{crp} assembly will be pipelined with the \dword{lem} and anode production and testing during these two production years with an initial  shift of three months. This is intended to constitute a large enough buffer between the \dwords{lem} and anodes to start the \dword{crp} assembly activities.

The assembly of a \dword{crp} in the clean room is a process that goes from the reception of the Invar frame to the  closing of the transport box. 
A detailed animation of the assembly steps and procedures for the construction of a \dword{pddp} \dword{crp} is available\footnote{\url{https://www.youtube.com/watch?v=jcnJjlU-Cyc&feature=youtu.be}.}.

This is divided in the following way:
\begin{enumerate}
\item Reception of the Invar frame in the clean room;
\item  Suspension of the Invar frame below the supporting structure;
\item  Adjustment of vertical position of the frame using lifting cranks;
\item  Assembly of the G10 frame on the optical table (proceeds in parallel);
\item  Positioning of the Invar frame 
above the G10 frame, and connection through \num{50} decoupling systems;
\item  Assembly of \dword{lem} and anodes (done from below);
\item  Connection of anodes  with jumpers;
\item  Weaving and soldering of the extraction grid on the special tooling (done in parallel), then storage on shelves, or direct installation on the \dword{crp}.
\end{enumerate}
Some of the steps are illustrated in Figure~\ref{fig:assembly} and real components and tooling for \dword{pddp} are shown in Figure~\ref{fig:assembly2}.

\begin{dunefigure}[\dword{lem} and anode connection and extraction grid module installation.]{fig:assembly}
{Some assembly steps: \dword{lem} and anode connection and extraction grid module installation.}
\includegraphics[width=0.3\textwidth]{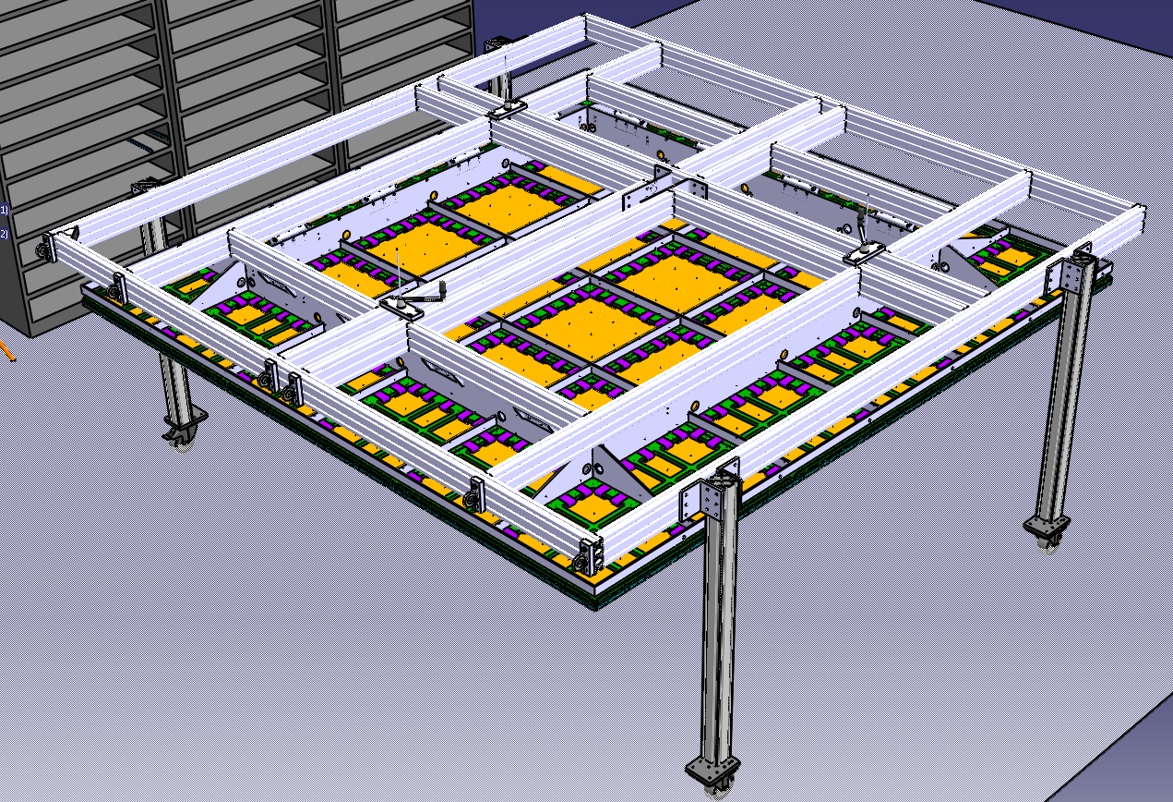}
\includegraphics[width=0.35\textwidth]{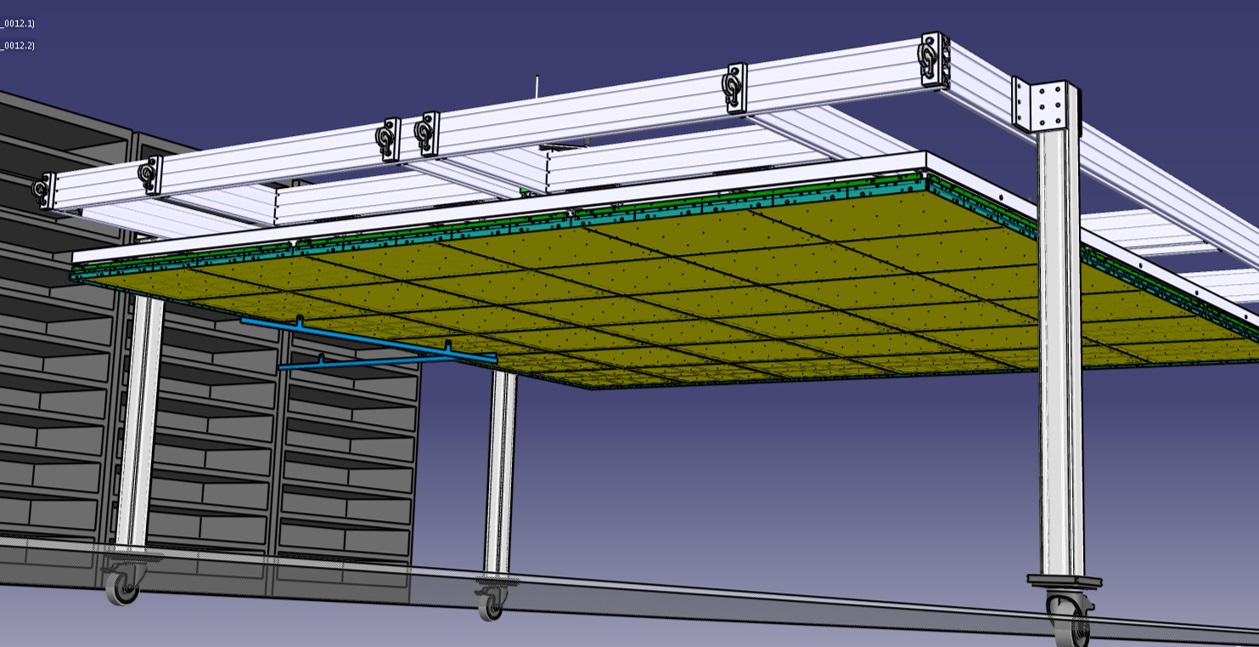}
\includegraphics[width=0.3\textwidth]{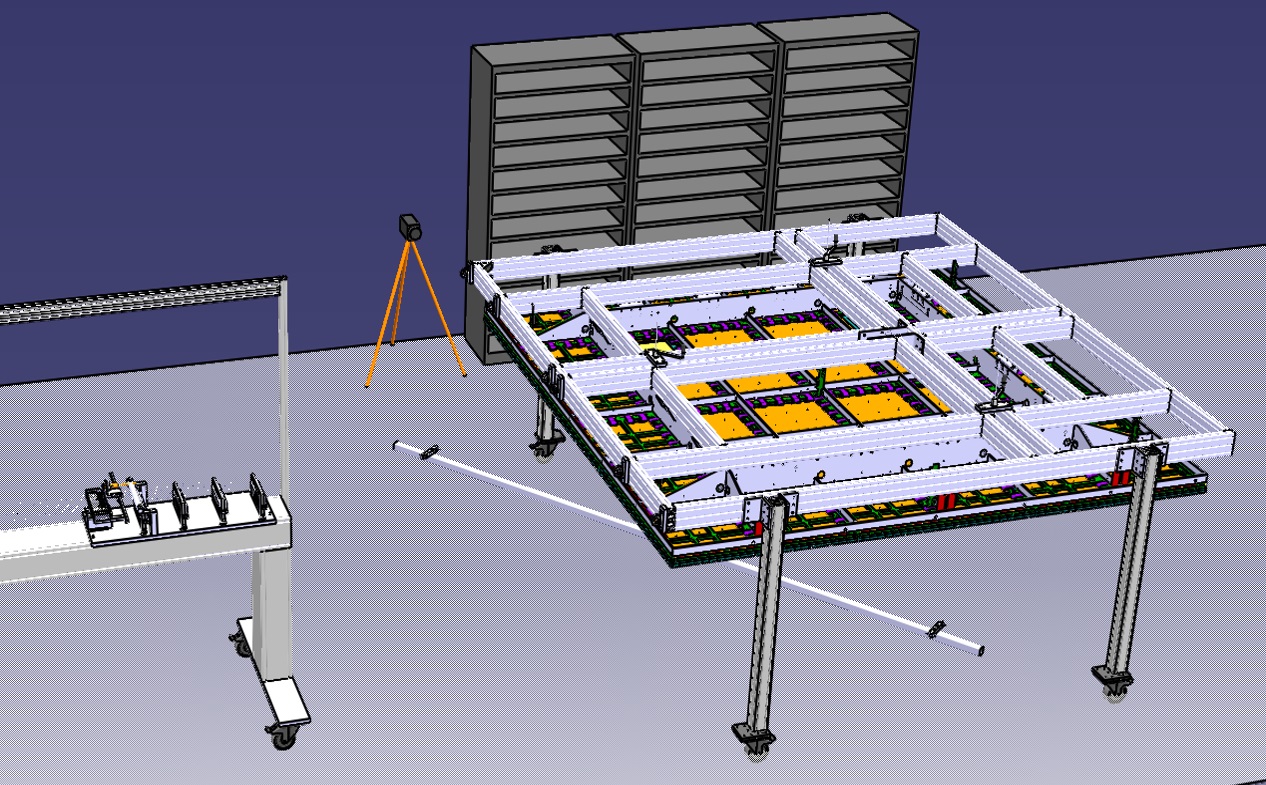}
\end{dunefigure}

\begin{dunefigure}[Elements for assembly of \dword{pddp} \dword{crp} at CERN]{fig:assembly2}
{Elements for assembly of \dword{pddp} \dword{crp} in clean room 185 at CERN: assembly structure and G10 subframes being prepared.}
\includegraphics[width=0.4\textwidth]{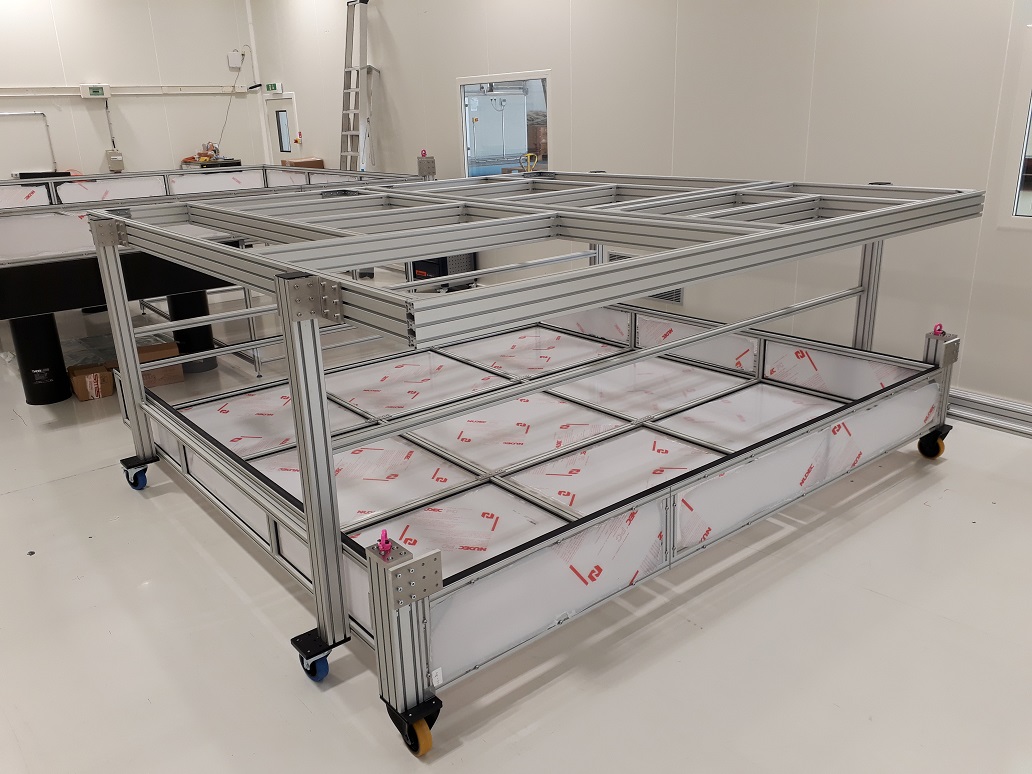}
\includegraphics[width=0.4\textwidth]{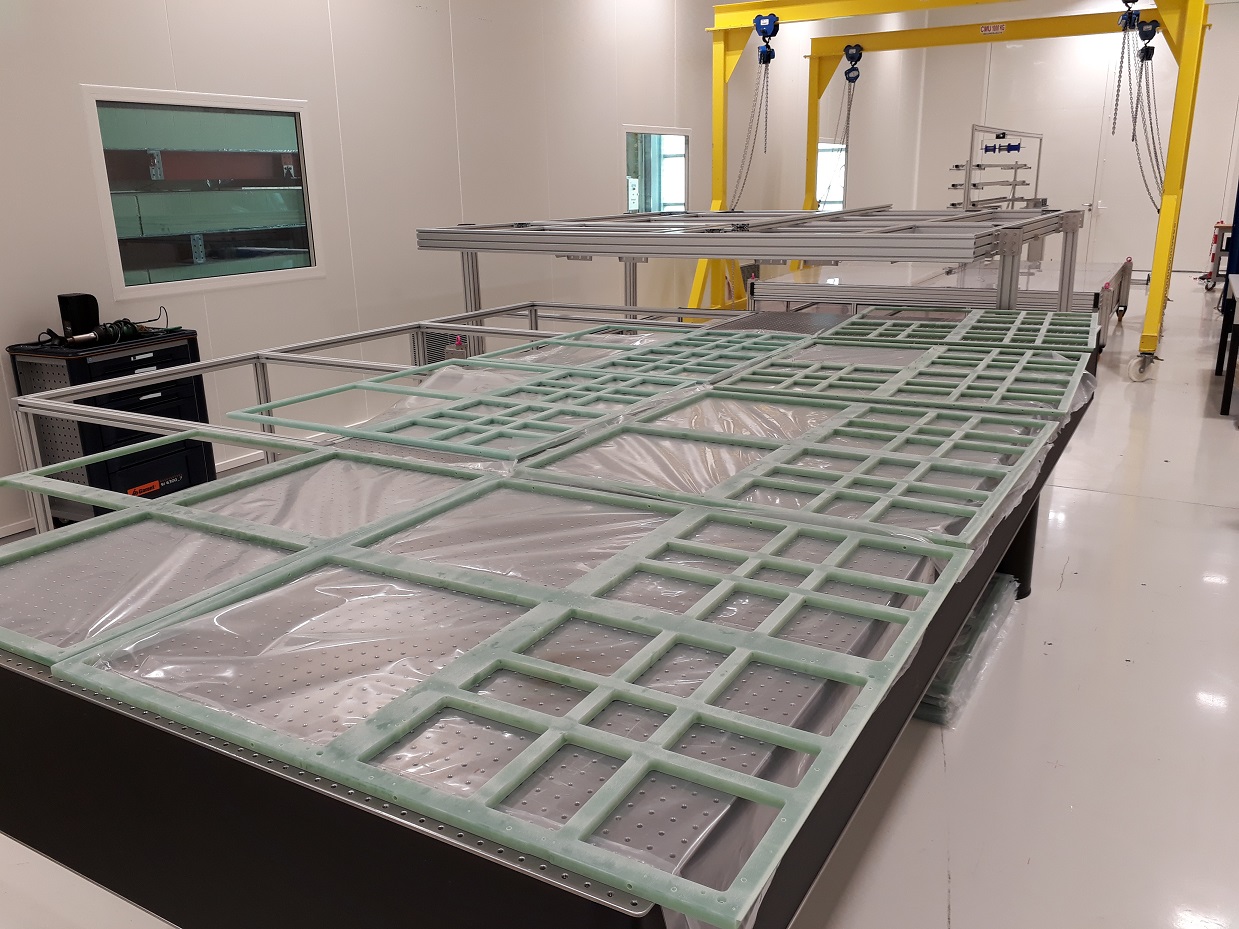}
\end{dunefigure}
Once ready, the planarity of the \dword{crp} is checked and adjusted using metrology and the decoupling systems.
All the components and instrumentation of the \dword{crp} are carefully checked, and the assembly is then packed in the transport box.


\section{Interfaces}
\label{sec:fddp-crp-intfc}

The main \dword{crp} interfaces  are with the elements connected directly to them. The documents that describe the interfaces are referenced.
\begin{itemize}
\item The readout electronics  (DocDB 6751), concerning the cabling of the signal cables on the bottom part of the cold 
flange of the signal chimneys; 
\item The \dword{cisc} (DocDB 6760),  concerning the power supplies for the \dword{lem} and extraction grid, the cameras, \dword{led} ribbons, temperature sensors, and distance meters; all these devices go to the \dword{crp} instrumentation \fdth{}s; 

\item Drift \dword{hv} (DocDB 6754), at the level of design requirements; this includes the system for maintaining the proper distance between the top-most field shaping profile to the extraction grid in order to maintain the proper extraction field and to protect the contact between the two systems.
\end{itemize}

\subsection{TPC Electronics}
\label{sec:fddp-crp-intfc-elec}

The interface with the \dual TPC electronics is at the level of the bottom flanges of the \dwords{sftchimney} where the flat cables bringing the signals from the anodes are plugged in during the \dword{crp} installation in the cryostat. The system for 
pulsing the anode strips is installed by the \dword{crp} consortium 
concurrently with the \dword{crp} installations. 
Calibration of the strips is then performed jointly by the two consortia.

\subsection{Instrumentation and HV Feedthrough Flanges}
\label{sec:fddp-crp-intfc-FT}
The \num{36} \dword{lem} modules of a \dword{crp} are supplied with \dword{hv} by \num{42} coaxial cables 
connecting the \fdth flange from the top of the cryostat to distribution boxes located on the \dword{crp} (Figure \ref{fig:CRP_DB}). Each cable (\num{20}\,AWG, \SI{50}{\ohm} Kapton\footnote{DuPont\texttrademark{} Kapton, polymide film,  E. I. du Pont de Nemours and Company,  \url{http://www.dupont.com/}.} insulated coaxial cable from ACCU-GLASS\footnote{ACCU-GLASS\texttrademark{}, \url{http://www.accu-glass.com/}.}) can sustain \SI{30}{kV} in air and has SHV plugs on each end in order to facilitate connections during the \dword{crp} installation.

Six coaxial cables are used for the top \dword{lem} electrodes and \num{36} cables for the bottom ones. Each of the cables for the top \dword{lem} electrodes is connected to a single distribution box that contains a PCB for distributing \dword{hv} individually to the six \dword{lem} modules through a \SI{500}{\mega\ohm} resistor in series (see Figure~\ref{fig:LEM_DB}). The \num{36} coaxial cables for the bottom  \dword{lem} electrodes are connected in groups of four to nine distribution boxes. 
%
For these, each \dword{hv} cable is directly linked to an individual bottom \dword{lem}  electrode. In total, \num{72} output coaxial cables rated for \SI{10}{kV} (\num{26}\,AWG, \SI{50}{\ohm} Kapton insulated coaxial cable from ACCU-GLASS) are used to supply \dword{hv} to both electrodes of all \dword{lem} modules of a \dword{crp}. The grounding of each coaxial line is ensured by the \fdth flange down to about \SI{5}{cm} from the \dword{lem} final connection point via the PCB located inside each of the \num{15} distribution boxes. In this way, contributions from the \dword{hv} lines to the electronic noise is limited as much as possible. In order to avoid electrical discharges inside the distribution boxes, the latter are filled with Arathane\footnote{Arathane\texttrademark{} advanced polyurethane based adhesives, Huntsman Advanced Materials , \url{http://www.huntsman.com/advanced\_materials/a/Home}.} epoxy glue. Finally, the connections to the \dword{hv} contacts of the \dword{lem} modules are done via  Deutsch 6862-201-22278 female connectors and each of them is covered by a heat-shrink sleeve. 

\begin{dunefigure}
[\dword{lem} \dword{hv} distribution boxes mounted on a \dword{crp}]
{fig:CRP_DB} 
{\dword{lem} \dword{hv} distribution boxes mounted on a \dword{crp}. }
  \includegraphics[width=.8\textwidth]{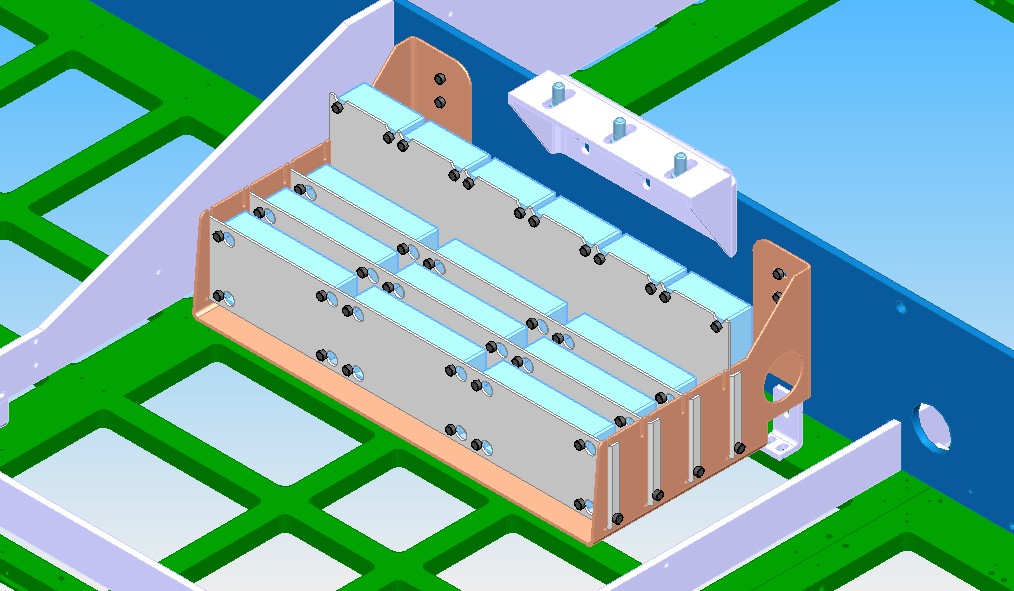}
\end{dunefigure}

\begin{dunefigure}
[\dword{hv} distribution box for \dword{lem} top electrodes before filling]
{fig:LEM_DB} 
{\dword{hv} distribution box for \dword{lem} top electrodes before filling with ARATHANE glue. }
  \includegraphics[width=.8\textwidth]{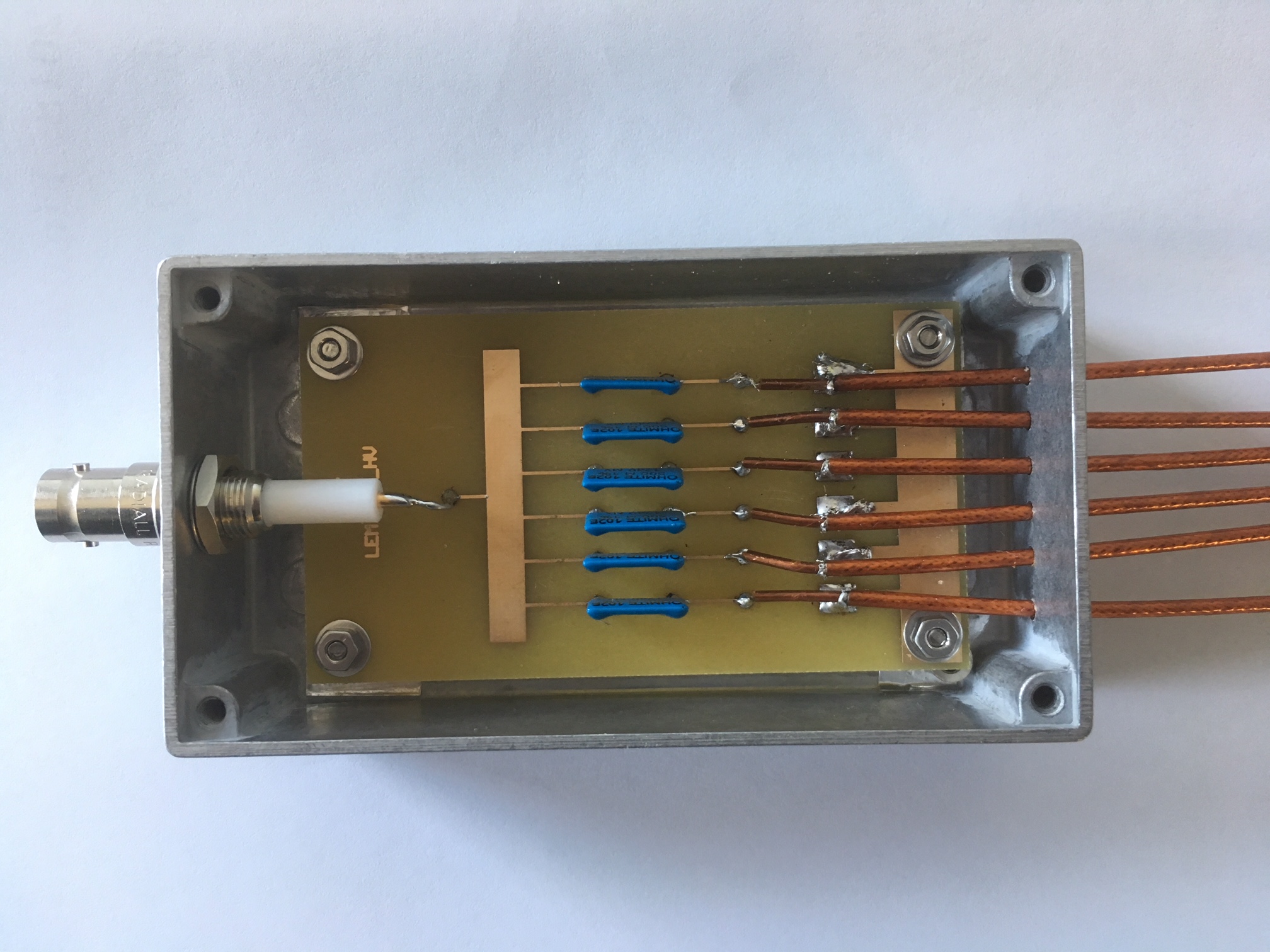}
\end{dunefigure}

\subsection{Cryostat and Detector Support Structure}
\label{sec:fddp-crp-intfc-support}

The cryostat includes the dedicated penetrations for the hanging system of each \dword{crp}. Each penetration has a diameter of \SI{60}{mm}. The layout of the cryostat penetrations   matches the position of the hanging points on the \dword{crp} supporting structure. The penetration flanges and \dword{crp} suspension step motors are provided by the \dword{crp} consortium.

\subsection{HV and Slow Controls}
\label{sec:fddp-crp-intfc-HV-slowcontrol}

The interface with the slow controls includes:

\begin{itemize}
\item The procurement and control of the power supplies for the HV and grid \dword{lem} biasing;
\item The control of the \dword{crp} temperature probes and level meters;
\item The procurement and control of the \dword{crp} pulsing system (external to the cryostat);
\item The procurement of the instrumentation and \dword{hv} \fdth flanges;
\item The control of the \dword{crp} suspension step motors.
\end{itemize}

\section{Installation, Integration and Commissioning}
\label{sec:fddp-crp-install}

The installation of the \dwords{crp} in the cryostat will be performed on a time span of nine  months, following the installation of the signal chimneys, which will take three months. The signal chimneys must be present in order to 
connect the \dword{crp} signal flat cables. The suspension, instrumentation and \dword{hv} flanges must also be  installed prior to \dword{crp} installation. The installation of these flanges can be performed in parallel with the installation of the signal chimneys. 

The production of the \dwords{crp} will be performed before the installation, and the \dwords{crp} will be stored in temporary storage boxes at the production centers. During installation the already produced \dwords{crp} will be moved to the transportation boxes in order to be shipped to the ITF, which is used as a delivery destination buffer, and then immediately moved underground for installation. Given the installation rate of eight \dwords{crp}/month a set of \num{30} transportation boxes will ensure a sufficient turnover among the production sites and the installation site.


The installation procedure 
plans for three two-person teams working in parallel inside the cryostat to 
install three \dwords{crp} at a time. It is assumed that 
approximately \num{1.3} weeks is needed for a team to prepare, survey, tune and cable a \dword{crp} in the  cryostat, and \num{35} working weeks for the overall installation of the \dptotcrp \dwords{crp}.

Once the three \dwords{crp} are ready to be lifted, three people are needed to manipulate the manual winches on the top of the cryostat for a few hours. This activity is expected to occur during the \num{1.3} week period just before the cabling activity for the connection to the flanges.


The sequence of operations needed underground per \dword{crp} is similar to the one foreseen for \dword{pddp}  and is the following: 
\begin{enumerate}
\item A \dword{crp} module in its transport box is brought to the entrance of the cryostat.
\item The box is hung from the side of the insertion rail, and guided through the \dword{tco}.
\item  Inside the cryostat the \dword{crp} is laid down horizontally, and rolled below the \dwords{sftchimney}.
\item The structure is suspended from temporary cables coming down from the chimney.
\item The transport box is dismounted and removed.
\item The \dword{crp} planarity is measured and tuned based on metrology survey.
\item The \dword{crp} is raised up with the manual winches, then the mechanical stop is assembled.
\item The \dword{crp} is lowered down  on the mechanical stop.
\item The cabling of the \dword{crp} patch panels to \dword{hv}, signal and slow control \fdth{}s is done.
\item The winch cable is disconnected  and the winch removed.
\item The bellows is compressed using special tooling.
\item The cable from the bellows is connected with a pin.
\item The compression tool is removed and the bellows attached.
\item The motor is inserted and screwed from the top.
\end{enumerate}

The assembly is then complete and operational.
The lateral and vertical alignment of the \dword{crp} is performed from the top of the cryostat with a SPFT translation mechanism, distance meter measurements and metrology.

The cabling activity is done  working at the nominal height position using elevating platforms. The access between two adjacent \dwords{crp} is done by staggering the altitude of the two modules by about \SI{20}{cm} allowing enough space to do the electrical connection. 

\subsection{Transport and Handling}
\label{sec:fddp-crp-install-transport}

The transportation boxes are sized (\num{3.5}$\times$\num{3.5}$\times$\SI{0.8}{m}) for the shafts and underground handling at \surf{}.
They are protected by a plastic layer that should be removed once the transportation box arrives at the clean area 
underground so that the box can be introduced into the cryostat in a clean state via the \dword{tco}. The transportation box is also essential for manipulating the \dword{crp} from the vertical orientation (i.e., for insertion in the \dword{tco}) to the horizontal orientation, required 
 prior to hanging it from the suspension system. Once the installation is complete the transportation boxes are wrapped again with a protective layer and shipped back to the production centers. 

\subsection{Calibration}
\label{sec:fddp-crp-install-calib}

The \dword{crp} calibration relies on a pulsing system that performs charge injection in the strips via \SI{1}{pF} capacitors. The pulse distribution system and related cabling, and the boards with the charge-injection capacitors are integrated  with the \dword{crp} during  \dword{crp} assembly. This system is connected via flat cables to the instrumentation flange. A pulse distribution system external to the cryostat (provided by the \dword{cisc} consortium) ensures the signals feeding to the charge injection system. 
The information from the pulse calibration is then combined
with that from the \dword{fe} electronics calibration and the analysis of the reconstructed charged tracks, enabling extraction of the overall calibration constants per channel.

\section{Quality Control}
\label{sec:fddp-crp-qc}

Several quality control 
procedures are applied at the level of the \dword{lem} and anode production, as described in more detail in their corresponding sections. The components of the \dword{hv} distribution system are also individually tested. Continuity tests are performed at the time of the \dword{crp} assembly. The \dword{crp} geometry is also systematically surveyed as well as the tensioning of the grid wires. It will also be possible to perform cold-box testing on a small subsample of the \dwords{crp} production using the infrastructure set up at CERN for \dword{pddp}.



\section{Safety}
\label{sec:fddp-crp-safety}

Safety must be a central feature of all tasks performed by the \dword{crp} consortium.  All aspects of \dword{crp} construction, installation, and commissioning will adhere to procedures established by the DUNE Technical Coordinator and relevant host institutions. 

The \dword{crp} installation and operation does not present particular safety issues apart from working at heights 
for the connection of the  \dword{crp} cabling to the \dwords{sftchimney} and the instrumentation and \dword{hv} \fdth{}s.
\cleardoublepage

@@ -0,0 +1,770 @@
\chapter{TPC Electronics}
\label{ch:fddp-tpc-elec}

\section{TPC Electronics System Overview}
\label{sec:fddp-tpc-elec-ov}

\subsection{Introduction}
\label{sec:fddp-tpc-elec-intro}

The aim of the \dword{dp} TPC electronics is to collect and digitize the signals from the 
\dwords{crp} (see chapter~\ref{ch:fddp-CRP}) and \dwords{pds} (see chapter~\ref{ch:fddp-pd}), which implements 
\dwords{pmt}. 
These two tasks are respectively accomplished by the \dword{cro} and \dword{lro} sub-systems.
The design of the system incorporates the components already developed for \dword{pddp} as a result of an R\&D activity started in 2006. One of the key objectives of this R\&D program has been the design of an electronics system that is easily scalable and cost-effective in order to meet the needs of the large-scale neutrino \dword{lar} detector.  

While a single \dword{dpmod} has a factor of \num{20} more readout (both charge and light) channels than \dword{pddp}, a simple scaling of the number of the components used in the prototype design is sufficient to meet these needs without necessitating any additional R\&D. A small-scale version of the TPC electronics system was used in the \dword{wa105} at CERN, a preliminary \dual \lartpc prototype with an active volume  of \SI[product-units=power]{3x1x1}{m} (\dword{crp} area of \SI[product-units=power]{3x1}{m}) that took data in the summer-fall of 2017. 
Operation of the \dword{wa105} validated the design choices and provided checks on various performance markers, e.g., noise. 

\begin{dunefigure}[Schematic layout of the \dword{dp} \dword{cro} sub-system]{fig:dpele-crosystem-sketch}
{Schematic layout of the \dword{dp} \dword{cro} sub-system.}
\includegraphics[width=0.6\textwidth]{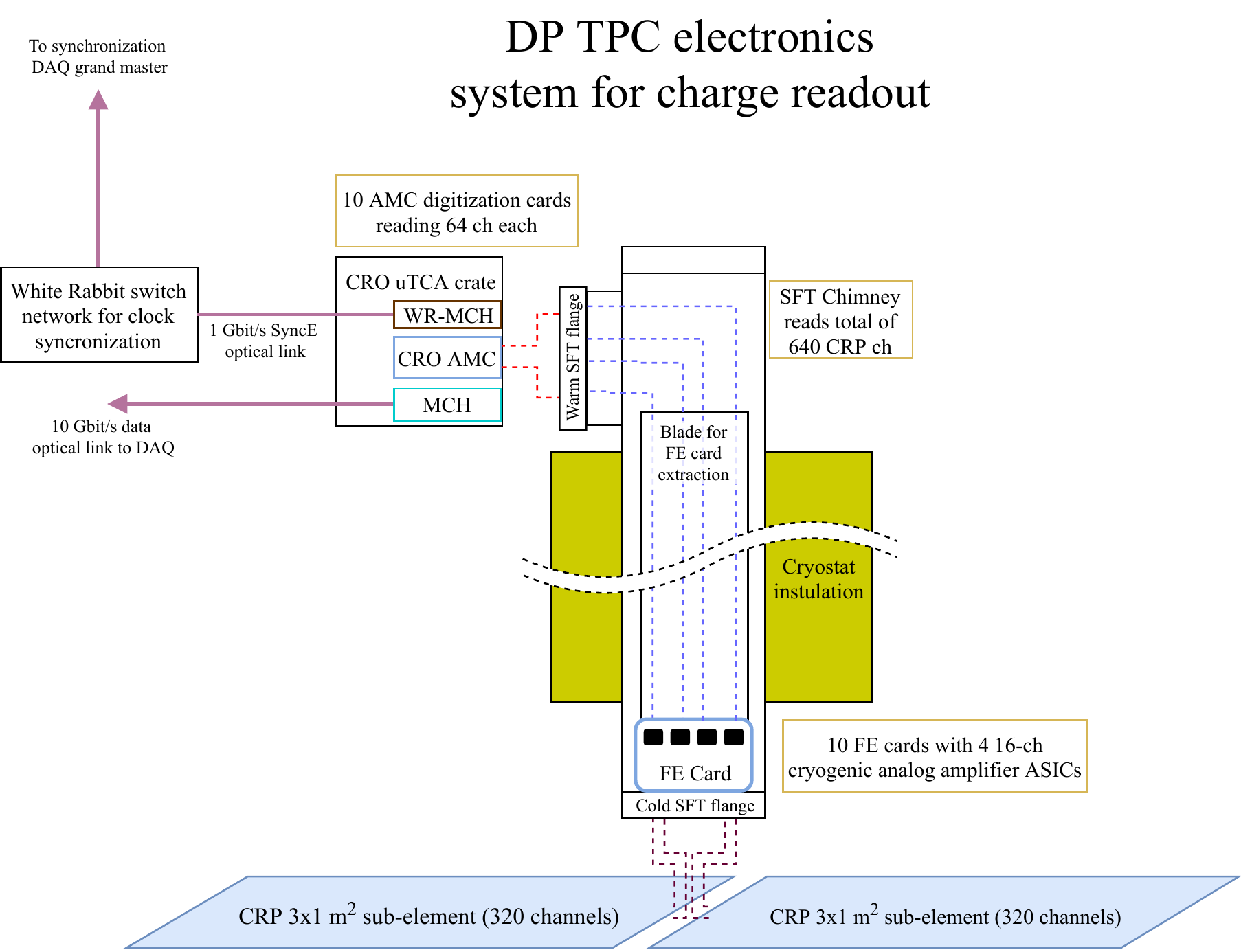}
\end{dunefigure}

The \dword{cro} electronics system, illustrated in the Figure~\ref{fig:dpele-crosystem-sketch}, is designed to provide continuous, non-zero-suppressed, and losslessly compressed digital signals by reading the charge collected on the \dwords{crp}' \SI{3}{m} long strips that are arranged in two collection views, all with a pitch of \SI{3.125}{mm}.
The system consists of 
the \dword{fe} analog electronics operating at cryogenic temperatures and digital electronics working in the warm environment outside of the cryostat.  The cryogenic \dshort{fe} analog electronics is based on an application-specific integrated circuit (\dword{asic}) chip with a large dynamic range (up to \SI{1200}{fC}) to cope with the charge amplification in the \dwords{crp}. The analog \dword{fe} cards are housed in dedicated \dwords{sftchimney} and are accessible from the outside even after the \dword{dpmod} is in operation, thus removing any significant risks associated with their long-term survivability. The \dwords{sftchimney} are approximately \SI{2.3}{m} long stainless steel pipes 
that traverse the entire insulation layer of the cryostat allowing placement of the \dword{fe} electronics close to the \dwords{crp} to minimize cable capacitance (noise).  In addition, their metallic structure shields the \dword{fe} cards 
from any interference from the warm digital electronics and ambient environment. The analog signals are digitized by \dwords{amc}, which are housed in the commercial \dword{utca} crates on top of the cryostat near the \dwords{sftchimney}. 

The \dword{cro} data are sampled at the rate of \SI{2.5}{MHz} with \SI{12}{bit} resolution.  
This frequency, traditionally used in \lartpc experiments, is well matched to the \SI{1}{\micro\second} pulse-shaping time of the \dword{fe} electronics and the detector response times determined by the electron drift velocity in the \lar. The corresponding sampling resolution along the drift coordinate is better than \SI{1}{\mm}. 

The \dword{lro} electronics system, illustrated in the Figure~\ref{fig:dpele-lrosystem-sketch},  collects and digitizes the signals from the \dword{pds}, which consists of \dword{tpb}-coated \num{8}\,in \dwords{pmt} (Hamamatsu\footnote{Hamamatsu\texttrademark R5912-02-mod, \url{http://www.hamamatsu.com/}.} R5912-02-mod) located beneath the TPC cathode. The \dword{lro} electronics 
facilitates the detection of the primary scintillation signals, which provide the absolute time reference for the interaction events. It 
also enables recording the light signals generated by photons from the so-called \textit{proportional scintillation component}, the light created by the electrons extracted and amplified in the gaseous phase. A fraction of this light can reach the \dwords{pmt} after traversing the entire detector volume.  
The \dword{lro} electronics, consisting of analog and digital stages, is housed in the \dword{utca} crates on top of the cryostat structure (similar to the \dword{cro} electronics). The \dword{lro} \dword{amc} card design shares a similar architecture with the \dwords{amc} for the charge readout. The \dword{lro} \dword{utca} crates are connected to specific \dword{lro} signal \fdth flanges on top of the cryostat (see chapter~\ref{ch:fddp-pd}).

\begin{dunefigure}[Schematic layout of the \dword{dp} \dword{lro} sub-system]{fig:dpele-lrosystem-sketch}
{Schematic layout of the \dword{dp} \dword{lro} sub-system.}
\includegraphics[width=0.6\textwidth]{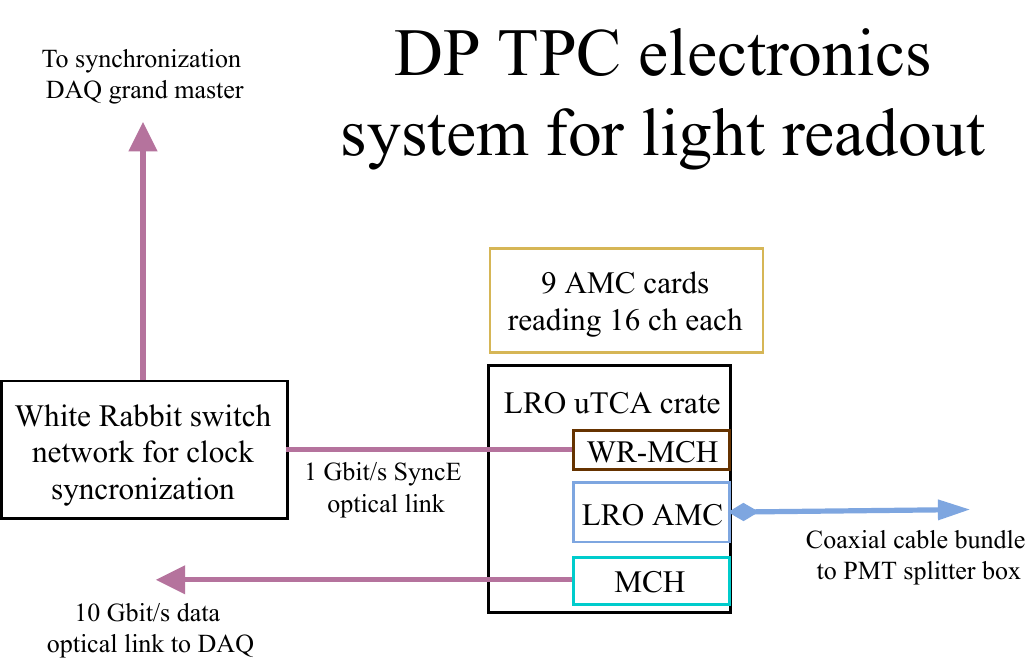}
\end{dunefigure}

Each \dword{utca} crate for either charge or light readout is connected to the \dword{daq} system via an optical fiber link that supports at least \SI{10}{Gbit/s}. 
Every crate also contains a 
\dword{wrmch} for the time synchronization of the digital electronics. This timing slave unit is connected via \SI{1}{Gbit/s} optical fiber to a master node that serves as a synchronization reference for all the connected slave nodes on the network. This system for the time synchronization is based on the commercially available components developed within the framework of the \dword{wr} project\footnote{\url{https://www.ohwr.org/projects/white-rabbit}.} with ad-hoc hardware and firmware development. The system performs automatic and continuous self-calibrations to account for any propagation delays and is able to provide sub-\si{\nano\s} accuracy for the timing synchronization.

\subsection{Design Considerations}
\label{sec:fddp-tpc-elec-des-consid}


The \dword{cro} electronics design covers the analog \dword{fe} cards containing pre-amplifier \dwords{asic} operating at cryogenic temperatures and digitization cards with the relevant system for their synchronization working in the warm environment outside of the cryostat. The system 
reads and digitizes signals from a total of \num{153600} channels (per one \dword{dpmod}) and is capable of continuously streaming the collected and losslessly compressed data to the \dword{daq} without any zero suppression. 
The design of the \dshort{cro} electronics system was developed to fit the following requirements:
\begin{itemize}
\item{The \dword{cro} electronics 
must measure signals of up to \SI{1200}{\femto\coulomb} without saturation; this has been optimized following \dword{mc} studies on the maximal occupancy per channel in shower events~\cite{WA105-TDR}. For a nominal \dword{crp} gain of \num{20}, a \dword{mip} signal is expected to be around \SI{30}{fC} -- the lowest limit that assumes a particle travelling horizontally with an azimuthal angle of \num{0} degrees -- giving 
a maximal operational range of up to \num{40} \dwords{mip}.}

\item{The electronic noise in the \dshort{cro} analog electronics is required to be \SI{< 2500}{e^{-}}. This condition can be derived from the requirement on the minimal \dword{s/n}, which should be greater than \num{5}:\num{1} once the charge attenuation is taken into account. Given the maximum drift distance of \SI{12}{\meter}, the largest attenuation factor due to electro-negative impurities assuming the \SI{3}{\milli\second} (minimally required) electron lifetime and the drift field of \SI{0.5}{\kilo\volt/\cm} is \num{0.08}. The smallest \dword{mip} signal with the \dword{crp} effective gain of \num{20} is therefore \SI{2.5}{\femto\coulomb} (\SI{15600}{e^{-}}).}

\item{The peaking time of the \dword{fe} analog amplifiers 
must be \SI{1}{\micro\second}. This requirement is driven by the need for optimal vertex resolution, determined in turn by the single track resolution and the power to separate two or more tracks that are close to one another.}

\item{The sampling frequency 
must be \SI{2.5}{\MHz} to match the peaking time of the \dword{fe} electronics.}

\item{The power dissipated by the \dword{fe} analog electronics 
must be below \SI{50}{\milli\watt/channel} in order to minimize the heat input to the cryostat volume.}

\item{The \dword{fe} analog electronics 
must be replaceable without the risk of contaminating the main \lar volume in order to guarantee the long-term reliability of the system.}

\item{The \dword{adc} resolution 
must be such that the noise is at the level of an \dword{adc}
given a dynamic range wide enough to match the response of the \dword{fe} amplifier. This can be achieved with a \num{12} bit \dword{adc}.}

\item{The digital electronics to be placed outside of the cryostat in the warm environment 
must be capable of adopting standard industrial components and solutions, to keep the costs lower and to benefit from  
technological evolution, e.g., higher network speeds.}
\end{itemize}
As described 
in subsequent sections, the achieved performance of the final system is significantly better than many of the listed requirements.  

The magnitude of the noise also 
has an effect on the quality of the lossless compression of the raw data. 
A compression factor of about \num{10} is achieved with the \rms noise level below \SI{1}{\dword{adc}}. To give an idea of the compression efficiency dependency on the noise level a compression factor of four is obtained with the noise at around \SI{1.5}{\dword{adc}} counts. 

The primary objective of the \dword{lro} system is to detect signals, from a minimum of one \phel on one \dword{pmt}, giving a precise timestamp that can be used in conjunction with the charge signals to determine the absolute event time ($T_0$). 
Precise measurements of \dword{lro} signal charge allow the continual monitoring of the \dword{pmt} gain at the single \phel level, and the determination of the number of photons in each scintillation event.  In addition, an \dword{adc}  continuously streams data, downsampled to \SI{400}{ns} as for the \dword{cro} signals,  which, amongst other items, allows measurements of the scintillation time profile. The \dword{lro} system also reads \num{20} channels from reference SiPM sensors from the \dword{pd} calibration system.

The cryogenic analog electronics for the \dword{cro} is housed in the dedicated \dwords{sftchimney}. 
Its design 
enables access to the \dword{fe} card for possible replacement without 
risk of contaminating the pure \lar in the main cryostat volume. The chimneys 
possess a cooling system that can control the temperature around the \dword{fe} cards to roughly \SI{110}{\kelvin} 
to reach their optimal noise level and 
that compensates for the heat input from the chimneys into the cryostat volume. 

The digital electronics for both charge and light readout is located in the warm environment on the top of the cryostat supporting structure and is therefore easily accessible. This fact removes any constraints associated with the accessibility and operation in cryogenic environments allowing for the usage of standard components and industrial solutions in the design. Digital electronics must be continuously and automatically synchronized to better than \SI{400}{\nano\s} to ensure the correct temporal alignment of the \dword{adc} samples from all of the readout channels. This is a minimal requirement dictated by the fact that the sampling rate is \SI{2.5}{\MHz}.  

Key parameters for the electronics system design are summarized in Table~\ref{tab:dpele-physicsparams}. 

\begin{dunetable}
[Parameters for the TPC electronics system design]
{lr}
{tab:dpele-physicsparams}
{Parameters for the  TPC electronics system design. The numbers are given for one \dword{dpmod}.}   
Parameter & Value  \\ \toprowrule
  \dword{cro} channels    &  \num{153600}            \\ \colhline
  \dword{cro} continuous sampling rate & \SI{2.5}{\MHz}\\ \colhline
  \dword{cro} \dword{adc} resolution & \num{12}\,bit           \\ \colhline
  \dword{cro} data compression factor   & \num{10}    \\ \colhline 
  \dword{cro} data flow  & \num{430}\,Gibit/s          \\ \colhline 
  \dword{lro} channels       & \num{720}               \\ \colhline
  \dword{lro} continuous sampling rate & \SI{2.5}{\MHz} \\ \colhline
  \dword{lro} \dword{adc} resolution & \num{14}\,bit            \\ \colhline
  \dword{lro} data compression factor  & \num{1}       \\ \colhline
  \dword{lro} data flow   & \num{24}\,Gibit/s          \\ 
\end{dunetable}

\subsection{Scope}
\label{sec:fddp-tpc-elec-scope}

The scope of the TPC electronics system covers the procurement, production, testing, validation, installation, and commissioning of all the components necessary to ensure the complete readout of the charge and light signals from a given \dword{dpmod}. This includes: 
\begin{itemize}
\item{Cryogenic analog \dword{fe} cards for charge readout;}
\item{\dword{amc} cards for charge and light readout;}
\item{The \dword{wrmch} cards for \dword{amc} clock synchronization;}
\item{\dword{utca} crates;}
\item{Switches for the \dword{wr} network;}
\item{\dwords{sftchimney};}
\item{Low-voltage power supplies, distribution, and filtering system for the \dword{fe} cards;}
\item{Flat cables connecting the \dword{fe} cards to the warm flange interface of the \dwords{sftchimney};}
\item{VHDCI cables connecting the warm flange interface of the \dwords{sftchimney} to \dwords{amc}.}
\end{itemize}


The total numbers for components to be procured to instrument one \dword{dpmod} are given in Table~\ref{tab:dpele-num-components}

\begin{dunetable}
[Numbers for \dual electronics components to procure]
{lr} {tab:dpele-num-components}
{Numbers for \dual electronics components to procure for one \dword{dpmod}.}
Name & Number  \\ \toprowrule
\dword{cro} cryogenic \dwords{asic} (\num{16} ch) & \num{9600} \\ \colhline
\dword{cro} cryogenic analog \dword{fe} cards (\num{64} ch) & \num{2400} \\ \colhline
\dword{cro} \dwords{amc} & \num{2400} \\ \colhline
\dwords{sftchimney} & \num{240} \\ \colhline
Flat cables for \dword{sftchimney} (\num{68} ch) & \num{2400} \\ \colhline
Flat cables for \dword{sftchimney} (\num{80} ch) & \num{2400} \\ \colhline
VHDCI cables (\num{32} ch) & \num{4800} \\ \colhline
\dword{lro} \dwords{amc} with analog \dword{fe} & \num{45} \\ \colhline
\dword{utca} crates & \num{245} \\ \colhline
\dword{wrmch} units & \num{245} \\ \colhline
WR switches (\num{18} ports) & \num{16} \\ 
\end{dunetable}

\section{TPC Electronics System Design}
\label{sec:fddp-tpc-elec-design}


The \dword{cro} \dword{fe} analog electronics is based on a cryogenic \dword{asic} chip with a large dynamic range (up to \SI{1200}{\femto\coulomb}) to accomodate the charge amplification in the 
\dword{crp}. The \dword{fe} cards read \num{64} \dword{crp} channels each. They are mounted in dedicated  \dwords{sftchimney} and are located within a short distance (\SI{<1}{\metre}) from each \dword{crp} to minimize the noise caused by long cables (large cable capacitance). The cards remain accessible throughout the \dword{dpmod} operation. Each \dword{sftchimney} 
accommodates \num{10} \dword{fe} analog cards, which corresponds to the readout of \num{640} \dword{crp} channels per chimney. There are, therefore, \num{240} \dwords{sftchimney} to be installed for the charge readout in a given \dword{dpmod}.   

The differential analog signals from the analog \dword{fe} cards, carried by the twisted-pair ribbon cables and routed via an interface flange of the \dwords{sftchimney}, are digitized by the \dword{amc} cards located 
outside of the cryostat. These cards are 
installed in \dword{utca} crates placed in the immediate vicinity of the \dwords{sftchimney} (one crate per chimney). 
In the design for \dword{pddp}, each \dword{amc} digitizes \num{64} channels, corresponding to reading one \dword{fe} analog card. Each \dword{utca}  contains \num{10} \dwords{amc} reading a total of \num{640} channels. However, an implementation with \dwords{amc} supporting a higher channel density is being investigated for cost-reduction purposes. 

The \dword{lro} \dword{fe} analog and digital electronics is based on a custom-built \dword{amc}. The card contains a \dword{catiroc} \dword{asic} \cite{Blin:2017}, which is used to determine 
the charge and start times of signals from each individual \dword{pmt} to high precision. In addition, a \SI{14}{bit} \SI{65}{MHz} \dword{adc} digitizes the data for continuous streaming of the \dword{pmt} signals. Each card reads up to \num{16} channels. A potential 
upgrade is to increase the channel density per card to \num{32} channels. The \dword{lro} cards are housed in five dedicated \dword{utca} crates located close to the \dword{pmt} instrumentation \fdth{}s.

\begin{dunefigure}[Top corner view of \dword{dpmod}]{fig:dpele-sft-chimney-pattern}
{Corner view of \dword{dpmod} showing the pattern of the \dwords{sftchimney} and \dword{utca} crates for charge readout above \dwords{crp}.}
\includegraphics[width=0.6\textwidth]{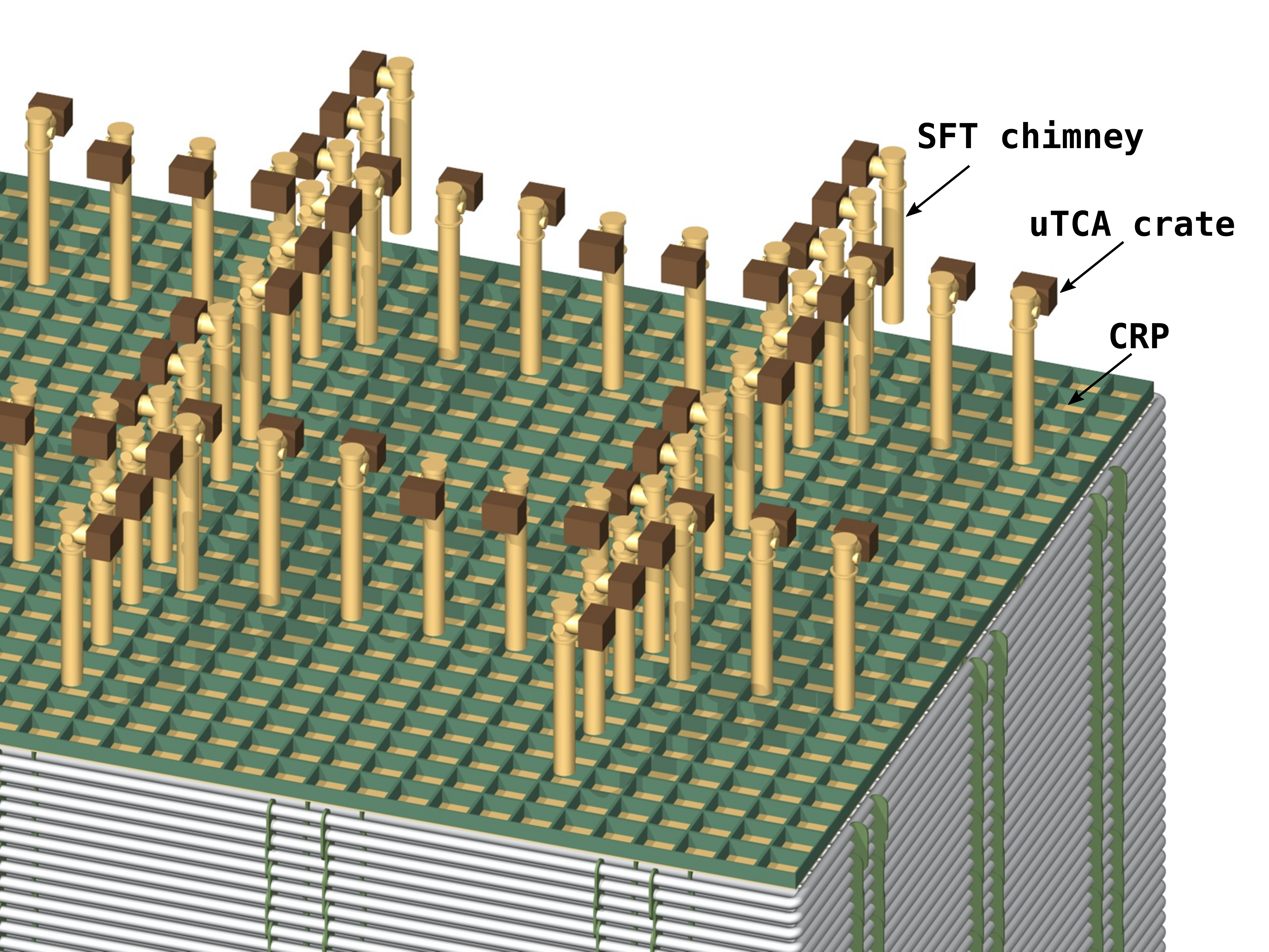}
\end{dunefigure}

Every \dword{utca} crate contains a network switch, \dword{mch}, via which the data are sent to the  \dword{daq} and to 
a (\dword{wrmch}) for 
time synchronization and trigger timestamp distribution to the \dwords{amc}. Both \dword{mch} and \dword{wrmch} require one optical fiber link each. 

The \dword{mch} switch streams the data from \dwords{amc} via a dedicated optical link. Currently \dword{pddp} uses \dword{mch} operating at \SI{10}{Gbit/s}. However, a move to \SI{40}{Gbit/s} links for the \dword{dpmod} implementation is under consideration because of the technology evolution with associated cost reductions and a possible increase in the channel density of each \dword{amc}.

The \dword{wrmch} time synchronization unit is based on the \dword{wr} system, which provides hardware and protocols for the network-based sub-\si{\nano\s} synchronization between a master and different slave nodes. The connection of the \dword{wrmch} to the \dword{wr} network is done via \SI{1}{Gbit/s} synchronous Ethernet optical link. \dword{wrmch} distributes the timing information for synchronization of the \dwords{amc} via the \dword{utca} backplane. In addition, this unit can be used to transmit triggers to the digitization units within the crate. This is achieved by sending it dedicated data packets containing trigger timestamp information. 

Figure~\ref{fig:dpele-sft-chimney-pattern} shows a corner view of the \dword{dpmod} illustrating the pattern of the \dwords{sftchimney} and the attached \dword{utca} crates above the \dwords{crp}. Each crate/\dword{sftchimney} collects signals from \SI{3}{\meter} long strips of two \SI[product-units=power]{1x3}{\meter} \dword{crp} segments. Each chimney completely 
penetrates the cryostat insulation layers (not shown in the figure). 

\begin{dunetable}
[Summary of some of the principal 
parameters of the TPC electronics system.]
{lr} {tab:dpele-numparts}
{Summary of some of the principal 
parameters  of the TPC electronics system for charge and light readout of a \dword{dpmod}.}
Name & Number  \\ \toprowrule
   \dword{cro} \dwords{sftchimney}/\dword{utca} crates              &  \num{240}   \\ \colhline
   \dword{cro} channels per \dword{sftchimney}/\dword{utca} crate & \num{640} \\ \colhline
   \dword{cro} cryogenic analog \dword{fe} cards per \dword{sftchimney}    &  \num{10}     \\ \colhline
   \dword{cro} \dwords{amc} per \dword{utca} crate                       & \num{10}      \\ \colhline
   \dword{lro} \dword{fe} cards  per \dword{utca} crate & \num{9} \\ \colhline
   \dword{lro} channels per \dword{utca} crate & \num{144} \\ \colhline
   \dword{lro} \dword{utca} crate                      & \num{5} \\ \colhline
   \dword{wrmch} per \dword{utca} crate                 & \num{1} \\ 
\end{dunetable}

A short summary of some of  the number of principal components and channel granularity in the design of the \dual electronics is provided in Table~\ref{tab:dpele-numparts}. 

\subsection{Cryogenic Analog Front-end Electronics}
\label{sec:fddp-tpc-elec-design-cryofe}

The cryogenic amplifier \dword{asic} is the main component of the \dword{fe} analog cards. Its design is based on the CMOS \SI{0.35}{\micro\meter} technology and is an outcome of an R\&D  activity started in 2006, which resulted in several versions of the cryogenic amplifier for both \single and \dual \dword{lartpc} detectors. Two principal versions of \dword{asic} chips have been produced for \dual \dword{lartpc} operation. In the first version the amplifiers have a constant gain in signal inputs of \numrange{0}{1200}\,\si{\femto\coulomb} (\numrange{0}{40} \dword{mip}). In the second, the amplifiers have a higher linear gain for signals up to \SI{400}{\femto\coulomb} (roughly \num{10} \dword{mip} signals) and a logarithmic response in the \numrange{400}{1200}\,\si{\femto\coulomb} range. This double-slope behavior is obtained by using a MOSCAP capacitor in the feedback loop of the amplifier that changes its capacitance above a certain signal threshold. The aim of this solution is to optimize the resolution for few-\dword{mip} charge depositions while preserving the overall large dynamic range of the amplifier.

\begin{dunefigure}[Cryogenic \dword{fe} \dword{asic} properties]{fig:dpele-fe-asic-prop}
{Cryogenic \dword{fe} \dword{asic} properties: amplifier response (left) and noise (right) at different temperatures measured at the output of differential buffer.}
\includegraphics[width=0.49\textwidth]{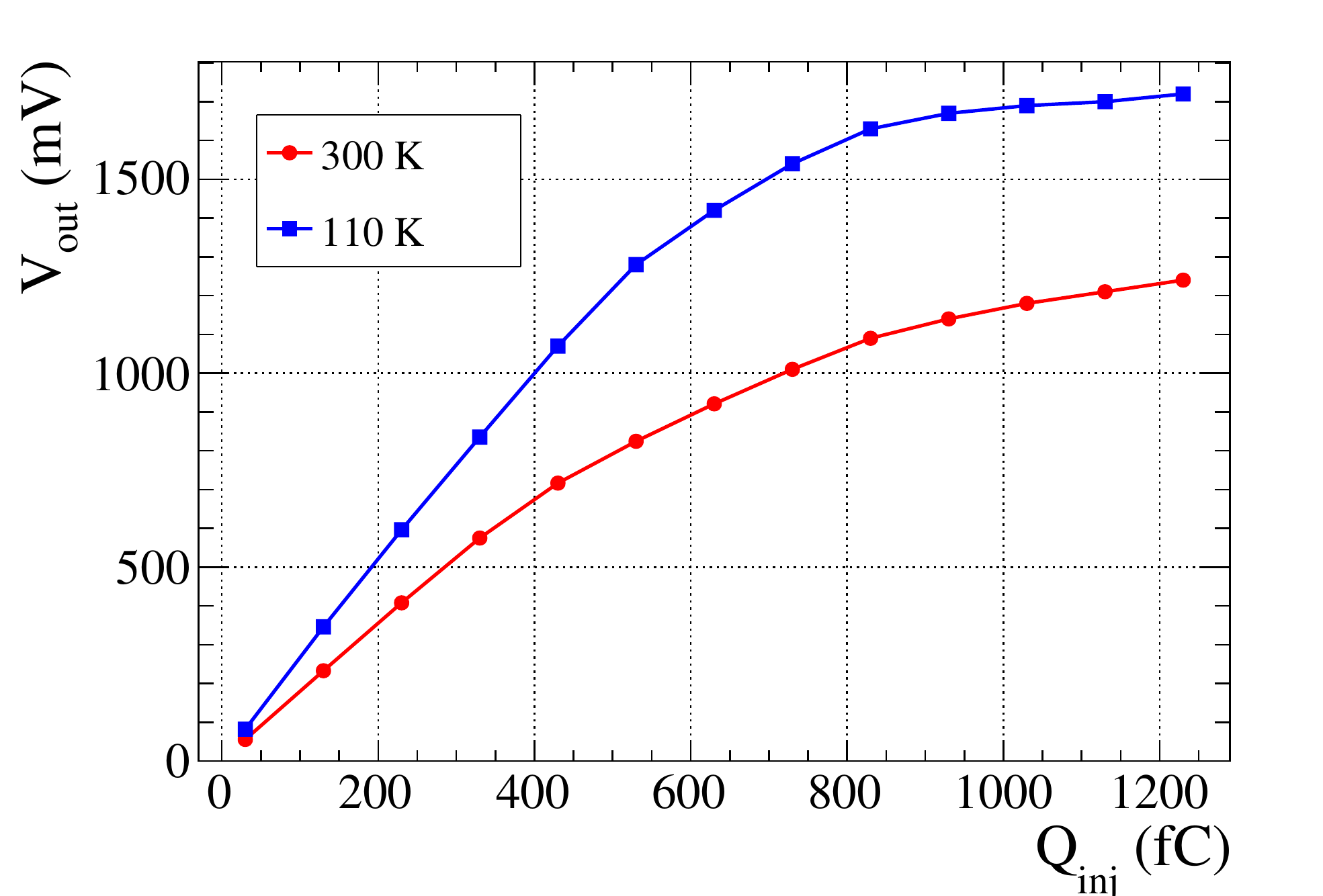}
\includegraphics[width=0.49\textwidth]{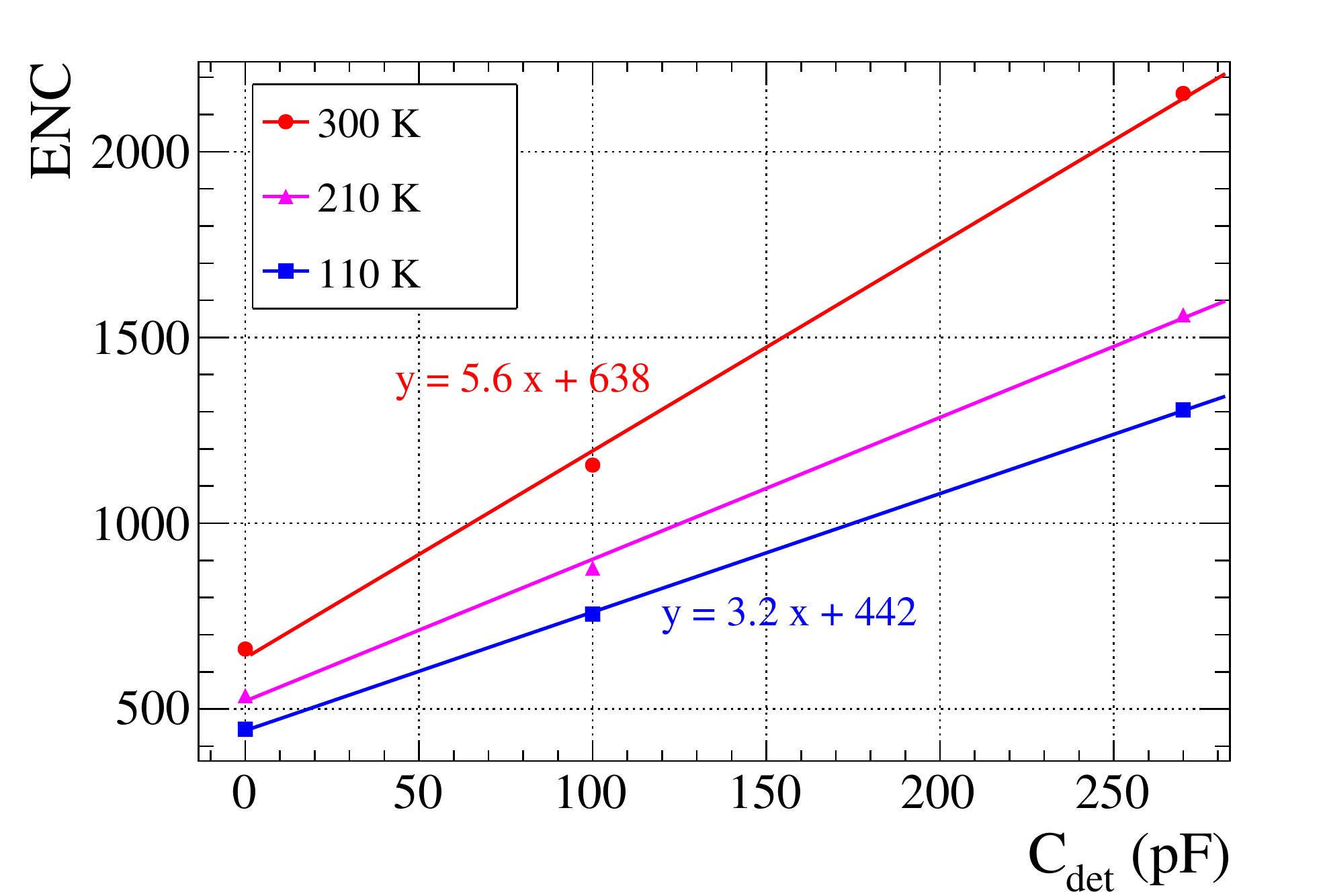}
\end{dunefigure}

The \dword{asic} version with the double-slope gain was selected for \dword{pddp} and has been adopted for the \dual TPC electronics design. The left plot in Figure~\ref{fig:dpele-fe-asic-prop} illustrates the response of this amplifier for different values of the injected charge, while that on the right shows the measured noise in units of Equivalent Noise Charge (ENC) as a function of the detector capacitance at different temperatures. 
For the \dword{crp} anode (detector) capacitance of \SI{480}{\pico\farad} per \SI{3}{\metre} strip, the expected noise is around \SI{2000}{e^{-}} at \SI{110}{\kelvin}. Each \dword{asic} contains \num{16} amplifier channels with differential line buffers and has a power consumption of \SI{11}{\milli\watt/channel}, surpassing the \SI{< 50}{\milli\watt/channel} requirement. 

\begin{dunefigure}[Image of an analog \dword{fe} card mounted on the extraction blade]{fig:dpele-fe-card-blade-image}
{Image of an analog cryogenic \dword{fe} card mounted on the extraction blade, which is a part of the \dword{sftchimney} sub-system.}
\includegraphics[width=0.7\textwidth]{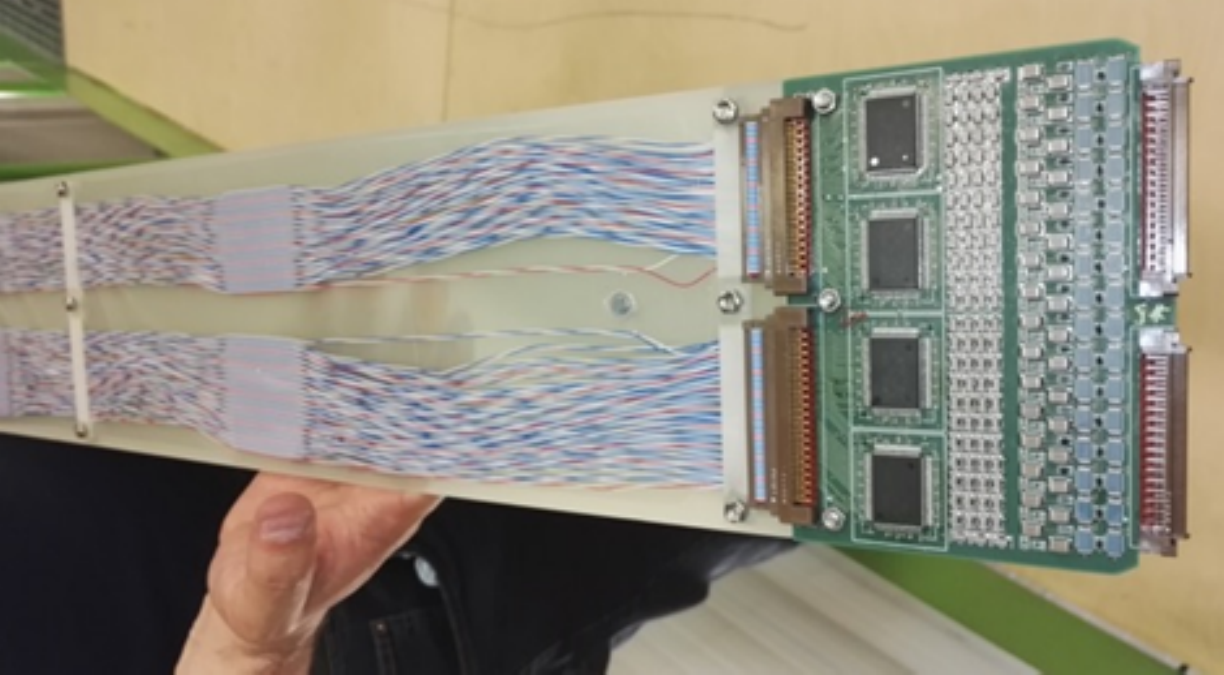}
\end{dunefigure}

Each cryogenic \dword{fe} card, shown in Figure~\ref{fig:dpele-fe-card-blade-image}, 
holds four \dword{asic} amplifier chips and a few passive discrete components. The input stage of each amplifier channel has a \SI{1}{\giga\ohm} resistor to ground followed by a \SI{2.2}{\nano\farad} decoupling capacitor; both are rated for \dword{hv} operation. The 
resistor grounds the \dword{crp} anode strips. A TVS diode\footnote{Bourns\texttrademark{} CDSOD323-T08LC TVS diode.} protects each input stage against discharges coming from the \dword{dpmod}. This component was selected after studying the performance of different devices for the electrostatic discharge protection by subjecting them systematically to discharges of a few \si{kV} with an energy similar to that of the \dwords{lem} in the \dword{crp}. The \dword{fe} cards also house the blocking capacitors for further filtering of the low-voltage power lines. These are first filtered at the power supply and transported via shielded cables.

\begin{dunefigure}[Noise measurements in the \dword{wa105} in different conditions]{fig:dpele-311-noise}
{Noise of measurements in the \dword{wa105} in different conditions. Left: warm with the slow control cables connected to the cryostat flanges (red points) and disconnected (black points). The horizontal dashed line in this plot indicates the noise level expected for \SI{3}{\meter} strips. Right: warm (red points) and cold (black points) with the slow control cables disconnected. The channels with negative (positive) channel number correspond to the strips of \SI{3}{\meter} (\SI{1}{\meter}).}
\includegraphics[width=0.45\textwidth]{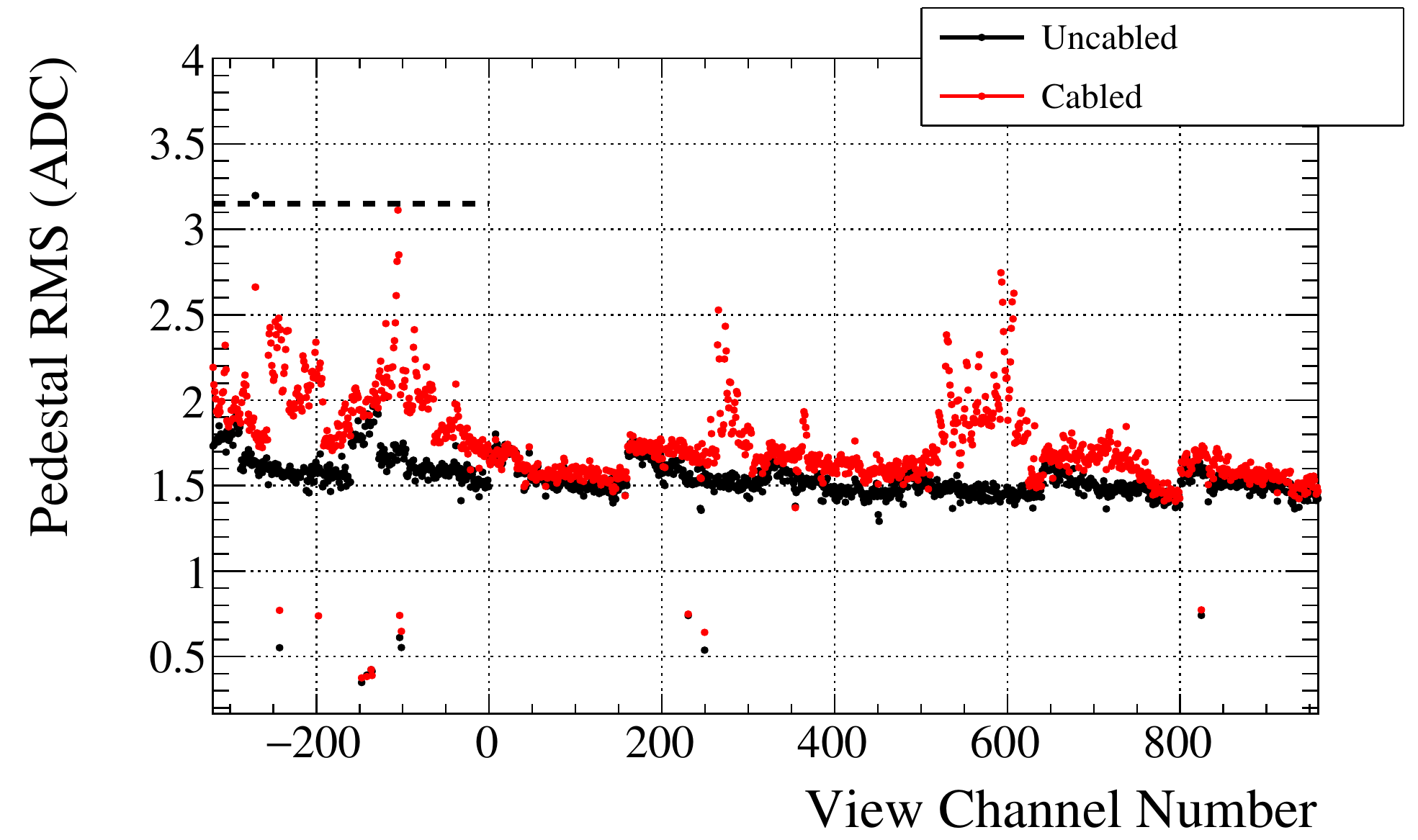}
\includegraphics[width=0.45\textwidth]{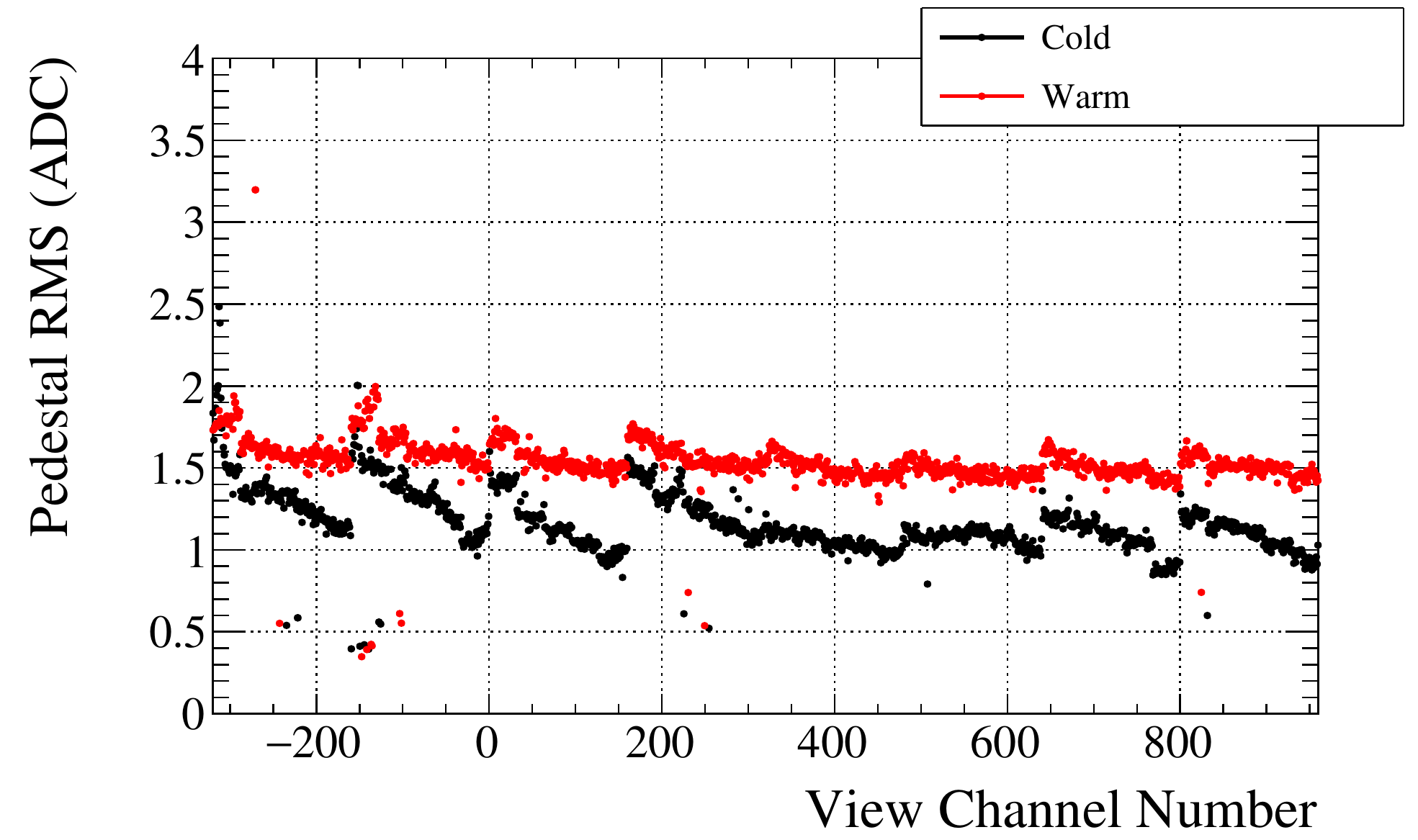}
\end{dunefigure}

Although the \dword{fe} amplifier \dwords{asic} are in a shielded environment provided by the chimneys, which act as Faraday cages,
interference from other equipment (via a noisy ground or ground loops) could significantly worsen the noise performance from the design target. Figure~\ref{fig:dpele-311-noise} shows 
results of 
noise measurements performed under different conditions in the \dword{wa105}. The channels reading \SI{3}{\meter} (\SI{1}{\metre}) long strips correspond to negative (positive) channel numbering in the plots and the \num{1} \dword{adc} count is equivalent to about \num{900} electrons. The left plot of Figure~\ref{fig:dpele-311-noise} shows the noise measurements performed at warm temperature with and without slow control cables (\dword{crp} \dword{hv}, \dword{crp} motors, level meters, temperature probes, etc.) connected. 
The noise is clearly affected by the grounding of the slow controls: the average value of the noise \rms is around \num{1.7} \dword{adc} with slow control cables connected, and decreases to about \num{1.5} \dword{adc} counts when they are removed. 
The grounding scheme of the \dword{wa105} does not correspond to the one 
planned for the \dword{dpmod}, where all electrical equipment is referred uniquely to the cryostat ground and is completely insulated from the external environment. 

One interesting feature, particularly visible with the cables disconnected, is 
the similarity of the noise measured for the channels connected to the \SI{1}{\meter} and \SI{3}{\meter} long strips in the \dword{wa105}. 
Given that the longer strips have 
three times the input capacitance of the shorter ones, the expected noise (see Figure~\ref{fig:dpele-fe-asic-prop}) for these should be larger by about a factor of two as indicated by the dashed line in the plot. In addition, the noise on the short strips is lower than expected (\num{1.5} \dword{adc}) for the \SI{160}{pF/m} strip input capacitance (\num{1.7} \dword{adc}). The reason for this behavior 
in the \dword{wa105} is still under investigation. 
Measurements have already shown that the capacitance of the \dword{crp} anode strips is not purely to ground, but rather it is driven by the inter-strip couplings, which creates a more complicated electrical network. 

Figure~\ref{fig:dpele-311-noise} (right) also shows a comparison of the noise measurements in the \dword{wa105} taken with the \dword{fe} electronics at warm (red points) and cold (black points) at around \SI{150}{\kelvin}. The slow control cables were disconnected in both cases. However, the cold measurements were performed with the recirculation pump active and the cathode \dword{hv} connected. 
The \rms noise averaged over all channels decreases by about 25\% from roughly \SI{1.5}{\dword{adc}} to \SI{1.1}{\dword{adc}} when the \dword{fe} analog cards are cold. For comparison, the expected signal for a \dword{mip} with the \dword{crp} gain of \num{20} should be around \SI{200}{\dword{adc}} counts. 

The overall grounding principle of the \dword{wa105} was based on 
using the cryostat as the ground reference. The low-voltage power supplies for the \dword{fe} analog electronics and the \dword{utca} crates were powered 
using insulating transformers to ensure that 
no other ground could interfere. 
In contrast, the design of the slow control system did not include any insulation transformers. This equipment was grounded to the building electrical network, thereby creating an interference with the ground of the cryostat. Stricter treatment of the ground connections, as 
implemented for \dword{pddp} and planned for the \dword{dpmod}, and a lower \dword{sftchimney} operating temperature of around \SI{110}{\kelvin} (from \SI{150}{\kelvin}) should help to further reduce the noise 
from the levels observed in the \dword{wa105}.

\subsection{Signal Feedthrough Chimneys}
\label{sec:fddp-tpc-elec-design-sft}

The \dwords{sftchimney} are designed to enable access to the \dword{fe} analog electronics for a potential repair or exchange while the \dword{dpmod} is filled and in operation. 
In addition, their metallic structure acts as a Faraday cage isolating the \dword{fe} \dwords{asic} from environmental interference.  Each \dword{sftchimney} hosts \num{10} analog cryogenic \dword{fe} cards (reading \num{640} channels in total).  
Details of the design are illustrated in Figure~\ref{fig:dpele-sft-chimney-design}. 

\begin{dunefigure}[Details of the \dword{sftchimney} design]{fig:dpele-sft-chimney-design}
{Details of the \dword{sft} chimney design.}
\includegraphics[width=0.8\textwidth]{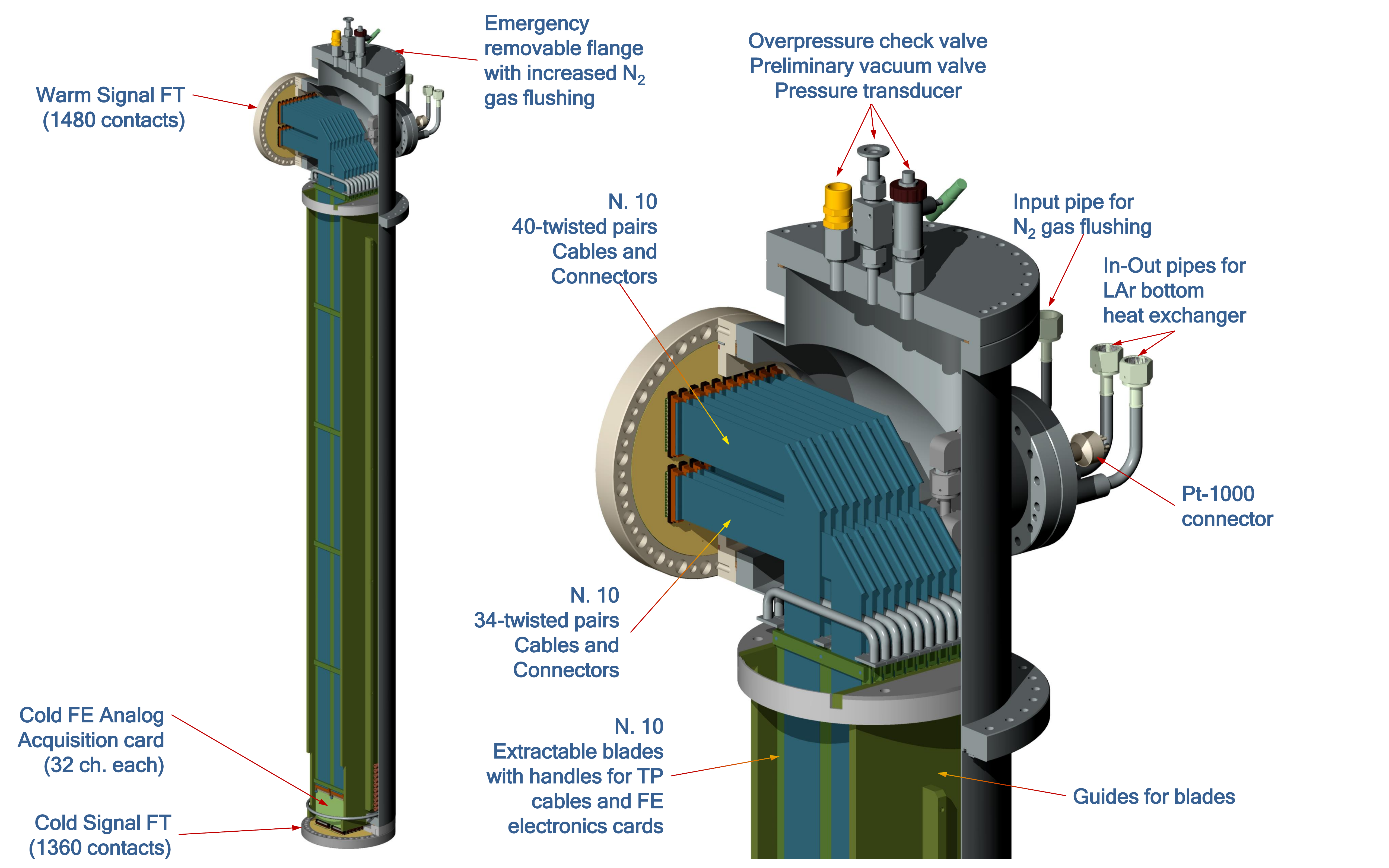}
\end{dunefigure}

The chimneys are closed at the bottom and top with vacuum-tight \fdth flanges whose function is to dispatch the signal and slow control lines. The \fdth at the bottom, the cold (signal) \fdth, isolates (ultra-high vacuum tightness standard) the inner volume of the \dword{dpmod} from the chimney volume and interconnects the signals from the \dword{crp} to the analog \dword{fe} cards. The \fdth at the top, the warm (signal) \fdth, seals the chimney from the outside environment. It also passes the low voltage and control lines to the \dword{fe} electronics inside and brings out the differential analog signal lines from the \dword{fe} amplifiers. 

The \dword{sftchimney} volume is filled with nitrogen gas at near-atmospheric pressure. The temperature inside the chimney can be adjusted using a heat exchanger copper coil cooled with \lar. It is located at the bottom close to the cold \fdth around the \dword{fe} cards. 
This cooling system 
mitigates the heat input to the main \dword{dpmod} volume and provides an optimal (lowest noise) operating temperature for the \dword{fe} electronics of around \SI{110}{K}. A pressure release valve, indicated in Figure~\ref{fig:dpele-sft-chimney-design}, protects the structure from an accidental overpressure in the inner volume. 

The expected heat input from a given \dword{sftchimney} is about \SI{20}{\watt}. This number includes the heat through the twisted-pair cables connected to the warm \fdth, the heat in the  \dword{sft} outer metallic tube, as well as the heat dissipation by the \dword{fe} cards. A total heat input from all \num{240} \dwords{sftchimney} is at the level of \SI{5}{\kilo\watt}. 

\begin{dunefigure}[Signal \fdth chimney cold flange]{fig:dpele-sft-cold-pcb}
{\dword{sftchimney} cold \fdth flange with one of the \dword{fe} cards mounted.}
\includegraphics[width=0.6\textwidth]{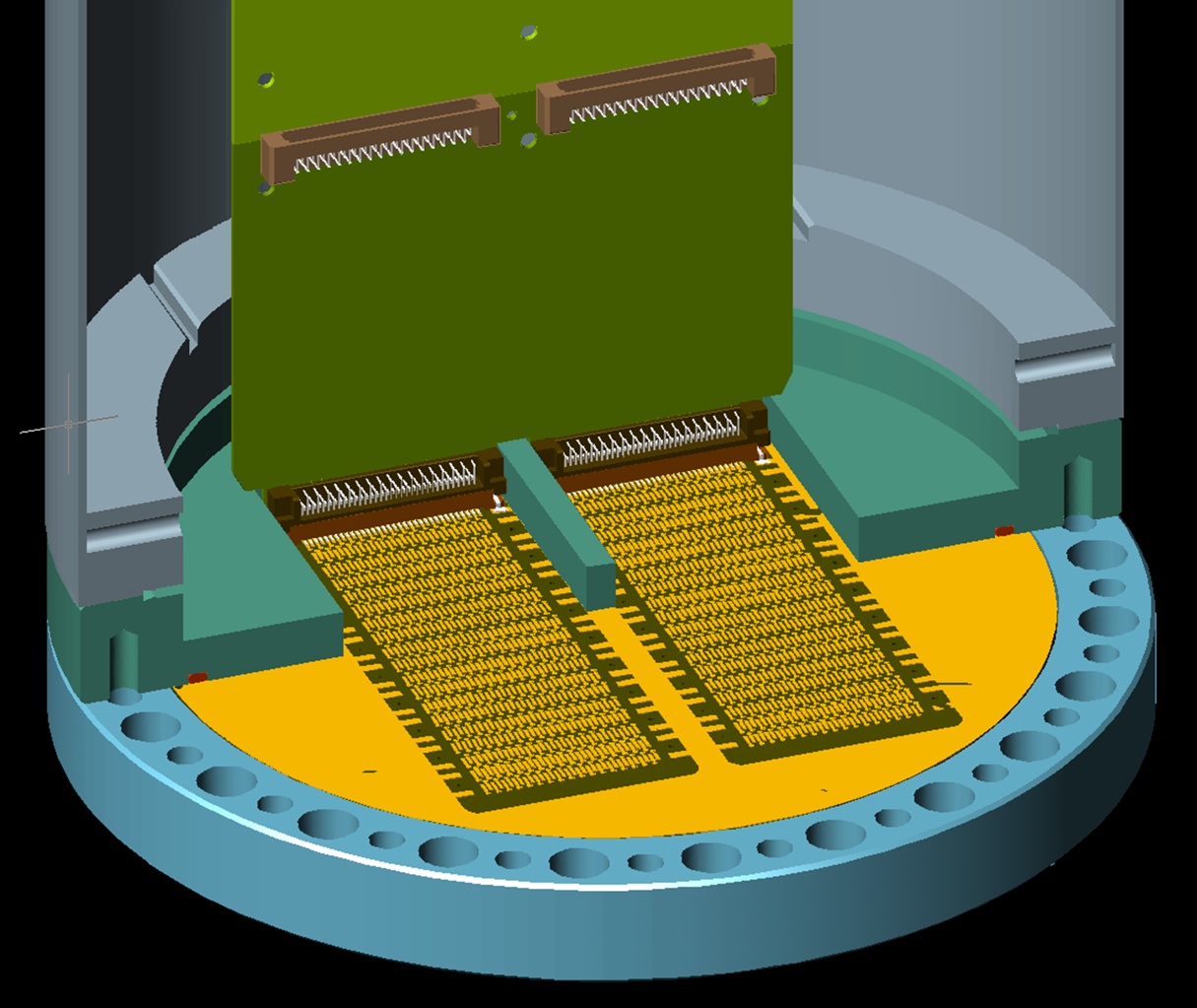}
\end{dunefigure}
The analog \dword{fe} cards are inserted directly onto the PCB of the cold \fdth of \dword{sftchimney}
(see Figure~\ref{fig:dpele-sft-cold-pcb}). The other side of the PCB (facing inside the cryostat) hosts the connectors for the flat cables coming from the \dword{crp} anodes.  The \dword{fe} cards are mounted on \SI{2}{\m} long blades made from FR4 that enable the insertion and extraction of the electronics, and also support the flat cables carrying signals, low voltages, and slow control to and from the warm flange interface.  The blades slide along the rails installed inside the chimney at opposite sides; 
these rails guide the \dword{fe} cards to their respective connectors on the cold \fdth. 

Prior to the commissioning of a \dword{dpmod}, the chimneys are evacuated via a dedicated ISO standard KF16 port (see Figure~\ref{fig:dpele-sft-chimney-design}) and then filled with nitrogen gas. This ensures the removal of 
any moisture that would otherwise condense around the \dword{fe} cards, once the \dword{dpmod} is filled with the \lar, damaging the electronics. To access the \dword{fe} cards once the \dword{dpmod} is cold, the stainless steel flange at the top of the \dword{sftchimney} (Figure~\ref{fig:dpele-sft-chimney-design}) must be removed. This procedure requires continuous flushing of nitrogen gas at slight over-pressure (with respect to atmospheric) in order to prevent the humid air entering. 
Once a chimney is opened, it is possible to extract the blades with the \dword{fe} cards after unplugging the flat cables (two per card) connected on the inner side of the warm flange (Figure~\ref{fig:dpele-sft-chimney-design}).

The procedure to access the \dword{fe} cards under cold operation was successfully tested during the operation of the \dword{wa105} detector. The 
top of the chimney was very close to room temperature, allowing manipulation of the cable connections on the warm \fdth flange without 
cryogenic gloves. The movement of the blades on the rails and the \dword{fe} card extraction and insertion did not indicate any mechanical problems 
due to the shrinking of various elements 
at lower temperatures.  The signals from the \dword{fe} cards that underwent 
this process were also checked and 
all channels functioned properly.


\subsection{Digital Advanced Mezzanine Card Electronics for Charge Readout}
\label{sec:fddp-tpc-elec-design-amc}
The 
 \dword{cro} \dword{amc} cards 
 read and digitize the data from the \dword{fe} amplifiers and 
 transmit them to the \dword{daq} system. The cards also include a last stage of analog shaping before the \dword{adc} input. The analog \dwords{fe} produce differential unipolar signals defined with respect to a baseline offset. 
Prior to the digitization, this offset is removed and the signals are subtracted in the analog input stage of the digital electronics.
Each card has eight \dword{adc} chips (Analog Devices, AD9257\footnote{Analog Devices\texttrademark{}, 
 \url{http://www.analog.com/media/en/technical-documentation/data-sheets/AD9257.pdf}.}, Table~\ref{tab:dpele-adc9257}), two dual-port memories (Integrated Device Technology IDT70T3339\footnote{Integrated Device Technology\texttrademark{} (IDT), \url{https://www.idt.com/document/dst/70t33391999-data-sheet}.}), and an \dword{fpga} 
 (Alterra\footnote{Alterra\texttrademark{}, \url{https://www.altera.com/products/fpga/cyclone-series/cyclone-v/overview.html}.} Cyclone V) on board. The \dword{fpga} provides a virtual processor (NIOS) that handles the readout and the data transmission. The choices for all of the components have been optimized with respect to the design requirements and technical criteria such as costs, chip footprint (small enough to fit on the \dword{amc}), power consumption, and ease of use (available functionality). 

\begin{dunetable}
[Main characteristics of \dword{adc} AD9257]
{lr} {tab:dpele-adc9257}
{Main characteristics of \dword{adc} AD9257.}
Item &   \\ \toprowrule
Channels & \num{8} \\ \colhline
Sampling & up to \SI{40}{MSPS} \\ \colhline 
Resolution & \SI{0.122}{\milli\volt} \\ \colhline
Dynamic range & \num{14} bit/ \SI{2.0}{\volt} \\ \colhline
Differential non-linearity & typical \num{\pm0.6} LSB\\ 
& with min. \num{-1.0} and max. \num{+1.7} LSB  \\ \colhline
Integral non-linearity & typical \num{\pm1.1}  LSB\\
& with min. \num{-3.1} and max. \num{+3.1} LSB  \\ 
\end{dunetable}

\begin{dunefigure}[Block diagram of \dword{amc}]{fig:dpele-amc-scheme}
{Block diagram of \dword{amc}.}
\includegraphics[width=0.9\textwidth]{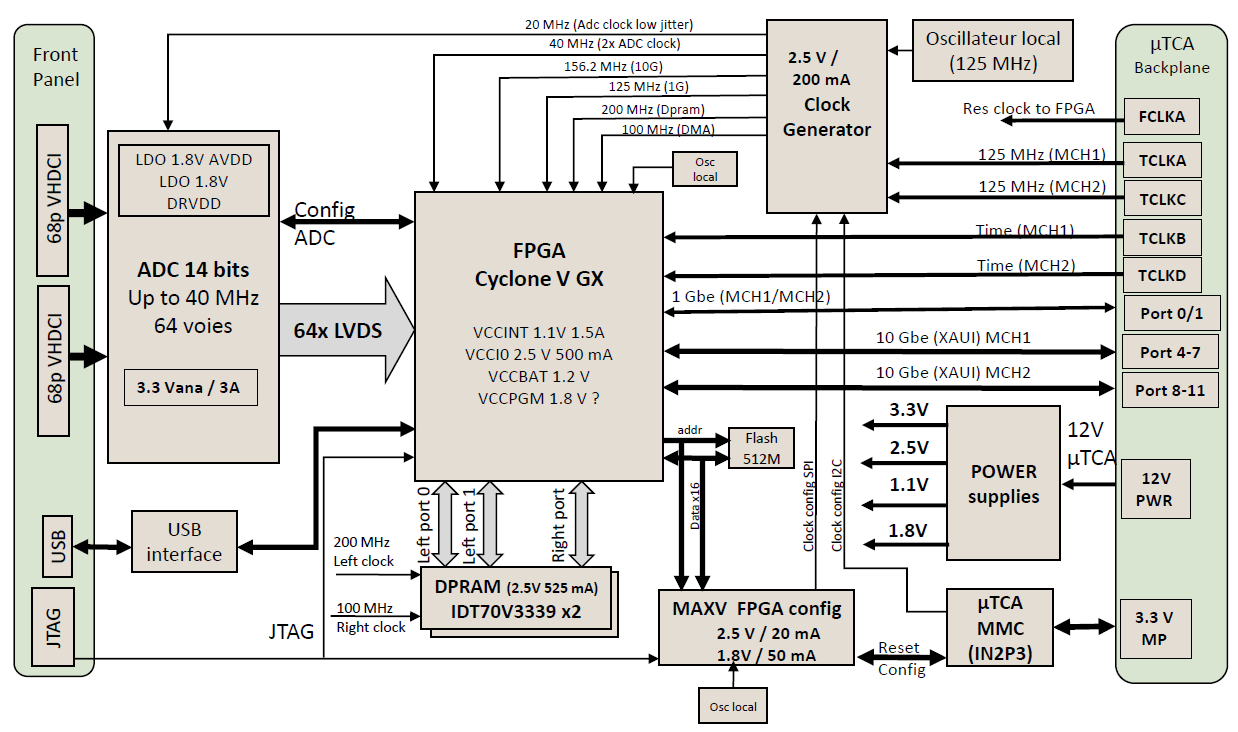}
\end{dunefigure}
Figure~\ref{fig:dpele-amc-scheme} shows block diagram of the \dword{amc} functionality. Each \dword{amc} generates a continuous compressed stream of \SI{2.5}{MSPS} \num{12}\,bit data per readout channel. The on-board ADCs operate at a rate of \SI{25}{\MHz} per channel. The data are down-sampled in the \dword{fpga} to \SI{2.5}{\MHz} by performing ten-sample averaging, which leads to further digital filtering of the noise. The data, consisting of only the  \num{12} most significant bits from each digitized \num{14}\,bit sample, are then compressed using an optimized version of the Huffman algorithm and organized in frames for transmission.  The frames contain the absolute timing information of the first data sample for reliability purposes. In the current design, each \dword{amc} has \num{64} channels and reads one analog \dword{fe} card.

The \dwords{amc} are housed in \dword{utca} crates and send their data via the \dword{mch} switch. The timing synchronization of \dwords{amc} is achieved via a \dword{wrmch} module (also housed in the crate) that is connected to the \dword{wr} network. In addition, a  \dword{wrmch} could also be used for triggered readout of \dwords{amc} by sending it dedicated packets containing trigger timestamp information over the \dword{wr} network.

In \dword{pddp}, \dwords{amc} are operated in the triggered mode, reading a \SI{4}{\milli\second} drift time window at a trigger rate of \SI{100}{Hz}, which is not far from 
continuous readout mode. The analog data are continuously digitized and buffered. It is possible to acquire a sub-sample of these data 
by providing \dword{amc} with a timestamp generated by an external trigger. The timestamp defines the start time for the data sequence to be read, while the length of the sequence is determined by the size of the drift window. In \dword{pddp} this length corresponds to \num{10000} \SI{400}{\nano\second} samples per full drift window (\SI{4}{\milli\second}). 
 Triggers (beam counters, cosmic-ray counters, \dwords{pmt} detecting the UV light, and starts of beam spills) are time stamped in a dedicated \dword{wr} slave node (\dword{wr}-TSN), an FMC-DIO mezzanine mounted on \dword{wr} SPEC carrier card, which runs a custom firmware and is hosted in a computer. The \dword{wr}-TSN is connected to the \dword{wrgm} for synchronization and for transmission of the trigger information. The timestamp data produced by the \dword{wr}-TSN are sent over the \dword{wr} network as Ethernet packets with a customized protocol. 

\subsection{Electronics for Light Readout}
\label{sec:fddp-tpc-elec-design-lro}
%
The \dword{lro} card is a \num{16} channel \dword{amc} containing one \num{16} channel \num{14} bit \SI{65}{\MHz} \dword{adc} (AD9249) and one \dword{catiroc} \dword{asic}. A block diagram of the prototype board used for \dword{pddp} is shown in Figure~\ref{fig:dpele-lro-scheme} and a photo in Figure~\ref{fig:dpele-lro-proto} . In this prototype a mezzanine board containing the \dword{asic} and \dword{adc} sits on a commercial mother board (Bittware S4 \dword{amc}\footnote{Bittware\texttrademark{}, Inc., \url{https://www.bittware.com/fpga/intel/boards/s4am/}.}) with a high specification \dword{fpga} (Alterra\footnote{Alterra\texttrademark{}, \url{https://www.altera.com/products/fpga/stratix-series/stratix-iv/overview.html}.} Stratix IV). In the final implementation for the \dword{dpmod}, the mezzanine is integrated with the layout of the \dword{amc} board developed for the charge readout.  
A proposed upgrade is a \num{32} channel card, 
which would reduce the number of cards required and increase the channel density to \num{352} channels per \dword{utca} crate.

\begin{dunefigure}[Block diagram of \dword{lro}]{fig:dpele-lro-scheme}
{Block diagram of \dword{lro} prototype.}
\includegraphics[width=0.8\textwidth]{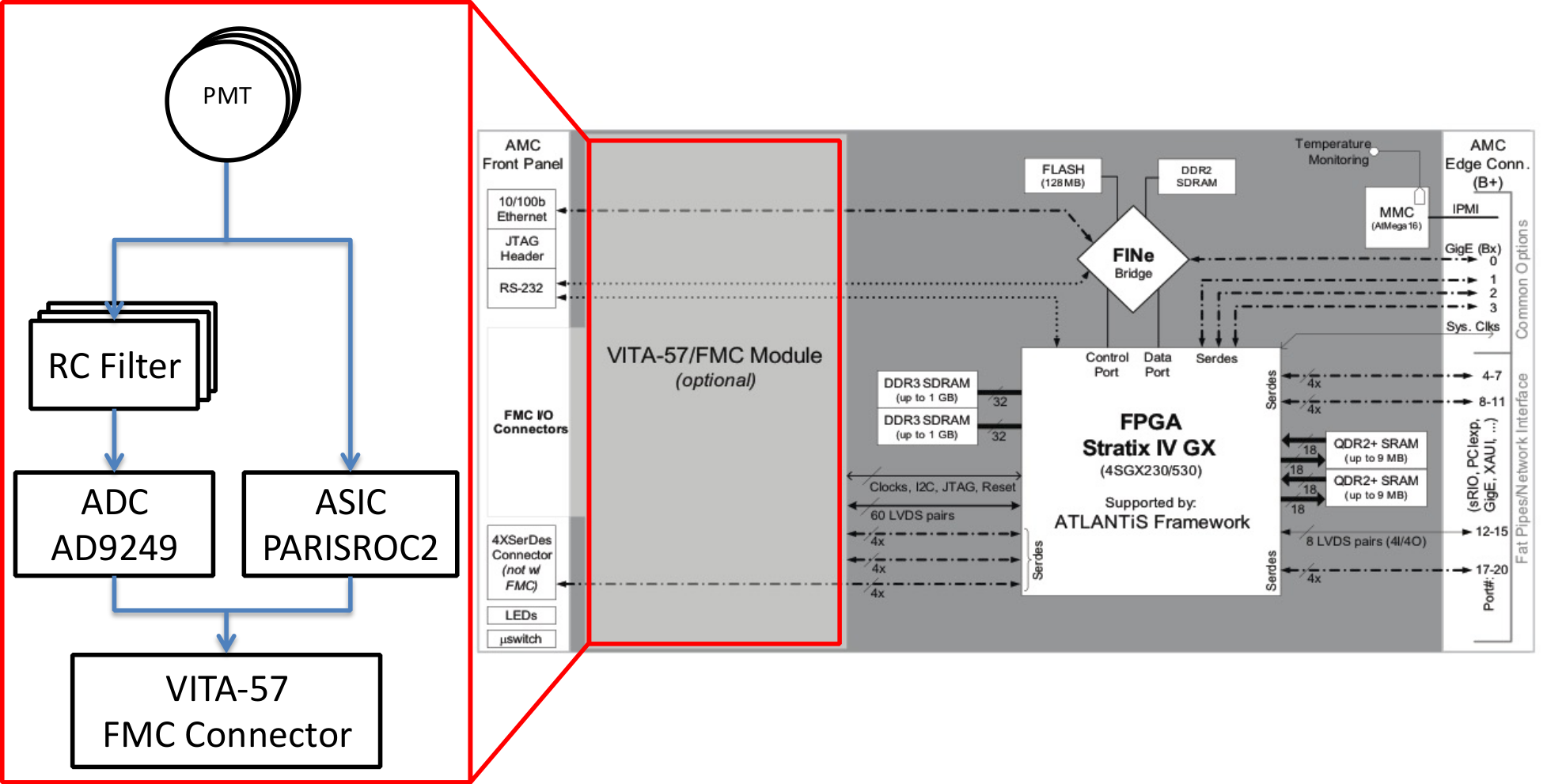}
\end{dunefigure}

\begin{dunefigure}[Photo of prototype \dword{lro} card]{fig:dpele-lro-proto}
{A photo of the \dword{lro} prototype.}
\includegraphics[width=0.7\textwidth]{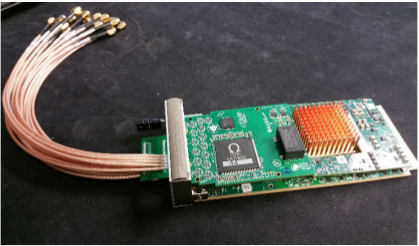}
\end{dunefigure}

The analog signals from each \dword{pmt} channel are split equally into two separate branches (see Figure~\ref{fig:dpele-lro-scheme}) . One path (waveform branch), through an anti-aliasing low-pass filter and the \num{14} bit \SI{65}{\MHz} \dword{adc} (AD9249), produces continuous digitization of the \dword{pmt} waveform data, which are down-sampled to \SI{2.5}{MHz} prior to the transmission to \dword{daq}. The other (\dword{catiroc} branch) is routed directly to the \dword{catiroc} \dword{asic} for precise measurements of pulse charge and timing. Both paths produce data continuously and independently.

\subsubsection{Waveform branch} 
The main characteristics of the \dword{adc} used for continuous digitization of the \dword{pmt} signals are shown in Table \ref{tab:dpele-adc9249}.
\begin{dunetable}
[Main characteristics of \dword{adc} AD9249]
{lr} {tab:dpele-adc9249}
{Main characteristics of \dword{adc} AD9249.}
Item &   \\ \toprowrule
Channels & \num{16} \\ \colhline
Sampling & \SI{65}{MSPS} \\ \colhline
Resolution & \SI{0.122}{\milli\volt} \\ \colhline
Dynamic range & \num{14} bit/ \SI{2}{\volt} \\ \colhline
Differential non-linearity & typical \num{\pm0.6} LSB\\
& with min. \num{-0.9} and max. \num{+1.6} LSB  \\ \colhline
Integral non-linearity & typical \num{\pm0.9}  LSB\\
& with min. \num{-3} and max. \num{+3} LSB  \\ 
\end{dunetable}

For normal operation, in the continuous mode, the digitized signals are down-sampled by the \dword{fpga} to a coarse \SI{400}{ns} sampling to match that of the \dword{cro} and limit the quantity of data streamed. 
The use of a higher specification \dword{adc}, with time-sampling of \SI{15.4}{ns}, allows for greater flexibility. 
For particular calibration runs, waveforms with finer time sampling could be read-out, allowing studies of, e.g., the \lar scintillation time profiles. 
During normal operation, as well, online pulse processing is possible within the \dword{fpga} using the finer time-sampled waveforms (before the down sampling; this would enable continuous measurements of quantities such as the rise and fall times of the pulses. Even at the coarse sampling rate of \SI{400}{ns}, studies of the \lar scintillation time profile are possible (given the long fall-time constant of $\sim$\SI{1500}{ns}) 
as is matching of the electroluminescence signal (also known as proportional scintillation light) to that of the charge signal.  Low light-level signals, 
as from single or a few \phel{}s, 
will show no time structure, but will consist of one sample several LSB above the baseline. 

\subsubsection{CATIROC branch} 

The \dword{catiroc} is a \num{16} channel \dword{asic} dedicated to measurement of charge and precision timing of negative-polarity \dword{pmt} signals~\cite{Blin:2017}. It auto-triggers on single \phel{}s and can sustain a high dark rate of up to \SI{20}{kHz/channel}. Charge measurements are possible over the range of \SI{160}{fC} to \SI{70}{pC} (corresponding to approximately to a range of \numrange{1}{400} \phel{}s with a \dword{pmt} gain of \num{1E6}). Timing measurements per channel can 
reflect an accuracy of \SI{200}{ps}.

\begin{dunefigure}[\dword{catiroc} \dword{asic}]{fig:dpele-lro-catiroc}
{Functional diagram of \dword{catiroc} \dword{asic}.}
\includegraphics[width=0.8\textwidth]{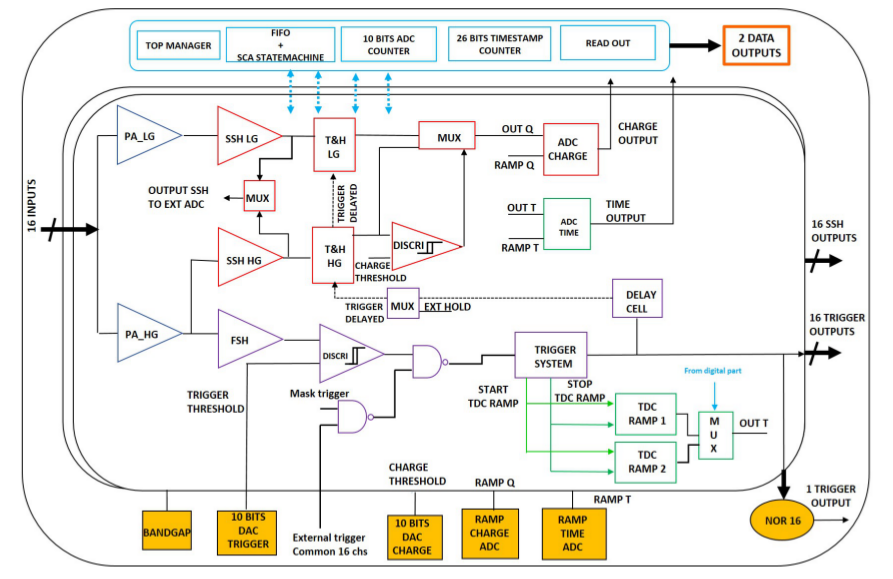}
\end{dunefigure}

Figure~\ref{fig:dpele-lro-catiroc} shows the schematic of the \dword{catiroc} \dword{asic}. Its main properties are summarized in Table~\ref{tab:dpele-catiroc}. The slow channel, from which precision charge and timing measurements are made, is formed by two variable-gain (\SI{8}{bit}) amplifiers followed by two variable slow shapers; one high gain for small signals, and one low gain for larger signals, and two track-and-hold stages. The slow shaper has a tunable shaping time (up to \SI{100}{ns}) and a variable gain.  If the high gain is saturated, corresponding to passing a predetermined threshold common to all \num{16} channels, the lower gain value is chosen. The chosen charge value is converted by an internal 10-bit Wilkinson \dword{adc} operating at \SI{160}{MHz}.  This slow channel operates in a ping-pong mode, with two capacitors to store the slow shaper signals, giving an effective buffer of 2 events. If both capacitors are full, a deadtime of \SI{5}{\micro\second} arises.

\begin{dunetable}
[Main characteristics of \dword{catiroc}.]
{lr} {tab:dpele-catiroc}
{Main characteristics of \dword{catiroc}.}
Item &   \\ \toprowrule
Number of channels & \num{16}\\ \colhline
Signal polarity & negative \\ \colhline
Timing & Timestamp: 26 bit counter at \SI{40}{MHz} \\
       & Fine time: resolution $<$\SI{200}{ps}\\ \colhline
Charge Dynamic Range & \SI{160}{\femto\coulomb} to \SI{100}{\pico\coulomb}\\ \colhline
Trigger & auto-trigger \\
        & Noise = \SI{5}{fC} Minimum threshold = \SI{25}{fC} (5$\sigma$)\\ \colhline
Digital & 10-bit Wilkinson \dword{adc} at 160 MHz \\ 
        & Read-out frame of 50 bits \\ \colhline
Outputs & \num{16} trigger outputs \\
        & NOR16 \\
        & \num{16} slow shaper outputs \\
        & Charge measurement over \num{10} bits \\
        & Time measurements over \num{10} bits \\ \colhline
Main Internal &  Variable preamplifier gain \\
Programmable  &  Variable shaping and gain \\
Features & Common trigger threshold \\
         & Common gain threshold \\ 
\end{dunetable}

The fast channel is used to auto-trigger the \dword{asic} and make the fine-timing measurement. It comprises a high gain preamplifier, fast shaper (shaping time \SI{5}{ns}) and discriminator with a \num{10} bit programmable threshold that is common to all \num{16} channels. The output of the discriminator is used for the two time-to-digital convertors to get the fine timing. A coarse timestamp could also be obtained from a \num{26} bit counter running at \SI{40}{MHz}.  Only the data from the triggered channels are digitized; their information is transferred to the internal memory, which is read by the external \dword{fpga}.

\subsection{Network-based $\mu$TCA Architecture}
\label{sec:fddp-tpc-elec-design-utca}

The digital electronics is based on \dword{utca} standard which offers an industrial solution with a very compact and easily scalable architecture to handle a large number of channels at low cost.  The standard (or related standards such as \dword{atca} or xTCA) is widely used in the telecommunication industry and is being adopted by the HEP community. The backplane of the \dword{utca} crates host high-speed serial links that support a variety of transmission protocols (Ethernet, PCI Express, SRIO, etc.). In addition, dedicated lanes are available for the distribution of the clock signals to all the boards hosted in the crate.  The Ethernet-based solution has been adopted for both the data and clock distribution in this design of the \dual electronics system for both charge and light readout. 

Each \dword{amc} for either charge or light readout plugged into the \dword{utca} is connected to the crate \dword{mch} board through the backplane serial links. The \dword{mch} provides the switch functionality that enables \dwords{amc} to communicate with each other or external systems through the \dword{mch} uplink interface. In the \dual electronics system design, \dword{mch} also manages the WR clock distribution. 

\begin{dunetable}
[Bandwidth requirements per $\mu$TCA crate.]
{lr}{tab:dp-utcabandwidth}
{Bandwidth requirements per \dword{utca} crate for continuous data streaming. A compression factor of 10 for the charger readout data is assumed }   
Parameter & Value  \\ \toprowrule
  \dword{cro} data rate  &  \SI{1.8}{Gibit/s}         \\ \colhline
  \dword{lro} data rate  &  \SI{4.7}{Gibit/s}            \\ \colhline
  Current \dword{mch} bandwidth & \SI{10}{Gibit/s}              \\ \colhline
  Upgradable \dword{mch} bandwidth & \SI{40}{Gibit/s}           \\ 
\end{dunetable}

In the current design, as used for \dword{pddp}, the \dword{mch} operates with a \SI{10}{Gbit/s} uplink. Given that a \dword{utca} crate hosts \num{10} \dwords{amc} for charge readout, the required bandwidth to stream the data to \dword{daq} is about \SI{1.8}{Gbit/s}. This assumes that the data exiting the \dwords{amc} are losslessly compressed with the compression factor \num{10}. The bandwidth required per crate link for streaming the \dword{lro} data is \SI{4.7}{Gbit/s}. The \SI{10}{Gbit/s} \dword{mch} is therefore sufficient to support these data rates. However, the technology is moving towards supporting the \SI{40}{Gbit/s} rates. In addition, the channel density per \dword{amc} could also be increased for cost optimization. For these reasons an upgrade to a \SI{40}{Gbit/s} \dword{mch} could be foreseen in the future. This would also imply that the optical links connecting the \dword{daq} system to \dword{utca} \dword{mch} should be operable at \SI{40}{Gbit/s}. A summary of the required and supported bandwidths per \dword{utca} crate for continous data streaming is provided in Table~\ref{tab:dp-utcabandwidth}.

\begin{dunefigure}[Instrumented \dshort{utca} crate from the \dword{wa105}]{fig:dpele-311-utca-image}
{Pictures of an instrumented \dword{utca} crate from the \dword{wa105}. The crate contains five \dword{amc} cards, correspondingly to the number of readout channels per the \dword{sftchimney}. The images below show the crate after the  cables are connected to the warm flange of the \dword{sftchimney}.}
\includegraphics[width=0.6\textwidth]{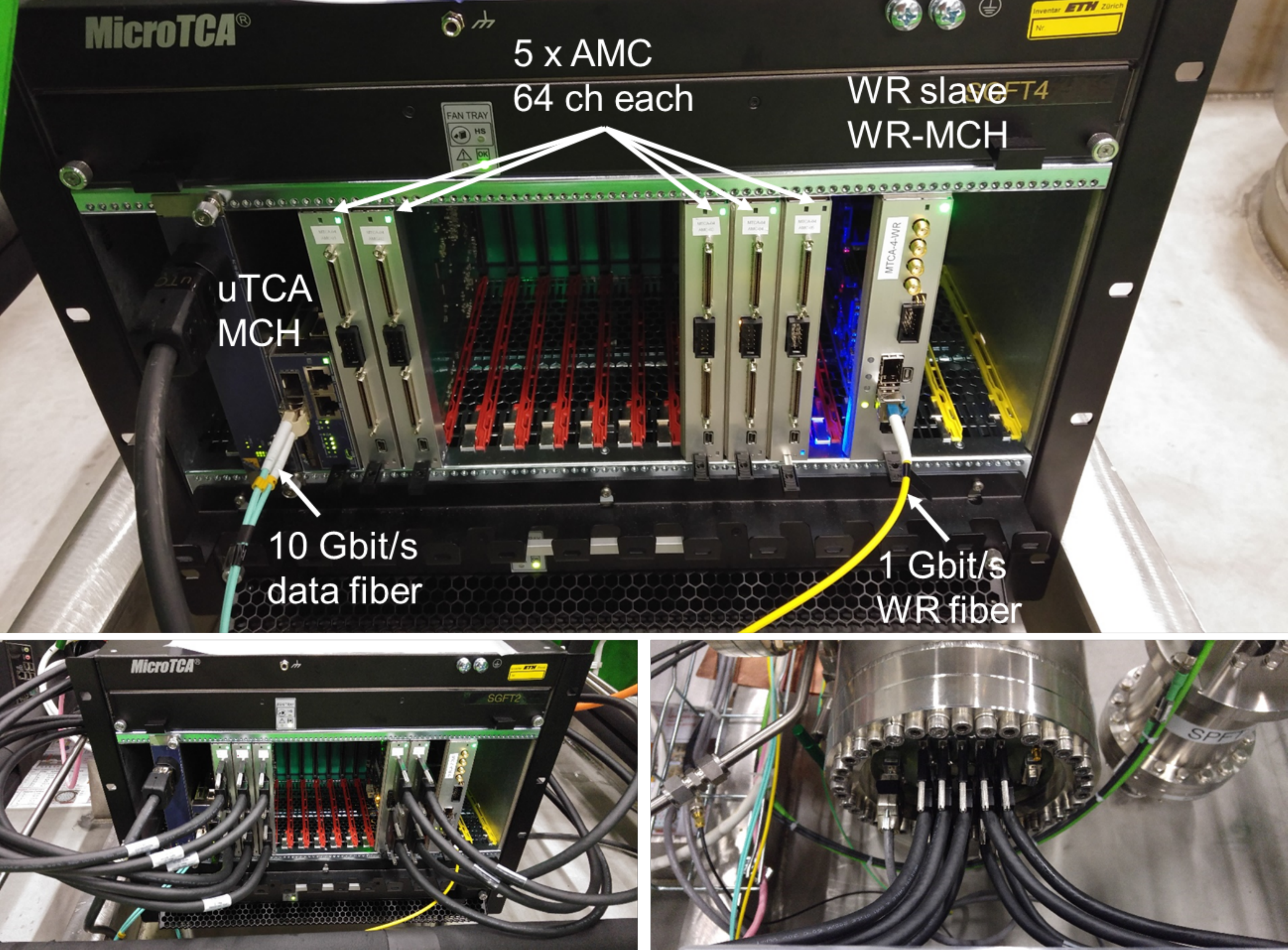}
\end{dunefigure}

As an illustration, Figure~\ref{fig:dpele-311-utca-image} shows pictures of one of the instrumented \dword{utca} crates used for the charge readout of the \dword{wa105} at CERN. In this detector each \dword{sftchimney} reads \num{320} channels, thus requiring only five \dwords{amc} per the \dword{utca} crate. The two optical fiber links, one (\SI{10}{Gbit/s}) for data and the other (\SI{1}{Gbit/s}) for clock and trigger timing distribution, are visible in the images.       

\subsection{Timing Distribution}
\label{sec:fddp-tpc-elec-wr}
The time synchronization system selected for the \dword{dpmod} utilizes a \dword{wr} network, which combines the synchronous \SI{1}{Gbit/s} Ethernet (SyncE) technology with the exchange of PTPV2 packets, to synchronize clocks of distant nodes to a common time. A high stability GPS disciplined oscillator (GPSDO) with  accuracy similar to that of an atomic clock provides a clock reference signal to be distributed over the physical layer interface of the \dword{wr} Ethernet network. The network topology is built using specially designed switches that have the standard IEEE802.1x Ethernet bridge functionality with an addition of \dword{wr}-specific extensions to preserve the clock accuracy. Time and frequency information are distributed to the nodes on the \dword{wr} network via optical fibers. The \dword{wr} protocol automatically performs dynamic self-calibrations to account for any propagation delays and keeps all connected nodes continuously synchronized to sub-ns precision. 

The sub-ns 
precision on the clock synchronization is not strictly needed for aligning samples in the different \dword{amc} digitization units, since the 
timing granularity on the data is \SI{400}{ns}. However, the \dword{wr} timing system offers readily available industrial components and the necessary protocols 
for synchronization with automatic calibration of delay propagation. 
R\&D on this timing distribution solution started in 2006; the final design for integrating this system, 
planned for the \dword{wa105}, \dword{pddp}, and the \dword{dpmod} readout, was completed in 2016. 

In the implementation specific to \dword{pddp}, a GPS-disciplined 
clock unit (Meinberg LANTIME M600\footnote{Meinberg\texttrademark{}, \url{https://www.meinbergglobal.com/english/products/advanced-1u-ntp-server.htm}.}) feeds \SI{10}{MHz} and \num{1}\,PPS reference signals to a commercial \dword{wr} switch (Seven Solutions WRS v3.4\footnote{Seven Solutions\texttrademark{}, \url{http://sevensols.com/index.php/products/white-rabbit-switch/}.}). The switch acts as grandmaster of the \dword{wr} network. It is connected via \SI{1}{Gbit/s} optical links to the dedicated \dword{wr} timestamping node (\dword{wr}-TSN) and the \dword{wr} end-node slave cards present within each \dword{utca} crate (\dword{wrmch}) keeping these synchronized to its reference time. The \dword{wrgm} also communicates through a standard Ethernet port with the LANTIME unit for its date and time synchronization via NTP. The \dword{wr}-TSN module recieves analog TTL-level trigger signals, generates their timestamps, and transmits them over the \dword{wr} network to the connected \dword{wrmch} units. This timestamp information is then used by \dwords{amc} to find the data frame corresponding to the trigger. 

\begin{dunefigure}[Picture of \dword{wr} slave node card]{fig:dpele-wrmch-image}
{Picture of the \dword{wr} slave node card (\dword{wrmch}) present in each \dword{utca} crate for time synchronization.The \dword{wr}-LEN mezzanine card is visible in the bottom right corner.}
\includegraphics[width=0.45\textwidth]{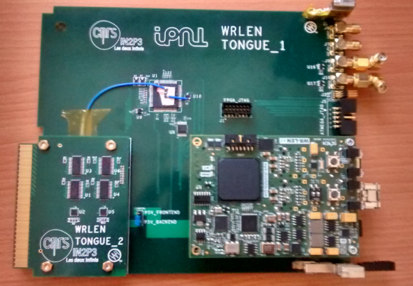}
\end{dunefigure}

The \dword{wrmch} card (Figure~\ref{fig:dpele-wrmch-image}) enables clock/timing/trigger distribution to \dwords{amc}. It communicates with them via dedicated lines in the backplane of the \dword{utca} crate using a customized data-frame protocol. The module contains a commercial WR slave node card, the \dword{wr} Lite Embedded Node (Seven Solutions OEM WR-LEN\footnote{Seven Solutions\texttrademark{}, \url{http://sevensols.com/index.php/products/oem-wr-len/}.}), as mezzanine card. WR-LEN runs on a customized firmware which also enables it to decode the trigger timestamp data packet received over the WR network.

\begin{dunefigure}[Architecture of \dword{wr} network]{fig:dpele-wrnet-layout}
{Architecture of WR network for time synchronization of digital readout electronics.}
\includegraphics[width=0.8\textwidth]{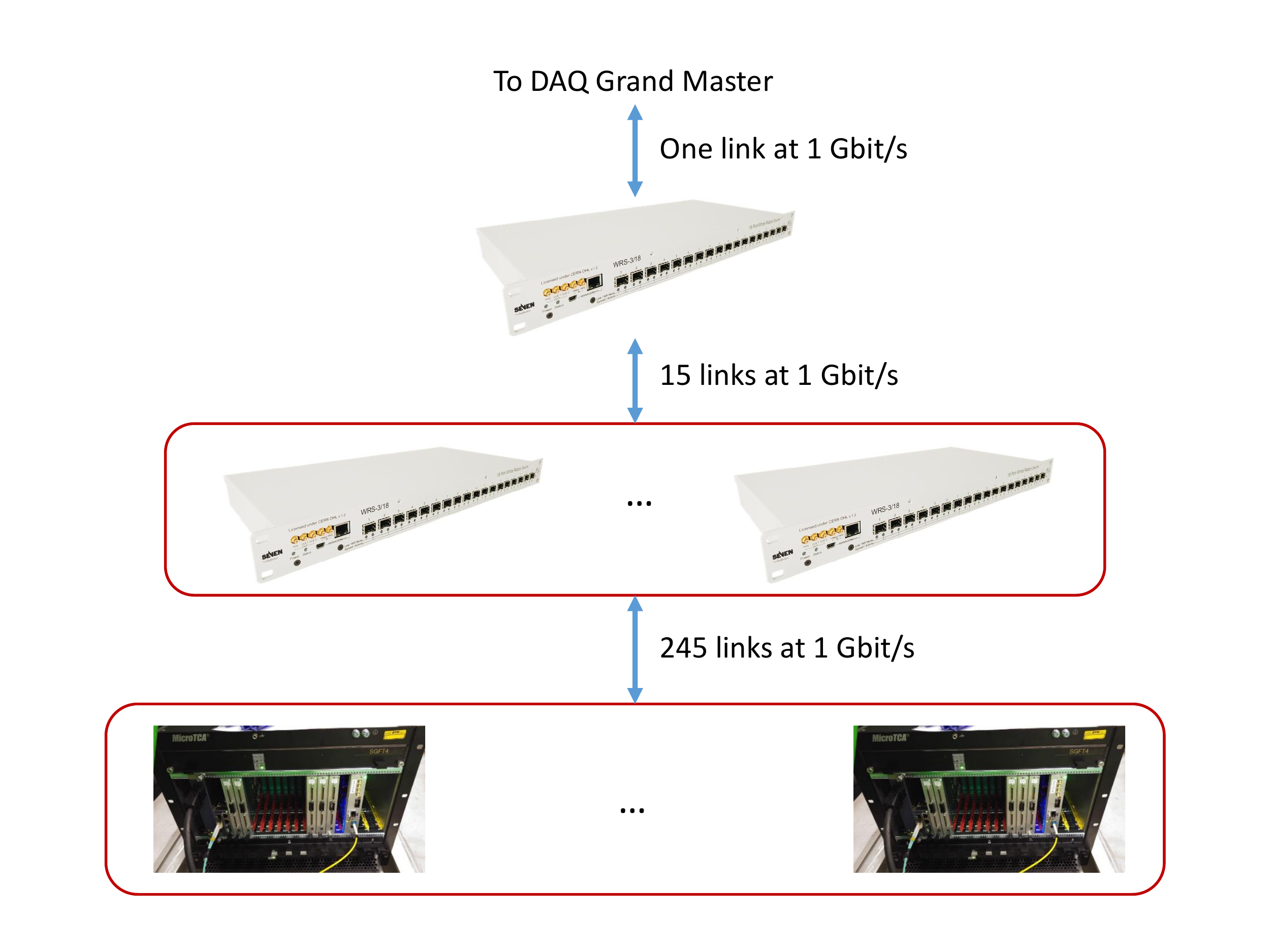}
\end{dunefigure}

The architecture of the \dword{wr} network layout for one \dword{dpmod} is illustrated Figure~\ref{fig:dpele-wrnet-layout}. It is built in a hierarchical structure from \num{16} \dword{wr} switches with \num{18} ports each,  chained with \SI{1}{Gbit/s} optical fibers. The switch at the top of the hierarchy interconnects the synchronization 
\dword{wrgm} from the \dword{daq} system with the \num{15} switches in the middle layer. These are in turn connected to the \dword{wrmch} slave nodes in each \dword{utca} crate (\num{245} in total for charge and light readout).

\section{Production and Quality Assurance}
\label{sec:fddp-tpc-elec-prod-assy}

\subsection{Cryogenic Analog FE Electronics}
\label{sec:fddp-tpc-elec-prod-fe}
The production of the cryogenic \dwords{asic} and analog \dword{fe} cards is envisioned to be split between several sites located in France and Japan at the moment. The delivered cards are then split between five institutions in France (IPNL), Japan (KEK, NITKC, IU), and USA (SMU), where they are tested for various performance parameters such as noise levels, dead channels, hot channels, gain and its uniformity across channels, etc., at both room and operating cold temperature. An appropriate and common database will be developed and populated with test results. 

\subsection{Signal Feedthrough Chimneys}
\label{sec:fddp-tpc-elec-prod-sft}
A number of items require manufacture in order to produce the \dwords{sftchimney}. These include 
\begin{itemize}
\item the PCB flanges for the warm and cold \fdth flange interfaces, 
\item the stainless steel pipe structure, 
\item the flanges containing the interfaces to the gas and liquid lines and slow control, 
\item the blades and railing, and 
\item the heat exchanger system. 
\end{itemize}
The flat cables that connect the \dword{fe} cards to the warm flange are commercially available products and are part of the \dword{sftchimney} procurement process. 

The 
manufactured components are delivered to 
designated institutions participating in the \dual electronics consortium where teams verify the signal continuity 
for both cold and warm flanges, then assemble them into \dwords{sftchimney} and test for leaks. 
They also check the blade insertion, 
 test the flat cables,  
 then, once verified, pack the assembled \dwords{sftchimney} 
 and ship them to SURF. 

\subsection{The Timing System and $\mu$TCA}
\label{sec:fddp-tpc-elec-prod-utca}

The timing system components, the  \num{16} \dword{wr} switches and the \num{245} \dword{utca} crates containing the power modules, carrier hubs (\dword{mch}), and fan units,  
are commercially available. The manufacturer takes the responsibility for the necessary quality control and quality assurance of these components, requiring no further testing on the part of the \dual electronics consortium. Once the components are delivered to the designated institutions, they can be sent to SURF for the installation. 

The commerical VHDCI signal cables (connecting the \dwords{amc} to the \dwords{sftchimney}) are procured and tested with the \dword{sftchimney} warm flanges.



\subsection{Charge Readout Electronics}
\label{sec:fddp-tpc-elec-prod-cro}
The production of the \dword{amc} cards for the charge readout as well as the \dword{wrmch} slave cards for synchronization is currently shared between four institutions (IPNL, KEK, NITKC, IU). The cards ordered and delivered to each respective institution are subjected to quality assurance tests agreed upon by all participants.  

\subsection{Light Readout Electronics}
\label{sec:fddp-tpc-elec-prod-lro}

The production of the \dword{lro} \dword{amc} cards is 
occurs in the same manner as the cards for \dword{pddp} since the number of cards to be produced and the channels to test are both small. The cards' electronic components, meeting required specifications, are purchased commercially. 
The project will be managed by a qualified engineer, working with a specialist in \dword{qa}.

The produced cards are 
delivered to 
designated consortium institutions. 
Upon delivery, teams conduct basic quality tests, including visual inspection and electrical testing, to ensure conformity of production. 
Another series of tests is performed on 
the cards to ensure their correct functionality and to evaluate their performance. Measurements include: linearity measurements (DNL and INL) of each \dword{adc} channel, and 
linearity of response of the \dword{asic}. The level of cross-talk on the \dword{asic} 
is also quantified.

A dedicated single-channel setup, with \dword{pmt} (Hamamatsu R5912-02-mod), and identical cabling and splitter as in the \dword{fd}, can be used to characterize the expected noise level of each channel, and response to single \phel{}s up to saturation. 
Multiple cards are operated in a \dword{utca} crate with the 
\dword{daq}.

After 
shipment to SURF and installation on-site, a small series of tests is performed with a pulse generator to verify the good working condition of the cards. Noise-level measurements are 
included in the integration effort.



\section{Interfaces}
\label{sec:fddp-tpc-elec-intfc}


The \dual electronics system interfaces to several other systems, starting with the \dword{crp} and the \dword{pd} systems.  The digitized data must in turn 
flow to the \dword{daq} via the optical links in each \dword{utca} crate. The \dwords{sftchimney} integrate into the cryostat structure and connect to the cryogenics and gas systems. The slow-control system takes on management of the low-voltage power supplies for the \dword{fe} analog electronics and \dword{utca} crates, and  monitors various sensors in the \dwords{sftchimney}. 
Table~\ref{tab:dpele-interfaces} provides the references to the relevant interface documents for each interface, stored in the DUNE document database (DocDB).

\begin{dunetable}
[Interface documents relevant to \dual TPC electronics system]
{lr}{tab:dpele-interfaces}{Interface documents relevant to \dual electronics system.}   
Interface document    & DUNE DocDB No. \\ \toprowrule
\dword{dp} TPC electronics to \dword{dp} \dword{crp} & 6751 \\ \colhline
\dword{dp} TPC electronics to \dword{dp} \dword{pd} & 6772 \\ \colhline
\dword{dp} TPC electronics to Joint \dword{daq} & 6778 \\ \colhline
\dword{dp} TPC electronics to Joint CISC & 6784 \\ \colhline
Facility Interfaces to \dword{dp} TPC electronics & 6982 \\ \colhline
Installation interfaces to \dword{dp} TPC electronics & 7009 \\ \colhline
Integration facility to \dword{dp} TPC electronics & 7036 \\ \colhline
Calibration to \dword{dp} TPC electronics & 7063 \\ \colhline
DUNE physics to \dword{dp} TPC electronics & 7090 \\ \colhline
Software and computing to \dword{dp} TPC electronics & 7117 \\ 
\end{dunetable}


\subsection{Electronics System to \dword{crp} and Photon Detection Systems}
\label{sec:fddp-tpc-elec-intfc-crppmt}

The cold \fdth flange of the \dwords{sftchimney} forms the interface between the \dword{crp} and the \dword{cro} electronics system. On the side facing the cryostat the flange PCB has \num{20} \num{68}\,pin connectors (KEL 8930E-068-178MS-F\footnote{KEL Corporation\texttrademark{}, \url{https://www.kel.jp/english/product/product_detail/?id=490\&pageID=3\%20}.}) for plugging the flat cables from the \dword{crp}. These are \num{68}\,channel twisted-pair flat cables, each carrying signals from \num{32} anode strips and are within the scope of the \dword{crp} system. Each analog \dword{fe} card reads \num{64} anode strips, i.e., 
signals from two KEL connectors. The order in which the cables are connected 
to the cold flange determines the mapping of the electronic channels to the physical location of the strips on the \dword{crp} and is 
coordinated carefully with the \dword{crp} consortium. 
Figure~\ref{fig:dpele-sft-cold-flange} shows two images of the cold \fdth from the \dword{wa105}.

\begin{dunefigure}[Images of the \dword{wa105} \dword{sft} cold \fdth]{fig:dpele-sft-cold-flange}
{Images of the \dword{wa105} \dword{sft} cold \fdth with the \dword{fe} cards inserted (right) and signal cables from \dword{crp} connected (left). The \dword{wa105} \dwords{sftchimney} read only \num{320} channels thus requiring \num{5} \dword{fe} cards.}
\includegraphics[width=0.8\textwidth]{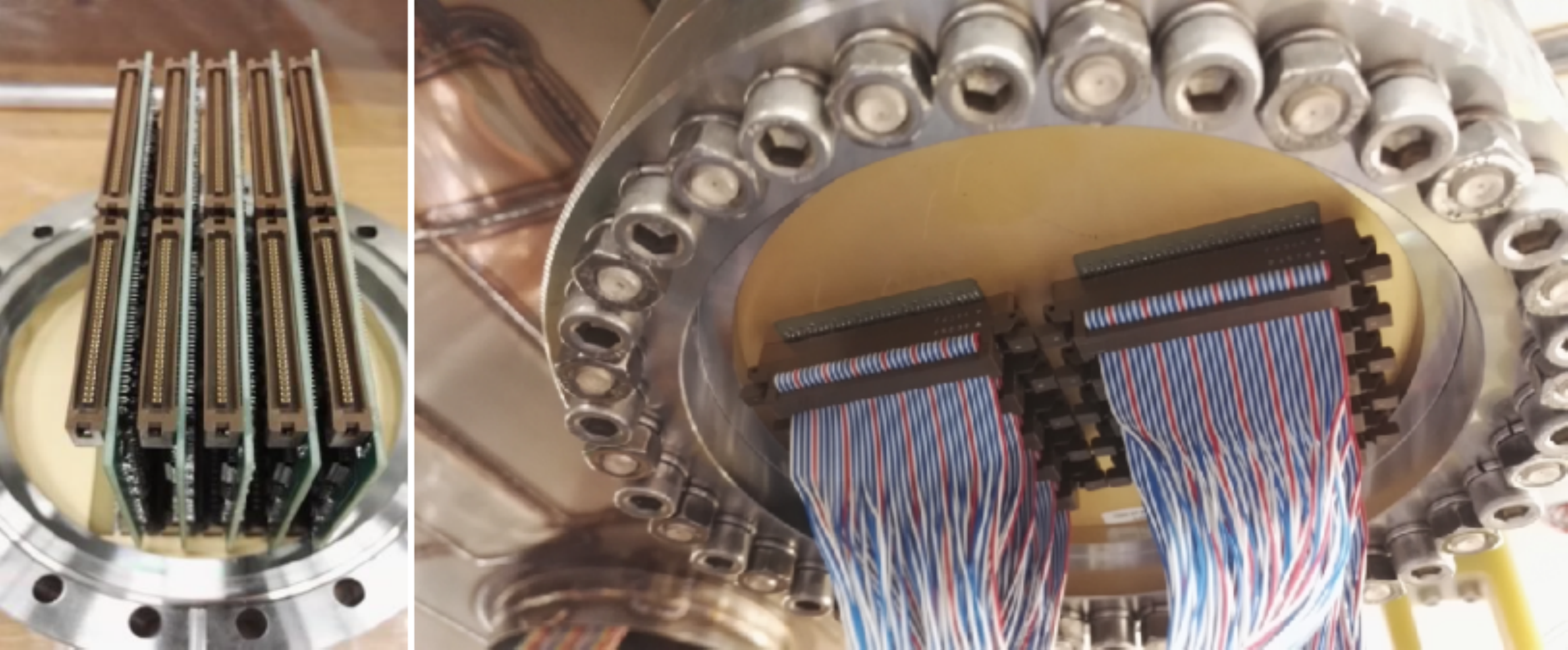}
\end{dunefigure}

The \dword{lro} electronics system is connected to the specific \dword{lro} signal \fdth flanges on top of the cryostat via coaxial cables, which are within the scope of the \dword{pds}.
The \dword{lro} electronics is designed for negative polarity \dword{pmt} signals, with the amplitude of single \phel{}s on the input of the card between \num{1} and \SI{10}{\milli\volt}. Assuming a typical \dword{pmt} gain of \num{1E6} (not accounting for attenuation of the signals), the Catiroc \dword{asic} can measure a range of \num{1} to \num{400} \phel{}s (\SI{160}{\femto\coulomb} to \SI{70}{\pico\coulomb}). The \dword{adc} samples from \SI{1}{\milli\volt} to \SI{1}{\volt} corresponding to \num{1} to \num{1000} \phel{}s, including the time response of the scintillator the range can increase to $\sim$\num{6000}. Increasing the gain of the \dword{pmt} to \num{1E7}, lowers the upper values by a factor of 10. The internal noise level of the \dword{catiroc} is below \SI{0.1}{\milli\volt}. The objective for the noise level of the \dword{adc} is for each channel to have the \rms noise level greater than \SI{0.5}{LSB}, aiming for \SI{1}{LSB} \SI{0.1}{\milli\volt}.

\subsection{Electronics System to DAQ System}
\label{sec:fddp-tpc-elec-intfc-daq}

The hardware interface between the \dual \dword{cro} and \dword{lro} electronics sub-systems and \dword{daq} has two components. 
The first interface is the \SI{10}{Gbit/s} optical fibers for data transfer between the \dword{utca} crates and the network interface of the \dword{daq} system. The second one is a \SI{1}{Gbit/s} optical fiber that connects the \dword{daq} \dword{wrgm} switch to the \dual electronics timing system.   

In the current design 
a given \dword{dpmod} would have \num{245} \SI{10}{Gbit/s} optical links for streaming the digitized data to the \dword{daq} from the \dword{cro} (\num{240} links) and \dword{lro} (\num{5} links) electronics housed in \dword{utca} crates on top of the cryostat structure.  In the current specifications, the fibers are multimode OM3 fibers \cite{om3fibers} with LC-LC connectors suitable for the transmission over distances of up to \SI{300}{\metre}.  They are provided by the \dword{daq} consortium. On the side of the \dword{utca} crate, the fibers are connected to an optical transceiver in the \dword{mch} (two SFP+XAUI links) \cite{natmch}.  On the \dword{daq}, they go to the level-1 machines of the trigger farm, or switches, depending on the network topology adopted in the \dword{daq} system design.

The \SI{1}{Gbit/s} link going from the \dword{wrgm} to the \dual electronics time distribution network serves to provide the synchronization to the reference clock common for the entire FD and derived from a GPSDO 
clock unit installed on the surface. The clock information is distributed to the \dword{wrmch} slave module in each \dword{utca} crate via a set of \dword{wr} switches. These switches and the interconnecting \SI{1}{Gbit/s} fibers form the timing sub-system of the \dual electronics system and are included in the design of the latter. The \dword{wr} synchronization protocol includes the automatic and continuous calibration of the propagation delays between the master and the connected slaves. This allows maintaining the overall synchronization between different nodes at sub-ns level. The \dword{wrgm} 
will be located either:
\begin{itemize}
\item{On the surface near the GPSDO. In this case, a single fiber connects it to the \dual timing system underground. 
The system automatically accounts for the incurred latency due to the extensive 
 fiber length.}
\item{Underground in the \dword{cuc}. In this case, calibration of the propagation delays between GPSDO and the \dword{wrgm} is performed manually, and a timing correction 
is applied to the data afterward.}
\end{itemize} 

The TPC electronics design assumes that the data are streamed continuously via the \SI{10}{Gbit/s} links to the \dword{daq}, where they are buffered until a trigger decision 
is made. The triggers are to be issued by processing the buffered data in some suitable sliding time window on the trigger farm machines. 
The window may be as long as \SI{10}{s} for \dword{snb}-triggered events.
The triggers determine whether the data contained in the buffers are to be written on disk. 

The software interface between the \dword{daq} and the electronics system
includes the tools for handling the data transmission and buffering, i.e.,  data formatting in \dword{udp} packets, compression and decompression, and exchange of the control packets.

\subsection{Electronics System to Cryostat and Cryogenics}
\label{sec:fddp-tpc-elec-intfc-cryo}

The interface point between the cryostat and the \dual electronics system is at the cryostat penetrations where the \dwords{sftchimney} are 
installed. Each penetration 
accommodates the chimney (of external diameter \SI{254}{\mm}). Each chimney has a CF-273 flange welded to its outer structure (see Figure~\ref{fig:dpele-sft-chimney-crosspipe}). After the chimney is inserted, this flange is in contact with the corresponding flange on the crossing (or penetration) pipe embedded in the cryostat structure to which it is eventually fastened. In order to avoid any leaks at this interface a CF-273 copper gasket is used to ensure the vacuum tightness.  

\begin{dunefigure}[Details of \dword{sftchimney} interface to the cryostat structure]{fig:dpele-sft-chimney-crosspipe}
{Details of \dword{sftchimney} interface to the cryostat structure.}
\includegraphics[width=0.7\textwidth]{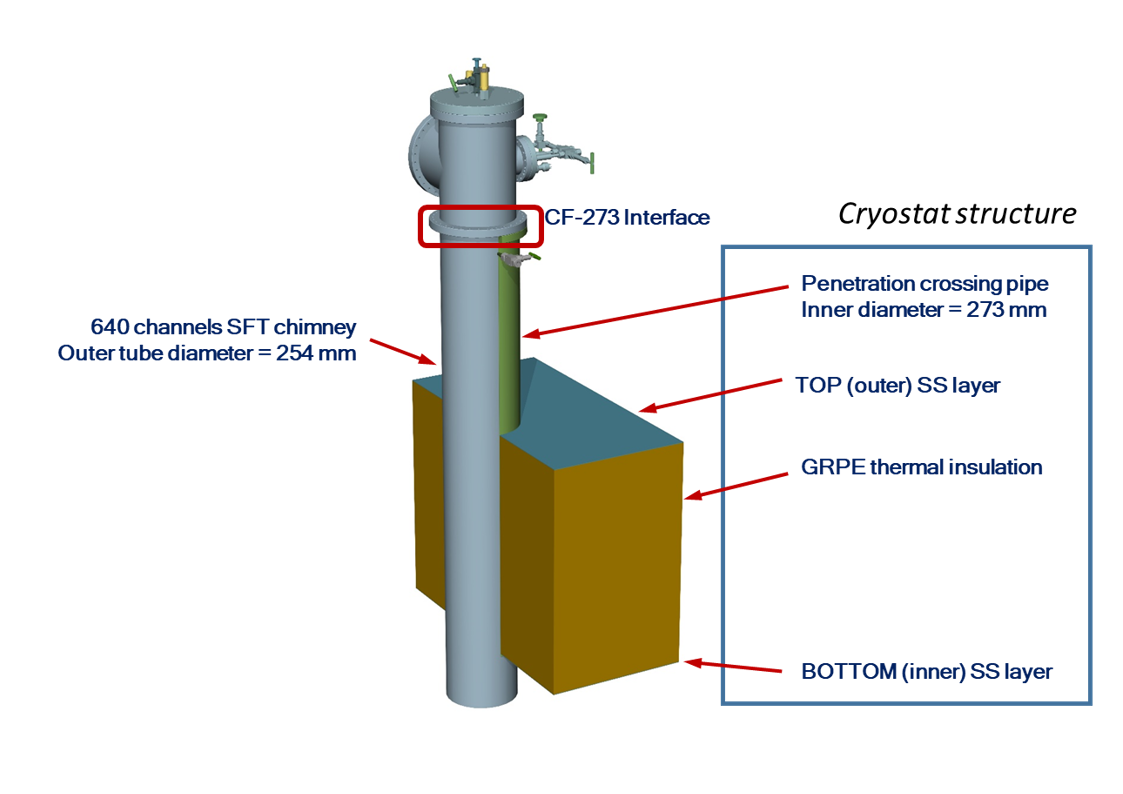}
\end{dunefigure}

Each chimney contains a heat exchanger copper coil cooled with \lar. There are two (inlet and outlet) stainless steel pipe connections with \SI{10}{\mm} and \SI{12}{\mm} inner and outer diameters, respectively, that need to be branched to the respective system for the \lar delivery and recirculation. In addition, 
a connection for nitrogen gas line with the same pipe dimensions as those for the \lar cooling, 
is used for filling the chimney after it is closed following the installation of the \dword{fe} electronics. The nitrogen line is also required for flushing the chimney in the case of an access to the \dword{fe} cards after the \dword{dpmod} is cooled for the operation. 

The \dword{utca} crates for charge readout 
are installed within a short \SI{<0.5}{\meter} distance from the \dwords{sftchimney} on top of the cryostat roof. The five \dword{utca} crates for the light readout are also placed on the roof of the cryostat at optimal locations defined by the routing of the \dword{pmt} signal cables. The required volume to accommodate the crates is roughly \SI[product-units=power]{60x50x40}{\cm}. 

\subsection{Electronics System to Slow Control System}
\label{sec:fddp-tpc-elec-intfc-sc}

The integration with the slow control of the low-voltage power supply system for the \dword{fe} cards and \dword{utca} crates is required to enable the remote management and monitoring (current consumption by \dwords{asic}, set voltage, etc.). In addition, the \dwords{sftchimney} contain several sensors that need to be monitored. These include a pressure transducer that measures the pressure inside the chimney and at least two temperature probes (PT1000) that monitor the gas temperature inside near the cold flange at the bottom and close to the warm flange at the top. The readout of the \dword{utca} crate information and the sensors in the \dwords{sftchimney} is part of the Slow Control system.

\section{Transport and Handling}
\label{sec:fddp-tpc-elec-install-transport}

The \dwords{sftchimney} are \SI{2350}{\mm} long 
and weigh \SI{180}{\kg}.  The cold and warm flanges are mounted on them at the assembly site, and the chimneys undergo leak-testing prior to shipping in wooden crates (approximate dimensions \SI[product-units=power]{2.5x0.5x0.5}{m}). 
Once at SURF the crates are moved underground and placed on the roof of the cryostat by the \dword{uit}. The 
\dual electronics consortium 
is then responsible for unpacking the crates and installing the \dwords{sftchimney}. 

The boxes containing the electronic cards and \dword{utca} crates are also handled by \dual electronics consortium personnel. These are expected to be lightweight and 
easy to carry. A box containing \num{100} \dword{amc} cards has a maximal dimesions of \SI{60}{\cm} and weighs less than \SI{10}{\kg}. 

\section{Installation, Integration and Commissioning}
\label{sec:fddp-tpc-elec-install}

The installation of the TPC electronics systems proceeds in several stages. In order to cable the \dwords{crp} to the \dwords{sftchimney}, 
the chimneys 
are installed first, prior to the start of the \dword{crp} installation inside the cryostat. 
Next the \dword{fe} cards 
are mounted on the blades and inserted. The installation of the digital electronics and \dword{utca} crates 
is postponed until all of the heavy work finishes on top of the cryostat in order to prevent 
damage to the fragile components (e.g., optical fibers)  
due to movement of material and traffic. 
Once the \dword{utca} crates are installed and all the digital cards are inserted, the 
\dwords{amc} are cabled to the warm flanges of the \dwords{sft} for the charge readout and are connected to the \dword{pmt} signal cables for the light readout. Finally, to complete the installation and integrate the system with the \dword{daq}, the \SI{10}{Gbit/s} and \SI{1}{Gbit/s} optical links to the \dword{daq} and \dword{wr} timing network are connected. At this stage the full system is ready for commissioning.

\subsection{SFT Chimneys}
\label{sec:fddp-tpc-elec-install-sft}

The installation of the \dwords{sftchimney} requires a compact gantry crane with 
movable supports along the length of the cryostat. The crane itself moves along the transverse direction. The crates containing the \dwords{sftchimney} are placed along the edges of the cryostat roof. An unpacked chimney is hoisted and transported to 
the appropriate penetration crossing pipe for installation. Once in place, the chimney is fastened to the flange on the crossing pipe. 
Enough overhead room to accommodate a chimney's \SI{2.4}{m} length is required
to allow to free movement of the chimney with the crane along the direction transverse to the beam axis. 

In parallel with the \dword{sftchimney} installation, the \dword{fe} cards are unpacked on top of the cryostat and mounted on the blades prior to their insertion in the chimneys.  
With \dwords{sftchimney} secured in the cryostat structure, the blades with mounted \dword{fe} cards 
are inserted prior to sealing the chimney.
At this stage, the \lar and gas nitrogen delivery pipes are already installed, and  
it is possible to make the connections with them. The pressure probes and temperature sensors are also connected to the slow control system.

\subsection{Digital $\mu$TCA Crates}
\label{sec:fddp-tpc-elec-install-utca}

The installation of the \dword{utca} crates with the digital electronics 
takes place in the final stage of the \dword{dpmod} installation to avoid damaging the fragile equipment. The crates are placed in their designated positions on the cryostat and connected to the power distribution network. The \dword{amc} cards and \dword{wrmch} modules are inserted in their slots. The VHDCI cables are then attached connecting the \dword{cro} \dwords{amc} to the warm flange interface of the \dwords{sftchimney}.  The fibers from the timing system are connected to \dword{wrmch}. 



\subsection{Integration within the DAQ}
\label{sec:fddp-tpc-elec-install-daq}
The integration of the \dual TPC electronics with the \dword{daq} system requires connecting the \SI{10}{Gbit/s} fiber links to each of \num{245} \dword{utca} crates. The connection of the timing system to the synchronization \dword{wrgm} is done via a single \SI{1}{Gbit/s} fiber link. 

The necessary software for the \dword{daq} to read and decode the data packets sent by each \dword{utca} crate would also be provided by the electronics consortium.   

\subsection{Integration with the Photon Detection System}
\label{sec:fddp-tpc-elec-install-pmt}
The cables carrying the \dword{pmt} signals from the splitter boxes 
are connected to the \dword{lro} analog electronics in each \dword{utca} crate. The position of the crates 
is optimized with respect to the layout of \dword{pmt} cables. In addition, the calibration system of the \dword{pds} 
is connected to specified inputs on the cards.



\subsection{Commissioning}
\label{sec:fddp-tpc-elec-comission}

The \dwords{sftchimney} are commissioned as a first step. This consists of evacuating and then filling them with nitrogen gas at slight overpressure. 
It is necessary to check the leak rate when the chimney is under vacuum and to monitor the nitrogen pressure once it is filled in order to verify that no damage occurred to the flange interfaces during installation.

The electronics system 
is commissioned after completing the installation of the \dword{utca} crates with the \dwords{amc}, and the timing system. 
The functionality of the full \dword{daq} system is not strictly required at this stage. The data from each crate 
is read with a portable computer connected to the crate \dword{mch} \SI{10}{Gbit/s} or \SI{1}{Gbit/s} interface. 
The non-functioning channels are identified by pulsing the \dword{crp} strips and the data quality is examined to ensure the correct functioning of the digital electronics and the temporal alignment of the data segments.   




\section{Risks and Vulnerabilities}
\label{sec:fddp-tpc-elec-risks}

The design of the \dual electronics system takes into account several risk factors:
\begin{itemize}
\item{\textbf{Obsolescence of electronic components over the period of experiment}: allocation of enough spares (preferably complete cards instead of components) should be sufficient to address this issue. }

\item{\textbf{Modification to \dword{fe} electronics due to evolution in design of \dwords{pd}}: Strict and timely follow-up of the \dword{fe} requirements from the \dual \dword{pds} is required.}

\item{\textbf{Damage to electronics due to \dword{hv} discharges or other causes}: 
The \dword{fe} cards should include suitable protection components. The TVS diodes used in the current design  have been sufficient to protect the electronics in the \dword{wa105}. 
 In addition, the cards are accessible and could be replaced if damaged. }
 
\item{\textbf{Overpressure in the \dwords{sftchimney}}: The \dwords{sftchimney} are equipped with safety valves that vent the excess gas in case of the sudden pressure rise. The overpressure threshold 
must be set low enough such that no significant damage could happen to the flanges. }

\item{\textbf{Leak of nitrogen inside the \dword{dpmod} via cold flange}: The chimney volume 
is  filled with argon gas instead of nitrogen.}

\item{\textbf{Mechanical problems with \dword{fe} card extraction due to insufficient overhead clearance}: 
This is addressed by imposing a requirement for LBNF to ensure enough overhead clearance to extract the blades from the \dwords{sftchimney}.}
\item{\textbf{Data flow increase due to inefficient compression caused by higher noise}: Currently there is a factor of \num{5} margin in the available bandwidth with \SI{10}{Gbit/s} \dword{mch}.} 
\item{\textbf{Damage to \dword{utca} crates due to presence of water on the roof of the cryostat}: This is addressed by imposing a requirement for LBNF to ensure that the top cryostat surface remains dry.}
\item{\textbf{Problems with the ventilation system of the \dword{utca} crates due to bad air quality}: Normal conditions similar to any industrial environment (e.g., at CERN or Fermilab) 
is expected to be sufficient 
for proper crate functioning. It is important to avoid liberation of large quantities of dust in the detector caverns at SURF.} 
\end{itemize}



\section{Organization and Management}
\label{sec:fddp-tpc-elec-org}

\subsection{Dual-Phase TPC Electronics Consortium Organization}
\label{sec:fddp-tpc-elec-org-consortium}

The \dual TPC electronics consortium 
consists of seven participating institutions from France (\num{3}), Japan (\num{3}), and the USA (\num{1}). The consortium leader is 
from IPNL, France, and the technical leader is 
from KEK, Japan. The consortium includes members from APC, IPNL, and LAPP in France; Iwate University, KEK, and NITKC in Japan; and SMU in the USA.


\subsection{Planning Assumptions}
\label{sec:fddp-tpc-elec-org-assmp}
The 
design of the \dual TPC electronics system 
largely on the elements that have already been developed and tested in the \dword{wa105}. Commissioning of the \dword{pddp} towards the end of 2018 should provide some additional information, but is not expected to affect the design of principal components. Some additional improvements related to the increase in the channel density supported by \dwords{amc}  
is possible for the purpose of further cost reduction. 

\subsection{WBS and Responsibilities}
\label{sec:fddp-tpc-elec-org-wbs}

The description of the \dword{wbs} including the assignments of the responsible institutions is documented in DUNE-doc-5594. 

\subsection{High-level Cost and Schedule}
\label{sec:fddp-tpc-elec-org-cs}

The key milestones are listed in Table~\ref{tab:dpele-milestones}. 
Table~\ref{tab:dpele-schedule} shows an extract from the international project schedule pertaining to the technical activities of this consortium. The detailed cost model has been developed based on the scaling of the costs for the electronics system of \dword{pddp}. It is provided in an addendum. 

\begin{dunetable}[\dual TPC electronics consortium key milestones]
{ll}
{tab:dpele-milestones}
{\dual TPC electronics consortium key milestones.}
Date & Milestone \\ \toprowrule
September 2018 & Number of \dword{lro} channels finalized \\ \colhline
November 2018 & Final routing for \dword{lro} \dwords{amc} for production \\ \colhline
March 2019 & Costing model for \dword{tdr} finalized \\ \colhline
March 2019 & Firmware for \dword{cro} \dwords{amc} finalized \\ \colhline
March 2019 & Commissioning of \dword{pddp} finished \\ \colhline
January 2023 & Start of component production and procurement \\ \colhline
July 2023 & \dword{utca} infrastructure components produced \\ \colhline
July 2023 & Components of \dword{wr} system delivered and validated \\ \colhline
January 2024 & \dword{sft} chimneys produced and tested \\ \colhline
January 2024 & Cryogenic \dword{fe} analog electronics produced and tested \\ \colhline
January 2024 & \dwords{amc} for \dword{cro} and \dword{lro} produced and tested \\ \colhline
August  2024 & Cryostat of the second detector module is ready \\ \colhline
November 2024 & \dword{sft} chimneys installed \\ \colhline
December 2024 & Cryogenic \dword{fe} electronics installed \\ \colhline
December 2024 & \dword{utca} crates and \dword{wr} network installed \\ \colhline
January  2025 & Installation of \dwords{amc} completed \\ \colhline
January  2025 & Commissioning of the \dual TPC electronics system \\ \colhline
August   2025 & Closure of the cryostat \dword{tco} \\
\end{dunetable}

\begin{dunetable}[\dual TPC electronics consortium schedule]
{p{0.6\linewidth}lll}
{tab:dpele-schedule}
{\dual TPC electronics consortium schedule.}
 Techincal activity  &  Days & Start date & End date \\ \toprowrule
Preparation of costing for \dword{tp} & \num{20} & 02/26/18 & 03/23/18 \\ \colhline
Initial development of installation schedule & \num{20} & 02/26/18 & 03/23/18 \\ \colhline
Further development of installation schedule & \num{145} & 09/03/18 & 03/22/19 \\ \colhline
Installation and commissioning of \dword{pddp} & \num{320} & 01/01/18 & 03/22/19 \\ \colhline
Finalization of the number of channels for \dword{lro} & \num{20} & 09/03/18 & 09/28/18 \\ \colhline
Implementation of routing for digital cards of \dword{lro} & \num{40} & 10/01/18 & 11/23/18 \\ \colhline
Preparation of final costing for \dword{tdr} & \num{85} & 11/26/18 & 03/22/19 \\ \colhline
Firmware development for charge readout cards & \num{145} & 09/03/18 & 03/22/19 \\ 
\end{dunetable}

\cleardoublepage

\chapter{High Voltage System}
\label{ch:fddp-hv}

\section{High Voltage System (HV) Overview}
\label{sec:fddp-hv-ov}

\subsection{Introduction}
\label{sec:fddp-hv-intro}

A \dword{lartpc} requires an equipotential cathode plane at \dword{hv} and a precisely regulated interior \efield to drive 
electrons from particle interactions to sensor planes.  In the case of the DUNE \dlong{dp} technology, 
this requires a horizontal cathode plane, held at negative \dword{hv}; a horizontal \dword{crp} in the gas phase as described in  Chapter~\ref{sec:fddp-crp-intro}; and formed sets of conductors at graded voltages surrounding the central drift volume, collectively called the \dlong{fc} as shown in Figure~\ref{fig:dune_dp_fd}. The \dword{fc} consists of continuous field shaping rings that provide voltage degradation in the vertical direction and forms one continuous active volume.

\begin{dunefigure}[\dword{dpmod} overview]{fig:dune_dp_fd}
{A cutaway showing an overview of a \dword{dpmod}, with the cathode plane and the \dword{pds} on the floor, the \tpcheight tall \dword{fc} modules surrounding the active volume, and the top view of the 
 \dwords{crp} showing the anode plane (the only portion visible from this angle).}
\includegraphics[width=0.8\textwidth]{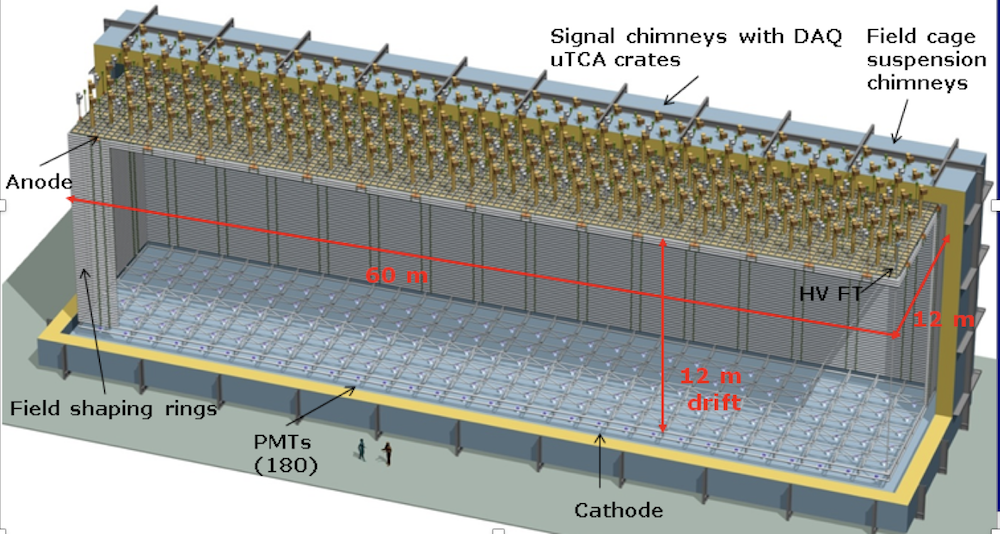}
\end{dunefigure}

The \dword{hv} consortium supplies the systems that operate at the nominal voltages to establish the uniform \SI{500}{\volt/\cm} \efield in the \dword{tpc} drift volume. As a result, its systems 
constitute a large fraction of the 
internal structures of the \dword{tpc}. 
Mechanical and structural concerns are taken into account, together with the electrical design to meet the requirements. 

The design presented in this chapter is primarily based on the \dword{pddp} design, which includes a set of basic elements (i.e., \dword{fc} sub-modules, cathode, and ground grid modules) that are deployed to build a TPC with a 6$\times$6$\times$6\,m$^3$ active volume. The extrapolation to the \dword{dpmod} structure with 
\tpcheight$\times$\dptpcwdth$\times$\dptpclen\,m$^3$ active volume 
requires some electrical and mechanical adaptations.  The size of each element is kept within a roughly 3$\times$3\,m$^2$ envelope, matching the size of the  \dword{crp} modules, which was optimized for underground transportation and assembly.


\subsection{Design Considerations}
\label{sec:fddp-hv-des-consid}

The \dword{hv} system is designed to meet the physics requirements of the DUNE experiment. This includes both physical requirements (e.g., an \efield 
that allows robust event reconstruction) and operational (e.g., 
avoidance of over-complication in order to maximize data collection efficiency). 
A collection of essential requirements for the \dword{hv} system is shown in Table~\ref{tab:hvphysicsreqs}.

\begin{dunetable}
[\Dword{hv} system requirements]{p{0.05\textwidth}p{0.2\textwidth}p{0.35\textwidth}p{0.15\textwidth}p{0.15\textwidth}}
{tab:hvphysicsreqs}
{\dword{hv} System Requirements}
No. & Requirement & Physics requirement driver & Requirement & Goal \\ \toprowrule
1 & Exceed minimum \efield TPC drift volume & Maintain adequate particle ID, which is impacted by slower drift speed and increased recombination, diffusion and space charge effects. & >\SI{250}{V/cm} &\SI{500}{V/cm} \\ \colhline
 2 & Do not exceed maximum \efield in \lar volume & Avoid damage to detector to enable data collection over long periods. & \SI{30}{kV/cm} & \dword{alara} \\  \colhline
3 & Minimize power supply ripple & Keep readout electronics free from external noise 
\\ \colhline
4 &  Maximize power supply stability & Maintain the ability to reconstruct data taken over long period.  Maintain high operational uptime to maximize experimental statistics. \\ \colhline
5 & Provide adequate decay time constant for discharge of the cathode plane and \dword{fc} as well as cathode plane resistive segmentation & Avoid damage to detector to enable data collection over long periods. Maintain high operational uptime to maximize experimental statistics. & \si{\giga\ohm} resistors per each connection of the $3\times3$\,m$^2$ cathode units  \\ \colhline
6 & Provide redundancy in all \dword{hv} connections & Avoid single-point failures in detector that interrupt data taking. & >2 voltage divider chains to distribute \dword{hv} to the \dword{fc} profiles & One voltage divider chain every four \dword{fc} modules\\ 
\end{dunetable}

\subsection{Scope}
\label{sec:fddp-hv-scope}
The scope of the \dword{hv} system 
includes the continued procurement of materials for, and the fabrication, testing, delivery, and installation of systems to generate, distribute, and regulate voltages so as to create a precision \efield within the \detmodule volume. 

The \dword{hv} system consists of components both exterior and interior to the cryostat. The \dword{hv} power supply is located external to the cryostat.  In the \dword{dpmod}, the \dword{hv} power supply is expected to be located on top of the \dword{hv} \fdth. The \dword{hv} is further distributed by interior components that form part of the TPC structure, as depicted in Figure~\ref{fig:dune-dp-hvs}.a.  These components are:

\begin{itemize}
\item power supply (similar to that shown in Figure~\ref{fig:dune-dp-hvs}b),
\item \dword{hv} \fdth (similar to that shown in Figure~\ref{fig:dune-dp-hvs}c),
\item \dword{hv} extender and voltage degrader,
\item cathode plane and \dword{gp}, 
\item \dlong{fc},
\item \dword{hv} return \fdth and resistor box  (similar to that shown in Figure~\ref{fig:dune-dp-hvs}d).
\end{itemize}

\begin{dunefigure}[HV system for a \dpmod ]
{fig:dune-dp-hvs}
{(a) Schematic overview of the VHV system for a \dpmod{}, 
(b) photo of the \SI{300}{\kV} Heinzinger power supply\footnote{Heinzinger\texttrademark{} PNChp 300000 power supply.}, (c) the VHV \fdth{}, and (d) the VHV return connection. (All photos from the \dword{wa105}.).}
\includegraphics[width=1.0\textwidth,height=1.2\textwidth]{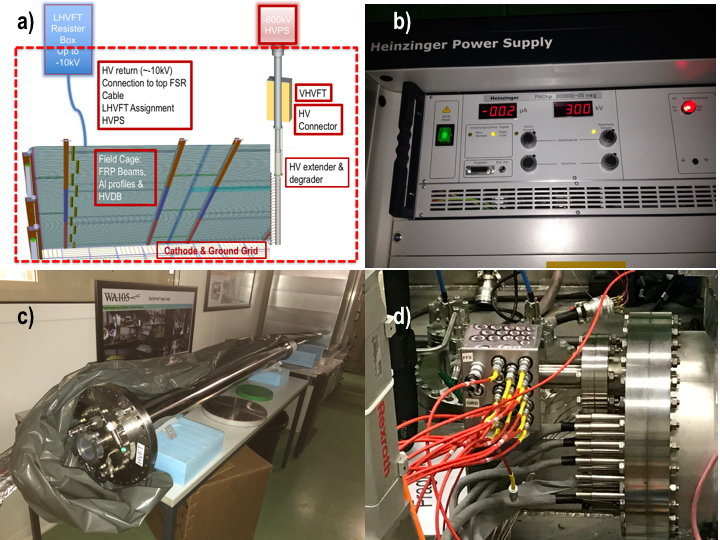}
\end{dunefigure}

\subsection{System Overview}

A \dpmod  has a single modular cathode plane that forms the bottom of the single
  \dptpcwdth (W) $\times$ \tpcheight (H) $\times$ \dptpclen (L) drift volume. It is constructed from eighty \SI{3}{\m} $\times$ \SI{3}{\m} contiguous units each consisting of a stainless steel tube frame holding a stainless steel grid. 
A similar but highly transparent \dword{gp} is placed just above the \dword{pds}, which is located near the bottom of the cryostat and below the cathode to shield it from the \dword{hv}. The cathode bias voltage of \dptargetdriftvoltneg is provided by an external \dword{hv} power supply through a \dword{hv} \fdth and a voltage extender unit that reaches the cathode.
 
The \dword{fc} surrounds the drift volume in a set of equidistant, stacked, horizontal, rectangular-shaped aluminum rings. Its function is to ensure a uniform drift field of \SI{500}{\V\per\cm}. This is accomplished by gradually decreasing the voltage over the \tpcheight height from the cathode voltage of \dptargetdriftvoltneg to \SI{-10}{\kV} at the top-most field shaping ring, to allow for electron extraction into the gas volume of the \dword{crp}. The \dword{fc} consists of aluminum field-shaping rings, made up of mechanically and electrically connected \SI{3}{m} long aluminum profiles. This is a cost-effective system for establishing the required equipotential surfaces. 

\section{HV System Design}
\label{sec:fddp-hv-design}

\subsection {High Voltage Power Supply and Feedthroughs}
The \dword{hv} delivery system consists of
\begin{itemize}
\item one power supply,
\item \dword{hv} cryogenic \fdth{}s,
\item \dword{hv} cryogenic extender.
\end{itemize}

To ensure the nominal \efield of \SI{500}{V/cm} over  the \dpmaxdrift drift distance, an external power supply must deliver \dptargetdriftvoltneg to  the cathode through one \dword{hv} cryogenic \fdth, with a maximum current draw of \SI{0.5}{\milli\ampere}.
At present such a power supply does not exist, but  Heinzinger, the industrial partner and leader in the production of \dword{hv} power supplies, is executing a vigorous R\&D program towards this goal, relying on the following facts:

\begin{itemize}
\item \dptargetdriftvoltpos power supplies are feasible, scaling from present industrial technology;
\item The same is possibly true for the \dword{hv} cryogenic \fdth, scaling to large diameter and longer size with respect to the present \SI{300}{\kV} prototypes;
\item The critical points of the \dword{hv} distribution are then the cable and its connectors on the power supply and on the \dword{hv}{}-\fdth. 
\end{itemize}

The joint R\&D program between DUNE and Heinzinger aims to eliminate cables and connectors, and build a power supply that can be connected directly on the top of the \dword{hv}{}-\fdth.  A sample schematic and some details are shown in Figure~\ref{fig:dune-dp-hvps-ft}. 
Heinzinger is motivated to pursue this effort because of possible new industrial applications.

\begin{dunefigure}[HV powersupply and \fdth for a \dpmod ]
{fig:dune-dp-hvps-ft}
{(a) Vertical cross section of the proposed VHV power supply inserted over the \SI{750}{kV} HVFT for a \dpmod{}, 
(b) Insertion detail of the proposed VHV power supply inserted over the \SI{750}{kV} HVFT. The female HDPE of the HVFT is indicated in green. The male plug of the HVPS, shown inserted, is a metallic conductor inserted in a HDPE insulating tube (indicated in yellow. The gap between male and female is filled, via a tube inside the HVPS (not indicated) by a silicone oil such as RHODORSIL 47 V1000. (c) Vertical cross section of the HVPS. The front panel is on the left. The \dword{hv} multiplication and regulation (not indicated) is in the beige region.}
\includegraphics[width=0.5\textwidth]{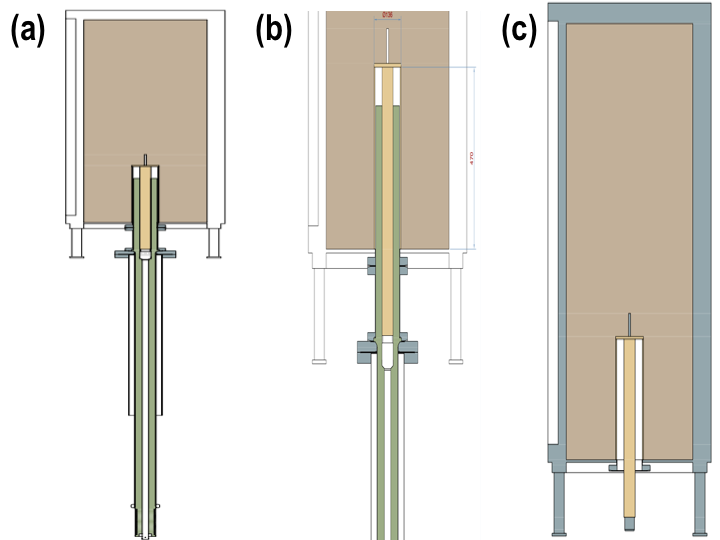}
\end{dunefigure}

Typical Heinzinger power supplies have ripples in the range of $\sim$\SI{30}{k\hertz} with an amplitude of \SI{0.001}{\%V_{nom}} $\pm$ \SI{50}{mV}. A low-pass RC filter designed to reduce the voltage ripple could be integrated into the output of the power supply.  It should be noted, however, that the required ripple suppression does not need to be as high as for the \dword{spmod} due to the \dword{dpmod}'s more effective shielding of the anodic structure, performed by the extraction grid and by the \dword{crp} signal amplification stage. 


The \dword{hv} \fdth is based on the same successful ICARUS design that was adopted in both \dword{pdsp} and \dword{pddp}.  In this design, the voltage is transmitted along a stainless steel center conductor on the warm exterior of the cryostat, where this conductor mates with a cable end.  Inside the cryostat, the end of the center conductor has a spring-loaded tip that  contacts a receptacle cup mounted on the cathode, from which point \dword{hv} is delivered to the \dword{fc}.  The center conductor of the \fdth is surrounded by Ultra-High Molecular Weight Polyethylene (UHMW PE).

To first order, the upper bound of operating voltage on a \fdth is set by the maximum \efield on the \fdth.  Increasing the insulator radius reduces the \efield.  For the target voltage, the \fdth uses a UHMW PE cylinder of at least \SI{15.2}{\cm} (\SI{6}\,in) diameter.  In the gas space and into at least \SI{15.2}{\cm} of the liquid, a tight-fitting stainless steel ground tube surrounds the insulator.  The ground tube has a CF-type 
flange of at least \SI{25.4}{\cm} (\SI{10}\,in) welded on for attachment to the cryostat.  A prototype\footnote{The prototype was manufactured by the company CINEL\texttrademark{} Strumenti Scientifici Srl.}  has been successfully tested up to \SI{-300}{\kV} in pure argon in a dedicated setup; two similar prototypes are currently being installed in \dword{pdsp} and \dword{pddp}.

\subsection{High Voltage Extender and Voltage Degrader}

Since the \dword{hv} has to be guided from the top of the cryostat to the cathode (\SI{12}{\m} below the \dword{lar} surface), an extension of the \dword{hv} \fdth is required, as shown in Figure~\ref{fig:dp-hvft-extender}a, b and c. The extender contains an inner conductor at \dptargetdriftvoltneg surrounded by an insulator. Since the extension runs the entire height of the drift volume, metallic rings (degrader rings) are installed on the periphery of the extension close to the field-shaping ring. Each degrader ring is electrically connected to the field shaping ring at the same height thus guaranteeing that the \efield in the \lar between the extender and the \dword{fc} remains at zero.

\begin{dunefigure}[\dual HVFT and extender]{fig:dp-hvft-extender}{Pictures of \dword{hv} \fdth and \dword{hv} extender-degrader; a) Overview of the \dword{hv} FT, \dword{hv} extender and the degrader chain, b) details of the top portion of the \dword{hv} extender and its connections to the field shaping rings, c) detail of the \dword{hv} extender and degrader connection to the bottom part of the \dword{fc}, including the connection to the cathode plane.}
\includegraphics[width=0.75\textwidth]{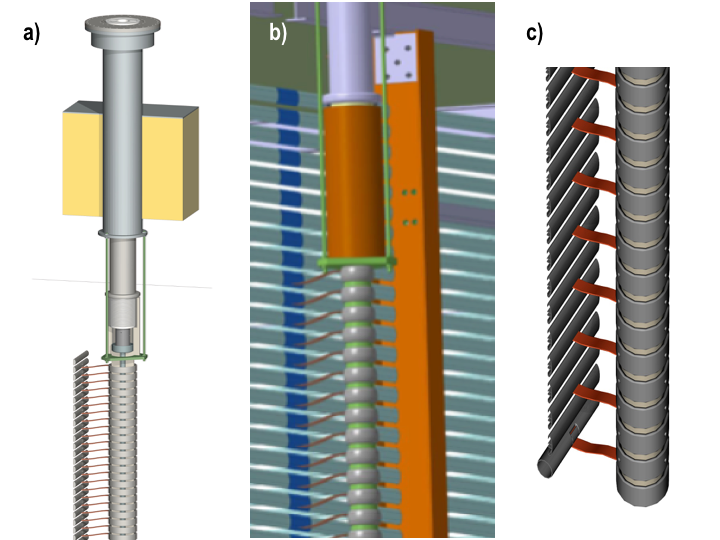}
\end{dunefigure}

\subsection{Cathode Plane}

The \dpmod{}'s cathode plane forms the  bottom of the single 
\dptpcwdth (W) $\times$ \tpcheight (H) $\times$ \dptpclen
drift volume and provides a constant potential surface at \dptargetdriftvoltneg{}.  It receives its \dword{hv} from the central conductor of the extender that carries the voltage from the power supply through the \dword{hv} \fdth.  

The cathode plane consists of eighty \SI{3}{\m} $\times$ \SI{3}{\m} modules. 
The cathode module design is based on the design used in  \dword{pddp}, for which the cathode consists of a stainless steel (316L) mechanical structure 
made from two types of tubes: 
an external frame made from \SI{60}{\milli\m} diameter tubes, and the internal portion is made from oval pipes of $\SI{20}{\milli\m}\times \SI{40}{\milli\m}$ with \SI{1.5}{mm} thickness, as shown in Figure~\ref{fig:dune-dp-cathode}. 
This frame is filled with smaller tubes of \SI{12}{\milli\m} diameter (not shown in the figure), forming a grid, to provide a uniform equipotential surface, while ensuring \SI{60}{\%} optical transparency. All diameters are optimized to guarantee a uniform potential across the cathode, to satisfy the maximum local field requirement of $\SI{30}{kV/\centi\m}$, and to minimize the local \efield{}s to ground. 
The cathode plane components are bolted together and fixed to the supporting  fiber-reinforced plastic (FRP) I-beams of the \dword{fc}. 
Modules of the same size as for \dword{pddp} are foreseen for the \dpmod{} cathode. Some mechanical modifications will be performed in order to guarantee the flatness over the longer distance of  \SI{12}{\m} compared to the  \SI{6}{\m} of  \dword{pddp}.  In addition, the shape of the conductors could undergo some slight modifications in order to lower the local \efield values based on the \dword{pddp} experience.

\begin{dunefigure}[Cutaway view of \dword{pddp} cathode]{fig:dune-dp-cathode}
{A cutaway view of \dword{pddp} cathode(Credit: ETHZ)}
\includegraphics[width=0.8\textwidth]{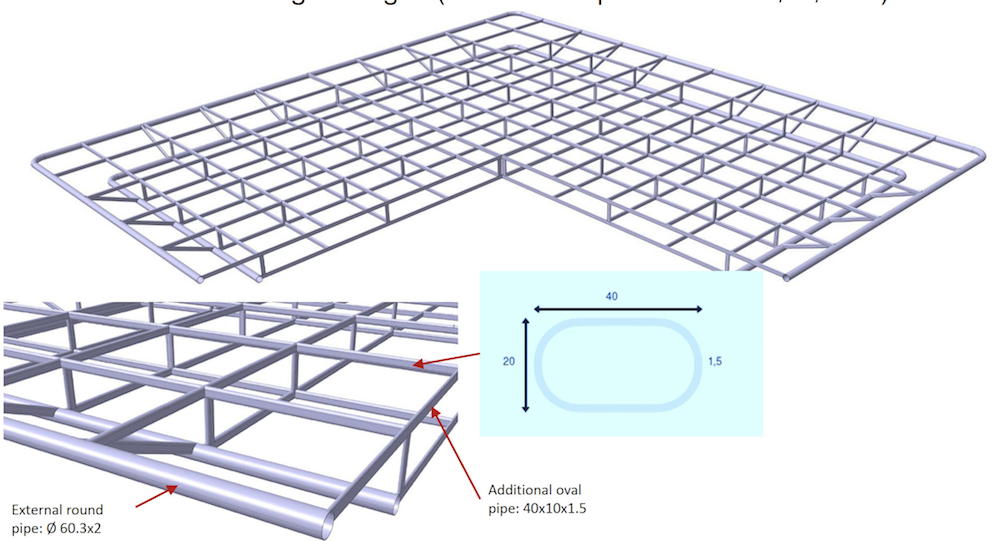}
\end{dunefigure}

\begin{dunefigure}[\dword{pddp} cathode \efield]{fig:dune-dp-cathode-field}
{Electric field map near the cathode, showing the maximum local field at <\SI{30}{kV/cm}, satisfying the requirement.} 
\includegraphics[width=0.7\textwidth]{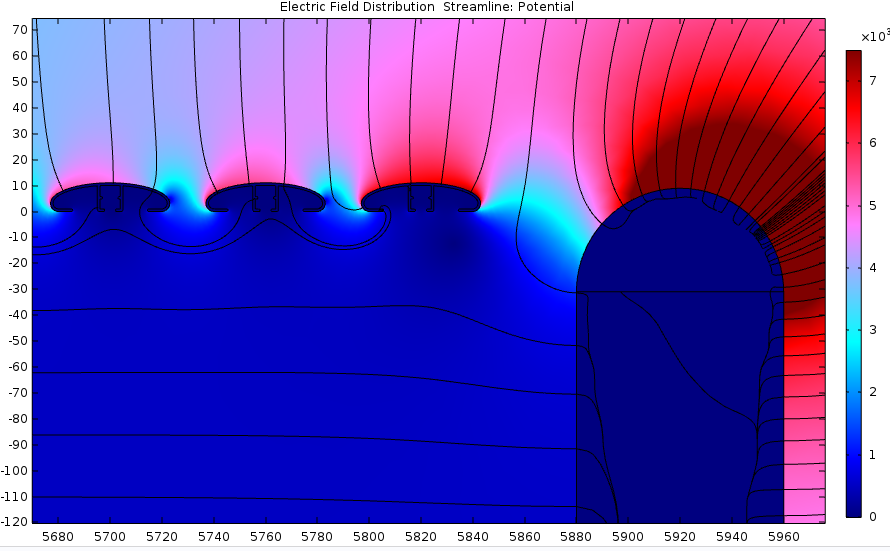}
\end{dunefigure}
The energy stored in the volume between the cathode plane and the \dword{gg} (which sits under the cathode and above the \dwords{pd}) is estimated to be about \SI{1.7}{\kilo\joule} over the \dptpcwdth $\times$ \dptpclen area, based on the cathode voltage and the distance of \SI{1}{\m} between the cathode and the \dword{gg} described in Section~\ref{sec:dp-hv-groundgrid}. 
A sudden discharge from the cathode frame to the cryostat membrane could cause severe damage to the membrane.
The modular construction of the cathode helps minimize this effect in case of discharge. During assembly in the cryostat, the cathode units are kept electrically insulated and connected to their adjacent neighbors through \si{\giga\ohm}  
resistors. Given the \SI{100}{\pico\farad} capacitance of each cathode unit, any discharge occurring in one unit will release at most 
\SI{21}{\joule} of stored energy while the discharge rate  
to the other units is slowed to the several-hundred-millisecond range.

Detailed calculations are in progress to determine the final shape and the size of the cathode and \dword{gg} frames to 
limit the maximum \efield to \SI{30}{\kV\per\cm}  
throughout the \lar volume (Figure~\ref{fig:dune-dp-cathode-field}), as per requirement 2 (Table~\ref{tab:hvphysicsreqs}).  Structural calculations are also in progress to verify the planarity of the cathode as it hangs on the \dword{fc} supports.
Value and voltage characteristics of the connecting resistors will also be defined according to results from dedicated simulations of the cathode electrical model.

\subsection{Ground Grid}
\label{sec:dp-hv-groundgrid}
The \dword{gg} is installed between the \dword{pd} and the cathode plane to shield the \pmt{}s from a discharge.  

The \dword{gg} consists of 316 L stainless tubes, as does the cathode. It is made of eighty \SI{3}{\m} $\times$ \SI{3}{\m} modules. Unlike the cathode, the \dword{gg} has a single layer supported by a set of feet resting on the membrane floor.
Detailed studies on the grid geometry are ongoing to ensure that the requirement on the maximum local field is satisfied.

\subsection{Field Cage}

\subsubsection{Mechanical Structure}
The field shaping rings of the \dword{fc} are made up of extruded roll-formed aluminum open profiles.  
The profiles are stacked horizontally and constructed into continuous rectangular rings that form the vertical sides of the drift volume. The aluminum profiles are attached to structural elements made of pultruded  FRP. 

FRP is non-conductive and strong enough to withstand the \dword{fc} loads in the temperature range of \num{-150}\,C to \num{23}\,C.
This material meets the  Class A International Building Code classification for flame spread and smoke development, 
as characterized by ASTM E84.

Each of these modules is composed of three 
styles of submodules of dimensions \SI{3}{\m} (W) $\times$ \SI{2}{\m} (H). These submodules are supported by two \SI{15.2}{cm} (\SI{6}{in}) FRP I-beams (with profile slots cut out) and two \SI{7.62}{cm} (\SI{3}{in}) cross bar I-beams that form a rectangular frame.

The \dword{fc} is modular, each module covering a vertical area of \SI{3}{\m} (W) $\times$ \tpcheight (H). 
There are two types of modules, straight section and corner, both types having the dimensions \SI{3}{\m} (W) $\times$ \tpcheight (H). A total of \num{40} straight section modules (i.e., with straight profiles) and eight corner section modules, with profiles bent \num{45} degrees at one corner to allow straight connections via clips at the corner, as shown in Figure~\ref{fig:dune-dp-fc-all}.c.  A photo of the \dword{pddp} \dword{fc} is shown in Figure~\ref{fig:dune-dp-fc-all}.d.

\begin{dunefigure}[\dual \dword{fc} parts]{fig:dune-dp-fc-all}
{\dword{fc} parts and connections for \dword{pddp}.  a. One \dword{pddp} \dword{fc} panel consisting of three submodules.  The \dword{dpmod} has virtually the identical structure, except for the number of middle submodules, which is four, b. A photo of the top module connection to the stainless steel I-beam and an inter-submodule connection, c. Aluminum clip connection at the corner and at the straight sections, d. \dword{hv} divider boards and their connections on \dword{pddp} \dword{fc}. }
\includegraphics[width=1.0\textwidth,height=1.0\textwidth]{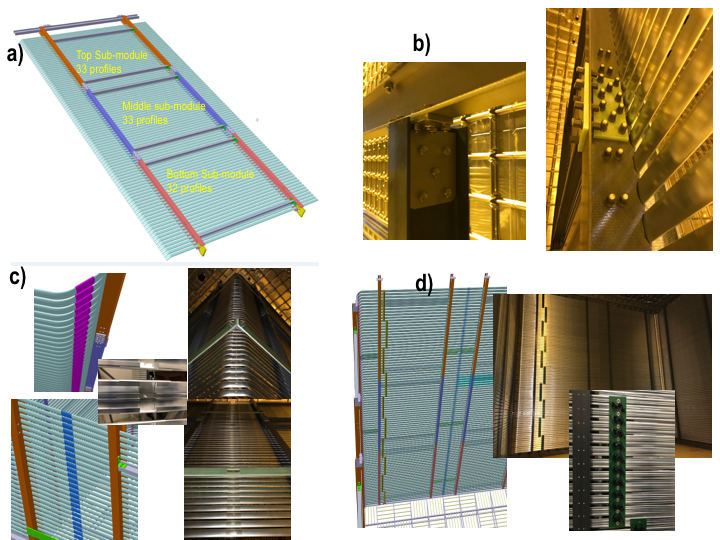}
\end{dunefigure}

Each \dword{fc} module is made of six submodules of three distinct types: top, bottom and middle. The dimensions for all submodules are \SI{3}{\m} (W) $\times$ \SI{2}{\m} (H).
Each module has one top submodule with \num{33} profiles, four middle submodules with \num{33} profiles each, and one bottom submodule with \num{32} profiles, for a total of \num{197} profiles per module. The voltage of the top-most field shaping ring  is \SI{-9}{\kV}. 
The top submodules make the mechanical connection to the ceiling of the cryostat from which the entire \dword{fc} hangs, as shown in Figure~\ref{fig:dp-fc-installation-connection}, left. The bottom modules make both mechanical and electrical connections to the cathode.

A railed rib runs along the length of the aluminum profiles at the center of the profile.  The rib provides mechanical strength and acts as the rail for the slip nuts that hold the profiles onto the supporting FRP frame. The rib also serves as the rail for the nuts that hold the \dword{hv} divider boards and the slip nuts for inter-module profile connections.  
All profiles at a given height are electrically connected via an aluminum clip screwed onto the slip nut across two neighboring profiles.  A resistive divider chain interconnects the aluminum profiles to provide a linear voltage gradient between the cathode and the top field-shaping ring.

The extruded aluminum profiles are mounted to the \SI{15.2}{\cm} (\SI{6}{in}) I-beam via two stainless screws and aluminum slip nuts in the center enforcement rail of the profile. The mounting is only on one of the \SI{15.2}{\cm} I-beams to allow contraction on either side of the profile. The top submodule has an extended \SI{15.2}{\cm} FRP beam with holes to connect it to the stainless steel I-beam hanging from the ceiling.  The bottom submodule has a cutout to hold the cathode plane onto it. The four middle submodules are symmetric, and thus interchangeable.
\Dword{fc} modules are horizontally interconnected with six G10 bars screwed to the vertical FRP I-beams at positions evenly distributed along the \tpcheight height. This interconnection guarantees the required alignment of the adjacent aluminum profiles for clipping. Given this interconnection, the full \dword{fc} can be considered as single skeleton that is mostly composed of FRP, which will shrink by about \SI{12}{cm} over \SI{60}{m} once the cryostat is filled with \lar. 
To compensate for this, the G10 bars are designed to be slightly longer than their final length in the \lar.

\subsubsection{Electrical Interconnections}

An aluminum clip connects the ends of each set of two end-to-end \dword{fc} profiles, forming continuous equipotential rectangular rings  \SI{144}{\m} long. 
Rows of \dwords{hvdb}, consisting of two resistors and a series of four surge-protection varistors in parallel, bridge the gap between the two neighboring stacked profile rings.   The total number of rows will be determined based on the redundancy and the current-limit requirements;  one row of \dwords{hvdb} is in principle sufficient to provide the required potential difference of \SI{3}{\kV} between neighboring rings, but more are desirable for redundancy.

The resistive chain for voltage division between the profiles provides a linear voltage gradient between the cathode and the top-most field shaping ring. It is critical as it determines the strength of the \efield between one profile and its neighbors, as well as between the profile and other surrounding parts, e.g., the grounded stainless steel membrane. The \efield needs to be kept well below \SI{30}{\kV\per\cm} at all points in the \lar bath to enable safe TPC operation.

The identified profile, Dahlstrom Roll Form \#1071\footnote{Dahlstrom Roll Form \#1071, Dahlstrom\texttrademark{}.}  is estimated to lead to \efield{}s of up to \SI{12}{\kV\per\cm},   
for the planned \dword{fc} configuration and operating voltage. Figure~\ref{fig:profile-e-field} illustrates results from an \efield calculation.

\begin{dunefigure}
[\efield map and equipotential contours of an array of roll-formed profiles]{fig:profile-e-field}
{\efield map (color) and equipotential contours of an array of roll-formed profiles biased up to \SI{-300}{\kV} and a ground clearance of about \SI{100}{\cm} in \dword{pddp} (CAD model)} 
\includegraphics[width=0.8\textwidth]{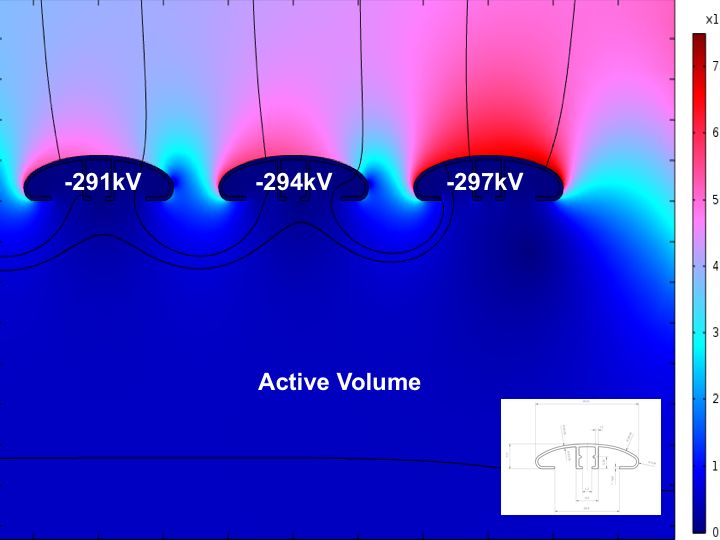}
\end{dunefigure}

Two distinct types of \dword{hv} divider boards are used for \dword{pddp}: \num{10} stage and \num{8} stage boards.  Each stage consists of two \SI{2}{\giga\ohm} resistors in parallel and four varistors of threshold voltage \SI{1.28}{kV} to protect the resistors in case of a sudden discharge.  The total expected current at \dptargetdriftvoltpos is therefore \SI{3}{\micro\ampere} per row.  Since multiple rows, which are connected in parallel, are used for redundancy, the expected current in the entire system is simply the number of rows times the current per each row.
For optimization purposes, one \dword{dpmod} \dword{hvdb} will connect \num{11} stages.  This enables  one \tpcheight tall module to be covered by fifteen \num{11} stage \dword{hvdb} and one \num{10} stage \dword{hvdb} at the bottom to make the final connection.

Figure~\ref{fig:dp-hvdb} shows one \dword{hvdb} board and  Figure~\ref{fig:dune-dp-fc-all}.d. shows the connection of one row of \dword{hvdb} along the entire height of the \dword{fc} as installed in \dword{pddp}. The  \dptpcwdth (W) $\times$ \dptpclen (L) ground plane consists of \num{80} unit planes (each \SI{3}{\m} $\times$ \SI{3}{\m}).  
The cathode plane is mounted to the bottom of the \dword{fc}, together forming one contiguous unit of field-providing structure.  The bottom-most \dword{hvdb} makes the connection between the cathode and the bottom-most field-shaping ring.
For redundancy, a total of two \dword{hvdb} rows are used in \dword{pddp}.  
The \dpmod will use one \dword{hvdb} chain every four \dword{fc} modules, providing an ample redundancy.  However, the final number of rows of \dword{hvdb} must take into account the impact of the underlying current through the \dword{fc} due to the particle interactions.

\begin{dunefigure}[\dual \dword{hvdb}]{fig:dp-hvdb}{\dword{pddp} \dword{hv} divider board (a) schematic circuit diagram, (b)photo of the top of the board, (c) photo of the bottom of the board}
\includegraphics[width=0.75\textwidth]{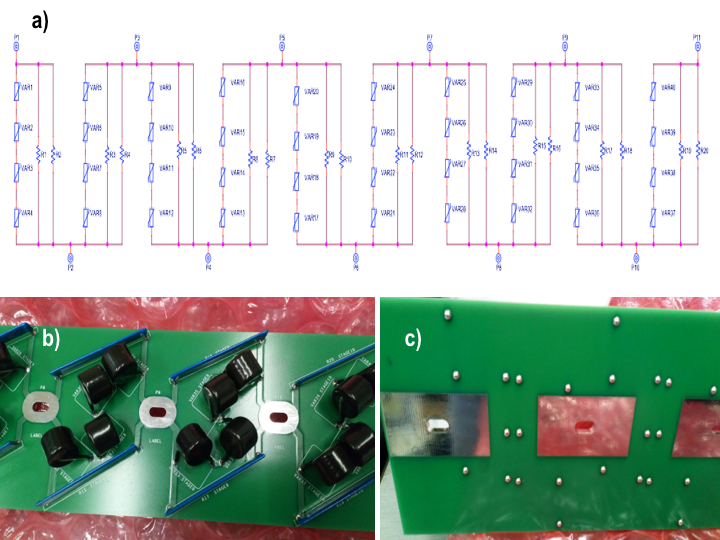}
\end{dunefigure}

\begin{dunefigure}[\dual \dword{fc} installation process and inter-module connection]{fig:dp-fc-installation-connection}{Left: Two submodules connected and hanging from the two sets of stainless steel cables on the ceiling.  The lifting wires are used to raise the module to its position as submodules are connected; the hanging wires keep the fully integrated module in its final position.  Right: Inter-submodule connections.  Each connection is made by two \num{1}{cm} thick G10 plates along the height of the I-beams and one \num{1}{cm} thick G10 plate on the flange.}
\includegraphics[width=0.75\textwidth]{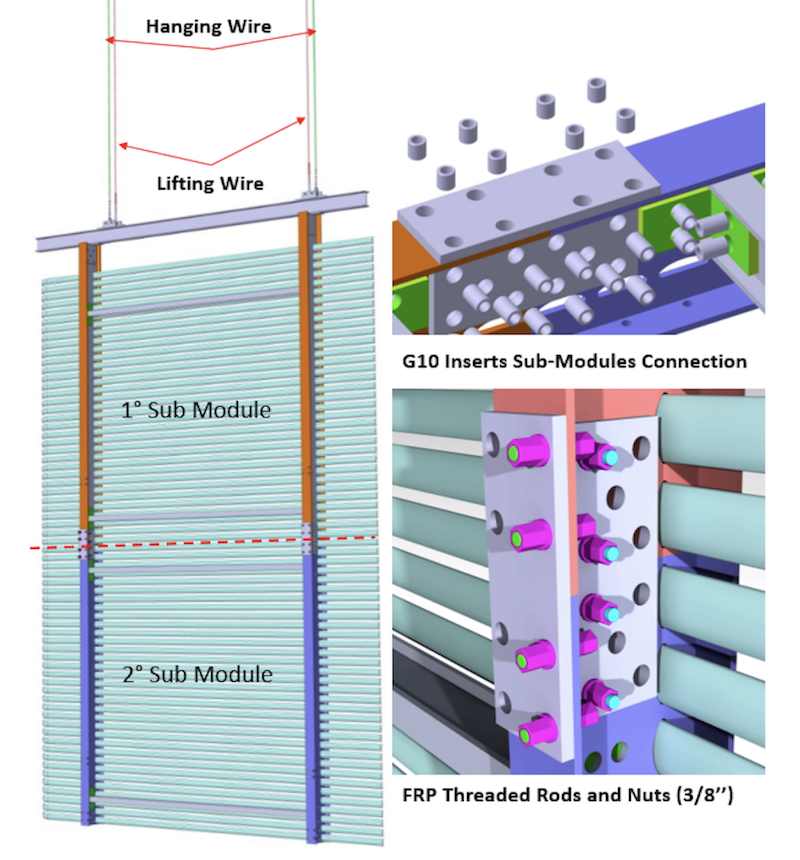}
\end{dunefigure}

\subsection{HV Return and Monitoring Devices}

In order to maintain the potential difference between the top-most field shaping ring and the extraction grid of the charge readout plane, either a dedicated \dword{hv} supply and an adjustable resistor chain is used in the \dword{hv} return outside of the cryostat.  This requires an independent \SI{10}{kV} \fdth similar to those developed for the extraction grid.

Multiple devices are planned for monitoring the \dword{hv}.  
\begin{itemize}
\item The Heinzinger units have typical sensitivities down to tens of nanoamperes with current readback capability.  The units are able to sample the current and voltage every few \SI{100}{\milli\s}.  
\item Inside the cryostat, so-called pick-off points near the anode will monitor the current through the \dword{hvdb} resistor chain.  

\item Additional pick-off points could be implemented on the \dword{gg} below the cathode to monitor possible stray currents.
\end{itemize}
\noindent

\section{Quality Assurance}
\label{sec:fddp-hv-qa}

{\bf Field Cage FRP Parts} Upon delivery of the FRP I-beams and other parts, all parts undergo a visual inspection process to look for defects, in particular those affecting structural integrity, such as cracks, air bubble holes, depression and flatness. The parts are sorted into three preliminary categories: category 0 (pass), category 1 (problematic but could be repairable) and category 2 (severe and unusable) .

The visual inspection is followed by critical dimension measurements to verify individual submodule assembly and module interconnection integrity.  These measurements focus on cross sectional dimensions for rods, plates and bars, and on the length, straightness, flatness and camber of the beams and flanges.
All measurements must satisfy the mechanical tolerance provided by the design drawings and the standard industry quality criteria.

The FRP I-beams and parts in category 1 undergo a repair process during the preparation stage through defibering, deburring and sanding.  Another set of measurements is performed after the repair to redetermine the part's category.

Those in category 2 are returned to the vendor for replacements.

{\bf \dword{hv} Divider Board}
All resistors used for board production are numbered and undergo a three-stage \dword{qa} test for final selection. Resistance is measured at voltages up to \SI{4}{kV} in \SI{500}{V} steps for each stage. The first stage is done at  room temperature, the second at LN temperature, and the third again at room temperature after warming up.  The measured values are histogrammed for the final selection, in order to look for groupings of values.   
It is more important that the resistance values be close to each other than that they be at any particular value (unless they are too far from the design values). The resistors for which the resistance values are within 1\% of each other are selected.  

All varistors for board production are also numbered for \dword{qa} purposes and undergo a three-stage \dword{qa} testing program for the final selection. Each stage is a clamping voltage measurement, first at room temperature, then at LN temperature, and again after warming back up up to room temperature.  
The measured values are histogrammed for the final selection; those for which the clamping voltage is closer to the design values are selected, to ensure proper protection.   

Once the electrical parts are mounted on the \dword{hvdb} by the vendor, they undergo a three-stage \dword{qa} testing program. Each stage involves a resistance measurement, first at room temperature, then at LN temperature, and again after warming back up up to room temperature.  The measured values are histogrammed for the final selection.
The boards can fall into three categories: 0. Pass, 1. Repairable and 2. Rejected.  
If at any testing stage the resistance is more than \SI{0.5}{\%} away from the mean, the part does not fall into category 0, pass.  Boards in category 1 are sent back to the vendor with the selected resistors so that the parts in the failed stages can be replaced. 

{\bf Aluminum Profiles and Clips}
The \dword{qa} testing of the aluminum profiles and clips is done on prototype production samples prior to full production.   The samples are visually inspected for their shape and their adherence to the design drawing, the surface smoothness, the surface-coating quality, and the smoothness of the bend (for the profiles with the \num{45}$^{\circ}$ bend).  Clip samples are visually inspected for their shape and their adherence to the drawing, their edge smoothness, their mechanical fitness, and how tightly they fit to the profiles.

{\bf Cathode, HVFT and the Extender} The \dword{qa} for other components is under development as part of the \dword{pddp} construction efforts.

{\bf Power Supply and Feedthrough} The power supply is tested extensively along with the controls and monitoring software.  
Capabilities to test include:
\begin{itemize}
\item Ramp and change the voltage; including rate change and pause capabilities and settings. 
\item Accept user-defined current limit.  This parameter sets the value of current at which the supply reduces the voltage output in order to stay below it.  The current limiting itself is done in hardware.
\item Accept a setting for the trip threshold current.  At this threshold the software would reduce the voltage output. 
In previous experiments, the trip function in software would set the output to \SI{0}{kV}. 
\item Record the current and voltage readback values with a user-defined frequency, and also to record any irregular current or voltage events. 
\end{itemize}

\section{Production and Assembly}
\label{sec:fddp-hv-prod-assy}

\subsection{Field Cage Submodule Frames}
\label{sec:fddp-hv-prod-assy-fc-frames}

All FRP parts that pass the \dword{qa}  process undergo the preparation process, which involves \dword{qc} at each step.  Each of the parts is first defibered, deburred, sanded to smooth the edges, and then varnished to
suppress any remaining fibers.  Once the part completes the preparation process, it is stored in a humidity- and temperature-controlled drying area for \numrange{24}{48} hours to fully cure the varnish.  Upon full curing, each module is laser-engraved with a unique part number and recorded in a \dword{qc}  database. 

At this point the submodule is pre-assembled on a table prior to packaging to ensure fitness of all parts, including inserts, screws and slip nuts.   When the fitness is verified, the submodule is disassembled and packaged into a \SI{2}{\cm}$\times$\SI{20}{\cm}$\times$\SI{2}{\m} compact package with the edges wrapped with a thick plastic layer to protect the parts from mechanical damage, e.g., from a fall.  This package includes all necessary parts to assemble the given submodule 
along with \numrange{10}{20}\% spare parts.  This ensures that each package is self-sufficient for the final assembly.  Each package is given a submodule number for further tracking purposes.  This submodule number stays with it throughout assembly at \surf and final installation so that the submodule can be assigned to its specific module and in its proper location within the module. 

These packages are shipped to \surf and transported underground 
for final assembly prior to the installation.   
The \SI{3}{\m} long aluminum profiles are produced and shipped to \surf separately for final assembly.  The final \dword{qc} of the profiles is conducted during submodule assembly.   Profiles with deep scratches and sharp protrusions that could cause excess charge concentration will be rejected. This process must be done underground at the time of submodule assembly because of the delicate nature of the coated surface and the smoothness requirement. 

Assembly of submodules from the parts is carried out inside the cryostat.   An assembly table with a precision alignment bar for rapid profile alignment is used for this task.   The package is opened on the table and the FRP frame is assembled with four L-brackets and sixteen \SI{5.61}{\cm} (\num{2.21}\,in) threaded rods through the G10 insert and eight \SI{5.77}{\cm} (\num{2.27}\,in) threaded rods through G10 inserts held by the plastic nuts.  These nuts are first hand-tightened and then set with a torque wrench. 

Once the frame is assembled, aluminum profiles are inserted through the corresponding slots on the \SI{15.2}{\cm} (\num{6}\,in) on either side.  One slip-nut 
is inserted onto the center rib rail prior to the profile insertion.  
Care must be taken to ensure that the profiles are not scratched during the insertion process. 

Once all profiles are inserted into the frame, each of the profiles is mounted onto the I-beam using two \SI{45}{\mm} M4 button head hex drive stainless steel screws into the aluminum slip nut. 
When all screws are hand-tightened as much as possible, the final torque is applied using a torque wrench.  The final alignment of the profiles is then made by hand, tightening only one or two screws, as necessary.
A wheel base made of unistrut bars is then mounted to the bottom of the submodule and the entire unit is placed in a designated area to await installation.   The wheel is clearly labeled with the submodule number for tracking purposes.

\subsection{Cathode Plane}
\label{sec:fddp-hv-prod-assy-cathode}


Each cathode module is assembled off-site and shipped to \surf in its transport container.   The containers are transported underground to the cryostat and opened in front of the cryostat.  The cathode module is then inserted into the cryostat for installation.

\subsection{Ground Plane} 
\label{sec:fddp-prod-assy-ground-grid}


Each \dword{gg} module is assembled off-site and shipped to \surf in its transport container.   The containers are transported underground to the cryostat and opened in front of the cryostat.  The \dword{gg} module is then inserted into the cryostat for installation. 

\subsection{Electrical Interconnections}
\label{sec:fddp-prod-assy-elec-connec}

The \dwords{hvdb} are mounted on the profiles as the submodules are built.  For optimization purposes, one \dword{dpmod} \dword{hvdb} connects \num{11} \dword{fc} stages.   
One \tpcheight tall module is covered by fifteen \num{11}-stage \dwords{hvdb}, and one \num{10}-stage \dword{hvdb} is at the bottom to make the final connection.   One row of \dwords{hvdb} is mounted every four \dword{fc} modules.  All connections and electrical functionality must be checked with a high-sensitivity electrometer.

\section{Interfaces to the HV System}
\label{sec:fddp-hv-intfc}

\subsection{Cryostat}
\label{sec:fddp-hv-intfc-to-cryostat}

Each \dword{fc} module is suspended and raised by two ropes hung from the cryostat roof through the \dword{fc} suspension \fdth{}s, by winches as in \dword{pddp}.  

Once an \dword{fc} module of the full height is completed for \dword{dpmod}, it will be hung 
from the cables attached to the final suspension hooks. The suspension \fdth{}s are then fully sealed. 
A possible improvement on the \dword{pddp} interface is to use remote-controlled electrical winches (as opposed to manual) to ensure synchronized lifting of the modules.

\subsection{Charge Readout Plane}
\label{sec:fddp-hv-intfc-to-crp}

No direct hardware interface exists between the \dword{hv} and \dword{crp} systems. The cryostat penetrations and \fdth{}s for the two systems are completely independent, as are their control electronics. The only interface envisaged includes a system to ensure the appropriate physical separation between the top-most field-shaping profile and the \dword{crp} extraction grid.

\subsection{Photon Detection System}
\label{sec:fddp-hv-intfc-to-pds}

No direct hardware interface exists between the \dword{hv} and \dword{pd} systems. The cryostat penetrations and \fdth{}s for the two systems are completely independent, as are their control electronics. The only interfaces envisaged include (1) maintenance of a safe minimum distance between the \dwords{pd} and the \dptargetdriftvoltneg cathode; and (2)
 definition of the \dword{pd} power dissipation limit (production of bubbles would compromise the \dword{hv} stability). The power dissipation depends on the final \dword{pd} density chosen, which is awaiting simulation results. Aspects of the cathode design are also awaiting simulation results to determine its impact on the \dword{pds}. 
These interfaces are effectively at the level of design requirements.

\section{Transport, Handling and Integration}
\label{sec:fddp-hv-install-transport}

The \dword{fc} FRP beams and G10 parts are shipped in a standard wooden crate.  Parts for each submodule are packed into one 
flatpack of dimensions \SI{0.2}{m}$\times$\SI{0.2}{m}$\times$\SI{2}{m} sealed in multiple layers of shrink wrap and a thick soft cushion of plastic for edge protection.   Due to the compact size of each submodule package and since a total of \num{530} or so submodules are expected, including 10\% spare, the total number of crates for FRP parts to be transported down the shaft will be small. Given the dimensions of the hoist cage, an optimal crate size could be \SI{1.2}{m}$\times$\SI{1.5}{m}$\times$\SI{2.3}{m}, each containing \num{30} flatpacks of FRP parts.


The extruded aluminum profiles for \dword{fc} are shipped separately in  standard wooden crates.  A total of \num{197} profiles are needed for each \tpcheight module. Therefore, the total number of profiles to be transported down underground would be of the order \num{18000}, including \SI{10}{\%} spares.  The same number of aluminum clips are also shipped in standard wooden crates and are transported underground separately.

The cathode modules of dimension  \SI{3}{m}$\times$\SI{3}{m} are assembled off-site and shipped to \surf in their transport containers. Given the paucity of storage space at the 4850L, the cathode plane is assembled as each unit cathode plane arrives at the cavern -- in its assigned order -- and is unwrapped. The cathode plane is assembled inside the cryostat by connecting the cathode units mechanically and electrically, taking into account the constraints imposed by the installation of the \dword{gg} modules and the \dwords{pmt}. 
%

The integration starts from  the cryostat end opposite to the \dword{tco}.  The field cage modules covering the 12 m long side of the cryostat are installed  first. This installation is followed by the CRP installation, which requires access to the cryostat floor. Once a 3$\times$12\,m$^2$ strip of CRP modules is installed, the corresponding field cage side modules are installed followed by the installation of the four cathode modules. The four ground grid modules are initially attached to the cathode modules allowing access for the removal of the false floor  and the subsequent installation and cabling of the photomultipliers below the cathode modules.  After the installation of the photomultipliers, the ground grid modules are lowered to their final positions. This installation sequence is repeated until 
the entire detector is complete. The field cage modules covering the endwall on the TCO side are installed before the installation of the last  $3 \times 12$ m$^2$ cathode strip.

The power supply, \fdth{}s and \dword{hv} extender are sent to \surf in standard shipping crates. Unwrapping requires clean areas and careful handling. Surfaces can be cleaned with alcohol and allowed to dry.

\section{Quality Control}
\label{sec:fddp-hv-qc}

The assembly, testing, transport, and installation procedures required to ensure adequate \dword{qc} of all \dword{hv} system components are being defined, tested and documented during the construction of \dword{pddp}.

The \dword{fc} submodules are assembled inside the cryostat on an assembly table with a precision alignment bar, as in \dword{pddp}.  
Each submodule-FRP part package is visually inspected for external damage and is opened carefully to avoid damage to the FRP parts.  
The bags of hardware are removed and set side.   The two \SI{15.2}{\cm} (\SI{6}\,in) I-beams, two \SI{7.62}{cm} (\SI{3}{in}) I-beams and connecting L-brackets are visually inspected for  damage during transport.  Once the FRP parts pass the inspection, they are assembled into the frame on the table.  


The aluminum profiles are visually inspected and felt by hand for severe scratches and any excess sharp points.  The profiles that pass are inserted into the profile slots on the FRP submodule frame, with one alignment-fixing slip nut inserted into the rail on the reinforcement rib, 
 and follow the assembly and alignment process described in Section~\ref{sec:fddp-hv-prod-assy-fc-frames}.  The alignment of the profiles is checked using a straight-edge along one end of the profile.  The submodules that pass are put on the wheel base and stored for installation.

The \dword{hv} divider boards are tested on-site for resistance of each stage at room temperature to ensure the integrity of the board and each electrical connection of the resistors before installation onto the submodule.

At \surf, cathode and \dword{gp} modules are checked for the required planarity and mechanical integrity. Also during the installation phase  the electrical continuity between the modules is checked.

The \fdth and the \dword{hv} extender are tested simultaneously at the 
testing facility  site (possibly at CERN or the \dword{itf}), 
preferably with the planned 
power supply.  To pass, the \fdth must hold the goal voltage (\dptargetdriftvoltneg{}) in ultra pure \lar (TPC-quality purity corresponding to a free electron lifetime, $\tau\geq$\SI{7}{\ms}) for at least \num{24} hours. The ground tube submersion and \efield environment of the test setup must be comparable to the real \dword{fc} setup or more challenging.  Additionally, the \fdth must be UHV-grade leak-tight.

Upon arrival at \surf the power supply used in the \dword{dpmod} \dword{hv} system is tested before installation, with  output voltages and currents checked on a known load. 

\section{Safety}
\label{sec:fddp-hv-safety}



%

%

In all phases of \dword{hv} system development of the \dpmod, including fabrication, installation, and operations, safety is the highest priority.  
As for \dword{pddp}, assembly, testing, transport, and installation procedures will be documented. Explicit attention is paid to the transferability of the \dword{pddp} procedures to the \dpmod; the most critical of these are noted in the preliminary \dword{pddp}  risk assessment document\footnote{CERN EDMS document number 1856841.}.
\fixme{make a citation later}


The structural and electrical designs for the \dword{dp} \dword{hv} system are based on designs that are vetted and validated in the \dword{pddp} construction, which is currently in its final phase of deployment at CERN. In parallel with the \dword{pddp} contruction and operation,  \dword{hv} tests at CERN are planned using 
a full-voltage and full-scale \dword{hv} \fdth, power supply, and monitoring system in dedicated \dword{hv} test facilities. This also provides an opportunity 
to complete full safety reviews. 

Operating the \dword{fc} at its full operating voltage produces a substantial amount of stored energy. The modular design of the cathode specifically addresses this safety concern: in the event of a power supply trip or other failure that unexpectedly drops the \dword{hv}, the charge stored in the segmented cathode structure limits the power dissipated. 
This design will be tested in \dword{pddp} at \SI{300}{kV} voltage over \SI{9}{\m$^2$} surfaces segmented in four cathode modules.  

Integral to the \dword{dp} \dword{fc} design, both in \dword{pddp} and the \dpmod, is the concept of pre-assembled modular panels of field-shaping conductors with individual voltage divider boards. The structural design and installation procedures used in \dword{pddp} were selected to be compatible with use at the \dword{fd} site and were vetted by project engineers, engineering design review teams, and CERN's safety engineers. Some revisions to these designs are expected based on lessons learned in installation and operations; these revisions will be reviewed both within the DUNE project and by \fnal EH\&S personnel. The overall design is on solid footing. 

Assembly of the \dword{fc} panels and resistor-divider boards does not present unusual  hazards. 
The \dword{hv} consortium will work closely with each assembly site to ensure that procedures meet both \fnal{}'s and institutional requirements for safe procedures, personal protective equipment, environmental protection, 
materials handling, and training. The vast majority of 
part fabrication will be carried out commercially, and shipping will be contracted through approved commercial shipping companies. Prior to approving a site as a production venue, the site will be visited and reviewed by an external safety panel to ensure best practices are in place and maintained. 

\section{Organization and Management}
\label{sec:fddp-hv-org}

\subsection{HV Consortium Organization}
\label{sec:fddp-hv-org-consortium}

At present, the \dword{hv} consortium is gathering 
institutions to participate in the design, construction and assembly of the \dword{hv} systems for both \dwords{spmod} and \dwords{dpmod}. 
The consortium needs to grow in the near future, and it hopes to attract new  institutions, in particular from EU to balance USA participation with additional international participants. 

\begin{dunetable}
[HV consortium participants]
{p{0.6\linewidth}}
{tab:hvconsortiumparticipants}
{HV Consortium Participants}   
 Institution  \\ \toprowrule
EU: CERN   \\ \colhline
USA: Argonne National Lab     \\ \colhline
USA: Brookhaven National Lab   \\ \colhline
USA: University of California (Berkeley)   \\ \colhline
USA: University of California (Davis)    \\ \colhline
USA: Fermi National Accelerator Lab   \\ \colhline
USA: University of Houston    \\ \colhline
USA: Kansas State University   \\ \colhline
USA: Lawrence Berkeley National Lab  \\ \colhline
USA: Louisiana State University   \\ \colhline
USA: South Dakota School of Mines and Technology    \\ \colhline
USA: Stony Brook University    \\ \colhline
USA: University of Texas (Arlington)   \\ \colhline
USA: Virginia Tech.  \\ \colhline
USA: College of William and Mary  \\

\end{dunetable}
The consortium management structure currently includes a consortium leader from CERN, a techincal lead from BNL, and \dword{tdr} editors from CERN and UTA.

In the current \dword{hv} consortium organization, each institution is naturally assuming the same responsibilities 
it had for \dword{pddp}. The consortium is organized into six working groups (WG) that are addressing the design and  R\&D phases, and the hardware production and installation.

\begin{itemize}
\item WG1. Design optimization for \single and \dual; assembly, system integration, detector simulation, physics requirements for monitoring and calibrations. 
\item WG2. R\&D activities and facilities.
\item WG3. \single{}-CPA: Procurement, in situ \dword{qc}, resistive panels, frame strips, electrical connections of CPA modules, \dword{qc}, assembly, shipment to assembly site / \dword{qc}.
\item WG4. \dual Cathode. 
\item WG5. \dword{fc} modules. 
\item WG6. \dword{hv} supply and filtering, \dword{hv} power supply and cable procurement, R\&D tests, filtering and receptacle design and tests. 
\end{itemize}

\noindent Merging of \single and \dual groups is envisaged for the working groups where synergies are being identified: \dword{hv} \fdth{}s, voltage dividers, aluminum profiles, FRP beams, and assembly infrastructures.

\subsection{Planning Assumptions}
\label{sec:fddp-hv-org-assmp}

The present baseline design for all elements of the \dword{hv} system for \dword{dpmod} strictly follows the \dword{pddp} design as it has been produced and is being assembled.  It is also assumed that no major issues in the \dword{hv} system operation of \dword{pddp} will be encountered and therefore that the basic \dword{hv} system concepts are sound.

However some design modifications and simplifications must be implemented to take into account the 
doubled drift distance, implying an increase in \dword{hv} delivery to the cathode from \SI{-300}{\kV} to \dptargetdriftvoltneg{}.

The \dual \dword{hv} system distribution and the related cathode structure still require intense R\&D, given the unprecedented value of the required \dword{hv} (\dptargetdriftvoltneg).
The related results could lead to revision of design details such as the shape of the cathode elements and of the 
\dword{gp} structures, the distance from the cryostat walls, the distance between the cathode and the 
\dword{gp} protecting the \dwords{pd}, and resistive connections of the cathode modules. It is important to ensure that the \efield intensity in the \lar is below the critical value of  about \SI{30}{\kV/\cm} everywhere and that the energy stored in the \dword{fc} is not  released catastrophically to the detector membrane. 

As for the \dword{spmod}, \dword{pdsp} serves as the test-bed for understanding and optimizing detector element assembly, installation sequence, and integration as well as requirements in 
human resources, space and tooling, and schedule. 


%
%
%

\cleardoublepage

\chapter{Photon Detection System}
\label{ch:fddp-pd}

\section{Overview}
\label{sec:fddp-pd-1}

\subsection{Introduction}
\label{sec:fddp-pd-1.1}

The \dword{dp} \dword{pds} primarily serves three purposes. It provides the trigger for non-beam events; it enables determination of the event absolute time for beam and non-beam events; and it enables calorimetric measurements. 
It is essential to ensure that the \dual \dword{pds} is optimized for the full DUNE physics program. In particular, low-energy signals like \dword{snb} neutrinos and multi-messenger astronomy, other low-energy signals, and proton decay, 
impose more stringent requirements on \dword{pds} performance than the primarily higher energy, beam-synchronous, neutrino oscillation physics. The final specifications of the system will be determined so as to achieve these physics requirements. 
A number of scientific and technical issues impact the \dual \dword{pds}  and \single \dword{pds}  in a similar way, and the consortia for these two systems cooperate closely.  See \voltitlespfd{}, Chapter 5 for details on the \single \dword{pds}.

This 
 chapter concentrates on direct projection of the \dword{pddp} design to the DUNE scale. The optimization and final design of the \dual \dword{pds} is driven by the \dword{pddp} \cite{protoDUNDP-tdr} data and simulation studies.

The chapter begins with an overview of the system in Section~\ref{sec:fddp-pd-1}. Section~\ref{sec:fddp-pd-2} describes the photosensors, namely \dwords{pmt} 
and the related \dword{hv} system, wavelength shifters and light collectors. The mechanics associated with the \dwords{pmt} is described in Section~\ref{sec:fddp-pd-3}, and the readout electronics in~\ref{sec:fddp-pd-4}. Section~\ref{sec:fddp-pd-5} details the photon calibration system to monitor the \dword{pmt} gain and stability. Then, the \dword{pd} performance is described in Section~\ref{sec:fddp-pd-6}, and the operations in Section~\ref{sec:fddp-pd-7}. Interfaces with other subsystems are described in Section~\ref{sec:fddp-pd-8}. Section~\ref{sec:fddp-pd-9} includes the installation, integration and commissioning plans. The \dword{qc} procedures are outlined in Section~\ref{sec:fddp-pd-10}. The main safety issues to consider are specified in Section~\ref{sec:fddp-pd-11}. To finish, the management and organization is described in Section~\ref{sec:fddp-pd-12}.

\subsection{Physics and the Role of Photodetection}
\label{sec:fddp-pd-1.2}

The main physics goals of the \dword{dpmod} are to register beam events 
and to operate outside of the beam spill as an efficient observatory for \dwords{snb} and proton decays. DUNE will also collect atmospheric neutrino and muon events, and will conduct searches for a number of exotic phenomena postulated by extensions of the standard model.  Expected or searched for signals can range in energy from a few \si{MeV} to many \si{GeV} and have characteristic time duration and topological features that challenge the performance of large noble liquid TPCs. 
A critical part of the \lartpc is the \dword{pds}, sensitive to light produced by interactions in argon \cite{Cuesta:2017nrs}. In \dual TPCs, the timing of prompt scintillation light (usually referred to as the S1 signal) in \lar is needed for time stamping of events and propagation of tracks in the detector. The extraction and amplification of drift electrons in the gas phase (usually referred to as the S2 signal) yields information on the drift time and amount of ionization charge, thus supplementing information from the charge readout on the anode plane. The interplay between the charge and light from an event enables 
pattern recognition and the measurement of energy of interactions.

Ionizing radiation in liquid noble gases leads to the formation of excimers in either singlet or triplet states, which decay radiatively to the dissociative ground state with characteristic S1 fast and slow lifetimes (fast is about \SI{6}{ns}, slow is about \SI{1.3}{$\mu$s} in \lar with the so-called second continuum emission spectrum peaked at the wavelength of approximately \SI{127}{nm}, \SI{126.8}{nm} with a full width at half maximum of \SI{7.8}{nm} \cite{Heindl}). This prompt and relatively high-yield (about \num{40000} photons per \si{MeV}) of \SI{127}{nm} scintillation light is exploited in a \lartpc to provide the absolute time ($t_0$) of the ionization signal collected at the anode, thereby providing the absolute value of the drift coordinate of fully contained events, as well as a prompt signal used for triggering purposes.

The secondary scintillation light S2 is produced in the gas phase of the \dword{dpmod} when electrons, extracted form the liquid, are accelerated in the \efield between the liquid phase and the anode. The secondary scintillation in the argon gas (i.e., the vapor phase) is the luminescence in gas caused by accelerated electrons in the \efield and in the \dword{lem} anode through Townsend amplification. For a given argon gas density, the number of photons of this S2 signal is proportional to the number of electrons, the \efield, and the length of the drift path in gas covered by the electrons. In an extraction field of \SI{2.5}{kV/cm} in gas, one electron generates about \num{75} photons. The time scale of S2 reflects the extraction time of original ionization in the liquid phase into the gas phase, thus for about \SI{1}{kV/cm} \efield, this time scale is of the order of hundreds of microseconds. The time between the occurrence of the primary scintillation light and the secondary scintillation light is equivalent to the drift time of the electrons from the ionization coordinate to the \lar surface.

The baseline design of the light collection system calls for \SI{20.3}{cm} (\SI{8}{in}) diameter cryogenic \dwords{pmt} distributed uniformly on the floor of the cryostat and electrically shielded from the bottom cathode plane. The proposed density of \dwords{pmt} and their arrangement follows the design of the \dword{pddp} detector. On the other hand, modeling and simulations of light collection both for \dword{pddp} and the DUNE \dwords{detmodule} are still ongoing. Therefore, even critical system parameters and their impact on the physics reach are still tentative. Results from the \dword{pddp} will provide the critical validation of simulations and will guide optimizations for the large \dword{dpmod}.

\subsection{Technical Requirements}
\label{sec:fddp-pd-1.3}

\Dwords{pmt} provide the sharpest time information of events in the \lartpc and in the gas phase of the extraction stage. Due to necessary wavelength-shifting of photons from the argon luminescence and shadowing by the cathode plane, the efficiency of light detection is challenging and requires careful mechanical, electrical, and optical designs. The \dual \dword{pd} consortium is presently in contact with \dword{pmt} manufacturers, including 
Hamamatsu Photonics~\footnote{Hamamatsu Photonics\texttrademark{}, \url{http://www.hamamatsu.com/resources/pdf/etd/LARGE_AREA_PMT_TPMH1286E.pdf}} in Japan, 
Electron Tubes Limited~\footnote{Electron Tubes Ltd\texttrademark{}, \url{http://www.electron-tubes.co.uk//}} in the USA and UK, and 
HZC~\footnote{HZC Photonics\texttrademark{}, \url{http://hzcphotonics.com/en_index.html}.} in China, 
to define optimal choice and configuration of \dwords{pmt} satisfying our requirements, tentatively summarized in Table~\ref{tab:dppd_t_1_3}. These requirements will be reviewed based on the physics needs. For this, simulations and \dword{pddp} results will be key.

\begin{dunetable}
[Summary of tentative requirements for the \dword{dpmod} \dword{pds}] 
{p{0.3\textwidth} p{0.25\textwidth} p{0.45\textwidth}}
{tab:dppd_t_1_3}
{Summary of tentative requirements for the \dword{dpmod} \dword{pds}. 
This table represents the baseline choice for the R5912-MOD20 \dword{pmt} as manufactured by Hamamatsu Photonics \cite{hamamatsu-5912} is made.}
\footnotesize
{\bf Feature}	& {\bf Goal}  	& {\bf Comment	}	\\
\toprowrule
\multicolumn{3}{l}{\bf Optical} \\ \specialrule{1.5pt}{1pt}{1pt}
Spectral response & \SI{127}{nm} 	& Wavelength shifters are required.				\\
Light yield & 2.5 \phel/\si{MeV} & For \dword{pddp}, final value depends on simulation results.\\
\toprowrule
\multicolumn{3}{l}{\bf Electronic} \\ \specialrule{1.5pt}{1pt}{1pt}
Minimum light signal & SPE & Required to perform the \dword{pmt} gain measurement. \\
Gain			& $\sim$\num{e6}-\num{e9} & Given by \dword{pmt} specifications.			\\
Noise (or signal/noise) &  <\SI{1}{mV} & To distinguish SPE from noise, depends on \dword{pmt} gain. \\
Timing resolution & few ns & To distinguish S1 from S2 component. \\
Power		& < \SI{0.2}{W/\dword{pmt}}& Used successfully in the  \dword{wa105}.			\\
\dword{adc} dynamic range & TBD & Depends on simulation results, see Section \ref{sec:fddp-pd-4.3}  \\ 
\toprowrule	
\multicolumn{3}{l}{\bf Electrical} \\ \specialrule{1.5pt}{1pt}{1pt}
 	\dword{hv} range		& \si{0-2500}{V}	& Individual cable per  \dword{pmt}.			\\
	\dword{hv} resolution	& \SI{1}{V}		    & Fine tuning of \dword{pmt} gain.			\\
	\dword{hv} noise		& <\SI{100}{mV}		& Extra filtering is required. \\
	\dword{hv} grounding	& Isolated		    & \dword{hv} outputs are floating, crate ground is independent of the return of the \dword{hv} channels.	\\
\dword{pmt} placement		& Isolated		    & \dwords{pmt} electrically shielded from the TPC cage.	\\
\toprowrule	
 \multicolumn{3}{l}{\bf Mechanical} \\ \specialrule{1.5pt}{1pt}{1pt}
    Temperature 	& Cryogenic	        & \dwords{pmt} operated in \lar and tested in liquid nitrogen. 	\\
	Pressure 		& \SI{+2}{bar}	    & The argon column is about \SI{10}{m} high.	\\		
\end{dunetable}

\Dwords{pmt} will be installed at a baseline density of one per \SI{1}{m^2}. The choice of the Hamamatsu R5912-MOD20 \dword{pmt} 
is assumed as the baseline plan. Extension of the \dword{pmt} light sensitivity region to the \lar light emission wavelength of \SI{127}{nm} requires use of a wavelength shifter. 
Therefore, in the baseline plan, the hemispherical windows of the \dwords{pmt} are 
coated with a thin layer of \dword{tpb}~\cite{tpb} for wavelength-shifting into the range suitable for R5912 \dword{pmt} photocathode sensitivity~\cite{hamamatsu-5912}. The \dwords{pmt} have to be rigidly anchored to the bottom of the cryostat. Different \dword{pmt} densities and placements along the walls are also being considered in simulations. The \dword{hv} signal cables are routed along the cryostat walls to \fdth{}s installed in the roof of the cryostat. Each \dword{pmt} is controlled individually so that its gain can be individually adjusted to match the \dword{fe} dynamic range and \dword{s/n} ratio. An average light yield of \SI{2.5}{\phel/MeV} is required based on \dword{pddp} simulations (see Section~\ref{sec:fddp-pd-6}). The precision of the light yield requirement will be established after the completion of the full DUNE simulations.

The cathode plane, described in Chapter~\ref{ch:fddp-hv}, 
 is placed at a height of about \SI{2}{m} above the bottom of the cryostat, and the \dword{pmt} plane is distant enough from the cathode plane, taking into account the high electrical rigidity of the \lar phase. In order to protect the \dwords{pmt}, the ground grid is installed and placed at an identical potential as the \dword{pmt} photocathode (\SI{0}{V}). The \dwords{pmt} are powered to values between \numrange{1.5}{2.0}\si{kV} such that the \dword{pmt} gain is $\sim$\numrange{e7}{e9}. Future developments on the quantification of possible \dword{tpb} dissolution in \lar will be encouraged and followed \cite{TPBdiss}.

\subsection{Detector Layout}
\label{sec:fddp-pd-1.4}

The \dword{pmt} plane is placed far enough below the cathode plane to be sufficiently electrically shielded. According to the baseline plan, the \dwords{pmt} are uniformly distributed across this plane with a density of \SI{1}{\dword{pmt}/m^2}, with a total of \dpnumpmtch \dwords{pmt} installed. Other \dword{pmt} configurations as determined by the simulations are also being considered. The \dwords{pmt} are individually mounted to the cryostat floor. The exact location of the \dwords{pmt} will be determined by the location of the other floor structures like the cryogenic piping. The outline of the \dword{dpmod} is shown in Figure~\ref{fig:dppd_3_1}.

\begin{dunefigure}[The \dword{dpmod} (partially open)]{fig:dppd_3_1}
{The \dword{dpmod} (partially open) with cathode, \dwords{pmt}, \dword{fc} and anode plane with chimneys.}
\includegraphics[width=0.95\textwidth]{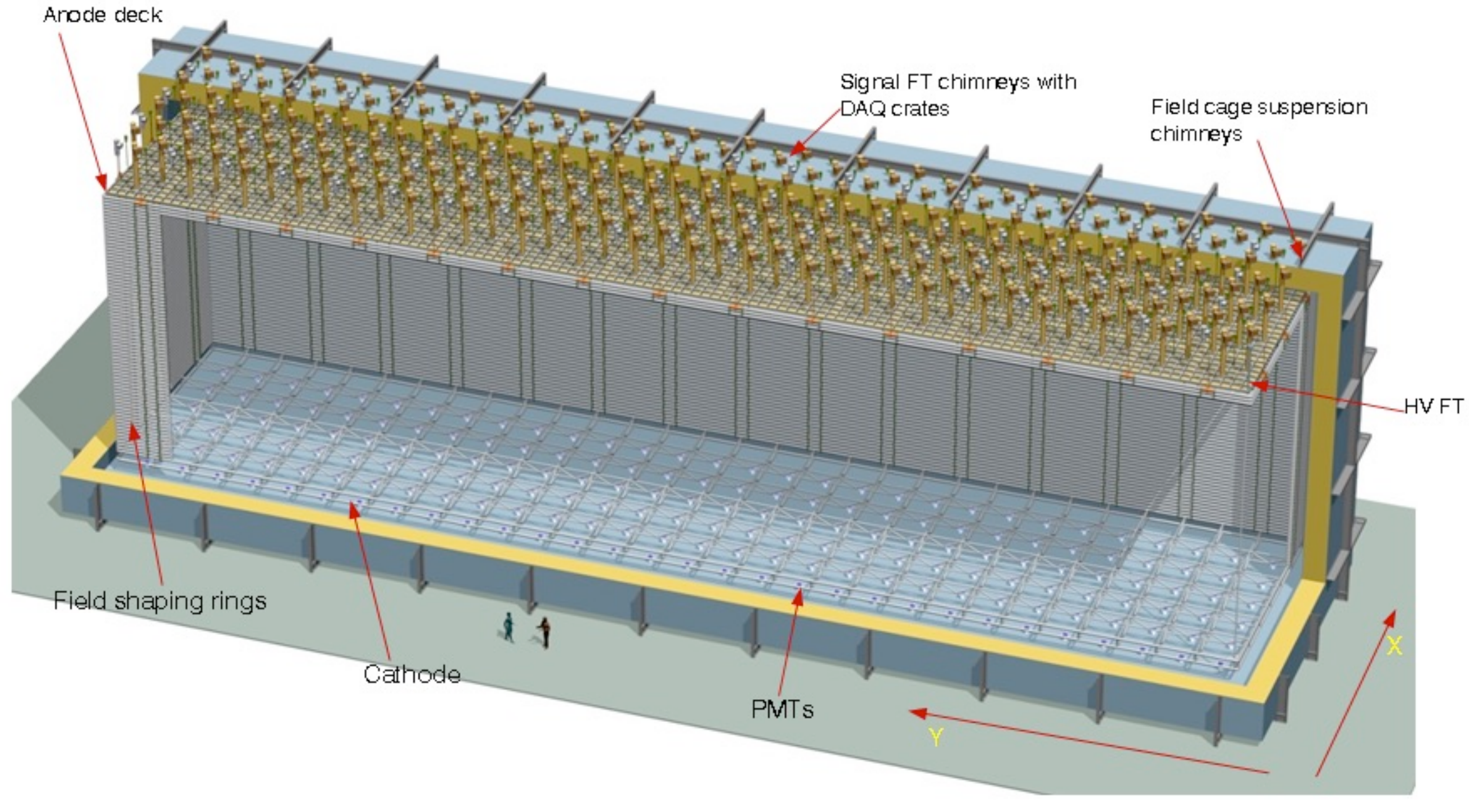}
\end{dunefigure}

Since few light sensors are directly sensitive to \SI{127}{nm}, a wavelength shifter is required. A \dword{tpb} coating directly on the \dword{pmt} is the default plan. Light collectors to increase the photons detected are under study. A single cable per \dword{pmt} carries power and signal, and splitters are placed outside the cryostat. A photon calibration system will be formed by external light sources and internal optical fibers.  

The cable trays from the side walls of the cryostat to the \dwords{pmt} carry the cables and calibration fibers. The cables and fibers are routed from the \fdth flanges at the top of the cryostat and  combined at the side wall trays. These side trays carry the \dword{hv}/signal cables in blocks of \num{24} \dwords{pmt} and four calibration fibers. Each block of \num{24} \dwords{pmt} in a \num{6}$\times$\SI{4}{m^2} area forms a sector of underground installation, with a total of \num{30} sectors.

\subsection{Operation Principles}
\label{sec:fddp-pd-1.5}

The physics program defines the operation principles of the \dword{dpmod}: the measurement of the neutrino oscillation parameters requires recording events based on an external trigger coming from the beam, while non-beam physics such as  \dwords{snb}, proton decay, or other exotic transitions events, require special trigger conditions, including the \dword{pds}. \Dword{pmt} calibration, which has to be performed regularly, presents another operation mode wherein  a hardware trigger provided by the calibration system starts the data recording. \\    

Thus, the operation modes are:
\begin{itemize}
\item External trigger: 
the typical case is when the beam generates a hardware trigger,  
but it also includes software-generated triggers for test data;
\item Non-beam physics trigger: the electronics based on the \dword{pds} signals provides the trigger for  \dword{snb}, proton decay events, etc.;
\item Calibration: during \dword{pds} calibrations, the trigger is provided by the light calibration system.
\end{itemize}

The external and non-beam physics triggers run in parallel to ensure that rare events such as  \dwords{snb} are recorded effectively.

\section{Photosensor System}
\label{sec:fddp-pd-2}

\subsection{Photodetector Selection and Procurement}
\label{sec:fddp-pd-2.1}

The \dword{pd} 
selected as baseline for the light-readout system is the Hamamatsu R5912-MOD20 \dword{pmt} as used in \dword{pddp}. The Hamamatsu R5912-MOD20, see Figure \ref{fig:dppd_2_1}, is an 8-inch, 14-stage, high gain \dword{pmt} (nominal gain of \num{e9}). In addition, this \dword{pmt} was designed to work at cryogenic temperature adding a thin platinum layer between the photocathode and the borosilicate glass envelope to preserve the conductance of the photocathode at low temperature. This particular \dword{pmt} has proven reliability on other cryogenic detectors. The same or similar \dwords{pmt} have been successfully operated in other \lar experiments like MicroBooNE \cite{microboone}, MiniCLEAN \cite{miniclean}, ArDM, ICARUS T600 \cite{icarus}, as well as in \dword{pddp} \cite{protoDUNDP-tdr}. Contacts with other manufacturers such as Electron Tubes Limited (UK) \cite{electrontubeslim} and HZC (China) \cite{hzc} are on-going to engage them in the program.

\begin{dunefigure}[Picture of the Hamamatsu R5912-MOD20 \dword{pmt}.]{fig:dppd_2_1}
{Picture of the Hamamatsu R5912-MOD20 \dword{pmt} \cite{hamamatsu-5912}.}
\includegraphics[width=0.2\textwidth]{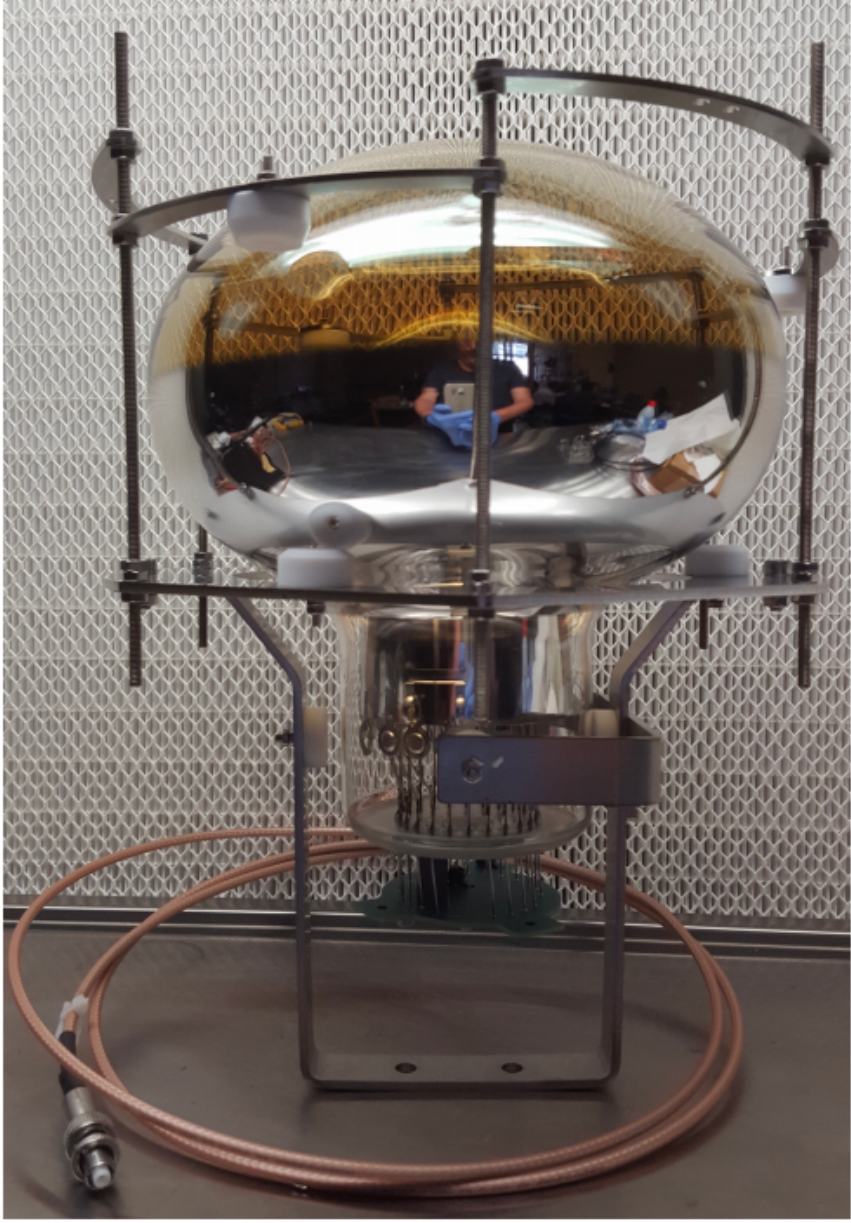}
\end{dunefigure}

As the baseline number of \dwords{pmt}, \dpnumpmtch + \num{80} spares, is high and several operations and tests have to be performed with them before the installation, the \dwords{pmt} have to be ordered with sufficient time in advance to complete the following planned operations: assembly of the voltage divider circuit, mounting on the support structure, testing at room and cryogenic temperatures, application of \dword{tpb} coating, packing and shipment. They are 
re-tested at \surf before installation (see Sections~\ref{sec:fddp-pd-9} and \ref{sec:fddp-pd-10}). Considering the large number of \dwords{pmt} required by \dual \dword{pd}, the purchase order 
must be completed at least two years ahead of installation. A staged or staggered order allowing us to receive a steady supply of \dwords{pmt} would be most convenient. 

\subsection{Photodetector Characterization}
\label{sec:fddp-pd-2.2}

Prior to installation, the most important parameters of the \dword{pmt} response have to be measured with two aims: first, to reject under-performing \dwords{pmt} and second, to store the characterization information in a database for later use during the \dword{dpmod} commissioning and operation.

The basic and most important parameters to characterize are the dark count rate versus voltage and the gain versus voltage. Both parameters must be measured at both room and  cryogenic temperatures. Prepulsing and afterpulsing are not expected to be an issue, but will be measured, too. 

From the mechanical point of view, the test setup requires a light-tight vessel filled with a cryogenic liquid (argon or nitrogen) plus the infrastructure for filling and operating the vessel with temperature and liquid-level controls. For \dword{pddp}, \num{10} \dwords{pmt} were tested at a time 
over the course of a week, as cryogenic tests of \dwords{pmt} 
require several days for the \dword{pmt} thermalization. Figure~\ref{fig:dppd_2_2a} shows the  \dword{pddp} \dwords{pmt} being installed in the testing vessel. 
Increasing the capacity of the vessel, and thus the number of \dwords{pmt} to test simultaneously, 
will reduce the characterization test duration.

\begin{dunefigure}[Picture of the \dwords{pmt} being installed in the testing vessel]{fig:dppd_2_2a}
{Picture of the \dwords{pmt} being installed in the testing vessel used for the \dword{pddp} \dwords{pmt}.}
\includegraphics[width=0.3\textwidth]{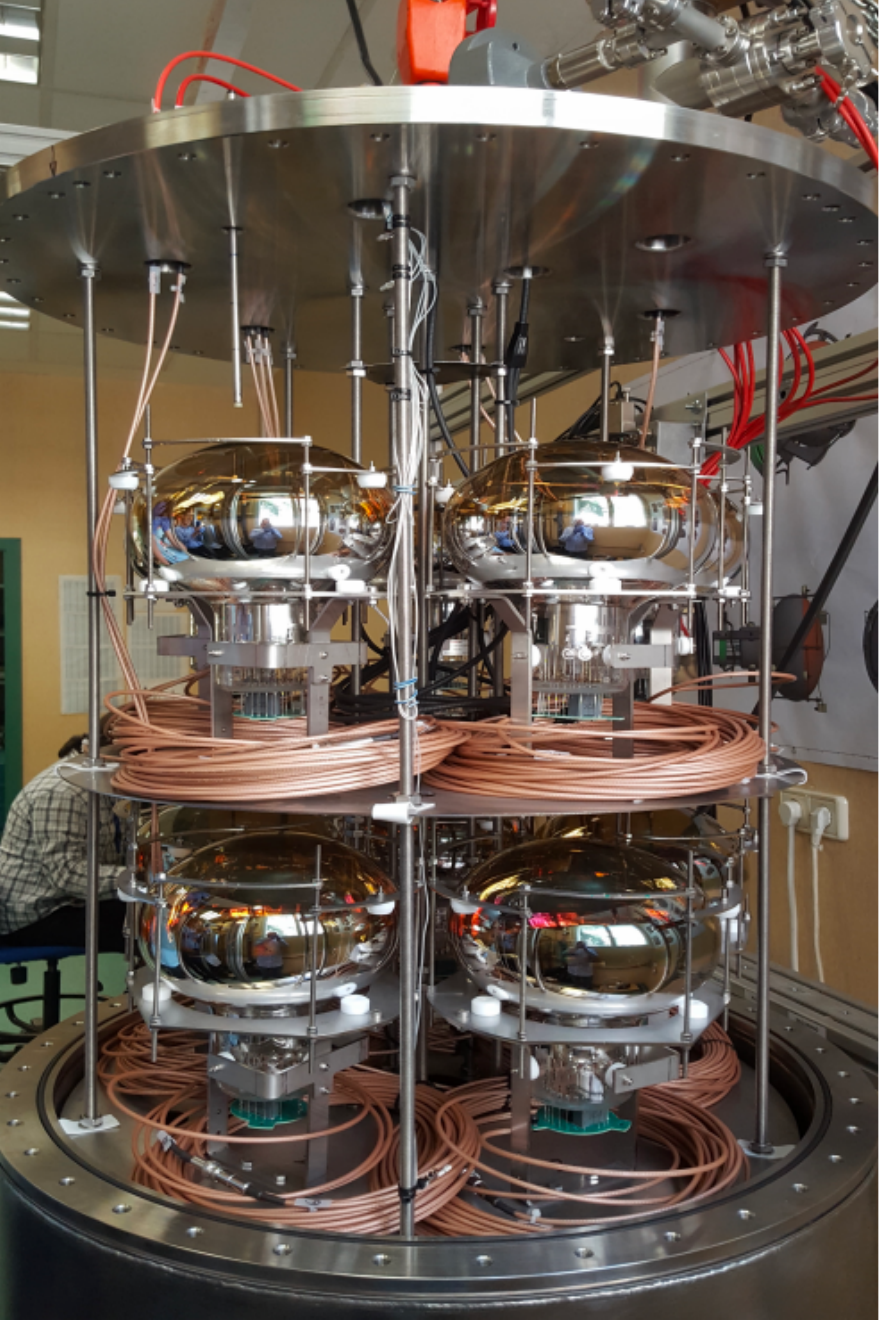}
\end{dunefigure}

Figure~\ref{fig:dppd_2_2b} shows the sketch of the envisaged setup for \dword{pmt} characterization tests. From the electronics point of view, the test setup requires a \dword{hv} power supply, a discriminator, a counter for the dark rate measurements, a pulsed light source, and a charge-to-digital or analog-to-digital converter for the \dword{pmt} gain versus voltage measurements. All those instruments must allow computer control to automate the data acquisition.

\begin{dunefigure}[Sketch of the setup for \dword{pmt} characterization tests.]{fig:dppd_2_2b}
{Sketch of the setup for \dword{pmt} characterization tests.}
\includegraphics[width=0.9\textwidth]{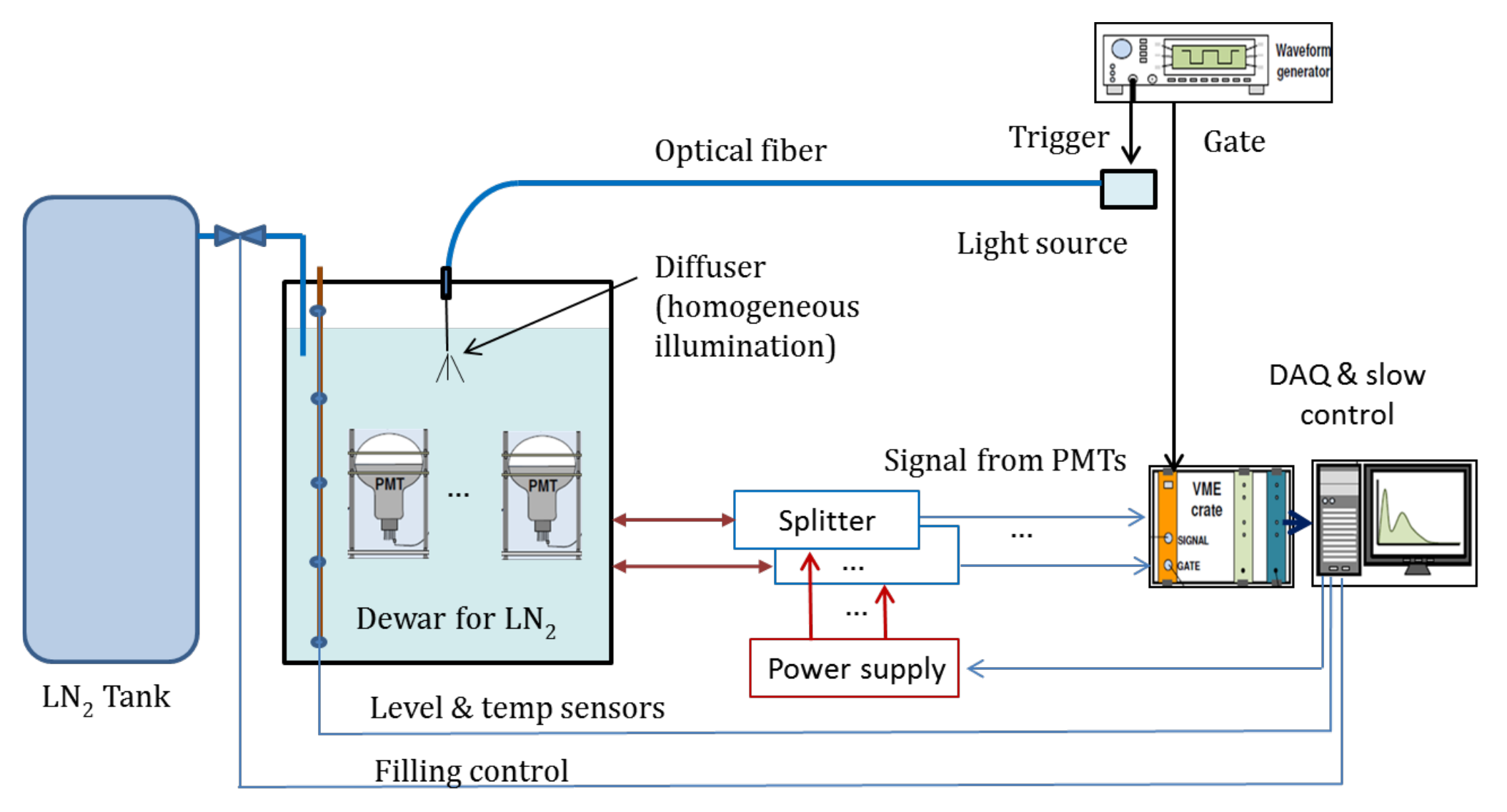}
\end{dunefigure}

\subsection{High Voltage System}
\label{sec:fddp-pd-2.3}

Based on the experience with the \dword{wa105} 
prototype, 
the A7030 power supply modules from CAEN~\footnote{CAEN\texttrademark{}, \url{http://www.caen.it/csite/CaenFlyer.jsp?parent=222}} 
form the baseline design of the \dword{pmt} \dword{hv} system. 
These modules provide up to \SI{3}{kV} with a maximal output current of \SI{1}{mA} and a common floating ground to minimize the noise. Module versions with \num{12}, \num{24}, \num{36}, or \num{48} \dword{hv} channels are available. The \dword{hv} polarity can be chosen for each module. According to the baseline \dword{pmt} powering scheme, modules with positive \dword{hv} polarity will be acquired for the experiment. Modules with \num{48} \dword{hv} channels and Radiall \num{52}\footnote{Radiall\texttrademark{}, \url{https://www.radiall.com/}.}  connectors are 
under consideration. The corresponding \dword{hv} cable connects the modules with the \dword{hv} splitters, described in Section~\ref{sec:fddp-pd-4.2}. This choice allows the design of a compact and most cost-effective system occupying only 
\num{1} to \num{2} racks. 
For \dpnumpmtch \dwords{pmt}, \num{15} A7030 modules (+ \num{2} spares) are needed. These \num{15} \dword{hv} modules will be installed in mainframes from CAEN.

Each \dword{pmt} is powered individually thus allowing the gain of all \dwords{pmt} to be equalized by adjusting the operating voltage. 
This will be sofware-controlled. The software must interface to the \dword{pmt} calibration system (Section~\ref{sec:fddp-pd-5}) in order to extract the calibration factors needed for the gain equalization.

\subsection{Wavelength Shifters}
\label{sec:fddp-pd-2.4}

The \dword{dpmod} \dword{pds} requires waveshifting the \si{127}{nm} scintillation photons into visible photons. 
Coating the \dword{pmt} windows with a thin film of \dword{tpb} 
has already been validated~\cite{tpb} and is adopted as the baseline plan. \dword{tpb} is a wavelength shifter with high efficiency for conversion of \lar scintillation \dword{vuv} photons into visible light, where the \dword{pmt} cathode is sensitive. The \dword{tpb} is deposited on the \dword{pmt} by means of a thermal evaporator which consists of a vacuum chamber with two copper crucibles (Knudsen cells) placed at the bottom of the chamber, following the sanding of the \dword{pmt} window. A \dword{pmt} is fixed at the top of the evaporator, with its window pointing downwards, on a rotating support in order to ensure a uniform coating. The crucibles, filled with the \dword{tpb}, are heated to \SI{220}{C}. At this temperature, the \dword{tpb} evaporates through a split in the crucible lid into the vacuum chamber, eventually reaching the \dword{pmt} window.

Several tests were performed to tune some parameters, e.g., the coating thickness (\dword{tpb} surface density) and the deposition rate. A \dword{pmt} mock up covered with mylar foils was used for the tests. A \dword{tpb} density of \SI{0.2}{mg/cm^2} -- the value where the \dword{pmt} efficiency is stable as a function of the density -- was chosen for \dword{pddp}. Efficiency measurements were performed using a \dword{vuv} monochromator by comparing the cathode current of a coated \dword{pmt} with the current value of a calibrated photodiode. As a result of the efficiency tests, about \SI{0.8}{g} of \dword{tpb} must be placed in the crucibles at each evaporation, in order to achieve the desired \dword{pmt} coating density. 
This value optimizes the quantity of \dword{tpb} used per evaporation while keeping 
the coating density fluctuations below \num{5}$\%$. With these specifications, two to four \dwords{pmt} can be coated per day at a single coating station. 
Multiple coating stations will be required in order to remain on schedule. 

\subsection{Light Collectors}
\label{sec:fddp-pd-2.5}

Although we are still lacking detailed physics simulations of photon collection in the full  \dwords{dpmod}, it can be generally argued that further optimization (i.e., cost 
balanced against physics reach) of light collection is desirable. In addition to  maximizing the overall light yield, another crucial figure of merit is the uniformity of the light collection efficiency within the full \dwords{dpmod} active volume. Geometrical acceptance effects, as well as light absorption processes at the detector boundaries and within the \lar itself, can greatly degrade the uniformity in response. Active detector regions close to the \dword{fc} and further away from the cathode are the most 
affected. As a result, differences of up to an order of magnitude in response throughout the 
active volume are not uncommon in a \lartpc.

In the case of a \lartpc, there are at least four main parameters for optimizing the light yield and the uniformity in response: (1) the number of \dwords{pmt} per unit area, (2) the placement of \dwords{pmt}, (3) the augmentation of \dwords{pmt} with additional light collectors, and (4) the choice of where and how the original \SI{127}{nm} photons can be wavelength-shifted. The most obvious direction for optimizing cost effectiveness are the latter two options. Detector components that are not strictly part of the \dword{pds} may also play a role in this optimization process, one relevant example being the transparency of the cathode plane. The options to use shifter-reflectors (Winston cones) to increase the effective area of individual \dword{pmt} windows, or to move shifting of light closer to the cathode and attaching wave guides coupled to the \dwords{pmt}, are under study.

Another promising and cost-effective option to increase both light yield and response uniformity is the use of \dword{tpb}-coated reflector foils covering the detector inner walls. This option is routinely used in \dual \lartpc{}s searching for dark matter, such as the ArDM~\cite{Boccone:2009zz} experiment. 
This is also under investigation for the \dword{spmod} concept, building on the experience already accumulated with the \lariat experiment, and the one to be gained with SBND. In the \dual case, up to four of the six inner faces of the TPC -- those corresponding to the \dword{fc} structure -- could be covered with dielectric foils. The same \dword{wls} used to coat the \dword{pmt} windows, \dword{tpb}, would be vacuum-evaporated on the foils. The shifted blue light emitted by the foils would then have a greater chance of reaching the \dword{pmt} windows compared to \SI{127}{nm} light, owing to the better reflective properties given by the combination of foils and blue light. To be adopted, 
this concept would first need to demonstrate satisfactory stability on the timescale of the experiment duration.

\section{Mechanics}
\label{sec:fddp-pd-3}


An individual \dword{pmt} mount has been designed and tested in the  \dword{wa105} 
prototype~\cite{Zambelli:2017dkg}. The same design is used for \dword{pddp} and is planned 
for the \dword{dpmod}. A \dword{pmt} with this mechanical structure is shown in Figure~\ref{fig:dppd_2_1}. The support frame structure is mainly composed of \num{304}L stainless steel with some small Teflon (PTFE) pieces assembled by A4 stainless steel screws that minimize the mass while ensuring the \dword{pmt} support to the cryostat membrane. The design 
takes into account the shrinking of the different materials during the cooling process so as to avoid breakage of the \dword{pmt} glass.
Over-pressure tests were carried out for \dword{pddp}, and further tests to ensure the correct performance under pressure will be carried out.


A uniform array of \dpnumpmtch cryogenic Hamamatsu R5912-MOD20 \dwords{pmt}, below the transparent cathode structure, is fixed on the membrane floor in the areas between the membrane corrugations. The arrangement of the \dwords{pmt} 
accommodates the cryogenic piping on the membrane floor, 
and other elements 
installed in this area. 

The mechanics for the attachment of the \dwords{pmt} has been carefully studied for \dword{pddp}. It must counteract the \dword{pmt} buoyancy while avoiding stress to the \dword{pmt} glass due to differentials in the thermal contraction between the support and the \dword{pmt} itself. The 
attachment is done via a stainless steel supporting base that could be point-glued to the membrane. The weight of the support and \dword{pmt} exceeds the buoyancy force of the system. Given the large standing surface of the stainless steel plate support basis, these supports also ensure stability against possible lateral forces acting on the \dwords{pmt} due to the liquid flow. Figure \ref{fig:dppd_3_2} depicts the \dword{pmt} together with its support base attached to the bottom of the cryostat.

\begin{dunefigure}[Cryogenic Hamamatsu R5912-MOD20 \dword{pmt} fixed on the membrane floor.]{fig:dppd_3_2}
{Cryogenic Hamamatsu R5912-MOD20 \dword{pmt} fixed on the membrane floor, with the optical fiber of the calibration system.}
\includegraphics[width=0.42\textwidth]{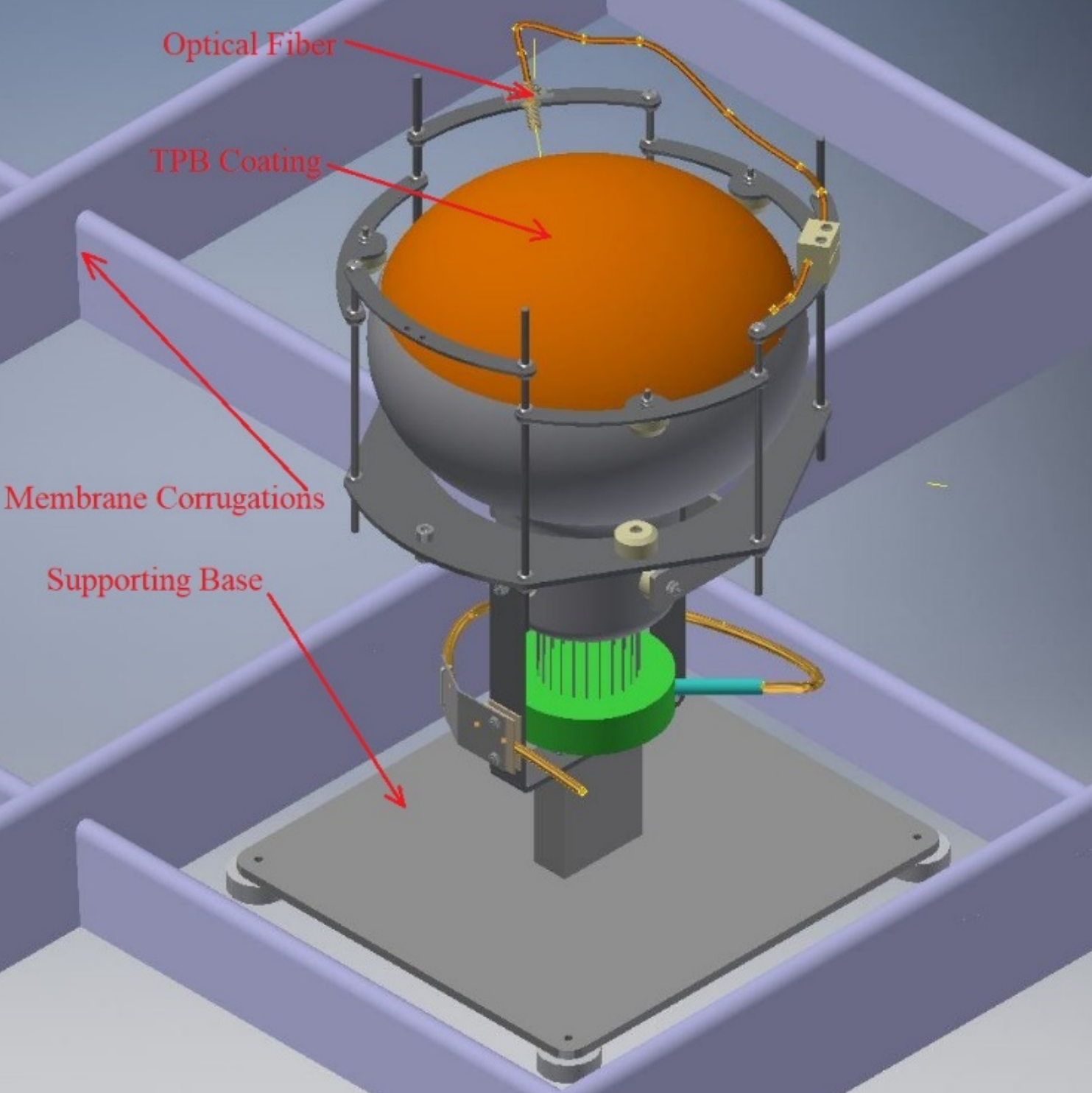}
\end{dunefigure}

\section{Readout Electronics}
\label{sec:fddp-pd-4}

\subsection{Photomultiplier High Voltage Dividers}
\label{sec:fddp-pd-4.1}

The \dword{pddp} \dword{pmt} power supply has a grounded cathode and positive \dword{hv} applied to the anode. A single cable for each \dword{pmt} carries both power and signal. This configuration, which requires half as many cables and \fdth{}s on the detector as would the negative voltage configuration, offers a clear advantage given the large number of \dwords{pmt} in the \dword{detmodule}. In addition, the cathode grounding shows fewer dark counts than the anode grounding scheme. Although a coupling capacitor must be used to separate the \dword{hv} from the \dword{pmt} signal, this signal and power splitting can be done externally, mitigating this drawback.  Figure~\ref{fig:dppd_4_1} shows the positive power supply and cathode grounding scheme.

\begin{dunefigure}[Positive power supply and cathode grounding scheme.]{fig:dppd_4_1}
{Positive power supply and cathode grounding scheme.}
\includegraphics[width=0.5\textwidth]{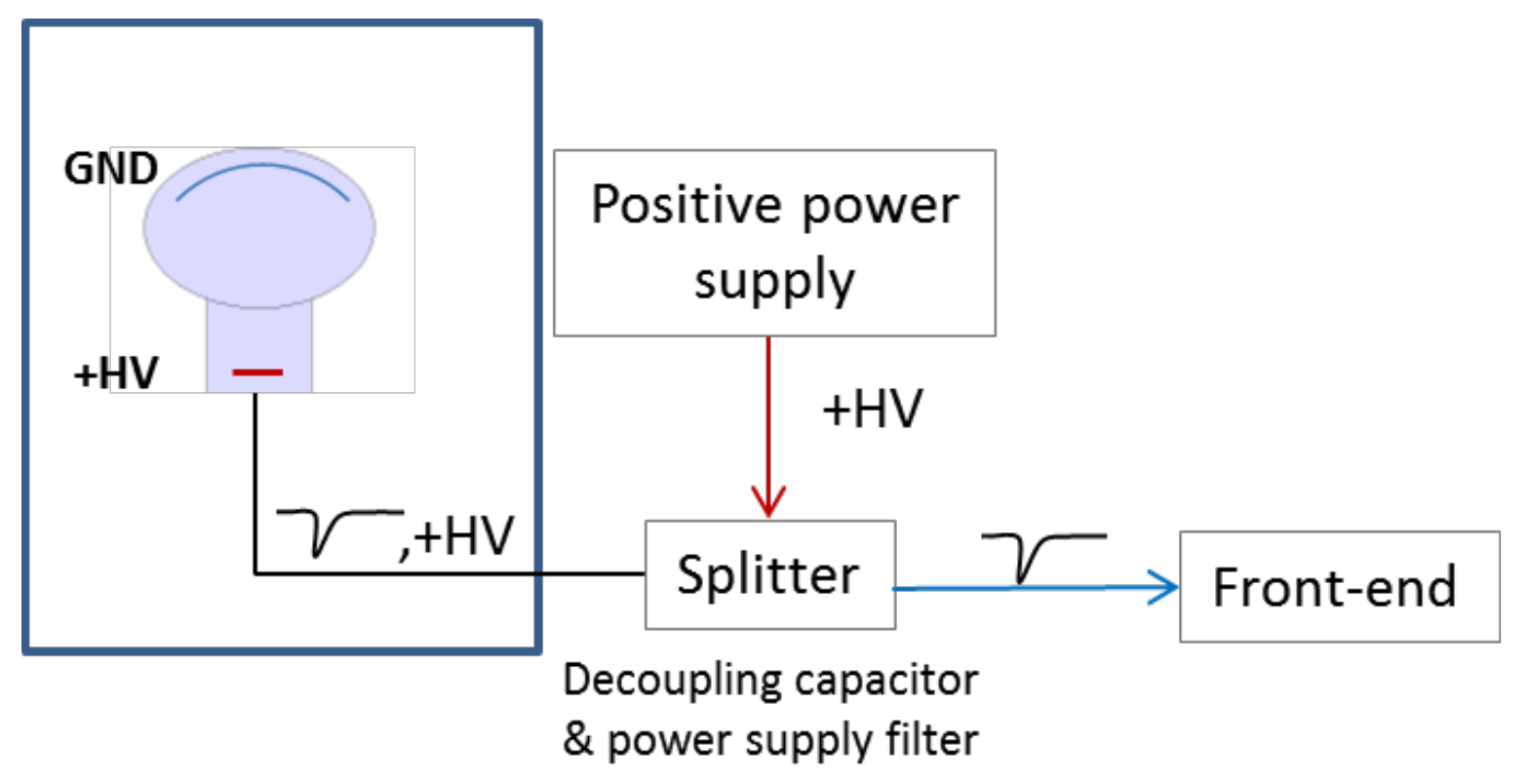}
\end{dunefigure}

The \dword{pmt} base circuit uses only resistors and capacitors. The components are carefully selected and tested to minimize the variations in their characteristics with temperature. The polarization current of the voltage divider (total circuit resistance) is chosen to meet the \dword{pmt} light linearity range and maximum power requirements. The dynodes' voltage ratio will follow the manufacturer recommendations for increased linearity range on the space-charge effect area (tapered divider). In addition, capacitors are added to the last stages to increase the \dword{pmt} linearity in pulsed mode. The precise values for the components have not been decided yet, as they depend on concrete requirements and also on the results from \dword{pddp}. The \dword{pddp} 
design is considered as the baseline solution.

For the connection between the \dword{pmt} base and the \fdth, the RG-303/U cable was selected for its low attenuation and its proven reliability in cryogenic environments. 
This cable is directly soldered to the \dword{pmt} base on one side, and it ends with an SHV connector on the other side for attachment to the flange. 

\subsection{High Voltage and Signal Splitters}
\label{sec:fddp-pd-4.2}

\Dword{hv} and signal splitters 
are used to separate the fast \dword{pmt} response signal from the positive \dword{hv} with capacitive decoupling. 
A low-pass filter between the \dword{hv} supply and the \dword{pmt} 
reduces the noise.

It is possible for radiated electromagnetic interference (EMI) picked up by the cables and conducted noise from the \dword{hv} power supply to be synchronous across many \dword{pmt} channels (i.e., coherent noise). This noise could add up to produce false detector triggers. Since the \dword{pmt} signal can be as low as a few \si{mV}, 
control of the EMI over the circuit is very important. The splitter \dword{hv} input filter is intended to reduce the EMI induced and conducted by the power supply cables. Enclosing each splitter channel in its own metallic grounded box reduces the EMI directly received in the splitter circuit and 
the cross-talk between different splitter channels.

Figure \ref{fig:dppd_4_2} shows a generic splitter circuit where R1 and C1 form the \dword{hv} input low-pass filter (with a cut-off frequency below \SI{60}{Hz}). The resistor R7 and the  \dword{led} are for safety purposes only, warning when \dword{hv} is applied to the splitter. The C4 capacitor splits the signal coming from the \dword{pmt} from the \dword{hv}, and R2 prevents the \dword{pmt} signal from going to ground through the C1 capacitor. R4 and R5 are zero \si{\ohm} optional resistors that allow some flexibility in the grounding configuration. Finally, R3 ensures the discharging of C4 if the splitter is not connected to the \SI{50}{\ohm} input at the \dword{daq} system. The RC constant of the capacitor C4 and the load (\SI{50}{\ohm}) must be as large as possible to minimize baseline oscillations due to the charge-discharge of the capacitor. Values of C4 between \SI{150}{nF} and \SI{300}{nF} have already been tested in the  \dword{wa105}. 

\begin{dunefigure}[Generic splitter circuit diagram.]{fig:dppd_4_2}
{Generic splitter circuit diagram.}
\includegraphics[width=0.75\textwidth]{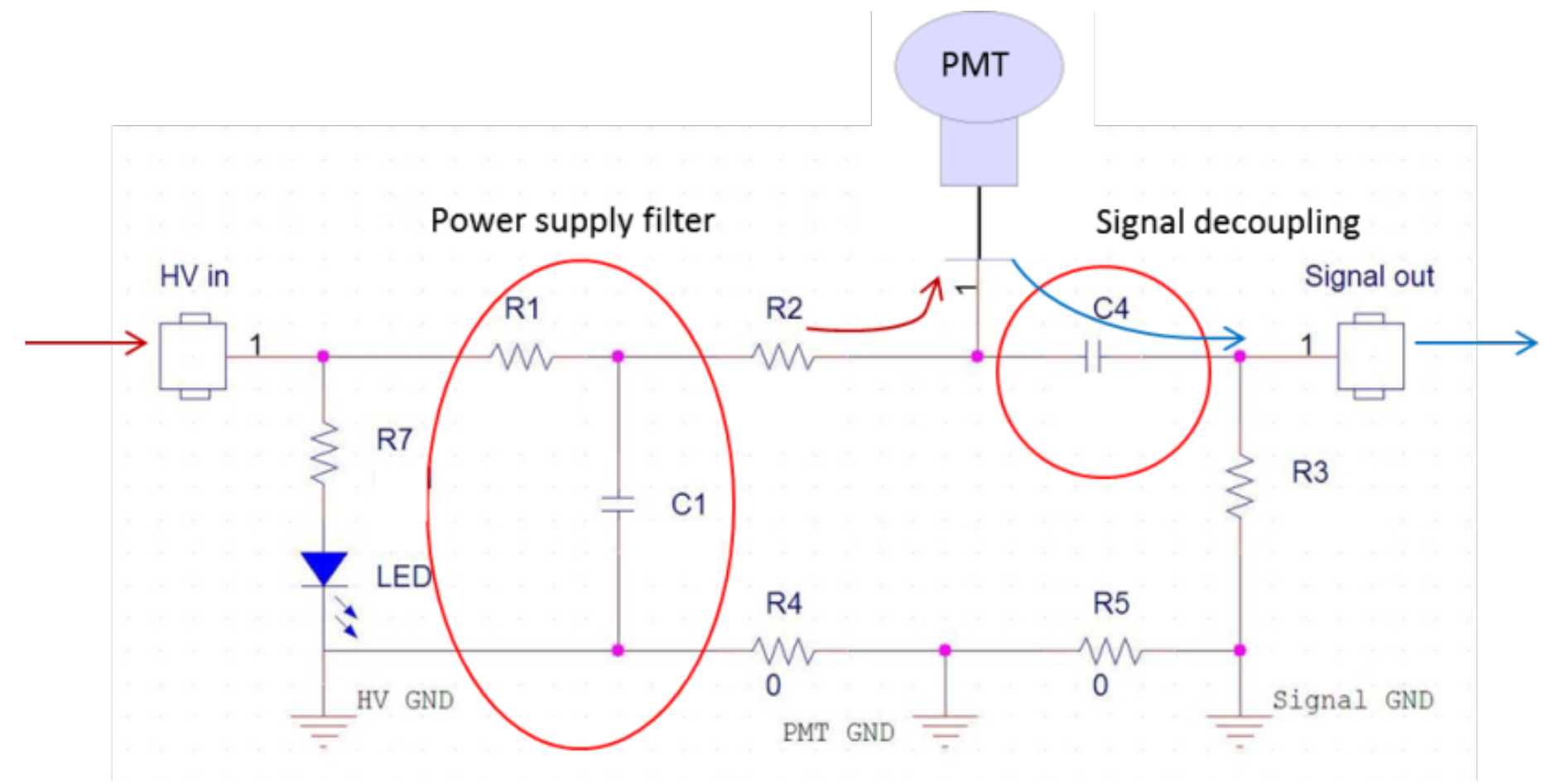}
\end{dunefigure}

For the connections between the \dword{hv} power supply and the splitters, and between the splitters and the cryostat \fdth{}s, the HTC 50-3-2 cables have been chosen as baseline. The HTC 50-3-2 has a similar attenuation 
as the RG-303/U (used inside the cryostat), but with a factor of \numrange{8}{10} lower cost. Both cables are attached on one side directly to the \dword{hv} splitter and have an SHV connector on the other end. An RG-58 cable 
terminated on the connector required by the \dword{fe} card provides the connection between the splitter and the \dword{fe}.

\subsection{Signal Readout Requirements}
\label{sec:fddp-pd-4.3}

In order to meet the physics requirements, the information that needs to be extracted from the \dword{pmt} signals is the following:

\begin{itemize}
\item S1 fast component shape, charge and timing;
\item S1 slow component shape;
\item S2 shape, charge and timing (distance from S1 and duration);
\item Single \phel (SPE) charge spectrum for gain calculation during \dword{pmt} calibration;
\item Trigger signal generation by the coincidence of several \dword{pmt} signals.
\end{itemize}

There is currently no estimate on the dynamic range of the light expected to reach the \dwords{pmt} in the \dword{dpmod}. Our calculations are based on the signals detected by the \dwords{pmt} in the \dword{wa105}, 
which has quite different dimensions from the \dword{dpmod}. However it is the only available reference 
until the \dword{pddp} and simulations are operational.

In general, the \dword{pmt} signal dynamic range goes from the \si{mV} level to several volts (over \SI{50}{\ohm} load). During the operation of the \dword{wa105}, \dword{pmt} signals larger than \SI{2}{V} were observed with \dword{pmt} gains around \num{e6}. 
Figure~\ref{fig:dppd_4_3_ab} shows the SPE waveforms (left, normalized) and amplitudes (right) for the \dword{wa105} at different voltages. The light levels in the \dword{dpmod} will have a larger dynamic range due to its large volume, therefore higher gains are required to see the far light signals. However, higher gains increase the output from closer light signals, requiring that the \dword{fe} electronics cover a large range of input voltages. To cover a dynamic range of \SI{10}{V} with a resolution below the \si{mV} level, \num{14} bits are necessary (least significant bit (LSV) $\sim$\SI{0.6}{mV}). For \SI{2}{V} of dynamic range \num{12} bits would be sufficient (LSB $\sim$\SI{0.5}{mV}). Results from \dword{pddp} and relevant simulations are needed to determine the required dynamic range.

\begin{dunefigure}[SPE waveforms and amplitudes from the WA105 at different voltages.]{fig:dppd_4_3_ab}
{SPE waveforms and amplitudes from the WA105 at different voltages.}
\includegraphics[width=0.47\textwidth]{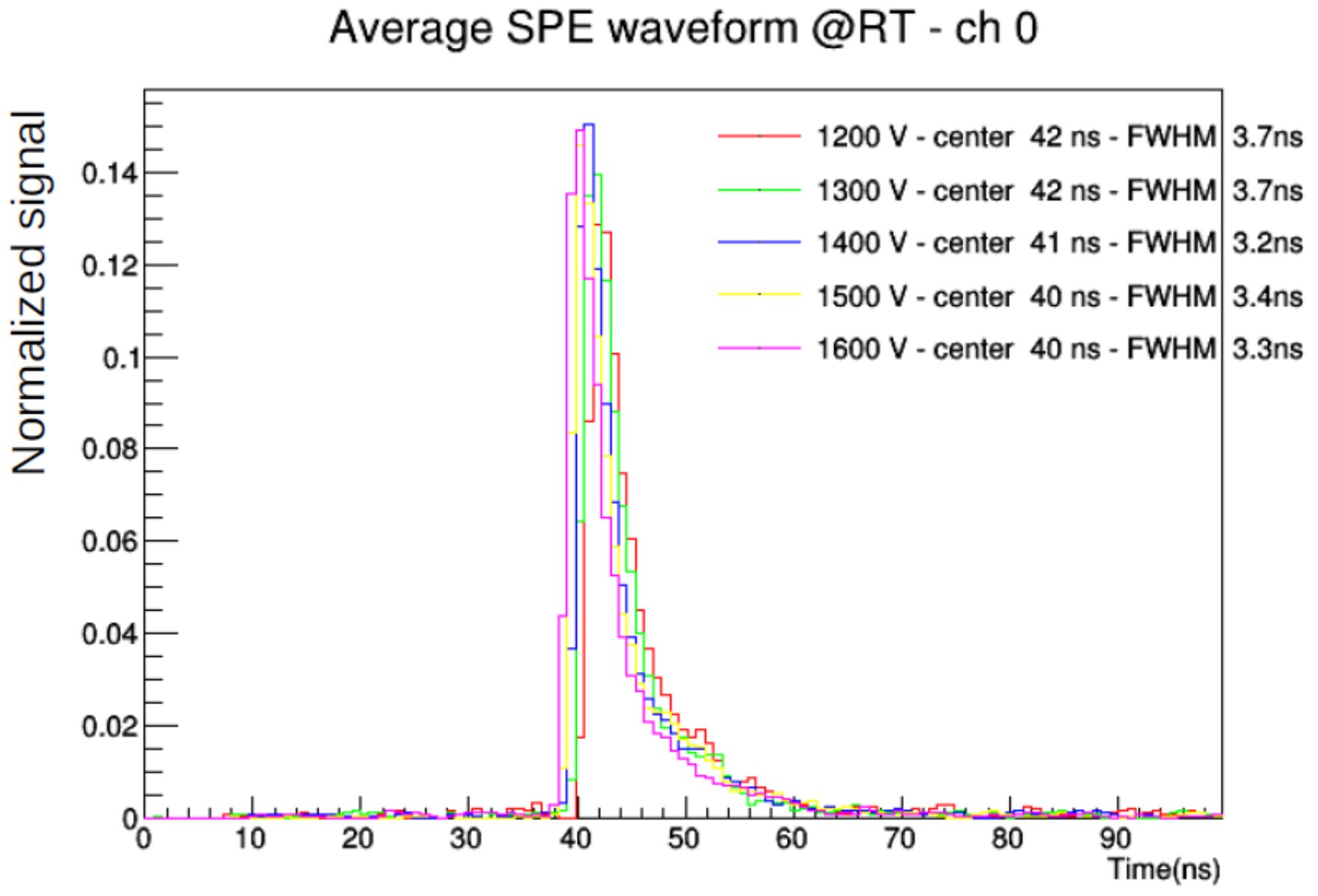}
\includegraphics[width=0.47\textwidth]{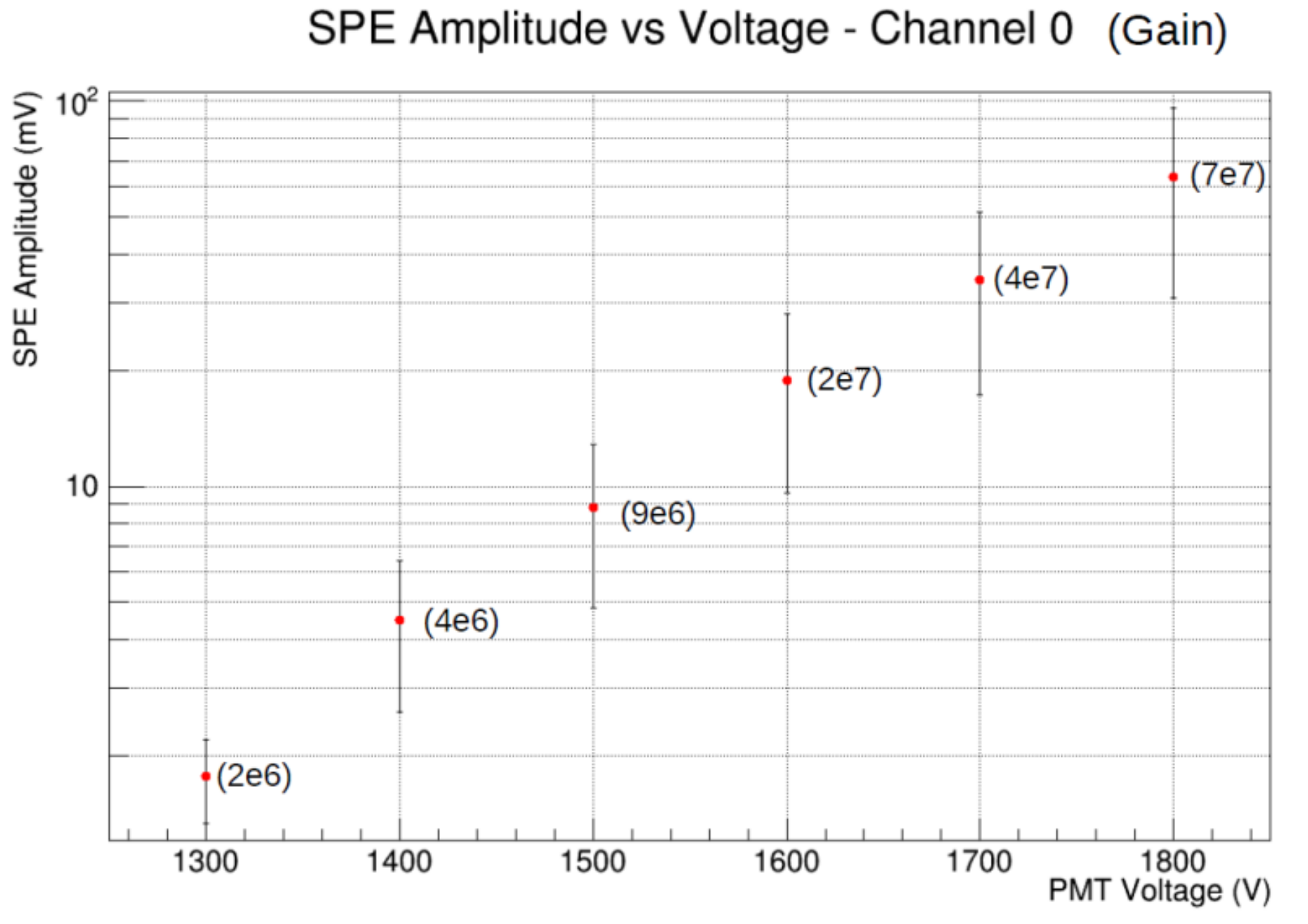}
\end{dunefigure}

To calculate the \dword{pmt} gains, the SPE charge measurement will be performed. Depending on the \dword{pmt} gain, the SPE amplitude varies from the \si{mV} level to hundreds of \si{mV}, as shown in Figure~\ref{fig:dppd_4_3_ab}. Due to the very long cables from the \dwords{pmt} to the \dword{fe} electronics, the noise into the cables could be high. If one considers a noise level around \SI{1}{mV},  the \dword{pmt} gain must be set to \num{e6} or higher in order to distinguish the SPE from noise. The average SPE pulse width is around \SI{3.5}{ns} full width at half maximum (FWHM). In order to digitize this signal to reconstruct it with fidelity, a sampling period on the order of \SI{}{ns} is required.

The sampling frequency also affects the time tagging precision. The time uncertainty due to the \dword{pmt} alone is around \SI{3}{ns} (transit time spread). Other factors, e.g., Rayleigh scattering, increase this uncertainty, as does the sampling period; therefore, the lower sampling frequency, the better. In the  \dword{wa105} \SI{4}{ns} sampling was used to digitize waveforms. 

The rate of the events observed in the \dword{wa105} was around \SI{300}{kHz} with the threshold at the SPE level. The rate at the \dword{dpmod}, is not yet known, but expected to be much larger despite the underground location. The time-tagging system needs to process events at high rates to ensure that no events are lost. Figure~\ref{fig:dppd_4_3_c} shows the event rates for different trigger thresholds observed in the \dword{wa105}.

\begin{dunefigure}[Event rates for different trigger thresholds observed in the \dword{wa105}.]{fig:dppd_4_3_c}
{Event rates for different trigger thresholds observed in the \dword{wa105} .}
\includegraphics[width=0.6\textwidth]{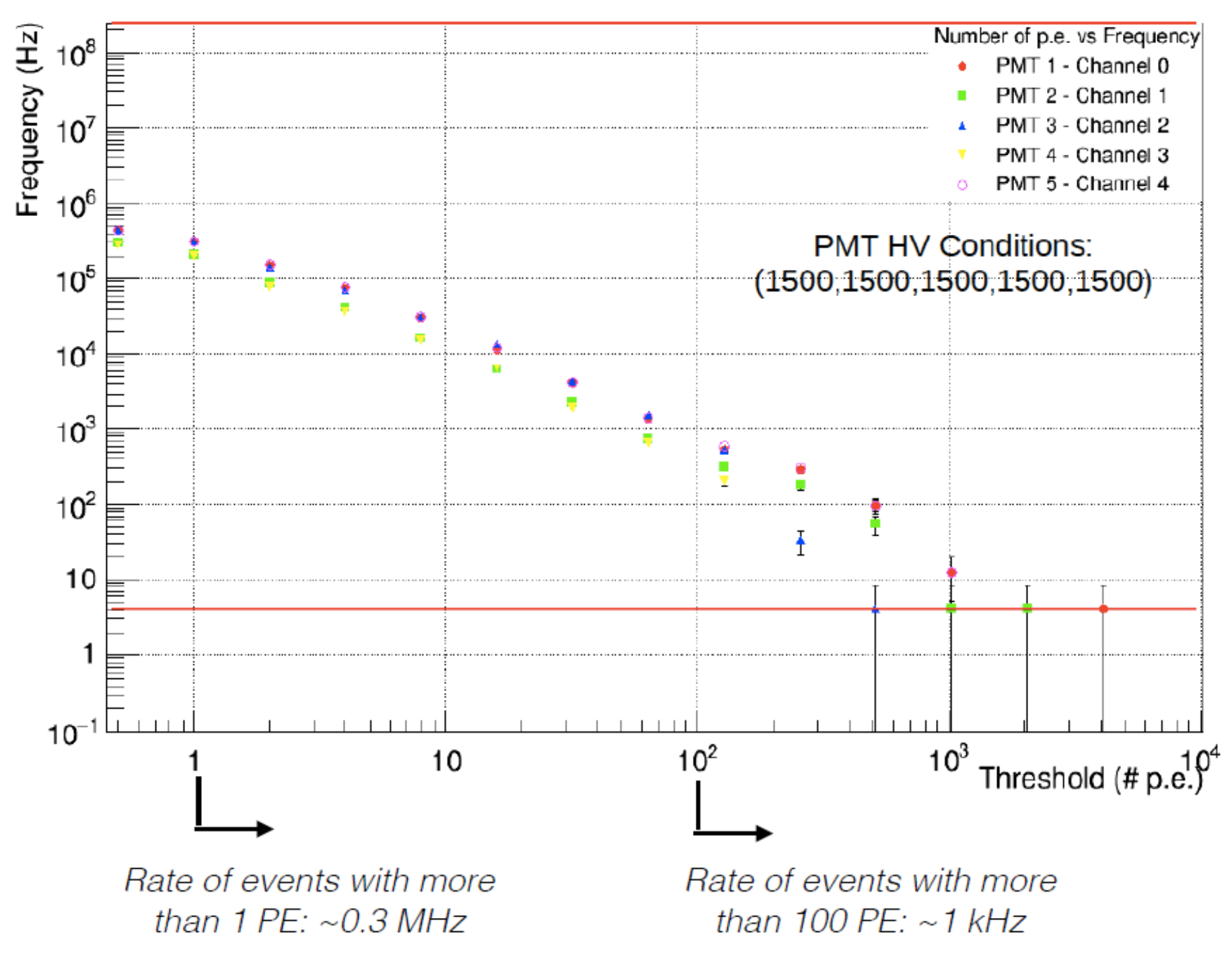}
\end{dunefigure}

The light signal has to be synchronized with the \dword{daq}. All the \dword{daq} electronics use the \dword{wr} protocol for synching. 
A dedicated \dword{wr} \dword{utca}~\cite{utca} slave node is on the light readout \dword{fe} electronics as sync receiver, distributing clocks to the different \dword{fe} cards.

\section{Photon Calibration System}
\label{sec:fddp-pd-5}

\subsection{System Design and Procurement}
\label{sec:fddp-pd-5.1}

A photon calibration system is required to be integrated into the \dword{dpmod} to calibrate the \dwords{pmt} 
installed in the \lar volume. The goal is to determine the \dword{pmt} gain and maintain the \dword{pmt} performance stability. A design similar to the one used in \dword{pddp} will be used although some R\&D measurements are planned to make it more effective, reduce the cost and mitigate issues related to the scaling.

In \dword{pddp}, an optical fiber is installed at each \dword{pmt} in order to provide a configurable amount of light (see Figure~\ref{fig:dppd_3_2}). The calibration light is provided by a blue  \dword{led} of \SI{460}{nm} using a Kapuschinski circuit as  \dword{led} driver; this is much less expensive than using a laser.
One \dword{led} is connected to one fiber that goes to one female optical \fdth from Allectra~\footnote{Allectra\texttrademark{}, \url{http://www.allectra.com/index.php/en/}.} 
In total,  six  \dwords{led} are placed in a hexagonal geometry. The direct light goes to the fiber, and the stray light to a \dword{sipm} used as a single reference sensor 
in the center. Fibers of length \SI{22.5}{m} (from Thorlabs $\phi$ \SI{800}{\micro\meter}, FT800UMT,~\footnote{Thorlabs\texttrademark{}, \url{https://www.thorlabs.com/thorproduct.cfm?partnumber=FT800UMT}.} 
 and stainless-steel jacket) are used inside the cryostat. Each of these fibers is attached to a \numrange{1}{7}-fiber bundle (from Thorlabs $\phi$ \SI{200}{\micro\meter}, FT200UMT~~\footnote{Thorlabs\texttrademark{}, \url{https://www.thorlabs.com/thorproduct.cfm?partnumber=FT200UMT}.} 
 stainless-steel jacket common end, and black jacket at split ends), so that one fiber is finally installed at each \dword{pmt}. A diagram of the \dword{pddp} photon calibration system is shown in Figure~\ref{fig:dppd_5_1}. 

\begin{dunefigure}[Diagram of the photon calibration system to be implemented in \dword{pddp}.]{fig:dppd_5_1}
{Diagram of the photon calibration system to be implemented in \dword{pddp}}
\includegraphics[width=0.4\textwidth]{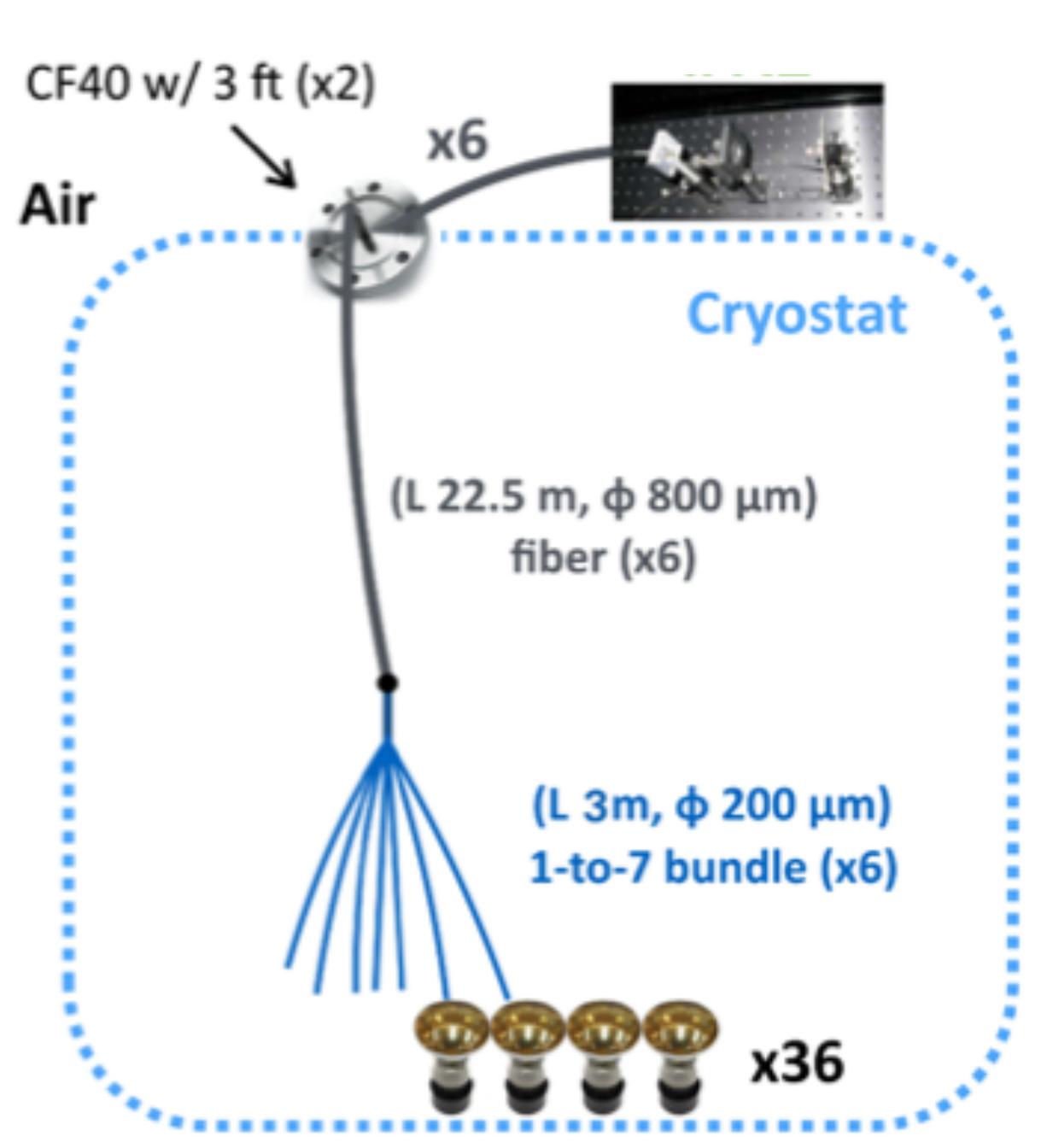}
\end{dunefigure}

Assuming the \dword{pddp} design for the DUNE \dword{dpmod} with \num{720} \dwords{pmt}, \num{120} bundles, \num{120} fibers, \num{120} light sources, \num{120} flange \fdth{}s, and \num{20} reference sensors are needed. The length of the fibers and bundles has to be calculated considering the exact position of the \fdth flanges. The number of flanges required to host \num{120} SMA \fdth{}s will depend on their size. However, alternatives to this design will be pursued with R\&D measurements in order to reduce the amount of fibers, study other options for the reference sensor, and increase the input light if necessary. In order to reduce the number of fibers, light diffusers can be used, so that one fiber can illuminate at least \num{4} \dwords{pmt}. For instance, a diffuser could be placed at the ground grid. 

\subsection{Validation Tests}
\label{sec:fddp-pd-5.2}

In order to validate the design, the most important result will come from the \dword{pddp} performance. In any case, since the fibers to be used in DUNE FD will be longer, dedicated calculations and measurements to confirm that sufficient light reaches the \dwords{pmt} will be performed. Also, alternative designs, will be validated in different laboratories. The possibility of using a diffuser can be tested in a vessel. The light source will also be validated by studying the different options in the lab. All these measurements will be performed at room temperature and in liquid nitrogen to test the behavior at cryogenic temperatures.

Once the design is fixed, basic characterization measurements will be performed on the fibers upon receiving them from the manufacturer. Those measurements will consist of providing light with a known source and measuring the output with a power meter. Measurements at cryogenic temperatures may not be needed at this point.

Finally, during the photon calibration system installation, each fiber and source will be re-tested to check that the expected light is arriving to each \dword{pmt} using a photodiode. A dedicated procedure will be designed with this purpose, similar to the one used in \dword{pddp}.

\section{Photon Detector Performance}
\label{sec:fddp-pd-6}

To define the \dword{pds} performance, a good understanding of the light generation is needed. For this, optical simulations and a good knowledge of the light properties are required. The DUNE experiment expects to record not only accelerator neutrino interactions, but also rare non-beam events such as \dwords{snb} or nucleon decays. In those cases, an internal trigger is required: an optimized light collection system is hence mandatory. This section describes the tools developed in the consortium for the light simulation in large detector volumes for these purposes.

The main feature of a \lartpc detector is to collect electrons produced by the energy loss of charged tracks when crossing the volume. This signal provides a high resolution \threed image of the event. The reconstructed topology and the amount of charge collected gives the characterization of the tracks (identification and energy). Together with the charge, scintillation light is also produced in \lar. There are many advantages to collect and exploit the scintillation signal. As only a fraction of the initial energy deposition is converted into electrons, the rest being emitted as photons, light collection can improve the calorimetry of the detector. The light signal can provide the $t_0$ of the event, which is a necessary observable for a proper reconstruction. The study of the slow component can give insights into the purity of the \lar. 

When energy deposition occurs, either the knocked argon atom gets excited or an electron is ejected. In the latter case, the probability of the ejected electron being recaptured by an argon ion depends on the drift field and on the amount of energy deposited. This case also produces an excited argon state. In either case, in order to decay to the ground state, the excited argon atom combines with another argon atom to form an excited excimer. A photon at \SI{127}{nm} is then emitted, returning  the excimer to the ground state. As the excimer can be formed in a singlet or triplet state, two time constants will be observed: the singlet at \SI{6}{ns}
and the triplet at \SI{1.3}{$\mu$s}. These principles are sketched in Figure~\ref{fig:dppd_6_0}.

\begin{dunefigure}[A sketch depicting the mechanism of light production in argon.]{fig:dppd_6_0}
{A sketch depicting the mechanism of light production in argon.}
\includegraphics[width=0.8\textwidth]{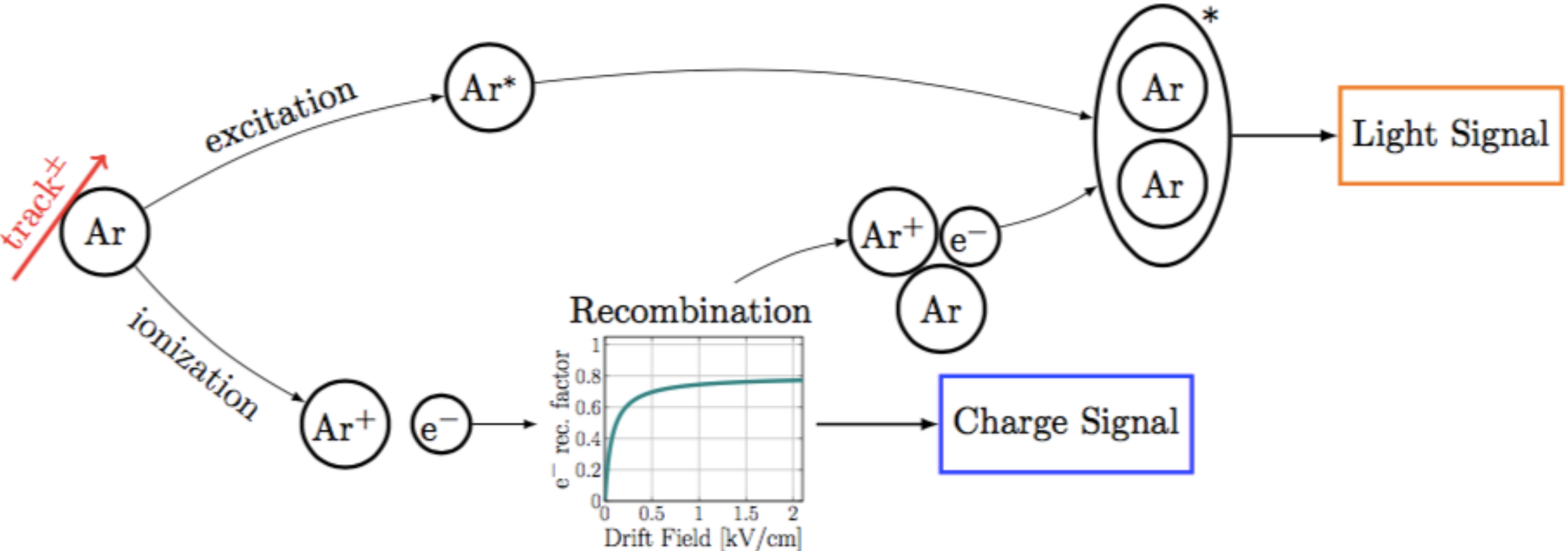}
\end{dunefigure}

In the \dual technology two light signals produced, one (S1) when a charged particle crosses the \lar volume and the second (S2) once the particle is above the liquid surface in the argon gas. 
As electrons drifting in the gas enter high field regions (such as the extraction field or the amplification field in the \dwords{lem}), their velocities increase and Townsend avalanches occur. This current of electrons produces electroluminescence light with the same wavelength and similar time structure as for the S1 signal. 
The S2 light is expected to be an irreducible background for the light studies in \dword{pddp}, since the detector is on the surface. Indeed, the S2 signal can last as long as the total drift time of the electrons: \SI{0.625}{ms} per meter of drift at a drift field of \SI{500}{V/cm}.

Table~\ref{tab:dppd_t_6_0} summarizes the default optical parameters chosen for the light simulation methods described in Section~\ref{sec:fddp-pd-6.1}. The \lar optical properties are the subject of significant discussions in the community, in particular regarding the \lar absorption length and the Rayleigh scattering length. The former affects the collected light yield  whereas the latter mostly affects its uniformity and timing resolution. The absorption and reflection of the \dword{vuv} photons on stainless steel (i.e., the drift cage, cathode, extraction grid and ground grid) and on copper (the \dword{lem} surfaces) are poorly known, 
largely because those reflection coefficients depend strongly on the polishing procedure. 

The measurement of the quantum efficiency of the \dwords{pmt} at \dword{vuv} wavelengths requires a specific setup operating in vacuum since \dword{vuv} photons are absorbed in air. The \dword{pmt} quantum efficiencies in the \dword{wa105} \dwords{pmt} were measured before and after the \dword{tpb} coating using a  \dword{led} that could emit light in the \numrange{200}{800}\,{nm} range. 

Finally, the electroluminescence gain $G$, defined as the number of S2 photons produced per extracted drifting electron, is also subject to discussion. Experimental measurements of $G$ have been performed in a setup quite similar to the amplification design of the \dual technology, although the measurements were made above the \dword{lem}~\cite{Monteiro:2012zz}.  In our case, the S2 photons are the ones leaving the \dword{lem} from below, where the number can be significantly lower.  


\begin{dunetable}
[Default optical parameters chosen for the light simulation methods]
{lcc p{0.8\textwidth}}
{tab:dppd_t_6_0}
{Default optical parameters chosen for the light simulation methods.  Below the thick line are presented some quantities used in our studies although they are not linked to the optical properties of the \lar.}
 & \dwords{vuv} photons & Shifted photons \\ 
 & $\lambda$ = \SI{127}{nm} & $\lambda$ = \SI{435}{nm}\\ \toprowrule
 Absorption length & \multicolumn{2}{c}{$\infty$} \\ \colhline
 Rayleigh scattering length & \SI{55}{cm} & \SI{350}{cm}\\ \colhline
 Absorption coefficients & \num{100}\% & \num{50}\% \\ \colhline
 \lar refractive index & \num{1.38} & \num{1.25}\\ \colhline
 \dword{pmt} quantum efficiency & \multicolumn{2}{c}{0.2 }\\ \colhline
 Electroluminescence gain & \num{300}\\ 
\end{dunetable}

To understand the performance of the \dword{pds}, it is important to take into account the following indicators:
\begin{itemize}
\item Overall detected light yield, in \phel{}s per \si{MeV} of deposited energy in \lar{};
\item Uniformity of the light yield across the entire \lartpc active volume;
\item Event time resolution extracted from the detected photon signal. 
\end{itemize}

In turn, these indicators directly affect the strategy and performance of the \dword{dpmod} trigger system (Section~\ref{sec:fddp-pd-7}), and determine whether the \dword{pd} technical design is sufficient to meet the DUNE physics goals. These higher-level studies will be available on the \dword{tdr} timescale. 

Our current understanding of these performance indicators is largely based on \dword{pddp} simulations and the current status of the simulation work is discussed in detail in Section~\ref{sec:fddp-pd-6.1} Work is focused on \dword{pddp} in a first phase, and will expand to the \dword{dpmod}. For a realistic \dword{pddp} geometry, an average light yield of \SI{2.5}{\phel/MeV} is expected across the entire active volume. This promising yield assumes \num{36} \SI{8}{in} \dwords{pmt} located below the \dword{pddp} cathode plane, averaging to one \dword{pmt} per \si{m$^2$}. On the other hand, spatial non-uniformities in the \dword{pd} response are found to be important and need to be modeled in detail. Variations as large as one order of magnitude both parallel to the drift direction (due to geometrical effects and absorption of light by \lar) as well as perpendicular to it (due to light absorption on detector boundaries) are obtained. 
The event time resolution due solely to light production and light propagation times, i.e., neglecting electronics and \dword{daq} effects for now, is expected to be of order $\mathcal{O}$(\SI{100}{ns}) and hence largely sufficient for our purposes. These initial low-level performance estimates will be refined with more realistic simulations and with \dword{pddp} data (Section~\ref{sec:fddp-pd-6.2}) in the future. They will also be extended to the full \dword{dpmod} geometry on the \dword{tdr} timescale.
\subsection{Simulations}
\label{sec:fddp-pd-6.1}

At zero drift field, when the electron recombination is maximum, roughly \SI{40000}{$\gamma$/MeV} are produced. At the nominal drift field of \SI{500}{V/cm}, then \num{24000}{$\gamma$/MeV} are generated. For reference, the energy deposited by a \dword{mip} track is \SI{2.12}{MeV/cm}. Given the size of the \dword{pddp} ($6\times6\times6$\,m$^3$) and the fact that it is located on surface, roughly \num{100} muons are expected to cross the fiducial volume during the \SI{4}{ms} time window of the \dword{daq}. With a full \dword{geant4}~\cite{geant4} simulation, it takes more than three hours to propagate all the photons emitted by a single \dword{mip} track crossing the \dword{pddp} detector. A full optical simulation is hence computationally prohibitive. Three simulation approaches are being explored to provide the light simulation needed for the design optimization of the \dword{dpmod}. 

\subsubsection{Generation of light maps}
\label{subsec:fddp-pd-6.1.1}

In this method, the photons are propagated in a full dedicated \dword{geant4} simulation only once. The main light characteristics needed for light studies (photon detection probability, called \textit{visibility} hereafter, and time profile) are stored in a map in a ROOT \cite{root} file format which can then be read by any other simulation program. This work was done first using LightSim, a dedicated software developed at LAPP, France. These maps have been adapted to be readable by \larsoft, where light maps are known as \textit{photon libraries}. Work to generate them directly in \larsoft is in progress, in particular for S2 light, which has not yet been simulated for \dual technology.

In the dedicated \dword{geant4}  code, special care has been taken to precisely describe all subdetector components that might affect the light propagation: \dword{lem} plates, extraction grid, \dword{fc} rings, the cathode and its supporting structure, and the ground grid above the \dwords{pmt}. The \lar fiducial volume is then divided into voxels of \SI{25}{cm^3} and \num{e8} photons are isotropically generated at the center of each voxel. The number of photons reaching each \dword{pmt}, and their arrival times are stored. The light map can then be built from these results. For each voxel and for each \dword{pmt}, the visibility is computed as: $w=N\gamma^{\textrm{collected}}/N\gamma^{\textrm{generated}}$. In order to be able to reproduce the time profile, each distribution is fit to a Landau function. From the fits, three parameters are extracted: the minimum time for photons to arrive to the \dword{pmt}, $t_0$; the peak of the distribution, $t_{\textrm{peak}}$ from the Landau most probable value (MPV); and the distribution spread, the $\sigma$ of the Landau function. These three parameters are stored in the light maps for each \dword{pd}. The same procedure is done for the gaseous phase, although the voxel size is smaller in height (only \SI{5}{mm}). In Figure~\ref{fig:dppd_6_1_1_ab}, two fitted time distributions are presented. As is visible, the shapes of the time distributions depend strongly on the source-to-\dword{pmt} distance. For close sources, the distributions are very sharp and the Landau description may not be the optimal function to use. On the other hand, for longer distances, the distribution is broader and the Landau fit reproduces the simulations quite well. 
For practical purposes, the Landau parametrization was used for all cases.

\begin{dunefigure}[Landau fits of the travel time distributions for a sources close to and far from the \dword{pmt}.]{fig:dppd_6_1_1_ab}
{Landau fits (red line) of the travel time distributions (black histogram) for a source close to (left) and far from (right) the \dword{pmt}.}
\includegraphics[width=0.45\textwidth]{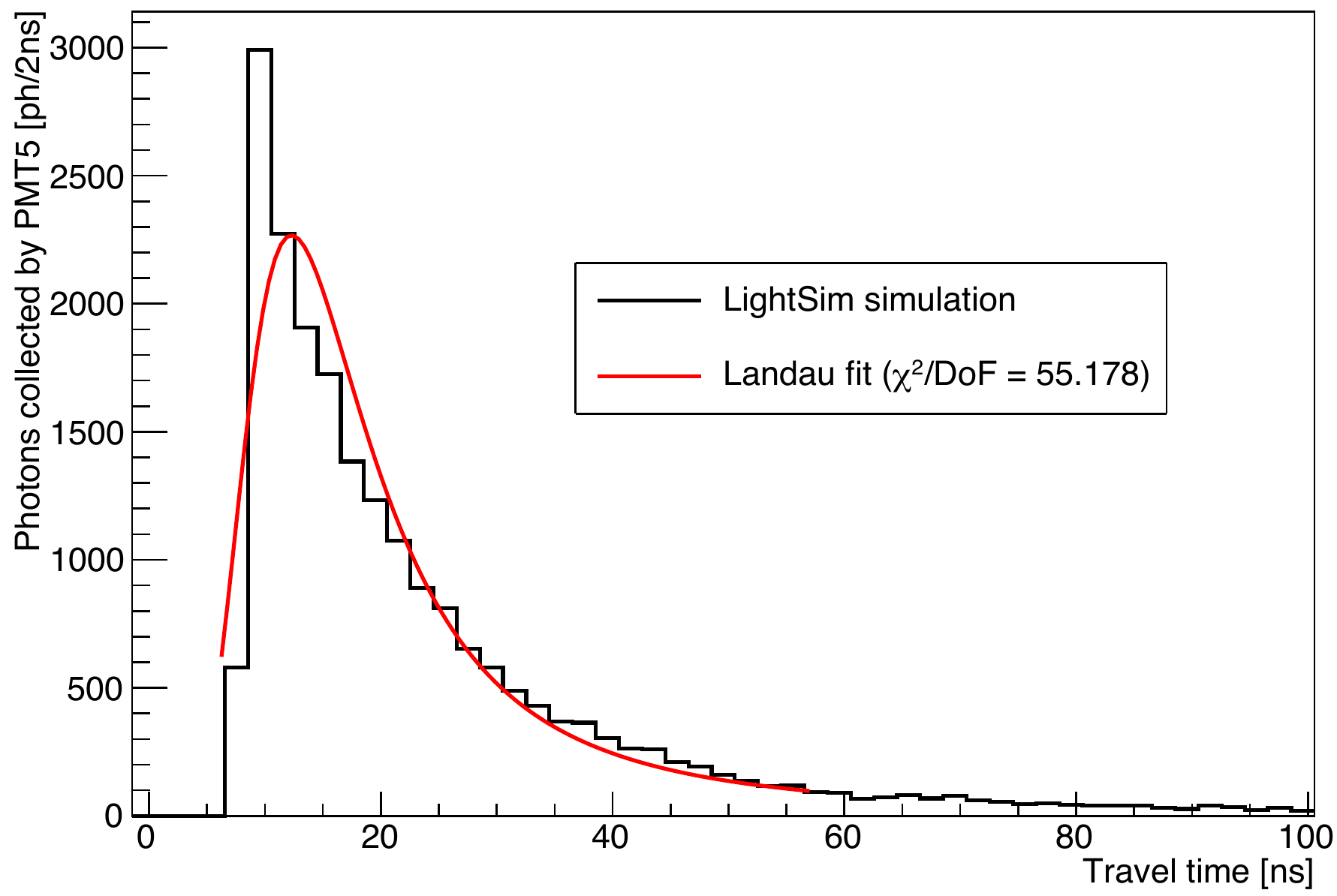}
\includegraphics[width=0.45\textwidth]{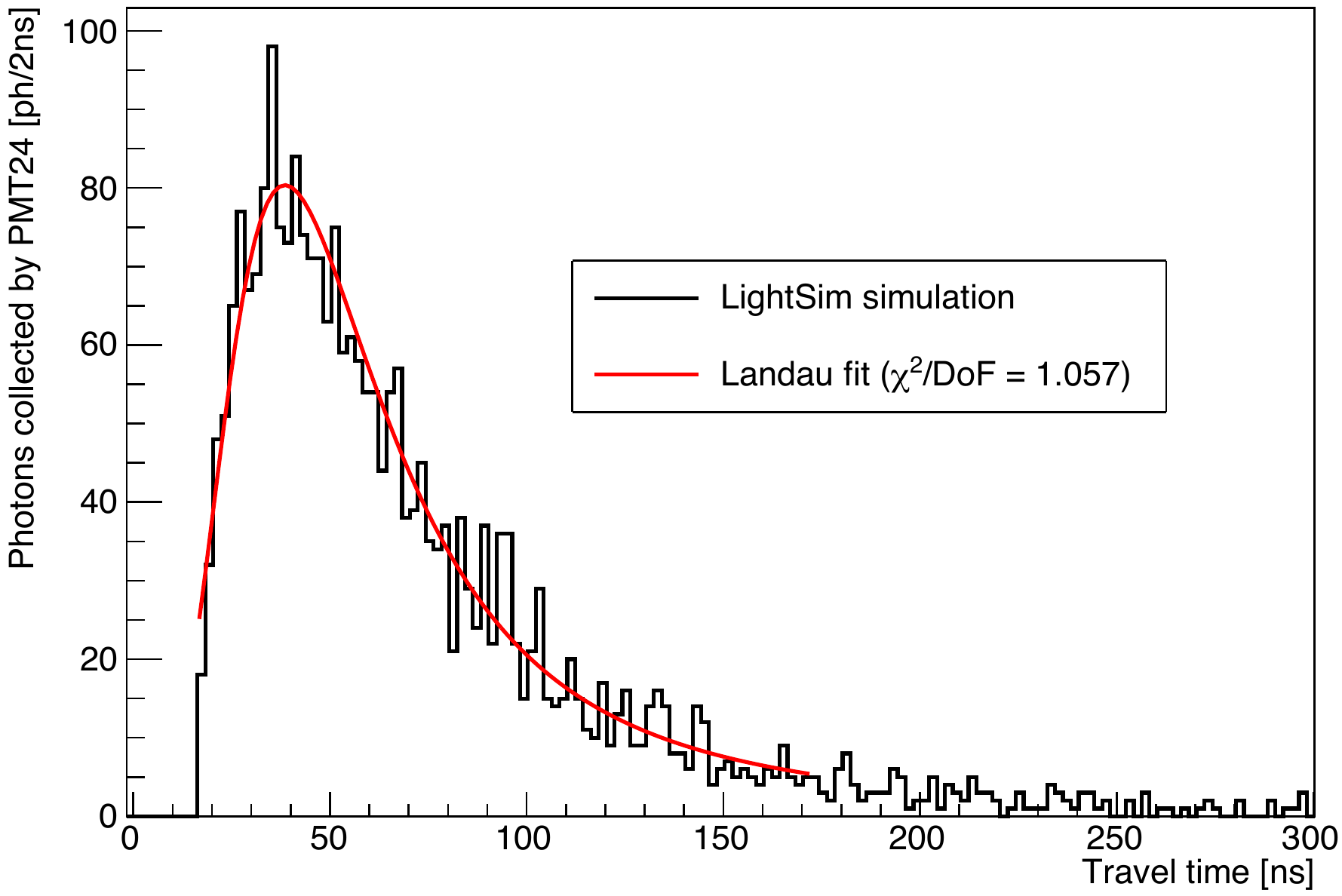}
\end{dunefigure}

As the map has been computed with discrete entries, an interpolation of the four light parameters ($w$, $t_0$, $t_{\textrm{peak}}$, $\sigma$) between the actual source position and the closest voxel centers is performed. An example of the distribution of the visibility 
and its \threed interpolation is presented in Figure~\ref{fig:dppd_6_1_1_cd}. The loss of photons due to the cathode and ground grid are visible. For the \dword{pddp} cathode and supporting structure design, and using the default optical parameters presented in Table~\ref{tab:dppd_t_6_0}, it has been shown that up to $\sim$\num{70}\% of the photons generated in the active volume are absorbed by those structures before reaching the \dword{pmt} array.

Table~\ref{tab:dppd_t_6_0} presents the light propagation parameters used during the generation of the light maps. 
After generation, it is possible to study the loss of photons due to absorption using an approximation of the probability that the medium absorbs the photon 
as: $p_{\textrm{abs}} = \exp(-\frac{D_{\textrm{travel}}}{\lambda_{\textrm{abs}}})$. For the study of other light propagation parameters (Rayleigh scattering and absorption on the stainless-steel and copper) new maps have to be generated.

It takes roughly three days of computing to generate the light maps for \dword{pddp}, even though only 
an eighth of the voxels need to be simulated, since the detector and the \dword{pmt} positioning are symmetric. Generating maps for larger volumes such as the \dword{dpmod}, where the maximum source-to-\dword{pmt} distance is around \SI{60}{m}, could be too time-consuming. Moreover, the light simulation for the \dword{dpmod} is foreseen to drive the optimization of the positioning of the \dwords{pmt} and will guide the studies of possible implementation of light reflectors. As most of the light propagation parameters in \lar are still subject to large uncertainties, these studies will have to 
consider various absorption and diffusion values. Therefore, it is crucial to find 
a faster way to 
simulate the light reliably, even at the cost of losing some precision.

\begin{dunefigure}[Evolution of the visibility seen by a central \dword{pmt}]{fig:dppd_6_1_1_cd}
{Evolution of the visibility seen by a central \dword{pmt} (see arrow) in \dword{pddp} as a function of different source positions in $x$ and $z$ ($y$ is set at \SI{0}{mm}). The position of the cathode and the ground grid are highlighted. Results are limited by the number of photons generated (\num{e7} photons per voxel), and voxels with less than \num{50} photons arriving to the \dword{pmt} are not taken into account. Left: discrete values from the maps, right: after \threed interpolation.}
\includegraphics[width=0.45\textwidth]{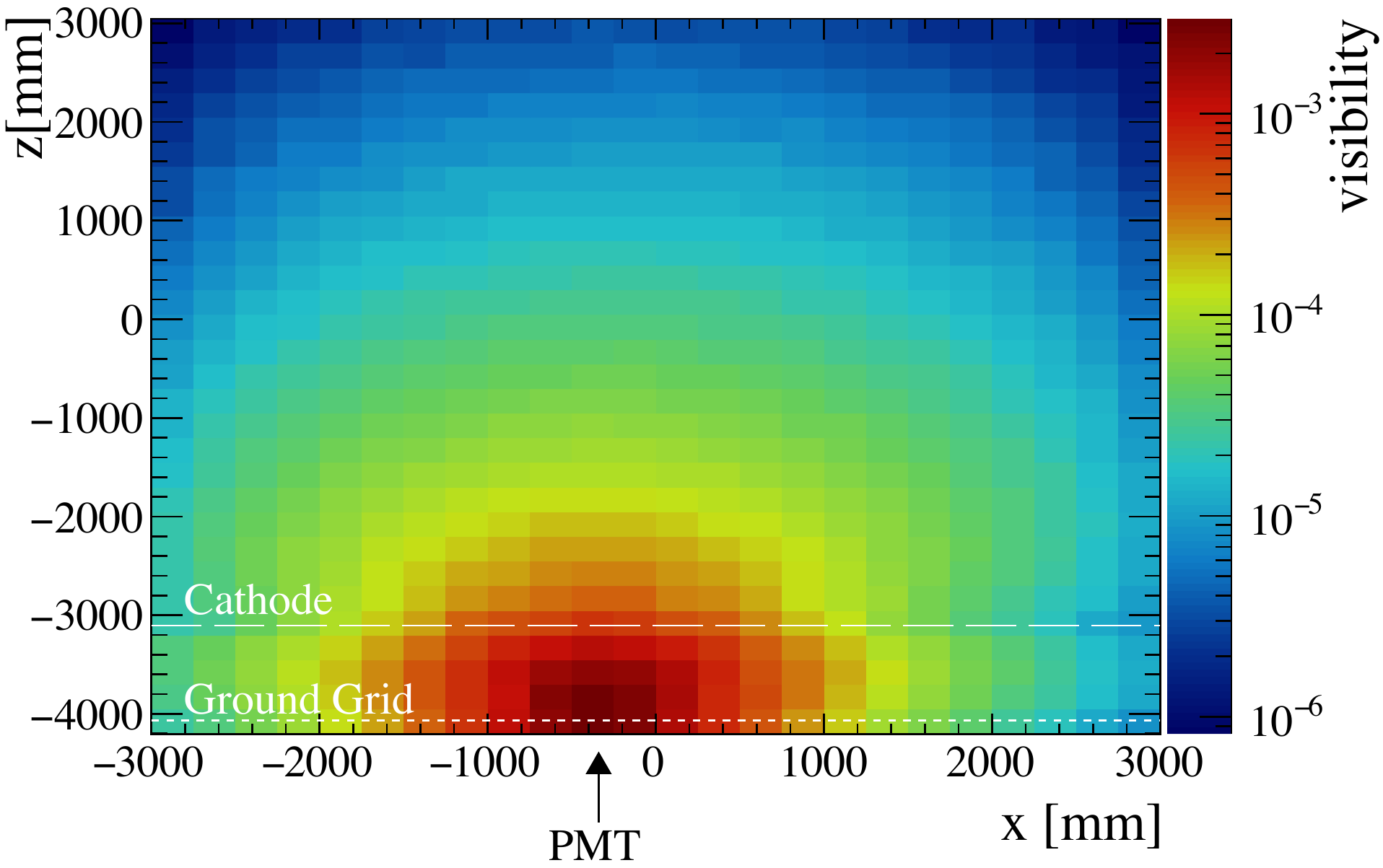}
\includegraphics[width=0.45\textwidth]{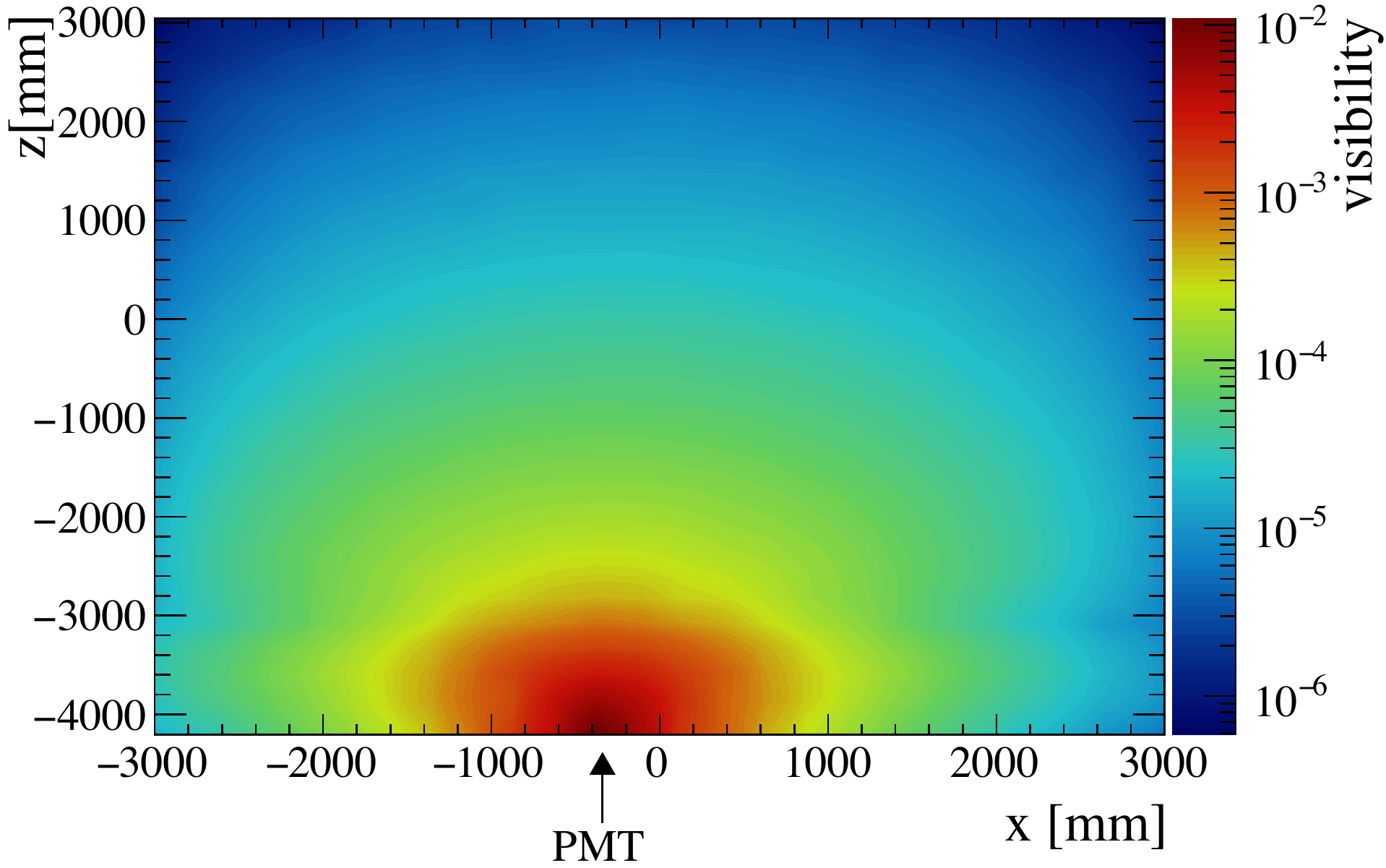}
\end{dunefigure}

\subsubsection{Parametrization from the Light Maps}
\label{subsec:fddp-pd-6.1.2}

Without considering the border effects where the photons are mostly absorbed,
the visibility and the time profile depend only on the source-to-\dword{pmt} distance.

This approach has been followed for the SBND~\cite{sbnd} light simulation and is under consideration for the \dword{dpmod} as well. In Figure~\ref{fig:dppd_6_1_2} the evolution of the visibility and the peak time as a function of the source-to-\dword{pmt} distance are shown. On the left, the borders are taken into account, and the visibility structure is quite complicated due to the complexity of the light simulation in a closed space. However, on the right, the same evolutions are presented only for voxels at least \SI{1}{m} from the active volume boundaries, and a clear correlation between distance and visibility is observed. As for the time distribution (here for the peak time, but the same goes for $t_0$ and $\sigma$ parameters), one can notice two different regimes when looking at all voxels, with a transition at a propagation distance of around  \SI{2}{m}. When considering only the central voxels, the evolution of the peak time is fully correlated with the propagation distance. The parametrization of the light propagation parameters as a function of the propagation distance is a promising option for very large volumes such as the \dword{dpmod}, at least for light sources far from the fiducial volume boundaries. 

This preliminary study is quite encouraging for the light simulation in the \dword{dpmod}, at least for light sources far from the fiducial volume boundaries. Since it is complicated to disentangle the effects due to the propagation and absorption parameters from the light maps, a careful dedicated study should be performed to get parametrization of the visibility and time distribution parameters as a function of the photon traveling distance. 

\begin{dunefigure}[Evolution of the visibility and peak time as a function of source-\dword{pmt} distance (preliminary)]{fig:dppd_6_1_2}
{Evolution of the visibility (top) and peak time (bottom) as a function of the source-\dword{pmt} distance as simulated in the \dword{pddp} geometry (preliminary study). On the left, all voxels are included, on the right only the voxels at least 1\,m away from the fiducial border are included.}
\includegraphics[width=0.75\textwidth]{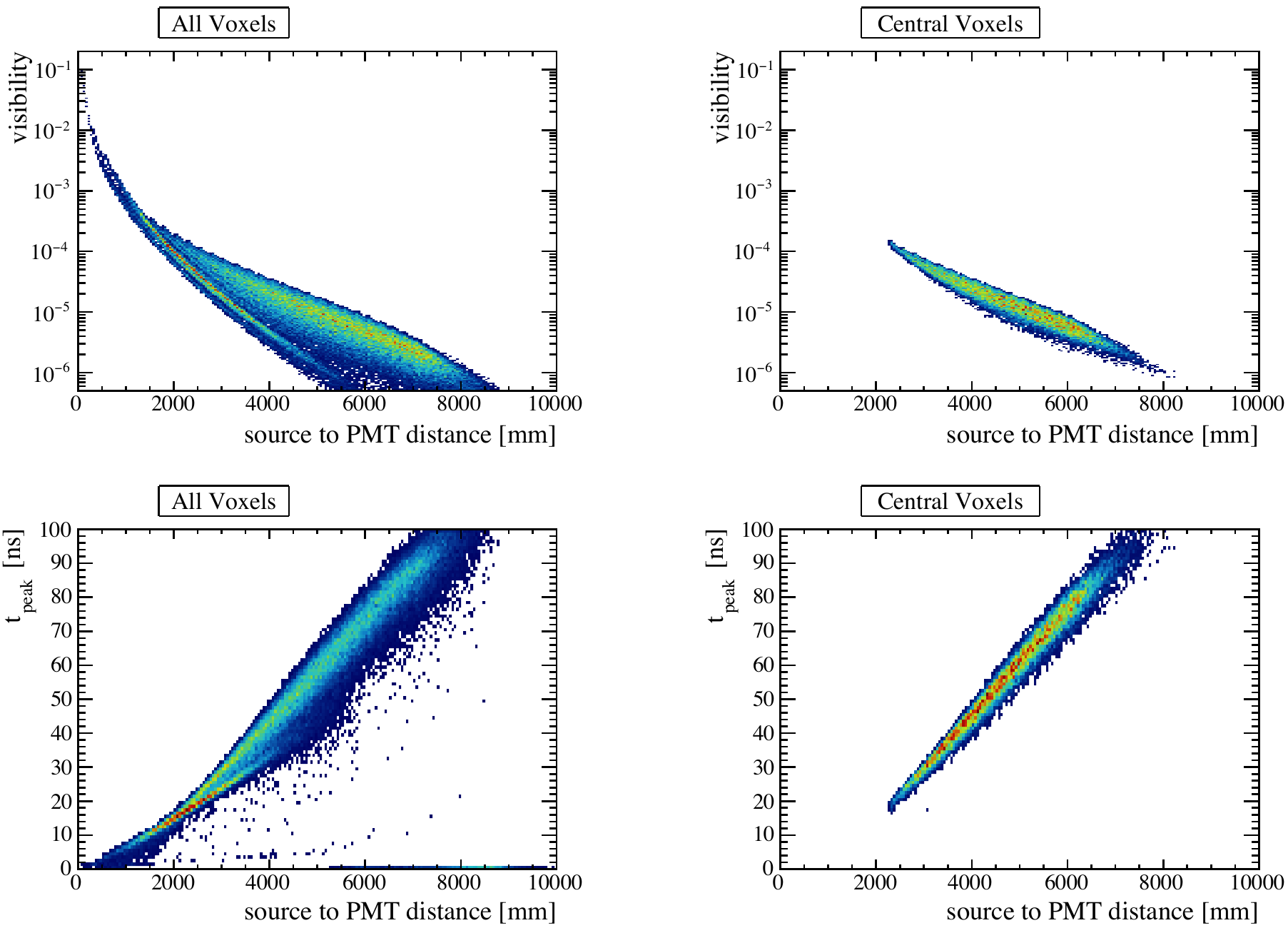}
\end{dunefigure}

\subsubsection{Analytical approach}
\label{subsec:fddp-pd-6.1.3}

The propagation of light in a uniform material such as \lar can be described by the Fokker-Planck diffusion equation:

$$\frac{\partial}{\partial t}p(x,y,z,t) = D\left[\frac{\partial^2}{\partial x^2}p(x,y,z,t) + \frac{\partial^2}{\partial y^2}p(x,y,z,t) + \frac{\partial^2}{\partial z^2}p(x,y,z,t)\right]$$ 

where $D$ is the diffusion coefficient. In an unbound medium, the Fokker-Planck equation is solved by the Green function:

\begin{eqnarray*}
G(\textbf{r}, t; \textbf{r}_0, t_0) &=& \frac{1}{[4\pi D c (t-t_0)^{3/2}]}\exp\left(-\frac{|\textbf{r}-\textbf{r}_0|^2}{4Dc(t-t0)}\right) \\
D &=& \frac{1}{3(\mu_A + (1-g)\mu_S)}
\end{eqnarray*}

where $\mu_A$ and $\mu_S$ are the absorption and scattering coefficients, respectively (both in units of \si{m$^{-1}$}), $g$ is the average scattering cosine ($g$ = \num{0.025}). In \lar with the default optical properties in Table~\ref{tab:dppd_t_6_0}, $D$ = \SI{18.8}{\cm}. In a bound medium, with full absorption of the photons by the \dword{fc} and \dwords{lem}, a few additional techniques have to be used to obtain a solution. With this method, it takes only a few \si{ms} to generate the photon density at a given \dword{pd} from a specific point source. From preliminary studies, relatively good agreement between analytical approach and full simulation has been found. In particular, the arrival time distributions of photons on the \dwords{pmt} are well reproduced. The only drawback is that one cannot easily implement or study a complicated geometry including regions that are semi-transparent to light. Hence, the visibilities generated by the two methods are not in agreement in the overall light yield, but have a very similar trend in terms of spatial dependences. Studies to improve the analytical method results are in progress since this approach could be extremely powerful for physics studies in the \dword{spmod}.

\subsubsection{Simulation of light yield}
\label{subsec:fddp-pd-6.1.4}
The light collected per \dword{pmt} can be simulated together with the charge for crossing tracks in a standard simulation code where a detailed description of the detector is not needed. At each step of the track propagation, the energy deposited is computed by \dword{geant4}. This energy is converted into number of electrons and photons produced. As for the light simulation, the number of photons reaching each \dword{pmt} and their time of arrival is now obtained from the light maps. As an example, the light yield from a uniform generation of \SI{10}{MeV} electrons in the active volume is shown in Figure~\ref{fig:dppd_6_1_4}. The number of \phel{}s/\si{MeV} shown is summed over all \dwords{pmt} and average over the $y$ axis ($z$ being the drift direction). One can notice the large spread in terms of light yield.

\begin{dunefigure}[Light yield in terms of \phel/\si{MeV} summed over all \dwords{pmt}]{fig:dppd_6_1_4}
{Light yield in terms of \phel/MeV summed over all \dwords{pmt} and averaged along the y-axis. The mean of all voxels gives a light yield of \SI{2.5}{\phel/MeV}, although the distribution is not uniform, in particular along the $z$ (drift) axis.}
\includegraphics[width=0.6\textwidth]{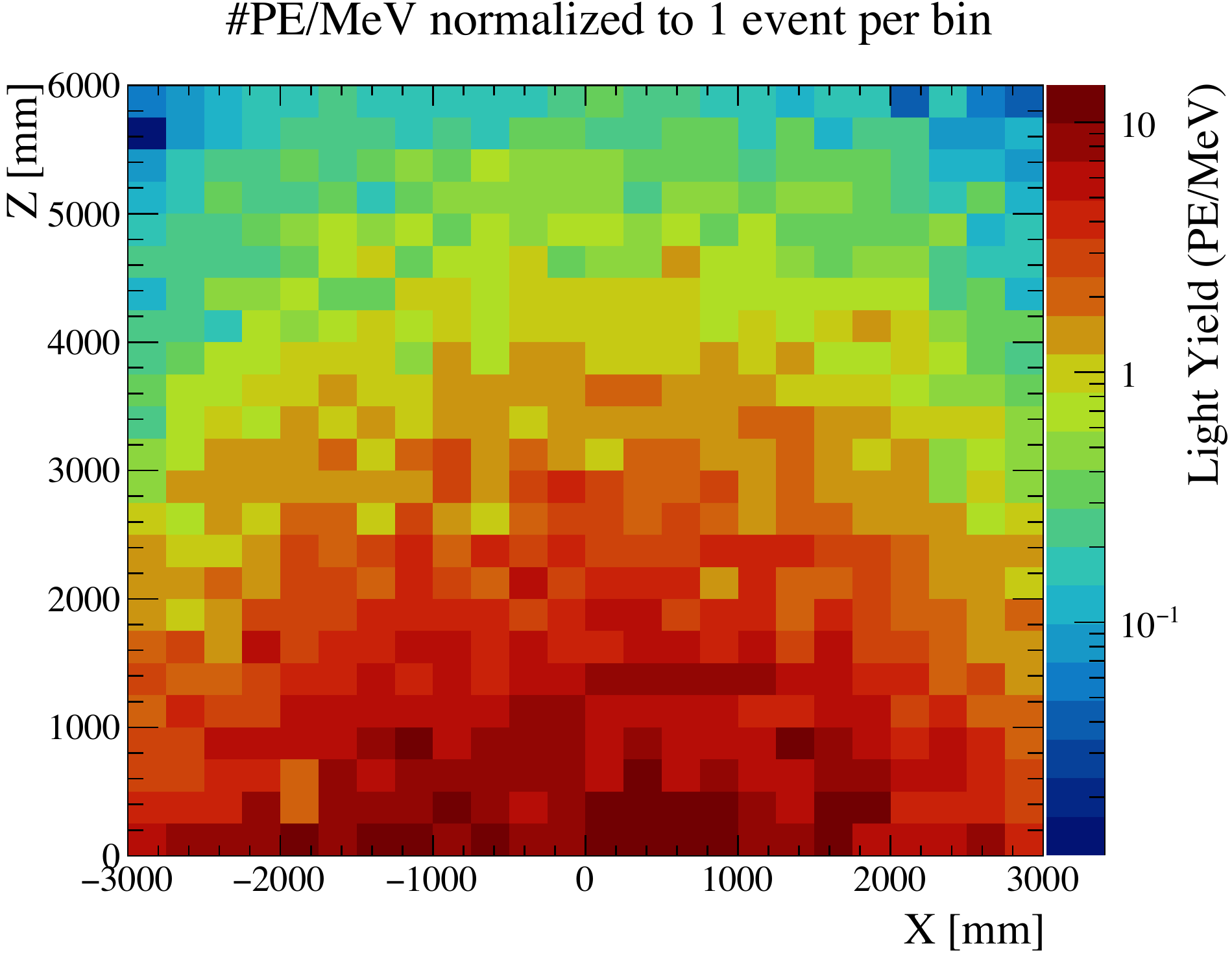}
\end{dunefigure}

For larger volumes such as the \dword{dpmod}, the light maps might be too big and the time spent accessing the four parameters might strongly reduce the speed of the simulation. Either the parametrization method or the analytical approach are foreseen to replace the current light map usage, the exact strategy is yet to be defined.

\subsection{Light Data in \dual Prototypes}
\label{sec:fddp-pd-6.2}

The  \dword{wa105} was operated from June to November 2017 with cosmic data. About \num{5} million light events were taken with various configurations. The study of the S1 light as a function of the drift field was performed. An example of an averaged waveform fitted to a fast and a slow scintillation components is shown in Figure~\ref{fig:dppd_6_2}. The amount of S2 light can be monitored as a function of the extraction and \dword{lem} amplification fields.

\begin{dunefigure}[Averaged waveform of the S1 light signal from the  \dword{wa105}]{fig:dppd_6_2}
{Averaged waveform of the S1 light signal taken with one \dword{pmt} from the  \dword{wa105}, fitted with a function (red line) that is the sum of a Gaussian, parametrized by $t_0$ and $\sigma$, and two exponential functions, with decay time constants $\tau_{fast}$ and $\tau_{slow}$, and normalization factors $A_{fast}$ and $A_{slow}$}
\includegraphics[width=0.7\textwidth]{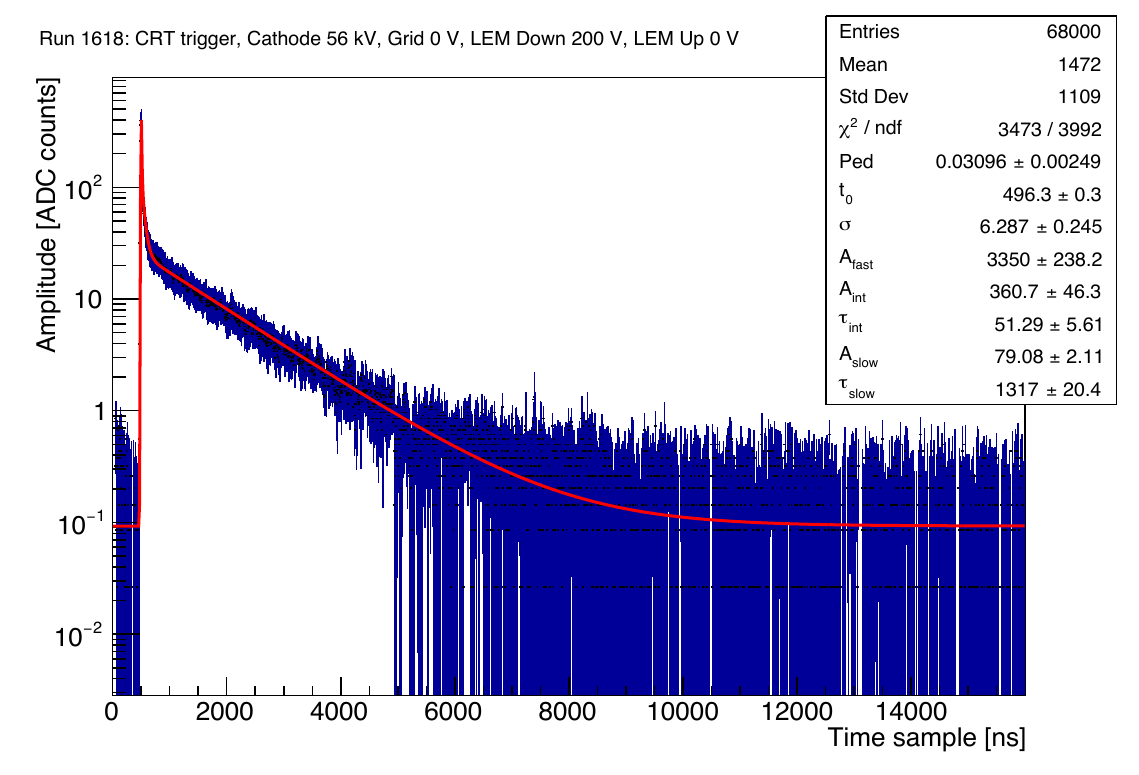}
\end{dunefigure}

Light maps have also been generated with the demonstrator geometry, and data-\dword{mc} comparisons are ongoing. The preliminary results look promising, although the statistics in each setting and the relatively small size of the detector still constitute a challenge to extract the entire optical properties of the \lar.

\subsection{Simulation of Physics Events}
\label{sec:fddp-pd-6.3}

A preliminary study to understand whether the \dual \dword{pd} technical design meets the experiment's physics requirements has been performed. In this study, event topologies of interest for DUNE physics have been simulated using \larsoft fast optical simulation tools.

The simulation framework used represents the current state of the art. It includes realistic models for the primary scintillation production yields in \lar, for Rayleigh scattering in \lar, for detector optical properties (such as \dword{fc} reflectivity and cathode transparency), for the density of \dwords{pmt} underneath the cathode, and for the quantum efficiency of the \dword{tpb}-coated \dwords{pmt} (taken to be \num{20}\%). On the other hand, these simulations do not yet include the full \dword{dpmod} geometry, but rather are performed in a \dword{pddp} geometry with the same average \dword{pmt} density as the one proposed here for the \dword{dpmod} (one \dword{pmt} per \si{m$^2$}). Relevant aspects such as secondary scintillation light emission in gaseous argon (a nuisance for event $t_0$ determination), light absorption in \lar, electronics effects, reconstruction effects, and background contributions coming from $^{39}$Ar decays are not accounted for either in this study. While more realistic simulation results including the above effects will be produced on the DUNE \dword{tdr} timescale, this preliminary study already provides a sense of the capabilities of the planned \dword{pd} design.

Figure~\ref{fig:dppd_6_3_1} shows the expected light yield for \dword{snb} neutrino CC interactions. As a representative \nue flux from a \dword{snb}, we assume a Fermi-Dirac distribution with $T$=\SI{3.5}{\MeV} temperature and no neutrino oscillation effects, yielding an average neutrino energy of about \SI{11}{MeV}. Low-energy \nue CC interactions throughout the entire \lartpc active volume are generated with the \larsoft-based Marley package. For the assumed \dword{snb} neutrino flux and for a single interacting neutrino (hence, after convoluting flux and cross-section effects), Marley expects about \SI{19}{\MeV} of energy deposited in the \lar active volume, primarily from the final state electron and from nuclear de-excitation gamma rays. The left panel of Figure~\ref{fig:dppd_6_3_1} shows a broad light yield distribution, averaging at about \num{50} detected \phel{}s per interaction and after summing all \dwords{pmt}. This is as expected from the light yield distributions per deposited energy shown in Figure~\ref{fig:dppd_6_1_4}. The right panel shows the fraction of \dword{snb} \nue CC interactions within the \lartpc active volume above a given \phel detection threshold, as a function of the \phel threshold. From the figure, we conclude that about a \SI{70}{\%} fraction of \dword{snb} \nue CC interactions would be seen by the \dword{pd}, if the detector threshold was set at \num{10}~\phel{}s on the sum of the \dword{pmt} charges.

\begin{dunefigure}[Response for simulated \dshort{snb} neutrino interactions  (\dword{pddp} geometry)]{fig:dppd_6_3_1}
{Photon detector response for simulated \dword{snb} neutrino interactions in the \dword{pddp} geometry. Left panel: distribution of detected \phel{}s per neutrino interaction, for \dword{snb} \nue CC interactions throughout the active volume. Right panel: fraction of \dword{snb} \nue CC interactions above \phel threshold, as a function of the \phel threshold.}
\includegraphics[width=0.44\textwidth]{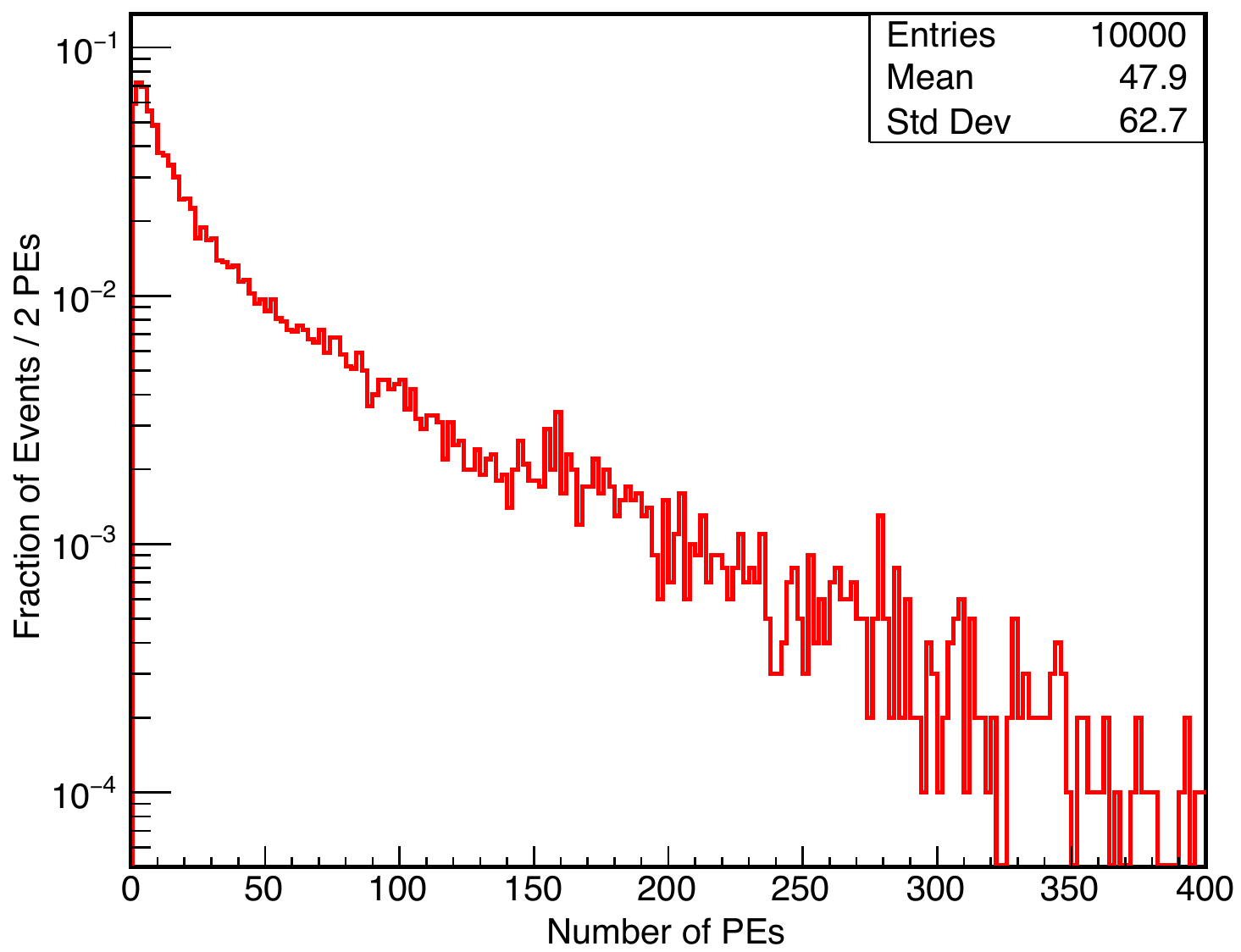} \hfill 
\includegraphics[width=0.44\textwidth]{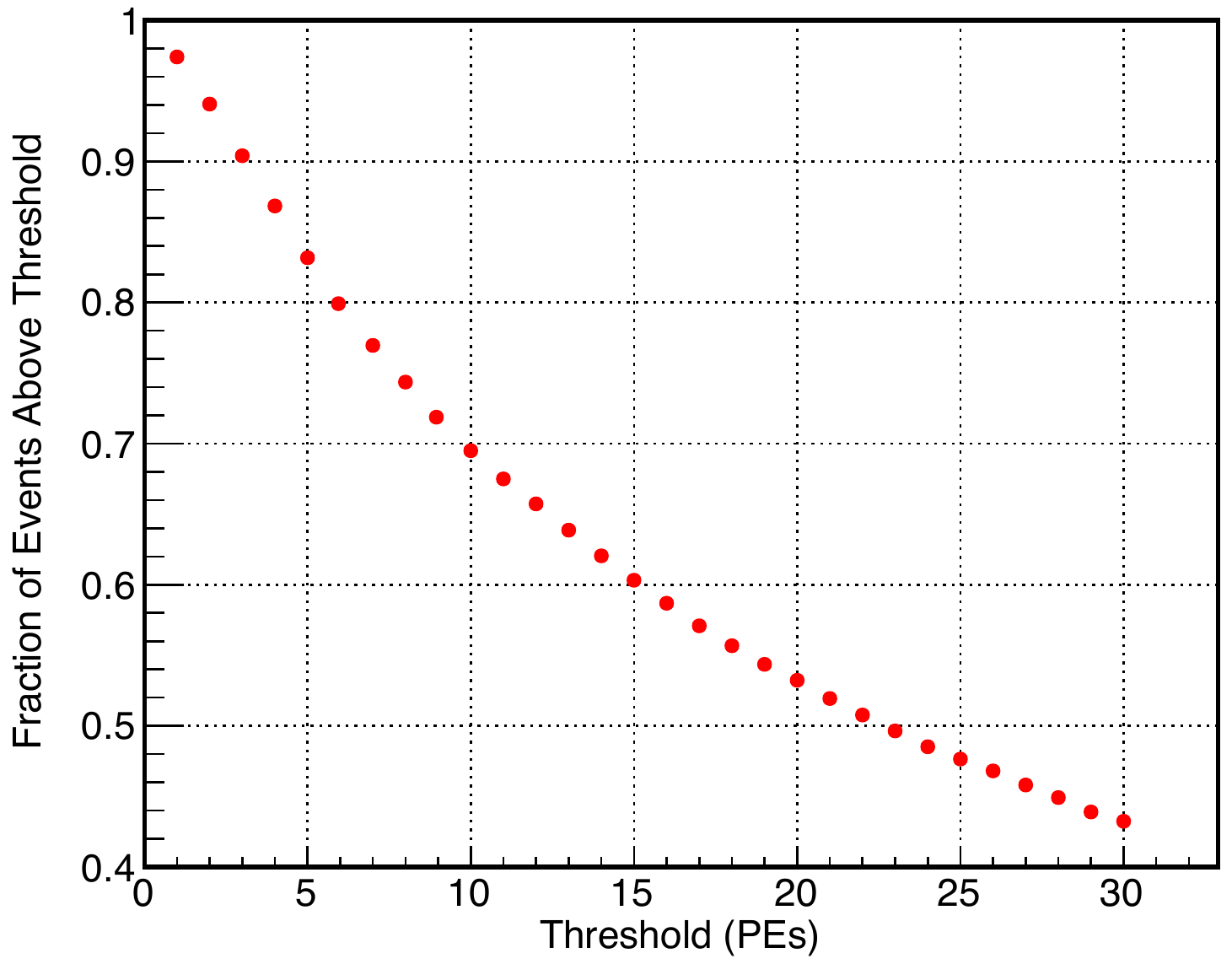} 
\end{dunefigure}

Figure~\ref{fig:dppd_6_3_2} shows the corresponding plots for a representative nucleon decay final state in DUNE, namely $p\to\bar{\nu}K^+$. Nucleon decay events are generated using \dword{genie}, accounting for both initial and final state nuclear effects in argon nuclei. Particles exiting the nucleus are then propagated in \lar using all relevant, \dword{geant4}-based, physics processes. The deposited energy per nucleon decay, of order \SI{300}{\MeV}, is much higher than the one per \dword{snb}  neutrino interaction. As a result, the expected light yield for $p\to\bar{\nu}K^+$ events throughout the active volume, shown in the left panel of Figure~\ref{fig:dppd_6_3_2}, averages to about 800~\phel{}s in this case. The right panel of Figure~\ref{fig:dppd_6_3_2} shows that about a 98\% fraction of 
\ptoknubar decays in the TPC active volume are expected to be seen by the \dword{pd}, for a \dword{pd} threshold of 10~\phel{}s on the \dword{pmt} charge sum.

\begin{dunefigure}[Response for simulated nucleon decays (\dword{pddp} geometry)]{fig:dppd_6_3_2}
{Photon detector response for simulated nucleon decays in the \dword{pddp} geometry. Left panel: distribution of detected \phel{}s per nucleon decay, for $p\to\bar{\nu}K^+$ decays throughout the active volume. Right panel: fraction of 
\ptoknubar decays above \phel threshold, as a function of the \phel threshold.}
\includegraphics[width=0.44\textwidth]{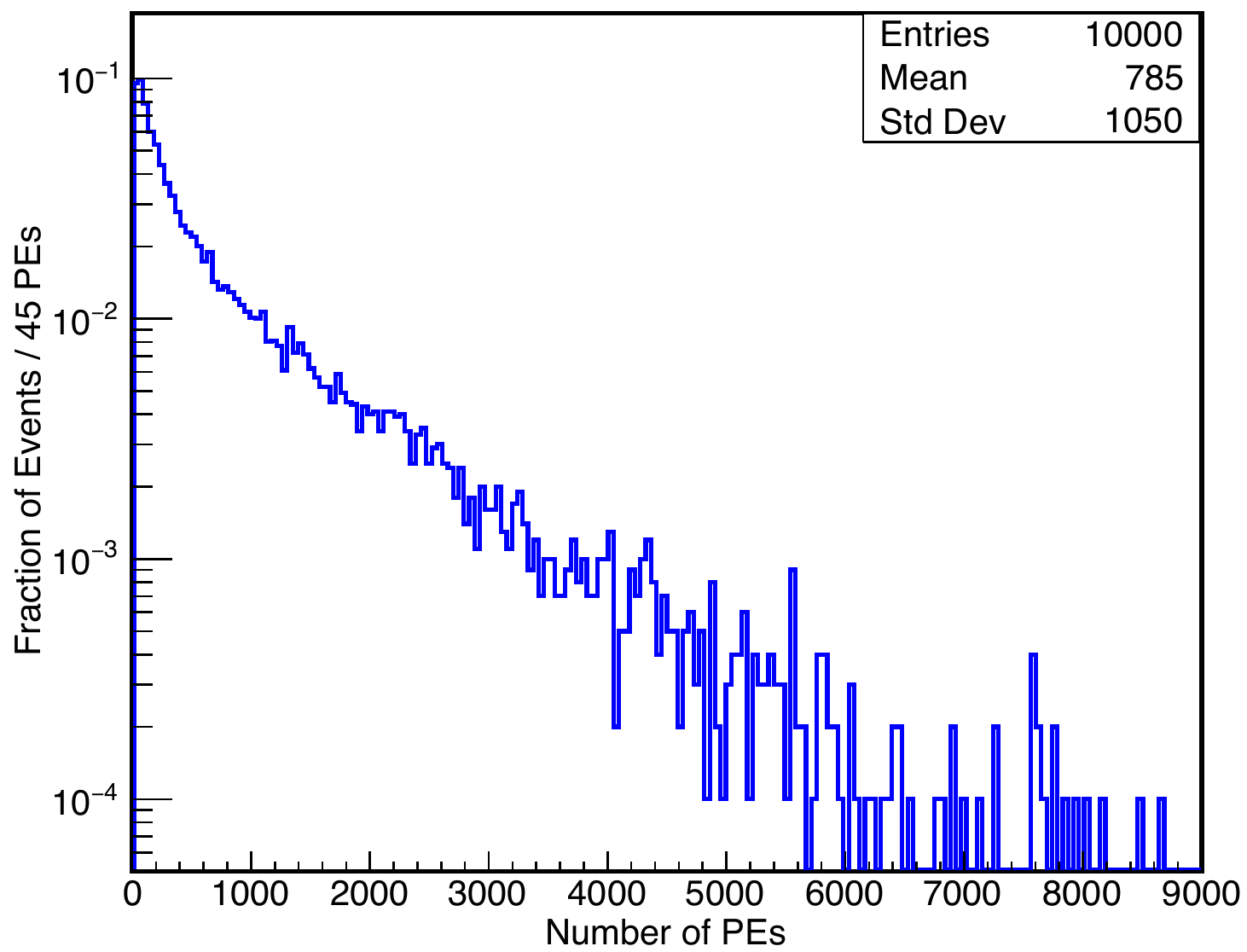} \hfill 
\includegraphics[width=0.45\textwidth]{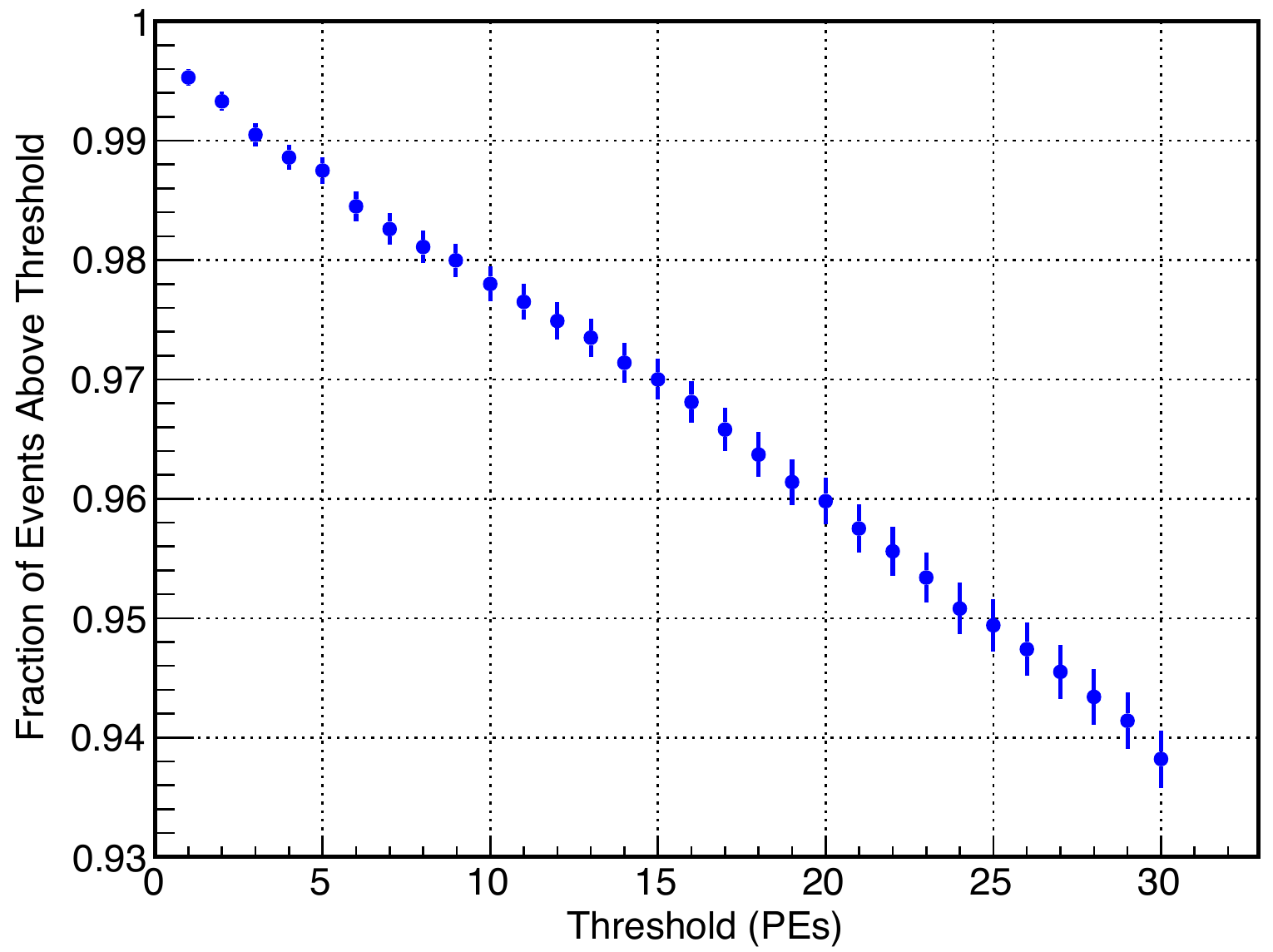} 
\end{dunefigure}

\section{Photon Detector Operations}
\label{sec:fddp-pd-7}

\subsection{Trigger Strategy}
\label{sec:fddp-pd-7.2}

As explained in Section~\ref{sec:fddp-pd-1.5}, the \dword{pds} operates in different acquisition modes. These modes include the external trigger 
(used for the beam events), the trigger for non-beam events, 
and the calibration mode. 

In the \lartpc there are different uses of the light signal: cosmic ray and track timing for the reconstruction; non-beam events trigger such as \dword{snb}, atmospheric neutrinos, and proton decay; and calorimetry, as the light and charge signal are anti-correlated. These physics studies imply different requirements in terms of dynamics of the electronics and data sampling, from a few \phel to a much higher level.

For the non-beam event trigger strategies, the requirements can be very different. In the event of a nearby (\SI{10}{kpc})  \dword{snb}, it is expected that a few thousands of neutrinos will homogeneously interact in the \dword{detmodule} for a period as long as $\sim$\SI{100}{s}. Hence, the  \dword{snb} trigger strategy is mostly driven by the energy threshold set for $\nu$ detection and its efficiency: \SI{30}{MeV} is sufficient for a galactic SN, \SI{5}{MeV} is needed for a burst in Andromeda. A high-efficiency trigger for proton decay events has to be designed considering the worst case scenario, e.g., the event happening at the top of the \dword{detmodule}, \SI{12}{m} away from the closest \dword{pmt}. In order to minimize the amount of spurious triggers, one can think of signal thresholds for a cluster of close-by \dwords{pmt}.

All these important studies will be further investigated once a reliable light simulation of the \dword{dpmod} is available. For the \dual  technology, the main light trigger concerns are the amount of light collectable for a photon traveling distance of \SI{12}{m} and the S1-S2 separation. The data that is collected in the \dword{pddp} will provide crucial inputs for the optimization of the \dword{dpmod} light-collection system and for the design of an efficient trigger strategy for rare non-beam events. 

The \dword{pds} trigger design is flexible so as to fulfill the different physics requirements.  
The light readout \dword{fe} board controls 
the \dword{pds} trigger generation. The trigger is decided based on the coincidence of several \dword{pmt} signals over a threshold during a time window. The number of \dwords{pmt} that contribute to the trigger, the signal threshold and the length of the coincidence time window will be programmable online in order to be adaptable to different physics cases.

\subsection{Data Quality Monitoring}
\label{sec:fddp-pd-7.3}

The \dwords{pmt} installed at the bottom of the tank will be operated for \numrange{10}{20} years with no possibility to access them. Monitoring tools to ensure data quality of the \dword{pds} will have to be developed to catch any malfunctioning detector before data analysis. For instance, the amount of dark noise and the stability of the \dword{pmt} response will have to be monitored over time. For the gain evolution, either studies of standard candles, e.g., from Michel electrons or average collected light produced by cosmic tracks, or with the dedicated calibration system are under consideration.

Monitoring tasks were performed during the six-month \dword{wa105} operation with no dedicated light calibration system. This and the forthcoming operation of the \dword{pddp} will again provide crucial input towards the \dword{pds} monitoring system in the \dword{dpmod}.

\section{Interfaces}
\label{sec:fddp-pd-8}

The \dword{pds} has several interfaces with other subsystems and the global DUNE systems. The interface documents related to \dword{dpmod} \dword{pds} are given in Table~\ref{tab:dppd_t_8}. Only some 
of the basic interfaces are summarized below. 

\begin{dunetable}
[\dual \dword{pd} interface documents]
{|l|c| p{0.8\textwidth}}
{tab:dppd_t_8}
{\dual \dword{pd} interface documents}

\dual \dword{pd} Interface Document & DUNE docdb number \\ \toprowrule
\dword{dpmod} electronics & 6772~\cite{bib:docdb6772} \\
\dword{dpmod} \dword{hv} & 6799~\cite{bib:docdb6799} \\
\dword{daq} & 6802~\cite{bib:docdb6802} \\
Cryogenic instrumentation and slow controls (CISC) & 6781~\cite{bib:docdb6781} \\ 
DUNE Physics & 7087~\cite{bib:docdb7087} \\
Software and Computing & 7114~\cite{bib:docdb7114} \\
Calibration & 7060~\cite{bib:docdb7060} \\
Integration and test facility (ITF) & 7033~\cite{bib:docdb7033}\\
Detector and Facilities (LBNF) Infrastructure & 6979~\cite{bib:docdb6979} \\
Installation & 7006~\cite{bib:docdb7006} \\
\end{dunetable}

\begin{itemize}

\item \dword{dpmod} electronics: The \dword{pds} shares the same \dword{fe} electronics standard as the charge readout, which is \dword{utca}-based \cite{utca}. Specifications of both \dword{pds} and \dword{fe} electronics will be determined by the simulations and \dword{pddp} data.

\item \dword{hv}: This interface includes the consideration of the distance between the cathode and the \dword{pmt} planes, power dissipation from the \dwords{pmt} and the combined impact on the simulations.

\item \dword{daq}: The hardware interface is mainly through optical fibers. \dword{dpmod} \dword{pds} provides trigger and data in continuous streaming;  the interface also includes the \dword{daq} software.

\item \dword{cisc}: The main interface points are the layout of the cryogenic instrumentation (e.g., purity monitors and light emitting system for the cameras) and the \dword{pmt} support structures and cabling; and the slow control and the \dword{pds} power supplies and calibration system.

\item DUNE physics: \dual \dword{pd} has interfaces with the overall physics requirements on energy and time together with classification of events, decay modes and neutrino flavors.

\item Software and Computing: This interface is on the development of simulation, reconstruction and analysis tools.

\item Calibration: The \dword{pds} is participating in the DUNE global calibration task force and will provide handles to allow global monitoring of the \dword{pmt} performance.

\item \dword{itf}: The operations at the \dword{itf} are described in Section~\ref{sec:fddp-pd-9.2}. The interface items can be summarized as shipping and receiving of the \dword{pds} components and basic testing and repairing at the facility. The interface includes recycling and returning the packaging materials.

\item Detector and Facilities (LBNF) Infrastructure: The \dword{pds} 
\dword{pmt} mounting structures stand on the cryostat membrane; cold cables are routed in cable trays to the ceiling \fdth flanges and racks and to cable trays on top of the cryostat. Other interfaces with the facility include access to conventional facilities and participation in the detector safety systems. 

\item Installation: This interface 
includes the transportation of the \dword{pds} components to and between underground areas, clean room activities and storage, and installation coordination with the other teams. 

\end{itemize}

\section{Installation, Integration and Commissioning}
\label{sec:fddp-pd-9}

\subsection{Transport and Handling}
\label{sec:fddp-pd-9.1}

The \dpnumpmtch \dwords{pmt} of the \dword{pds} are shipped from various locations following base and cable assembly for the \dword{tpb} coating at the \dword{itf}. 
They are shipped in boxes of \num{24}, for a total of \num{30} deliveries.
The \dwords{pmt} are individually wrapped with special wrapping materials. 

The \dwords{pmt} are placed in modular shock-absorbing assemblies inside the boxes. 
The boxes have integrated pallets for easy handling and short-distance transportation, and a limited amount of inclination for the assemblies is safely allowed. The \dwords{pmt} reach the \dword{itf} by air and ground transportation. Each box has a dedicated bar code  visible on each side. This bar code is also associated with the shipping documents. 

\subsection{Integration and Testing Facility Operations}
\label{sec:fddp-pd-9.2}

The \dword{itf} receives the \dword{pmt} boxes and manages a shipping and delivery database. The received status of the boxes is available to the \dual \dword{pd} consortium as the boxes arrive at the \dword{itf}. The \dword{pds} characteristics database managed by the \dual \dword{pd} consortium is updated accordingly to reflect the received status of the contents of the boxes. Each \dword{pmt} assembly gets an identifying bar code that is directly connected to the \dword{pds} characteristics database. This database stores the \dword{pmt} serial number, the base board serial number, special information about \dword{tpb} coating and assembly if any, and performance and calibration characteristics. This database  forms the basis of the operations database, providing the initial calibration values. It also stores information about the \dword{itf} tests and underground installation and commissioning tests.

The \dword{tpb} coating is performed at the \dword{itf} in the coating stations, after which  the \dwords{pmt} are placed back in their boxes and dedicated testing electronics are connected to the \dword{pmt} cables soldered to the \dword{pmt} bases. The test electronics enables connecting several \dwords{pmt} at a time. The tests include basic functionality checks of both the \dwords{pmt} and the base boards to assess the performance after transportation. No detailed performance characteristics are measured at the \dword{itf}. The tests are performed in a dedicated room with light and climate control. Once the performance of the \dwords{pmt} in a box is validated, the box is closed with the original cover. Before closing, additional quality checks on the shock-absorbing assemblies are made.

The preparation of the \dword{pmt} boxes for underground transportation includes installing holding and lifting fixtures to the top and sides 
that allow crane operation. The boxes are delivered to the surface station by ground transportation with 
appropriate updates to the shipping database.

\subsection{Underground Installation and Integration}
\label{sec:fddp-pd-9.3}

Once the \dword{pmt} boxes are underground, the same top and side covers are opened as at the \dword{itf}. The \dwords{pmt} are carried to the underground storage area in sub-units of the modular shock-absorbing assemblies. The underground storage area for the \dword{pds} is expected to be sufficiently large to store at least \num{30} \dwords{pmt}, enabling continuous installation operations.

The removal of the individual \dword{pmt} wrappings is done in the clean room. \Dwords{pmt} together with their base boards undergo visual inspection by the \dword{pds} installation supervisor. Once signed-off, the installation can proceed with multiple \dwords{pmt} at a time by multiple teams. Cabling is carried out in parallel and relevant database updates are made in situ. The installation time management is done in coordination with the cathode and \dword{fc} installation groups.

The bundles of cables are routed through the cable trays along the cryostat walls from the \dword{pds} flanges. Following the mechanical mounting of the \dwords{pmt} to the cryostat floor, the \dword{pmt} cables are 
connected to the cables coming from the flanges. In parallel, the calibration fibers are installed and routed through cable trays.

\subsection{Commissioning}
\label{sec:fddp-pd-9.4}

The commissioning of the \dword{pds} is performed in partitions. The size of a single partition will mainly be determined by the \dword{daq} and the \dword{hv} systems. The \dword{daq} and \dword{hv} partitions are commissioned, including the relevant control systems, prior to the connection of the \dwords{pmt} to these systems. Once the physical sector corresponding to a partition is installed, the \dwords{pmt} are powered up, and basic functionality and performance checks are performed. These include pedestal data taking, i.e., recording event data with external periodic triggering, and tests with the calibration system where the data taking is triggered in synchronization with a light source, as described in Section \ref{sec:fddp-pd-5}.

As a result of the commissioning tests, the basic performance characteristics of the \dwords{pmt}, e.g., the dark count rate and gain, are measured in their final places. Installation-related issues are identified and eliminated at this stage. A commissioned sector 
becomes a part of the overall detector and can join the global calibration data taking and commissioning.

\section{Quality Control}
\label{sec:fddp-pd-10}

 \subsection{Production and Assembly}
 \label{sec:fddp-pd-10.1}
 
The \dword{qc} performed at the different institutions labs includes reception of \dwords{pmt} from the manufacturer and execution of the \dword{qc} tests to accept or return the \dwords{pmt} according to the acceptance and rejection criteria.

\begin{itemize}
\item The \dword{pmt} support structure design is validated by immersing its mounted \dword{pmt} in cryogenic temperatures and at an over-pressure equivalent to 
a depth of \SI{12}{m} in \lar{}. 
\item Design validation tests are carried out to confirm that the \dword{pmt} base design fulfills the specifications at room and cryogenic temperatures. A cable with SHV connector is soldered to each \dword{pmt} base to facilitate the different base and \dword{pmt} tests and the final \dword{pmt} connection during the installation. The \dword{pmt} bases are labeled (on the cable) in order to keep track of them. After production of the \dword{pmt} base boards they are individually tested before mounting to the \dword{pmt} to verify that components are correctly mounted. Later they are cleaned and tested at maximum voltage in an argon gas environment to confirm that there are no sparks on these (worst case conditions).
After mounting the bases on the \dwords{pmt} they are tested again in argon gas at maximum voltage to confirm that there are no sparks due to bad soldering.
\item All the light readout units (\dword{pmt} + base + support) are tested and characterized in liquid nitrogen in order to check their performance at cryogenic temperature and to obtain a database with the most important parameters from each \dword{pmt} (gain versus voltage, dark counts, etc.). The \dword{pmt} base number attached to each \dword{pmt} is also included in the database. 
\item The wrapping materials and techniques are studied with one fully assembled light readout unit. The handling, transportation and installation scenarios are carefully studied and the transportation box design is validated. The transport box and \dword{pmt} wrapping must  ensure complete darkness. 
\item The light output of the \dwords{led} and the fibers' light transmission from the photon calibration system are measured with a power meter.
\end{itemize}

 \subsection{Post-Factory Installation}
 \label{sec:fddp-pd-10.2}
 
Upon receipt at \dword{itf}, the \dwords{pmt} go through verification measurements in order to identify any items damaged during transport.  Gain versus voltage and dark current values are compared with those obtained before transportation.

The \dword{tpb} coating is also performed at the \dword{itf}. The first few samples undergo microscopic examination and surface uniformity tests, and the coating procedure is validated. The production \dwords{pmt} are randomly sampled for basic coating \dword{qc}.

After  transport from the \dword{itf} to \surf the \dwords{pmt} are tested again before installation to confirm that no damage occurred during the last stage of transportation. During the installation, the \dwords{pmt} database is updated with the position in the \dword{detmodule} of each \dword{pmt} (identified by its serial number and base number). After installation, the full connection from the \dword{fe} to the \dwords{pmt} is checked. The \dword{fe} channel and splitter number connected to each \dword{pmt} are also included in the \dword{pmt} database. 
Once it is possible to fully darken the \dword{detmodule}, voltage can be applied to the \dwords{pmt} to test the signal with a scope or, if available, with the \dword{fe} electronics.

\section{Safety}
\label{sec:fddp-pd-11}

Safety is the highest priority at all stages of \dual \dword{pd} operations. Since DUNE is an international project, the international safety regulations will be followed closely 
and safety documents will be prepared accordingly.

The main risks at the production and testing sites are electrocution, exposure to excessive heat, chemicals and cryogenics, and heavy lifting. Detailed procedures will be developed by the relevant institutes and approved by the \dual \dword{pd} consortium. Contents of the electrical safety rules will range from utilizing regular power equipment to handling \dwords{pmt} for testing. The chemical and heat exposure hazards only concern the sites where the \dword{tpb} coating is performed. The heavy-lifting risks concern mainly 
the \dword{pmt} delivery boxes.

The \dword{itf} \dual \dword{pd} safety procedures will be developed in a similar way. The main hazards at this site are electrocution and heavy lifting. Also, due to the quantity and frequency of shipments from all other subsystems, tripping and operating in a limited space will be considered.

The underground operation and installation safety rules will 
follow the general facility rules on e.g., working in confined spaces, oxygen deficiency hazard and emergency procedures. The \dual \dword{pd}-specific safety rules are particularly related to lifting the boxes and their contents 
for installation and working at heights for cabling.

\section{Management and Organization}
\label{sec:fddp-pd-12}

The \dual \dword{pd} consortium was formed in 2017 and it is composed of eleven institutes from France, Peru, Spain, UK and USA. The charge of the \dual \dword{pd} consortium is to plan and execute the construction, installation and commissioning of the \dword{dpmod} \dword{pds}.

\subsection{Consortium Organization}
\label{sec:fddp-pd-12.1}

The current \dual \dword{pd} consortium Leader (CL) is 
 from CIEMAT (Spain) and the Technical Lead (TL) is 
 from LAPP (France). They are members of the DUNE Technical Board and they represent the consortium to the overall DUNE collaboration. The CL is responsible for the subsystem deliverables and for the effective management of the consortium. The TL acts as the overall project manager and is the interface to the International Project Office (IPO); he is responsible for monitoring and reporting on progress with respect to the agreed schedule and for issues related to interface documentation.

The institutions participating in the consortium are responsible for the design or construction of a particular subsystem. It is hoped that the national groups within the consortia will be able to approach relevant funding agencies with a specific construction-phase proposal, such that a likely funding line can be established in or before 2019. The \dual \dword{pd} consortium is open to any new institution willing to join the current effort.

The current institutions participating in the \dual \dword{pd} consortium are LAPP (France); PUCP (Peru); IFAE, CIEMAT, and IFIC (Spain); UCL (UK); and ANL, Duke U., U. of Iowa, SDSMT, and UTA (USA).


The \dual \dword{pd} consortium is divided into four working groups: photosensors and electronics, calibration system, mechanics and integration, and simulation and physics. The corresponding current WG convener institutions are:
\begin{itemize}
\item WG1: Photosensors and Electronics -  CIEMAT 
\item WG2: Calibration System -  CIEMAT 
\item WG3: Mechanics and Integration - U. of Iowa 
\item WG4: Sim. \& Phys. - 
Duke U., 
IFIC, 
LAPP

\end{itemize}


\subsection{Planning Assumptions}
\label{sec:fddp-pd-12.2}

The optimization and final design of the \dual \dword{pd} system will be driven by:
\begin{enumerate}
\item \dword{pddp} data (expected by beginning of 2019)
\item Simulation studies (in progress)
\end{enumerate}

\dword{pddp} operation and data analysis are fundamental steps to understanding whether the current \dword{pds} design considered as baseline, based on cryogenic \dwords{pmt} with \dword{tpb} coating, is able to provide $t_0$ for non-beam events, background rejection and triggering on non-beam events. These data will be used to tune the \dword{mc} simulations and extrapolate the performance of the system to the \dword{dpmod}. 

Simulations are needed to determine and optimize the \dual \dword{pd} system to meet the physics requirements in terms of:
\begin{itemize}
\item light collection efficiency,
\item number of channels,
\item photosensor requirements,
\item dynamic range of readout electronics and timing resolution, and 
\item trigger strategy on non-beam events.
\end{itemize}

The performance requirements on the \dword{pds} will be provided by the DUNE physics working group. Alternate designs for aspects of the \dword{dpmod}'s baseline 
\dword{pds}~\cite{cdr-vol-4} 
 will be developed based on the compatibility of \dword{pddp} data and \dword{mc} light simulation results with the DUNE physics requirements.

\subsection{WBS and Responsibilities}
\label{sec:fddp-pd-12.3}

The \dual \dword{pd} consortium has developed a detailed breakdown of deliverables and responsibilities included in the overall DUNE collaboration \dword{wbs}~\cite{bib:docdb5594} coordinated by the IPO. The main deliverables are based on the \dword{pddp} \dword{pds}  and are divided into seven topics. These are listed along with the participating institutions: 

\begin{enumerate}
\item Management \dual \dword{pds} (includes milestones and review dates) \textit{- LAPP, CIEMAT }
\item Physics and Simulations \textit{- Duke, LAPP, IFIC, SDSMT, CIEMAT, PUCP, UCL, Texas-Austin}
\item Design, Engineering, R\&D and validation tests \textit{- Iowa, CIEMAT, IFIC, UCL, Texas-Austin, IFAE, SDSMT}
\item Production Setup (includes tooling) \textit{- UCL}
\item Production (includes component production, assembly, testing, and \dword{qc}) \textit{- Iowa, CIEMAT, IFAE, IFIC, UCL, Texas-Austin, Duke, SDSMT, LAPP}
\item Integration (contributions to activities at global integration facility) \textit{- SDSMT}
\item Installation (contributions to activities at \surf) \textit{- CIEMAT, IFIC, SDSMT, Iowa}
\end{enumerate}

\subsection{High-Level Cost and Schedule}
\label{sec:fddp-pd-12.4}

The cost of the baseline \dual  \dword{pds} will be defined in a separate document. \fixme{need a ref to it?}

The \dual \dword{pds} consortium's main activities during the next \num{16} months are focused on developing the \dword{tdr}. The main high-level milestones are detailed in Table~\ref{tab:dppd_t_12_5} for this period. The plan for the activities in the post-\dword{tdr} period is summarized in Table~\ref{tab:dppd_t_12_6}.

\begin{dunetable}
[Pre-\dword{tdr} key milestones]
{|l|l| p{0.8\textwidth}}
{tab:dppd_t_12_5}
{Pre-\dword{tdr} key milestones}

Milestone & End date \\ \toprowrule
Simulations and physics: 
Implementation of \dual optical & \\
simulation in \larsoft for \dword{pddp} & 08/2018 \\ \colhline
Simulations and physics: Optimization of the & \\
\dword{dpmod} performance to fulfill the physics requirements and & \\
definition of a trigger strategy & 05/2019 \\ \colhline
Photosensors: Components selection and final design & 03/2019 \\ \colhline
\dword{pmt} calibration system design and selection of components & 03/2019 \\ \colhline
Cabling definition and design of flange & 03/2019 \\ \colhline
Design review in light of \dword{pddp} calibration data & 03/2019 \\ \colhline
\dword{qc} plan & 06/2018 \\ \colhline
Identification of Interfaces & 06/2018 \\ \colhline
Integration, installation and commissioning plans & 12/2018 \\ \colhline
\dword{dpmod} \dword{tdr} & 06/2019 \\ 
\end{dunetable}

\begin{dunetable}
[Post-\dword{tdr} key milestones]
{|l|l|l| p{0.8\textwidth}}
{tab:dppd_t_12_6}
{Post-\dword{tdr} key milestones}

Milestone & Start date & End date \\ \toprowrule
\textbf{\dword{pmt} preparation and installation} (can be done in batches) & & \\ \colhline
\dword{pmt} procurement procedure and production & 01/2021 & 12/2022 \\ \colhline
\dword{pmt} base design and manufacturing & 01/2022 & 12/2022 \\ \colhline
\dword{pmt} support structure production and assembly & 08/2022 & 01/2023 \\ \colhline
\dword{pmt} characterization - \num{10} \dwords{pmt}/week (two facilities) & 02/2023 & 12/2023 \\ \colhline
\dword{tpb} coating (two facilities similar to that for CERN ICARUS) & 01/2024 & 12/2024 \\ \colhline
Splitter production and tests & 05/2024 & 12/2024 \\ \colhline
\textbf{Installation at \surf} & & \\ \colhline
\dword{pmt} cable and fiber routing in cryostat from flange to bottom & & \\
                  (depends on \dword{fc} and flange installation) & 09/2024 & 09/2024 \\ \colhline
\dword{pmt} testing, installation in cryostat and cabling (\num{72} \dwords{pmt}/month) & 10/2024 & 07/2025 \\ \colhline
\dword{pmt} support installation on the membrane & & \\
                  (in parallel by sector with \dword{pmt} installation) & 10/2024 & 07/2025 \\ \colhline
Splitter installation & & \\
                  (in parallel with \dword{pmt} installation to test cabling and connections) & 10/2024 & 07/2025 \\ \colhline
\textbf{Light calibration system} & & \\ \colhline
Fibers, light source tests and procurement & 06/2023 & 05/2024 \\ \colhline
Fiber calibration system installation & & \\
                  (in parallel with \dword{pmt} installation with validation test) & 09/2024 & 07/2025 \\ 
\end{dunetable}

\cleardoublepage

\chapter{Data Acquisition System}
\label{ch:fddp-daq}

\section{Data Acquisition (DAQ) System Overview}
\label{sec:fd-daq-ov}

\metainfo{DP/SP shared.  Georgia Karagiorgi and Dave Newbold. 2 Pages - largely
  generic but some highlighting of SP-specifics. 
  Focus on describing to HEP but non-DAQ expert. 
  Include how design is resilient in the face of potential
  uncertainties such as excess noise or the need to reduce drift HV
  (just two examples, maybe there are more).}

\subsection{Introduction}
\label{sec:fd-daq-intro}

The DUNE \dword{fd} \dword{daq} system must enable the readout,
triggering, processing and distribution to permanent storage of data
from all \dwords{detmodule}, which includes both their electrical
\dword{tpc} and optical \dword{pds} signals.  
The final output data must retain, with very high efficiency and low
bias, a record of all activity in the detector that pertains to the
recognized physics goals of the DUNE experiment. 
The practical constraints of managing this output requires that the
\dword{daq} achieve these goals while reducing the input data volume by almost four
orders of magnitude.

The current generation of \dword{lartpc} \dwords{daq}, such as used in
\dword{protodune} and \microboone, produce data spanning a fixed window of
time that is chosen based on the acceptance of an external trigger. 
The DUNE \dword{daq} faces several major challenges beyond those of the
current generation. 
Foremost, it must accept data from about two orders of magnitude more
channels and from that data it must form its own triggers.
This self-triggering functionality requires immediate processing of
the full-stream data from a large portion of all TPC channels with a
throughput of approximately one terabyte per second per
\dword{detmodule}. 
From this data stream, triggers must be raised based on two very
different patterns of activity. 
The first is activity 
localized in a small region of one
\dword{detmodule}, such as due to beam neutrino interactions or the
passage of relatively rare cosmic-ray muons. 
This activity tends to correspond to a relatively large deposition of
energy, around \SI{100}{\MeV} or more. 
The second pattern that must lead to a trigger is lower energy activity
dispersed in both time and spatial extent of the \dword{detmodule}, such as due to a 
\dword{snb}.

The 
\dword{daq} must also contend with a higher order of
complexity compared to the current generation. 
The \dword{fd} is not monolithic but ultimately will consist
of four \dwords{detmodule} each of \nominalmodsize fiducial mass. 
Each module will 
implement somewhat different 
technologies and the
inevitable asymmetries in the details of how data are read out from
each must be absorbed by the unified \dword{daq} at its front end. 
Further, each \dword{detmodule} is not monolithic but has at least one
layer of divisions, here generically named \dwords{detunit}. 
For example, the \dword{sp} \dword{detmodule} has \dwords{apa} each
providing data from a number of \dwords{wib} and the \dword{dp} \dword{detmodule} has
\dword{cro} and \dword{lro} units associated with specific electronics
crates.
In each \dword{detmodule}, there are on the order of \num{100} \dwords{detunit}
(\num{150} for \dword{sp} and \num{245} for \dword{dp}) and each unit has a
channel count that is of the same order as that of an entire \lartpc
detector of the current generation.
The DUNE \dword{daq}, composed of a cohesive collection of \dword{daq} instances
called
\dwords{daqpart}, must run on a subset of all possible
\dwords{detunit} for each given \dword{detmodule}. 
Each instance effectively runs independently of all the others, however
some instances indirectly communicate through the exchange of
high-level trigger information. 
This allows, for example, each \dword{detmodule} to take data in
isolation. It also allows for all \dwords{detmodule} to contribute to forming and
accepting global \dword{snb} triggers, and to simultaneously run small portions -- consisting of a few \dwords{detunit} -- separately in
order to debug problems, run calibrations or 
 perform other activities while not interfering with nominal data taking in order to maintain high uptime.

Substantial computing hardware is required to provide the processing
capability needed to identify such activity while keeping up with the
rate of data.
The nature of various technical, financial and physical constraints
leads to the need for much of the computing hardware 
required for this processing
to reside underground, near the \dwords{detmodule}. 
In such an environment, power, cooling, space, and access is far more
costly than in typical data centers. 

Past \lartpc and \dword{lbl} neutrino detectors have successfully
demonstrated external triggering using information related to their beam. 
The DUNE \dword{fd} \dword{daq} will accept external information on recent
times of Main Injector beam spills from \fnal. 
This will assure triggering with high efficiency to capture activity
pertaining to interactions from the produced neutrinos. 

However, even if the DUNE experiment were interested only in 
neutrinos from 
beam spills, an external beam
trigger alone would not be sufficient. 
Absent any other information, such a trigger must inevitably call for
the readout of all possible data from the \dword{fd} 
over at least one \lartpc drift time.
This would lead to an annual data volume approaching an exabyte
($10^{18}$ bytes), the vast majority of which would consist of just noise. 
This entire data volume would have to be saved to permanent storage
and then processed offline in order to get to the signals.

DUNE's physics goals of course extend beyond beam-related interactions, including
cosmic-ray muons, which provide an important
source of detector calibration, and atmospheric neutrino interactions,
which give a secondary source from which to measure neutrino
properties. 
Taken together, 
recording their activity will
dominate the data rate.
The \dword{daq} must also record data with sensitivity to rare interactions
(both known and hypothetical) such as nucleon decay, other baryon
number violating processes (such as neutron-antineutron oscillation),
and interactions from the products of \dwords{snb} as well as possibly
being able to observe isolated low-energy interactions from solar
neutrinos and diffuse supernova neutrinos. 

Some of these events, while rare in themselves, produce patterns of
activity that can be mimicked by other higher-rate backgrounds, particularly
in the case of \dwords{snb}. 
While the exact processes involved in \dwords{snb} are not fully understood,
it is expected that a prolonged period of activity of many tens of
seconds will occur over which their neutrino interactions may be
observed. 
Individually, these interactions will be of low energy (relative to
that of beam neutrino interactions, for example), and will be spread
over time and over the bulk of the \dwords{detmodule}. 
Because of their signature and their importance, special attention is
required to first ascertain that a \dword{snb} may be occurring and to save as
much data as possible over its duration.

Thus the \dword{daq} must greatly reduce the full-stream of its input data
while using the data itself to do so. 
It must do this efficiently both in terms of recording essentially all
activity important to the physics goals of DUNE and in terms of a rate of data output 
that is manageable.  
To perform these primary duties the \dword{daq} 
provides run
control, configuration management, monitoring of both its processes
and the general health of the data, and a user interface for these activities.

\subsection{Design Considerations}
\label{sec:fd-daq-des-consid}

The different \dwords{detmodule} vary in terms of their
readout technology and schemes, timing systems, channel counts and data
throughput and format.
These aspects determine the nature of the digital data input
to the \dword{daq}. 
The design of the \dword{daq} strives to contain the unique layers that adapt
to the variation in the \dwords{detmodule} toward its front end in
order to allow as many of its back end components to remain as identical across
the \dwords{detmodule} as possible. 
In particular, the \dword{daq} must present a unified interface to the
ultimate consumer of its data, DUNE offline computing.
It must also accept and process the data from a variety of other
sources including the accelerator, various calibration systems
(including laser, \dword{ce}, \dwords{pd}, and potentially
others) as well as trigger sources external to DUNE.
The modular nature of the DUNE \dword{fd} implies that the \dword{daq} instances running
on each module must also exchange trigger information. 
In particular, exchanging module-local \dword{snb} trigger information
will allow higher efficiency for this important physics.
The \dword{daq} must be optimized for the above while also retaining
the flexibility to scale to handle risks such as excess noise,
changes in \dword{hv}, cut network connectivity and other issues that could arise. 

\begin{dunefigure}[DAQ overview]{fig:daq-overview}
  {The high-level, \textit{nominal} design for the DUNE \dword{fd} \dword{daq} in
    terms of data (solid) and trigger (dashed) flows between one
    \dword{daqfrag} \dword{fe} and the trigger processing and event
    building back end for one \dword{daqpart}. 
    Line thickness indicates relative bandwidth requirements.
    Blue indicates where the full data flow for the \dword{daqfrag} is
    concentrated to one endpoint.
    Green indicates final output of normally triggered (non-\dword{snb}) data.
    Red indicates special handling of potential \dword{snb}. 
    Each detector module has specialized implementation of some of these
    high level components, particularly toward the upstream \dword{fe}
    as described in the text. 
    The grayed boxes are not in the \dword{daq} scope.
  }
  \includegraphics[width=\textwidth{}, trim={1cm 0 1cm 0},clip]{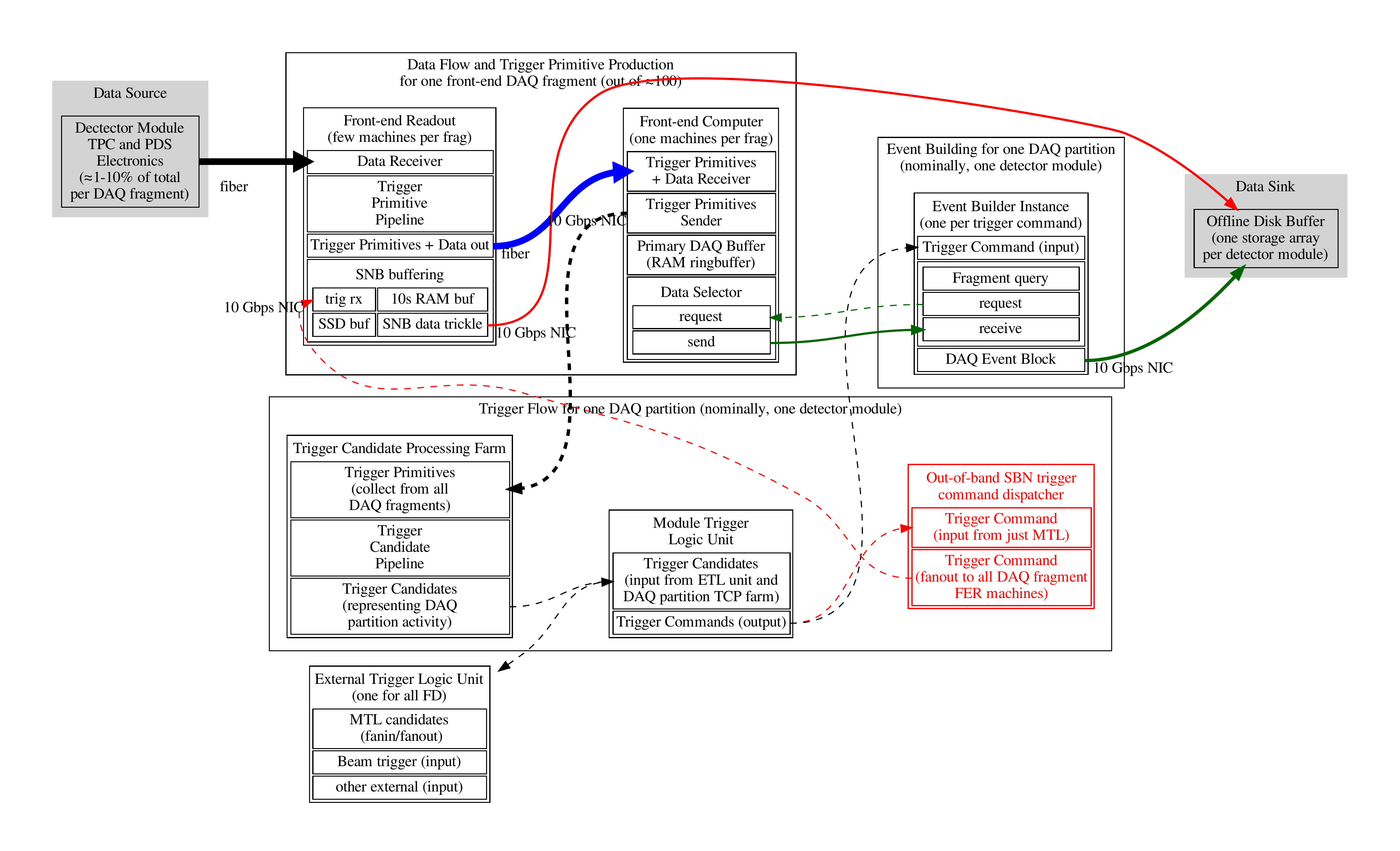}%
\end{dunefigure}

\begin{dunefigure}[\dword{daq} overview]{fig:daq-overview-alt}
  {The high-level, alternate design for the DUNE \dword{fd}\dword{fd} \dword{daq} in
    terms of data (solid) and trigger (dashed) flows between one
    \dword{daqfrag} \dword{fe} and the trigger processing and event
    building back end for one \dword{daqpart}. 
    Line thickness indicates relative bandwidth requirements.
    Blue indicates where the full data flow for the \dword{daqfrag} is
    concentrated to one endpoint.
    Green indicates final output.
    Note, except for a longer readout, \dword{snb} is handled
    symmetric to normal data.
    Each detector module has specialized implementation of some of
    these high level components, particularly toward the upstream
    front-end as described in the text. 
    The grayed boxes are not in the \dword{daq} scope.
  }
  \includegraphics[width=\textwidth{}, trim={1cm 0 1cm 0},clip]{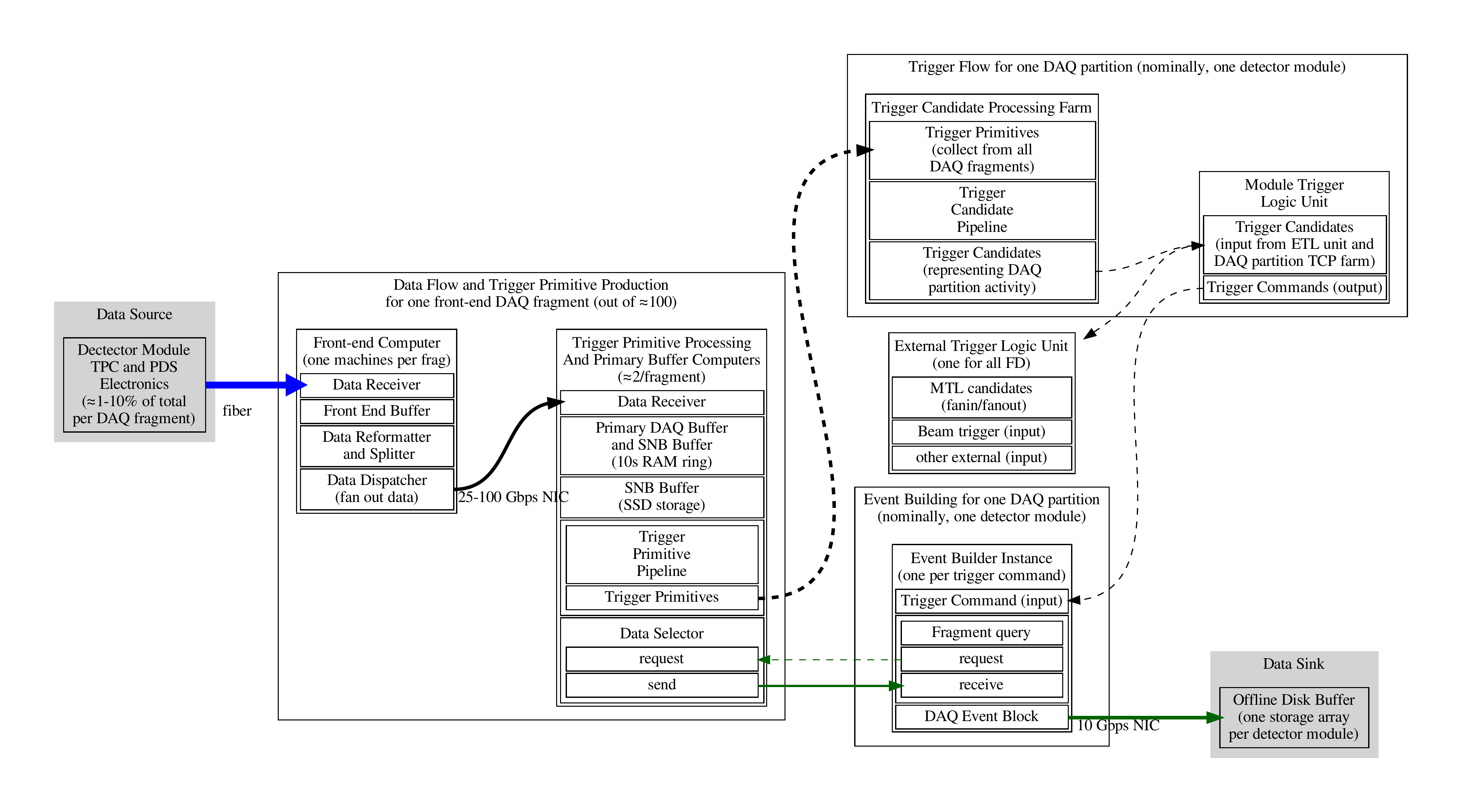}%
\end{dunefigure}

Currently, two major variations for the DUNE \dword{daq} are under consideration. 
The eventual goal is to reduce this to a single high-level design
which will service both \single and \dual \dwords{detmodule} and be reasonably
expected to support the third and forth modules to come.
The first design, designated in this proposal as \textit{nominal}, is
illustrated in a high-level way in terms of its data and trigger flow
in Figure~\ref{fig:daq-overview}. 
The second design, designated as the alternate, is similarly
illustrated in Figure~\ref{fig:daq-overview-alt}. 
The two variants differ largely at their \dwords{fe} in terms of the
order in which they buffer the data received from the \dword{detmodule} 
electronics and use it to form \dwords{trigprimitive}. 
They also differ in how they treat triggering and data flow due to a
potential \dword{snb}. 
As their \dwords{fe} are also sensitive to differences between the
\dword{detmodule} electronics, this further variation for each general
design is described below in Sections~\ref{sec:fd-daq-fero}
and~\ref{sec:fd-daq-fetp} for the \dword{detmodule}  specific to this volume.

At this general high level, the two designs are outlined. 
For both, the diagrams are 
centered on one \dword{daqfrag}
\dword{fe}, which is a portion of the entire \dword{daqpart} servicing a
\dword{detmodule} that has one \dword{fec} accepting about
\numrange{10}{20}~\si{\Gbps} of data (uncompressed rate) from some integral
number of \dwords{detunit}. 
Each of the participating \dwords{daqfrag} do the following: 
\begin{itemize}
\item Accept TPC and \dword{pds} data from the \dwords{detunit} associated with the \dword{daqfrag}.
\item Produce and emit a stream of per-channel \dwords{trigprimitive}.
\item Buffer the full data stream long enough for the \dword{trigdecision} to complete (at least \snbpretime as driven by \dword{snb} requirements).
\item Accept data selection requests and return corresponding \dwords{datafrag}.
\end{itemize}

All participating \dwords{daqfrag} in the particular \dword{daqpart}
(i.e., the \dword{daq} instance) servicing a portion of one \dword{detmodule}
communicate with one trigger processing and event building system.
The trigger processing system must:
\begin{itemize}
\item Receive the stream of per-channel \dwords{trigprimitive} from all \dwords{daqfrag}.
\item Correlate the primitives in time and spatially (across channels), and otherwise use them to form higher-level \dwords{trigcandidate}.
\item Exchange \dwords{trigcandidate} with the \dword{etl}.
\item From them form \dwords{trigcommand}, each of which describes a
  portion of the data in time and a channel to be read out, such that
  no two trigger commands overlap.
\item Dispatch these commands as required (in general to the \dword{eb}).
\end{itemize}
The event building system is responsible for performing the following actions:
\begin{itemize}
\item Accept a trigger command and allocate one \dword{eb} instance to dispatch it.
\item 
Interpret and execute the command by making
  data selection requests to referenced \dwords{daqfrag}.
\item 
Accept the returned \dword{datafrag} from each
  \dword{daqfrag} and combine them into a \dword{rawevent}.
\item Write the result to the \dword{diskbuffer}, which is the boundary
  shared with DUNE offline computing.
\end{itemize}

The nominal and the alternate \dword{daq} designs differ largely in where
the \dword{trigprimitive} and \dword{snb} buffering exist. 
The \textit{nominal} design places these functions in machines
comprising a \dword{daqfer}, which is upstream of the \dword{fec}. 
This then requires the \dword{snb} data and trigger handling to be
different than that for normal (non-\dword{snb}) data. 
When a \dword{snb} trigger command is raised it is forwarded to the
\dword{daqoob} which sends it down to the \dwords{daqfer}. 
After the \dword{snb} data is dumped to \dwords{ssd} it is
``trickled'' out via a path separate from the normal data to the
\dword{diskbuffer}. 
The \textit{alternate} design, on the other hand, places these
functions downstream of the \dword{fec} in trigger processing and data
buffering nodes.
The RAM of these nodes is used to provide the primary \dword{daq} buffer for
normal triggering as well as the deeper buffers needed for \dword{snb}. 
This design handles the \dword{snb} data somewhat
symmetrically with normal data. 
When an \dword{eb} makes a request for \dword{snb} data, it differs
only in its duration, spanning tens of seconds of instead just a few
milliseconds. 
The \dword{fe} buffering nodes, instead of directly attempting to
return the full \dword{sbn} data immediately, streams it to local
\dword{ssd} storage. 
From that storage, the data is sent to the \dword{eb} as low
priority (i.e., also trickled out).
Since the \dword{mtl} ensures no overlapping commands, the buffer
nodes may service subsequent requests from post-dump data that is 
in the RAM buffer.
Since each trigger command is handled by an individual \dword{eb}
instance, the trickle proceeds asynchronously with respect to any subsequent
trigger command handled by another \dword{eb} instances.

Further description of 
these designs
is
given in Section~\ref{sec:fd-daq-design}.

The most critical requirements for the DUNE \dword{fd} \dword{daq} 
are summarized in Table~\ref{tab:daqrequirements}.

\begin{dunetable}[Important requirements on the \dword{daq} system design]
{p{0.2\textwidth}p{0.6\textwidth}}
{tab:daqrequirements}
{Important requirements on the \dword{daq} system design}   
Requirement  & Description \\ \toprowrule
Scalability & The DUNE \dword{fd} \dword{daq} shall be capable of receiving and
buffering the full raw data from all four \dwords{detmodule} \\ \colhline 
Zero deadtime & The DUNE \dword{fd} \dword{daq} shall operate without deadtime under
\textit{normal} operating conditions \\ \colhline
Triggering & The DUNE \dword{fd} \dword{daq} shall provide full-detector triggering
functionality as well as self-triggering
functionality; the data selection shall maintain high efficiency to
physics events while operating within a total bandwidth of \offsitepbpy
for all operating \dwords{detmodule} \\ \colhline
Synchronization & The DUNE \dword{fd} \dword{daq} shall provide synchronization of
different \dwords{detmodule} to within \SI{1}{$\mu$s}, and of different subsystems
within a module to within \SI{10}{ns}\\ 
\end{dunetable}

The input bandwidth and processing needs of the \dword{daq} are expected to be
dominated by the rate of data produced by the TPC system of each
\dword{detmodule}.
These rates vary between the modules and their estimations are summarized in
Table~\ref{tab:daq-input-bandwidth}.
\begin{dunetable} [Pre-trigger data rates from the  \dword{fd} TPCs and
  into \dword{daq} front end.]
  {lll} {tab:daq-input-bandwidth} {The parameters governing the
    pre-trigger data rate from units of each \dword{detmodule} TPC
    \dwords{ce} and the aggregate throughput into the \dwords{fec} of
    the \dword{daq} \dwords{daqfrag}. 
    Compression is an estimate and will be reduced if excess noise is
    introduced.  
  }
  Parameter & \dlong{sp} & \dlong{dp} \\
  \colhline
  TPC unit & \dword{apa} & \dword{cro} crate \\ \colhline
  Unit multiplicity & \num{150} & \num{240} \\ \colhline
  Channels per unit & \num{2560} (\num{800} collection) & \num{640} (all collection) \\ 
  ADC sampling & \SI{2}{\MHz} & \SI{2.5}{\MHz} \\
  ADC resolution & \num{12}\,bit & \num{12}\,bit \\ \specialrule{1.5pt}{1pt}{1pt}
  Aggregate from \dword{ce} & \SI{1440}{\GB/\s} & \SI{576}{\GB/\s} \\
  Aggregate with compression & \SI{288}{\GB/\s} (5$\times$) & \SI{58}{\GB/\s} (10$\times$)  \\
  \colhline
\end{dunetable}

The ultimate limit on the output data rate of the DUNE \dword{fd} \dword{daq} is
expected to be provided by the available bandwidth to the tape,
disk and processing capacity of \fnal. 
An ample guideline has been established that places this limit at
about \offsitepbpy or \offsitegbps.
Extrapolating to four \dwords{detmodule}, this requires a \dword{daq} data reduction factor of almost four orders of magnitude. 
This is achieved through a simple self-triggered readout strategy.

An overestimate of the annual triggered but uncompressed data volume
for one \nominalmodsize  \dword{detmodule} is summarized in
Table~\ref{tab:daq-data-rates}. 
It assumes a very generous and simple trigger scheme whereby the data
from the entire \dword{detmodule} is saved for a period longer than
two drift times around the trigger time.
This essentially removes any selection bias at the cost of
recording a substantial amount of data that will simply contain noise.
Detailed trigger efficiency studies still remain to be performed. 
Initial understanding indicates that trigger efficiency should be near
\SI{100}{\%} for localized energy depositions of at least \SI{10}{\MeV}. 
Sub-\si{\MeV} signals can be ascertained from noise in existing \lartpc{}s
so the effective trigger threshold may be even lower with high
efficiency. 
Of course, data rates rise quickly when the threshold drops into the
range of an \si{\MeV}. 
Additional simulation and use of early data will be used to better
optimize this threshold.

\begin{dunetable} [Uncompressed data rates for one \dword{spmod}.]
  {p{0.30\textwidth}p{0.13\textwidth}p{0.4\textwidth}}
  {tab:daq-data-rates} {Anticipated annual, uncompressed data rates
    for a single \dword{spmod}. The rates for normal (non-\dword{snb} triggers)
    assume a readout window of \SI{5.4}{\ms}. 
    For planning purposes these rates are assumed to apply to a  
    \dword{dpmod} as well, which has a longer readout time but fewer channels. 
    In reality, application of lossless compression is expected
    to provide as much as a $5\times$ reduction in data volume for the \dword{spmod}
    and as much as $10\times$ for the \dword{dpmod}.}   
  Event Type  & Data Volume \si{\PB/year} & Assumptions \\ \toprowrule
  Beam interactions & \num{0.03} & \num{800} beam and \num{800} dirt muons; \SI{10}{\MeV} threshold in coincidence with beam time; include cosmics\\ \colhline
  Cosmics and atmospherics & \num{10} &  \SI{10}{\MeV} threshold, anti-coincident with beam time \\ \colhline
	 Front-end calibration & \num{0.2} & Four calibration runs per year, \num{100} measurements per point \\ \colhline
 Radioactive source calibration & \num{0.1} & Source rate $\le$\SI{10}{Hz}; single fragment readout; lossless readout \\ \colhline
 Laser calibration & \num{0.2} & \num{1e6} total laser pulses, lossy readout \\ \colhline
 Supernova candidates & \num{0.5} & \num{30} seconds full readout, average once per month \\ \colhline
 Random triggers & \num{0.06} & \num{45} per day\\ \colhline
 Trigger primitives & $\le$\num{6} &  All three wire planes; \num{12} bits per primitive word; \num{4} primitive quantities; $^{39}$Ar-dominated\\ \colhline
\end{dunetable}

The data volume estimates also assume that any excess noise beyond
what is expected due to intrinsic electronics noise will not lead to
an increase in trigger rates. 
If, for example, excess noise occurs such that it frequently mimics
more than about \SI{10}{\MeV} of localized ionization, this would
lead to an increase in various types of triggers and subsequently more
data.
However, at the same time, these estimates do not take into account
that some amount of lossless compression of the TPC data will be
achieved. 
In the absence of excess noise it is expected that a compression
factor of at least $5\times$ can be achieved with the \single data and up
to $10\times$ may be achieved with the \dual data, although the actual  
factor achieved will ultimately depend on
the level of excess noise experienced in each \dword{detmodule}. 
Studies using data from the DUNE \dword{35t} and early \microboone
running have shown that a compression factor of at least $4\times$ can
be expected even in the case of rather high levels of excess noise.

One category that will be particularly sensitive to excess noise is the trigger primitives. 
As discussed further in Section~\ref{sec:fd-daq-fetp}, their primary
intended use is as transient objects produced and consumed locally
and directly by the \dword{daq} in the \dword{trigdecision} process. 
However, as their production is expected to be dominated by $^{39}$Ar
decays (absent excess noise) they may carry information that proves
very useful for calibration purposes. 
Future studies with simulation and with early data will determine 
 the most feasible methods to exploit this data. 
These may include committing all or a portion to permanent storage or
potentially developing processes that can summarize their data while
still retaining information salient to calibration.

Finally, it is important to note that early data will be used to
evaluate other selection criteria. 
It is expected that efficient and bias-free selections can be
developed and validated that save a subset of the entire
\dword{detmodule} for any given trigger type. 
For example, a cosmic-muon trigger command for a \dword{spmod} will indicate which \dwords{apa}
contributed to its formation (i.e., which ones had local ionization activity). 
This command can then direct reading out these \dwords{apa}, possibly also
including their neighbors, while discarding the data from all other
\dwords{apa}. 
This may reduce the estimated \SI{10}{\PB/year} for cosmics and
atmospherics by an order of magnitude. 
A similar advanced scheme can be applied to the 
\dword{dpmod} by retaining data for the given readout window from
only the subset of \dword{cro} crates (and again, potentially their
nearest neighbors) that contributed to the formation of the given
trigger.

\subsection{Scope}
\label{sec:fd-daq-scope}


The nominal scope of the \dword{daq} system is illustrated in
Figure~\ref{fig:daq-overview} by the white boxes. 
It includes the continued procurement of materials for, and the
fabrication, testing, delivery and installation of the following
systems:

\begin{itemize}
\item \dword{fe} readout (nominal design) or trigger farm (alternate
  design) hardware and firmware or software development for
  \dword{trigprimitive} generation.
\item \dword{fe} computing for hosting of \dword{daqdr}, \dword{daqbuf} and \dword{daqds}.
\item Back-end computing for hosting \dword{mtl}, \dword{eb} and the \dword{daqoob} processes.
\item External trigger logic and its host computing.
\item Algorithms to generate trigger commands that perform data selection.
\item Timing distribution system.
\item \dword{daq} data handling software including that for receiving and building 
  events.
\item The \dword{om} of \dword{daq} performance and data content.
\item Run control software, configuration database, and user interface
\item Rack infrastructure in the \dword{cuc} for readout
  electronics, \dword{fe} computing, timing distribution, and data
  selection.
\item Rack infrastructure on surface at \surf for back-end computing.
\end{itemize}

\section{DAQ Design}
\label{sec:fd-daq-design}
\metainfo{16 Pages.  This section is mostly DP/SP shared but with some subsections broken out.  This file is \texttt{far-detector-dual-phase/chapter-fddp-daq/design.tex}}

\subsection{Overview}
\label{sec:fd-daq-overview}

The design for the \dword{daq} has been driven by finding a cost-effective solution that satisfies the requirements. Several design
choices have 
been made and two major variations remain to
be studied. 
From a hardware perspective, the \dword{daq} design follows a standard HEP
experiment design, with customized hardware at the upstream, feeding
and funnelling (merging) and moving the data into computers. 
Once the data and triggering information are in computers, a
considerable degree of flexibility is available;  the processing
proceeds with a pipelined sequence of software operations, involving
both parallel processing on multi-core computers and switched
networks. The flexibility allows the procurement of computers and
networking to be done late in the delivery cycle of the DUNE
\dwords{detmodule}, to benefit from increased capability of commercial devices
and falling prices.

Since DUNE will operate over a number of decades, the \dword{daq} has been
designed with upgradability in mind. 
With the fall in cost of serial links, a guiding principle is to
include enough output bandwidth to allow all the data to be passed
downstream of the custom hardware.
This allows the possibility for a future very-fast farm of computing
elements to accommodate new ideas in how to collect the DUNE data. 
The high output bandwidth also gives a risk mitigation path in case
the noise levels in a part of the detector are higher than specified
and higher than tolerable by the baseline trigger decision mechanism;
it will allow additional data processing infrastructure to be added
(at additional cost).

Digital data will be collected from the TPC and \dword{pd}
readout electronics of the \single and \dual
\dwords{detmodule}. 
These categories of data sources are viewed as essentially four types
of \dwords{submodule} within the \dword{daq} and follow the same overall
data collection scheme as shown for the nominal design in
Figure~\ref{fig:daq-overview} and for the alternate design in
Figure~\ref{fig:daq-overview-alt}. 
The readout is arranged to allow making a \dword{trigdecision} 
in a hierarchical manner. 
Initial inputs are formed at the channel level, then combined at the
\dword{detunit} level and again 
combined at the
\dword{detmodule} level.
In addition, the \dword{trigdecision} process combines 
information at this level that may come from the other \dwords{detmodule} as well as
information from sources external to the \dword{daq}. 
This hierarchical structure in forming and consuming triggers 
allows safeguards to be developed so that any problems in one cavern or
in one \dword{detunit} of one \dword{detmodule} need not overwhelm the
entire \dword{daq}.
It also allows a \dword{snb} to be recorded in all
operational parts of the detector while others may be down for
calibration or maintenance.

Generally speaking, the \dword{daq} consists of data flow and trigger flow.
The trigger flow involved in self-triggering originates from
processing a portion of the data flow. 
The trigger flow is then consumed back by the \dword{daq} in order to govern
what portion of the data flow is finally written out to permanent
storage. 
The nominal and alternate designs differ in where in the data flow
the trigger flow originates. 

In both designs, a single \dword{daqfrag} associates an integral
number of \dwords{detunit} with one \dfirst{fec}.
This fragment forms one conceptual unit of the \dword{fe} \dword{daq}.
The processing on a \dword{fec} is kept minimal such that each has a
throughput limited by I/O bandwidth. 
The recently released PCIe v4 doubles the bandwidth from the prior
version and thus we assume that $\approx$\SI{20}{\GB/\s} throughput (out of
a theoretical \SI{32}{\GB/\s} max) can be achieved based on tests
using PCIe v3.
In principle then, this allows one \dword{fec} to accept the data
from: two (if uncompressed) or ten (if $5\times$ compressed) of the
\num{150} \single \dwords{apa}, ten of the \num{240} \dual \dword{cro} crates
given their nominal $10\times$ compression or the uncompressed data
from all five \dword{dp} \dword{lro} crates.

In the nominal design, the data enters the \dword{daq} via the fragment's
\dfirst{daqfer} component.
In the \dword{sp} the \dword{daqfer} consists of eight \dwords{rce}
and in the \dword{dp} it consists of a number of \dword{bow}
computers, (see Section~\ref{sec:fd-daq-fero} in each respective \dword{detmodule} volume).
The \dword{daqfer} is responsible for accepting that data and from it
producing channel level \dwords{trigprimitive}.
It is also responsible for forwarding compressed data and the
primitives to the \dfirst{daqdr} in the corresponding \dword{fec}.
The \dword{daqfer} is also responsible for supplying transient memory
(RAM) and non-volatile buffer in the form of \dword{ssd} sufficient
for \dword{snb} triggering and readout.
The \dword{daqdr} accepts the full data stream and transfers it to the
\dlong{daqbuf} of its \dword{daqfrag}. 
There it is held awaiting a query from the \dfirst{eb}. 
When the \dword{eb} receives a \dword{trigcommand} it uses the
included information to query all appropriate \dwords{daqds} and from
their returned \dwords{datafrag} an \dword{rawevent} is built and
written to file on the \dword{diskbuffer}. 
From there the data becomes responsibility of the offline group to
transfer to \fnal for permanent storage and further processing.

In the alternate design, the data is accepted directly by the
\dfirst{daqdr} in a \dword{fec} from the detector electronics
for the particular \dword{detmodule}.
The data then flows into the \dword{daqbuf} and the portion required
for forming trigger primitives is dispatched to the trigger computers
of the fragment for the production of \dwords{trigprimitive}.
Current \dword{ssd} technology may allow \dword{ssd} to be directly mounted to the
\dword{fec} to provide for the \dword{snb} dump buffer. 
Another solution, which puts less pressure on write throughput, is to
distribute the \dword{ssd} for the \dword{snb} dumps to the trigger computers. 
In order to supply enough CPU for trigger primitive pipelines it is
expected that at least two hosts per \dword{fec} will be needed.
While their CPUs are busy finding trigger primitives, their I/O
bandwidth will be relatively unused and thus they provide synergistic,
cost-effective hosting for the \dword{ssd}s.

Regardless of where the \dwords{trigprimitive} are produced in either
the nominal or alternate design, they are further processed at the
\dword{daqfrag} level to produce \dwords{trigcandidate}. 
At this level, they represent possible activity localized in time and
by channel to a portion of the overall \dword{detmodule}.
The \dwords{trigcandidate} emitted by all \dwords{daqfrag} are sent to
the \dfirst{mtl} associated with the \dword{daqpart}.
There, they are time ordered and otherwise processed to form
\dwords{trigcommand}.
At this level they represent activity localized across the
\dword{detmodule} and over some period of time.

The \dword{daqpart} (or \dword{daq} instance) just introduced is the cohesive
collection of \dword{daq} parts. 
One \dword{daqpart} operates essentially independently from any other,
and there is typically one per \dword{detmodule}. 
In some cases multiple \dwords{daqpart} may operate simultaneously in
a \dword{detmodule}, such as when some fraction of \dwords{detunit}
are undergoing isolated testing or calibration.

Each \dword{trigcommand} is consumed by a single \dword{eb} instance
in order to query back to the \dwords{daqfrag} of its \dword{daqpart}
as described above.
In addition, the \dword{mtl} of one module is exchanging messages in
the form of \dwords{trigcandidate} with the others. 
For example, one module may raise a local \dword{snb}
\dword{trigcandidate} and forward it to all other modules.
Each module is also emitting candidates to sinks and accepting them
from sources of external trigger information.

The exact implementation of some of these high-level functions,
particularly those near the \dword{fe}, depends on the particular
\dword{detmodule}. 
The required specialization and in general, more implementation-level
details are described in the following sections.
Subsequent description proceeds toward the \dword{daq} back end including
processes handling dataflow, triggering, event building and data
selection.

\subsection{Front-end Readout and Buffering}
\label{sec:fd-daq-fero}

\metainfo{Giles Barr \& Giovanna Miotto \& Brett Viren, this is DP-specific.  This file is \texttt{far-detector-dual-phase/chapter-fddp-daq/design-fero.tex}}

\begin{dunefigure}[\dual \dword{fe} \dword{daq} fragment]{fig:daq-readout-buffering-baseline}
  {Illustration of data (solid arrows) and trigger (dashed) flow for
    two \dual \dword{fe} \dword{daq} fragments. 
    One servicing eight of \num{240} \dword{cro} crates and the other servicing all
    five \dword{lro} crates.  
    Black arrows indicate nominal flow and red indicate special flow
    for handling of potential \dword{snb}.} 
  \includegraphics[width=0.95\textwidth,clip,trim=1cm 0 1cm 0]{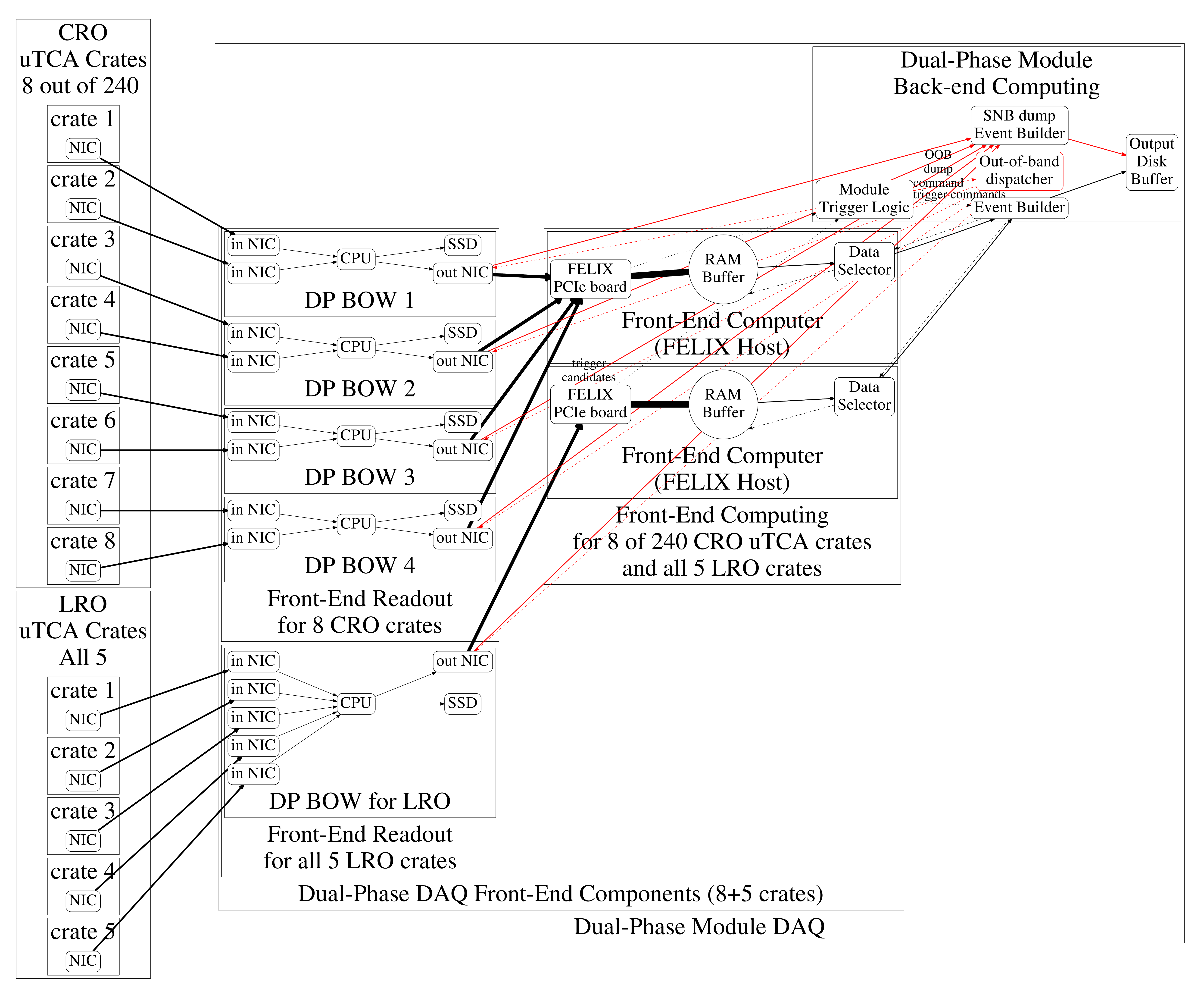}%
\end{dunefigure}

Figurere~\ref{fig:daq-readout-buffering-baseline} illustrates
\dual-specific implementation of the nominal, generic \dword{daq} front-end
\dword{daqfrag} design shown in Figure~\ref{fig:daq-overview}. 
The \dword{cro} crates deliver data on \dword{udp} to \dword{bow} computers
which implement the \dword{daqfer} duties.
The \dword{cro} data is delivered to the \dword{daq} \dword{bow} computers with lossless
compression applied and so before the \dwords{trigprimitive} can be
produced the data must be decompressed. 

In order to save on cost, power, space, cooling, etc., a number of \dword{cro}
crate data streams can be aggregated into one \dword{fec}. 
The \dword{cro} data stream is sent using \dword{udp} which is not expected to provide
reliable transport of high-throughput data if a network switch
intervenes. 
Thus each \dword{cro} requires a corresponding \SI{10}{\Gbps} \dword{nic} in the receiving
\dword{bow} computer.
With the expected 10$\times$ lossless compression factor the nominal
output from each \dword{cro} crate will be \SI{2}{Gbps}. 
If noise levels are higher than expected the throughput will increase,
however even very noisy data should be compressible enough to fit into
the \SI{10}{Gbps} bandwidth. 
Noise will be better understood from \dword{pddp} data, and studies
are needed to optimize the number of \dword{cro} crates per \dword{bow}
computer.

After processing, the input data is sent out, along with the
corresponding primitives to a \dword{felix} board in a \dword{fec}. 
Similar to the argument above about noise, bandwidth and \dword{bow} computer
multiplicity, the number of \dword{bow} data streams that can be aggregated
into each \dword{felix} board requires additional study. 
The current generation of \dword{felix} boards have been tested with a
throughput to system RAM of \SI{10}{\GB/\s}. 
The next generation is expected to at least double that. 
With the caveat that these tests did not receive data on \dword{udp}, a nominal
\SI{20}{\GB/\s} throughput is assumed possible for the next-generation
PCIe v4 boards. 
Given this assumption and that the expected noise levels are achieved,
then based solely on bandwidth, as many as \num{80} \dword{cro} data streams might be
aggregated into a single next-generation \dword{felix} board. 
Figure~\ref{fig:daq-readout-buffering-baseline} indicates the
aggregation of eight \dword{cro} streams, which represents the multiplicity
required to deal with very high noise. 
Future studies are needed to determine 
which of these two extremes to favor for optimization of the design,
but the basic design itself is
fairly elastic between them.

The \dword{lro} crates are nominally not involved in self-triggering although
the data from all five \dword{lro} crates are assumed to flow through a \dword{lro} \dword{bow}. 
The data are then sent to a
single \dword{felix} board. 
Given the expected data rates, that \dword{felix} board must ingest about
\SI{25}{\GB/\s}.

The second duty of the \dword{daqfer} in the nominal design is to
provide non-volatile storage for receiving full-stream data dumps when
an \dword{snb} dump trigger command is issued. 
With today's technology, individual \dwords{ssd} can write at about \SI{2.5}{GB/s}. 
Up to four \dwords{ssd} have been placed on a \num{1}-lane PCIe v3 board and have
achieved about three times this speed. 
This would allow an individual \dword{bow} computer to aggregate \numrange{10}{30} \dword{cro}
data streams before the \dwords{ssd} become a bottleneck. 
The five \dword{lro} data streams, each producing \SI{5}{Gbps}, could together be
streamed to a two modern \dwords{ssd}. 
The \dword{bow} computers must also have sufficient RAM to hold
pre-\dword{snb}-trigger data of about \num{10} seconds.

The alternate design (not diagrammed), which corresponds to
Figure~\ref{fig:daq-overview-alt}, deletes the layer containing \dword{bow}
computers and directly connects the \dword{udp} streams from the \dword{cro} and \dword{lro}
crates to the \dword{felix} boards in the \dwords{fec}. 
The \dword{cro} (compressed) data is buffered on the \dword{felix} host computer RAM
and distributed to the trigger farm for decompression and \dword{trigprimitive} and \dword{trigcandidate} processing. 
The remaining part of the \dword{trigdecision} is as in the nominal
design except that the \dword{snb} dump data stream is handled symmetrically
with the normal triggered data.

\subsection{Front-end Trigger Primitive Generation}
\label{sec:fd-daq-fetp}

In the nominal design, the \dword{bow} computers decompress the
\dword{cro} data stream and execute algorithms to search for per-channel
localized activity above some threshold based on a recent measure of
the noise. 
These \dwords{trigprimitive} are then sent out along with the original,
compressed \dword{cro} data to the \dword{fec} associated with the \dword{bow}.

Initial studies have shown some promise that this type of \dwords{trigprimitive} pipeline can be implemented on commodity CPUs and keep up
with the data. 
The studies made use of \dword{sp} signal and noise simulation but
given the relatively higher signal-to-noise ratio expected in the \dword{dpmod}
data, coupled with fewer total channels, these studies should
be applicable and even better performance may be expected. 
With that said, more realistic studies using \dword{dp}-specific
simuilations and with \dword{pddp} data are needed.

Most of the same technical triggering issues apply in the nominal and
the alternate designs (generically diagrammed in
Figure~\ref{fig:daq-overview-alt}), as both call for deploying \dword{trigprimitive} pipelines on commodity CPUs. 
The main difference is that the \dword{bow} hosts become trigger farm
hosts. 
They move from being upstream and inline of the data flow before the
\dword{fec} to being downstream and receiving the data after it has
been buffered. 
Their \dwords{trigcandidate} no longer have to be inserted into the
stream and stripped out by the \dword{fec} and instead are sent
directly to the \dword{mtl}. 
Additional study is required to understand the multiplicity of trigger
processing computers and how that might scale as one considers
different multiplicities of \dword{cro} \dword{utca} crates with respect to each
\dword{fec}.

In both designs, the \dword{lro} data is currently not considered for
triggering as the \dword{cro} triggering is expected to be more
efficient. However, using light information for triggering has not yet
been ruled out. 
Future studies may indicate additional benefit to including
\dword{lro} information in triggering and both the nominal and
alternate design can elastically accommodate this increased scope
although possibly at the cost of either more \dword{bow} or trigger
processor computers.
As described elsewhere, the \dword{lro} data flow is handled by
separate \dwords{daqfrag} from those that service the \dwords{cro}. 
Thus, if any \dwords{trigprimitive} and \dwords{trigcandidate} are to
be formed from \dword{lro} data they would come from these separate
fragments and be combined in the \dword{mtl} as peers of the
\dword{cro} and external candidates.

\subsection{Dataflow, Trigger and Event Builder}
\label{sec:fd-daq-hlt}

\metainfo{Giles Barr \& Josh Klein \& Giovanna Miotto \& Kurt Biery \& Brett Viren.  This is a DP/SP shared section.  It's file is \texttt{far-detector-generic/chapter-generic-daq/design-flow.tex}}

In the general data and trigger flow diagrams for the nominal
(Figure~\ref{fig:daq-overview}) and alternate
(Figure~\ref{fig:daq-overview-alt}) designs, the dataflow, trigger and
event builder functions take as input data from the \dword{detmodule}
electronics and culminate in files deposited to the \dword{diskbuffer} for
transfer to permanent storage by offline computing processes.  
The continuous, uncompressed data rate of the input from one
\dword{detmodule} is on the order of \SI{1}{\TB/\s}. 
The final output data rate, for all \dwords{detmodule} operating at
any given time is approximately limited to \offsitegbyteps. 

To accept this high data-inflow rate and to apply the substantial
processing needed to achieve the required reduction factor, which is on the 
order of \num{1000}, the \dword{daq} follows a distributed design.
The units of distribution for the front end of the \dword{daq} must match up
with natural units of the \dword{detmodule} providing the data. 
This unit is called the \dword{daqfrag} and each accepts input at a
rate of about \numrange{10}{20}\,\si{\GB/\s}. 
The exact choice maps to some integral number of physical
\dword{detmodule} units (e.g., \dword{sp} \dwords{apa} or \dword{dp}
\dwords{cro} and \dwords{lro}).

As described in the previous sections, the nominal and alternate
designs differ essentially in the order and manner in which the
\dword{snb} buffering occurs and the \dwords{trigprimitive} are
formed. 
The overall data flow, higher level triggering and building of
\textit{event} data blocks for final writing are conceptually very similar.
This processing begins with the data being received by the
\dword{felix} PCIe board hosted in the \dword{fec}. 
The \dword{felix} board performs a DMA transfer of the data into the
\dword{daqbuf} for the \dword{daqfrag}, which resides in the
\dword{fec} host system RAM.  
This buffer is sized to hold ten seconds of data assuming the maximum 
uncompressed input rate associated with the fragment.
While data is being written to the buffer, a delayed portion is
also being read in order to dispatch it for various purposes.
Any and all requests to further dispatch a subset of this data from
the \dword{daqbuf} must arrive within this buffer time.
In the nominal design, the only dispatching will be from a request
made by an \dword{eb} (described more below) upon receipt of a
\dword{trigcommand}. 
In the alternate design, a suitable fraction of the data is also
dispatched via high bandwidth (at least \numrange{25}{50}\,\si{\Gbps} simplex, less
if data is compressed at this stage) network connections to a trigger
farm so that \dwords{trigprimitive} may be formed. 
Whether the primitives are formed in this manner or extracted from the
stream sent by the \dword{daqfer} (as in the nominal design) these
trigger primitives from one \dword{daqfrag} are collectively sent for
further processing in order to be combined across channels and  to then 
produce \dwords{trigcandidate}. 
These are finally combined for one \dword{detmodule} in the
\dword{mtl}. 
It is in the \dword{mtl} where \dwords{trigcandidate} from additional
sources are also considered, as described in
section~\ref{sec:fd-daq-sel}.

In both the nominal and alternate designs the dispatch of data
initiated by normal (non-\dword{snb}) \dwords{trigcommand} is
identical. 
This dispatch, commonly termed \textit{event building} involves collection
of data spanning an identical and continuous period of time from
multiple \dwords{daqbuf} across the \dword{daq}.
As introduced above, each \dword{trigcommand} is consumed by an
\dword{eb} process. 
It  uses fragment address information in the \dword{trigcommand} to
query the \dword{daqds} process representing each referenced
\dword{daqfrag} and accepts the returned a \dword{datafrag}.
In the exceptional case that the delay of this request is so large
that the \dword{daqbuf} no longer contains the data, then an error
return is supplied and recorded by the \dword{eb} in place of the
lost data. 
Such failures lead to indicators displayed by the detector operation
monitoring system.
The \dword{eb} finally assembles all responses into a
\dword{rawevent} and writes it to file on the \dword{diskbuffer} where
it becomes the responsibility of DUNE offline computing.

The \dword{daqds} and \dword{eb} services are implemented using
 the general-purpose \fnal data acquisition framework \dword{artdaq} for
distributed data combination and processing. 
It is designed to exploit the parallelism that is possible with
modern multi-core and networked computers, and has been used in \dword{protodune} and other experiments.
The \dword{artdaq} framework is the principal architecture that will be used for the DUNE \dword{daq} back-end computing.
The authors of \dword{artdaq} have accommodated DUNE-specific 
requests for feature additions. Also, a number of libraries have been developed based on
existing parts of \dword{artdaq} used to handle incoming data from data
sources. 
It is likely that future DUNE extensions will be made by one of these two
routes.

Unlike the dispatch of data initiated by a normal \dwords{trigcommand},
a command formed to indicate the possibility of a \dword{snb} is
handled differently between the nominal and alternate designs. 
Such a command is interpreted to save all data from all channels
for a rather extended time of \snbtime starting from \snbpretime
before the time associated with the \dword{trigcommand}. 
As no data selection is being performed, given the required bandwidth, special buffering to nonvolatile storage, in the form of \dword{ssd}, is required.  
Today's technology supplies individual \dword{ssd} in the M.2 expansion card form factor,
which supports individual write speeds up to \SI{2.5}{\GB/\s}. 
The two designs differ as to the location of and data source for these
buffers.

In the nominal design, these \dwords{ssd} reside in the \dword{daqfer}
as described in Section~\ref{sec:fd-daq-fero}. 
In that location, due to larger granularity of computing units, the
data rate into any one \dword{ssd} is within the quoted write
bandwidth. 
However, and as shown in Figure~\ref{fig:daq-overview}, the data and
trigger flow for \dword{snb} in the nominal case takes a special
path. 
Instead of an \dword{eb} consuming the \dword{trigcommand} as
described above, it is sent to the \dfirst{daqoob}, which dispatches
it to each \dword{daqfer} unit hosting an \dword{ssd}. 
This component is used 
to immediately free up the \dword{mtl}
to continue to process normal triggers.
When the command is received, each host must begin to stream data from
its local RAM, supplying at least \snbpretime of buffer to the
\dword{ssd}, and continue until the full \snbtime has elapsed. 
While it is performing this dump it must continue to form
\dwords{trigprimitive} and pass them and the full data stream to the
connected \dword{fec}.

In the alternate design the same \dword{daqbuf} provides the
\snbpretime of pre-trigger \dword{snb} buffering. 
As in the nominal case, it must rely on fast, local \dword{ssd}
storage to sink the dump. 
Current \dword{ssd} technology allows four M.2 \dword{ssd} devices to
be hosted on a PCIe board. 
Initial benchmarks of this technology show that such a combination can
achieve \SI{7.5}{\GB/\s} write bandwidth, which is short of linear
scaling. 
To support the maximum of \SI{20}{\GB/\s}, three such boards would be
required.
The alternate design presents a synergy between the need to dump
high-rate data and the need to provide CPU to form the
\dwords{trigprimitive}. 
With current commodity computing hardware it is expected that each
\dword{fec} will need to be augmented with about two computers in the trigger
farm. 
These trigger processors will need to accept the entire \dual and
three-eighths of the \single data stream from their \dword{daqfrag}. 
If they instead accept the entire stream, they can also provide
RAM buffering and split up the data rate, which must be sunk to
\dword{ssd} buffers.

In both designs, the data dumped to \dword{ssd} may contain precious
information about a potential \dword{snb}. 
It must be extracted from the buffer, processed and either discarded
or saved to permanent storage. 
The requirements on these processes are not easy to determine.
The average period between actual \dwords{snb} to which DUNE is
sensitive is measured in decades. 
However, to maintain high efficiency for capturing such important
physics, the thresholds will be placed as low as feasible, limited
only by the ability to acquire, validate and (if validated) write out the
data to permanent storage. 
Notwithstanding, the (largely false positive) \dword{snb} trigger rate is
expected to be minuscule relative to normal triggers.
Understanding the exact rate requires more study, including using
early data, but for planning purposes it is simply assumed that one
whole-detector data dump will occur per month on average.
Using the \dword{spmod} as an example, and choosing the
nominal time span for the dump to be \snbtime{}, about \spsnbsize of
uncompressed data would result.
In the nominal \dword{sp} \dword{daq} design, this dump would be spread over
\num{600} \dword{ssd} units leading to \SI{75}{\GB} per \dword{ssd} per dump.
Thus, typical \dwords{ssd} offer storage to allow any given dump to be
held for at least one half year before it must be purged to assure
storage is available for subsequent dumps.
If every dump were to be sent to permanent storage, it would represent
a sustained \SI{0.14}{\Gbps} (per \dword{detmodule}), which is a small
perturbation on the bandwidth supplied throughout the \dword{daq} network. 
Saved to permanent storage this rate integrates to \SI{0.5}{\PB/\year}, 
which while substantial, is a minor fraction of the total data budget.
The size of each dump is still larger than is convenient to place into
a single file, so the \dword{snb} event-building will likely differ from
that for normal triggers in that the entire dump is not held in a
single \dword{rawevent}. 
Finally, it is important to qualify that these rates assume
uncompressed data. 
At the cost of additional processing elements, lossless compression
can be expected to reduce this data rate by \numrange{5}{10}\,$\times$ or
alternatively allow lower thresholds that lead to the same factor of 
more
dumps.
Additional study is required to optimize the costs against the expected
increase in sensitivity.


\subsection{Data Selection Algorithms}
\label{sec:fd-daq-sel}

\metainfo{Josh Klein \& Brett Viren.  This is a shared DP/SP section.  It's file is \texttt{far-detector-generic/chapter-generic-daq/design-sel.tex}}

Data selection follows a hierarchical design. 
It begins with forming \dword{detunit}-level \dwords{trigcandidate}
inside the \dword{daqfrag} \dword{fe} computing using channel-level
\dwords{trigprimitive}. 
These are then used to form \dword{detmodule} \dwords{trigcommand}
in the \dword{mtl}.
When executed, they lead to readout of a small subset of the total
data. 
In addition, \dwords{trigcandidate} are provided to the
\dword{mtl} from external sources such as the \dword{etl} in order to
indicate external events such as beam spills, or \dword{snb} candidates
detected by the other \dwords{detmodule}. 
In addition to supplying triggers to \dword{snews}, triggers from
\dword{snews} or other cosmological detector sources such as LIGO and
VIRGO can be accepted in order to possibly record low-energy or
dispersed activity that would not pass the self-triggering. 
The latency of arrival for these sources must be less than the nominal
\snbpretime buffers used to capture low-level early \dword{snb}
activity.
A \dword{hlt} may also be active within the \dword{mtl}. 
The hierarchical approach is natural from a design standpoint and 
it allows for vertical slice testing and running multiple
\dwords{daqpart} simultaneously during commissioning of the system or
when debugging of individual \dwords{detunit} is required.

As discussed in Sections~\ref{sec:fd-daq-fero}
and~\ref{sec:fd-daq-fetp}, \dwords{trigprimitive} are generated in
either in \dwords{daqfer} (in the nominal design) or in trigger
processing computers (in the alternate design). 
In both designs, and for both \dword{sp} and \dword{dp}
\dwords{detmodule}, only data from TPC collection channels (three-eighths of \single and all of \dual channels) feed
the self-triggering, as their waveforms directly supply a measure of
ionization activity without computationally costly signal processing.
The \dwords{trigprimitive} contains summary information for each 
channel, such as the time of any threshold-crossing pulse, its
integral charge, and time over threshold. 
A channel with an associated \dword{trigprimitive} is said to be
\textit{hit} for the time spanned by the primitive. 
Trigger primitives from one \dword{detunit} are then further processed
to produce a \dword{trigcandidate}. 
The candidate represents a cluster of hits across time and
channel, localized to the \dword{detunit}.
The candidates from all \dwords{daqfrag} are passed to the
\dword{mtl}.

The \dword{mtl} arbitrates between various trigger types, determines
trigger priority and ultimately the time range and detector coverage
for a \dword{trigcommand}, which it emits back to the \dwords{fec}.
The \dword{mtl} assures that no \dwords{trigcommand} are issued that 
overlap in time or in detector channel space.
It also may employ a \dword{hlt} to reduce or aggregate triggers into
fewer \dwords{trigcommand} so as to optimize the subsequent readout. 
For example, aggregating many small readouts into fewer but larger
ones may allow for more efficient processing.   This can be particularly
important during periods of high-rate activity due to e.g., various
backgrounds or instrumental effects.

When activity leads to the formation of a \dword{trigcommand} this
command is sent down to the \dwords{fec} instructing which slice of
time of its buffered data should be saved. 
The \dword{trigcommand} information is saved along with this data. 
At the start of DUNE data taking, it is anticipated that for any given
single-interaction trigger (a cosmic-ray track, for example), waveforms
from all channels in the \dword{detmodule} will be recorded over a one
\dword{readout window} (nominally, \spreadout for \dword{sp} and
\dpreadout for \dword{dp}, chosen to be two drift times  plus an
extra \SI{20}{\%}). 

Such an approach is clearly very generous in terms of the amount of
data saved, but it ensures that associated low-energy physics (such as
captures of neutrons produced by neutrino interactions or cosmic rays)
are recorded without any need to fine-tune \dword{detunit}-level
triggering, and does not depend on the noise environment across
\dwords{detunit}. 
In addition, the wide \dword{readout window} ensures that the data of
all associated activity is recorded.
As generous as it is, it is estimated that this \dword{readout window}
will not produce an unmanageable volume of data.
As shown in Table~\ref{tab:daq-data-rates}, the uncompressed selected
data from the \dword{spmod} will fill about half of the
nominal annual data budget. 
The longer \dual drift and its fewer channels will give approximately the
same data rate. 
However, once a modest amount of lossless compression is applied, the
nominal data budget can be met. 
Early running will allow experience to be gained and more advanced
data selection algorithms to be validated allowing the \dword{daq} to discard
the many \dwords{datafrag} in each trigger consistent
with just electronics noise. 
This has the potential for a reduction of at least another factor of ten.

Other trigger streams -- calibrations, random triggers, and prescales
of various trigger thresholds -- are also generated at the
\dword{detmodule} level, and filtering and compression can be applied
based upon the trigger stream. 
For example, a large fraction of random triggers may have \dword{zs}
applied to their waveforms, reducing the data volume substantially, as
the dominant data source for these will be $^{39}$Ar events.
Additional signal-processing can also be done on particular trigger
streams if needed and if the processing is available, such as fast
analyses of calibration data.

At the \dword{detmodule} level, a decision can also be made on whether
a series of interactions is consistent with an \dword{snb}. 
If the number of \dword{detunit}-level, low-energy
\dwords{trigcandidate} exceeds a threshold for the number of such
events in a given time, a trigger command is sent from the \dword{mtl}
back to the \dwords{daqfer}, which store up to \SI{10}{\s} of full
waveform data. 
That data is then streamed to non-volatile storage to allow for
subsequent analysis by the \dword{snb} working group, perhaps as an
automated process. 
If not rejected, it is sent out of the \dword{daq} to permanent offline
storage.

In addition, the \dword{mtl} passes \dwords{trigcandidate} up to a
detector-wide \dword{etl}, which among other functions, can decide
whether, integrated across all modules, enough \dwords{detunit} have
detected interactions to qualify as an \dword{snb}, even if within a
particular module the threshold is not exceeded. 
\Dwords{trigcandidate} from the \dword{etl} are passed to the
\dword{mtl} for dispatch to the \dwords{fec} (or \dwords{daqfer} in the
case of \dword{snb} dump commands in the nominal design). 
That is, to the \dword{mtl}, an \dword{externtrigger} looks like just
one more \textit{external} trigger input.

\Dword{detunit} level \dwords{trigcandidate} are generated within
the context of one \dword{daqfrag}, specifically in each \dword{fec}. 
The trigger decision is based on the number of nearby channels
hit in a given fragment within a time window (roughly \SI{100}{\micro\s}),
the total charge collected in these adjacent channels, and possibly the
union of time-over-threshold for the \dwords{trigprimitive} in the
collection plane.
Studies show that even for low-energy events (roughly
\SIrange{10}{20}{\MeV}) the reduction in radiological backgrounds is
extremely high with such criteria.
The highest-rate background, $^{39}$Ar, which has an overall rate of
\SI{10}{MBq} within a \nominalmodsize volume of argon, has an endpoint 
of \SI{500}{keV} and requires significant pileup in both space and time to get near
a \SI{10}{\MeV} threshold.
One important background source is $^{42}$Ar, which has a \SI{3.5}{MeV}
endpoint and an overall rate of \SI{1}{kBq}. 
$^{222}$Rn decays via a 
 \SI{5.5}{MeV} kinetic energy $\alpha$  and is
also an important source of background.
The radon decays to $^{218}$Po, which within a few minutes leads to a
\SI{6}{MeV} kinetic energy $\alpha$, and ultimately to a $^{214}$Bi daughter (many
minutes later), which has a $\beta$ decay with its endpoint near \SI{3.5}{MeV}  kinetic energy. 
The $\alpha$ ranges are short, resulting in charge being collected on one or two anode wires at most,
but the charge deposit can be large, and therefore the charge
threshold must be well above the $\alpha$ deposits plus any
pileup from $^{39}$Ar and noise.

At the level of one \dword{detunit}, two kinds of local
\dwords{trigcandidate} can be generated.
One is a high-energy trigger that indicates local ionization
activity corresponding to more than than \SI{10}{\MeV}. 
The per-channel thresholds on total charge and time-over-threshold
will be optimized to achieve at least \SI{50}{\%} efficiency at this energy
threshold, with efficiency increasing to \SI{100}{\%} via a turn-on curve
that ensures at least \SI{90}{\%} efficiency at \SI{20}{\MeV}. 
The second type of trigger candidate 
generated is for
low-energy events between \SI{5}{\MeV} and \SI{10}{\MeV}. 
In isolation, these candidates do not lead to formation of a
\dword{trigcommand}. 
Rather, at the \dword{detmodule} level they are combined, time
ordered and their aggregate rate compared against a threshold based on
fluctuations due to noise and backgrounds in order to form an
\dword{snb} \dword{trigcommand}.

The \dword{mtl} takes as input \dwords{trigcandidate} (both low-energy
and high-energy) from the participating \dwords{daqfrag}, as well as
\dword{externtrigger} sources, such as the \dword{etl}, which includes
global, detector-wide triggers, external trigger sources such as
\dword{snews}, and information about the time of a \fnal beam
spill. 
The \dword{mtl}  also generates \dwords{trigcommand} for internal
consumption, such as random triggers and calibration triggers (for
example, telling a laser system to fire at a prescribed time). 
The \dword{mtl} can also generate \dwords{trigcommand} from a
prescaled fraction of trigger types that otherwise do not generate
such commands on their own. 
For example, a prescaled fraction of single, low-energy trigger
commands could be allowed to generate a trigger command, even though
those candidates normally only result in a trigger command when
aggregated (i.e., as they would be for an \dword{snb}).

The \dword{mtl} is also responsible for checking candidate triggers
against the current \dword{rc} trigger mask: in some runs, for
example, we may decide that only random triggers are accepted, or that
certain \dwords{trigcandidate} streams should not be considered
because their \dwords{daqfrag} have been producing unreasonably large
rates in the recent past (such as may be due to noise spikes, flaky
hardware or buggy software).
In addition, the \dword{mtl} counts low-energy trigger candidates,
and based upon their number and distribution over a long time interval
(e.g., \SI{10}{\s}), decides to generate an \dword{snb} trigger command.
The trigger logic will be optimized to record the data due to at least
\SI{90}{\%} of all Milky Way supernovae, and studies of simple low-energy
trigger criteria show that a much higher efficiency can likely be
achieved.

The \dword{hlt} can also be applied at this level, particularly if
there are unexpectedly higher rates from instrumental or low-energy
backgrounds that require some level of reconstruction or pattern
recognition. 
An \dword{hlt} might also allow for efficiently triggering on
lower-energy single interactions, or allow for better sensitivity for
supernovae originating outside the Milky Way galaxy, by employing a weighting scheme to
individual \dwords{trigcandidate} -- higher-energy
\dwords{trigcandidate} receiving higher weights. 
Thus, for example, two \dwords{trigcandidate} consistent with
\SI{10}{\MeV} interactions in \SI{10}{\s} might be enough to create a
\dword{snb} candidate trigger, while a hundred \SI{5}{\MeV} trigger
candiates in \SI{10}{\s} might not.
Lastly, the \dword{hlt} can allow for dynamic thresholding; for
example, if a trigger appears to be due to a cosmic-ray muon, the
threshold for single interactions can be lowered (and possibly
prescaled) for a short time after that to identify spallation
products. 
In addition, the \dword{hlt} could allow for a dynamic threshold after
a \dword{snb}, to extend sensitivity beyond the \SI{10}{\s}
\dword{snb} \dword{readout window}, while not increasing the data
volume associated with \dword{snb} candidates linearly. 

All low-energy \dwords{trigcandidate} are also passed upwards to the
\dword{etl} so that they may be integrated across all \SI{10}{\kton}
\dwords{detmodule} in order to determine that a \dword{snb} may be
occurring. 
This approach increases the sensitivity to trigger on \dwords{snb} by
a factor of four (for \SI{40}{\kton}), thus extending the burst
sensitivity to a distance twice as far as for a single \nominalmodsize
\dword{detmodule}.

The \dword{mtl} is also responsible for including in the
\dword{trigcommand} a global timestamp built from its input
\dwords{trigcandidate}, and information on what type of trigger was
created. 
Information on \dwords{trigcandidate} is also kept, whether or not
they contribute to the formation of a \dword{trigcommand}. 
As described above, the \dword{readout window} for nominal
\dwords{trigcommand} (those other than for \dword{snb} candidates)
is somewhat more than two times the maximum drift time. 
Further, a nominal readout spans all channels in a \dword{detmodule}. 
The \dword{mtl} is also responsible for sending the trigger commands
that tell the \dwords{daqfer} to stream all data from the past
\snbpretime and for a total of \snbtime in hopes to catch
\dwords{snb}.
This command may be produced based on \dwords{trigcandidate} from
inside the \dword{mtl} itself or it may be produced based on an external
\dword{snb} \dword{trigcandidate} passed to the \dword{mtl} by the
\dword{etl}.

\subsection{Timing and Synchronization}
\label{sec:fd-daq-timing}

\metainfo{David Cussans \& Kostas Manolopoulos.  This is a \dual-specific section. The file is \texttt{far-detector-dual-phase/chapter-fddp-daq/design-timing.tex}}

The timing distribution is integrated into the TPC electronics design
and is described in section \ref{sec:fddp-tpc-elec-intfc-daq} and
\ref{sec:fddp-tpc-elec-wr} using a \dfirst{wr} network over
synchronous \SI{1}{\Gbps} Ethernet (SyncE), using Precision Time Protocol-version 2 (PTPV2) packets.

The synchronization between the \dword{fd} and the beam pulses from
the LBNF beamline is done by a single overall system for \single 
and \dual  and is described in the following section.

\subsubsection{Beam timing}
\label{sec:fd-daq-design-beamtiming}

The neutrino beam is produced at the Fermilab accelerator complex in
spills of \SI{10}{\us} duration. 
A \dword{sls} at the Far Detector site will locate the time periods in
the data when beam could be present, based on network packets received
from Fermilab containing predictions of the GPS-time of spills soon to
occur or absolute time stamps of recent spills. 
Experience from MINOS and \nova shows that this can provide beam
triggering with high reliability with some small fraction of late or
dropped packets.
To improve reliability further, the system outlined here contains an extra layer
of redundancy in the prediction process. 
Several stages of prediction based on recent spill behavior will be applied, aiming
for an accuracy of better than 10\% of a readout
time (sub-\si{\ms}) in time for the data to be selected from
the \dword{daq} buffers. 
Ultimately, an offline database will match the actual time of the
spill with the data, thus removing any reliance on real-time network
transfer for this crucial stage of the oscillation measurements. The
network transfer of spill-timing information is simply to ensure that a
correctly located and sufficiently wide window of data is considered
as beam data. This system is not required, and is not designed to
provide signals accurate enough to measure neutrino time-of-flight.

The precision to which the spill time can be predicted at \fnal
improves as the acceleration process of the protons producing the
spill in question advances.  The spills currently occur at intervals
of \SI{1.3}{\s}; the system will be designed to work with any interval, and
to be adaptable in case the sequence described here changes.  For
redundancy, three packets will be sent to the far detector for each
spill.  The first is approximately \SI{1.6}{\s} before the spill-time, which
is at the point where a \SI{15}{\Hz} booster cycle is selected; from this
point on, there will be a fixed number of booster cycles until the
neutrinos and the time is subject to a few ms of jitter.  The second
is about \SI{0.7}{\s} before the spill, at the point where the main injector
acceleration is no longer coupled to the booster timing; this is
governed by a crystal oscillator and so has a few \si{\us} of jitter.
The third will be at the so called `\texttt{\$74}' signal generated before the beamline kicker magnet fires
to direct the protons at the LBNF target; this doesn't improve the
timing at the Far Detector much, but serves as a cross check for
missing packets.  This system is enhanced compared to that of
MINOS-\nova, which only use a third of the above timing signals.  The
reason for the larger uncertainty in the time interval from \SI{1.6}{\s} to
\SI{0.7}{\s} is that the booster cycle time is synchronized to the
electricity supply company's \SI{60}{\Hz}, which has a variation of about
1\%.

Arrival-time monitoring information from a year of MINOS data-taking
was analyzed, and it was found that 97\% of packets arrived within
\SI{100}{\ms} of being sent and 99.88\% within \SI{300}{\ms}.

The \dword{sls} will therefore have estimators of the GPS-times of
future spills, and recent spills with associated data contained in the
\dwords{daqbuf}. These estimators will improve in precision as
more packets arrive.  The \dword{daq} will use data in a wider window than
usual, if, at the time the trigger decision has to be made, the
precision is lower 
due to missing or late packets.  From the
MINOS monitoting analysis, this 
expected to be very rare.

\subsection{Computing and Network Infrastructure}
\label{sec:fd-daq-infra}

\metainfo{Kurt Biery \& Babak Abi.  This is a DP/SP shared section.  It's file is \texttt{far-detector-generic/chapter-generic-daq/design-compnet.tex}}

The computing and network infrastructure that will be used in each
of the four \dwords{detmodule} is similar, if not identical.
It supports the buffering, data selection, event
building, and data flow functionality described
above, and it includes computing elements that consist of servers that:

\begin{itemize}
\item buffer the data until a \dword{trigdecision}
  is received;
\item host the software processes that
  build the data fragments from the relevant
  parts of the detector into complete events;\fixme{whole detector or module?}
\item host the processes that make the
  \dword{trigdecision};
\item host the data logging processes and
  the disk buffer where the data is written;
\item host the real-time \dlong{dqm} processing;
\item host the control and monitoring processes.
\end{itemize}

The network infrastructure that connects these computers has the following components:

\begin{itemize}
\item subnets for transferring triggered data from the buffer
  nodes to the event builder nodes; these need to
  connect underground and above-ground computers;
\item a control and monitoring subnet that connects all
  computers in the \dword{daq} system and all \dword{fe}
  electronics that support Ethernet communication; this
  sub-network must connect to underground and
  above-ground computers;
\item a subnet for transferring complete events from the
  event builder servers to the storage servers; this subnet
  is completely above-ground.
\end{itemize}

\subsection{Run Control and Monitoring}
\label{sec:fd-daq-tcm}

\metainfo{Giovanna Miotto \& Jingbo Wang.  THis is a DP/SP shared section.  The file is \texttt{far-detector-generic/chapter-generic-daq/design-ctrmon.tex}}


The online software constitutes the backbone of the DUNE \dword{daq}
system and provides control, configuration and monitoring of the data taking in
a uniform way.
It can be subdivided logically into four subsystems: the run control,
the management of the \dword{daq} and \dword{detmodule} electronics configuration, the monitoring, and the non-physics data archival.
Each of these subsystems has a distinct 
function, but their
implementation will share underlying technologies and tools.

In contrast to experiments in which data taking sessions, i.e., runs, are
naturally subdivided into time slots by external conditions (e.g., a
collider fill, a beam extraction period), the DUNE experiment aims to
take data continuously.
Therefore, a classic run control with a coherent state machine and a
predefined and concurrently configured number of active detector and
\dword{daq} elements does not seem adequate. 

The DUNE online software is thus structured according to the
architecture principle of loose coupling: each component has as
little knowledge as possible of other components.
While the granularity of the back-end \dword{daq} components may match the 
individual software processes, for the front-end \dword{daq} a minimum
granularity must be defined, balancing fault tolerance and
recovery capability against the requirement of data consistency.
The smallest independent component is called a \dword{daqfrag}, which
is made up of the \dwords{detunit} associated with a single
\dlong{fec}.
In the nominal design, this corresponds to two \dword{sp} \dwords{apa}
and about ten \dword{dp} \dword{cro} crates.

The concept of a \textit{run} represents a period of time in which the same
\dword{fe} elements are active or the same data selection criteria
are in effect (possibly with maximum lengths for offline processing
reasons). 
More than just orchestrating data taking, the run control
provides the mechanisms allowing \dword{daq} applications to publish
their availability, subscribe to information, and exchange messages. 
In addition, the online software provides a configuration service
for \dword{daq} elements to store their settings and a conditions
archive, keeping track of varying detector electronics settings and
status.

Another important aspect of the online software is the monitoring
service.
Monitoring can be subdivided into two main domains: the monitoring of
the data taking operations (rates, number of \dwords{datafrag}
in flight, error flags, application logs, network bandwidth, 
computing and network infrastructure) and the monitoring of the
physics data.
Both are essential to the success of the experiment and must be 
designed and integrated into the \dword{daq} system from the start.
In particular, for such a large and distributed system, the information sharing and archival system is very important, as are 
scalable and easily accessible data visualization tools, which will evolve during the lifetime of the experiment.
The online software provides the glue that holds the
\dword{daq} applications together and enables 
data taking.
Its architecture guides the
approach to \dword{daq} application design and also shapes the view
that the operators will have of the experiment.

\section{Interfaces} 
\label{sec:fd-daq-intfc}

\metainfo{5 Pages.  This is a shared SP/DP section.  If there are assymetries simply describe both detector modules.}

\fixme{Include an image of each interface in appropriate section.  Can maybe refer to Figure~\ref{fig:daq-overview} but it currently lacks some of the interfaces.}

\subsection{TPC Electronics}
\label{sec:fd-daq-intfc-elec}

Details about the interfaces between the \dword{daq} and the \dword{dp}
\dlong{cro} TPC electronics are documented in~\cite{bib:docdb6778}.
\fixme{Add references to bib.}

In the case of the \dword{dp} \dword{detmodule}, signals from the
\dlong{cro} electrodes are amplified and then digitzed by \dword{amc}
boards residing in 240 \dword{utca} crates. 
Each crate produces \SI{2.5}{\MHz} samples from 640 channels on a
\SI{10}{\Gbps} optical fiber. 
Data is losslessly compressed and sent via \dword{udp} producing a
stream with an expected average throughput of \SI{2}{\Gbps}.
The \dword{daq} consortium will be responsible for acquiring the fibers while
the \dual electronics consortium will be responsible for their
installation on the cryostat roof down to their connection to the
\dword{utca} crates.

\fixme{Need some statement about the white rabbit fiber?}

\subsection{PD Electronics}
\label{sec:fd-daq-intfc-photon}

Details about the interfaces between the \dword{daq} the \dword{dp}
\dlong{lro}~\cite{bib:docdb6802} are documented. 

\fixme{Add references to bib.}

For the \dword{dp} \dword{lro}, the signals are digitized in 14 bits
at \SI{65}{\MHz} and then downsampled to \SI{2.5}{\MHz}. 
The data is then handled largely symmetrically with the \dword{cro}
data except that only five \dword{utca} crates are required and their
average uncompressed output is expected to fill \SI{5}{\Gbps} on
\SI{10}{\Gbps} fiber optical connections. 
The installation of these fibers is similar to that described above for the
\dwords{cro} fibers.

\subsection{Offline Computing} 
\label{sec:fd-daq-intfc-fnal-cmptg}

The interface between the \dword{daq} and offline computing is
described in~\cite{docdb-7123}.
The \dword{daq} team is responsible for reducing the data volume
to the level that is agreed upon by all interested parties, and the
raw data files are transfered from \surf to \fnal using a
dedicated network connection.
A disk buffer is provided by the \dword{daq} on or near the \surf
site to hold 
several days worth of data 
so that the
operation of the experiment is not 
affected if there happens to
be a network disruption between \surf and \fnal.

During stable running, the data volume produced by the
\dword{daq} systems of all four \dwords{detmodule} will be no larger
than \offsitepbpy.
The maximum data rate is expected to be independent of the number of
\dwords{detmodule} that are operational.
During the construction of the second, third, and fourth
\dwords{detmodule}, the extra rate per \dword{detmodule} will be used
to gather data to aid in the refinement of the data selection
algorithms.
During commissioning, the data rate is expected to be higher than
nominal running and it is anticipated that  
a data volume corresponding to (order) one year will be necessary to commission a \dword{detmodule}.

The disk buffer at \surf is planned to be \SI{300}{\TB} in size.
The data link from \surf to \fnal will support \surffnalbw
(\offsitepbpy corresponds to about \offsitegbps).
The offline computing team is responsible for developing the
software to manage the transfer of files from \surf to \fnal.
The \dword{daq} team is responsible for producing a reference
implementation of the software that is used to access and decode the
raw electronics data.
The offline group is also responsible for providing the framework
for real-time \dword{dqm}. 
This monitoring is distinct from the \dfirst{om}.
Developing the payload jobs that run various algorithms to
summarize the data is the joint responsibility of the \dword{daq}, offline,
reconstruction and other groups.
The \dword{dqm} system includes a visualization system that can be
accessed from the Internet and shows specifically where operation shifts are
performed.

\subsection{Slow Control}
\label{sec:fd-daq-intfc-sc}
\label{sec:fd-daq-intfc-sc}

The \dword{cisc} systems monitor detector hardware and conditions not
directly involved in taking the data described above.
That data is stored both locally (in \dword{cisc} database servers in the
\dword{cuc}) and offline (the databases will be replicated back to \fnal)
in a relational database indexed by timestamp.
This allows bi-directional communications between the \dword{daq} and \dword{cisc} by
reading or inserting data into the database as needed for non
time-critical information.  

For prompt, time sensitive status information such as \textit{run is in
progress} or \textit{camera is on}, a low-latency software status register
is available on the local network to both systems.

There is no hardware interface. However, several racks of \dword{cisc} servers are in the counting room of the \dword{cuc}, and rack monitors in \dword{daq} racks are read out into the \dword{cisc} data stream.

Note that life and hardware safety-critical items will be hardware
interlocked 
according to \fnal standards, and fall outside the scope of this interface.

\subsection{External Systems} 
\label{sec:fd-daq-intfc-ext}

\fixme{Need to receive information on beam spills (Giles) , SNEWS (Alec).}


The \dword{daq} is required to save data based on external triggers, e.g., when a pulse of beam neutrinos  arrives at the \dword{fd}; or upon notice of an interesting astrophysical event by \dword{snews}~\cite{snews} or LIGO. This could involve going back 
to save data that has already been buffered (see Section~\ref{sec:fd-daq-fero}), or changing the trigger or zero suppression criteria for data taken during the interesting time period.

\subsubsection{Beam Trigger} 

The method for predicting and receiving the time of the beam spill is described in
Section~\ref{sec:fd-daq-design-beamtiming}.
Once that time is known to the \dword{daq}, a high-level trigger can be issued
to ensure that the necessary full data can be saved from the buffer
and saved as an event.

\subsubsection{Astrophysical Triggers} 

\dfirst{snews} is a coincidence
network of neutrino experiments that are individually sensitive to
an \dword{snb} 
observed from a core-collapse
supernova somewhere in our galaxy.
While DUNE must be sensitive to such a burst on its own, and 
is expected to be able to contribute to the coincidence network (Section~\ref{sec:fd-daq-sel}) via a TCP/IP socket, the capability to save data based on other observations provides an additional opportunity to ensure capture of this rare and valuable data. 
A \dword{snews} alert is formed when two or more neutrino experiments
report a potential \dword{snb} signal within \SI{10}{\s}.
A script running on the \dword{snews} server at BNL, provided by a given experiment that wishes to receive an alert,  sends out a message with the earliest time in the coincidence.
The latency from the neutrino burst is set by the response time of the
second fastest detector to report to \dword{snews}. This could be as
short as seconds, but could be tens of seconds.
At latencies larger than \SI{10}{\s}, full data might not be
available, but selected data is expected to be manageable. 


Other astrophysical triggers are available to which DUNE alone is unlikely to have sensitivity, except in rare cases, or if the triggers are taken as an ensemble. 
 Among these are gravitational wave triggers 
 (the details are being worked
out during the current LIGO shutdown), and high-energy photon
transients, most notably gamma ray bursts.
In fact, the use of network sockets on the timescale of seconds
enabled cooperation between LIGO, VIRGO, the Gamma Ray Coordinates
Network (GCN)~\footnote{Described in detail at
  \url{https://gcn.gsfc.nasa.gov/gcn_describe.html}}, and a number of
automated telescopes to make the discovery that \textit{short/hard} gamma ray bursts are caused by colliding neutron stars~\cite{kilonova}.

\section{Production and Assembly}
\label{sec:fd-daq-prod-assy}

\subsection{DAQ Components}

The \dword{fd} \dword{daq} system comprises the classes of components listed below. In each case, we outline the production, procurement, \dword{qa}, and \dword{qc} strategies.

\subsubsection{Custom Electronic Modules}

Custom electronic modules, specified and designed by the \dword{daq}
consortium, are used for two functional components in the \dword{daq}
\dword{fe}. 
The first is to interface the \dword{detmodule} electronics to the \dword{daq} \dword{fec} systems, which are likely to be based on the \dword{felix}
PCIe board.
The other is for real-time data processing (particularly for the
\dword{spmod}), which will likely be based on the
\dword{cob} \dshort{atca} blade.
\Dword{pdsp} currently implements both designs, and new designs optimized according to
DUNE requirements will be developed.
It is possible that we will make use of commercially-designed hardware
in one or other of these roles. \dword{daq} consortium institutes have
significant experience in the design and production of high-performance digital electronics for previous experiments.
Our strategy is therefore to carry out design in-house, manufacturing
and \dword{qa} steps in industry, and testing and \dword{qc} procedures at a number of
specialized centers within the DUNE collaboration. 
Where technically and economically feasible, modules will be split
into subassemblies (e.g., carrier board plus processing
daughter cards), allowing production tasks to be spread over more
consortium institutes.

DUNE electronic hardware will be of relatively high performance by commercial standards, and will contain high-value subassemblies such as large \dwords{fpga}. Achieving a high yield will require significant effort in design verification, prototyping and pre-production tests, as well as in tendering and vendor selection. The production schedule is largely driven by these stages and the need for thorough testing and integration with firmware and software before installation, rather than by the time for series hardware manufacture. This is somewhat different from the majority of other DUNE \dword{fd} components.

\subsubsection{Commercial Computing}

The majority of procured items will be standard commercial computing equipment, in the form of compute and storage servers. Here, the emphasis is on correct definition of the detailed specification, and the tracking of technology development, in order to obtain the best value 
during the tendering process. Computing hardware will be procured in several batches, as the need for \dword{daq} throughput increases during the construction period. 

\subsubsection{Networking and data links}

The data movement system is a combination of custom optical links (for data transmission from the cryostats to the \dword{cuc}) and commercial networking equipment. The latter items will be procured in the same way as other computing components. The favored approach to procurement of custom optics is purchase of pre-manufactured assemblies ready for installation, rather than 
on-site fiber preparation and termination. Since transmission distances and latencies in the underground area are not critical, the fiber run lengths do not need to be of more than a few variants. It is assumed that fibers will not be easily accessible for servicing or replacement during the lifetime of the experiment, meaning that procurement and installation of spare \textit{dark} fibers (including, if necessary, riser fibers up the \surf hoist shafts) is necessary.

\subsubsection{Infrastructure}

All \dword{daq} components will be designed for installation in \SI{48.3}{cm} (standard \SI{19}{in}) rack infrastructure, either in the \dword{cuc} or above ground. Standard commercial server racks with local air-water heat exchangers are likely to be used. These items will be specified and procured within the consortium, but will be pre-installed (along with the necessary electrical, cooling and safety infrastructure) under the control of \dword{tc} before \dword{daq} beneficial occupancy.

\subsubsection{Software and firmware}

The majority of the \dword{daq} construction effort will be invested in the production of custom software and firmware. Based on previous experiments, these projects are likely to use tens to hundreds of staff-years of effort, and will be significant projects even by commercial standards, mainly due to the specialized skills required for real-time software and firmware. A major project management effort is required to guide the specification, design, implementation and testing of the necessary components, especially as developers will be distributed around the world. Use of common components and frameworks across all areas of the \dword{daq} is mandatory. Effective \dword{daq} software and firmware development has been a demonstrated weakness of several previous experiments, and substantial work is required 
 in the next two years to put in place the necessary project management regime.

\subsection{Quality Assurance and Quality Control}

High availability is a basic requirement for the \dword{daq}, and this rests upon three key principles:

\begin{itemize}
	\item A rigorous \dword{qa} and \dword{qc} regime for components (including software and firmware);
	\item Redundancy in system design, to avoid single points of failure;
	\item Ease of component replacement or upgrade with minimal downtime.
\end{itemize}

The lifetime of most electronic assemblies or commercial computing components will not match the \dunelifetime lifespan of the DUNE experiment. It is to be expected that essentially all components will therefore be replaced during this time. Careful system design will allow this to take place without changes to interfaces. However, it is intended that the system  run for at least the first three to four years without substantial replacements, and \dword{qa} and \dword{qc}, as well as spares production, will be steered by this goal. Of particular importance is adequate burn-in of all components before installation underground, and careful record-keeping of both module and subcomponent provenance, in order to identify systematic lifetime issues during running.

\subsection{Integration testing}

Since the \dword{daq} will use subcomponents produced by many different teams,
integration testing is a key tool in ensuring compatibility and
conformance to specification. This is particularly important in the
prototyping phase before the design of final hardware. Once
pre-production hardware is in hand, an extended integration phase will
be necessary in order to perform final debugging and performance tuning
of firmware and software. In order to facilitate ongoing optimization in
parallel with operations, and compatibility testing of new hardware or
software, we envisage the construction of one or more permanent
integration test stands at \dword{daq} institutions. These will be in locations
convenient to the majority of consortium
members, i.e., at major labs in Europe and the USA. A temporary \dword{daq}
integration and testing facility near \surf will also be required as
part of the installation procedure.

\section{Installation, Integration and Commissioning}
\label{sec:fdsp-daq-install}

\subsection{Installation}
\label{sec:fdsp-daq-install-transport}

The majority of \dword{daq} components will be installed in a dedicated and partitioned area of the \dword{cuc} as shown in Figure~\ref{fig:daq-install-controlroom}, starting as soon as the consortium has beneficial occupancy of this space. 
The \dword{cf} is responsible for running fiber from the \dword{spmod}'s \dwords{wib} to the \dword{daq}, and from the \dword{daq} to the surface. This is currently projected to take place eighteen months before \dwords{apa} are installed in the \dword{spmod}, allowing time for final component acceptance testing in the underground environment, and to prepare the \dword{daq} for detector testing and commissioning. Some \dword{daq} components (event builder, storage cluster and WAN routers, plus any post-event-builder processing) will be installed above ground.

A total of \SI{500}{kVA} of power and cooling will be available to run the computers in the counting room. 
Twelve \SI{48.3}{cm} (standard \SI{19}{in}) server racks (of up to 58U height) per module have initially been allocated for each detector module, with two more each for facilities and \dword{cisc}. An optimized layout, including the necessary space for hardware installation and maintenance, plus on-site spares, will be developed once the \dword{daq} design is finalized. The racks will be water cooled with local air-to-water heat exchangers. To allow expanded headroom for initial testing, development, and commissioning throughput, the full complement of rack infrastructure and network equipment for four \dwords{detmodule} will be installed from the start.  

\begin{dunefigure}[CUC control room layout]{fig:daq-install-controlroom}
  {Floor plan for the \dword{daq} and control room space in the \dword{cuc}.  The \dword{daq}
    Room has space for at least \num{52} racks of servers and routers.
    Fiber from the \dwords{wib} in the detector caverns enter in the upper
    right of this room, terminate in a breakout panel, and are
    distributed to the \dwords{rce} in these racks, then to \dword{felix} servers (also
    in this room) as outlined in
    Figure~\ref{fig:daq-readout-buffering-baseline}.  Fibers to the
    surface enter this room from the lower left.}
\includegraphics[width=0.8\textwidth]{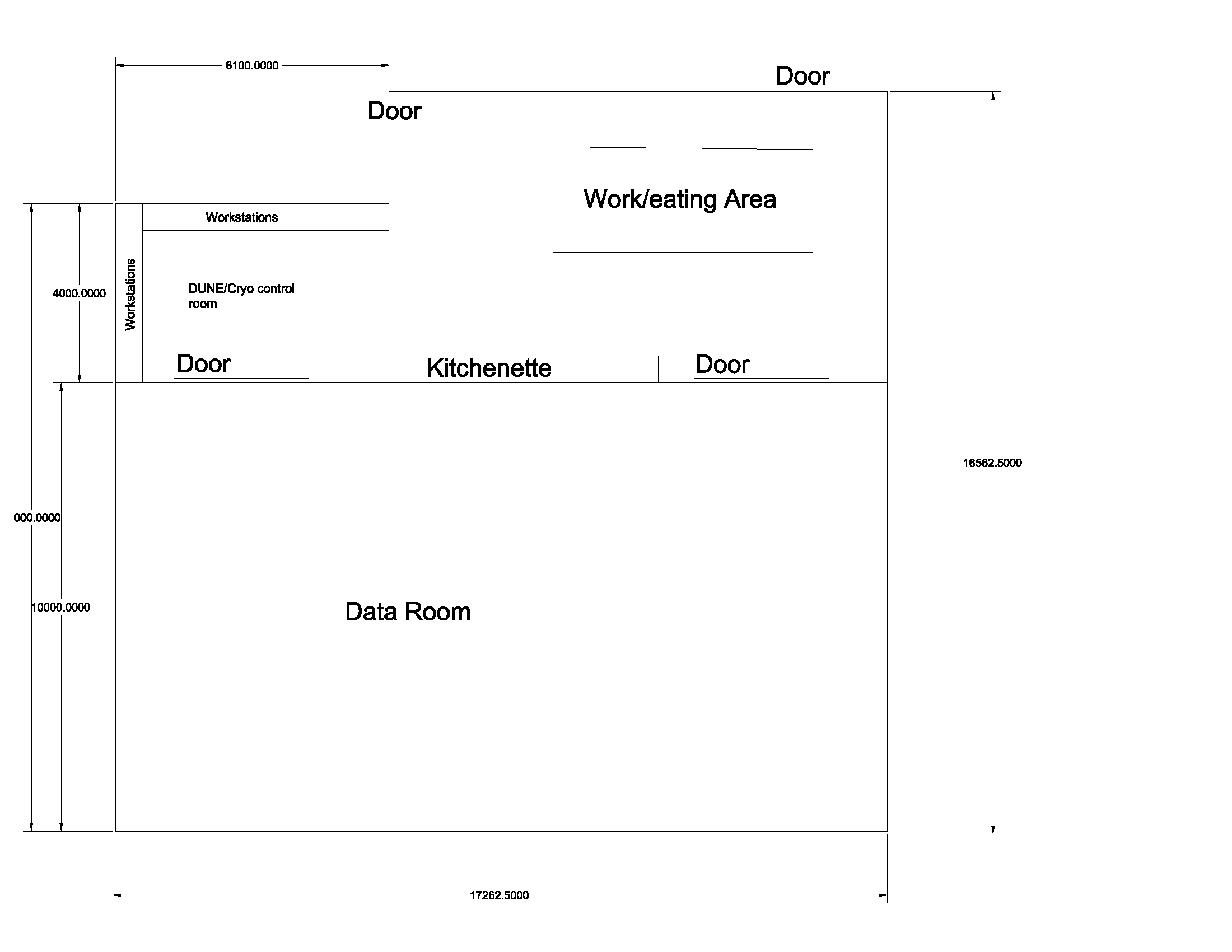}
\end{dunefigure}

The counting room is similar to a server room at a university or national lab in terms of the need for cleanliness, ventilation, fire protection, drop flooring, and access control. Networking infrastructure and fiber breakout will take up some of the rack space, but 
very little of the power budget. Power to individual machines and crates must to be controlled remotely via power distribution units, since it is desirable to minimize \dword{daq} workers' presence underground if there is work that can be done from the surface or remotely.  Some uninterruptible power supply (UPS) capacity is needed to allow for an orderly shutdown of computers, but only networking equipment requires longer-duration backup power, this is to enable remote recovery from short-term power failures.  

\subsection{Integration with Detector Electronics}
\label{sec:fdsp-daq-install-transport}

Basic technical integration with detector electronics will take place before installation, during a number of integration exercises in the preceding years. We anticipate that the consortium will supply and support small-scale instances of the \dword{daq} system for testing of readout hardware at the production sites, based on prototype or pre-production hardware. Full-scale \dword{daq} testing will have been completed with artificial data sources during internal integration. The work to be done during installation is therefore essentially channel-by-channel verification of the final system as it is installed, on a schedule allowing for any rectifying work to be carried out on the detector immediately (i.e., the \dword{daq} must gather and present data in effectively real time). This implies the presence of a minimal but sufficient functional \dword{daq} system before detector installation commences, along with the timing and fast control system, and the capability to permanently record data for offline analysis. However, it does not require triggering, substantial event building or data transfer capacity. The \dword{daq} installation schedule is essentially driven by this requirement.

In addition, the data pipeline from event builder, via the storage buffer and WAN, to the offline computing facilities, must be developed and tested. We anticipate this work largely happening at \fnal in parallel with detector installation, and the full-scale instances of these components being installed at \surf in preparation for start of data-taking.

\subsection{Commissioning}
\label{sec:fdsp-daq-commissioning}

System commissioning for the \dword{daq} comprises the following steps:

\begin{itemize}
	\item Integration with detector subsystems of successive \dwords{detmodule};
	\item Final integration and functional testing of all \dword{daq} components;
	\item Establishment of the necessary tools and procedures to achieve high-efficiency operation;
	\item Selection, optimization and testing of trigger criteria;
	\item Ongoing and continuous self-test of the system to identify actual or imminent failures, and to assess performance.
\end{itemize}

Each of these steps will have been carried out at the integration test stands before being used on the final system. The final steps are to some extent continuous activities over the experiment lifetime, but which require knowledge of realistic detector working conditions before final validation of the system can take place. We anticipate that these steps will be carried out during the cryostat filling period, and form the major focus of the \dword{daq} consortium effort during this time.

\section{Safety}
\label{sec:fd-daq-safety}

Two overall safety plans will be followed by the \dword{fd} \dword{daq}. General work underground will comply with all safety procedures in place for working in the detector caverns and \dword{cuc} underground at \surf. \Dword{daq}-specific procedures for working with racks full of electronics or computers, as defined at \fnal, will be followed, especially with respect to electrical safety and the fire suppression system chosen for the counting room. For example, a glass wall between the server room space and the other areas in Figure~\ref{fig:daq-install-controlroom} will be necessary to prevent workers in the server room from being unseen if they are in distress, and an adequate hearing protection regime must be put in place.

There are no other special safety items for the \dword{daq} system not already covered by the more general safety plans referenced above. The long-term emphasis is on remote operations capability from around the world, limiting the need for physical presence at \surf, and with underground access required only for urgent interventions or hardware replacement.

\section{Organization and Management}
\label{sec:fd-daq-org}

At the time of writing, the \dword{daq} consortium comprises \num{30} institutions, including universities and national labs, from five countries. Since its conception, the \dword{daq} consortium has met on roughly a weekly basis, and has so far held two international workshops dedicated to advancing the  \dword{fd} \dword{daq} design. The current \dword{daq} consortium leader is 
from U. Bristol, UK.

Several key technical and architectural decisions have been made in the last months, that have formed an agreed basis for the \dword{daq} design and implementation presented in this document.

\subsection{DAQ Consortium Organization}
\label{sec:fd-daq-org-consortium}

The DUNE \dword{daq} consortium is currently organized in the form of five active
Working Groups (WG) and WG leaders:
\begin{itemize}
\item Architecture, current WG leaders are from: U. Oxford and CERN;
\item Hardware, current WG leaders are from: U. Bristol and SLAC;
\item Data selection, current WG leader is from: U. Penn.;
\item Back-end, current WG leader is from: \fnal;
\item Integration and Infrastructure, current WG leader is from: U. Minnesota Duluth.
\end{itemize}

During the ongoing early stages of the design, the architecture and hardware WGs have been holding additional meetings focused on aspects of the design related to architecture solutions and costing. In parallel, the \dword{daq} Simulation Task Force effort, which was in place at the time of the consortium inception, has been adopted under the data selection WG, and simulation studies have continued to inform design considerations. This working structure is expected to remain in place through at least the completion of the \dword{tp}. During the construction phase of the project we anticipate a new organization, built around major subsystem construction and commissioning responsibilities, and drawing also upon expertise build up during the \dword{protodune} projects.

\subsection{Planning Assumptions}
\label{sec:fd-daq-org-assmp}

The \dword{daq} planning is based the assumption of a \dword{spmod} first, followed by a \dword{dpmod}. The schedule is sensitive to this assumption, as the \dword{daq} requirements for the two module types are quite different. Five partially overlapping phases of activity are planned (see Figure~\ref{fig:daq-schedule}):

\begin{itemize}
	\item A further period of R\&D activity, beginning at the time of writing, and culminating in a documented system design in the \dword{tdr} around July 2019;
	\item Production and testing of a full prototype \dword{daq} slice of realistic design, culminating in an engineering design review;
	\item Preparation and fit out of the \dword{cuc} counting room with a minimal \dword{daq} slice, in support of the first module installation;
	\item Production and delivery of final hardware, computing, software and firmware for the first module;
	\item Production and delivery of final hardware, computing, software and firmware for the second module.
\end{itemize}

This schedule assumes beneficial occupancy of the \dword{cuc} ounting room by end of the first quarter of 2022, and the availability of facilities to support an extended large-scale integration test in 2020 (e.g., CERN or \fnal). We assume the availability of resources for installation and commissioning of final \dword{daq} hardware (e.g., surface control room and server room facilities) from around the first quarter of 2023, and the \dword{itf} from the second quarter of 2022. The majority of capital resources for \dword{daq} construction will be required from the second quarter of 2022, with a first 
portion of funds for the minimal \dword{daq} slice from the first quarter of 2021.



\subsection{High-level Cost and Schedule}
\label{sec:fd-daq-org-cs}

The high-level \dword{daq} schedule, which is based upon the current DUNE \dword{fd} top-level schedule, is shown in Figure~\ref{fig:daq-schedule}.

\begin{dunefigure}[\dword{daq} high-level schedule]{fig:daq-schedule}
  {\dword{daq} high-level schedule}
\includegraphics[width=1.0\textwidth]{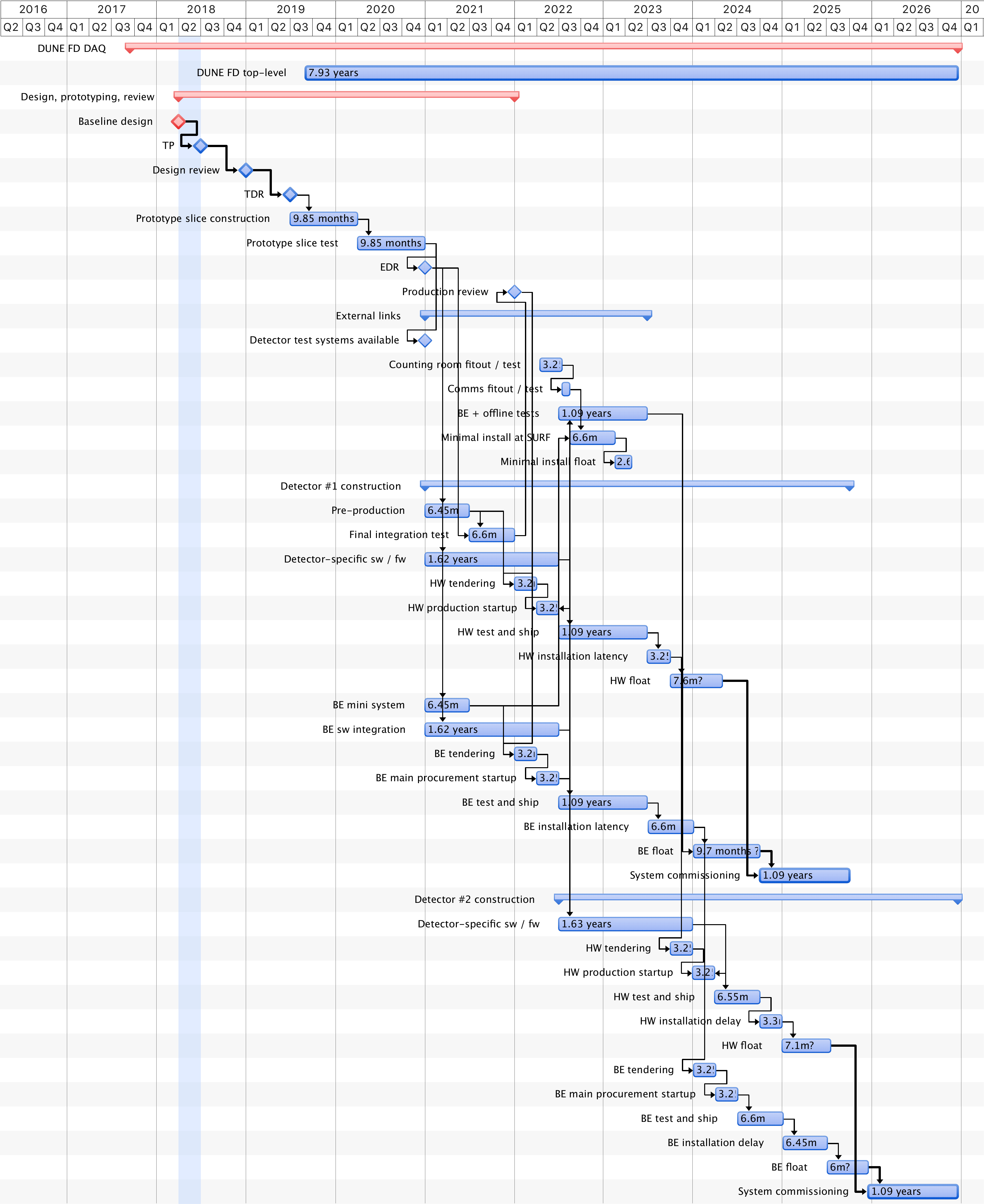}
\end{dunefigure}



\cleardoublepage

\chapter{Slow Controls and Cryogenics Instrumentation}
\label{ch:fddp-slow-cryo}

\section{Slow Controls and Cryogenics Instrumentation Overview}
\label{sec:fddp-slow-cryo-ov}



\subsection{Introduction}
\label{sec:fddp-slow-cryo-intro}


The \dword{cisc} system provides 
comprehensive monitoring for all \dword{detmodule} components as well as for the \lar quality and behavior, both being crucial
to guarantee high-quality of the data. Beyond passive monitoring, \dword{cisc} also provides a control system for some of the detector components. 
The structure of the \dword{cisc} consortium is quite complex. The subsystem chart
for the \dword{cisc} system in Figure~\ref{fig:sp-slow-cryo-subsys} shows the distinct 
 cryogenic instrumentation and slow controls branches. 
 
 The cryogenic instrumentation includes a set of devices 
to monitor the quality and behavior of the \lar volume in the cryostat interior, ensuring the correct functioning of
the full cryogenics system and the suitability of the \dword{lar} for good quality physics data. These devices are 
purity monitors, temperature monitors, gas analyzers, \lar level monitors, and cameras with their associated
light emitting system.

\begin{dunefigure}[CISC subsystems]{fig:sp-slow-cryo-subsys}
{\dword{cisc} subsystem chart}
\includegraphics[width=0.5\textwidth,trim=20mm 30mm 30mm 70mm,clip]{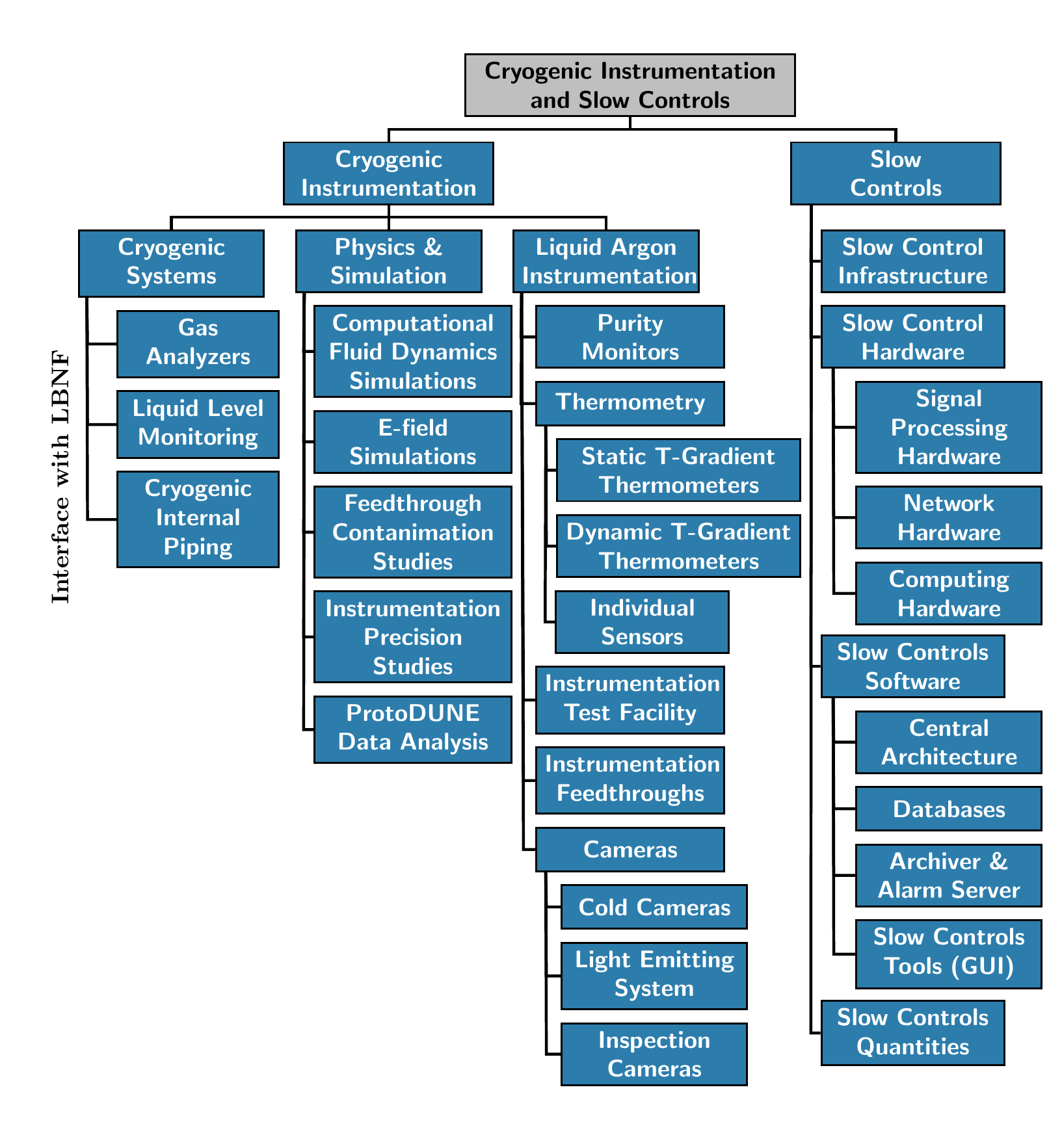}  
\end{dunefigure}

Cryogenic instrumentation also requires significant physics and
simulation work such as \efield simulations and cryogenics modeling
studies using \dfirst{cfd}. \efield simulations
are required to identify desirable locations for instrumentation
devices in the cryostat so that they are not in regions of high \efield and
that their presence does not induce large field distortions. \dshort{cfd}
simulations are needed to understand the expected temperature,
impurity and velocity flow distributions and guide the placement and
distribution of instrumentation devices inside the cryostat.

From the organizational point of view, the 
cryogenic instrumentation has been divided into three main parts: (1) cryogenics systems, which includes all components directly related to the external cryogenics system, such as
liquid level monitoring, gas analyzers and internal cryogenic piping -- all having substantial interfaces with LBNF; (2)  physics and simulation; and (3) \lar  instrumentation, which includes all
other instrumentation devices. 

The second branch of \dword{cisc} is the slow controls system, in charge of monitoring and controlling most detector elements, including power supplies, electronics, racks, instrumentation devices, and  
calibration devices, etc. It includes four main components: hardware, infrastructure,
software, and firmware. The slow controls hardware and infrastructure consists of
networking hardware, signal processing hardware, computing hardware, and relevant
rack infrastructure. The slow controls software and firmware are needed for
signal processing, alarms, archiving, and control room displays.

Two other systems have been included by the DUNE management as part of the \dword{cisc} consortium, a test facility for the instrumentation devices and the cryogenic piping inside the cryostat. Those are included inside the cryogenic instrumentation branch.


\subsection{Design Considerations}
\label{sec:fddp-slow-cryo-des-consid}


For all \lar instrumentation devices, \dword{pddp} designs are
considered as the baseline, and requirements for most design
parameters are extrapolated from \dword{protodune}. Hence a critical step for
the \dword{cisc} consortium is to analyze data from \dword{protodune} once data sets become  available
to validate the instrumentation designs and understand their
performance. For example, noise induced by instrumentation devices on the readout electronics  can confuse the event reconstruction; the tolerable noise level from this source is a crucial design parameter that should be evaluated in \dword{protodune}.

Some of the common design considerations for
instrumentation devices include stability, reliability and longevity, 
such that the devices can survive for a period of at least \dunelifetime{}s.  Since it is uncommon for any device
to have such a long lifetime, provisions are made in the overall
design to allow replacement of devices where possible.

As for any other element inside the cryostat, 
the \efield on the instrumentation devices is 
required to be less than \SI{30}{kV\per\cm},
so that the risk of dielectric breakdown in \dword{lar} is minimized.
This requirement imposes stringent constraints on the location and mechanical 
design of some devices. Electrostatic simulations  
will be performed to compute the expected field on the boundaries of 
instrumentation devices and to design the appropriate \efield shielding
in the case the field approaches the limit. 

Another common consideration for all instrumentation devices is their support structure
design, which is expected to be substantially different from the one used in \dword{protodune}.

For slow controls, the system needs to be 
robust enough to support a large number of monitored variables and a broad range of
monitoring and archiving rates. It must be capable 
interfacing 
with a large number of systems to establish two-way communication for
control and monitoring.  
Table~\ref{tab:dp-cisc-requirements} shows
some of the important \dword{cisc} system design requirements and parameters.

There are several aspects specific to the \dual design impacting the \dword{cisc} system design requirements:
\begin{itemize}
\item At the level of the cryogenic instrumentation, additional care is needed in order to monitor the gas phase above the liquid level. The temperature and the pressure of the gas phase affect the gas density, and consequently, the \dword{lem} gain calibration. The gas pressure must be accurately monitored. In proximity to the liquid surface the temperature gradient of the gas 
is measured with an array of temperature probes with a vertical pitch of about \SI{1}{cm}. Each \dword{crp} is also equipped with \num{36} thermometers to sample the temperature across its structure.

\item The \dword{crp}-specific instrumentation also includes: 

\begin{itemize}
\item the pulsing system for charge injection in the anode strips,
\item the precision level meters implemented only on the \dwords{crp} located at the cryostat borders, and 
\item the measurement of the \dword{lem}-grid capacitance allowing to know \fixme{which provides} the position of the \dword{crp} with respect to the liquid level for all \dwords{crp}, \fixme{what makes the measurement?}
\item the control of the stepping motors, \fixme{control mechanism?} which allows positioning each \dword{crp} parallel to the liquid level (keeping the extraction grid immersed in the liquid and the \dwords{lem} in the gas phase).
\end{itemize}

\item The slow control system generates and controls the \dword{hv} for biasing the \dwords{lem} (\num{41} channels/\dword{crp}) and the extraction grid  (\num{1} channel/\dword{crp}), at maximum voltages of  \SI{-5}{kV/channel} and \SI{-10}{kV/channel}, respectively;
\item The requirements related to the \dword{pds} include the generation and control of \dword{hv} biasing for the \dwords{pmt} (up to \SI{-3}{kV}) and the control of the calibration of the  \dwords{pmt},  performed with a  light distribution system with a common light source and a network of optical fibers;
\item The \dword{fe} electronics requires the control of the \dword{utca} crates, the control of the  analog \dword{fe} cards, the control of the \dword{lv} and of the charge injection system connected to the pre-amplifiers mounted on the \dword{fe} cards;
\item The \dual design also enables surveying from the cryostat roof the position of reference points connected to the \dword{crp} suspension system, to ensure proper \dword{crp} alignment. This aspect needs as well to be integrated in the alignment scheme. \fixme{IS integrated?}
\end{itemize}

\begin{dunetable}
[Important design requirements on the DP CISC system design]
{p{0.22\textwidth}p{0.17\textwidth}p{0.32\textwidth}p{0.19\textwidth}}
{tab:dp-cisc-requirements}
{Important design requirements on the \dual \dword{cisc} system design}   
Design Parameter
 & Requirement
 & Motivation
 & Comment \\ \toprowrule
Electron lifetime measurement precision
 & $<\SI{1.4}{\%}$ at \SI{3}{ms}
 & Per DUNE-FD Task Force\,\cite{fdtf-final-report}, needed to keep the bias on the charge readout in the \dshort{tpc} to below \SI{0.5}{\%} at \SI{3}{ms}
 & Purity monitors do not directly sample \dshort{tpc}: see Section~\ref{sec:fdgen-slow-cryo-purity-mon}
\\  \colhline
Thermometer precision
 & $<\SI{5}{mK}$
& Driven by \dshort{cfd} simulation validation; based on \dword{pdsp} design
& Expected \dword{protodune} performance \SI{2}{mK}
\\ \colhline
Pressure meters precision (\dual)
 & <\SI{1}{mbar}
& To measure the pressure (density) of the gas phase; based on \dword{pddp} design
&  \dword{wa105} / \dword{pddp} design $<\SI{1}{mbar}$
\\ \colhline
Thermometer density
 & \(>2/\si{m}\) (vert.), \(\sim\)~\SI{0.2}{m} (horiz.)
 & Driven by \dshort{cfd} simulation.
 & Achieved by design. 
\\ \colhline
Thermometer density gas phase (\dual)
 & \(>1/\si{cm}\) (vert.), \(\sim\)~\SI{1}{m} (horiz.)
 & Vertical array of thermometers with finer pitch close to the liquid level to measure the temperature gradient in the gas phase.
 & Achieved by design. 
\\ \colhline
 Thermometer density \dword{crp} structure (\dual)
 & \num{36} thermometers on each \dword{crp}
 & Monitoring the temperature across the \dword{crp}  structure.
 & Achieved by design. 
\\ \colhline
Liquid level meters precision (\dual)
 & \(<\SI{1}{mm}\)
&  Maintain constant \dword{crp} alignment with respect to the liquid surface
& \dword{wa105} / \dword{pddp} design \SI{0.1}{mm}
\\  \colhline
Cameras
 & \multicolumn{3}{p{0.64\textwidth}}{--- multiple requirements imposed by interfaces: see Table~\ref{tab:fdgen-cameras-req} ---}
 \\ \colhline
Cryogenic Instrumentation Test Facility cryostat volumes
 & 0.5 to \SI{3}{m^3}
& Based on filling costs and turn around times
& Under design
\\  \colhline
 Max.\ \efield on instrumentation devices
 & \(<\SI{30}{kV/cm}\)
 & The mechanical design of the system should be such that \efield is below this value, 
 to minimize the risk of dielectric breakdown in \dword{lar}
 & \dword{protodune} designs based on electrostatic simulations
\\ \colhline
 Noise introduced into readout electronics
 & Below significant levels
 & Keep readout electronics free from external noise, which confuses event reconstruction
 & To be evaluated at \dword{protodune}
\\ \colhline
Total no.\ of variables
 & \numrange{50}{100}\si{k}
& Expected number based on scaling past experiments; requires robust base software model that can handle large no. of variables.
& Achievable in existing control systems; DUNE choice in progress.
\\  \colhline
Max.\ archiving rate per channel
 & \SI{1}{Hz} (burst), \SI{1}{\per\minute} (avg.)
& Based on expected rapidity of interesting changes; impacts the base software choice; depends on data storage capabilities
& Achievable in existing control system software; DUNE choice in progress.
\\
%
%
%
\end{dunetable}

\subsection{Scope}
\label{sec:fddp-slow-cryo-scope}


As described above, and shown schematically in Figure~\ref{fig:sp-slow-cryo-subsys},
the scope of the \dword{cisc} system spans a broad range of activities.  In the
case of cryogenics systems (gas analyzers, liquid level monitors and
cryogenic internal piping), LBNF provides the needed expertise and
is responsible for the design, installation, and commissioning activities
while the \dword{cisc} consortium provides the resources as
needed. In the case of \lar Instrumentation devices (purity monitors,
thermometers, cameras and light-emitting system; and their associated \fdth{}s) and instrumentation
test facility, \dword{cisc} is responsible from design to commissioning in
the far \dwords{detmodule}.

From the slow controls side, \dword{cisc}  provides control and monitoring of
all detector elements that provide information on the health of the
\dword{detmodule} or conditions important to the experiment.
The scope of systems that slow controls includes is listed below:

\begin{itemize}
\item {\bf Slow Controls Base Software and Databases}: provides the central tools needed to develop control and monitoring for various detector systems and interfaces.
  \begin{itemize}
  \item Base input/output software;
  \item Alarms, archiving, display panels, and operator interface tools; 
  \item Slow controls system documentation and operations guidelines.
  \end{itemize}
\item {\bf Slow Controls for External Systems}: export data from systems external to the detector and provide status to operators and archiving.
  \begin{itemize}
  \item Beam status, cryogenics status, \dword{daq} status, and facilities systems status;
  \item For the systems above, import other interesting monitoring data as needed (e.g., pumps data from cryogenics system, heaters data from facility systems, etc.);
  \item Building controls, detector hall monitoring, and ground impedance monitoring; 
  \item Interlock status bit monitoring (but not the actual interlock mechanism).
  \end{itemize}
\item {\bf Slow Controls for Detector Hardware Systems}: develop software interfaces for detector hardware devices.
  \begin{itemize}
  \item Monitoring and control of all power supplies; 
  \item Full rack monitoring (rack fans, thermometers and rack protection system);
  \item Instrumentation and calibration device monitoring (and control to the extent needed);
  \item Power distribution units monitoring, computer hardware monitoring;
  \item \dword{hv} system monitoring through cold cameras;
  \item Detector components inspection through warm cameras.
  \end{itemize}
\end{itemize}

In terms of slow controls hardware, \dword{cisc}  develops, installs and
commissions any hardware related to rack monitoring and control. While
most power supplies might only need a cable from the device to an
Ethernet switch, some power supplies might need special cables (e.g., 
GPIB or RS232) for communication. The \dword{cisc} consortium is responsible for
providing such control cables.

The \dword{dpmod} has historically defined the scope of its slow controls system in a different way from that of the \dword{spmod}.  This chapter respects that historic definition and includes two systems within \dword{cisc} for now that could be taken on by other consortia at a later date.  These are: \dword{hv} biasing for the \dword{lem} and extraction grid, and  \dword{hv} biasing for the \dual \dword{pds}; and (2)  a calibration system for the \dwords{crp}.

In addition to the listed activities, \dword{cisc} also has activities that span
outside the scope of the consortium and require interfacing with other
groups. This is discussed in Section~\ref{sec:fdgen-slow-cryo-intfc}.



\section{Cryogenics Instrumentation}
\label{sec:fdsp-cryo-instr} 
\label{sec:fddp-cryo-instr} 
\label{sec:fdgen-cryo-instr} 

Instrumentation inside the cryostat must ensure that the condition of the \dword{lar} is adequate for operation of the \dshort{tpc}.
This instrumentation includes devices to monitor the impurity level of the argon, e.g., the purity monitors, which provide high-precision electron lifetime measurements,
and gas analyzers to ensure that the levels of atmospheric contamination drop below certain limits during the cryostat purging, cooling and filling.
The cryogenics system operation is monitored by temperature sensors deployed in vertical arrays and at the top and bottom of the detector, providing a 
detailed \threed temperature map that can help to predict the \dword{lar} purity across the entire cryostat. The cryogenics instrumentation also includes \lar level monitors and
a system of internal cameras to help in locating sparks in the cryostat and for overall monitoring of the cryostat interior. 
As mentioned in the introduction, cryogenics instrumentation requires simulation work to identify the proper location for these devices inside the cryostat and
for the coherent analysis of the instrumentation data. 

Figure~\ref{fig:sp-slow-cryo-ports} shows the current map of cryostat ports for the \dword{spmod}, highlighting the ones assigned to instrumentation devices,
as well as the preliminary location for some of these devices. Vertical temperature profilers are located behind the \dwords{apa} ($T_S$) and behind the east end wall ($T_D$).
They are complemented by a coarser \twod grid of sensors at the top and bottom of the cryostat (not shown in the figure). Purity monitors and level meters are planned
in each detector side, behind the two front end walls. Inspection cameras will use some of the multipurpose instrumentation ports, but their exact locations are yet to be decided.

\begin{dunefigure}[Cryostat ports]{fig:sp-slow-cryo-ports}
{Cryostat ports and preliminary location of some instrumentation devices. }
\includegraphics[width=0.95\textwidth]{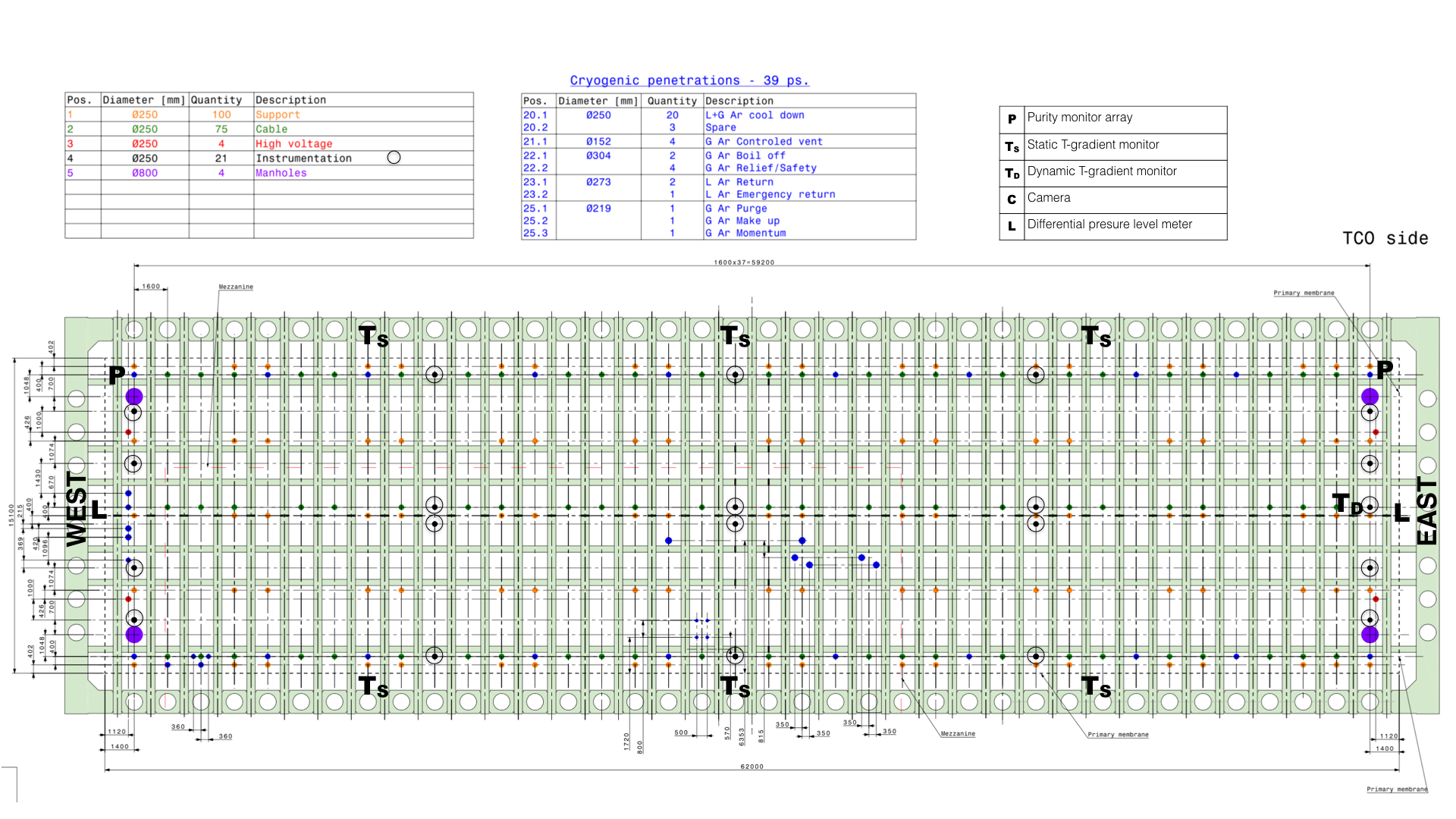}
\end{dunefigure}


For \dual, additional care must be taken to monitor the gas phase above the liquid level as the temperature and pressure of argon gas affect gas pressure, and consequently the \dword{lem} calibration.  The gas pressure must be monitored to \SI{1}{mbar} precision and accuracy. In proximity to the liquid surface the temperature gradient of the  gas has to be measured with an array of temperature probes with a vertical pitch of about \SI{1}{cm}, and with a coarser \SI{5}{cm} pitch up to the cryostat roof. Each \dword{crp} is also equipped with \num{36} thermometers to sample the temperature across its structure.

\subsection{Fluid Dynamics Simulations}
\label{sec:fdgen-slow-cryo-cfd}

Proper placement of purity monitors, thermometers, and liquid level monitors within the \dword{detmodule} requires knowledge of how \dword{lar} behaves within the cryostat in terms of its fluid dynamics, heat and mass transfer, and distribution of impurity concentrations. 
Fluid motion within the cryostat is driven primarily by small changes in density from thermal gradients, although pump flow rates and inlet and outlet locations also contribute. 
Heat sources include exterior heat from the surroundings, interior heat from the electronics, and heat flow through the pump inlet.

The fluid flow behavior can be determined through simulation of \dword{lar} flow within the detector using ANSYS CFX\footnote{ANSYS\texttrademark{}, \url{https://www.ansys.com/products/fluids/ansys-cfx}.}, a commercially available \dfirst{cfd} code. Such a model must include proper definition of the fluid characteristics, solid bodies and fluid-solid interfaces, and a means for measuring contamination, while still maintaining reasonable computation times. 
Although simulation of the \dword{detmodule} presents challenges, there exist acceptable simplifications for accurately representing the fluid, the interfacing solid bodies, and variations of contaminant concentrations. Because of the magnitude of thermal variation within the cryostat, modeling of the \dword{lar} is simplified through use of constant thermophysical properties, calculation of buoyant force through use of the Boussinesq Model (using constant a density for the fluid with application of a temperature dependent buoyant force), and a standard shear stress transport turbulence model. Solid bodies that contact the \dword{lar} include the cryostat wall, the cathode planes, the anode planes, the ground plane, and the \dword{fc}. As in previous \dshort{cfd} models of the DUNE \dword{35t} and \dword{protodune} by South Dakota State University (SDSU)\cite{docdb-5915}, the \dword{fc} planes, anode planes, and \dword{gp} can be represented by porous bodies. Since impurity concentration and electron lifetime do not impact the fluid flow, these variables can be simulated as  passive scalars, as is commonly done for smoke releases~\cite{cfd-1} in air or dyes released in liquids.

Significant discrepancies between real data and simulations can have potential impacts on detector performance, as simulation results contribute to decisions about where to locate sensors and monitors, as well as definitions of various calibration quantities. However, methods of mitigating such risks include well established convergence criteria, sensitivity studies, and comparison to results of previous \dshort{cfd} simulation work by SDSU and \fnal. Additionally, the simulation will be improved with input from temperature measurements and validation tests. 

\begin{dunefigure}[\dshort{cfd} example]{fig:cfd-example}
  {Distribution of temperature on a plane intersecting an inlet (right) and halfway between an inlet and an outlet (left), as predicted by SDSU \dshort{cfd} simulations \cite{docdb-5915}. (See Figure\ \ref{fig:cfd-example-geometry} for geometry.)}
  \includegraphics[height=0.4\textwidth]{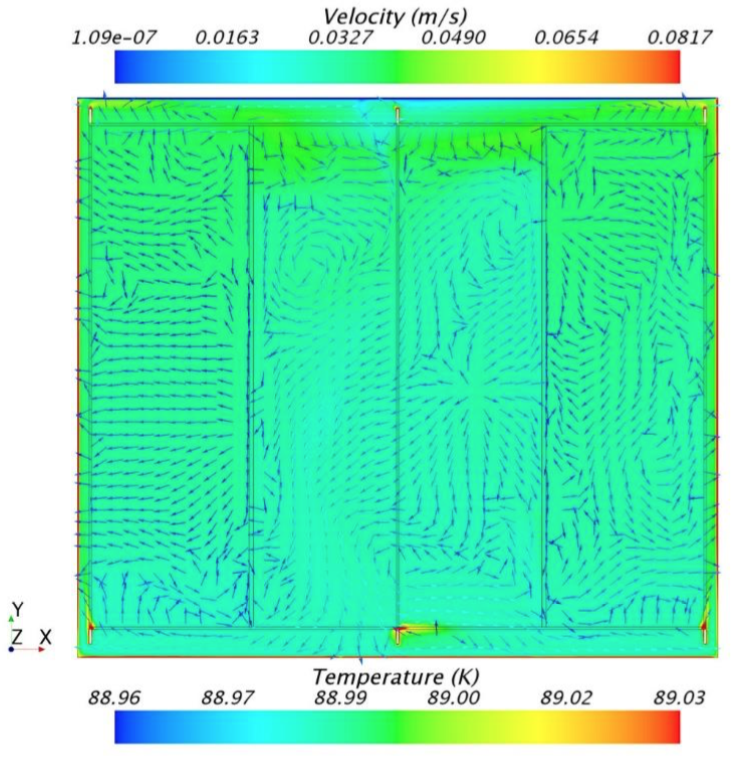}
  \includegraphics[height=0.4\textwidth]{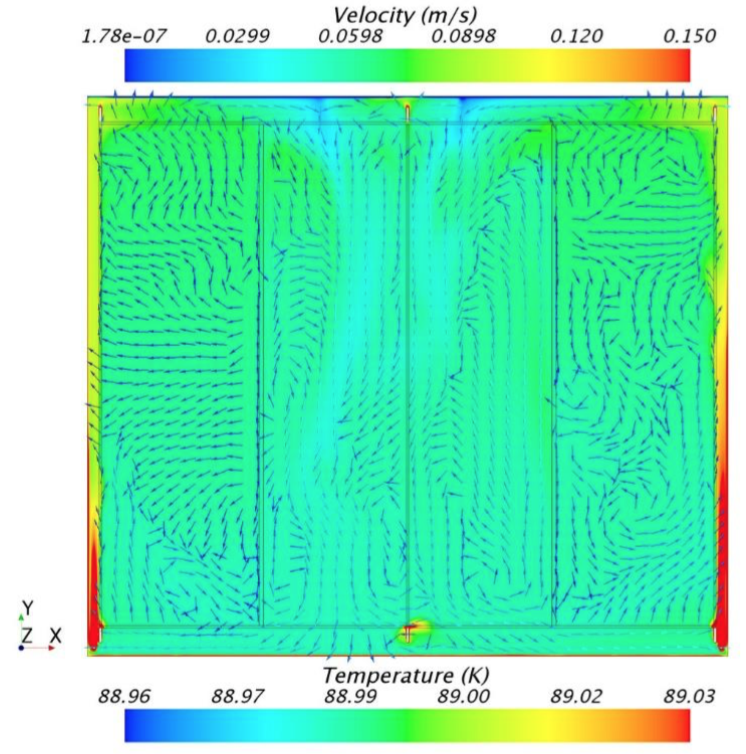}
\end{dunefigure}

Figure~\ref{fig:cfd-example} shows an example of the temperature
distribution on a plane intersecting a \dword{lar} inlet and at a
plane halfway between an inlet and an outlet; the geometry used for
this simulation is shown in Figure~\ref{fig:cfd-example-geometry}.
Note the plume of higher temperature \dword{lar} between the walls and
the outer \dword{apa} on the inlet plane.  The current locations of instrumentation in
the cryostat as shown in Figure~\ref{fig:sp-slow-cryo-ports} were
determined using the temperature and impurity distributions from these
previous simulations.

\begin{dunefigure}[\dshort{cfd} example geometry]{fig:cfd-example-geometry}
  {Layout of the \dshort{tpc} within the cryostat (top) and positions of
    \dword{lar} inlets and outlets (bottom) as modeled in the SDSU
    \dshort{cfd} simulations \cite{docdb-5915}.
    The Y axis is vertical and the X axis is parallel to the \dword{tpc}
    drift direction.
    Inlets are shown in green and outlets are shown in red.}
  \includegraphics[width=0.8\textwidth]{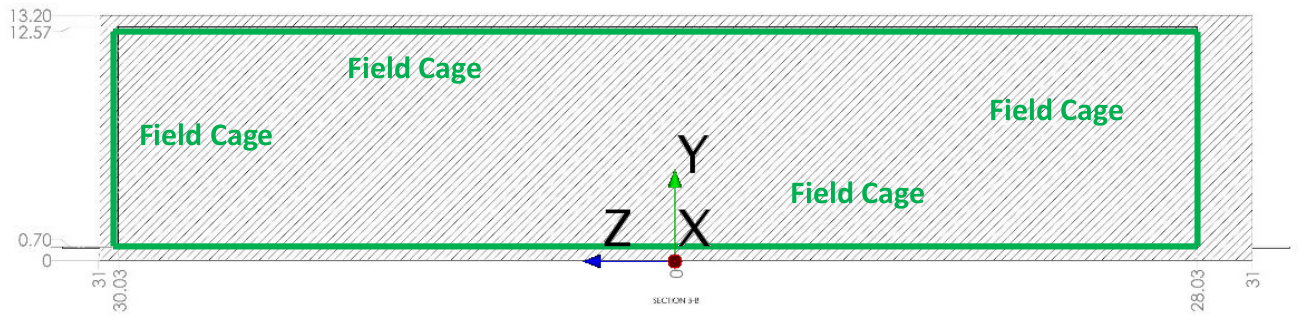}
  \includegraphics[width=0.8\textwidth]{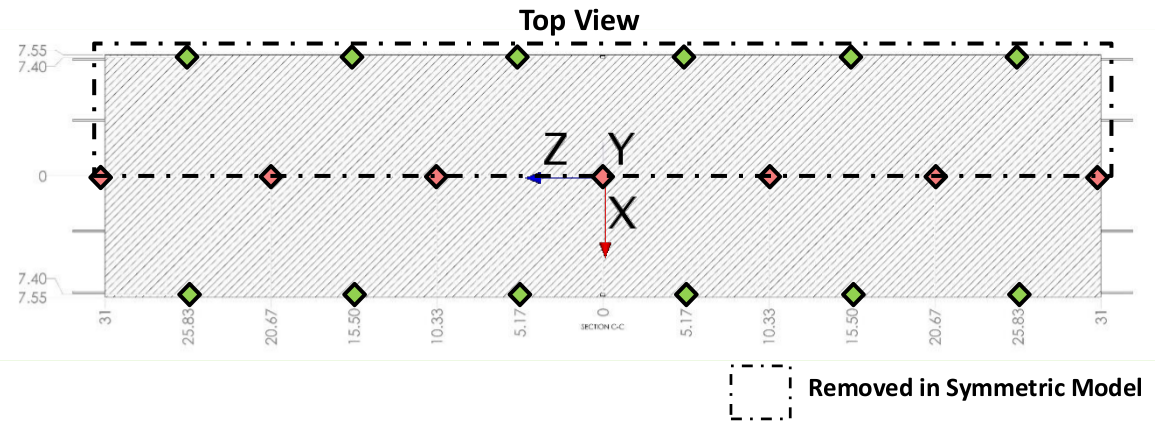}
\end{dunefigure}

The initial strategy for the future \dword{cfd} simulation effort is to understand the performance of \dword{protodune} cryogenics system and model the \dwords{fd} to derive requirements for instrumentation devices.
The following is a prioritized set of studies planned to help drive the requirements for other systems:
\begin{enumerate}
\item Review the DUNE \dword{fd} cryogenics system design and verify the current implementation in simulation; this is important to ensure that the model represents what will be built.
\item Model the \dword{pddp}  liquid and gas regions with the same precision as the \dword{fd}. Presently only the liquid model exists. The liquid model is needed to interpret the thermometer data, and the gas model is needed to understand how to place thermometers in the ullage and verify the design of the gaseous argon purge system.
\item Perform a \dword{cfd} study to determine the feasibility of a wier for \dual; this helps to determine if it can be used to clean the \lar surface before the extraction grid is submerged in the \dword{dpmod}.
\item Verify the \dword{spmod} \single \dword{cfd} model in simulation performed by LBNF; this defines the requirements for instrumentation devices (e.g., thermometry).
\item Model the \dword{pddp} liquid and gas regions with the same precision as the \dword{fd}. \fixme{same as a previous bullet}
\end{enumerate}


\subsection{Purity Monitors} 
\label{sec:fdgen-slow-cryo-purity-mon}
\label{sec:fdsp-slow-cryo-purity-mon} 
\label{sec:fddp-slow-cryo-purity-mon} 
A fundamental requirement of a \dword{lar} \dshort{tpc} is that ionization electrons drift over long distances in \dword{lar}. Part of the charge is inevitably lost due to the presence of electronegative impurities in the liquid. To keep such loss to a minimum, purifying the \dword{lar} during operation is essential, as is the monitoring of impurities.

Residual gas analyzers are an obvious choice when analyzing argon gas and can be exploited for the monitoring of the gas in the ullage of the tank. Unfortunately, commercially available and suitable mass spectrometers have a detection limit of \num{\sim10}\dword{ppb}, whereas DUNE requires a sensitivity down to the \dword{ppt} level. Instead, specially constructed purity monitors measure \lar purity in all the phases of operations, and enable the position-dependent purity measurements necessary to achieve DUNE's physics goal. 

Purity monitors also serve to mitigate \lar contamination risk.  
The large scale of the \dwords{detmodule} increases the risk of failing to notice a sudden unexpected infusion of contaminated \lar being injected back into the cryostat.   
If this condition were to persist, it could cause irreversible contamination to the \dword{lar} and terminate useful data taking.  Strategically placed purity monitors mitigate this risk. 

Purity monitors are placed inside the cryostat, but outside of the detector \dshort{tpc}, as well as outside the cryostat within the recirculation system before and after filtration. 
Continuous monitoring of  the \dword{lar} supply lines to the \dword{detmodule} provides a strong line of defense against contaminated \lar. Gas analyzers (described in Section~\ref{sec:fdgen-slow-cryo-gas-anlyz}) provide a first line of defense against contaminated gas.  Purity monitors inside the \dword{detmodule} provide a strong defense against all sources of contamination in the \lar volume and contamination from recirculated \lar. 
Furthermore, multiple purity monitors measuring lifetime with high precision at carefully chosen points can provide key inputs to \dshort{cfd} models of the detector, such as vertical gradients in impurity concentrations.

Purity monitors have been deployed in the ICARUS and \microboone detectors and in the \dword{35t} detector at Fermilab. In particular during the first run of the \dword{35t}, two out of four purity monitors stopped working during the cooldown, and a third was intermittent. It was later found out that this was due to poor electrical contacts of the resistor chain on the purity monitor. A new design was then implemented and successfully tested in the second run. 
The \dword{pdsp} and \dword{pddp} employ purity monitors based on the same design principles. \dword{pdsp} utilizes a string of purity monitors similar to that of the \dword{35t}, enabling measurement of the electron drift lifetime as a function of height.  A similar system design is exploited in the DUNE \dword{fd}, with modifications made to accommodate the instrumentation port placement relative to the purity monitor system and the requirements and constraints coming from the different geometric relations between the \dshort{tpc} and cryostat. 

\subsubsection{Physics and Simulation}

A purity monitor is a small ionization chamber that can be used to independently  infer the effective free electron lifetime in the \lartpc.  The operational principle of the purity monitor consists of generating a known electron current via illumination of a cathode with UV light, followed by collecting at an anode the current that survives after drifting a known distance.  The  attenuation of the current can be related to the electron lifetime.
The electron loss can be parameterized as
\(N(t) = N(0)e^{-t/\tau},\)
where $N(0)$ is the number of electrons generated by ionization, $N(t)$ is the number of electrons after drift time $t$, and $\tau$ is the electron lifetime.

For the \dword{spmod}, the 
\fixme{max?} drift distance is \spmaxdrift and the \efield is \SI{500}{\volt\per\centi\meter}. Given the drift velocity at this field of approximately \SI{1.5}{\milli\meter\per\micro\second}, the time to go from cathode to anode is around \SI{\sim2.4}{\milli\second} \cite{Walkowiak:2000wf}.
The \dword{lar} \dshort{tpc} signal attenuation, \([N(0)-N(t)]/N(0)\), is to be kept less than \SI{20}{\percent} over the entire drift distance \cite{fdtf-final-report}. The corresponding electron lifetime is $2.4/[-\ln(0.8)] \simeq \SI{11}{ms}$.

For the \dword{dpmod}, the maximum drift distance is \dpmaxdrift{}, therefore the requirement on the electron lifetime is much higher.

The 
\dword{35t} at Fermilab was instrumented with four purity monitors. The data taken with them during the first part of the second phase is shown in Figure~\ref{fig-35t-prm} and clearly shows the ability to measure the electron lifetime between \SI{100}{\micro\second} and \SI{3.5}{\milli\second}.  

\begin{dunefigure}[Electron lifetimes measured in the purity monitors in the \dword{35t}]{fig-35t-prm}
  {The measured electron lifetimes in the four purity monitors as a function of time at Fermilab \dword{35t}.}
  \includegraphics[width=0.6\textwidth]{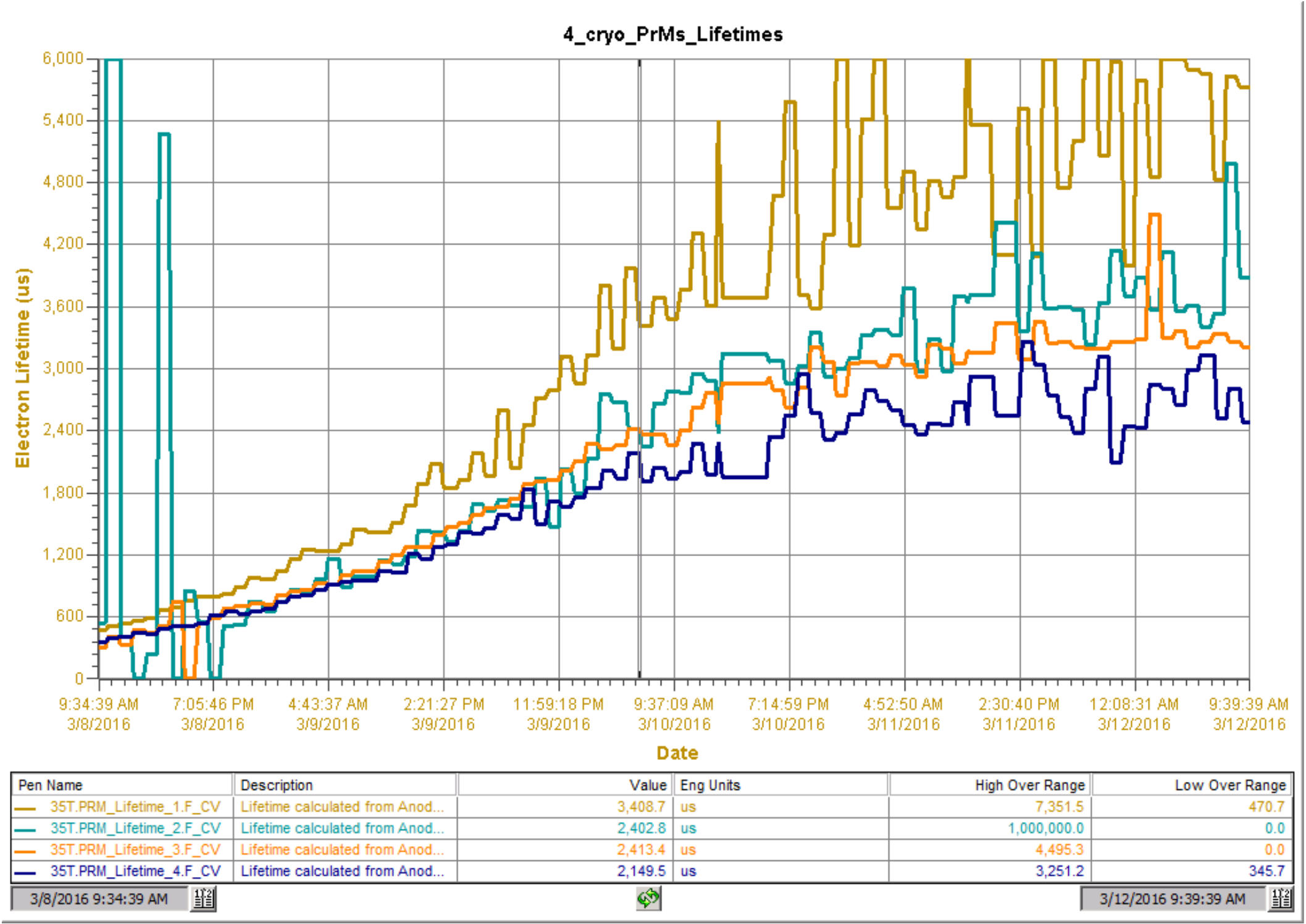}%
\end{dunefigure}

\subsubsection{Purity Monitor Design}

The basic design of a purity monitor is based on those used by the ICARUS experiment (Figure~\ref{fig:prm})\cite{Adamowski:2014daa}. It is a double-gridded ion chamber immersed in the \lar volume.   The purity monitor consists of four parallel, circular electrodes: a disk holding a photocathode, two grid rings (anode and cathode), and an anode disk. The cathode grid is held at ground potential. The cathode, anode grid, and anode are electrically accessible via modified vacuum grade high-voltage \fdth{}s and separate bias voltages held at each one.  
The anode grid and the field shaping rings are connected to the cathode grid by an internal chain of \SI{50}{\mega\ohm} resistors to ensure the uniformity of the \efield{}s in the drift regions. A stainless mesh cylinder is used as a Faraday cage to isolate the purity monitor from external electrostatic backgrounds. 

The purity monitor measures the electron drift lifetime between its anode and cathode. The electrons are generated by the purity monitor's UV-illuminated gold photocathode via the photoelectric effect. As the electron lifetime in \lar is inversely proportional to the electronegative impurity concentration, the fraction of electrons generated at the cathode that arrive at the anode ($Q_A/Q_C$) after the electron drift time $t$ gives a measure of the electron lifetime $\tau$:
\( Q_A/Q_C \sim e^{-t/\tau}.\)
%

It is clear from this formula that the purity monitor reaches its sensitivity limit once the electron lifetime becomes much larger than the drift time $t$. For $\tau >> t$ the anode to cathode charge ratio becomes $\sim\,1$. But, as the drift time is inversely proportional to the \efield, by lowering the drift field one can in principle measure any lifetime no matter the length of the purity monitor (the lower the field, the lower the drift velocity, i.e., the longer the drift time). 
In practice, at very low fields it is hard to drift the electrons all the way up to the anode. Currently, specific sensitivity limits for purity monitors with a drift distance of the order of $\sim$\SI{20}{\centi\meter} are still to be determined in a series of tests. If the required sensitivity is not achieved by these ``short'' purity monitors, longer ones may be developed.

\begin{dunefigure}[Purity monitor diagram]{fig:prm}
  {Schematic diagram of the basic purity monitor design~\cite{Adamowski:2014daa}.}
  \includegraphics[width=0.5\textwidth]{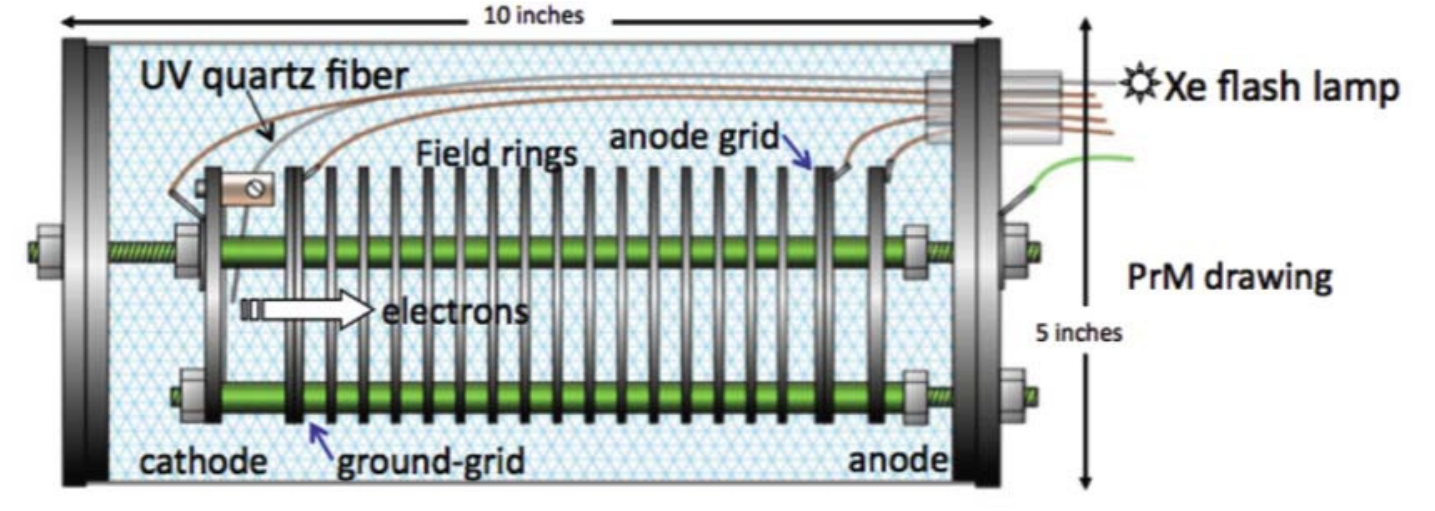}
\end{dunefigure}

The photocathode that produces the \phel{}s is an aluminum plate coated with \SI{50}{\angstrom} of titanium and \SI{1000}{\angstrom} of gold and attached to the cathode disk. A xenon flash lamp is used as the light source in the baseline design, although this could potentially be replaced by a more reliable and possibly submersible light source in the future, perhaps LED driven. The UV output of the lamp is quite good around $\lambda=$ \SI{225}{\nano\meter}, which is close to the work function of gold (\SIrange{4.9}{5.1}{\eV}). Several UV quartz fibers are used to carry the xenon UV light into the cryostat to illuminate the gold photocathode.   Another quartz fiber is used to deliver the light into a properly biased photodiode outside of the cryostat to provide the trigger signal for when the lamp flashes. 

\subsubsection{Electronics, DAQ and Slow Controls Interfacing}
The purity monitor electronics and \dword{daq} system consist of \dword{fe} electronics, waveform digitizers, and a \dword{daq} PC.  The block diagram of the system is shown in Figure~\ref{fig:cryo-purity-mon-diag}.

The baseline design of the \dword{fe} electronics is the one used for the purity monitors at the \dword{35t}, LAPD, and \microboone. The cathode and anode signals are fed into two charge amplifiers contained within the purity monitor electronics module.
The amplified outputs of the anode and cathode are recorded with a waveform digitizer that interfaces with a \dword{daq} PC.

\begin{dunefigure}[Purity monitor block diagram]{fig:cryo-purity-mon-diag}
  {Block diagram of the purity monitor system.}
  \includegraphics[width=0.7\textwidth]{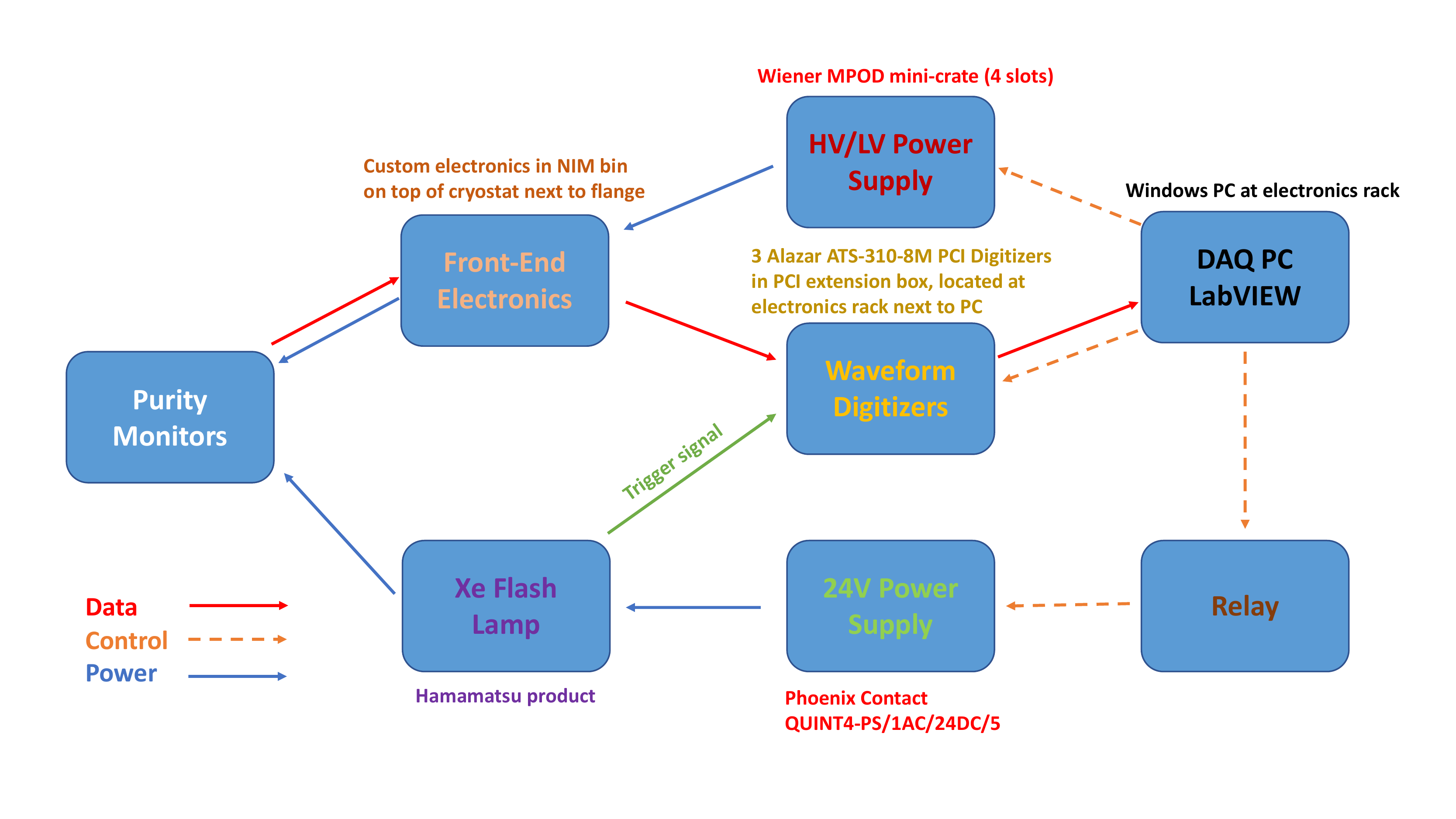}%
\end{dunefigure}

A custom LabVIEW application running on the \dword{daq} PC is developed and 
executes two functions: it controls the waveform digitizers and the power supplies, and it monitors the signals and key parameters. The application configures the digitizers to set the sampling rate, the number of waveforms to be stored in the memory, pre-trigger data, and a trigger mode. A signal from a photodiode triggered by the xenon flash lamp is directly fed into the digitizer as an external trigger to initiate data acquisition. The LabVIEW application automatically turns on the xenon flash lamp by powering a relay at the start of data taking and then turns it off when finished.
The application continuously displays the waveforms and important parameters, such as measured electron lifetime, peak voltages, and drift time of electrons in the purity monitors, and shows these parameters over time.

The xenon flash lamp and the \dword{fe} electronics are installed close to the purity monitor flange, to reduce light loss through the optical fiber and prevent signal loss. Other pieces of equipment are mounted in a rack separate from the cryostat. They distribute power to the xenon flash lamp and the \dword{fe} electronics, as well as collect data from the electronics. The slow control system communicates with the purity monitor \dword{daq} software and has control of the \dword{hv} and \dword{lv} power supplies of the purity monitor system. As the optical fiber has to be very close to the photocathode (less than \SI{0.5}{\milli\meter}) for efficient \phel extraction, no interference with the \dword{pds} is expected. Nevertheless light interference will be evaluated more precisely at \dword{protodune}.

Conversely the electronics of purity monitors may induce noise in the \dshort{tpc} electronics, largely coming from the current surge in the discharging process of the main capacitor of the purity monitor xenon light source when producing a flash.  This source of noise can be controlled by placing the xenon flash lamp inside its own Faraday cage, allowing for proper grounding and shielding; the extent of mitigation will be evaluated at \dword{protodune}.
If an unavoidable interference problem is found to exist, then software can be implemented to allow the \dword{daq} to know if and when the purity monitors are running and to veto purity monitor measurements in the event of a \dword{snb} alert or trigger.

\subsubsection{Production and Assembly}
\label{sec:PrMon-Production-Assembly}
Production of the individual purity monitors and their assembly into the string that gets placed into the \dword{detmodule} cryostat follows the same methodology that is being developed for \dword{protodune}.  Each of the individual monitors is fabricated, assembled and then tested in a smaller test stand.  After confirming that each of the individual purity monitors operates at the required performance, they are assembled together via the support tubes used to mount the system to the inside of the cryostat such that three purity monitors are grouped together to form one string, as shown in Figure~\ref{fig:PrMon-SystemString}.
Each monitor is assembled as the string is built from the top down, and in the end 
three individual purity monitors 
hang from a single string.  The assembly of the string concludes once the purity monitors are each in place, but with the Faraday cages removed and the \dword{hv} cables and optical fibers yet to be run.  This full string assembly is then 
shipped to the \dword{fd} site for installation into the cryostat.

\begin{dunefigure}[Purity monitor string]{fig:PrMon-SystemString}
  {Design of the purity monitor string that will contain three purity monitors.}
  \includegraphics[width=\textwidth]{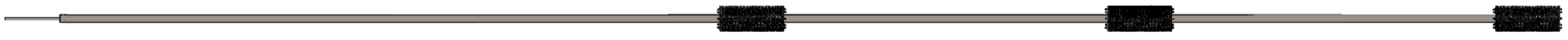}
\end{dunefigure}



\subsection{Thermometers}
\label{sec:fdgen-slow-cryo-therm}

A detailed \threed temperature map is important to monitor the correct functioning of the cryogenics system and the \lar uniformity.
Given the complexity and size of purity monitors, those can only be installed on the cryostat sides to provide a local measurement of
the \lar purity. While a direct measurement of the \lar purity across the entire cryostat is not viable, a sufficiently detailed \threed temperature map
can be used to predict the \lar purity using \dword{cfd} simulations. Measuring the vertical temperature profile is especially important since this is closely related to
the \lar recirculation and uniformity. 

High-precision temperature sensors are distributed near the TPC walls in two ways:
(1) in high-density (\(>2\) sensors/\si{m}) vertical arrays (the T-gradient monitors), and (2) in coarser ($\sim$ 1 sensor/\SI{5}{m}) \twod arrays 
at the top and bottom of the detector, which are the most sensitive regions (the individual sensors).   

Since temperature variations inside the cryostat are expected to be very small ($\SI{0.02}{K}$, see Figure~\ref{fig:cfd-example}), to properly measure the \threed temperature map 
sensors must be cross-calibrated to better than $\SI{0.005}{K}$. Most sensors are calibrated in the laboratory, prior to installation,
as described in Section~\ref{sp-cisc-thermom-static-t}.  This is in fact the only viable method for sensors in areas where the available space is restricted: on the long sides of the detector
(behind the \dwords{apa} for SP, and behind the lateral 
\dword{ewfc} for \dual) and top/bottom of the detector.
Given the precision required and the unknown longevity of the sensors (which could require a new calibration after some time), a complementary method
is used for T-gradient monitors behind the front endwalls, at least for the \dword{spmod}.
In those areas there is sufficient space for a movable system that can be used to cross-calibrate in situ
the temperature sensors. 

In the baseline design for all three systems mentioned above, three elements are common: sensors, cables and readout system.
Platinum sensors with \SI{100}{\ohm} resistance, PT100 series, produced by Lakeshore\footnote{Lakeshore\texttrademark{}, \url{https://www.lakeshore.com/Pages/Home.aspx}.},
are adequate for the temperature range of interest, \SIrange{83}{92}{K}, since in this range those sensors have $\sim\SI{5}{mK}$ reproducibility 
 and absolute temperature accuracy of \SI{100}{mK}.
In addition, using four-wire readout greatly reduces the issues related to the lead resistance, any parasitic resistances,
connections through the flange, and general electromagnetic noise pick-up. The Lakeshore PT102 sensors
have been previously used in the \dword{35t} and \dword{pdsp} detector,
giving excellent results. For the inner readout cables a custom cable made by Axon\footnote{Axon\texttrademark{}, \url{http://www.axon-cable.com/en/00_home/00_start/00/index.aspx}} is the baseline. It consists of four AWG28 teflon-jacketed copper wires forming two twisted pairs, with a metallic external shield
and an outer teflon jacket.
The readout system is described below in  Section~\ref{sec:fdgen-slow-cryo-therm-readout}. 



Another set of lower-precision sensors is used to monitor the filling of the cryostat in its initial stage. Those sensors are epoxied into the cryostat bottom membrane with
a density to be determined, not to exceed one sensor every \SI{5}{m}. 
Finally, the inner walls and roof of the cryostat are instrumented with the same type of sensors in order to monitor their temperature during cooldown and filling.
The baseline distribution has three vertical arrays of sensors epoxied to the membrane: one behind each of the two 
front \dwords{ewfc} and the third one in the middle of the cryostat
(behind the \dwords{apa} for \single and behind the lateral 
\dwords{ewfc} for \dual). 


\subsubsection{Static T-gradient Monitors}
\label{sp-cisc-thermom-static-t}

Several vertical arrays of high-precision temperature sensors cross-calibrated in the laboratory are installed near the lateral walls
(behind the \dwords{apa} for \single and behind the lateral 
\dwords{ewfc} for \dual). 
For the \dword{spmod}, since the electric potential is zero behind the \dwords{apa}, no \efield shielding is required, simplifying enormously the mechanical design.
This does not apply for the \dword{dpmod}, for which the proper shielding must be provided.

Sensors are cross-calibrated in the lab using a well controlled environment and a high-precision readout system, described below in Section~\ref{sec:fdgen-slow-cryo-therm-readout}. 
Although the calibration procedure will certainly improve, the one currently used for \dword{pdsp} is described here.
Four sensors are placed as close as possible (such that identical temperature can be assumed for all of them) inside a small cylindrical aluminum capsule,
which protects the sensors from thermal shocks and helps in minimizing convection.
One of the sensors acts as reference while the other three are 
calibrated. Five independent calibrations
are performed for each set of three sensors, such that the reproducibility of each sensor can be computed. For each calibration 
the capsule is introduced in a \threed printed polylactic acid (PLA) box of size \(9.5\times9.5\times\SI{19}{cm^3}\), with two concentric independent volumes of \lar
and surrounded by a polystyrene box with \SI{15}{cm} thick walls. A small quantity of \lar is used to slowly
cool down the capsule to $\sim\SI{90}{K}$, avoiding thermal shocks that could damage the sensors.
Then the capsule is immersed in  \lar such that it penetrates
inside, fully covering the sensors. Once the temperature stabilizes to the 1-\SI{2}{mK} level (after 5-15 minutes) measurements are taken. Then the capsule is taken out from \lar
and exposed to room temperature until it reaches \SI{200}{K}. As mentioned above, this procedure is repeated five times, before going to the next set of three sensors.  
As shown in Figure~\ref{fig:Trepro} a reproducibility (\rms of the mean offset in the flat region) of $\sim \SI{2}{mK}$ has been achieved in the \dword{pdsp} design.  

The baseline design for the mechanics of the \single system consists of two stainless strings anchored at top and bottom corners of the cryostat
using the available M10 bolts (see Figure~\ref{fig:sensor-support}). One of the strings is used to route the cables while the other,
separated by \SI{340}{mm}, serves as support for temperature sensors.
Given the height of the cryostat, the need of intermediate anchoring points is under discussion. For the \dword{dpmod} no baseline design exists yet,
since additional complications due to the required \efield shielding must be taken into account. Figure~\ref{fig:sensor-support} shows the baseline design of the
PCB support for temperature sensors, with an IDC-4 male connector. It has a size of $52\times \SI{15}{mm^2}$. Each four-wire cable from the sensor to the flange has an IDC-4 female connector
on the sensor side; on the other side, it is directly soldered into the inner pins of male SUBD-25 connectors on the flanges. The CF63 side ports on the \dword{dss}/cryogenic ports are 
used to extract the cables. 

\begin{dunefigure}[Cryostat bolts and temperature sensor support]{fig:sensor-support}
  {Left: bolts at the bottom corner of the cryostat. Right: Lakeshore PT102 sensor mounted on a PCB with an IDC-4 connector.}
  \includegraphics[height=0.2\textwidth]{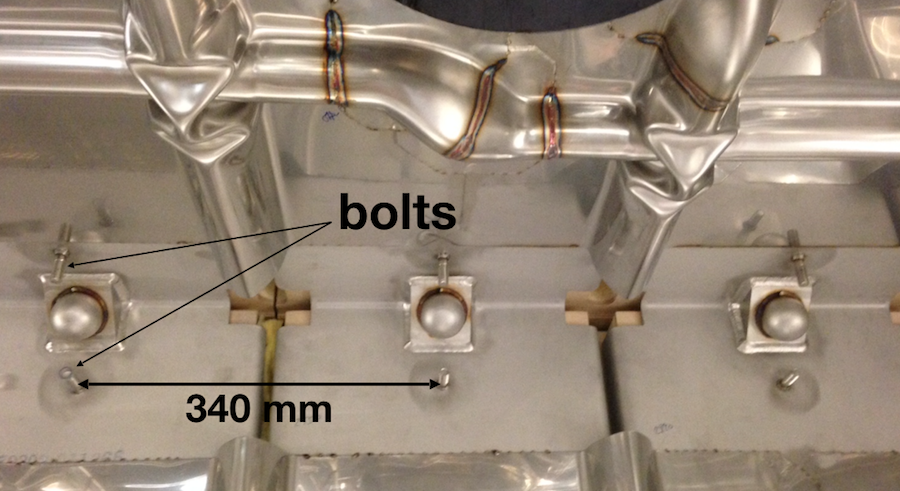}%
    \hspace{1cm}%
  \includegraphics[height=0.2\textwidth]{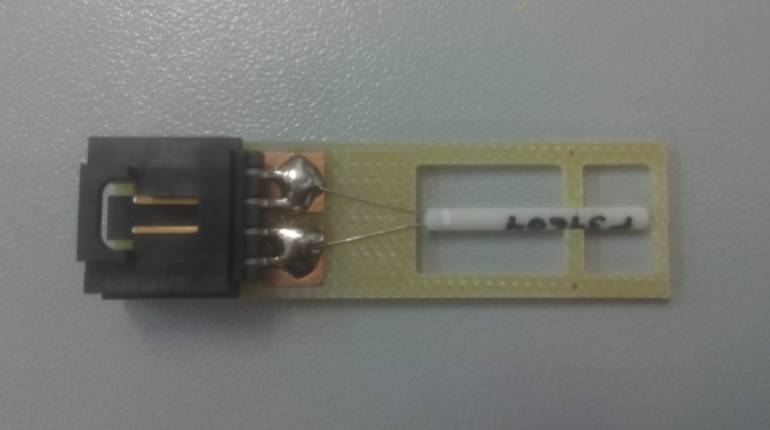}%
\end{dunefigure}

\begin{dunefigure}[Temperature sensor resolution and reproducibility]{fig:Trepro}
  {Temperature offset between two sensors as a function of time for five independent inmersions in \lar. The reproducibility of those sensors,
    defined as the RMS of the mean offset in the flat region, is $\sim \SI{2}{mK}$,
    The resolution for individual measurements, defined as the RMS of one of the offsets in the flat region, is better than \SI{1}{mK}.}
  \includegraphics[width=0.5\textwidth]{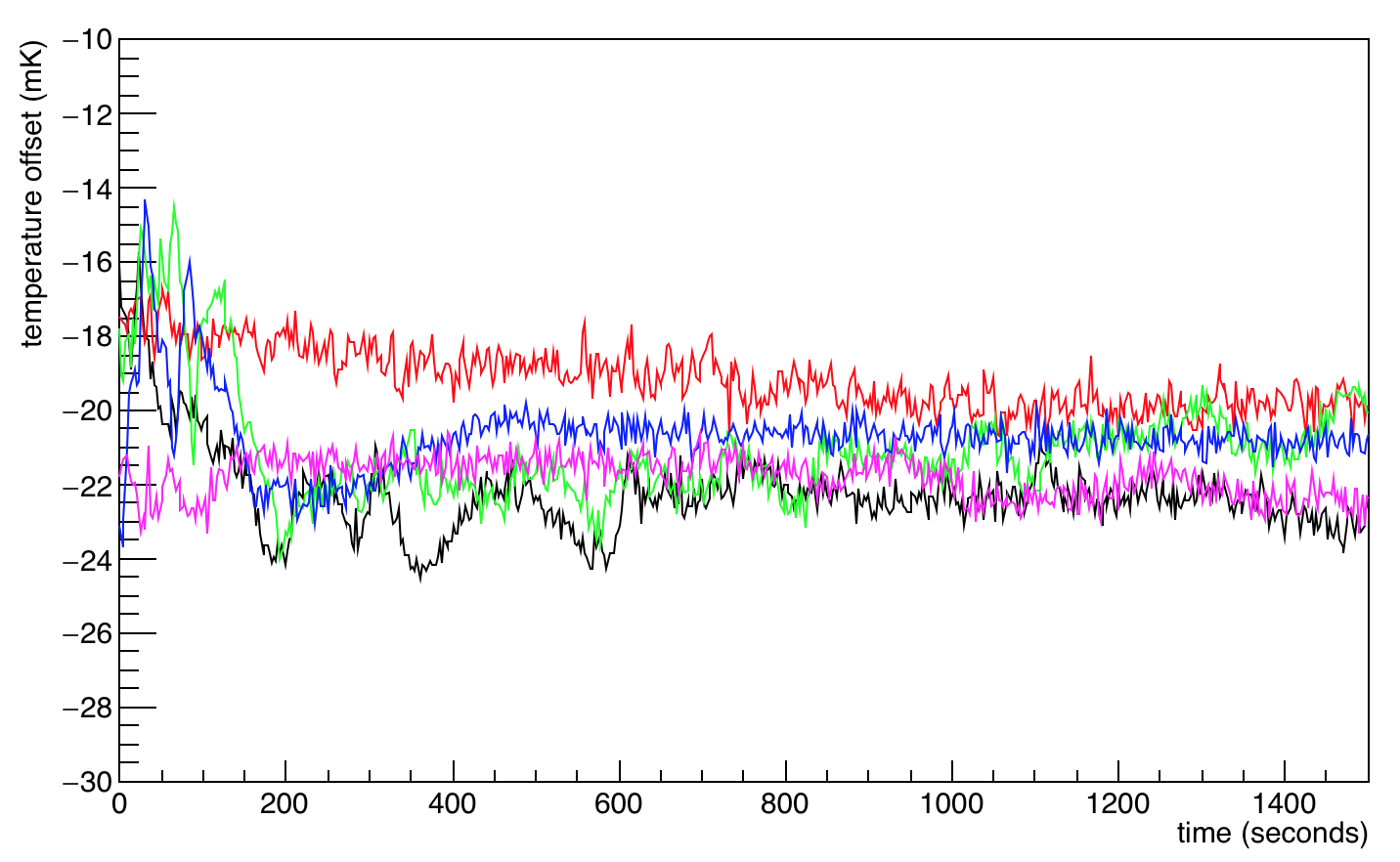}%
\end{dunefigure}

\subsubsection{Dynamic T-gradient Monitors}

The dynamic temperature monitor is a vertical array of high precision temperature sensors with the goal of measuring vertical temperature gradient with precision of few \si{mK}. The design of the system is driven by two factors:
\begin{itemize}
\item
A few-\si{mK} uncertainty in the measured vertical temperature profile over the entire detector height is required in order to monitor \lar purity and provide useful feedback of efficiency of cryogenic recirculation and purification.
\item
Simulations of the cryogenic recirculation predict very slow change in temperature at meter scale except at the bottom and top of the cryostat. Thus, sensors are placed every \SI{50}{cm} along the \dword{detmodule} height with increased frequency in the first \SI{50}{cm}, closest to the bottom of the cryostat and the last \SI{50}{cm}, closest to the top of the cryostat, where spacing between sensors is reduced to \SI{10}{cm}.
 \end{itemize}


 
 In order to address concerns related to possible differences in the sensor readings prior to 
 and after installation in a \dword{detmodule}, a dynamic temperature monitor allows cross-calibration of sensors in situ. 
 \fixme{difference in voltage, or differences in the sensor reading that may happen? I think the latter... (anne)}
 Namely, this T-gradient monitor  can move vertically while installed in the \dword{detmodule}, which allows for precise cross-calibration between the sensors in situ at predefined locations, as well as in between them. \fixme{not clear}
 The procedure for cross-calibrations is the following: the temperature reading is taken at the lowest position with all sensors. The stepper motor then moves the carrier rod up \SI{50}{cm},  
 putting all sensors in the previous location of their neighbor that was \SI{50}{cm} above them. 
 Then the second reading is taken. In this manner, except for the lowest position we have temperature measurement at each location with two adjacent sensors, and by linking the temperature offsets between the two readings at each location, temperature readings from all sensors are cross-calibrated in situ, cancelling all offsets due to electromagnetic noise or any parasitic resistances that may have prevailed despite the four point connection to the sensors that should cancel most of the offsets. These measurements are taken with very stable current source, which ensures high precision of repeated temperature measurements over time. The motion of the dynamic T-monitor is stepper motor operated, 
delivering measurements with high spatial resolution. 

\subsubsection{Dynamic T-gradient Monitor Design}

A dynamic T-gradient monitor consists of three distinct parts: a carrier rod on which sensors are mounted, an enclosure above the cryostat housing the space that allows vertical motion of the carrier rod 1.5\,m above its lowest location, and the motion mechanism. The motion mechanism consists of a stepper motor connected to a gear and pinion through a ferrofluidic dynamic seal. The sensors have two pins that are soldered to a printed circuit board (PCB). Two wires are soldered to the common soldering pad for each pin, individually.   There is a cutout in the PCB around the sensor that allows free flow of argon for more accurate temperature reading.  Stepper motors typically have very fine steps allowing high-precision positioning of the sensors.  Figure~\ref{fig:fd-slow-cryo-dt-monitor-overview} shows the overall design of the dynamic T-gradient monitor with the sensor carrier rod, enclosure above the cryostat and the stepper motor mounted on the side of the enclosure. The enclosure consists of two parts connected by a six-way cross flange. One side of this cross flange is used to for the signal wires, another side is used as a viewing window, while the two other ports are spares. Figure~\ref{fig:fd-slow-cryo-sensor-mount} (left) shows the mounting of the PCB board on the carrier rod and mounting on the sensor on the PCB along with the four point connection to the signal readout wires. Finally, Figure~\ref{fig:fd-slow-cryo-sensor-mount} (right) shows the stepper motor mounted on the side of the rod enclosure. The motor is kept outside, at room temperature, and its power and control cables are also kept outside.
 \fixme{above pgraph needs some work}

\begin{dunefigure}[Dynamic T-gradient monitor overview]{fig:fd-slow-cryo-dt-monitor-overview}
  {An overview of the dynamic T-gradient monitor.}
 \includegraphics[width=0.11\textwidth,angle=90]{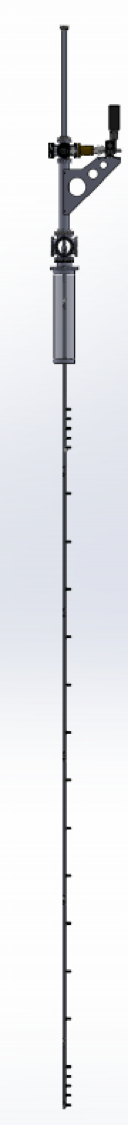}
\end{dunefigure}
\begin{dunefigure}[Sensor-cable assembly for dynamic T-gradient monitor]{fig:fd-slow-cryo-sensor-mount}
  {Left: Sensor mounted on a PCB board and PCB board mounted on the rod. Right:
    The driving mechanism of the dynamic T-gradient monitor. It consists of a stepper motor driving the pinion and gear linear motion mechanism. }
  \includegraphics[width=0.40\textwidth]{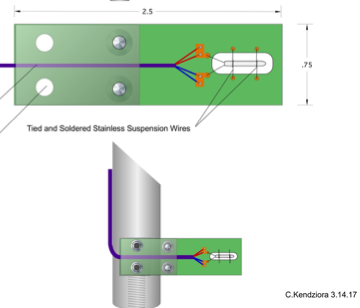}
  \hspace{3cm}%
  \includegraphics[width=0.12\textwidth]{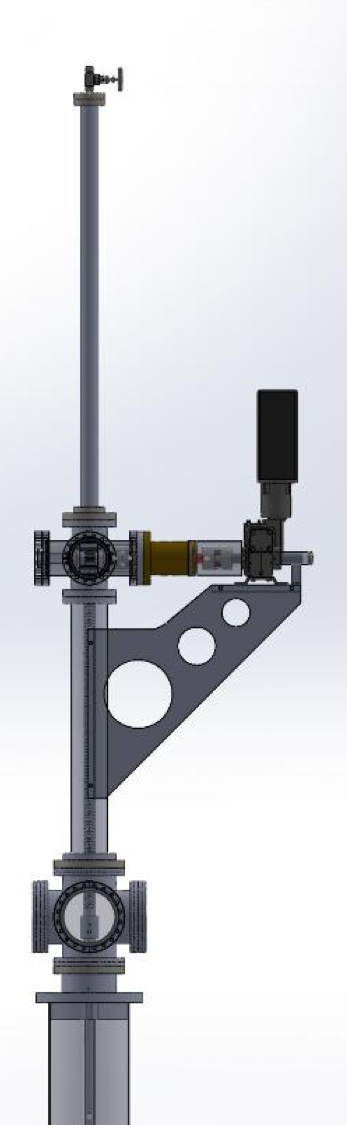}
\end{dunefigure}

\subsubsection{Individual Temperature Sensors}

T-Gradient monitors provide a vertical temperature profiling outside the TPCs. They are complemented by a coarser \twod array at the top and bottom of the
detector. Sensors, cables and readout system are the same as for the T-gradient monitors. 

In principle, a similar distribution of sensors is used at top and bottom.
Following the \dword{pdsp} design, bottom sensors use the cryogenic pipes as a support structure, while top sensors are anchored to the \dwords{gp}.
Teflon pieces (see Figure~\ref{fig:cable-support}) are used to route cables from the sensors to the CF63 side ports on \dword{dss}-cryogenics ports, which are used to extract the cables.
The PCB sensor's support, cables and connection to the flanges are the same as for the static T-gradient monitors. 

\begin{dunefigure}[Cable supports for individual temperature sensors]{fig:cable-support}
  {Left: support for two cables on ground planes. Right: Supports for three cables  mounted on cryogenics pipes using split clamps}
  \includegraphics[width=0.3\textwidth]{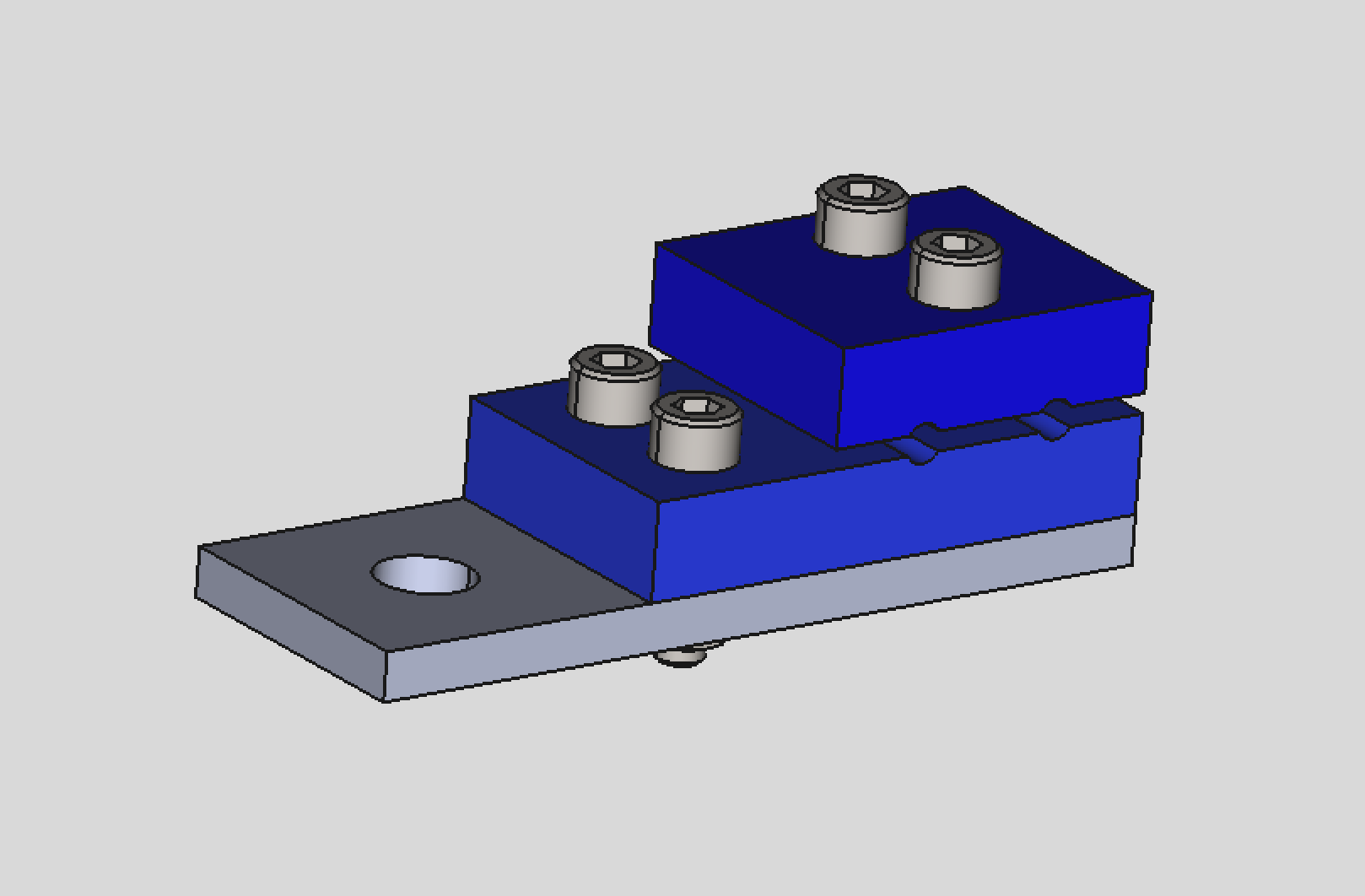}
  \includegraphics[width=0.315\textwidth]{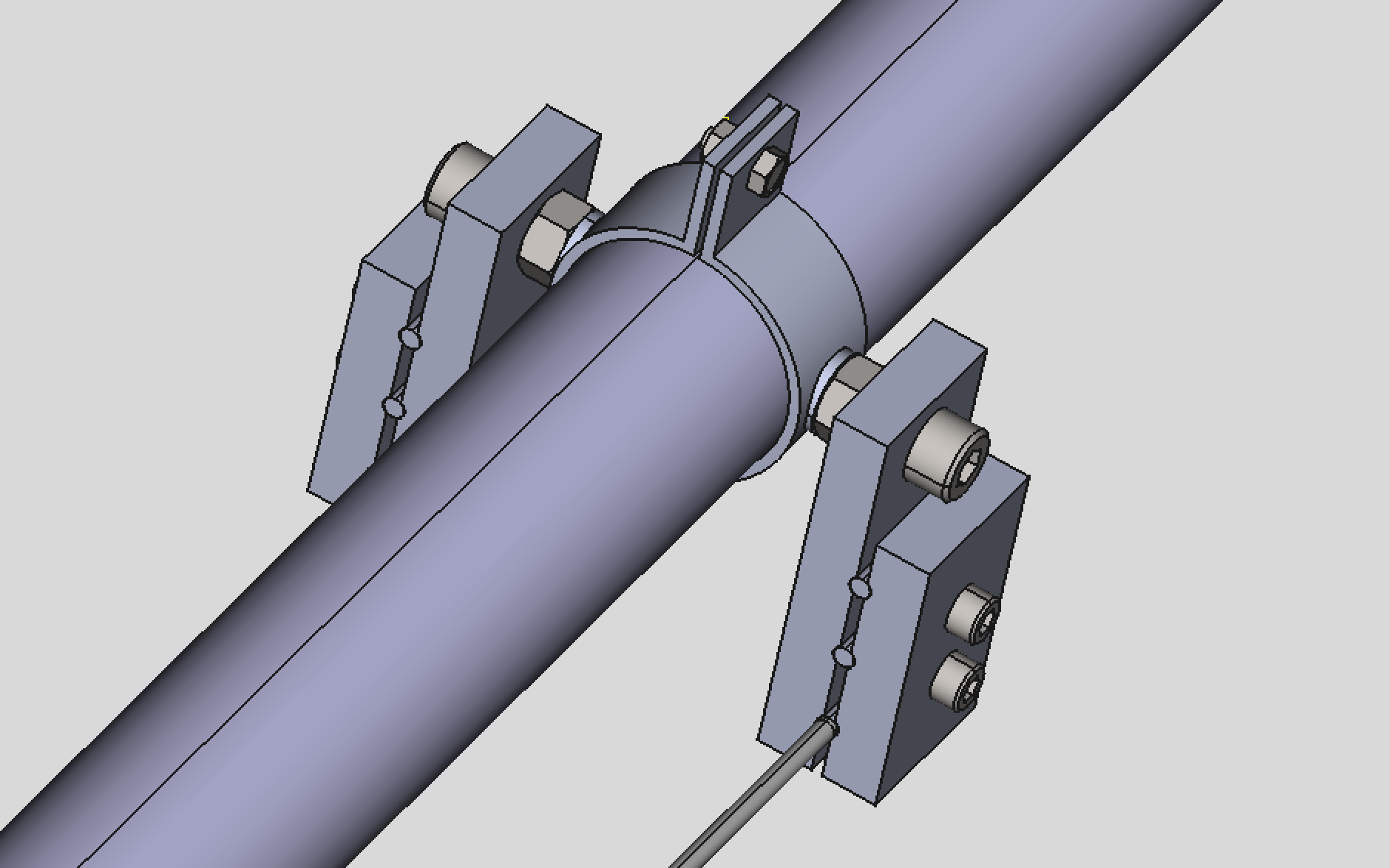}
\end{dunefigure}

\subsubsection{Readout System for Thermometers}
\label{sec:fdgen-slow-cryo-therm-readout}

A high precision and very stable system is required to achieved the design precision of $<\SI{5}{mK}$.
The proposed readout system is the one used in \dword{pdsp}, which is based on a variant of an existing mass PT100 temperature readout system developed at
CERN for one of the LHC experiments. The system consists of three parts:
\begin{itemize}
\item An accurate current source for the excitation of the temperature sensors, implemented by a compact electronic circuit using a high-precision voltage reference from Texas Instruments~\footnote{Texas Instruments\texttrademark{}, \url{http://www.ti.com/}.};
\item A multiplexing circuit based on an Analog Devices ADG707\footnote{Analog Devices\texttrademark{},\url{http://www.analog.com/media/en/technical-documentation/data-sheets/ADG706_707.pdf}.} multiplexer electronic device;
\item A high-resolution and accuracy voltage signal readout module based on National Instruments~\footnote{National Instruments\texttrademark{}, \url{http://www.ni.com/en-us.html/}.} NI9238, which has \SI{24}{bit} resolution over a \SI{1}{V} range.
  This module is inserted in a National Instruments compact RIO device that distributes the temperature values to the main slow control software
  through the standard OPC UA protocol. The Ethernet \dword{daq} also includes the multiplexing logic.
\end{itemize}

The current mode of operation averages over \num{2000} samples taken every second for each sensor. 
As inferred from Figure~\ref{fig:Trepro} the system has a resolution better than
\SI{1}{mK}, the \rms of one of the offsets in the stable region.


\subsection{Liquid Level Monitoring}
\label{sec:fdgen-slow-cryo-liq-lev}

The goals for the level monitoring system are basic level sensing when filling, and precise level sensing during static operations. 

For filling the \dword{detmodule} the differential pressure between the top of
the detector and known points below it can be converted to depth with
the known density of \lar.  The temperatures of \dwords{rtd} at known
heights may also be used to determine when the cold liquid has reached
each \dword{rtd}.

During operation, the purpose of liquid level monitoring is twofold:
the cryogenics system uses it to tune the \lar flow and 
the \dword{detmodule} uses it to guarantee that the top \dwords{gp} are always
submerged (otherwise the risk of dielectric breakdown is high).
Two differential pressure level meters are installed as part of
the cryogenics system, one on each side of the \dword{detmodule}.  They 
have a precision of \SI{0.1}{\%}, which corresponds to \SI{14}{mm} at the
nominal \lar surface.  This precision is sufficient for the single
phase detector, since the plan is to keep the \lar surface at least \SI{20}{cm} above the \dwords{gp} (this is the value used for the \dword{hv}
interlock in \dword{pdsp}); thus, no additional level meters are
required for the single phase. However, in the \dual \lar
system the surface level should be controlled at the millimeter level,
which can be accomplished with capacitive monitors. Using the same
capacitive monitor system in each detector reduces design differences
and provides a redundant system for the \single.  Either system
could be used for the \dword{hv} interlock.

Table \ref{tab:fdgen-liq-lev-req} summarizes the
requirements for the liquid level monitor system.

\begin{dunetable}
[Liquid level monitor requirements]
{p{0.45\linewidth}p{0.50\linewidth}}
{tab:fdgen-liq-lev-req}
{Liquid level monitor requirements}   
Requirement & Physics Requirement Driver \\ \toprowrule
 Measurement accuracy (filling) \(\sim \SI{14}{mm}\) & Understand status of detector during filling \\ \colhline
 Measurement accuracy (operation, \dual) \(\sim \SI{1}{mm}\) & Maintain correct depth of gas phase. (Exceeds \single requirements) \\ \colhline
 Provide interlock with \dword{hv} & Prevent damage to \dword{detmodule} from \dword{hv} discharge in gas \\
\end{dunetable}

Differential pressure level meters will be purchased from commercial sources.
Installation methods and positions will be determined as part of the
cryogenics internal piping plan.  Sufficient redundancy will be designed in
to ensure that no single point of failure compromises the level measurement.

Multiple capacitive level sensors are deployed along the top of
the fluid to be used during stable operation and checked against each
other.

During operations of the \dword{wa105}, the cryogenic programmable logic controller (PLC) continuously checked the measurements from one level meter on the charge readout plane (\dword{crp}) in order to regulate the flow from the liquid recirculation to maintain a constant liquid level inside the cryostat. Continuous measurements from the level meters around the drift cage and the \dword{crp} illustrated the stability of the liquid level within the \SI{100}{\micro\meter} intrinsic precision of the instruments. The observation of the level was complemented by live feeds from the custom built cryogenic cameras, hereby providing qualitative feedback on the position and flatness of the surface.

In the \dword{dpmod} each \dword{crp} is suspended with three ropes actuated by step motors and can be individually adjusted with respect to the liquid level in terms of parallelism and distance. The \dword{crp} has a gap of 1 cm in between the bottom of the surface of the \dwords{lem} and the extraction grid. The \dword{crp} positioning requirements include its parallelism to the liquid surface and the liquid level situated across this gap, namely the extraction grid is required to be immersed and the \dwords{lem} are required to not be wet.  The distance of the \dword{crp} with respect to the liquid level is required to be measured with the accuracy of 1 mm. The slow control system takes care of the control of the step motors, which control the \dword{crp} suspensions. The positions of the motors (and of the corresponding suspension points) can be surveyed from the cryostat roof with a typical 0.2 mm accuracy. By reading the controls of the motors and by knowing the results of this survey, the position of the \dwords{crp} with respect to the liquid can be predicted in absolute terms, provided that the liquid level is measured also with respect to the same reference system by the cryostat level meters. 

The horizontal alignment of the \dwords{crp} can be measured as well thanks to the survey of the reference points on the suspension feedthroughs. The relative position of a \dword{crp} with respect to the liquid level can be measured with high-accuracy level meters mounted on the \dword{crp} itself. These level meters are multi-parallel plate capacitors with the capacitance changing as a function of the liquid level between the plates. These level meters can be attached to the \dword{crp} borders and they guarantee a measurement accuracy at the level of 0.1 mm. It is possible to install these level meters only on the borders of the \dwords{crp}, which are at the periphery of the detection surface and not on the borders where the \dwords{crp} are side by side. In addition to the installed level meters, the liquid height in the extraction region of all \dwords{crp} can be inferred by measuring the capacitance between the grid and the bottom electrode of each \dword{lem}. Averaging over all 12 \dwords{lem}, the measured values of this capacitance typically ranged  from 150 pF with the liquid below the grid to around 350 pF when the \dwords{lem} are submerged. This  method offers the potential advantage of monitoring the liquid level in the \dword{crp} extraction region with a 50$\times$50 \si{cm$^2$}  granularity and can be used for the \dword{crp} level adjustment in the future large-scale detectors where, due to the space constraints, placement of the level meters along the \dword{crp} perimeter is not possible.

\subsection{Gas Analyzers}
\label{sec:fdgen-slow-cryo-gas-anlyz}

 Gas analyzers are commercially produced modules that measure trace quantities of specific gases contained within a stream of carrier gas. The carrier gas for DUNE is argon, and the trace gases of interest are oxygen ($\text{O}_2$), water ($\text{H}_2\text{O}$), and nitrogen ($\text{N}_2$). Oxygen and water impact the electron lifetime in \dword{lar}, while $\text{N}_2$ impacts the efficiency of scintillation light production. In the \dword{lar} environment, these trace gases represent contaminants that need to be kept at levels below \SI{0.1}{ppb}.
The argon is sampled from either the argon vapor in the ullage or from the \dword{lar} by the use of small diameter tubing run from the sampling point to the gas analyzer. Typically the tubing runs from the sampling points are connected to a switchyard valve that is used to route the sample points to the desired gas analyzers. Figure~\ref{fig:GA-switchyard} is a photo of such a switchyard.

\begin{dunefigure}[Gas Analyzer switchyard]{fig:GA-switchyard}
  {A Gas Analyzer switchyard that routes sample points to the different gas analyzers.}
  \includegraphics[width=0.35\textwidth]{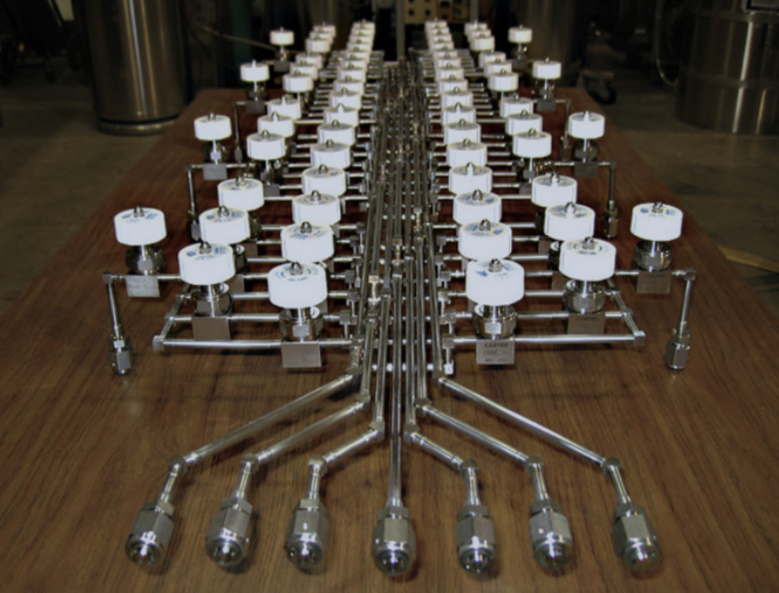}%
\end{dunefigure}

Gas analyzers can be used to:

\begin{enumerate}
\item Eliminate the air atmosphere from the cryostat after detector installation to levels low enough to begin cooldown is an argon piston purge followed by a recirculation of the remaining argon gas through the filtration system. This process is described more fully in Section~\ref{sec:fdgen-slow-cryo-install}. Figure~\ref{fig:GA-purge} shows the evolution of the $\text{N}_2$, $\text{O}_2$, and $\text{H}_2\text{O}$ levels from gas analyzer data taken during the purge and recirculation stages of the DUNE \dword{35t} 
phase 1 run.

\item Track trace $\text{O}_2$ and $\text{H}_2\text{O}$ contaminants from the $\>$tens of ppb to the hundreds of ppt. This is useful when other means of monitoring the impurity level (e.g., purity monitors, or \dshort{tpc} tracks) are not yet sensitive. Figure~\ref{fig:GA-O2} shows an example plot of the $\text{O}_2$ level at the beginning of \dword{lar} purification from one of the later \num{35}\si{t} 
\dword{hv} runs.

\item Monitor the tanker \dword{lar} deliveries purity during the cryostat-filling period. This allows tracking the impurity load on the filtration system and rejecting any deliveries that are out of specifications. Likely specifications for the delivered \dword{lar} are in the \SI{10}{ppm} range per contaminant.

\end{enumerate}

\begin{dunefigure}[Gas analyzer purge]{fig:GA-purge}
  {Plot of the $\text{O}_2$, $\text{H}_2\text{O}$, and $\text{N}_2$ levels during the piston purge and gas recirculation stages of the \num{35}\si{t} 
  phase 1 run.}
  \includegraphics[width=0.65\textwidth]{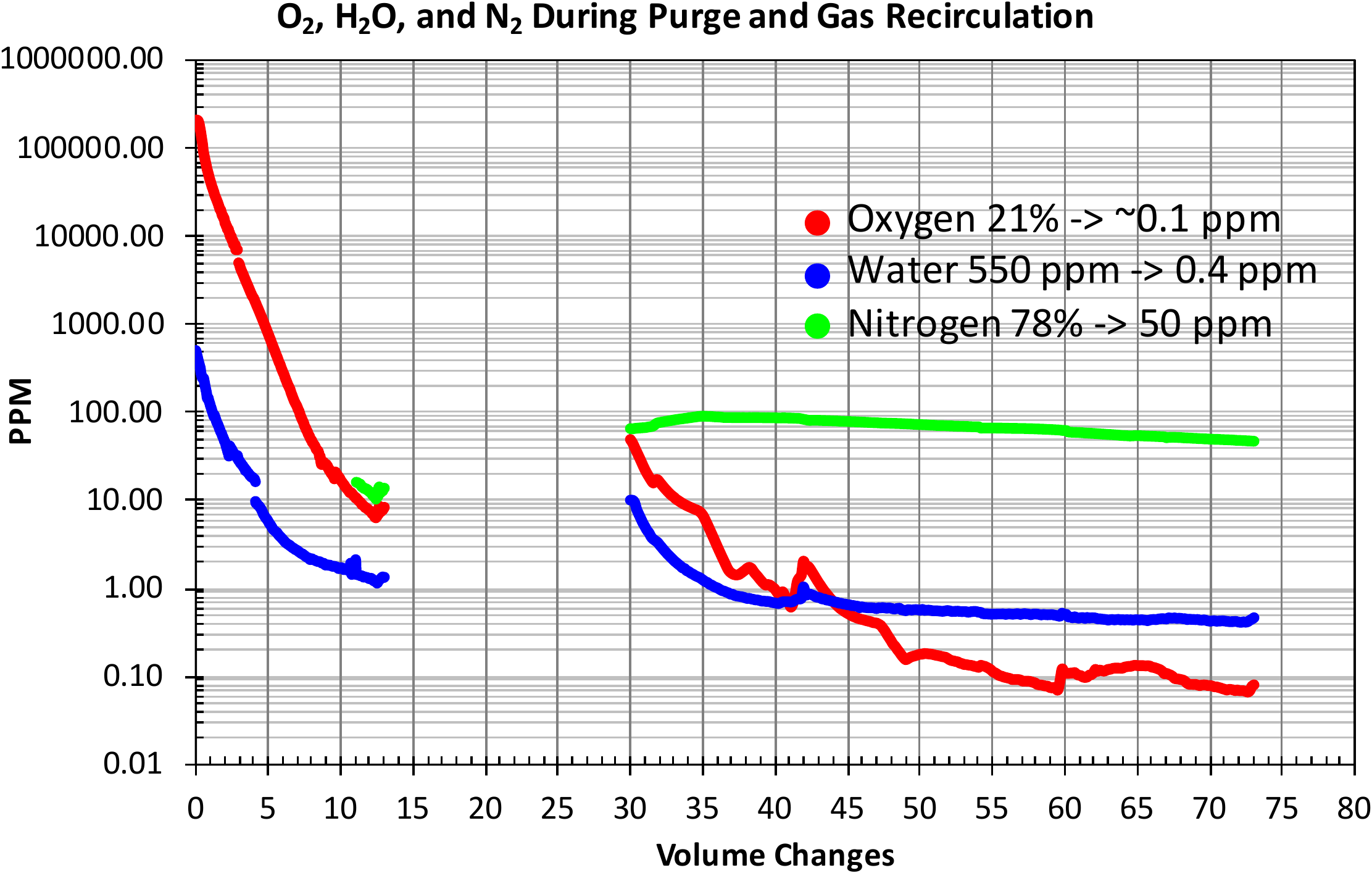}%
\end{dunefigure}

\begin{dunefigure}[Gas analyzer $O_2$ level after \dword{lar} filling]{fig:GA-O2}
  {$\text{O}_2$ as measured by a precision $\text{O}_2$ analyzer just after the \dword{35t} 
  was filled with \dword{lar}, continuing with the \dword{lar} pump start and beginning of \dword{lar} recirculation through the filtration system. As the gas analyzer loses sensitivity, the purity monitor is able to pick up the impurity measurement. Note that the purity monitor is sensitive to both $\text{O}_2$ and $\text{H}_2\text{O}$ impurities giving rise to its higher level of impurity.}
  \includegraphics[width=0.7\textwidth]{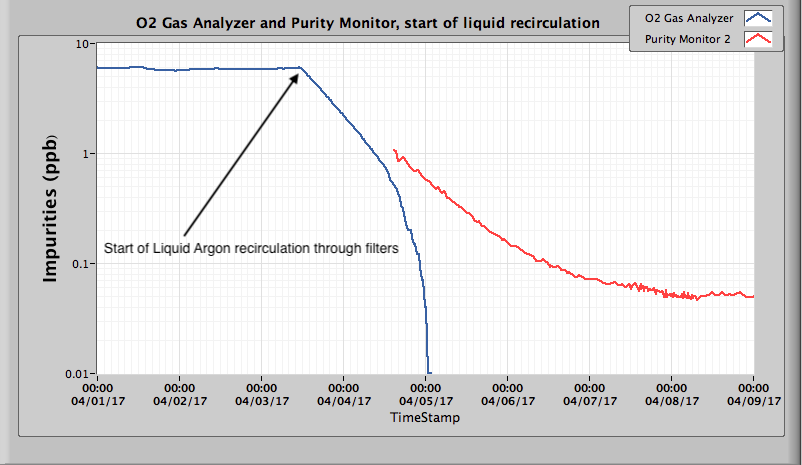}%
\end{dunefigure}

As any one gas analyzer covers only one contaminant species and \numrange{3}{4} orders of magnitude of range, multiple units are needed both for the three contaminant gases and to cover the ranges that are seen between the cryostat closure to the beginning of \dshort{tpc} operations:
\SI{20}{\percent} to $\lesssim 100$~ppt for $\text{O}_2$,
\SI{80}{\percent} to $\lesssim 1$~ppm for $\text{N}_2$, and
$\sim \SI{1}{\percent}$ to $\lesssim 1$~ppb for $\text{H}_2\text{O}$.
Since the total cost of these analyzers exceeds $\SI{100}[\textdollar]{k}$, it is useful to be able to  sample more than a single location or cryostat with the same gas analyzers. At the same time, the tubing run lengths from the sample point should be as short as possible in order to keep the response of the gas analyzer timely. This puts some constraints on the sharing of devices since, for example, the argon deliveries are at the surface, perhaps necessitating a separate surface gas analyzer.


\subsection{Cameras}
\label{sec:fdgen-slow-cryo-cameras}

Cameras provide direct visual information about the state of the
\dword{detmodule} during critical operations and when damage or unusual
conditions are suspected.  Cameras in the \dword{wa105} allowed spray from cool-down
nozzles to be seen and the level and state of the \lar to be
observed as it covered the \dword{crp} \cite{Murphy:20170516}.  A camera was
used in the Liquid Argon Purity Demonstrator
cryostat\cite{Adamowski:2014daa} to study \dword{hv} discharges in
\lar, and in EXO-100 during operation of a TPC
\cite{Delaquis:2013hva}.  Warm cameras viewing \lar from a distance
have been used to observe \dword{hv} discharges in \lar in
fine detail \cite{Auger:2015xlo}.  Cameras are commonly used during
calibration source deployment in many experiments (e.g., the
\kamland ultra-clean system \cite{Banks:2014hra}).

In DUNE, cameras are used to verify the stability, straightness,
and alignment of the hanging TPC structures during cool-down and
filling; to ensure that there is no bubbling near the \dwords{gp}
(\single) or \dwords{crp} (\dual); to inspect the
state of movable parts in the \dword{detmodule} (calibration devices, dynamic
thermometers) as needed; and to closely inspect parts of the TPC as
necessary following any seismic activity or other unanticipated
occurrence.  These functions are performed using a set of fixed
\textit{cold} cameras permanently mounted at fixed points in the cryostat
for use during filling and commissioning, and a movable, replaceable
\textit{warm} inspection camera that can be deployed through any free
instrumentation flange at any time throughout the life of the
experiment.  Table \ref{tab:fdgen-cameras-req} summarizes the
requirements for the camera system.

\begin{dunetable}
[Camera system requirements]
{p{0.45\linewidth}p{0.50\linewidth}}
{tab:fdgen-cameras-req}
{Camera system requirements}   
 Requirement & Physics Requirement Driver \\ \toprowrule
 \multicolumn{2}{l}{\bf General} \\ \specialrule{1.5pt}{1pt}{1pt}
 No component may contaminate the \lar{}. & High \lar purity is required for TPC operation. \\ \colhline
 No component may produce bubbles in the liquid argon if the \dword{hv} is on. & Bubbles increase risk of \dword{hv} discharge. \\ \colhline
 No point in the camera system shall have a field greater than \SI{15}{kV/cm} when the drift field is at nominal voltage. & Fields must be well below \SI{30}{kV/cm} to avoid risk of \dword{hv} discharge.\\ \colhline
The camera system shall not produce measurable noise in any detector system. & Low noise is required for TPC operation. \\ \colhline
 Cameras provide the viewing functionality as agreed upon with the other subsystems for viewing, as documented in the ICDs with the individual systems. \\ \toprowrule
\multicolumn{2}{l}{\bf Cold cameras}\\ \specialrule{1.5pt}{1pt}{1pt}
Minimize heat dissipation when camera not in operation. & Do not generate bubbles when \dword{hv} is on. \\ \colhline
Longevity must exceed \num{18} months. & Cameras must function throughout cryostat filling and detector commissioning. \\ \colhline
Frame rate \(\geq\SI{10}{\per s}\). & Observe bubbling, waves, detritus, etc. \\ \toprowrule
\multicolumn{2}{l}{\bf Inspection cameras}\\ \specialrule{1.5pt}{1pt}{1pt}
Keep heat transfer to \lar low when in operation. & Do not generate bubbles, some use cases may require operation when \dword{hv} is on. \\ \colhline
Deploy without exposing \lar to air. & Keep \lar free of N2 and other electronegative contaminants. \\ \colhline
Camera enclosure must be replaceable. & Replace broken camera, or upgrade, throughout life of experiment. \\ \colhline
{\bf Light emitting system} \\ \colhline
No emission of wavelengths shorter than \(\SI{400}{nm}\) & Avoid damaging \dword{tpb} waveshifter. \\ \colhline
Longevity must exceed \num{18} months. & Lighting for fixed cameras must function throughout cryostat filling and detector commissioning. \\ \colhline
\end{dunetable}

The following sections describe the design considerations for the cold
and warm cameras and the associated lighting system.  The same basic
design may be used for both the single and dual phase detectors.

\subsubsection{Cryogenic Cameras (Cold)}

The fixed cameras 
monitor the following items during filling:
\begin{itemize}
\item Positions of corners of \dword{apa} or \dword{crp}, \dword{cpa} or cathode, \dwords{fc}, \dwords{gp} (\SI{1}{mm} resolution);
\item Relative straightness and alignment of \dword{apa}/\dword{crp}, \dword{cpa}/cathode, and \dword{fc} (\(<\sim\SI{1}{mm}\));
\item Relative position of profiles and endcaps (\SI{0.5}{mm} resolution);
\item State of \lar surface: e.g., the presence of bubbling or debris.
\end{itemize}

There are published articles and unpublished presentations describing
completely or partially successful operation of low-cost,
off-the-shelf \dword{cmos} cameras in custom enclosures immersed in cryogens.
(e.g., EXO-100: \cite{Delaquis:2013hva}; DUNE \dword{35t} test
\cite{McConkey:2016spe}; \dword{wa105}: \cite{Murphy:20170516}.)  Generally
it is reported that such cameras show poor performance and ultimately
fail to function below some temperature of order \SIrange{150}{200}{K}, but some report that their cameras recover fully after
being stored (not operated) at temperatures as low as \SI{77}{K} and
then brought up to minimum operating temperature.

However, as with photon sensors, experience has also shown that it is
non-trivial to ensure reliable and reproducible mechanical and
electrical integrity of such cameras in the cryogenic environment 
(e.g., \cite{McConkey:2016spe} and
\cite{Valencia-Rodriquez:20180130}).  Off-the-shelf cameras and camera
components are generally only specified by the vendors and original
manufacturers for operation down to \SI{-40}{\celsius} or \SI{-50}{\celsius}.
In addition, many low-cost cameras use digital interfaces not intended
for long distance deployment, such as USB (\(2\sim\SI{5}{m}\)) or CSI (circuit
board scale), leading to signal degradation and noise problems.

The design for the DUNE fixed cameras uses an enclosure based on
the successful EXO-100 design\cite{Delaquis:2013hva}, which was also
used successfully in
LAPD (see Figure~\ref{fig:gen-fdgen-cameras-enclosure}). The enclosure is
 connected to a stainless steel gas line to allow it to be
flushed with argon gas at low enough pressure to prevent
liquification, using the same design as the gas line for the beam plug
tested in the \dword{35t} \dword{hv} test and in \dword{protodune}.  A thermocouple in the
enclosure allows temperature monitoring, and a heating element
provides temperature control.  The camera transmits its video
signal using either a composite video signal over shielded coax or
Ethernet over optical fiber.  Most importantly, the DUNE \dword{cisc}
consortium must work with vendors to design camera circuit boards that
are robust and reliable in the cryogenic environment.

\begin{dunefigure}[A camera enclosure]{fig:gen-fdgen-cameras-enclosure}
  {CAD exploded view of vacuum-tight camera enclosure suited for cryogenic applications from \cite{Delaquis:2013hva}.
    (1) quartz window, (2 and 7) copper gasket, (3 and 6) flanges, (4) indium wires, (5) body piece, (8) signal \fdth.
  }
  \includegraphics[width=0.6\textwidth]{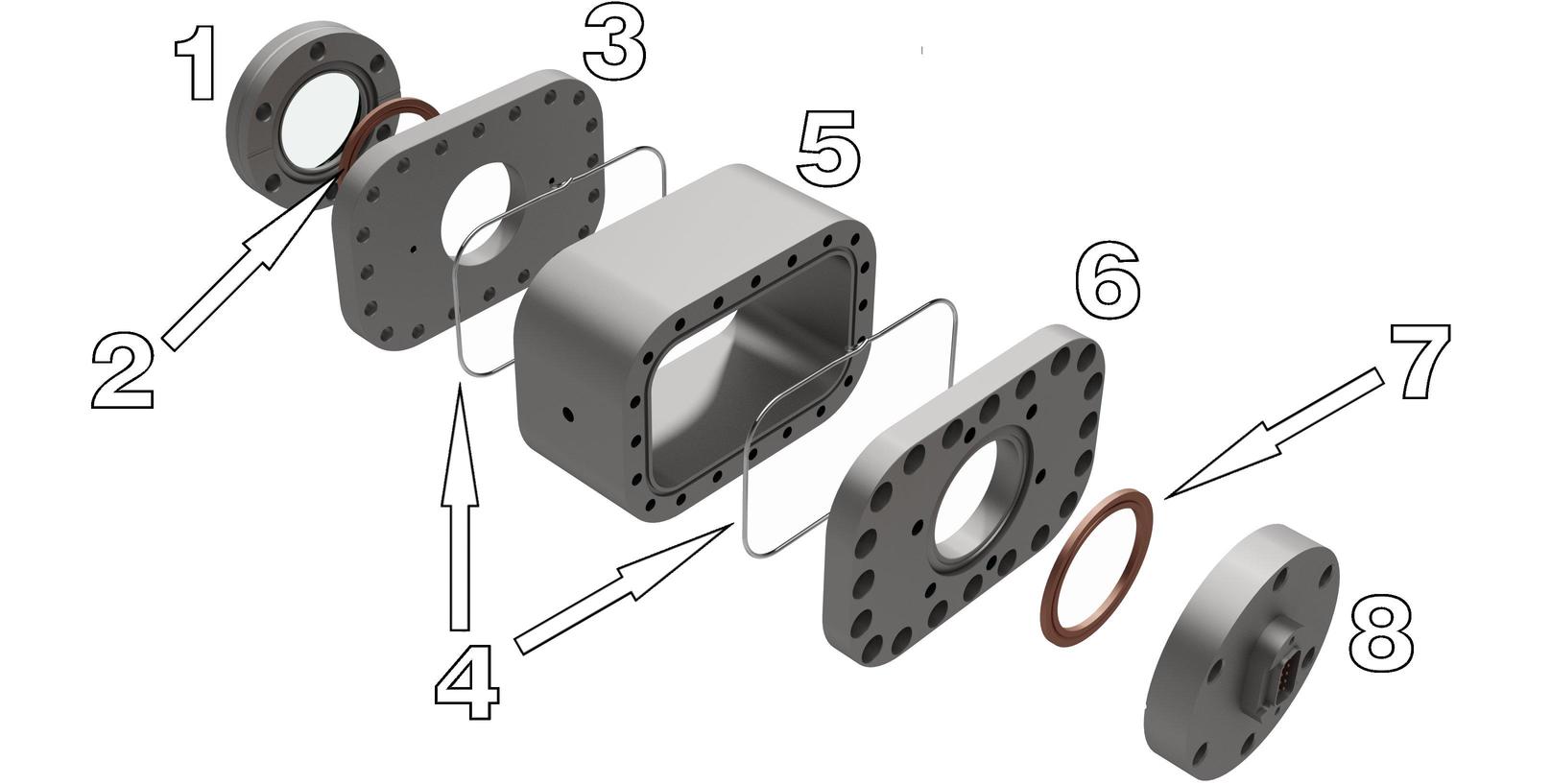}%
\end{dunefigure}

\subsubsection{Inspection Cameras (Warm)}

The inspection cameras are intended to be as versatile as possible.
However, the following locations have been identified as likely
to be of interest:
\begin{itemize}
\item Status of \dword{hv} \fdth and cup;
\item Status of \dword{fc} profiles, endcaps (\SI{0.5}{mm} resolution);
\item $y$-axis deployment of calibration sources;
\item Status of thermometers, especially dynamic thermometers;
\item \dword{hv} discharge, corona, or streamers on \dword{hv} \fdth, cup, or \dword{fc};
\item Relative straightness and alignment of \dword{apa}/\dword{crp}, \dword{cpa}/cathode, and \dword{fc} (\SI{1}{mm} resolution);
\item Gaps between \dword{cpa} frames (\SI{1}{mm} resolution);
\item Relative position of profiles and endcaps (\SI{0.5}{mm} resolution);
\item Sense wires at top of outer wire planes in \single \dword{apa} (\SI{0.5}{mm} resolution).
\end{itemize}

Unlike the fixed cameras, the inspection cameras need operate only as
long as inspection lasts, as the camera can be replaced in case of failure.  It
is also more practical to keep the cameras continuously \textit{warm}
(above \SI{-150}{\celsius}) during deployment; this offers 
more options for commercial cameras, e.g., 
the same model camera used successfully to observe discharges
in \lar from \SI{120}{cm} away \cite{Auger:2015xlo}.

\begin{dunefigure}[Inspection camera design]{fig:gen-fdgen-cameras-movable}
  {An overview of the inspection camera design.}
  \includegraphics[width=0.3\textwidth]{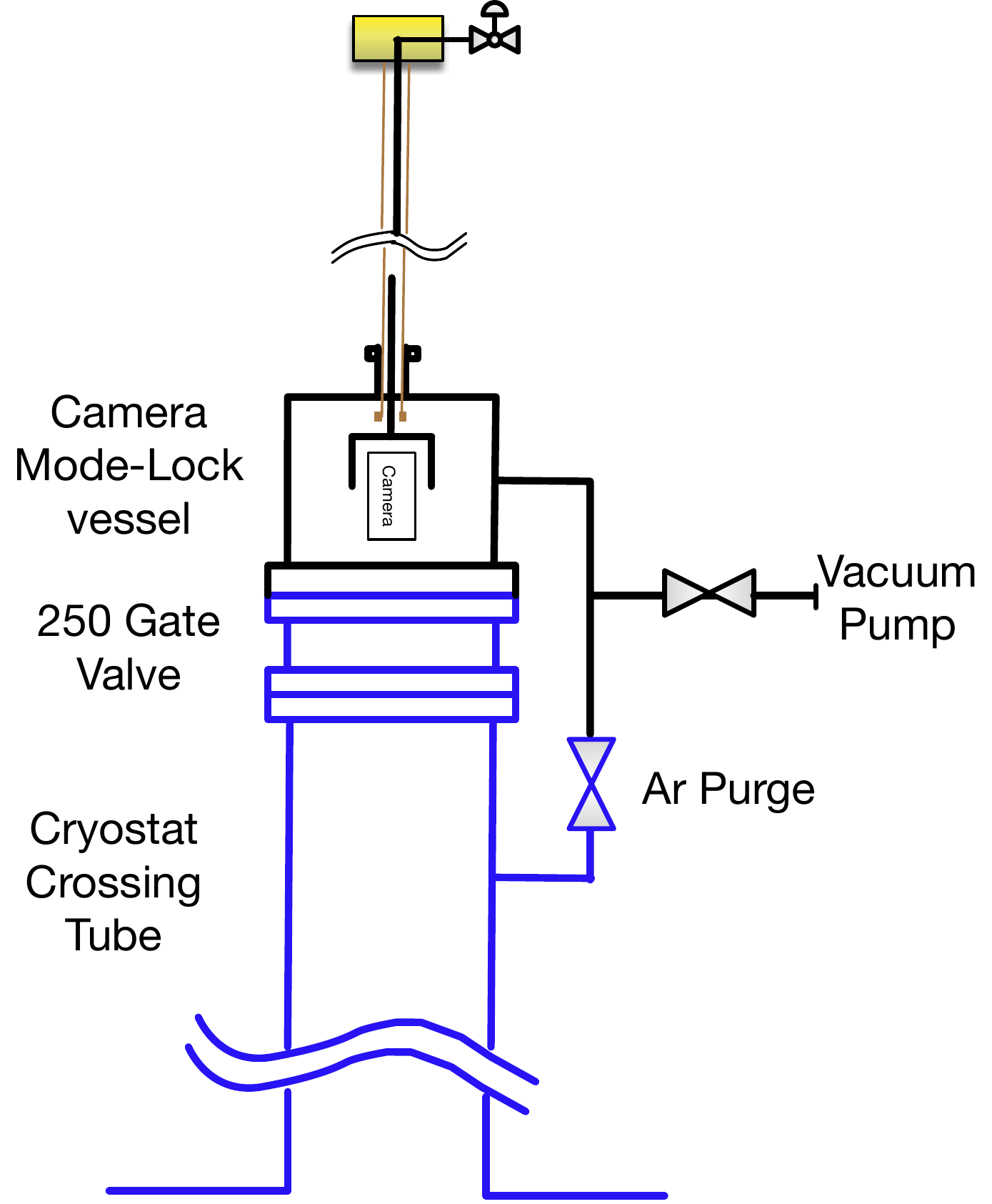}%
\end{dunefigure}

The design for the inspection camera system employs the same basic
enclosure design as for cold cameras, but mounted on an insertable
fork using a design similar to the dynamic temperature probes. See
Figure~\ref{fig:gen-fdgen-cameras-movable} and
Figure~\ref{fig:fd-slow-cryo-sensor-mount}.  The entire system is sealed to
avoid contamination with air. In order to avoid contamination, the
camera can only be deployed through a \fdth equipped with a gate
valve and a purging system, similar to that used for the vertical axis
calibration system at \kamland~\cite{Banks:2014hra}. The entire system
is  purged with pure argon gas before the gate valve is opened.

Motors above the flange allow rotation and vertical movement of the fork. 
 A chain drive system, with motor
mounted on the end of the fork, allows tilting of the camera assembly, 
creating a point-tilt mount with vertical motion capability.
Taking into account the room above the cryostat flanges and the
thickness of the cryostat insulation, a vertical range of motion of
\SI{1}{m} inside the cryostat is achievable.
The motors for rotation and vertical motion are located outside the sealed
volume, coupled mechanically using ferrofluidic seals, thus reducing
contamination risks and allowing for manual rotation of the vertical
drive in the event of a motor failure.  A significant protyping and
testing effort is needed to finalize and validate this design.

\subsubsection{Light-emitting system}
The light-emitting system is based on \dwords{led} with the capability of illuminating the interior with selected
wavelengths (IR and visible) that are suitable for detection by the
cameras.  Performance criteria for the light-emission system are based
on the efficiency of detection with the cameras, in conjunction with
adding minimal heat to the cryostat. The use of very high-efficiency
\dwords{led}  
helps reduce heat generation; as an
example, one \SI{750}{nm} \dword{led} has a specification of
\SI{32}{\%} conversion of electrical input power to light.

While data on the performance of \dwords{led} at cryogenic temperatures is sparse,
some studies related to NASA projects~\cite{Carron:2017zzz} 
indicate that \dword{led} efficiency increases with reduced temperature,
and that the emitted wavelengths may change, particularly for blue \dwords{led}.
The wavelength changes cited would have no impact on illumination, however, since
in order  to avoid degradation of wavelength-shifting materials in the \dword{pds},
such short wavelength \dwords{led} would not be used.

\begin{dunefigure}[\dword{led} chain for illumination]{fig:gen-cisc-LED}
  {Suggested \dword{led} chain for lighting inside the cryostat, with
    dual-wavelength and failure-tolerant operation.}
\includegraphics[width=0.7\textwidth]{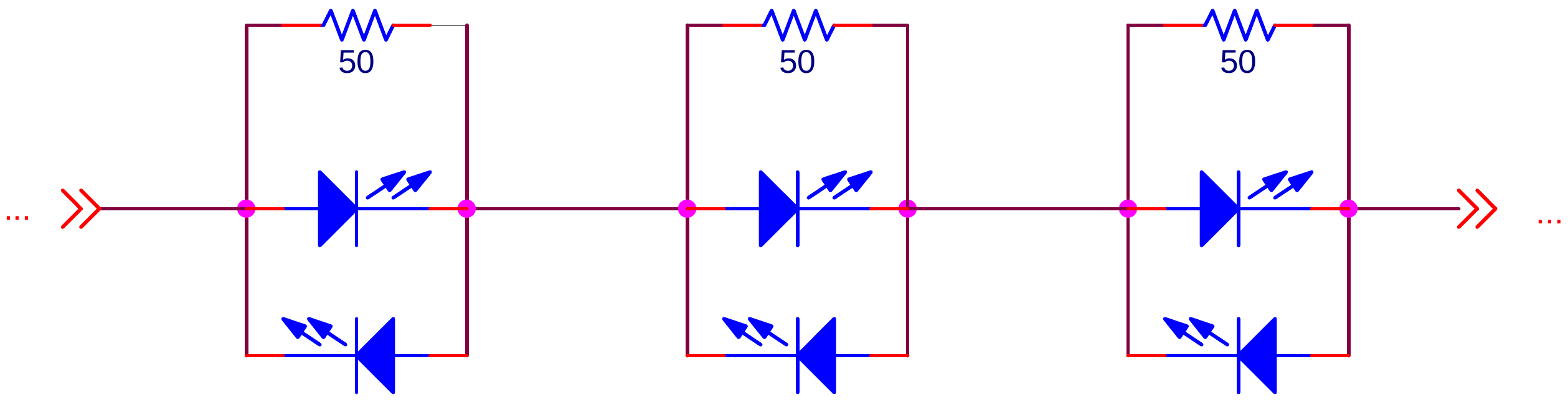}
\end{dunefigure}

\fixme{shoulds and woulds...}
A \textit{chain} of \dwords{led} should be connected in series and driven with a
constant-current circuit. It would be advantageous to pair each
\dword{led} in parallel with an opposite polarity \dword{led} and a resistor
(see Figure~\ref{fig:gen-cisc-LED}).
This allows two different wavelengths of illumination with a single installed
chain (by changing the direction of the drive current) and 
continued use of an \dword{led} chain even if individual \dwords{led} fail.

The \dwords{led} should be placed as a \textit{ring light} around the outside of each
camera lens, pointing in the same direction as the lens, to 
illuminate the part of the \dword{detmodule} within the field of
view of the camera. Commercially available \dwords{led} can be obtained with
a range of angular spreads, and can be matched to the needs of the
cameras without additional optics.


\subsection{Cryogenics Test Facility}
\label{sec:fdgen-slow-cryo-test-facil}
The cryogenics test facility is intended to provide the access to a small ($<$ \num{1} ton) to intermediate ($\sim$ \num{4} tons) volumes of purified TPC-grade \lar{}. Hardware that needs liquid of purity this high include any device intending to drift electrons for millisecond time periods. Not all devices require purity this high, but some may need a relatively large volume to provide the needed prototyping environment. Of importance is a relatively fast turn-around time of approximately a week for short prototyping runs.

Figure~\ref{fig:CryTest-Blanche} shows the Blanche test stand cryostat at \fnal.

\begin{dunefigure}[CryTest Blanche Test]{fig:CryTest-Blanche} 
  {Blanche Cryostat at \fnal. This cryostat holds $\sim 0.75$ tons of \lar{}.}
  \includegraphics[width=0.35\textwidth]{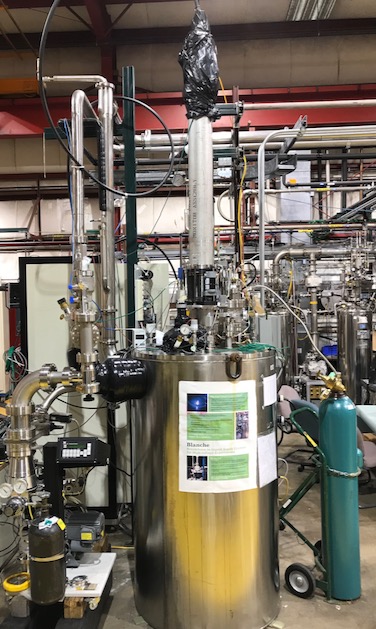}%
\end{dunefigure}


\subsection{Cryogenic Internal Piping}
\label{sec:fdgen-slow-cryo-int-piping}
\label{sec:fdsp-slow-cryo-int-piping}
\label{sec:fddp-slow-cryo-int-piping}


The cryogenic internal piping comprises several manifolds to
distribute the liquid and gaseous argon inside the cryostat during all
phases (e.g., gaseous purge, liquid distribution, cool down) and
various pipe stands to return argon to the outside (e.g.,
boil-off gaseous argon).  Vacuum-insulated pipe stands are needed to
transition from inside to outside in a way that does not affect the
purity and does not introduce a significant heat load.

LBNF has the expertise for engineering design and installation of the
detector internal piping, while the \dword{cisc} consortium has the expertise
on the physics requirements, the relevant risk registries, and the
interfaces with other detector systems. Ultimate responsibility for
costing the internal cryogenic piping system also lies with the \dword{cisc}
consortium. It is important for these two groups to interact closely
to ensure that the system enables achievement of the physics, 
avoids interference with other detector systems, and mitigates
risks.

DUNE has formed a cryogenics systems working group with conveners from
both the \dword{cisc} consortium and LBNF. This group has both LBNF and
\dword{cisc} members and provides an official forum where we interface and
establish the final design.

The initial design for the cryogenic internal piping calls for some
\SI{750}{m} of pipe per cryostat for purging and filling, laid out as
shown in Figure~\ref{fig:fd-slow-cryo-int-piping}-Left, and 20 flange-pipes assemblies, as the one shown
on the right pannel of Figure~\ref{fig:fd-slow-cryo-int-piping}, with a CF DN250 flange penetrated by two $\sim$ \SI{2.2}{m} long pipes.

\begin{dunefigure}[Cryogenic internal piping]{fig:fd-slow-cryo-int-piping}
  {Left: Cryogenic internal piping for purging (red) and filling (blue). Right: Cool-down pipes, \lar in blue (vacuum jacketed) and gaseous argon in red. }
  \includegraphics[height=0.3\textwidth]{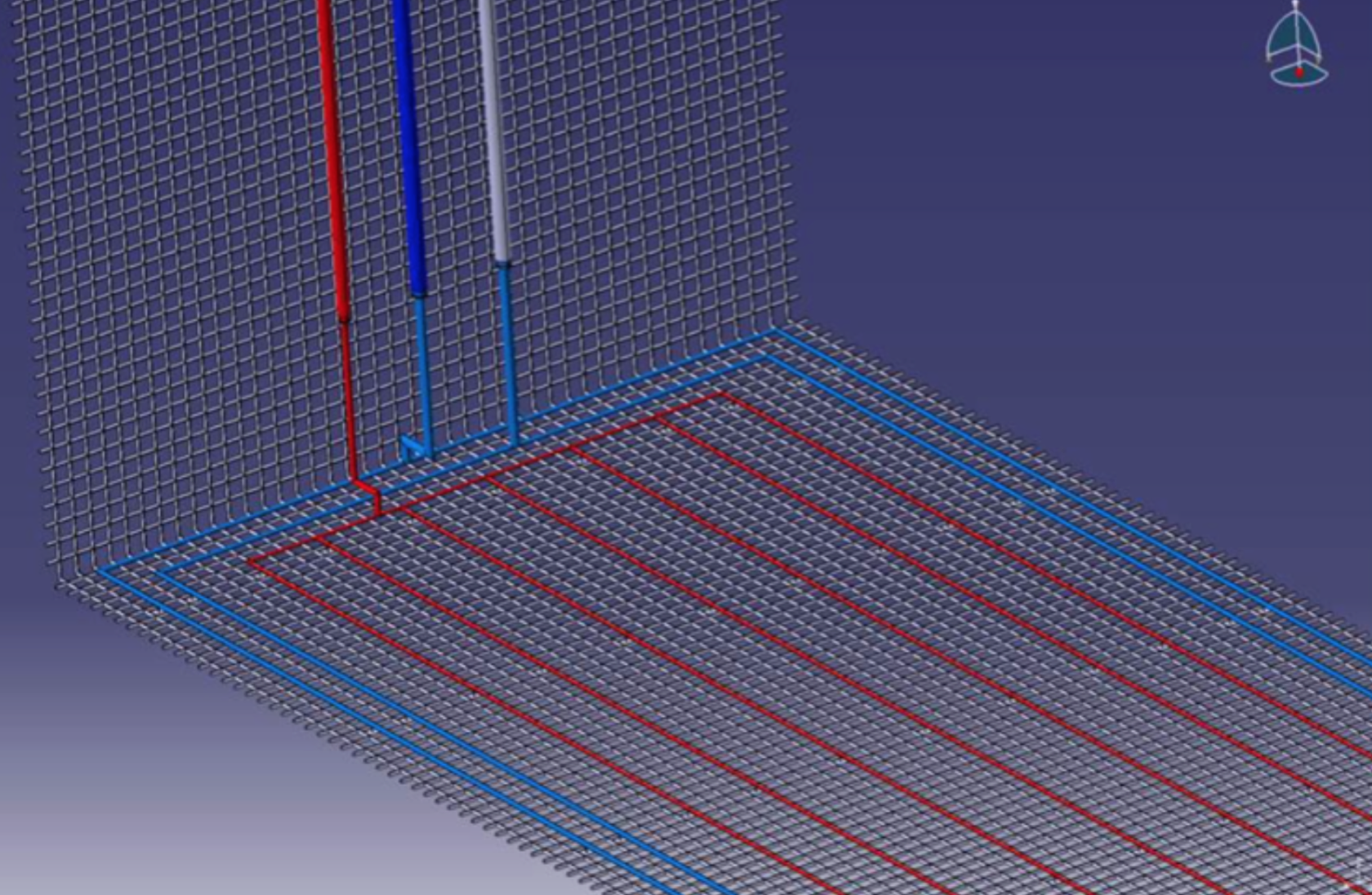}
  \includegraphics[height=0.3\textwidth]{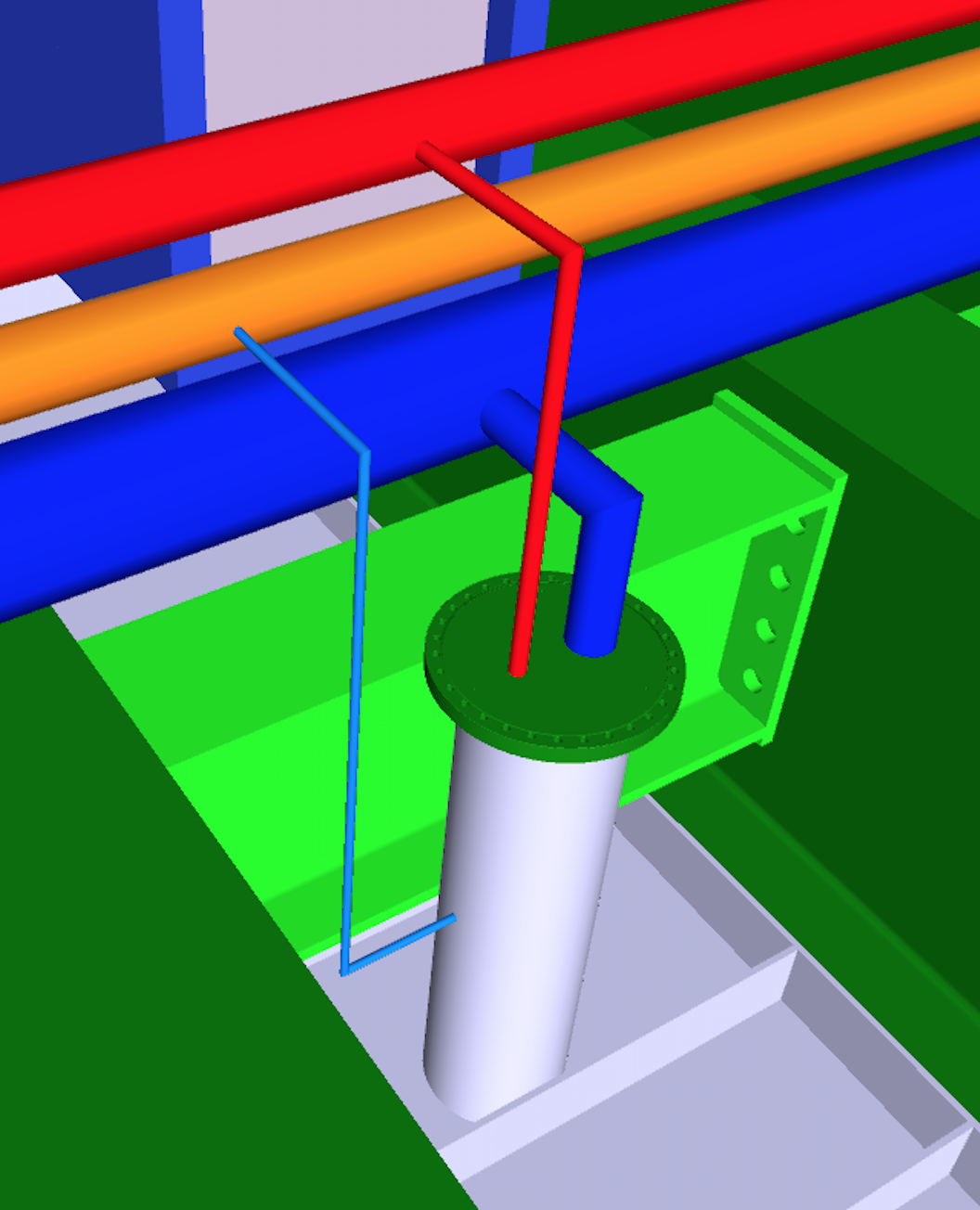}
\end{dunefigure}

\section{Slow Controls}
\label{sec:fddp-slow-cryo-ctrl}


The slow controls system collects, archives, and displays data from
a broad variety of sources, and provides real time alarms and
warnings for detector operators. Data is acquired via network
interfaces.  Figure \ref{fig:gen-slow-controls-diagram} shows the
connections between major parts of the slow controls system and other
systems.  

\begin{dunefigure}[Slow Controls connections and data]{fig:gen-slow-controls-diagram}
{Typical Slow Controls system connections and data flow}
\includegraphics[width=0.7\textwidth]{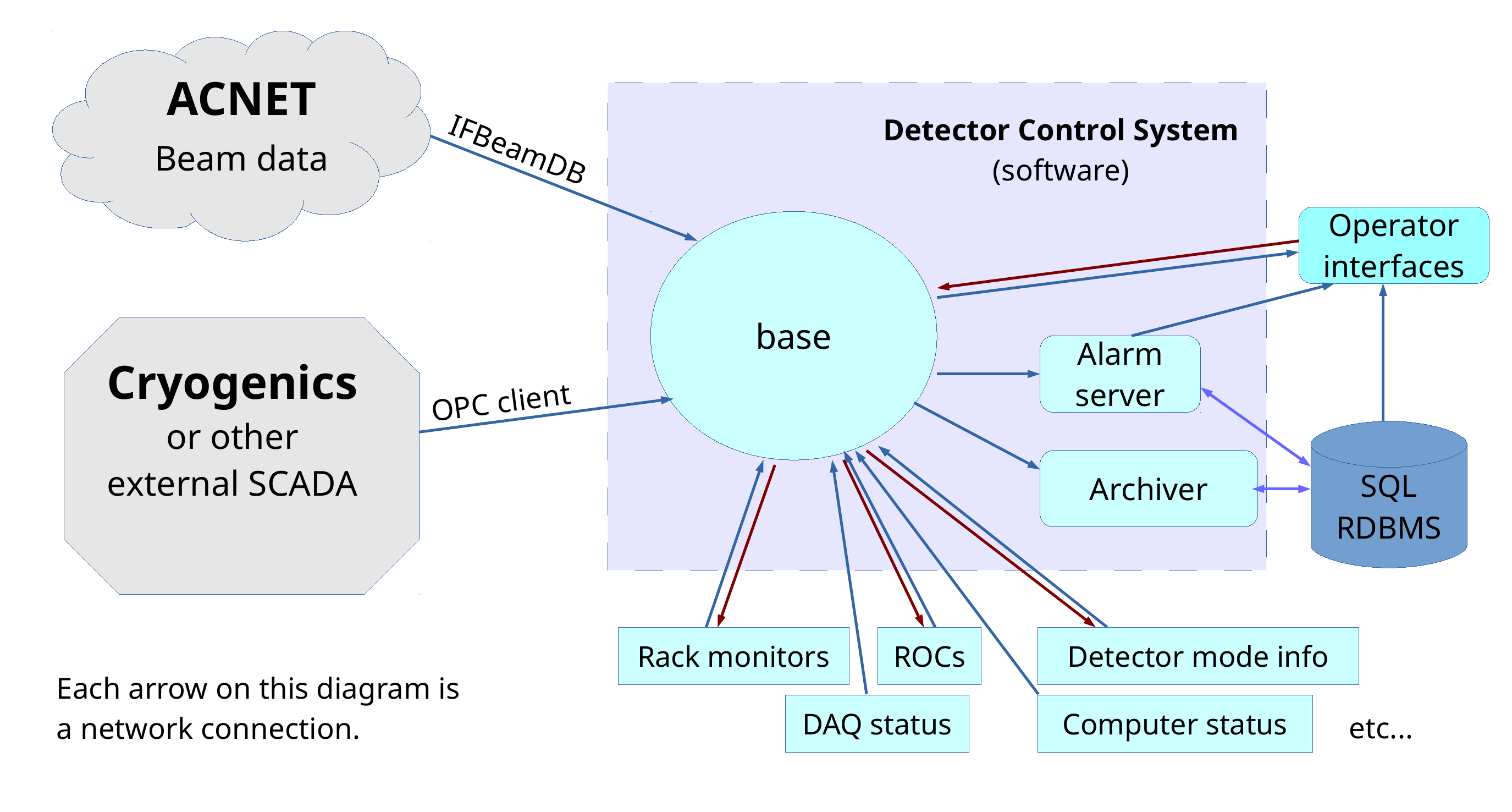}
\end{dunefigure}

\subsection{Slow Controls Hardware}
\label{sec:fddp-slow-cryo-hdwr}

The slow controls will always require a small amount of dedicated network and
computing hardware as described below.  It also relies on common
infrastructure, as described in
Section~\ref{sec:fddp-slow-cryo-slow-infra}.

In the current concept of the \dual detector, \dword{hv} biasing for the \dword{lem}, extraction grid, and \dword{pds} also falls within the scope of slow controls.  This hardware could be assigned to different consortia in the future.  For now its description is provided here.

\subsubsection{Slow Controls Network Hardware}
\label{sec:fddp-slow-cryo-slow-network}
The slow controls data originates from the cryogenic instrumentation discussed in
Section~\ref{sec:fdgen-cryo-instr} and from other systems,
and is collected by software running on servers
(Section~\ref{sec:fddp-slow-cryo-slow-compute})
housed in the underground data room in the \dlong{cuc},
where data is archived in a central \dword{cisc} database.
The instrumentation data is transported over
conventional network hardware from any sensors located in the cryogenic
plant.  However, the readouts that are located in the racks on top the
cryostats have to be careful about grounding and noise.  Therefore, each
rack on the cryostat has a small network switch that sends
any network traffic from that rack to the \dshort{cuc} via a fiber transponder.
This is the only network hardware specific to slow controls;
network infrastructure requirements are described in
Section~\ref{sec:fddp-slow-cryo-slow-infra}.

\subsubsection{Slow Controls Computing Hardware}
\label{sec:fddp-slow-cryo-slow-compute}

Two servers (a primary server and a replicated backup) suitable for the needed relational database discussed
in Section~\ref{sec:fddp-slow-cryo-sw} are located in the \dshort{cuc} data
room, with an additional
two servers to perform \dword{fe} monitoring interface services: for
example, assembling dynamic \dword{cisc} monitoring web pages from the adjacent
databases.  Any special purpose software, such as iFix or EPICS, would
also run here. It is expected that one or two more servers will accommodate
these programs.
Replicating this setup on a per-module basis would allow for easier
commissioning and independent operation, accommodate different module
design (and the resulting differences in database tables), and ensure
sufficient capacity.  Including four sets of networking hardware, this
would fit tightly into one rack or very comfortably into two.

\subsubsection{High Voltage Biasing}
\label{sec:fddp-slow-cryo-slow-hvbias}

The \dword{hv} biasing for a \dword{lem} requires \SI{-5}{kV/channel}, with \num{41} channels per \dword{crp}.  The extraction grid requires \SI{-10}{kV/channel}, with  one channel per \dword{crp}. The  \dword{lem} biasing system requires a current measurement sensitivity of \SI{50}{pA}. The biasing of the photomultipliers for the \dword{pds} needs of order \num{1000} channels operating up to a voltage of \SI{-3}{kV}.

\subsubsection{CRP and PDS Calibration}
\label{sec:fddp-slow-cryo-slow-crppdscalib}

Calibration of the charge readout system will be performed by charge injection onto the anode strips of the \dwords{crp} and into the preamplifiers. The slow controls system must include a high-accuracy pulser connected to the \dword{daq} system for time-tagging of the pulse triggers. The pulser will be configurable as a function of the different calibration runs (pulsing of the strips or pulsing of the amplifiers) acting as common pulse source. This will be connected to a distribution system that allows dispatching the pulses to the \dword{crp} anode strips or to the signal chimneys hosting the analog \dword{fe} cards.The distribution system will also include the possibility of selecting specific channels to which the pulses will be distributed. The \dword{pds} will also require a calibration system constituted by a common light source constantly monitored with reference photodetectors and a network of optical fibers distributing the light to each photomultiplier.


\subsection{Slow Controls Infrastructure}
\label{sec:fddp-slow-cryo-slow-infra}

The total number of slow controls quantities and the update rate are low enough
that the data rate will be in the tens of kilobytes per second range
(Section~\ref{sec:fddp-slow-cryo-quant}), placing minimal requirements
on the local network infrastructure.
Network traffic out of \surf to \fnal will be primarily database calls
to the central \dword{cisc} database: either from monitoring applications, or from
database replication to the offline version of the \dword{cisc} database.  This
traffic is of a low enough bandwidth that the proposed general purpose
links both out of the mine and back to \fnal can accommodate it.

Up to two racks of space and appropriate power and cooling are
available in the \dshort{cuc}'s \dword{daq} server room for \dword{cisc} usage.
Somewhat less space than that is currently envisioned, as described in
\ref{sec:fddp-slow-cryo-slow-compute}.

\subsection{Slow Controls Software}
\label{sec:fddp-slow-cryo-sw}


The slow controls software includes the following components in order 
to provide complete monitoring and control of detector subsystems:
\begin{itemize}
 \item{Control systems base} that performs input and output operations
  and defines processing logic, scan conditions, alarm conditions,
  archiving rate, etc.;
 \item{Alarm server} that monitors all channels and sends alarm
  messages to operators; 
 \item{Data archiver} that performs automatic sampling and storage of
  values for history tracking;
 \item{Integrated operator interface} that provides display panels for
  controls and monitoring.
\end{itemize}

An additional requirement for the software is the ability to indirectly
interface with external systems (e.g., cryogenics control
system) and databases (e.g., beam database) to export data into
slow controls process variables (or channels) for archiving and status
displays. This allows integrating displays and warnings into one
system for the experiment operators, and 
provides integrated
archiving for sampled data in the archived database. In this case, one
can imagine an input output controller (IOC) running on a central \dword{daq}
server provides soft channels for these data.
Figure~\ref{fig:gen-slow-controls-diagram} shows a typical workflow of a
slow controls system.

In terms of key features of the software, a highly evolved software is
needed that is designed for managing real-time data exchange, scalable
to large number of channels and high bandwidth if needed. The software
should be well documented, supported, and known to be reliable. The base
software should also allow easy access of any channel by name. The
archiver software should allow data storage in an SQL database with
adjustable rates and thresholds such that one can easily retrieve data
for any channel by using channel name and time range. Among the key
features, the alarm server software should remember state, support
arbitrary number of clients and provide logic for delayed alarms and
acknowledging alarms. As part of the software, a standard naming
convention for channels is followed to aid dealing with large
number of channels and subsystems.

\subsection{Slow Controls Quantities}
\label{sec:fddp-slow-cryo-quant}


The final set of quantities to monitor will ultimately be determined
by the needs of the subsystems being monitored, as documented in
appropriate  interface control documents (ICDs), and continually revised based on operational
experience.  The total number of quantities to monitor has been very
roughly estimated by taking the total number of quantities monitored
in \microboone and scaling by the detector length and the number of
planes, giving a number in the range of \numrange{50}{100}k.
Quantities are expected to update on average no faster than once per minute.
The subsystems
to be monitored include the 
cryogenic instrumentation
described in this chapter, the other detector systems, and relevant
infrastructure and external devices. Table \ref{tab:gen-slow-quant}
lists the kind of quantities expected from each system.

\begin{dunetable}
[Slow controls quantities]
{p{0.3\textwidth}p{0.6\textwidth}}
{tab:gen-slow-quant}
{Slow controls quantities}
System & Quantities \\ \toprowrule
\multicolumn{2}{l}{\bf Detector Cryogenic Instrumentation } \\ \specialrule{1.5pt}{1pt}{1pt}
Purity monitors & Anode and cathode charge, bias voltage and current, flash lamp status, calculated electron lifetime \\ \colhline
Thermometers & Temperature, position of dynamic thermometers \\ \colhline
Liquid level & Liquid level \\ \colhline
Gas analyzers & Purity level readings \\ \colhline
Cameras & Camera voltage and current draw, temperature, heater current and voltage, lighting current and voltage \\ \colhline
Cryogenic internal piping & \fdth gas purge flow and temperature \\ \toprowrule
\multicolumn{2}{l}{\bf Other Detector Systems } \\ \specialrule{1.5pt}{1pt}{1pt}
\dword{hv} systems & Drift \dword{hv} voltage, current; end-of-field cage current, bias; ground plane currents \\ \colhline
TPC electronics & Voltage and current to electronics \\ \colhline
\dword{pd} & Bias, current, electronics \\ \colhline
\dword{daq} & Warm electronics currents and voltages; run status; \dword{daq} buffer sizes, trigger rates, data rates, GPS status, etc.; computer and disk health status; other health metrics as defined by \dword{daq} group \\ \colhline
\dword{crp} / \dword{apa} & Bias voltages and currents \\ \toprowrule
\multicolumn{2}{l}{\bf Infrastructure and external systems } \\ \specialrule{1.5pt}{1pt}{1pt}
Cryogenics (external) & Status of pumps, flow rates, inlet and return temperature and pressure (via OPC or similar SCADA interface) \\ \colhline
Beam status & Protons on target, rate, target steering, beam pulse timing (via IFBeamDB) \\ \colhline
Near detector & Near detector run status (through common slow controls database) \\ \colhline
Racks power and status & PDU current and voltage, air temperature, fan status if applicable, interlock status (fire, moisture, etc.) \\
\end{dunetable}

\subsection{Local Integration}
\label{sec:fddp-slow-cryo-slow-loc-integ}


The local integration for the slow controls consists entirely of software
and network interfaces with systems outside of the scope of the \dword{detmodule}.
This includes the following:
\begin{itemize}
\item Readings from the LBNF-managed external cryogenics systems, for status of pumps, flow rates, inlet and return temperature and pressure, which are implemented via OPC or a similar SCADA interfaces;
\item Beam status, such as protons on target, rate, target steering, beam pulse timing, which are retrieved via IFBeamDB;
\item Near Detector status, which can be retrieved from a common slow controls database.
\end{itemize}

This integration occurs after both the slow controls and non-detector
systems are in place.  The LBNF-\dword{cisc} interface is managed by the
Cryogenics Systems working group described in Section
\ref{sec:fdgen-slow-cryo-int-piping}.  IFBeamDB is already well established.
An internal near-detector--\dword{fd} working group may be established
to coordinate detector status exchange between near and far sites interfacing.


\section{Interfaces}
\label{sec:fdgen-slow-cryo-intfc}


The \dword{cisc} consortium interfaces with all other consortia, task forces (calibration), working groups (physics, software/computing) and technical coordination.
This section provides a brief summary; further details can be found in references~\cite{bib:docdb6745}-\cite{bib:docdb7018}.

There are obvious interfaces with detector consortia since \dword{cisc} provides full rack monitoring (rack fans, thermometers and rack protection system),
interlock status bit monitoring (not the actual interlock mechanism) and monitoring and control for all \pwrsupps. The  \dword{cisc} consortium must maintain close contacts with all other consortia to ensure that specific hardware choices have acceptable slow controls (SC)  solutions.  
Also, installation of instrumentation devices interferes with other devices and must be coordinated with the respective consortia.  
On the software side \dword{cisc} must define, in coordination with other consortia, the quantities to be monitored and controlled by slow controls and the corresponding alarms,
archiving and GUIs. 



A major interface is the one with the cryogenics system. As mentioned in Section~\ref{sec:fdgen-slow-cryo-purity-mon} purity monitors and gas analyzers are essential
to mitigate the liquid argon contamination risk. The appropriate interlock mechanism to prevent the cryonenics system from irreversible contamination
must be designed and implemented. 

Another important interface is the one with the \dword{hv} system \cite{bib:docdb6787} since several aspects related with safety must be taken into account. 
For all instrumentation devices inside the cryostat, electric field simulations are needed to guaranty proper shielding is in place.
Although this is a \dword{cisc} responsibility, input from \dword{hv} is crucial.
During the deployment of inspection cameras, generation of bubbles must be avoided when \dword{hv} is on, as it can lead to discharges.

There are also interfaces with the \dword{pds}~\cite{bib:docdb6730}. Purity monitors and the light-emitting system for cameras both emit light that might damage \dwords{pd}.
Although this should be understood and quantified, \dword{cisc} and the \single \dword{pds} may have to define the necessary hardware interlocks
that avoid turning on any other light source accidentally when \dwords{pd} are on.

The \dword{daq}-\dword{cisc} interface~\cite{bib:docdb6790} is described in Section~\ref{sec:fd-daq-intfc-sc}. 
\dword{cisc} data is stored both locally (in \dword{cisc} database servers in the
\dword{cuc}) and offline (the databases are replicated back to \fnal)
in a relational database indexed by timestamp.
This allows bidirectional communications between \dword{daq} and \dword{cisc} by
reading or inserting data into the database as needed for non-time-critical information.



\dword{cisc} also interfaces with the beam and cryogenics group since at least the status of these systems must be monitored.


Assuming that the scope of software \& computing \dword{swc} group includes scientific computing support to project activities, there are substancial interfaces with that group~\cite{bib:docdb7126}. 
The hardware interfaces resposibility of the \dword{swc} include networking installation and maintenance,
maintenance of SC servers  and any additional computing hardware needed by instrumentation devices.
\dword{cisc} provides the needed monitoring for power distribution units (PDUs). Regarding software interfaces the \dword{swc} group  provides:
(1) SC database maintenance, (2) API for accessing the SC database offline,
(3) UPS packages, local installation and maintenace of software needed by \dword{cisc}, and (4) \dword{swc} creating and maintaining computer accounts on production clusters. 
Additionally  \dword{cisc} provides the required monitoring and control of \dword{swc} quantities including alarms, archiving, and GUIs, where applicable.

The \dword{cisc} consortium may have additional hardware interfaces with the not-as-yet formed calibration consortium \cite{bib:docdb7072}. Indeed, since 
the shared ports are multi-purpose to enable deploying various devices,
both \dword{cisc} and calibration must interact in terms of flange design and sharing space around the ports. Also, \dword{cisc} might use calibration ports to extract cables from \dword{cisc} devices. 
At the software level, \dword{cisc} is responsible for calibration device monitoring (and control to the extent needed) and 
monitors the interlock bit status for laser and radioactive sources.

\dword{cisc} indirectly interfaces to physics through the shared devices. One specific need for physics is to extract 
instrumentation or slow controls data to correlate high-level quantities to low-level or calibration data.
This requires tools to extract data from the slow controls database (see \dword{cisc}-\dword{swc} interface document \cite{bib:docdb7126}).
A brief list of what \dword{cisc} data is needed by physics is given in the \dword{cisc}-Physics interface document \cite{bib:docdb7099}. 

Interfaces between \dword{cisc} and technical coordination are detailed in the corresponding interface documents for the facility \cite{bib:docdb6991}, installation \cite{bib:docdb7018}
and integration facility~\cite{bib:docdb7045}.

\fixme{Interfaces with technical coordination has to be expanded}








\section{Installation, Integration and Commissioning}
\label{sec:fdsp-slow-cryo-install}    
\label{sec:fddp-slow-cryo-install} 
\label{sec:fdgen-slow-cryo-install} 

\subsection{Cryogenics Internal Piping}
\label{sec:fdgen-slow-cryo-install-pipes}

The installation of internal cryogenic pipes occurs soon after the cryostat is completed or towards the end of the cryostat completion, depending on how the cryostat work proceeds. A concrete installation plan will be developed by the company designing the internal cryogenics. It depends on how they address the thermal contraction of the long horizontal and vertical runs. We are investigating several options, which each have different installation sequences. 
All involve delivery and welding together of prefabricated spool pieces inside the cryostat, and vacuum insulation of the vertical lines. The horizontal lines are bare pipes. 

The cool-down assemblies are installed in dedicated cool-down \fdth{}s at the top, arranged in
two rows of ten each in the long direction of the cryostat. Each one features a \lar line connected to a gaseous argon line via a mixing nozzle and a gaseous argon line with spraying nozzles. The mixing nozzles generate droplets of liquid that are circulated uniformly inside the cryostat by the spraying nozzles. They are prefabricated at the vendor's site and delivered as full pieces, 
then mounted over the \fdth{}s.

The current \threed model of the internal cryogenics is developed and archived at CERN as part of the full cryostat model. CERN is currently responsible for the integration of the detector cavern: cryostat, detector, and proximity cryogenics in the detector cavern, including cryogenics on the mezzanine and main \lar circulation pumps.

The prefabricated spool pieces and the cool down nozzles undergo testing at the vendor  before delivery. The installed pieces are helium leak-checked before commissioning, but no other integrated testing or commissioning is possible after the installation, because the pipes are open to the cryostat volume. The internal cryogenics are commissioned once the cryostat is closed.

\subsection{Purity Monitors}
\label{sec:fdgen-slow-cryo-instal-pm}

The purity monitor system is built in a modular way, such that it can be assembled outside of \dword{detmodule} cryostat.  The assembly of the purity monitors themselves occurs outside of the cryostat and includes everything described in the previous section.  The installation of the purity monitor system can then be carried out with the least number of steps inside the cryostat.  The assembly itself is transported into the cryostat with the three individual purity monitors mounted to the support tubes but before installation of \dword{hv} cables and optical fibers. The support tube at the top and bottom of the assembly is then mounted to the brackets inside the cryostat that could be attached to the cables trays or the detector support structure. \fixme{brackets attached to trays or DSS depending on SP vs DP?} In parallel to this work, the \dword{fe} electronics and light source can be installed on the top of the cryostat, along with the installation of the electronics and power supplies into the electronics rack.  

Integration begins by running the \dword{hv} cables and optical fibers to the purity monitors, coming from the top of the cryostat.  The \dword{hv} cables are attached to the \dword{hv} \fdth{}s with enough length to reach each of the respective purity monitors.  The cables are run through the port reserved for the purity monitor system, along cable trays inside the cryostat until they reach the purity monitor system, and are terminated through the support tube down to each of the purity monitors.  Each purity monitor has three \dword{hv} cables that connect it to the \fdth, and then along to the \dword{fe} electronics.  The optical fibers are run through the special optical fiber \fdth, into the cryostat, and guided to the purity monitor system either using the cables trays or guide tubes.  Whichever solution is adopted for running the optical fibers from the \fdth to the purity monitor system, it must protect the fibers from accidental breakage during the remainder of the detector and instrumentation installation process.  The optical fibers are then run inside of the purity monitor support tube and to the respective purity monitors,  terminating at the photocathode of each. 

Integration  continues with the connection of the \dword{hv} cables between the \fdth and the system \dword{fe} electronics, and then the optical fibers to the light source.  The cables connecting the \dword{fe} electronics and the light source to the electronics rack are also run and connected at this point.  This allows for the system to turn on and the software to begin testing the various components and connections.  Once it is confirmed that all connections are successfully made, the integration to the slow controls system is made, first by establishing communications between the two systems and then transferring data between them to ensure successful exchange of important system parameters and measurements.  

The purity monitor system is formally commissioned 
once the cryostat is purged and a gaseous argon atmosphere is present.  At this point the \dword{hv} for the purity monitors is ramped up without the risk of discharge through the air, and the light source is turned on.  Although the drift electron lifetime in the gaseous argon is very large and therefore not measurable with the purity monitors themselves, comparing the signal strength at the cathode and anode gives a good indication of how well the light source is generating drift electrons from the photocathode and whether they drift successfully to the anode. 

\subsection{Thermometers}
\label{sec:fdgen-slow-cryo-instal-th}

Individual temperature sensors on pipes and cryostat membrane are installed prior to any detector component, right after the installation of the pipes.
First, all cable supports are anchored to pipes. Then each cable is routed individually starting from the sensor end (with IDC-4 female connector but no sensor)
towards the corresponding cryostat port. Once a port's cables are routed, 
they are cut to the same length such that they can be properly soldered
to the pins of the SUBD-25 connectors on the flange. 
To avoid damage, the sensors are installed at a later stage, just before unfolding the bottom \dwords{gp}.

Static T-gradient monitors are installed before the outer \dwords{apa}, 
after the installation of the pipes
and before the installation of individual sensors. This proceeds in several steps: (1) installation of the two stainless steel strings to the bottom and top corners of the cryostat,
(2) tension and verticality checks, (3) installation of cable supports in one of the strings, (4) installation of sensor supports in the other string, (5) cable routing starting from
the sensor end towards the corresponding cryostat port, (6) cutting all cables at the same point in that port, and (7) soldering cable wires to the pins of the SUBD-25 connectors on the flange. Then, at a later stage, just before moving corresponding \dword{apa} into its final position, (8) the sensors are plugged into IDC-4 connectors. 

For the \single{}, individual sensors on the top \dword{gp} must be integrated with the \dwords{gp}. For each \dshort{cpa} (with its corresponding four \dshort{gp} modules)
going inside the cryostat, cable and sensor supports are anchored to the \dshort{gp} threaded rods as soon as possible.
Once the \dshort{cpa} is moved into its final position and its top \dword{fc} is ready to be unfolded, sensors on those \dwords{gp} are installed. Once unfolded, cables 
exceeding the \dshort{gp} limits can be routed to the corresponding cryostat port either using neighboring \dwords{gp} or \dshort{dss} I-beams. 
\fixme{prev pgraph needs work}

Dynamic T-gradient monitors are installed after the completion of the detector.
The monitor comes in several segments with sensors and cabling already
in place. Additional slack is provided at segment joints to ease the
installation process. Segments are fed into the flange one at a 
time. The segments being fed into the \dword{detmodule} are held at the top
with a pin that prevents the segment from sliding in all the way. Then the next
segment is connected. The pin is removed, and the
segment is pushed down until the next segment top is held with the
pin at the flange. Then this next segment is installed. The
process  continues until the entire monitor is in its place
inside the cryostat. Use of a crane is foreseen to facilitate the process.
Extra cable slack at the top is provided again in order to ease  the
connection to the D-Sub standard connector flange and to allow  vertical movement of the
entire system. Then, a four-way cross flange with electric \fdth{}s on
one side and a window on the other side. The wires are connected to
the D-sub connector on the electric flange \fdth on the side. On the
top of the cross, a moving mechanism is then installed with a crane.
The pinion is connected to the top segment. The moving mechanism will
come reassembled with motor on the side in place and pinion and gear
motion mechanism in place as well. The moving mechanism enclosure  is then connected to top part of the cross and this completes the
installation process of the dynamic T-gradient monitor.
\fixme{prev pgraph needs work}

Commissioning of all thermometers proceeds in several steps. Since in the first stage only cables are installed,
the readout performance and the noise level inside the cryostat are
tested with precision resistors. Once sensors are installed the entire chain is checked again at room temperature.
The final commissioning phase occurs during and after cryostat filling.

\subsection{Gas Analyzers}
\label{sec:fdgen-slow-cryo-install-ga}
 
Prior to the piston purge and gas recirculation phases of the cryostat commissioning, the gas analyzers are installed near the 
tubing switchyard. This minimizes tubing runs and is  convenient for switching the sampling points and gas analyzers. Since each is a standalone module, a single rack with shelves, should be adequate to house the modules.

Concerning the integration, the gas analyzers typically have an analog output (\numrange{4}{20} \si{mA} or \numrange{0}{10}\si{V}) that maps to the input range of the analyzers. They also usually have a number of relays that indicate the scale they are currently running. These outputs can be connected to the slow controls for readout. However, using a digital readout is preferred since this directly gives the analyzer reading at any scale. Currently there are a number of digital output connections, ranging from RS-232, RS-485, USB, and Ethernet. At the time of purchase, one can choose the preferred option, since the protocol is likely to evolve. The readout usually responds to a simple set of text commands. Due to the natural time scales of the gas analyzers, and lags in the gas delivery times (depending on the length of the tubing runs), sampling on timescales of a minute most likely is adequate. \fixme{``sampling on timescales of a minute...'' otherwise `minute' may strike reader as pronounced ``minoot''}

Before the beginning of the gas phase of the cryostat commissioning, the analyzers must be brought online and calibrated. Calibration varies for the different modules, but often requires using argon gas with both zero contaminants (usually removed with a local inline filter) for the zero of the analyzer, and argon with a known level of the contaminant to check the scale. Since the start of the gas phase  begins with normal air, the more sensitive analyzers are valved off at the switchyard to prevent overloading their inputs and potentially saturating their detectors. As the argon purge and gas recirculation progress, the various analyzers are valved back in when the contaminant levels reach the upper limits of the analyzer ranges. 

\subsection{Liquid Level Monitoring}
\label{sec:fdgen-slow-cryo-install-llm}

%
%

Multiple differential pressure level monitors are installed in the
cryostat, connected both to the side
penetration of the cryostat at the bottom and to dedicated
instrumentation ports at the top.

The capacitance level sensors are installed at the top of the
cryostat in coordination with the \dword{tpc} installation.  Their
placement relative to the upper ground plane (single phase) or
\dshort{crp} (dual phase) is important as these sensors will be used for a
hardware interlock on the \dword{hv}, and, in the case of the \dword{dpmod}, to measure the \lar level at the millimeter level as required
for \dual operation.
Post installation in situ testing of the capacitive level sensors can be
accomplished with a small dewar of liquid.

\subsection{Cameras and Light-Emitting System}
\label{sec:fdgen-slow-cryo-install-c}

Fixed camera installation is in principle simple, but involves a
considerable number of interfaces. Each camera enclosure has 
threaded holes to allow bolting it to a bracket. A mechanical
interface is required with the cryostat wall, cryogenic internal
piping, or \dword{dss}. Each enclosure is attached
to a gas line for maintaining appropriate underpressure in the fill
gas; this is an interface with cryogenic internal piping. Each camera has a
cable for the video signal (coax or optical), and a multiconductor
cable for power and control, to be run through cable trays to flanges
on assigned instrumentation \fdth{}s.

The inspection camera is designed to be inserted and removed on any
instrumentation \fdth equipped with a gate valve at any time
during operation.  Installation of the gate valves and purge system
for instrumentation \fdth{}s falls under cryogenic internal
piping.

Installation of fixed lighting sources separate from the cameras would
require similar interfaces as fixed cameras.  However, the current
design has lights integrated with the cameras, which do not require separate
installation.

\subsection{Slow Controls Hardware}
\label{sec:fdgen-slow-cryo-install-sc-hard}

Slow controls hardware installation includes multiple
servers, network cables, any specialized cables needed
for device communication, and possibly some custom-built rack
monitoring hardware. The installation sequence is interfaced and
planned with the facilities group and other consortia. The network
cables and rack monitoring hardware are common across many racks
and are installed first as part of the basic rack installation, 
led by the facilities group. The installation of
specialized cables needed for slow controls and servers is done
after the common rack hardware is installed, and will be coordinated
with other consortia and the \dword{daq} group respectively.

\subsection{Transport, Handling and Storage}
\label{sec:fdgen-slow-cryo-install-transport}

Most instrumentation devices are shipped to \surf in pieces and mounted in situ. 
Instrumentation devices are in general small except the support structures for purity monitors and T-gradient monitors,
which will cover the entire height of the cryostat. Since the load on those structures is relatively small
 (\(<\SI{100}{kg}\)) they can be fabricated in parts of less than \SI{3}{m},
which can be easily transported to \surf. These parts are also easy to transport down the shaft and through the tunnels.
All instrumentation devices except the dynamic T-gradient monitors, which are introduced into the cryostat through a dedicated cryostat port, 
\fixme{above something?}
can be
moved into the cryostat without the crane.

Cryogenic internal piping needs special treatment given the number of pipes and their lengths.
Purging and filling pipes will be most likely pre-assembled by the manufacturer as much as possible, using the largest  
size that can be shipped and transported down the shaft. Assuming \SI{6}{m} long sections,
pipes could be grouped in bunches of \numrange{10}{15} pipes and stored in five pallets or boxes of about \SI{6.2}{m} $\times$ \SI{0.8}{m} $\times$ \SI{0.5}{m}. 
These would be delivered to the site, stored, transported down to the detector cavern,
 and stored again before they are used.
Depending on when they are installed, they could be stored inside the cryostat itself or in one of the drifts. 
Cool-down pipes are easier to handle. They could be transported in \num{20} boxes of \SI{2.2}{m} $\times$ \SI{0.6}{m} $\times$ \SI{0.6}{m}, although
there is room for saving some space using a different packaging scheme. 
Once in the cavern they could be stored on top of the cryostat.







\section{Quality Control}
\label{sec:fdgen-slow-cryo-qc}
A series of tests should be done by the manufacturer and the institute in charge of the device assembly. The purpose of  \dword{qc} is to ensure that the equipment is capable of performing its intended function. The \dword{qc} includes post-fabrication tests and also tests to run after shipping and installation. In case of a complex system, the whole system performance will be tested before shipping. 
Additional \dword{qc} procedures can be performed at the \dword{itf} and underground after installation if possible. The planned tests for each subsystem are described below.  
\fixme{the organization of prev pgraph is confusing}

\subsection{Purity Monitors}
\label{sec:fdgen-slow-cryo-qc-pm}

The purity monitor system undergoes a series of tests to ensure the performance of the system.  This  starts with testing the individual purity monitors in vacuum after each one is fabricated and assembled.  This test looks at the amplitude of the signal generated by the drift electrons at the cathode and the anode.  This ensures that the photocathode is able to provide a sufficient number of photoelectrons for the measurement 
with the required precision, and that the field gradient resistors are all working properly to maintain the drift field and hence transport the drift electrons to the anode.  A follow-up test in \lar is then performed for each individual purity monitor, ensuring that the performance expected in \lar is met.  

The next step 
is to assemble the entire system and make checks of the connections along the way.  Ensuring that the connections are all proper during this time reduces the risk of having issues once the system is finally assembled and ready for the final test.  
The assembled system is placed into the shipping tube, which serves as a vacuum chamber, and tested. 
If an adequately sized \lar test facility is available, a full system test can be performed at \lar temperature prior to installation. 
\fixme{and what if no \lar facility exists there? Orig sentence: If there is a \lar test facility with the height or length required for the full purity monitor system and it is available for use, then a final full system test would be made there ensuring that the full system operates in \lar and achieves the required performance.}


\subsection{Thermometers}
\label{sec:fdgen-slow-cryo-qc-th}

\subsubsection{Static T-Gradient Thermometers}
\label{sec:fdgen-slow-cryo-qc-thst}

Three type of tests are carried out at the production site prior to installation. First, the mechanical rigidity of the system is tested such that swinging is minimized (< \SI{5}{cm})
to reduce the risk of touching the \dwords{apa}. This is done with a \SI{15}{m} stainless steel string, strung horizontal anchored to two points; its tension is controlled and measured. 
Second, 
all sensors are calibrated in the lab, as explained in Section~\ref{sec:fdgen-slow-cryo-therm}.
The main concern is the reproducibility of the results since sensors could potentially change their resistance (and hence their temperature scale)
when undergoing successive immersions in \lar{}. In this case the \dword{qc} is given by the calibration procedure itself since five independent measurements
are planned for each set of sensors. Sensors with reproducibility (based on the \rms of those five measurements) beyond the requirements (\SI{2}{mK} for \dword{pdsp}) are discarded.  
The calibration serves as \dword{qc} for the readout system (similar to the final one) and of the PCB-sensor-connector assembly. Finally, the cable-connector assemblies are tested: sensors must measure the expected values with no additional noise introduced by the cable or connector. 

If the available \lar test facility has sufficient height or length to test a good portion of the system, an integrated system test is conducted there ensuring that the system
operates in \lar and achieves the required performance. Ideally, the laboratory sensor calibration will be compared with the in situ calibration
of the dynamic T-gradient monitors by operating both dynamic and static T-gradient monitors simultaneously.   

The last phase of \dword{qc} takes place after installation. 
The verticality of each array is checked and the tensions in the horizontal strings are adjusted as necessary.
Before soldering the wires to the flange, the entire readout chain is tested with temporary SUBD-25 connectors. 
This allows testing the sensor-connector assembly, the cable-connector assembly and the noise level inside the cryostat.
If any of the sensors gives a problem, it is replaced. If the problem persists, the cable is checked and replaced if needed.

\subsubsection{Dynamic T-Gradient Thermometers}
\label{sec:fdgen-slow-cryo-qc-thdy}

The dynamic T-gradient monitor consists of an array of high-precision temperature sensors mounted on a vertical rod. The rod can move vertically in order to perform cross-calibration of the temperature sensors in situ. Several tests are foreseen to ensure that the dynamic T-gradient monitor delivers vertical temperature gradient measurements with precision at the level of a few \si{mK}.

\begin{itemize}
\item
Before installation, temperature sensors are tested in LN to verify correct operation and to set the baseline calibration for each sensor with respect to the absolutely caibrated reference sensor. 
\item
Warm and cold temperature readings are taken with each sensor after mounting on the PCB board and soldering 
the readout cables.
\item
The sensor readout is taken for all sensors after the cold cables are connected to electric \fdth{}s on the flange and the warm cables outside of the cryostat are connected to the temperature readout system.
\item 
The stepper motor is tested before and after connecting to the gear and pinion system.
\item
The fully assembled rod is connected to the pinion and gear, and moved with the stepper motor on a high platform many times to verify repeatability, possible offsets and uncertainty in the positioning. Finally, by repeating the test a large number of times, the sturdiness of the system will be verified.
\item
The full system is tested after installation in the cryostat: both motion and sensor operation are tested by checking 
sensor readout and vertical motion of the system.
\end{itemize} 

\subsubsection{Individual Sensors}
\label{sec:fdgen-slow-cryo-qc-is}

The method to address the quality of individual precision sensors is the same as for the static T-gradient monitors.
The \dword{qc} of the sensors is part of the laboratory calibration. After mounting six sensors with their corresponding cables, a
temporary SUBD-25 connector will be added and the six sensors tested at room temperature. All sensors should work and give values within specifications.  
If any of the sensors gives problems, it is replaced.  If the problem persists the cable is checked and replaced if needed.

\subsection{Gas Analyzers}
\label{sec:fdgen-slow-cryo-qc-ga}

The gas analyzers will be guaranteed by the manufacturer. However, once received, the gas analyzer modules are checked for both \textit{zero} and the \textit{span} values using a gas-mixing instrument. This is done using two gas cylinders with both a zero level of the gas analyzer contaminant species and a cylinder with a known percentage of the contaminant gas. This should verify the proper operation of the gas analyzers. When eventually installed at \surf, this process is repeated before the commissioning of the cryostat. It is also important to repeat the calibrations at the manufacturer-recommended periods over the gas analyzer lifetime.

\subsection{Liquid Level Monitoring}
\label{sec:fdgen-slow-cryo-qc-llm}

The differential pressure level meters will require \dword{qc} by the manufacturer.
While the capacitive 
sensors can be tested with a modest sample of \lar in the lab,
the differential pressure level meters require testing over a greater range.  While they do not
require testing over the whole range,  lab tests  in \lar 
done over a meter or two can ensure operation
at cryogenic temperatures.  Depth tests can be accomplished using a
pressurization chamber with water.


\subsection{Cameras}
\label{sec:fdgen-slow-cryo-qc-c}

Before transport to \surf, each cryogenic camera unit (comprising the enclosure, camera, and internal thermal control and monitoring) is checked for correct operation of all operating features, for recovery from \SI{87}{K} non-operating mode, for no leakage, and for physical defects. Lighting systems are similarly checked for operation. Operations tests will include verification of correct current draw, image quality, and temperature readback and control. The movable inspection camera apparatus is inspected for physical defects, and checked for proper mechanical operation before shipping. A checklist is completed for each unit, filed electronically in the DUNE logbook, and a hard copy sent with each unit. 

Before installation, each fixed cryogenic camera unit is inspected for physical damage or defects and checked in the cryogenics test facility  for correct operation of all operating features, for recovery from \SI{87}{K} non-operating mode, and for no contamination of the \lar{}. Lighting systems are similarly checked for operation. Operations tests include correct current draw, image quality, and temperature readback and control. After installation and connection of wiring, fixed cameras and lighting are again  checked for operation. The movable inspection camera apparatus is inspected for physical defects and, after integration with a camera unit, tested in facility for proper mechanical and electronic operation and cleanliness, before installation or storage. A checklist is completed for each \dword{qc} check and filed electronically in the DUNE logbook. 

\subsection{Light-emitting System}
\label{sec:fdgen-slow-cryo-qc-les}

The complete system is checked before installation to ensure the functionality of the light emission. 
\fixme{to ensure it meets requirements?}
Initial testing of the light-emitting system (see Figure~\ref{fig:gen-cisc-LED}) is done by first
measuring the current when a low voltage (\SI{1}{V}) is applied, to check
that the resistive \dword{led} failover path is correct. Next, measurement
of the forward voltage is done with the nominal forward current applied, to
check that it is within \SI{10}{\%} of the nominal forward voltage drop of
the \dwords{led}, that all of the \dwords{led} are illuminated, and that each of the
\dwords{led} is visible over the nominal angular range. If the \dwords{led} are
infrared, a video camera with IR filter removed is used for the
visual check. This procedure is then duplicated with the current
reversed for the \dwords{led} oriented in the opposite direction.  

These tests are duplicated during
installation to make sure that the system has
not been damaged in transportation or installation. However, once
the \dwords{led} are in the cryostat a visual check could be difficult or impossible.

\subsection{Slow Controls Hardware}
\label{sec:fdgen-slow-cryo-qc-sc-hard}

Networking and computing systems will be purchased commercially, requiring \dword{qa}. However, the new servers are tested after delivery to confirm no damage during shipping. The new system is allowed to \textit{burn in} overnight or for a few days, 
running a diagnostics suite on a loop. This should turn up anything that escaped the manufacturer's \dword{qa} process.

The system can be shipped directly to the underground, 
\fixme{maybe ``once the system arrives on-site, it can be transported down to the 4850L''}
where an on-site
expert peforms the initial booting of systems and basic
configuration. Then the specific configuration information is pulled over
the network, after which others may log in remotely to do the final
setup, minimizing the number of people underground.


\section{Safety}
\label{sec:fdgen-slow-cryo-safety}


Several aspects related to safety must be taken into account for the different phases of the \dword{cisc} project, including R\&D, laboratory calibration and testing, mounting tests and installation. 
The initial safety planning for all phases is reviewed and approved by safety experts as part of the initial design review, and always prior to implementation. 
All component cleaning, assembly, testing  and installation procedure documentation includes a section on safety concerns
relevant to that procedure, and is reviewed during the appropriate pre-production reviews.

Areas of particular importance to \dword{cisc} include:
\begin{itemize}
\item Hazardous chemicals (e.g., epoxy compounds used to attach sensors to cryostat inner membrane) and cleaning compounds:
  All chemicals used are documented at the consortium management level, with an MSDS (Material safety data sheet) and approved handling and disposal plans in place.

\item Liquid and gaseous cryogens used in calibration and testing: LN and \lar are used for calibration and testing of most of the instrumentation devices.
  Full hazard analysis plans will be in place
  \fixme{are being developed?}  at the consortium management level for all module or
  module component testing involving cryogenic hazards, and these safety plans will be reviewed in the appropriate pre-production and production reviews

\item \dword{hv} safety:  Purity monitors operate at $\sim$\SI{2000}{V}. Fabrication and testing plans will demonstrate compliance with local
  \dword{hv} safety requirements at the particular institution or lab where the testing or operation is performed, and this compliance will be reviewed as part of the standard review process.


\item Working at heights: Some aspects of the fabrication, testing and installation of \dword{cisc} devices require working at heights. This is the 
  case of T-gradient monitors and purity monitors, which are quite long. 
  Temperature sensors installed near the top cryostat membrane and cable routing for all instrumentation devices
  require working at heights as well. The appropriate safety procedures including lift and harness training will be designed and reviewed. 
  
\item Falling objects: all work at height comes with associated risks of falling objects. The corresponding safety procedures, including the proper helmet ussage 
  and the observation of  well delimited safety areas, will be included in the safety plan. 
\end{itemize}


\section{Organization and Management}
\label{sec:fdgen-slow-cryo-org}

\subsection{Slow Controls and Cryogenics Instrumentation Consortium Organization}
\label{sec:fdgen-slow-cryo-org-consortium}


The organization of the \dword{cisc} consortium is shown in
Figure~\ref{fig:gen-slow-cryo-org}. The \dword{cisc} consortium board is
currently formed from institutional representatives from \num{17} institutes. 
The consortium leader
acts as the spokesperson for the consortium and is responsible for the
overall scientific program and management of the group. The technical
leader of the consortium is responsible for the project management for
the group.  Currently five working groups are envisioned in the
consortium (leaders to be appointed):



\begin{description}
 \item[Cryogenics Systems] gas analyzers, liquid level
  monitors and cryogenic internal piping; CFD simulations.
 \item[\lar Instrumentation] purity monitors, thermometers,
   cameras and lightemitting system, and instrumentation test facility;
   feedthroughs; \efield simulations;
   instrumentation precision studies;
   \dword{protodune} data analysis coordination efforts.
 \item [Slow Controls Base Software and Databases]  Base software, alarms and archiving databases, and monitoring tools;
   variable naming convention and slow controls quantities.
 \item [Slow Controls Detector System Interfaces] Signal processing software and hardware interfaces (e.g., power supplies);
   firmware; rack hardware and infrastructure.   
 \item [Slow Controls External Interfaces] Interfaces with external detector systems (e.g., cryogenics system, beam, facilities, \dword{daq}).
\end{description}

\begin{dunefigure}[\dword{cisc} consortium organization]{fig:gen-slow-cryo-org}
{\dword{cisc} consortium organizational chart}
\includegraphics[width=0.6\textwidth,trim=20mm 80mm 30mm 70mm,clip]{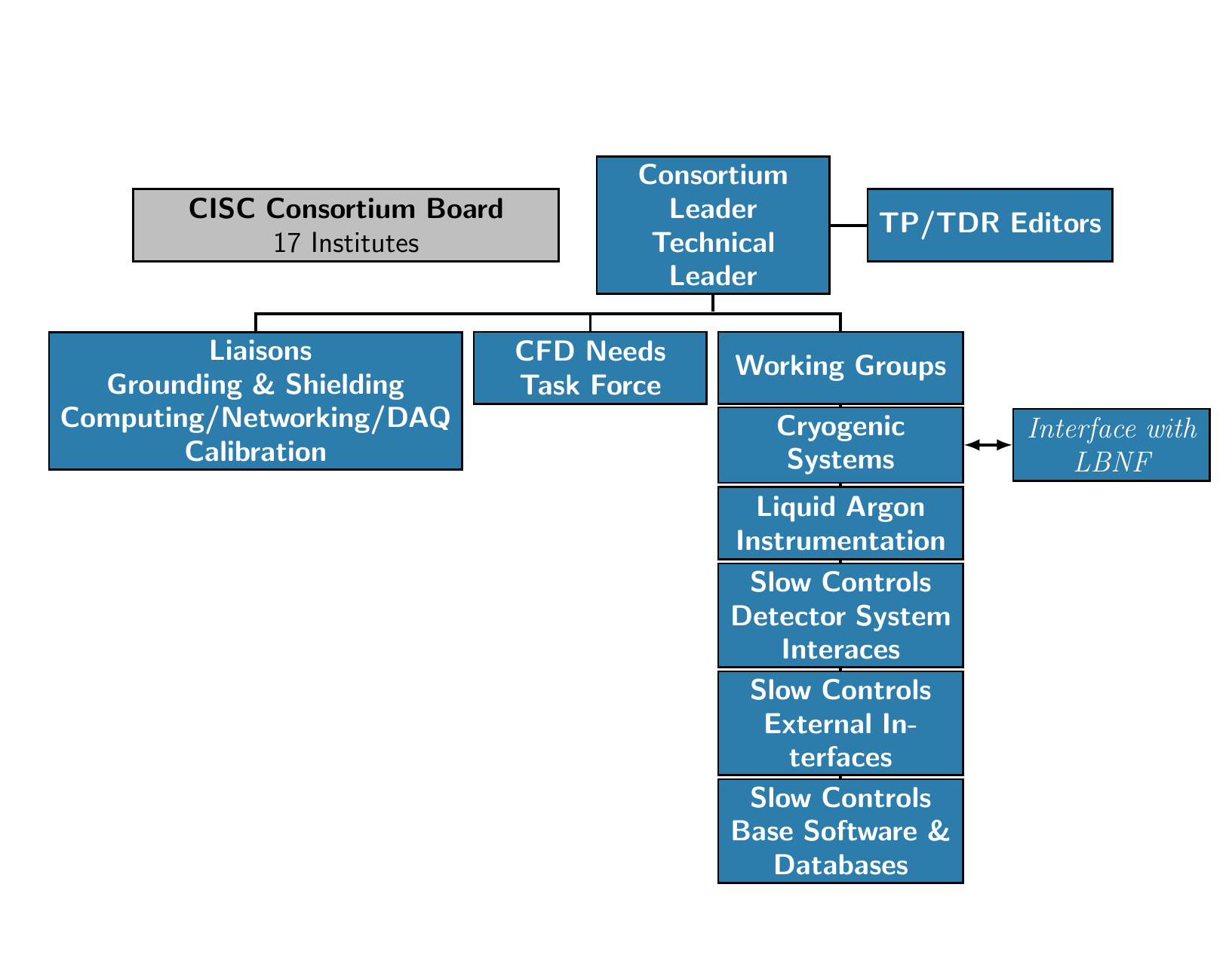}  
\end{dunefigure}

Additionally, since the \dword{cisc} consortium broadly interfaces with other
groups, liaisons have been identified for various roles as listed in
Figure~\ref{fig:gen-slow-cryo-org}. A short-term focus group was
recently formed to understand the needs for cryogenics modeling for the
consortium.  Currently members from new
institutes are added to the consortium based on consensus from the
consortium board members.


\subsection{Planning Assumptions}
\label{sec:fdgen-slow-cryo-org-assmp}

The slow controls and cryogenic instrumentation is a joint effort for \single and \dual{}.
A single slow controls system will be implemented to serve both \single and \dual{}.

Design and installation of cryogenics systems (gas analyzers, liquid level monitoring, internal piping) is coordinated with LBNF, with the consortium providing resources, and effort and expertise provided by LBNF.
\dword{protodune} designs for \lar instrumentation (purity monitors, thermometers, cameras, test facility) provide the basis for DUNE designs. Design validation, testing, calibration, and performance will be evaluated through \dword{protodune} data.


\subsection{High-level Schedule}
\label{sec:fdgen-slow-cryo-org-cs}

Table \ref{tab:fdgen-slow-cryo-schedule} shows key milestones on
the path to  commissioning of the first two DUNE detector modules.

\begin{dunetable}
[Key \dword{cisc} milestones]
{p{0.15\linewidth}p{0.60\linewidth}}
{tab:fdgen-slow-cryo-schedule}
{Key \dword{cisc} milestones leading towards commissioning of the first two DUNE detector modules.}   
Date & Milestone \\ \toprowrule
Aug.\ 2018 &	Validate instrumentation designs using data from ProtoDUNE  \\ \colhline
Jan.\ 2019 &	Complete architectural design for slow controls ready \\ \colhline
Feb.\ 2019 &	Full final designs of all cryogenic instrumentation devices ready \\ \colhline
Feb.\ 2023 &	Installation of Cryogenic Internal Piping for Cryostat 1 \\ \colhline
Apr.\ 2023 &	Installation of support structure for all instrumentation devices for Cryostat 1 \\ \colhline
Oct.\ 2023 &	All Instrumentation devices installed in Cryostat 1 \\ \colhline
Feb.\ 2024 &	All Slow Controls hardware and infrastructure installed for Cryostat 1  \\ \colhline
May 2024 &	Installation of Cryogenic Internal Piping for Cryostat 2 \\ \colhline
July 2024 &	Installation of support structure for all instrumentation devices for Cryostat 2 \\ \colhline
Jan.\ 2025 &	All Instrumentation devices installed in Cryostat 2 \\ \colhline
Apr.\ 2025 &	All Slow Controls hardware and infrastructure installed for Cryostat 2 \\ \colhline
July 2025 &	Full Slow controls systems commissioned and integrated into remote operations \\
\end{dunetable}


\cleardoublepage

\chapter{Technical Coordination}
\label{ch:fdsp-coord}

The \dword{tc} team is responsible for detector integration
and installation support. 
The \dword{dune} collaboration consists of a large number of
institutions distributed throughout the world. They are supported by a
large number of funding sources and collaborate with a large number of
commercial partners. Groups of institutes within \dword{dune} form
consortia that take complete responsibility for construction of their
system.  \dword{dune} has empowered several consortia (currently nine)
with the responsibility to secure funding and design, fabricate,
assemble, install, commission and operate their components of the
\dword{dune} \dword{fd}. There are three consortia focusing
exclusively on the \dword{spmod}: \dword{apa},
\single \dword{ce} and \single \dword{pds}. There are three
focusing exclusively on the \dword{dpmod}: 
\dword{crp}, \dual \dword{ce} and \dual \dword{pds}. There are
three joint consortia: \dword{hv}, \dword{daq} and \dword{cisc}. Other consortia may
be formed over time as concepts more fully emerge, such as a
\dword{fd} calibration system and various aspects of the \dword{nd}.
\dword{dune} \dword{tc}, under the direction of the
Technical Coordinator, has responsibility to monitor the
technical aspects of the detector construction, to integrate and
install the \dwords{detmodule} and to deliver the common projects. The
\dword{dune} \dword{tc} organization is shown in Figure~\ref{fig:TC_orgchart}.

\begin{dunefigure}[Organization of \dword{tc}]{fig:TC_orgchart}
  {Organization of \dword{tc}. This organization
 oversees the construction of the \dword{fd}, both \single and
 \dual, and the \dword{nd}.}
\includegraphics[width=0.85\textwidth,trim=20mm 90mm 30mm 90mm,clip]{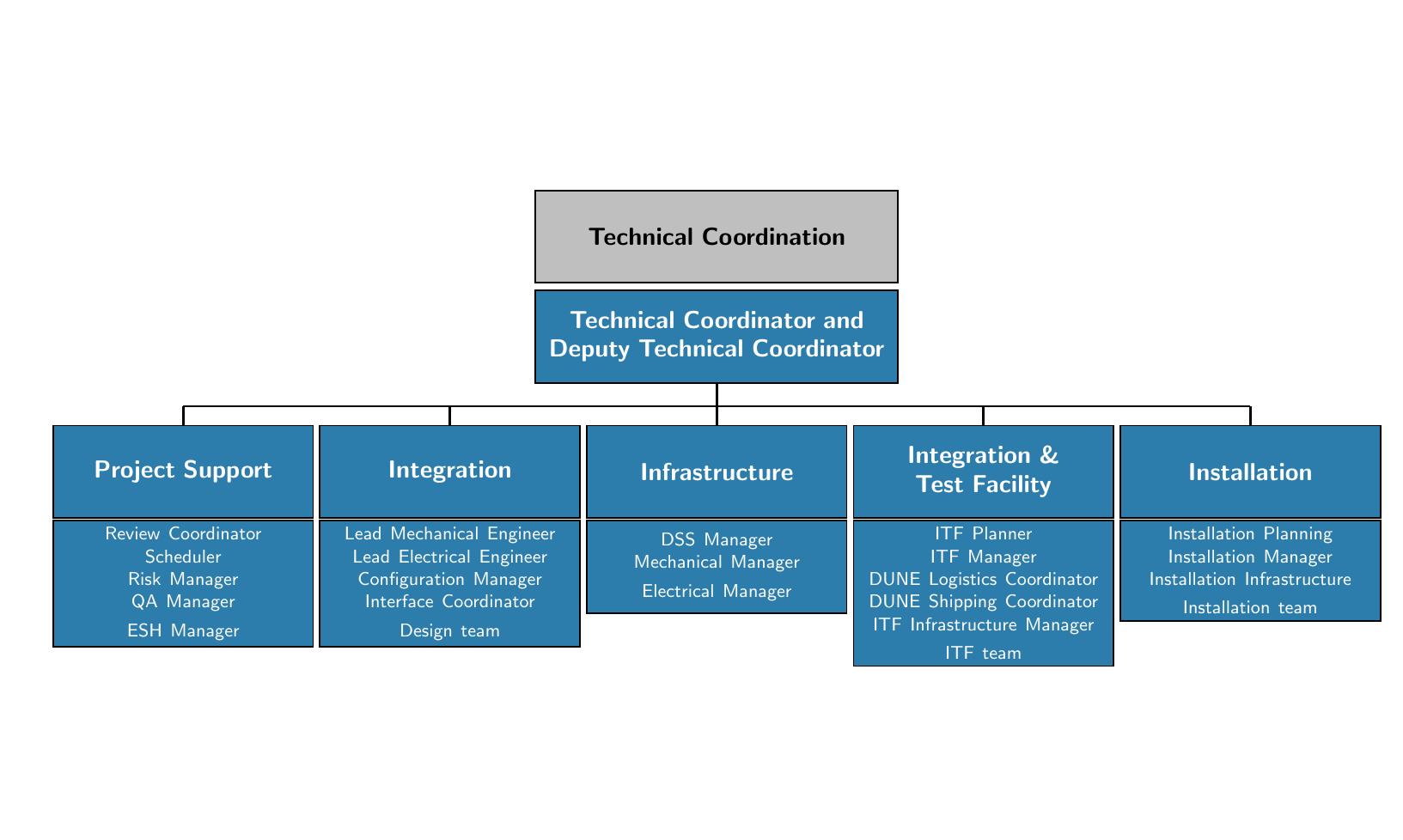}
\end{dunefigure}

The \dword{tc} organization staffing will grow over time as the project
advances. \Dword{tc} will provide staffing for teams underground at \surf, at
integration facilities, and at the near site at \fnal, in addition to
the core team distributed among collaborating institutions.

The \dword{dune} Project consists of a \dword{fd} and a
\dword{nd}. The \dword{nd} is at a pre-conceptual state; as the
conceptual design and organization emerges, it will become part of the
\dword{dune} Project. Currently the \dword{dune} Project consists of
the \dword{dune} \dword{fd} consortia and \dword{tc}.  The
\dword{dune} Project is moving towards a \dword{tdr} for
the \dword{fd}, both \single and \dual options, in 2019. It is
expected that a Conceptual Design Report for the \dword{nd} will be
prepared at the same time. 

The \dword{fd} components will be shipped
from the consortia construction sites to the \dword{itf}. 
\Dword{tc} will
evaluate and accept consortia components either at integration
facilities or the installation site and oversee the integration of
components as appropriate. The scope of the \dword{fd} integration
and installation effort is shown graphically in
Figure~\ref{fig:TC_flow}.

\begin{dunefigure}[Flow of components from the consortia to the \dword{fd}.]{fig:TC_flow}
  {Flow of components from the consortia to the \dword{fd}.}
 \includegraphics[width=\textwidth]{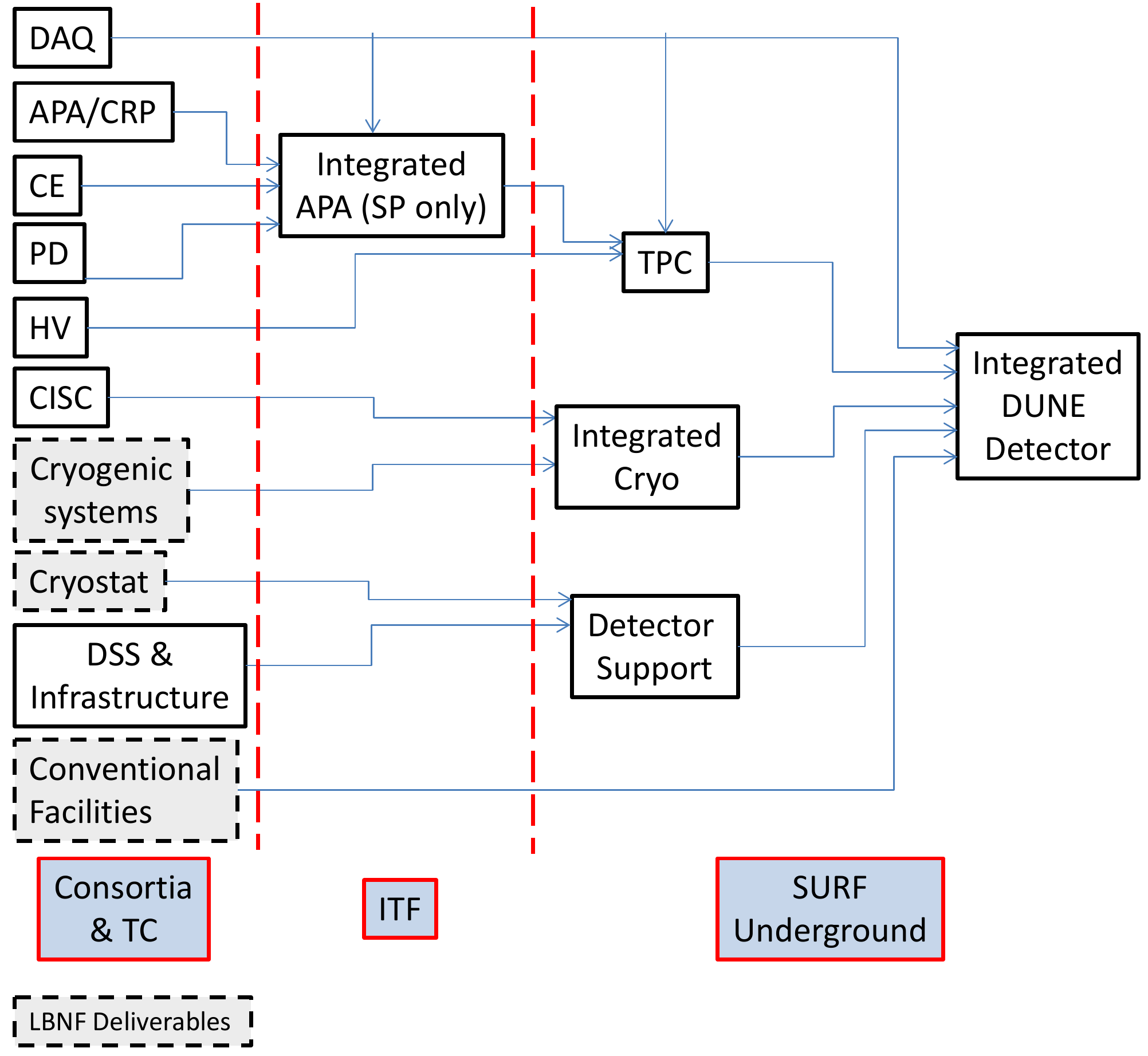}
\end{dunefigure}

\Dword{tc} interacts with the consortia via three main areas: project
coordination, integration, and installation.  Construction of the
\dword{dune} \dword{fd} requires careful technical coordination due to
its complexity.  Given the horizontal nature of the consortia
structure and the extensive interdependencies between the systems, a
significant engineering organization is required to deliver
\dword{dune} on schedule and within specifications and funding
constraints.

The responsibilities of \dword{tc} include:
\begin{itemize}
  \item management and delivery of all common projects;
  \item development and monitoring of the consortia interfaces;
  \item configuration control of all interface documents, drawings and envelopes;
  \item installation of detectors at the near and far sites;
  \item logistics for detector integration and installation at the near and far sites;
  \item survey of the detector;
  \item primary interface to \dword{lbnf} for conventional facilities, cryostat and cryogenics;
  \item primary interface to the host laboratory for infrastructure and operations support;
  \item development and tracking of project schedule and milestones;
  \item review of all aspects of the project;
  \item recording and approving all project engineering information, including: documents, drawings and models;
  \item project work breakdown schedules;
  \item project risk register;
  \item \dword{dune} engineering and safety standards, including grounding and shielding;
  \item monitoring of all consortia design and construction progress;
  \item \dword{qa} and all \dword{qa} related studies and documents;
  \item \dword{esh} organization and all safety related studies and documents.
\end {itemize}

\dword{dune} \dword{tc} interacts with \dword{lbnf} primarily through the
\dword{lbnf}/\dword{dune} systems engineering organization. \dword{tc}
provides the points of contact between the consortia and \dword{lbnf}.
\Dword{tc} will work with the \dword{lbnf}/\dword{dune} Systems Engineer to
implement the \dword{lbnf}/\dword{dune} Configuration Management Plan
to assure that all aspects of the overall \dword{lbnf}/\dword{dune}
project are well integrated. \Dword{tc} will work with \dword{lbnf} and the
host laboratory to ensure that adequate infrastructure and operations support
are provided during construction, integration, installation,
commissioning and operation of the detectors. The \dword{lbnf}/\dword{dune}  systems
engineering organization is shown in Figure~\ref{fig:DUNE_SE_org}.

\begin{dunefigure}[LBNF/DUNE systems engineering organizational structure.]{fig:DUNE_SE_org}
  {LBNF/DUNE systems engineering organizational structure.}
   \includegraphics[width=\textwidth,trim=20mm 105mm 30mm 110mm,clip]{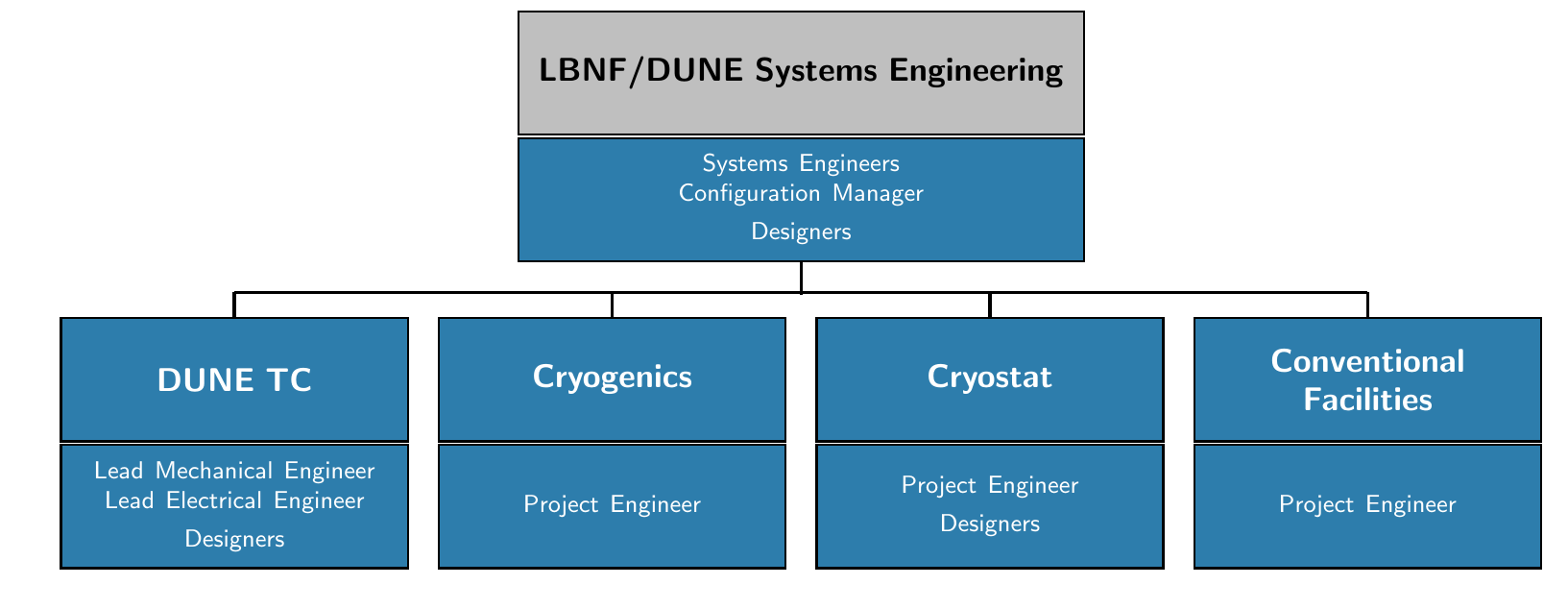}
\end{dunefigure}

Proper integration of the \dwords{fd} within the supporting
facilities and infrastructure at \surf is a major engineering task.
The \dword{lbnf}/\dword{dune} Systems Engineer is responsible for the
interfaces between the major \dword{lbnf} and \dword{dune} systems
(conventional facilities, cryostats, cryogenics systems and
\dwords{detmodule}). The \dword{lbnf}/\dword{dune} systems engineering team
includes several engineers and designers with responsibility for
maintaining computer aided design (CAD) models. \Dword{dune} \dword{tc}
supports an engineering team that works directly with the
\dword{lbnf}/\dword{dune} systems engineering team to ensure that the
detector is properly integrated into the overall system.

\dword{tc} has been working with the \dword{lbnf}/\dword{dune} systems engineering team to
define requirements from \dword{dune} for the conventional facilities final
design for the detector chambers, \dword{cuc}, drifts
and utilities. \Dword{tc} is representing the interests of the \dword{dune} detector
in the conventional facilities (CF) design. This includes refining the
detector installation plan to understand how much space is needed in
front of the \dwords{tco} of the cryostats
and therefore of the size of the chambers. \dword{tc} continues to refine the
detector needs for utilities in the detector caverns and the \dword{cuc} 
where the \dword{daq} will be housed.

Physics requirements on \dword{tc} include cleanliness in the cryostats,
survey and alignment tolerances, and grounding and shielding
requirements. The cleanliness requirement is for ISO 8 (class
100,000), which will keep rates from dust radioactivity below those of
the inherent $^{39}$Ar background. The alignment tolerances are driven
by physics requirements on reconstructing tracks. Grounding and
shielding are critical to enable this very sensitive, low-noise
detector to achieve the required \dword{s/n}. The physics
requirements for \dword{lbnf} and \dword{dune} are maintained in
DocDB-112.

\section{Project Support}
\label{sec:fdsp-coord-supp}

As defined in the \dword{dune} Management Plan (DMP), the \dword{dune}
Technical Board (TB) generates and recommends technical decisions to the 
collaboration executive board (EB) (see Figure~\ref{fig:TB_org}).
\begin{dunefigure}[DUNE Technical Board.]{fig:TB_org}
  {DUNE Technical Board.}
 \includegraphics[width=0.9\textwidth,trim=20mm 110mm 30mm 110mm,clip]{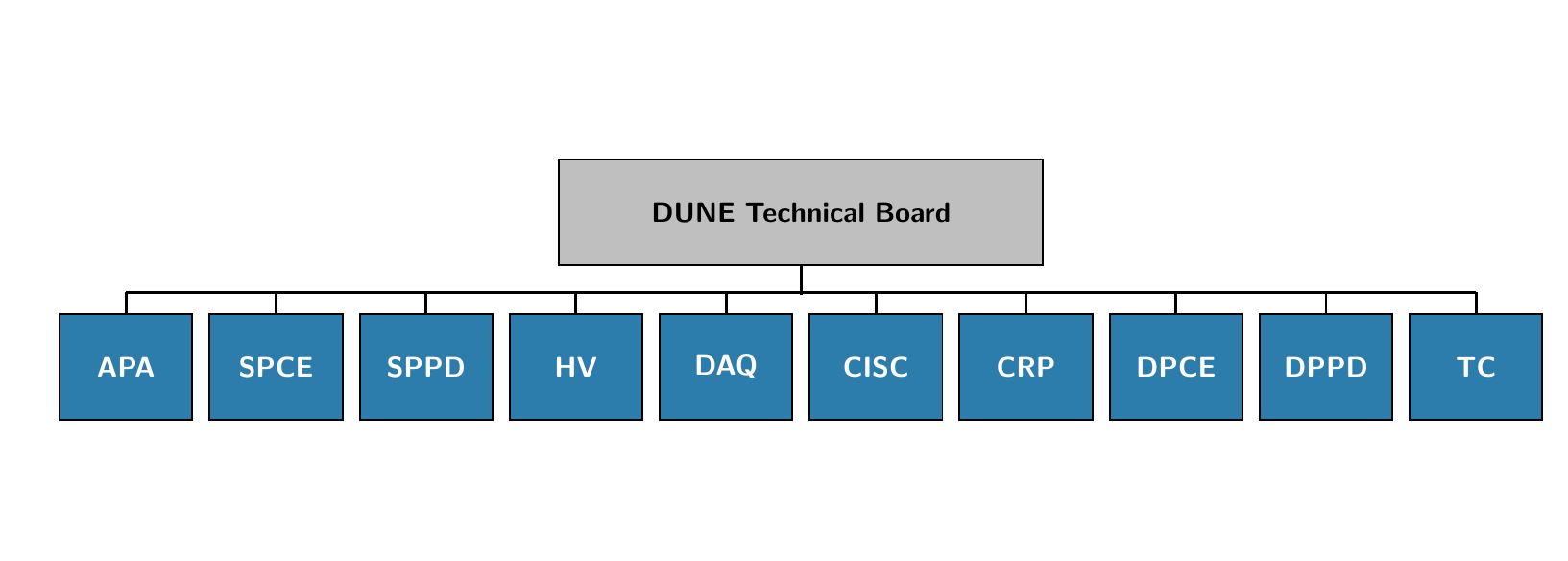}
\end{dunefigure}

It consists of all consortia scientific and technical leads. It meets
on a regular basis (approximately monthly) to review and resolve any
technical issues associated with the detector construction. It reports
through the EB to collaboration management. The \dword{dune} TB
is chaired by the technical coordinator. \dword{dune} collaboration
management, including the EB, is shown in Figure~\ref{fig:DUNE_org}. The
\dword{tc} engineering team also meets on a regular basis (approximately monthly)
to discuss more detailed technical issues. \Dword{tc} does not have
responsibility for financial issues; that will instead be referred to
the EB and Resource Coordinator (RC).

\begin{dunefigure}[DUNE management organizational structure.]{fig:DUNE_org}
  {DUNE management organizational structure.}
 \includegraphics[width=\textwidth]{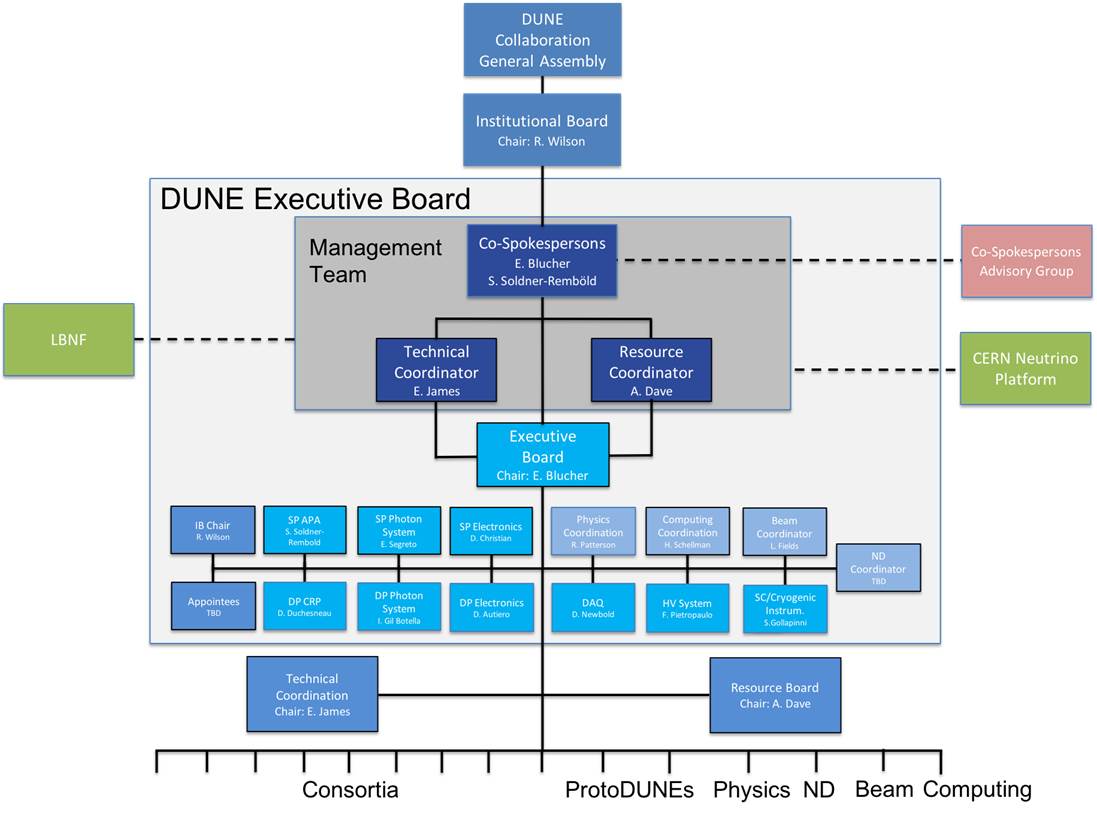}
\end{dunefigure}

\Dword{tc} has several major project support tasks that need to be accomplished:
\begin{itemize}
  \item Assure that each consortium has a well defined and complete
    scope, that the interfaces between the consortia are sufficiently
    well defined and that any remaining scope can be covered by \dword{tc}
    through \dword{comfund} or flagged as missing scope to the EB and RC. In
    other words, assure that the full detector scope is
    identified. Monitor the interfaces and consortia progress in
    delivering their scope.
  \item Develop an overall project \dlong{ims}
    that includes reasonable production schedules, testing plans and a
    well developed installation schedule from each consortium. Monitor
    the \dword{ims} as well as the individual consortium schedules.
  \item Ensure that appropriate engineering and safety standards are
    developed and agreed to by all key stakeholders and that these
    standards are conveyed to and understood by each
    consortium. Monitor the design and engineering work.
  \item Ensure that all \dword{dune} requirements on \dword{lbnf} for
    conventional facilities, cryostat and cryogenics have been clearly
    defined and understood by each consortium. Negotiate scope
    boundaries with \dword{lbnf}. Monitor \dword{lbnf} progress on
    final conventional facility design, cryostat design and cryogenics
    design.
  \item Ensure that all technical issues associated with scaling from
    \dword{protodune} have sufficient resources to converge on
    decisions that enable the detector to be fully integrated and
    installed.
  \item Ensure that the integration and \dword{qc} processes for each
    consortium are fully developed and reviewed and that the
    requirements on an \dword{itf} are well defined.
\end{itemize}

\Dword{tc} is responsible for technical quality and schedule and is not
responsible for consortia funding or budgets.  \Dword{tc} will try to help
resolve any issue that it can, but will likely have to push all
financial issues to the TB, EB and RC for resolution.

\Dword{tc} maintains a web page\footnote{\url{https://web.fnal.gov/collaboration/DUNE/DUNE\%20Project/\_layouts/15/start.aspx\#/}.}
with links to project documents. \Dword{tc} maintains repositories of project
documents and drawings. These include the \dword{wbs}, schedule, risk
register, requirements, milestones, strategy, detector models and
drawings that define the \dword{dune} detector.

\subsection{Schedule}
\label{sec:fdsp-coord-controls}

A series of tiered milestones are being developed for the \dword{dune}
project. The Tier-0 milestones are held by the spokespersons and host
laboratory director. Three have been defined and the current milestones and
target dates are:
\begin{enumerate}
\item Start main cavern excavation \hspace{2.1in} 2019
\item Start \dword{detmodule}~1 installation \hspace{2.1in} 2022
\item Start operations of \dword{detmodule}~1--2 with beam \hspace{1in} 2026
\end{enumerate}
These dates will be revisited at the time of the \dword{tdr} review.  Tier-1
milestones will be held by the technical coordinator and \dword{lbnf} Project
Manager and will be defined in advance of the \dword{tdr} review. Tier-2
milestones will be held by the consortia.

A high level version of the \dword{dune} milestones from the \dword{ims}
can be seen in Table~\ref{tab:DUNE_schedule}. 

\begin{dunetable}
[Overall \dword{dune} Project Tier-1 milestones.]
{p{0.84\linewidth}p{0.14\linewidth}}
{tab:DUNE_schedule}
{Overall \dword{dune} Project Tier-1 milestones.}
Milestone & Date   \\ \toprowrule
RRB Approval of Technical Design Review                       & 09/02/2019 \\ \colhline 
Beneficial Occupancy of Integration Test Facility             & 09/01/2021 \\ \colhline 
Construction of steel frame for Cryostat \#1 complete         & 12/17/2021 \\ \colhline 
Construction of Mezzanine for Cryostat \#1 complete           & 01/17/2022 \\ \colhline 
Begin integration/testing of Detector \#1 components at ITF   & 02/01/2022 \\ \colhline 
Beneficial Occupancy of Central Utility Cavern Counting room  & 04/16/2022 \\ \colhline 
Construction of steel frame for Cryostat \#2 complete         & 07/01/2022 \\ \colhline 
Construction of Mezzanine for Cryostat \#2 complete           & 08/01/2022 \\ \colhline 
\textbf{Beneficial occupancy of Cryostat \#1}                 & \textbf{12/23/2022} \\ \colhline 
Cryostat \#1 ready for TPC installation                       & 05/01/2023 \\ \colhline 
Begin integration/testing of Detector \#2 components at ITF   & 11/01/2023 \\ \colhline 
\textbf{Beneficial occupancy of Cryostat \#2}                 & \textbf{03/01/2024} \\ \colhline 
Begin closing Temporary Construction Opening for Cryostat \#1 & 05/01/2024 \\ \colhline 
Cryostat \#2 ready for TPC installation                       & 08/01/2024 \\ \colhline 
Cryostat \#1 ready for filling                                & 10/01/2024 \\ \colhline 
Begin closing Temporary Construction Opening for Cryostat \#2 & 07/18/2025 \\ \colhline 
\textbf{Detector \#1 ready for operations}                    & \textbf{10/01/2025} \\ \colhline 
Cryostat \#2 ready for filling                                & 12/05/2025 \\ \colhline 
\textbf{Detector \#2 ready for operations}                    & \textbf{12/18/2026} \\
\end{dunetable}

\Dword{tc} will maintain the \dword{ims} that links all consortium schedules
and contains appropriate milestones to monitor progress. The \dword{ims}
is envisioned to be maintained in MS-Project~\footnote{MicroSoft\texttrademark{} Project.} as it is expected
that many consortia will use this tool. It is currently envisioned as
three levels of control and notification milestones in addition to the
detailed consortium schedules. The highest level contains external
milestones, with the second level containing the key milestones for \dword{tc}
to monitor deliverables and installation progress, and the third level
containing the inter-consortium links. The schedules will go
under change control after agreement with each consortium on the
notification milestone dates and the \dword{tdr} is approved.

In addition to the overall \dword{ims} for construction and
installation, a schedule of key consortia activity in the period
2018--19 leading up to the \dword{tdr} has been developed.

To ensure that the \dword{dune} detector remains on schedule,
\dword{tc} will monitor schedule statusing from each consortium, will organize
reviews of schedules and risks as appropriate.  As schedule problems
arise \dword{tc} will work with the affected consortium to resolve the
problems. If problems cannot be solved, \dword{tc} will take the issue to the
TB and EB.

A monthly report with input from all consortia will be published by
\dword{tc}. This will include updates on consortium technical progress and
updates from \dword{tc} itself.

\subsection{Risk}
\label{sec:fdsp-coord-risk}

The successful operation of \dword{protodune} will retire a great many
potential risks to \dword{dune}. This includes most risks associated with the
technical design, production processes, \dword{qa}, integration
and installation. Residual risks remain relating to design and
production modifications associated with scaling to \dword{dune}, mitigations
to known installation and performance issues in \dword{protodune}, underground
installation at \surf and organizational growth.

The highest technical risks include: development of a system to
deliver \SI{600}{kV} to the \dual cathode; general delivery of the
required \dword{hv}; cathode and \dword{fc} discharge to the cryostat
membrane; noise levels, particularly for the \dword{ce}; 
number of dead channels; lifetime of components surpassing \dunelifetime{}; 
\dword{qc} of all components; verification of improved \dword{lem}
performance; verification of new cold  \dword{adc} and  \dword{coldata} performance;
argon purity; electron drift lifetime; \phel light yield;
incomplete calibration plan; and incomplete connection of design to
physics. Other major risks include insufficient funding, optimistic
production schedules, incomplete integration, testing and installation
plans.

Key risks for \dword{tc} to manage include the following:
\begin{enumerate}
    \item Too much scope is unaccounted for by the consortia and falls
      to \dword{tc} and \dword{comfund}.
    \item Insufficient organizational systems are put into place to
      ensure that this complex international mega-science project,
      including \dword{tc}, \fnal as host laboratory, \surf, DOE and all international
      partners continue to successfully work together to ensure
      appropriate rules and services are provided to enable success of
      the project.
  \item Inability of \dword{tc} to obtain sufficient personnel resources so as to
    ensure that \dword{tc} can oversee and coordinate all of its
    project tasks.  While the USA has a special responsibility towards
    \dword{tc} as host country, it is expected that personnel resources will
    be directed to \dword{tc} from each collaborating country. Related to this
    risk is the fact that consortium deliverables are not really
    stand-alone subsystems; they are all parts of a single \dword{detmodule}. This
    elevates the requirements on coordination between consortia.
\end{enumerate}

The consortia have provided preliminary versions of risk analyses that
have been collected on the \dword{tc} webpage. These are being developed into
an overall risk register that will be monitored and maintained by \dword{tc}
in coordination with the consortia.

\subsection{Reviews}
\label{sec:fdsp-coord-reviews}

\Dword{tc} is responsible for reviewing all stages of detector development
and works with each consortium to arrange reviews of the design
(\dword{pdr} and \dword{fdr}), production (\dword{prr} and
\dword{ppr}) and \dword{orr} of their system.  These
reviews provide input for the TB to evaluate technical decisions.
Review reports are tracked by \dword{tc} and provide guidance as to
key issues that will require engineering oversight by the \dword{tc}
engineering team. \Dword{tc} will maintain a calendar of \dword{dune}
reviews.

\Dword{tc} works with consortia leaders to review all detector designs,
with an expectation for a \dword{pdr}, followed by a \dword{fdr}.  All
major technology decisions will be reviewed prior to down-select.  \Dword{tc}
may form task forces as necessary for specific issues that need more
in-depth review.

Start of production of detector elements can commence only after
successful \dwords{prr}. Regular production progress
reviews will be held once production has commenced. The \dwords{prr}
will typically include review of the production of \textit{Module 0}, the
first such module produced at the facility. \Dword{tc} will work with
consortium leaders for all production reviews.

\Dword{tc} is responsible to coordinate technical documents for the LBNC
Technical Design Review.

\subsection{Quality Assurance}
\label{sec:fdsp-coord-qa}

The \dword{lbnf}/\dword{dune} \dword{qap} outlines the \dword{qa} requirements
for all \dword{dune} consortia and describes how the requirements
shall be met. The consortia will be responsible for implementing a
quality plan that meet the requirements of the
\dword{lbnf}/\dword{dune} \dword{qap}.  The consortia implement the
plan through the development of individual quality plans, procedures,
guides, \dword{qc} inspection and test requirements and travelers~\footnote{The
  traveler is a document that details the fabrication and inspection
  steps and ensures that all steps have been satisfactorily
  completed.} and test reports.  In lieu of a consortium-specific quality
plan, 
the consortia may work under the \dword{lbnf}/\dword{dune}
\dword{qap} and develop manufacturing and \dword{qc} plans, procedures and
documentation specific to their work scope.  The technical coordinator 
and consortia leaders are responsible for
providing the resources needed to conduct the Project successfully,
including those required to manage, perform and verify work that
affects quality.  The \dword{dune} consortia leaders are responsible
for identifying adequate resources to complete the work scope and to
ensure that their team members are adequately trained and qualified to
perform their assigned work.

The consortia work will be documented on travelers and applicable test
or inspection reports. Records of the fabrication, inspection and
testing will be maintained. When a component has been identified as
being in noncompliance to the design, the nonconforming condition
shall be documented, evaluated and dispositioned as one of the following: use-as-is (does
not meet design but can meet functionality as is), rework (bring into
compliance with design), repair (will be brought into meeting
functionality but will not meet design) or scrap. For nonconforming
equipment or material that is dispositioned as use-as-is or repair, a
technical justification shall be provided allowing for the use of the
material or equipment and approved by the design authority.

The \dword{lbnf}/\dword{dune} \dword{qam} reports
to the \dword{lbnf} Project Manager and \dword{dune} Technical
Coordinator and provides oversight and support to the consortium
leaders to ensure a consistent quality program.
\begin{enumerate}
  \item The \dword{qam} will plan reviews as independent assessments to assist
    the technical coordinator in identifying opportunities for
    quality and performance-based improvement and to ensure compliance
    with specified requirements.
  \item The \dword{qam} is responsible to work with the consortia in
    developing their \dword{qa} and \dword{qc} Plans.
  \item The \dword{qam} will review consortia \dword{qa} and \dword{qc} activity, including
    production site visits.
  \item The \dword{qam} will participate in consortia design reviews, conduct
    \dwords{prr} prior to the start of production,
    conduct \dwords{ppr} on a regular basis (as
    described in Section~\ref{sec:fdsp-coord-reviews}), and perform
    follow-up visits to consortium facilities prior to shipment of
    components to ensure all components and documentation are
    satisfactory.
\item The \dword{qam} is responsible for performing assessments at the
  \dword{itf}, the Far Site and the Near Site to
  ensure the activities performed at these locations are in accordance
  with the \dword{lbnf}/\dword{dune} \dword{qa} Program and applicable procedures,
  specifications and drawings.
\end{enumerate}

\subsubsection{Document Control}
\label{sec:fdsp-coord-document}

\Dword{tc} maintains repositories of project documents and drawings in two
document management systems.  DUNE Project documents will be stored in
the DUNE DocDB~\footnote{\url{https://docs.dunescience.org}}. DUNE drawings
will be stored in 
\dword{edms}~\footnote{\url{https://edms.cern.ch/ui/\#!master/navigator/project?P:1637280201:1637280201:subDocs}.}.
\Dword{tc} will maintain approved versions of \dword{qa}, \dword{qc} and testing plans,
installation plans, engineering and safety standards in the DUNE
DocDB.

Consortia have developed initial interface, risk, schedule and \dword{wbs}
documents that will be put under change control and managed by the \dword{tc}
engineering team along with the consortia involved. These
are currently in DocDB and will likely go under change control later
in 2018, although they will continue to be developed through the \dword{tdr}.

Thresholds for change control are described in the
\dword{lbnf}/\dword{dune} \dword{cmp}. The control process is further
described in Section~\ref{sec:fdsp-coord-integ-config}.

\subsection{ESH}
\label{sec:fdsp-coord-esh}

The \dword{dune} \dword{esh} program is described in the
\dword{lbnf}/\dword{dune} \dword{ieshp}. This plan is maintained by
the \dword{lbnf}/\dword{dune} \dword{esh} Manager, who reports to the
\dword{lbnf} Project Manager and the technical coordinator. The
\dword{esh} manager is responsible to work with the consortia in
reviewing their hazards and their \dword{esh} plans.  The \dword{esh}
Manager is responsible to review \dword{esh} at production sites,
integration sites and at \surf. It is expected that the \dword{esh}
reviews will be conducted as part of the \dword{prr} and \dword{ppr}
process described in Section~\ref{sec:fdsp-coord-reviews}.

A strong \dword{esh} program is essential to successful completion of
the \dword{lbnf}/\dword{dune} Project at \fnal, collaborating laboratories and
universities, the \dword{itf}, and \surf. \dword{dune} is committed to ensuring
a safe work environment for 
collaborators at
all institutions and to protecting the public from hazards associated
with construction and operation of \dword{lbnf}/\dword{dune}. In
addition, all work will be performed in a manner that preserves the
quality of the environment and prevents property damage. Accidents and
injuries are preventable and it is important that we work together to
establish an injury-free workplace.

To achieve the culture and safety performance required for this
project, it is essential that \dword{dune} ensure that procedures
are established to support the following \dword{esh} policy
statements:
\begin{itemize}
  \item Line managers are responsible for environmental stewardship
    and personal safety at the \dword{dune} work sites.
  \item Line managers, supported by the \dword{lbnf}/\dword{dune},
    \fnal, other collaborating laboratories, and \surf \dword{esh}
    organizations, will provide consistent guidance and enforcement of
    the \dword{esh} program that governs the activities of workers at each
    site where work is being performed.
  \item Incidents, whether they involve personal injuries or other
    losses, can be prevented through proper planning. All
    \dword{lbnf}/\dword{dune} work is planned.
  \item Workers are involved in the work planning process and
    continuous improvement, including the identification of hazards
    and controls.
  \item Working safely and in compliance with requirements is vital to
    a safe work environment. Line managers will enforce disciplinary
    policies for violations of safety rules.
  \item Each of us is responsible for  our own safety and for that of
    our co-workers. Together we create a safe work environment.
  \item A strong program of independent audits, self-assessments and
    surveillance will be employed to periodically evaluate the
    effectiveness of the \dword{esh} program.
  \item Any incidents that result or could have resulted in personal
    injury or illness, significant damage to buildings or equipment,
    or impact of the environment, will be investigated to determine
    corrective actions and lessons that can be applied to prevent
    recurrence. \dword{lbnf}/\dword{dune} encourages open reporting of
    errors and events.
\end{itemize}

To achieve the culture and safety performance required for this
project, it is essential that \dword{esh} be fully integrated into the
project and be managed as tightly as quality, cost, and schedule.

\section{Integration Engineering}
\label{sec:fdsp-coord-integ-sysengr}

The major aspects of detector integration focus on the mechanical and
electrical connections between each of the detector systems. This
includes verification that subassemblies and their interfaces are
correct (e.g., \dword{apa} and \single \dword{pds}). A second major area is in the support of
the 
\dwords{detmodule} and their interfaces to the cryostat and cryogenics. A
third major area is in assuring that the \dwords{detmodule} 
can be installed ---
that the integrated components can be moved into their final
configuration. A fourth major area is in the integration of the 
\dwords{detmodule} with the necessary services provided by the conventional
facilities.

\subsection{Configuration Management}
\label{sec:fdsp-coord-integ-config}

The \dword{tc} engineering team will maintain
configuration data in the appropriate format for the management of the
detector configuration. The consortia are responsible for providing
engineering data for their subsystems to \dword{tc}. The
\dword{tc} engineering team will work with the \dword{lbnf} project team to
integrate the full detector data into the global \dword{lbnf}
configuration files. Appropriate thresholds for tracking and for
change control will be established.

For mechanical design aspects, the \dword{tc} engineering team
will maintain full \threed CAD models of the \dwords{detmodule}. Appropriate level
of detail will be specified for each type of model. Each consortium
will be responsible for providing CAD models of their detector
components to be integrated into overall models. The \dword{tc} engineering
team will work with the \dword{lbnf} project team to integrate the
full \dword{detmodule} models into a global \dword{lbnf} CAD model which will
include cryostats, cryogenics systems, and the conventional
facilities. These will include models using varying software
packages. The \dword{tc} engineering team will work directly with the
consortium technical leads and their supporting engineering teams to
resolve any detector component interference or interface issues with
other detector systems, detector infrastructure, and facility
infrastructure.

For electrical design aspects, the \dword{tc} engineering team will maintain
high-level interface documents that describe all aspects of required
electrical infrastructure and electrical connections. All consortia
must document power requirements and rack space
requirements. Consortia are responsible for defining any cabling that
bridges the design efforts of two or more consortia. This agreed-upon
and signed-off interface documentation shall include cable
specification, connector specification, connector pinout and any data
format, signal levels and timing. All cables, connectors, printed
circuit board components, physical layout and construction will be
subject to project review. This is especially true of elements which
will be inaccessible during the project lifetime. Consortia shall
provide details on \lar temperature acceptance testing and
lifetime of components, boards, cables and connectors.

At the time of the release of the \dword{tdr}, the
\dword{tc} engineering team will work with the consortia to produce
formal engineering drawings for all detector components.  These
drawings are expected to be signed by the consortia technical leads,
project engineers, and Technical Coordinator.  Starting from that
point, the \dwords{detmodule} models and drawings will sit under formal change
control as discussed in Section~\ref{sec:fdsp-coord-qa}.  It is
anticipated that designs will undergo further revisions prior to the
start of detector construction, but any changes made after the release
of the \dword{tdr} will need to be agreed to by all of
the drawing signers and an updated, signed drawing produced. The major
areas of configuration management include:
\begin{enumerate}
  \item \threed model,
  \item Interface definitions,
  \item Envelope drawings for installation,
  \item Drawing management.
\end{enumerate}

\Dword{tc} will put into place processes for configuration
management.  Configuration management will provide \dword{tc}
and engineering staff the ability to define and control
the actual build of the detector at any point in time and to track any
changes which may occur over duration of the build as well as the
lifetime of the project as described in
Section~\ref{sec:fdsp-coord-qa}.

For detector elements within a cryostat, configuration management
will be frozen once the elements are permanently sealed within the
cryostat.  However, during the integration and installation process of
building the \dwords{detmodule} within the cryostat, changes may need to occur.
For detector elements outside the cryostat and accessible, all
repairs, replacements, hardware upgrades, system grounding changes,
firmware and software changes must be tracked.

Any change will require revision control, configuration
identification, change management and release management.

{\bf Revision Control}\\ Consortia will be responsible for providing
accurate and well documented revision control.  Revision control
will provide a method of tracking and maintaining the history of
changes to a detector element.  Each detector element must be clearly
identified with a document which includes a revision number and
revision history.  For mechanical elements, this will be reflected by
a drawing number with revision information.  For electrical elements,
schematics will be used to track revisions.  Consortia will be
responsible for identifying the revision status of each installed
detector element. Revisions are further controlled through maintenance
of the documents in DocDB and \dword{edms}.

{\bf Configuration Identification}\\
Consortia are responsible for providing unique identifiers or part
numbers for each detector element.  Plans will be developed on how
inventories will be maintained and tracked during the build.  Plans
will clearly identify any dynamic configuration modes which may be
unique to a specific detector element.  For example, a printed circuit
board may have firmware that affects its performance.

{\bf Change Management}\\
\dword{tc} will provide guidelines
for formal change management.  During the beginning phase of the
project, drawings and interface documents are expected to be signed by
the consortium technical leads, project engineers, and technical
coordinator.  Once this initial design phase is complete, the detector
models, drawings, schematics and interface documents will be under
formal change control.  It is anticipated that designs will undergo
further revisions prior to the start of detector construction, but any
changes made after the release of the \dword{tdr} will
need to be agreed to by all drawing signers and an updated signed
drawing produced.

{\bf Release Management}\\
Release management focuses on the delivery of the more dynamic aspects
of the project such as firmware and software.  Consortia with
deliverables that have the potential to affect performance of the
detector by changing firmware or software must provide plans on how
these revisions will be tracked, tested and released.  The
modification of any software or firmware after the initial release,
must be formally controlled, agreed upon and tracked.

\subsection{Engineering Process and Support}
\label{sec:fdsp-coord-integ-engr-proc}

The \dword{tc} organization will work with
the consortia through its \dword{tc} engineering team to ensure the proper
integration of all detector components.  The \dword{tc} engineering team will
document requirements on engineering standards and documentation that
the consortia will need to adhere to in the design process for the
detector components under their responsibility.  Similarly, the
project \dword{qa} and \dword{esh} managers will document \dword{qc} and safety
criteria that the consortia will be required to follow during the
construction, installation, and commissioning of their detector
components, as discussed in sections~\ref{sec:fdsp-coord-qa}
and~\ref{sec:fdsp-coord-esh}.

Consortia interfaces with the conventional facilities, cryostats, and
cryogenics are managed through the \dword{tc}
organization.  The \dword{tc} engineering team will work with the
consortia to understand their interfaces to the facilities and then
communicate these globally to the \dword{lbnf} Project team.  For conventional
facilities the types of interfaces to be considered are requirements
for bringing needed detector components down the shaft and through the
underground tunnels to the appropriate detector cavern, overall requirements for
power and cooling in the detector caverns, and the requirements on
cable connections from the underground area to the surface.
Interfaces to the cryostat include the number and sizes of the
penetrations on top of the cryostat, required mechanical structures
attaching to the cryostat walls for supporting cables and
instrumentation, and requirements on the global positioning of the
\dwords{detmodule} within the cryostat.  Cryogenics system interfaces include
requirements on the location of inlet and output ports, requirements on
the monitoring of the \lar both inside and outside the
cryostat, and grounding and shielding requirements on piping attached to
the cryostat.

\dword{lbnf} will be responsible for the design and construction of the
cryostats used to house the \dwords{detmodule}.  The consortia are required to
provide input on the location and sizes of the needed penetrations at
the top of the cryostats.  The consortia also need to specify any
mechanical structures to be attached to the cryostat walls for
supporting cables or instrumentation.  The \dword{tc} engineering
team will work with the \dword{lbnf} cryostat engineering team to understand
what attached fixturing is possible and iterate with the consortia as
necessary.  The consortia will also work with the \dword{tc} engineering
team through the development of the \threed CAD model to understand the
overall position of the \dwords{detmodule} within the cryostat and any issues
associated with the resulting locations of their detector components.

\dword{lbnf} will be responsible for the cryogenics systems used to purge,
cool, and fill the cryostats.  It will also be responsible for the
system that continually re-circulates \lar through filtering
systems to remove impurities.  Any detector requirements on the flow
of liquid within the cryostat 
will be developed by the consortia and
transmitted to \dword{lbnf} through the \dword{tc} engineering team.  Similarly,
any requirements on the rate of cool-down or maximum temperature
differential across the cryostat during the cool-down process will 
be specified by the consortia and transmitted to the \dword{lbnf} team.

The engineering design process is defined by a set of steps taken to
fulfill the requirements of the design.  By the time of the \dword{tdr}, 
all design requirements must be fully defined and
proposed designs must be shown to meet these requirements.  Based on
prior work, some detector elements may be quite advanced in the
engineering process, while others may be in earlier stages.  Each
design process shall closely follow the engineering steps described
below.

{\bf Development of specifications}\\ Each consortium is responsible
for the technical review and approval of the engineering
specifications.  The documented specifications for all major design
elements will 
include the scope of work, project milestones,
relevant codes and standards to be followed, acceptance criteria and
specify appropriate internal or external design reviews.
Specifications shall be treated as controlled documents and cannot be
altered without approval of the \dword{tc}
team.  The \dword{tc} engineering team will participate in and help facilitate
all major reviews as described in
Section~\ref{sec:fdsp-coord-reviews}.  Special TB reviews
will be scheduled for major detector elements.

{\bf Engineering Risk Analysis}\\ Each consortium is responsible for
identifying and defining the level of risk associated with their
deliverables as described in Section~\ref{sec:fdsp-coord-risk}.
\Dword{tc} will work with the consortia,
through its \dword{tc} engineering team, to document these risks in a risk
database and follow them throughout the project until they are
realized or can be retired.

{\bf Specification Review}\\
The \dword{dune} \dword{tc} organization and project engineers
shall review consortium specifications for overall compliance with the
project requirements.  Consortia must document all internal reviews
and provide that documentation to \dword{tc}.
Additional higher-level reviews may be scheduled by \dword{tc}
as described in Section~\ref{sec:fdsp-coord-reviews}.

{\bf System Design}\\
The system design process includes the production of mechanical
drawings, electrical schematics, calculations which verify compliance
to engineering standards, firmware, printed circuit board layout,
cabling and connector specification, software plans, and any other
aspects that lead to a fully documented functional design.  All
relevant documentation shall be recorded, with appropriate document
number, into the chosen engineering data management system and be
available for the review process.

{\bf Design Review}\\ The design review process is determined by the
complexity and risk associated with the design element.  For a simple
design element, a consortium may do an internal review.  For a more
complex or high risk element a formal review will be scheduled.
\dword{dune} \dword{tc} will facilitate the review,
bringing in outside experts as needed.  In all cases, the result of
any reviews must be well documented and captured in the engineering
data management system.  If recommendations are made, those
recommendations will be tracked in a database and the consortia will
be expected to provide responses. All results of these engineering
reviews will be made available for the subsystem design reviews
described in Section~\ref{sec:fdsp-coord-reviews}.

{\bf Procurement}\\ The procurement process will include the
documented technical specifications for all procured materials and
parts.  All procurement technical documents are reviewed for
compliance to engineering standards and \dword{esh} concerns.
\dword{dune} \dword{tc} will assist the consortia in working with
their procurement staff as needed.

{\bf Production and assembly}\\ During the implementation phase of the project,
the consortia shall provide the Technical Coordinator with updates on
schedule.  A test plan will be fully developed which will allow for
verification that the initial requirements have been met. This is part of
the \dword{qa} plan that will be documented and followed as described
in Section ~\ref{sec:fdsp-coord-qa}.

{\bf Testing and Validation\\} The testing plan documented in the
above step will be followed and results will be well documented in
consultation with the \dword{qam}.  The Technical Coordinator and \dword{tc}
engineering team will be informed as to the results and whether the
design meets the specifications.  If not, a plan will be formulated to
address any shortcomings and presented to the Technical Coordinator.

{\bf Final Documentation\\} Final reports will be generated that
describe the as built equipment, lessons learned, safety reports,
installation procedures, testing results and operations procedures.

\section{Detector Infrastructure}
\label{sec:fdsp-coord-infrastructure}

\dword{tc} is responsible for delivering the common infrastructure for
the \dword{fd}. This infrastructure is typically equipment that is used
by many groups. This may include: the electronics racks on top of the
cryostat with power, cooling, networking and safety systems, cable
trays, cryostat crossing tubes, and flanges, detector safety systems
and ground monitoring and isolation transformers.

\dword{tc} is responsible for the systems to support the
detectors. For the \dword{dpmod} this may consist of a cable
winch system similar to \dword{pddp}.  In the case of the
\dword{spmod} the installation group also provides the
\dword{dss}, which supports the detector and is needed to bring
equipment into the correct location in the cryostat.

\section{The Integration and Test Facility}
\label{sec:fdsp-coord-integ-test}

\dword{dune} \dword{tc} is responsible for interfacing with the
\dword{lbnf} logistics team at \surf to coordinate transport of all
detector components into the underground areas.  Due to the limited
availability of surface areas at \surf for component storage, nearby
facilities will be required to receive, store and ship materials to
the Ross Shaft on an as-needed basis. A team within the \dword{tc}
organization will 
develop and execute the plan for
receiving detector components at a surface facility and transporting
them to the Ross Shaft in coordination with the on-site \dword{lbnf}
logistics team.  The surface facility will require warehouse space
with an associated inventory system, storage facilities, material
transport equipment and access to trucking.  Basic functions of this
facility will be to receive the detector components arriving from
different production sites around the world and prepare them for
transport into the underground areas, incorporating re-packaging and
testing as necessary. As a substantial facility will 
be
required, it can also serve as a location where some detector
components are integrated and undergo further testing prior to
installation.

The logistics associated with integrating and installing the
\dwords{fd} and their associated infrastructure present
a number of challenges. 
These include the size and
complexity of the detector itself, the number of sites around the
world that will be fabricating detector and infrastructure components,
the necessity for protecting components from dust, vibration and shock
during their journey to the deep underground laboratory, and the lack
of space on the surface near the Ross Shaft. To help
mitigate the associated risks, \dword{dune} plans to establish an
\dword{itf} somewhere in the vicinity of \surf. This facility
and its associated staff will need to provide certain functions and
services connected to the \dword{dune} \dword{fd} integration and
installation effort and will have other potential roles that are still under
consideration.  The areas to which the \dword{itf} will and could
contribute are the following:
\begin{itemize}
  \item {\bf Transport buffer:} The \dword{itf} needs to provide
    storage capacity for a minimum one-month buffer of detector
    components required for the detector installation process in the
    vicinity of \surf and be able to accept packaging materials
    returned from the underground laboratory.
  \item {\bf Longer-term storage:} The \dword{itf} needs to provide
    longer-term storage of detector components that need to be
    produced in advance of when they are to be installed and for which
    sufficient storage space does not exist at the production sites.
  \item {\bf Re-packaging} The \dword{itf} needs to have the capability
    to re-package components arriving from the various production
    sites into boxes that can be safely transported through the shaft
    into the underground areas.
  \item {\bf Component fabrication:} It could be convenient to fabricate some
  components 
  in the vicinity
    of \surf at the \dword{itf}, taking advantage of
    undergraduate science and engineering students from the South
    Dakota School of Mines and Technology (SDSMT).
  \item {\bf Component integration:} Integration of detector components
  received from different production sites 
  that must be done prior to transport underground,
    such as the installation of \dwords{pd} and 
    electronics on the \dwords{apa}, could be done
    prior to re-packaging at the \dword{itf}.    
  \item {\bf Inspection, testing and repair:} In cases where
    components are re-packaged for transport to the underground area,
    \dword{itf} support for performing tests on these components to
    ensure that no damage has occurred during shipping is likely
    desirable.  In addition, components integrated at the \dword{itf}
    would require additional testing prior to being re-packaged.  The
    \dword{itf} could provide facilities needed to repair some of the
    damaged components (others would likely need to be returned to
    their production sites).
  \item {\bf Collaborator support:} The host institution of the \dword{itf}
    will need to 
    provide support for a significant number
    of \dword{dune} collaborators involved in the above activities
    including services such as housing assistance, office space,
    computing access, and safety training.
  \item {\bf Outreach:} The host institution of the \dword{itf} would be
    ideally situated for supporting an outreach program to build upon
    the considerable public interest in the experiment that exists
    within South Dakota.
\end{itemize}

 The facilities project (\dword{lbnf}) will provide the cryostats that 
 house the \dwords{detmodule} and the cryogenics
systems that support them.  Additional large surface facilities in the vicinity
of \surf will be required to stage the components of these
infrastructure pieces prior to their installation in the underground
areas.  The requirements for these facilities, in contrast to the
\dword{itf}, are relatively straight-forward and can likely be met by
a commercial warehousing vendor, who would provide suitable storage
space, loading and unloading facilities, and a commercial inventory
management and control system.  It is currently envisioned that
operation of the \dword{lbnf} surface facilities and the \dword{dune}
\dword{itf} will be independent from one another.  However, the
inventory systems used at the different facilities will need to 
ensure proper coordination of all
deliveries being made to the \surf site.

A reasonable criterion for the locations of these surface staging
facilities is to be within roughly an hour's drive of \surf.
Population centers in South Dakota within this radius are Lead,
Deadwood, Sturgis, Spearfish, and Rapid City.  The \dword{lbnf}
surface facilities, which are not expected to have significant
auxiliary functionality, would logically be located as near to the
\surf site as possible.  In the case of the \dword{dune}
\dword{itf}, however, locating the facility in Rapid City to take
advantage of facilities and resources associated with the South Dakota
School of Mines \& Technologies (SDSMT) has a number of potential
advantages.  SDSMT and the local business community in Rapid City
have expressed interest in incorporating the \dword{dune} \dword{itf}
into an overall regional development program.

\subsection{Requirements}

The leadership teams associated with the \dword{dune} 
consortia taking responsibility for the different \dword{fd}
subsystems have provided input to \dword{tc} on which potential
\dword{itf} functions and services would be applicable to their
subsystems.  The consortia have also made preliminary assessments of
the facility infrastructure requirements that would be necessary in
order for these functions and services to be provided for their
subsystems at the \dword{itf}.  An attempt to capture preliminary,
global requirements for the \dword{itf} based on the information
received from the consortia results in the following:
\begin{itemize}
  \item Warehousing space on the order of \SI{2800}{m^2} (\num{30000} square feet); driven by
    potential need to store hundreds of the larger detector components
    needed to construct the TPCs.  The provided space will need to
    maintain some minimal cleanliness requirements (e.g.; no insects)
    and be climate-controlled within reasonable temperature and
    humidity ranges.
  \item Crane or forklift coverage throughout the warehousing
    space to access components as needed for further processing or
    transport to \surf.
  \item Docking area to load trucks with detector components being
    transported to \surf and to receive packaging materials
    returned from the site.
  \item Smaller clean room areas within the warehousing space to
    allow for the re-packaging of detector components for transport
    underground.  Re-packaging of larger components
    will require local crane coverage within the appropriate clean room
    spaces.
  \item Dedicated space for racks and cabinets available for dry-air
    storage and testing of electronics components.  Racks and cabinets
    must be properly connected to the building ground to avoid
    potential damage from electrostatic discharges.
  \item Climate-controlled dark room space for the handling and
    testing \dword{pd} components.
  \item Dedicated clean room on order of \SI{1000}{m^2} (\num{10000} square feet) to facilitate
    integration of electronics and \dwords{pd} on \dwords{apa}.
    The
    integrated \dwords{apa} are tested in cold boxes supported by cryogenics
    infrastructure.  Clean room lighting must be UV-filtered to avoid
    damaging the \dwords{pd}.  The height of the clean room must
    accommodate crane coverage needed for movement of the \dwords{apa} in and
    out of the cold boxes.  It will also be necessary to have
    platforms for installation crews to perform work at heights within
    different locations in the clean room.
  \item Access to machine and electronics shops for making simple
    repairs and fabricating unanticipated tooling.
  \item Access to shared office space for up to \num{30} collaborators
    contributing to the activities taking place at the \dword{itf}.
    Assistance for identifying temporary housing in the vicinity of
    the \dword{itf} for the visiting collaborators.
\end{itemize}

\subsection{Management}

Overall management of the \dword{itf} is envisioned to be
responsibility of one or more of the collaborating institutions on
\dword{dune}.  If the \dword{itf} is located in Rapid City, SDSMT
would be a natural choice due to its physical proximity, connections
to the local community and ability to provide access to resources and
facilities that would benefit planned \dword{itf} activities.  Initial
discussions with SDSMT representatives have taken place, in which
they have expressed a clear interest for hosting the \dword{itf}.
Additional discussions will be needed to understand the details of the
\dword{itf} management structure with in the context of a more
finalized set of requirements for the facility.

\subsection{Inventory System}

Effective inventory management will be essential for all aspects of
\dword{dune} detector development, construction, installation and
operation.  While its relevance and importance go beyond the
\dword{itf}, the \dword{itf} is the location at
which \dword{lbnf}, \dword{dune} project management, consortium
scientific personnel and \surf operations will interface.  We therefore
will develop standards and protocols for inventory management as part
of the \dword{itf} planning.  A critical requirement for the project
is that the inventory management system for procurement, construction
and installation be compatible with future \dword{qa}, calibration and
detector performance database systems.  Experience with past large
detector projects, notably \nova, has demonstrated that the capability
to track component-specific information is extremely valuable
throughout installation, testing, commissioning and routine operation.
Compatibility between separate inventory management and physics
information systems will be maintained for effective operation and
analysis of \dword{dune} data.

\subsection{ITF Infrastructure}

\dword{tc} is responsible for providing the common support
infrastructure at the \dword{itf}. This includes the cranes and
forklifts to move equipment in the \dword{itf}, 
any
cryostats and cryogenics systems to enable cold tests of consortium-provided 
components, 
the cleanrooms and cleanroom equipment
to enable work on or testing of consortia components, and UV-filtered
lighting as needed. This also includes racks and cable trays.

\section{Installation Coordination and Support}
\label{sec:fdsp-coord-install}


Installation Coordination and Support 
(also called simply \textit{Installation}) is
responsible for coordinating the detector installations, providing
detector installation support and providing installation-related
infrastructure. The installation group management responsibility is
shared by a scientific lead and a technical lead that report to the
Technical Coordinator. The 
group responsible for activities in the underground areas is referred to as
the \dword{uit}. The \dword{itf} group, which delivers equipment to the
Ross Shaft, and the \dword{uit}, which receives the equipment
underground, need to be in close communication and work closely
together.

Underground installation is in general responsible for coordinating
and supporting the installation of the \dwords{detmodule} and providing
necessary infrastructure for installation of the experiment. While the
individual consortia are responsible for the installation of their own
detector equipment, it is essential that the detector installation be
planned as a whole and that a single group coordinates the
installation and adapts the plans throughout the installation
process. The \dword{uit} has responsibility for overall coordination
of the installations. In conjunction with each consortium the
\dword{uit} makes the installation plan that describes how the
\dwords{detmodule} are to be installed. The installation plan is used to define
the spaces and infrastructure that will be needed to install them.
The installation plan will also be used to define the
interfaces between the Installation group and the consortium
deliverables.

\subsection{UIT Infrastructure}

The installation scope includes the infrastructure needed to install
the \dword{fd} such as the cleanroom, a small machine shop, special
cranes, scissor lifts, and access equipment.  Additional equipment
required for installation includes: rigging equipment, hand tools,
diagnostic equipment (including oscilloscopes, network analyzers, and
leak detectors), local storage with some critical supplies and some
personal protective equipment (PPE). The \dword{uit} will also provide
trained personnel to operate the installation infrastructure. The
consortia will provide the detector elements and custom tooling and
fixtures as required to install their detector components.

\subsection{Underground Detector Installation}
\label{sec:fdsp-coord-undergd}

For the \dwords{detmodule} to be installed in safe and efficient
manner, the efforts of the individual consortia must be coordinated
such that the installation is planned as a coherent process. The
interfaces between the individual components must be understood
and the spaces required for the installation process planned and
documented. The installation planning must take into account the
plans and scope of the \dword{lbnf} effort and the individual plans of
the nine consortia. By working with the \dword{lbnf} team and the
members of the consortia responsible for building and installing their
components, a joint installation plan and schedule, taking into account
all activities and needs of all stakeholders, can be developed. Although
the organization of the installation effort is still evolving, 
an installation coordinator will be the equivalent of a scientific lead for this effort.

One of the primary early responsibilities of the \dword{uit} is to
develop and maintain the \dword{dune} installation plan and the
installation schedule. This installation plan 
describes the installation process in sufficient detail to demonstrate
how all the individual consortium installation plans mesh and it 
gives an overview of the installation process. The installation plan
is used by the \dword{uit} to define the underground infrastructure
needed for detector installation and the interfaces it has with respect  to 
the consortia. The \dword{uit} is responsible for reviewing and
approving the consortium installation plans. Approved installation
plans, engineering design notes, signed final drawings, and safety
documentation and procedures are all prerequisites for the \dword{prr}. 
Approved procedures, safety approval, and
proper training are all required before the \dword{uit} performs
work. During the installation phase the installation leadership 
coordinates the \dword{dune} installation effort and adapts the schedule
as needed. The installation coordinator, together with management, will also
resolve issues when problems occur.

The installation infrastructure to be provided by the \dword{uit}
includes: the underground ISO 8 (or class \num{100000}) clean room
used for the installation; cranes and hoists (if they are not
delivered by \dword{lbnf}); and scissor lifts, aerial lifts, and the common
work platforms outside the cryostat. The \dword{uit} will have
responsibility for operating this equipment and assisting the
consortia with activities related to rigging, material transport, and
logistics. Each consortium is responsible for the installation of
their own equipment, so the responsibility of the installation group is
limited, but the material handling scope is substantial. To support
the installation process, an installation floor manager will lead a
trained crew with the main responsibility of transporting the
equipment to the necessary location and operating the cranes, hoists,
and other common equipment needed for the installation. It is expected
that the installation crew will work with the teams from the various
consortia but will mainly act in a supporting function. The
\dword{uit} floor manager will be responsible for supervising the
\dword{uit} crew, but the ultimate responsibility for all detector
components remains with the consortia even while the underground
team is rigging or transporting these components.  This will be
critical in the case where any parts are damaged during transport or installation,
as the consortia need to judge the necessary actions. 
\fixme{judge the situation and determine the necessary actions?}
For this reason,
a representative or point of contact (POC) from the consortia must be
present when any work is performed on their equipment. The consortium
is responsible for certifying that each installation step is properly
performed.

The \dword{uit} acts as the primary point of contact with
\dword{lbnf} and \surf from the time the components reach the Ross
headframe until the equipment reaches the experimental cavern. If
something goes wrong, \surf calls the \dword{uit} leader who then
contacts the responsible party. The consortia are responsible for
delivering to the \dword{uit} all approved procedures and specialized
tooling required for transport. The \dword{uit} leader acts as a point
of contact if the \dword{lbnf} or \surf team has questions or difficulties
with the underground transport.  The \dword{uit} receives the
materials from \dword{lbnf} and \surf at the entrance to the \dword{dune}
excavations. The \dword{uit} then delivers the equipment to the
required underground location.

In an effort to get an early estimate of the equipment required to
install the detectors the \dword{uit} has developed a preliminary
installation plan that outlines the installation process. At present
the installation plan consists of a \threed model of the cryostat in the
excavations. The \dword{spmod} elements are inserted in the
model and a proposal for how they are transported, assembled, and
inserted into the cryostat has been conceptually developed and expressed in a series of images
some of which are shown in Figures~\ref{fig:Install-seq} and~\ref{fig:cpa-fc-unpack-assy}. 
Conceptual designs of the infrastructure needed to support
the transport and assembly are also included in the model. See Figures~\ref{fig:Install-ISO-Top} and~\ref{fig:Install-TopView}. With this
as a tool, the proposed installation sequence can be iterated with the
consortia to converge on a baseline installation plan. A similar
process will be followed for the \dword{dpmod} once the base
configuration for the \dword{sp} installation is agreed upon. The
\dword{uit} has focused initially on the \dword{spmod} as the
\dword{sp} components are larger and the installation process more
complex. 

\begin{dunefigure}[APA and CPA installation steps]{fig:Install-seq}
  {Top row from left:  crated \dword{apa} rotating to vertical position;  crated vertical \dword{apa} placed in cart; \dword{apa} panels moved to fixture using the under-bridge crane. Bottom row: series showing \dword{cpa} panels uncrated and moved to fixture. }
\includegraphics[width=.9\textwidth]{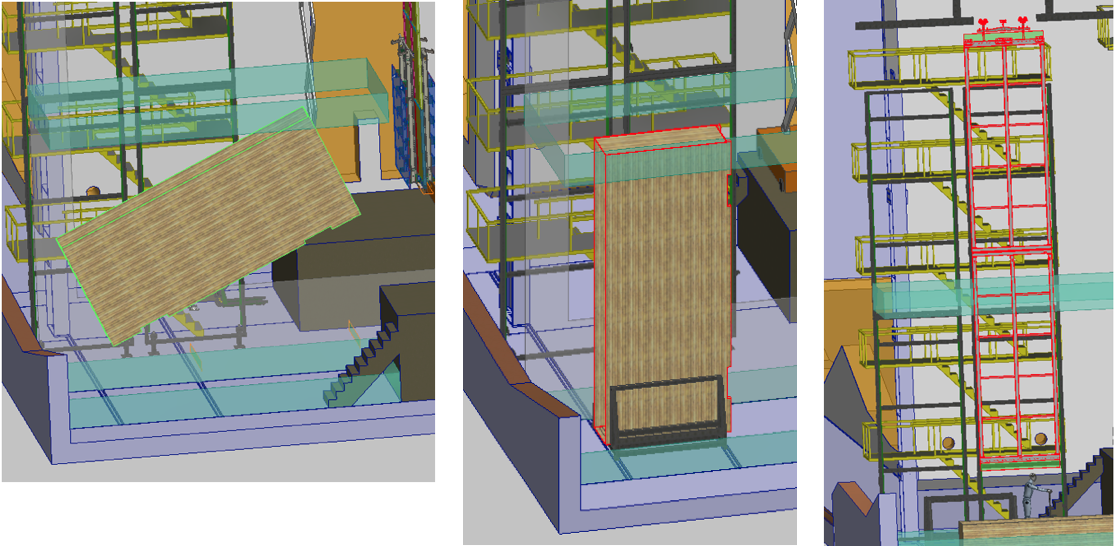}
\includegraphics[width=.9\textwidth]{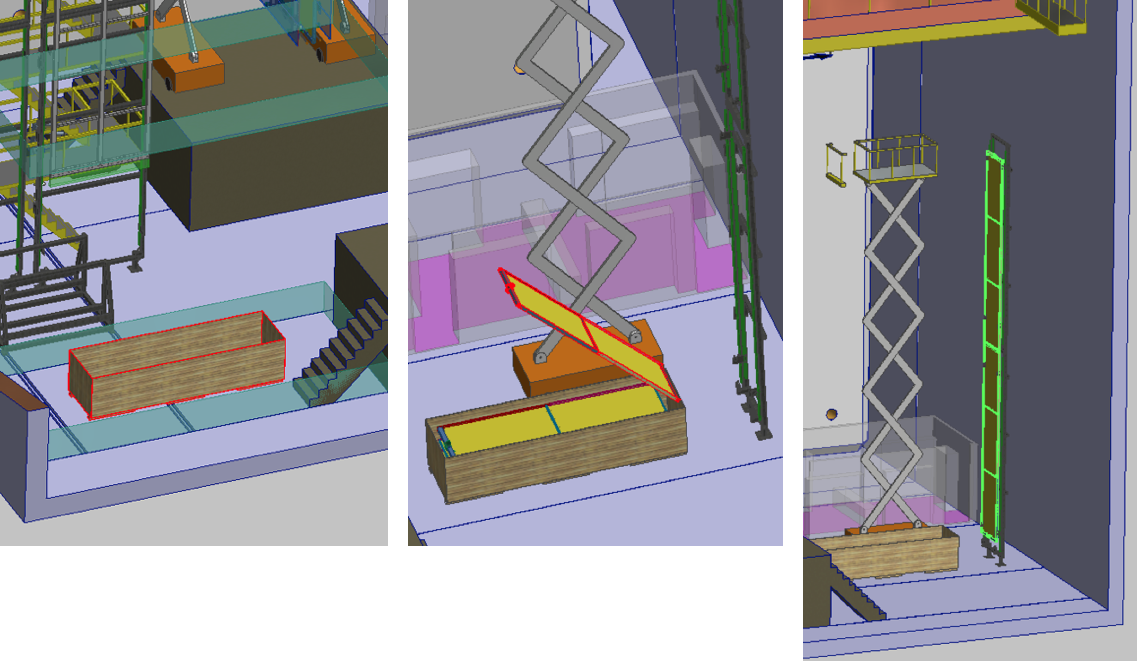}
\end{dunefigure}

\begin{dunefigure}[\dword{cpa} and \dword{fc} unpacking and assembly]{fig:cpa-fc-unpack-assy}
  {On the left, the assembled \dword{cpa} panel is placed onto the north \dword{tco} beam. On the right, the (green) \dword{fc} panels (already lowered into \dword{sas} and moved into the clean room) are installed as the \dword{cpa} array hangs under the \dword{tco} beam. }
\includegraphics[width=.9\textwidth]{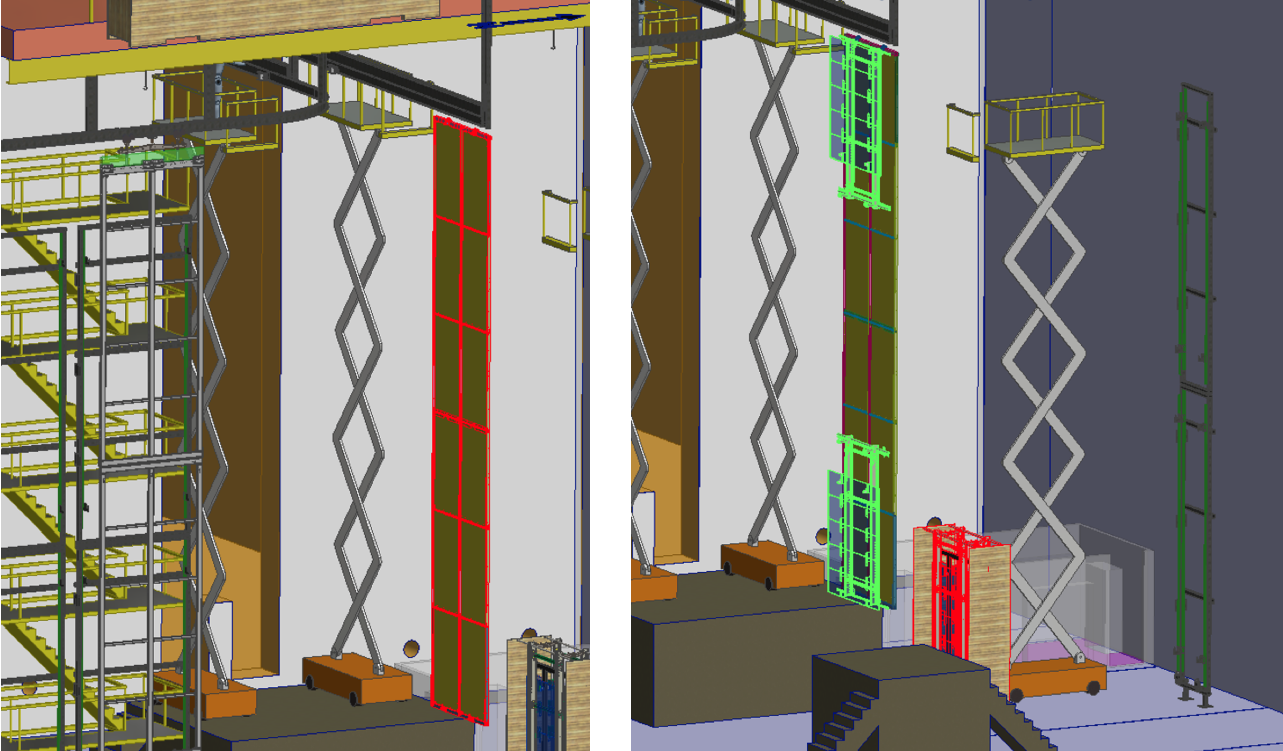}
\end{dunefigure}

\begin{dunefigure}[\threed model of underground area showing installation infrastructure]{fig:Install-ISO-Top}
  {\threed model of the underground area showing the infrastructure to install the \dword{spmod} in cryostat~1. The most significant features are presented including the \dword{apa} and \dword{cpa} assembly areas, the region around the \dword{tco} for materials entering the cryostat,  the changing room, the region for the materials air lock, (\dword{sas}), 
  and the means of egress.}
\includegraphics[width=.9\textwidth]{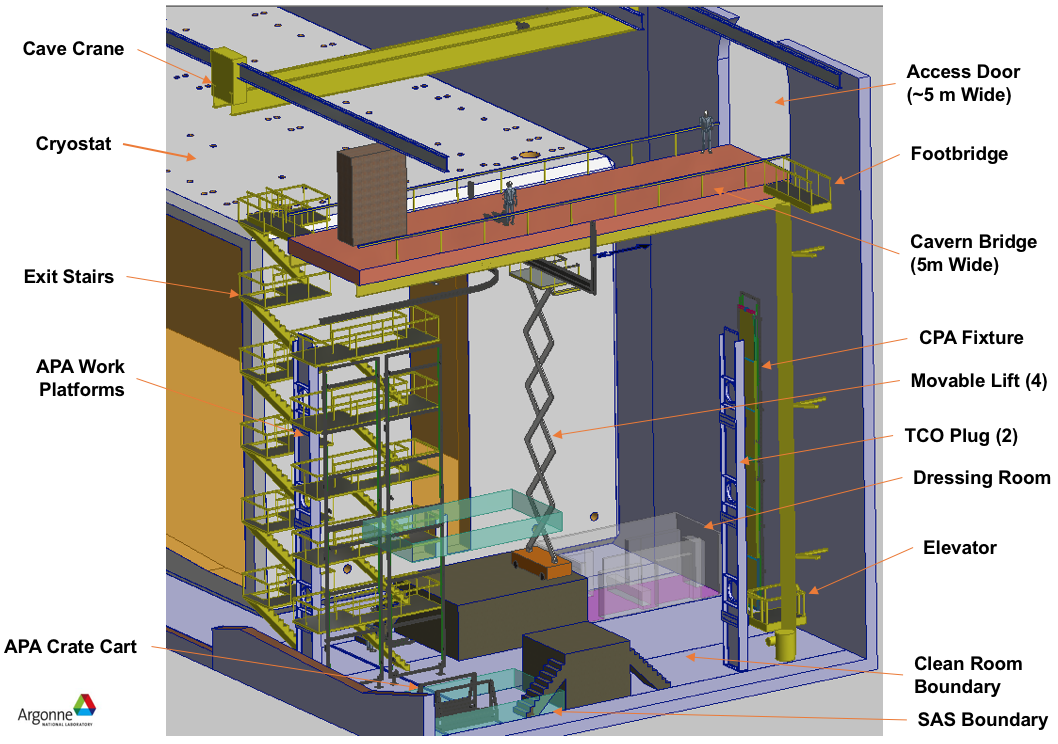}
\end{dunefigure}

\begin{dunefigure}[Section view of the \threed model showing layout]{fig:Install-TopView}
  {Section view of the \threed model showing layout, looking down on the installation area from below the bridge. Areas shown, left to right,  are the cryostat and \dword{tco}, the platform in front of the \dword{tco}, the dressing area, the \dword{apa} and \dword{cpa} assembly area (directly under the bridge), and the stairs and elevator. The lower right corner of the region is used as the materials air lock.}
\includegraphics[width=.9\textwidth]{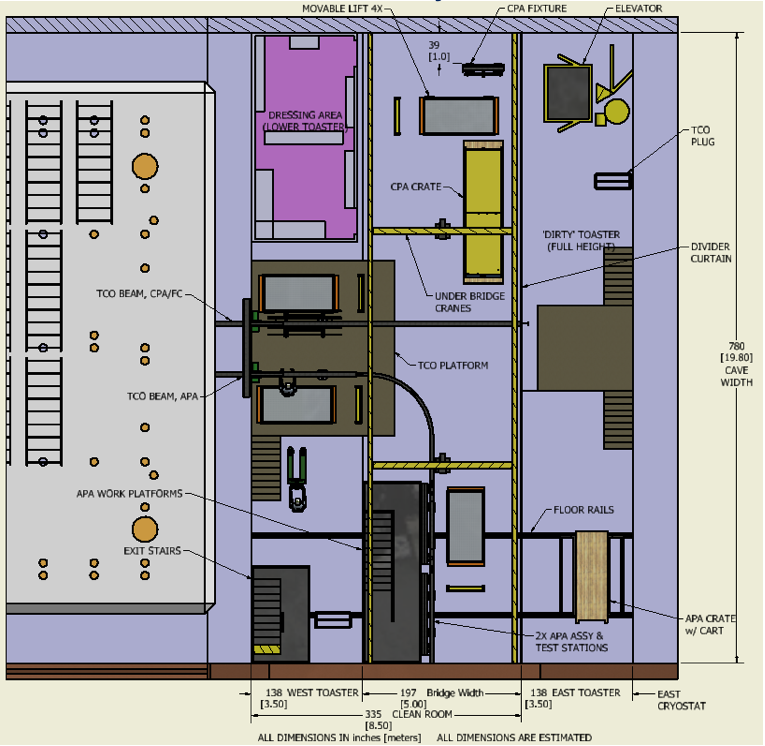}
\end{dunefigure}

In the current installation plan, \dword{dune} will take
ownership of the different underground areas at different times. The
surface data room and the underground room in the \dword{cuc} are available
significantly before the collaboration has access to the cryostats; 
the optical fibers between the surface and underground will be in
place even earlier. This will allow a \dword{daq} prototype to be developed
and tested early. The installation of the \dword{daq} hardware can also be
finished before the start of detector installation if desired, so the
\dword{daq} will not be on the critical path.  When the collaboration receives
access to Cryostat~1 the steel work for Cryostat~2 will be
finished and the work on installing the membrane will have
started. Excavation will be complete.  For planning purposes it is
assumed that the first \dword{detmodule} will be \dword{sp} and the second
\dword{dp}. The first step in the \dword{sp} installation is to
install the cryogenics piping and the \dword{dss}. As this piping will
require welding and grinding, it is a dirty process and must be
complete before the area can be used as a clean room. When this is
complete the cryostat can be cleaned and the false floor
re-installed. The clean infrastructure for installing the \dword{detmodule},
including the clean room, work platforms, scaffolding, the
fixturing to hold the detector elements during assembly, and all the
lifts need to be set up. Once the infrastructure is in place and the
area is clean, the installation of the main elements can start. The
general layout of the installation area showing the necessary space
and equipment is shown in Figure~\ref{fig:Install-seq}. 

The \dword{spmod}  is installed by first installing the west endwall or
endwall~1 (see Figure~\ref{fig:endwall}).

\begin{dunefigure}[End view of \dword{spmod} with \dword{ewfc} in
  place]{fig:endwall}
  {End view of \dword{spmod} with \dword{ewfc} in
  place, with one row of \dwords{apa} and \dwords{cpa}.}
\includegraphics[width=0.6\textwidth]{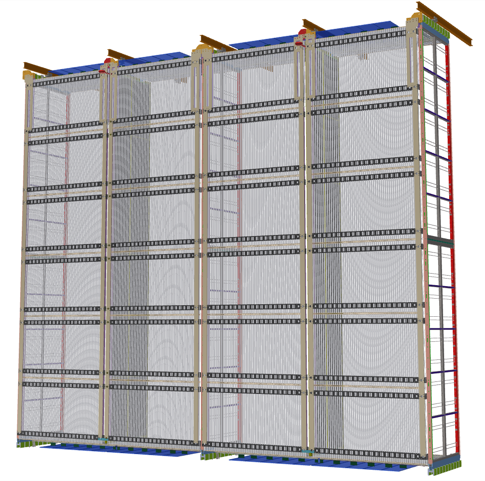}
\end{dunefigure}

The \dwords{apa} and \dwords{cpa} with top and bottom \dword{fc} panels are
installed next. The plan is to install six \dwords{apa} and four
\dwords{cpa} per week, which is enough to complete one of the \num{25}
rows every week. Additional time is built into the schedule to take
into account that the installation will be slower at the beginning and
some re-work may be needed. By building west-to-east, complete rows can
be finished and tested before moving to the next row. This reduces the
risk of finding a fault after final \dword{fc} deployment and cabling,
which would require dismantling part of the \dword{detmodule}. Some of the steps
needed to install the \dword{apa} and \dword{cpa} modules outside the
cryostat are also shown in Figure~\ref{fig:Install-seq}.  The middle three
panels show how the \dword{apa} needs to be handled in order to rotate
it and mount it to the assembly frame. After two \dwords{apa} are
mounted on top of each other, the cabling for the lower \dwords{apa}, and the
\dword{ce} and \dword{pd} cables can be installed. The
lower three panels show how the \SI{2}{m} \dword{cpa} sub-panels are
removed form the transport crates and assembled on a holding frame. Once
the \dword{cpa} module is assembled the \dword{fc} units can be
mounted. Finally, once the \dwords{apa} and \dwords{cpa} are installed,
the endwall~2 can be installed. A high-level summary of the schedule
is shown in Figure~\ref{fig:Install-Schedule}.

\begin{dunefigure}[High-level installation schedule]{fig:Install-Schedule}
  {High-level installation schedule.}
 \includegraphics[width=\textwidth]{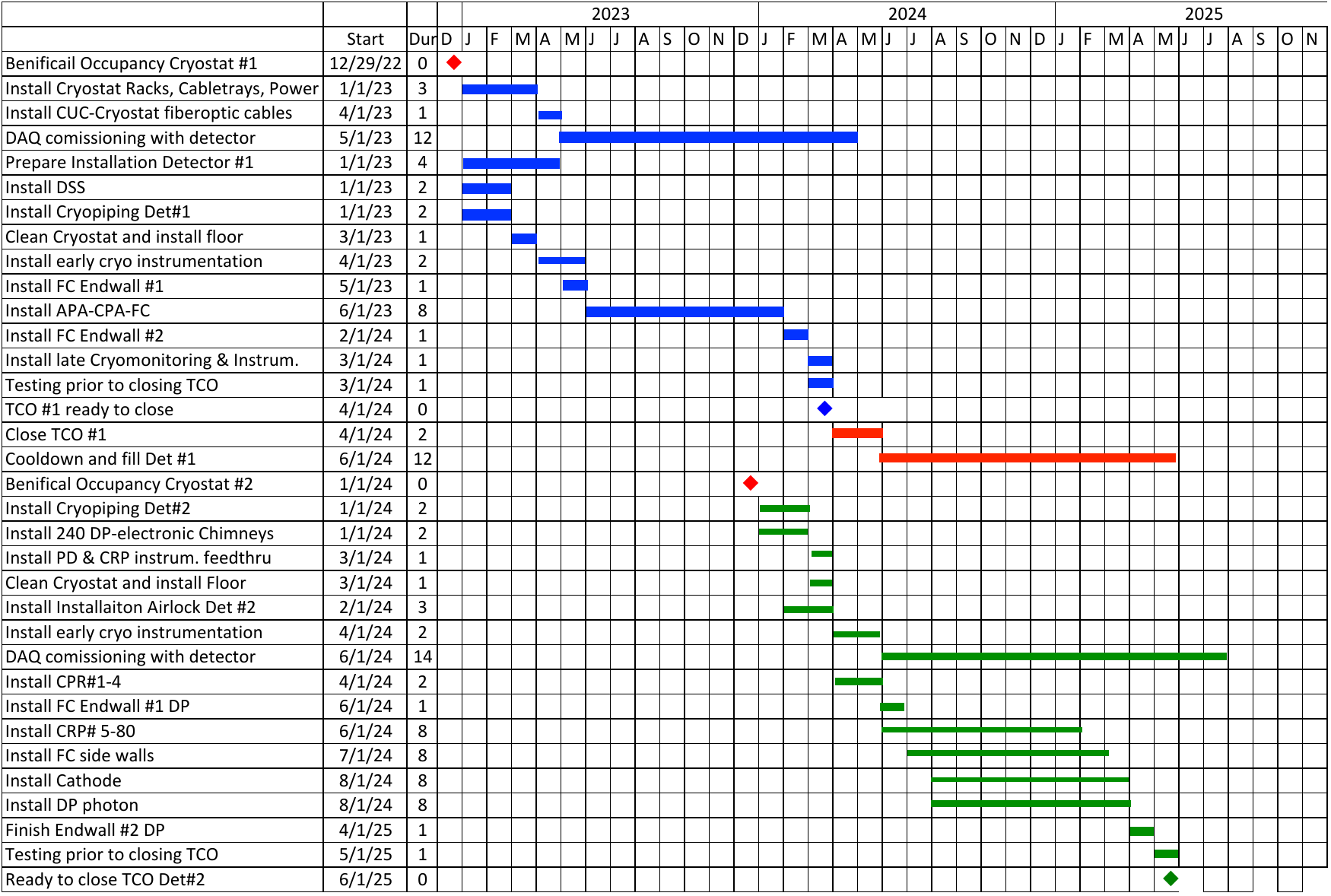}
\end{dunefigure}

As is seen in the installation schedule the second cryostat becomes
available four months before the first \dword{detmodule} installation is
complete. In this period, installation work for both \dwords{detmodule} 
proceeds in parallel. Like the \dword{spmod}, the first step is
the installation of the cryogenics piping, followed by a thorough cleaning
and installation of the false floor. While this piping is being
installed, the \dword{dp} chimneys for the electronics along with the
\dword{pds} and \dword{crp} instrumentation \fdth{}s can also be installed. Since the
chimneys are installed into the roof of the cryostat,  this work is
performed well away from the final installation work on the first
\dword{detmodule} so there should be no conflicts. Once the first \dword{detmodule} is
installed work on setting up the second \dword{detmodule}'s installation
infrastructure can begin. This work includes moving the cranes and
work platforms along with moving the walls of the clean room so that the
second cryostat is clean. The air filtration to the cryostat is also
moved to the second cryostat.  Since much of the work for the \dword{dp}
installation will be performed inside the cryostat, in principle, outside the cryostat a
clean room area smaller than that for the \dword{spmod}  would suffice. 
However, for
planning purposes, it will not be completely clear what type of \dword{detmodule}
will be installed in the second cryostat until fairly late. Therefore the \dword{uit}
will plan to provide a sufficiently large area outside the cryostat to
accommodate either detector technology.  
The much smaller clean room for a \dword{dpmod} 
could be installed just
outside the \dword{tco}. The installation process inside the \dword{dpmod} will
proceed east-to-west. At the start of the TPC installation the first
four \dwords{crp} -- comprising the first row -- will be installed. 
The left panel in Figure~\ref{fig:CRP-Install} shows two \dwords{crp}
being installed near the roof of the cryostat and the right panel
shows one of the \dwords{crp} in a transport box being moved into the cryostat.
Once the first \dword{crp} row is installed and tested, 
the first \dword{ewfc} can be installed.
In general rows of \dwords{crp} will be installed, followed by rows of \dword{fc} modules, followed by the cathode installation at the bottom of the \dword{detmodule}, followed by 
\dwords{pd} under the cathode plane. 
At the end of the installation, the second \dword{ewfc} is
installed and a final testing period for the full \dword{detmodule} is
foreseen. The \dword{dpmod} installation sequence is shown in green in
Figure~\ref{fig:Install-Schedule}.

\begin{dunefigure}[Image of the \dual CRPs being installed in
  the \dword{dpmod}]{fig:CRP-Install}
  {Left: Image of the \dword{dp} \dwords{crp} being installed in
  the \dword{dpmod}, showing the connection from the \dword{crp} to the
  electronics readout chimney. Right: Image of the \dword{crp} being
  inserted into the cryostat using a transport beam.  The \dual \dword{fc} modules will be inserted in a similar fashion.}
\includegraphics[width=0.45\textwidth]{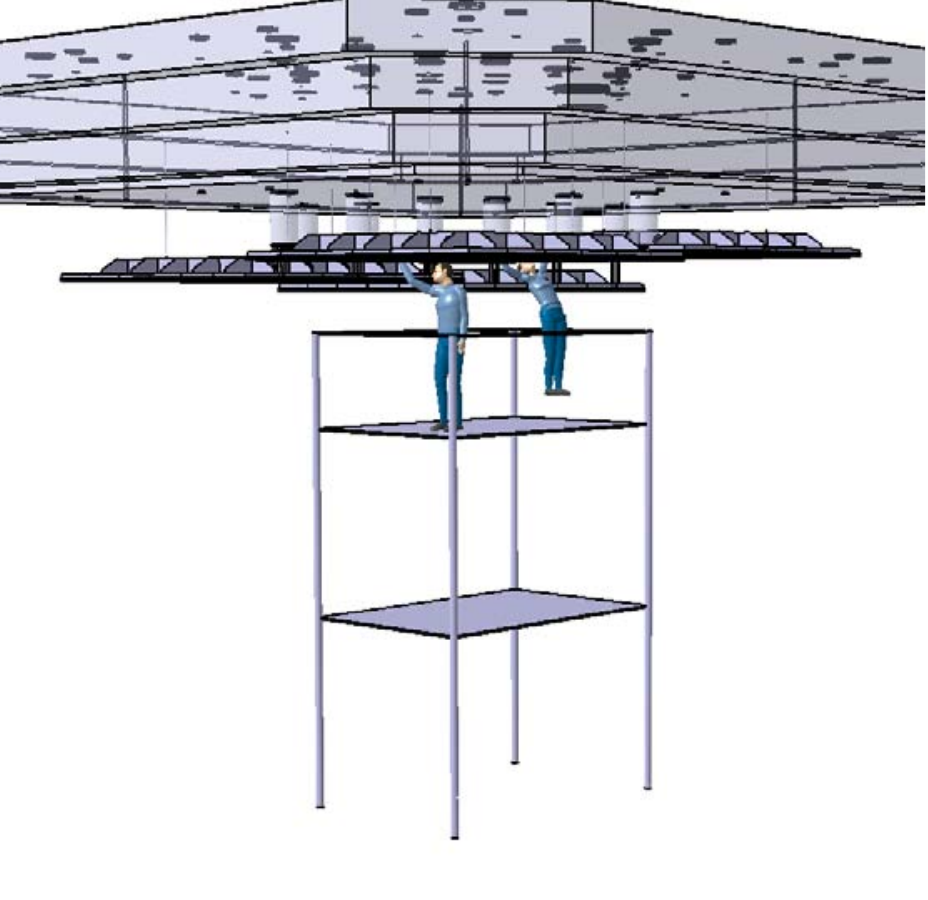}
\includegraphics[width=0.45\textwidth]{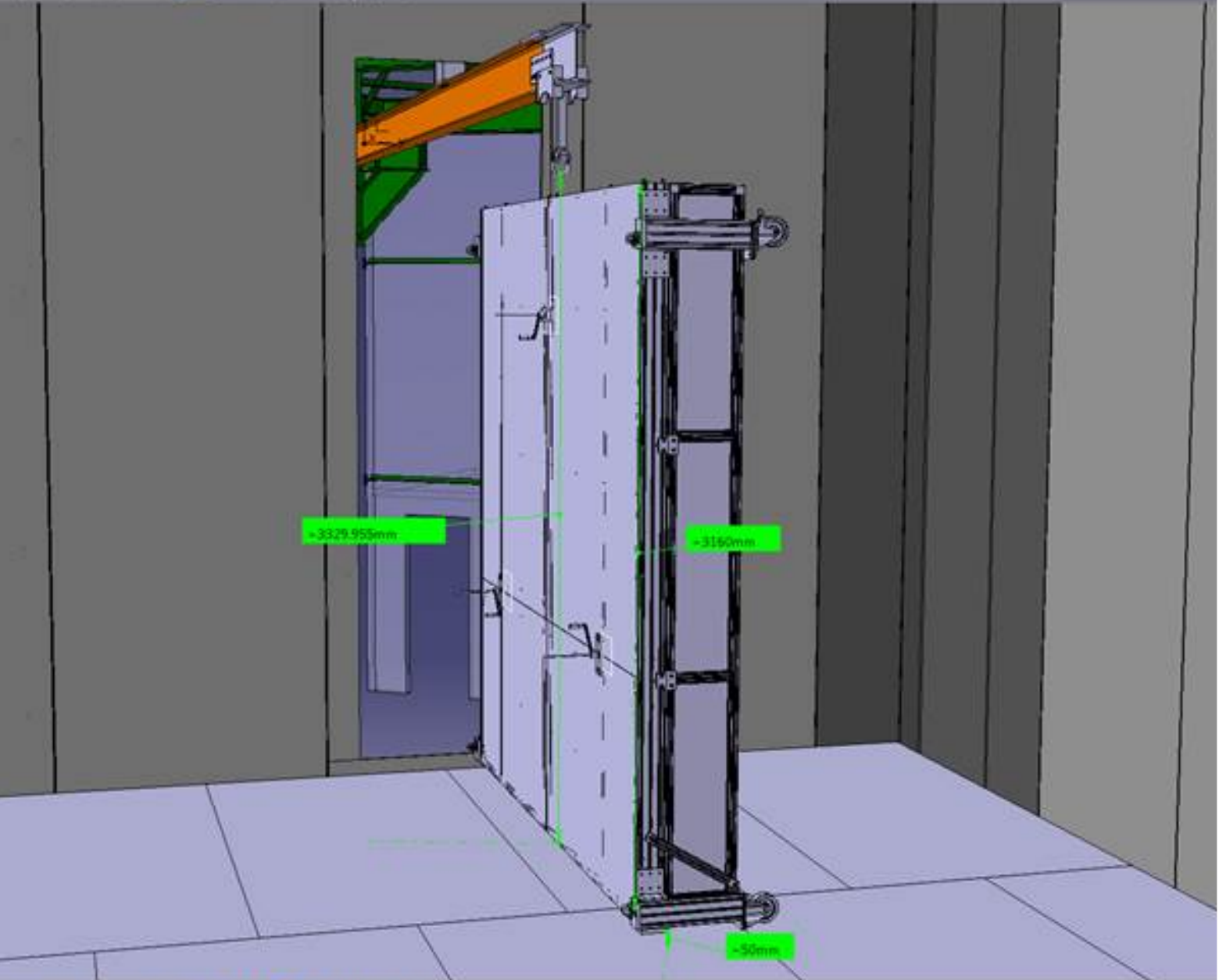}
\end{dunefigure}

Prior to the \dword{tdr}, mutually agreed upon installation plans must
be approved. These will set the schedule for the installation and will
determine the planning for staffing and budget. Having good estimates
for the time needed and having enough experience to ensure that the
interfaces are understood and the procedures are complete is important
for accurate planning. The experience at \dword{protodune} will be
very important as the \dword{protodune} installation establishes the
procedures for handling all the detector elements and in many cases
gives accurate estimates for the time needed. However, in the case of
the \dword{spmod}, many of these procedures need to revised or
newly developed. For example, the \dword{spmod} will be twice as high as
\dword{pdsp}, so two \dwords{apa} need to be assembled together
and a totally different cabling scheme is needed. Testing the
cabling must be done prior to the \dword{tdr} 
in order to
ensure the design is viable. The \dword{dp} will also need to develop
installation procedures as the \dword{dpmod} 
will have a significantly different \dword{fc} and cathode plane. 

By definition, the installation  is on the critical path, making it vital
that the work be performed efficiently and in a manner that has low
risk. In order to achieve this, a prototype of the installation
equipment for the \dword{spmod}  will be constructed at Ash
River (the \nova neutrino experiment \dword{fd} site in Ash River, Minnesota, USA), and the installation process tested with dummy detector
elements. It is expected that the setup will be available at the time
of the \dword{tdr}, but any lessons learned will need to be implemented and
tested after this. In the period just prior to the start of
installation, the Ash River setup will be used as a training ground for
the \dword{uit}.

\subsection{Preparation for Operations}

After the \dwords{detmodule} are installed in the cryostats there remains a lot
of work before they can be operated. First the \dword{tco}
must be closed. This requires bringing back the cryostat manufacturer. 
First the missing panel with the steel beams
and steel panel are installed to complete the cryostat's outer
structural hull. Then the remaining foam blocks and membrane panels
are installed from the inside using the roof access holes 
to enter the cryostat. 

In parallel, the \lar pumps are installed at
the ends of the cryostat and final connections are made to the
recirculation plant. Once the pumps are installed, the cryostat is
closed, and everything is leak tested, the cryogenics plant can be
brought into operation. First the air inside the cryostat is purged by
injecting pure argon gas at the bottom 
at a rate such
that the 
cryostat volume is filled uniformly but faster than the diffusion
rate. This produces a column of argon gas that rises through the volume 
and sweeps out the air. This process is referred to as the \textit{piston
purge}. When the piston purge is complete the cool-down of the \dword{detmodule}
can begin. Misting nozzles inject a liquid-gas mix into the cryostat
that cools the detector components at a controlled rate. 

Once the detector is
cold the filling process can begin. Gaseous argon stored at the surface 
at \surf is brought down the shaft and is re-condensed underground. The \lar then flows through filters to remove any H$_2$O or O$_2$ and
flows into the cryostat. Given the very large volume of the cryostats
and the limited cooling power for re-condensing, it is  
expected to take \num{12} months to fill the first \dword{detmodule} and \num{14} months to
fill the second. During this time the detector readout electronics
will be on monitoring the status of the detector. 
Once the
\dword{detmodule} is full, the drift high voltage can be carefully ramped up and
data taking can begin.

\cleardoublepage



\cleardoublepage
\printglossaries

\cleardoublepage
\cleardoublepage
\renewcommand{\bibname}{References}

\bibliographystyle{utphys} 
\bibliography{common/tdr-citedb}

\end{document}